**British Automobiles, Aging Theory, and the Death of Complex Machines**

Saul Justin Newman*[1,2,3]

1 Centre for Longitudinal Studies, University College London

2 Institute of Population Ageing, University of Oxford

3 University College Oxford, University of Oxford

* Correspondence to saul.j.newman@gmail.com

**Abstract**

Machines provide a longstanding model for how organisms accumulate damage, age, and die. However, the large-scale observation and analysis of complex machine populations under real-world conditions is routinely missing from this framework. Here, we analyze survival and repair patterns in sixty-five million complex machines to reveal fundamental challenges to our theories of biomechanical aging. We measure the reliability, survival, and mechanical failure rates of every privately registered used vehicle in Britain from 2005-2021, using comprehensive samples from 397 million mandatory annual inspections and billions of accompanying repair records. These data reveal that vehicle survival patterns are not a fixed outcome of mechanical reliability or accumulated physical 'wear-and-tear' but display non-aging and anti-aging patterns of survival. These patterns are robust to multiple reanalyses and remain after correcting for diverse external and mechanical predictors of mortality rate using survival forests. These findings challenge the perception of aging as an inevitable and cumulative physical phenomenon, complicate our longstanding comparison of organisms to machines, and highlight exciting new pathways to study the evolution of death in complex systems.

**Main Text**

**Introduction**

Machines have long provided a mathematical and intellectual template for biological systems(1–3) and theories of aging(1, 4). These mechanistic theories of aging often assume that biological aging is caused by the irreversible accumulation of damage – to information systems(5, 6) or physical parts(7–10) – in a way that is directly analogous to a mechanical process of accumulating 'wear-and-tear'. The broad view, that damage accumulation causes aging, remains one of the few dominant ideas in aging science(11).

Central to this assumption is the idea that complex systems progressively accumulate damage over time and, as a result, suffer an accelerating probability of failure. The physical 'wearing out' of parts or molecules is seen as a cause of aging in both organic and inorganic systems: a process evidenced by nonlinear increases in the probability of death, or complete mechanical failure, with age. Modern reliability theory has placed this comparison on a theoretical footing(12, 13) and linked machine survival directly to mathematical models of human ageing patterns(1, 4). In humans and other organisms, age-specific mortality rates approximate the Weibull distribution(6, 14): a distribution with an accelerating rate of failure or mortality that also approximates simulated data for complex machines, and more limited observational data from complex mechanical systems such as cars(14). In turn, this concordance in survival patterns is seen as further supporting the idea that models of physical reliability in machines are a suitable template for understanding the aging of organisms(14).

These biomechanical frameworks have heavily shaped debates on evolutionary trade-offs that involve organisms engaging in hypothetical repairs(7) and preventative

maintenance(5, 10, 15) to prevent ageing. However, this comparison has remained largely theoretical. Notably, work on the biomechanical basis of aging has assumed that evolutionary trade-offs(7), and behaviours such as damage accumulation and preventative maintenance(10), can be informed by mechanical systems without substantial analyses of survival in real-world machine populations. That is, nobody seems to have tested substantial theories against equally substantial observations of machine survival.

In the absence of such data, key questions on individual-level machine longevity seem not to have been tested 'in the wild'. It remains unclear whether mechanical systems always age, what ageing rate diversity exists in observed mechanical systems, what is the empirical importance of investment in repairs or preventative maintenance, and even if observed mechanical systems follow the theoretical distributions assumed by biologists.

In addition to these theoretical considerations, machine longevity and aging rates have a central pragmatic role in economics, personal financial choices, and climate change(16). Vehicles constitute one of the most significant ongoing costs to trade balances and household budgets, with vehicle replacement, repair, and maintenance costs directly impacted by vehicle longevity and ageing. The lifespan and ageing of vehicle fleets also deeply impact projections of air pollution(17) and future carbon emissions(16): the longevity, migration(18, 19), degradation(20, 21), and replacement rate of vehicles directly limit global efforts to decarbonise private transport(16).

To remedy both these theoretical and practical issues, we obtained data on the lifespan and repair rates of sixty-five million vehicles – all the private cars, motorcycles, and vans of Great Britain – observed in 'wild' conditions. Every privately owned vehicle in Great Britain undergoes a legally mandated annual roadworthy inspection called the MOT, producing a rolling annual census of all vehicles that records billions of required repairs, and a suite of vehicle data for every vehicle in the country. Our analysis of these data provides a unique insight into the behaviour, life-course, and ageing of complex systems – which exhibit proto-biological patterns of survival and behaviour – and reveal surprising patterns of longevity and reliability in complex machines. These records also reveal striking issues with our understanding of ageing in complex systems and, by extension, our biomechanical theories of aging.

**Results**

The privately-owned British vehicle population (Table S1) is subjected to a rolling annual census, the 'MOT' roadworthiness inspection, where any failing vehicle components are monitored and repaired by independent mechanics who adhere to strict legal standards of reporting and testing. During these mandatory roadworthiness tests, vehicles are tracked longitudinally using two independent systems of unique identifier (the Vehicle Identification Number or VIN; and number plates) and registered at a known date to postcodes throughout Great Britain. These data were used to generate summary statistics and life table data to capture measuring the reliability and longevity of all 6,281 common makes and models of vehicles driven in the UK over a 17-year period, including the longest-lived and most mechanically reliable vehicles (available in Dataset S1; and at drivedictionary.com).

The life table and survival data of these vehicles – provided in the Supplementary Information and at drivedictionary.com – demonstrates a rich diversity. Population pyramids of cars and motorbikes reveal shortfalls in productive output – corresponding with recessions and world wars – that display persistent signals of disruption in population structures (Fig 1a). Population pyramids also highlighted the long tail of survival times in older and therefore higher-polluting vehicles. Some 2.8 million vehicles (4.2%) in the study predated introduction of the EURO-1 standard, for example, which required catalytic converters. Of these, 1.2 million (1.9%) vehicles were still driving in 2021, and vehicles with no emissions-reduction are standards surviving and emitting pollution for over a century (Fig 1a).

Longitudinal lifetables constructed from these data revealed highly surprising patterns of population structure and mortality rates (Fig 1a-c) that contradict accepted wisdom on the aging patterns of machines. For most vehicles, the age-specific probability of death did not follow the Weibull-approximated 'bathtub' curve(14) – typical of most vertebrate species and indicative of biological ageing(22) – for either chronological (Fig 1b) or physical age (Fig 1c). Neither did the Weibull model provide an adequate or accurate model of aging mechanical systems. Indeed, the survival curve of complex mechanical systems did not resemble biological survival curves at all (Fig 1b-c; Fig S1a-b).

Vehicles initially did undergo a general log-linear acceleration of mortality rates with age (Fig 1b), but almost all observed makes and models then exhibited a deceleration in mortality profiles from around age 15 onward, before a dramatic long-term reversal in mortality patterns. Beginning around age 18, the probability of death declined with age: a process of progressive 'anti-aging' in which the probability of death became lower, at a log-linear rate, as the vehicle became older (Fig 1b; Dataset S1). This inversion of typical aging patterns was observed regardless of whether vehicle ages were measured by years (Fig 1b) or accumulated mileage (Fig 1c). That is, old vehicles (above 120,000 miles or over 18 years) often become increasingly long-lived and likely to survive, appearing to 'age in reverse' with increasing time or mileage. These patterns were not caused by under-reporting or by recent distortions to the roadworthiness testing rules during the COVID pandemic (Fig S2). Censoring data to allow for as much as a decade in reporting lag, back to 2011, still reproduced anomalous 'anti-aging' patterns of age-specific mortality (Fig S2; Software S1).

Reflecting these unexpected survival patterns, age-specific vehicle mortality exhibited a poor fit to a theoretical Weibull distribution under quantile-quantile regression within vehicle cohorts (Fig 2). When fitting the Weibull distribution to the most common vehicle types (N$\geq$10,000), most observed mortality patterns failed to exceed the minimum threshold required to support a Weibull distribution as the underlying distribution of mortality (N = 2342 makes and models; adjusted R$^2$ < 0.9 in 63.4% of populations). Instead, as sample sizes (Fig 2a) and observation times (Fig 2b) increased, the fit between the best-fit Weibull distribution and the observed data became progressively worse.

The unexpected diversity and shape of vehicle survival patterns violated fundamental assumptions of traditional survival models. As a result of this discrepancy, Cox proportional hazards models aimed at predicting vehicle survival all failed the Cox proportional hazards z-test, for both the omnibus test and for every individual parameter (p < 2x10$^{-16}$; Supplementary Information). Even when tested under diverse sample- and cohort-based stratification model structures, aimed at rescuing these models, every coefficient and omnibus test failed the test for proportionality in the make-model-year stratified Cox models of vehicle survival (p < 2x10$^{-16}$; Supplementary Information). Fitting Cox proportional hazards models to individual make-model cohorts – across all 392 common (N $\geq$ 10,000) class-4 vehicles, the top twenty 3-3.5T vans, and the top twenty 200cc+ motorbikes – did not address this persistent issue. Only nine of these 432 models (2.1%; Software S1) satisfied the z-proportional hazards omnibus test[23].

A machine learning approach was implemented to address the challenge of non-proportional hazards and the broad violation of assumptions in the Cox survival model caused by vehicle mortality patterns. Survival forests are a promising method for predicting survival patterns under non-proportional hazards[24], and can even outperform the Cox model[24, 25] when hazards are proportional across strata. Survival

forests(26) were therefore used to predict mortality in a sample drawn from 8.8 million common makes and models of vehicle that underwent a vehicle inspection during 2011: a census year in the middle of our survey period that generated contemporary social data available at the postcode level (Table S2; Supplementary Information). However, after using survival forests to predict mortality rates, and after correcting for past breakdown rates and social and geographic diversity, the mortality rate remained steadfastly unlike biological patterns of aging (Fig S3).

Further surprising patterns arose when examining mechanical failures, captured through billions of encoded vehicle repair records during the MOT roadworthy inspections. The rate of non-fatal mechanical failures is an indicator of vehicle frailty and wear-and-tear but, rather than mirroring the log-linear increases seen in vehicle mortality rates (Fig 1b,c; Fig S1) and aging organisms, mechanical failure rates increased linearly with age before flattening out at age fifteen (Fig 3a). Beyond age fifteen, the rate of mechanical failure progressively fell with age (Fig 3a). This general pattern was also reflected to a lesser degree in mechanical failures per driven mile, with increasing mechanical wear from longer driving distances associated with slight falls in mortality risk (Fig 3b). In addition, these late-life improvements in mortality risk did not seem to be a result of any baseline heterogeneity in vehicle reliability or mechanical failures(27). Correction for past heterogeneity in breakdown and repair rates, using survival forest models detailed above, did not explain, or eliminate anomalous patterns such as 'inverted ageing' or late-life survival improvements (Supplementary Information; Fig S4a,g; Fig S3). When oversampling for older vehicles and predicting mortality profiles using survival forests, to better approximate late-life patterns, estimated reductions in old-age mortality risk became even more pronounced (Fig 4g; Software S1). Neither was this pattern explained by the broad diversity within vehicle makes and models: age-specific rates of mechanical failure (Fig 3c) often had very weak concordance to age-specific mortality rates (Fig 3d).

The nonlinear and often counter-intuitive connection between mortality and physical wear-and-tear was also visible at the fleet level when comparing vehicles of a fixed age, say five (Fig S4a-b) or ten (Fig S4c-d) years old. Despite hardening thresholds on emissions(28, 29) and increases in the stringency of roadworthy test criteria, the rate of major roadworthy testing failures gradually fell over time (Fig S4a, c) indicating ongoing improvements in the mechanical reliability of vehicles. Vehicle mortality rates observed in the same cohorts, however, displayed marked nonlinear trends that were largely uncoupled from the gradual improvement in mechanical reliability rates (Fig S4b, d).

Social factors in vehicle use also appeared to play a role in variation between undifferentiated mechanical systems. In a process known as 'badge engineering', mechanically identical vehicles – from the same factory floor and with the same engine – are routinely sold as cosmetically different models with minor changes to trim levels and body shape. Such vehicles are mechanically indistinguishable at baseline, differing only in customer behaviour and the economic cost of the car. For example, Volkswagen group manufactured the Audi A3, VW Golf, and Skoda Octavia on the same factory floor using the same engines. These models carry the same World Manufacturer's Index and engine code, traceable through the Vehicle Identification Number (VIN). These physically undifferentiated vehicles had widely divergent lifespans (measured in miles or years; Fig S5): a pattern repeated in other re-badged or badge-engineered vehicles.

A case-by-case examination of vehicle diversity revealed many other unexpected patterns of aging, and inverted or anti-aging behaviours, in machine systems. For example, vehicles with the lowest mechanical failure rates were dominated by fuel-efficient vehicles widely used as taxis: the Ford Galaxy, Mercedes-Benz E class, and especially Toyota Prius models all achieved markedly low rates of failure per mile (at 100,000 miles) and per year across multiple models (Supplementary Information). This may represent either the rational choice of taxi companies to select the most mechanically reliable vehicles or, non-exclusively, the strikingly different driving patterns and maintenance-repair schedules taxis experience over their life-course (Fig 3c-d). Motorcycles were generally less mechanically reliable and shorter-lived than cars or small vans, especially per mile (Fig S1a), while purpose-built taxis (which have a legally constrained use as taxis and a legally mandated upper lifespan limit) displayed markedly different age-specific mortality patterns than privately owned cars or vans.

A closer examination of survival forest models revealed a detailed understanding of the statistical relationships between mortality rates, social correlates, and physical or chronological age (Fig 4; Software S1). This approach prompted a wide array of questions. For example, after modelling a wide diversity of social and physical predictors of vehicle mortality in a million-vehicle sample (Table S2; Software S1) the marginal effect of chronological age remained highly nonlinear and did not approximate increasing mortality risk or the Weibull function (Fig 4a). Rather than an increasing number of repairs causing an accelerating failure rate, as widely predicted under reliability models of aging(12), a greater number of repaired vehicle failures at baseline did not predict higher mortality rates once the vehicle had undergone five major repairs (Fig 4b).

Social variables were also predictive of vehicle mortality independent of geographic or physical vehicle states: for example, vote share for major parties in the 2010 elections predicting marginal vehicle survival (Fig 4c), even after correction for numerous spatial, geographic, and physical wear-and-tear indicators (Table S2; Dataset S1; Supplementary Information). The marginal effect of categorical variables like engine propulsion were examined, independently of other factors (Fig 4d-e). In contrast with marketing claims of higher mechanical reliability, electric cars achieved unremarkable reliability and age-specific mortality rates (Supplementary Information; Fig 3c-d) and suffered a small penalty under survival forest models (Fig 4e). While the two common vehicles in our database were an insufficient sample for more general conclusions, the comparable reliability and lifespan of electric vehicles should perhaps be viewed in historical terms: electric cars are an emergent, complex, and rapidly improving technology, yet by 2012 they had already achieved near-parity with legacy diesel and petrol systems.

These marginal effects revealed social variables that were informative of vehicle mortality, independent of measured geographic and mechanical patterns (Fig 3a-b; Fig 4). However, when evaluated by variable importance metrics(30) geographic diversity and physical predictors – often confounded by, driven by, or linked to social factors – remained the strongest indicators of vehicle mortality (Fig 4f).

Across a range of accurate survival forest models (Fig S6a) the odometer reading at baseline, engine capacity, location, and chronological age of the vehicle were dominant predictors of vehicle mortality (Fig 4f; Fig S6b-d), even though these factors strongly diverged from theoretical expectations that mortality should accelerate with increasing age or mileage. This stark gap between theory and observation remained even when constructing predictive models oversampling for high-mileage (Fig 4g; Fig S6a,c) or older (Fig S6a-b) vehicles, or when including other vehicle types like motorcycles (Fig S6d), and examining the resultant marginal patterns of mortality risk.

In these targeted over-sampled models, increasing wear and tear was even more strongly associated with lower adjusted mortality risk after correcting for hundreds of predictors of vehicle mortality, aligning with in the observational data on age-specific mortality (Fig 1b-c; Fig S1; Fig S3) and posing a deep challenge to our understanding of complex aging systems.

**Discussion**

These results yield pragmatic pieces of evidence on vehicle survival and reliability patterns. Some findings affect simple personal choices that could have been informed by independent government testing. Previously, the public have been left to guess at vehicle reliability, despite entire populations of vehicles being exhaustively tested for mechanical reliability on an annual basis by their own governments. This situation is somewhat remarkable given the time, resources, legislation, and care devoted to ensuring the reliability and roadworthiness of vehicles. Independent reliability data appear to be generated globally and available nowhere. Our data provided a partial remedy to this shortfall (Supplementary Information).

Vehicle survival data often contrasted markedly with marketing claims by major companies: such as the idea that SUVs or expensive vehicles, for example, are particularly reliable (Dataset S1). Yet these data also touch on substantially important problems in climate science and public health.

Even without tracing the fate of higher-polluting exported vehicles(19), British vehicle data highlight an unexpectedly long tail of vehicle survival times. Such long survival times generate pollution and CO2 for decades longer than the fleet averages often used for climate modelling(31), with direct repercussions for future estimates of air pollution(32) and carbon emissions(33). If vehicle longevity is not constrained by physical degradation but is more reflective of personal preferences and economic forces(34), we risk a situation where the changeover to low-carbon transport is not limited by the rate of aging, or the capacity to buy clean transport, but by our more limited capacity to change the cultural preferences of global vehicle owners.

Added to these immediate cultural and scientific questions are more philosophical and theoretical concerns. Our intellectual model of 'organisms as machines' has enjoyed a lengthy history in aging theory(2, 3) whilst supported by a surprisingly limited amount of observational data. The similarity between vehicle survival curves and human survival curves, for example, has long been used as evidence that human or even general biological aging patterns resemble the decay of complex mechanical systems. Yet this connection was based on the truncated survival curves of less than ten vehicle types(14): the only large analysis of machine aging that, somehow, ignored the reality of vehicles surviving for over a century and stopped observing vehicles around age sixteen(14).

Our analysis enjoys a considerable advantage over previous comparisons of machines and organisms(2, 3) – not in brains or logic, but in data and computation. Unlike philosophical or theoretical comparisons of machines and organisms(2, 14), we observed tens of millions of complex non-biological systems in the wild across a broad range of ages. These data suggest our model of 'organisms as a machine' has been long oversimplified, not because organisms are more complex than machines, but because machines display more complex behaviours than our theoretical models ever suggested.

The aging of vehicles appears far more diverse than a simple damage-accumulation model of aging. A linear accumulation of 'wear and tear' does not result in an accelerating rate of aging or mortality, and a higher rate of physical damage is routinely associated with a falling risk of failure and death. Even identical mechanical systems respond differentially to human behaviours and environments, in a way that substantially modifies lifespan and slows or reverses physical degradation and aging. These findings reinforce a growing suspicion amongst reliability engineers that the Weibull function does not hold in many complex systems(35–37), and that a substantial update to reliability theory is required.

A possible cause of this gap between theory and observation is that a vital theoretical distinction appears overlooked, or more often ignored, when comparing ageing machines to aging organisms. Aging resulting from accumulated damage, or wear and tear, arises as a theoretical property of redundant, non-repairable systems(35, 36). Vehicles and organisms are, however, repairable on almost all levels of organisation, and in repairable systems the shape of the survival curve can take arbitrary functional forms(35, 36).

Generalized renewal processes(38), for example, can have unexpected or counterintuitive effects on system mortality rates. Generalised repair processes initially did not allow replacement of parts with a 'better-than-new' repair(38) – as the accuracy of repair (the restoration factor $q$) was theoretically bounded between a minimal repair and a 'good-as-new' repair. However, in real systems a 'better-than-new' repair is possible under a wide range of conditions(37) – a possibility increasingly modelled in engineering theory(37) – with the result that mortality rates can decline with age if this condition is met.

Another related possibility is that, under generalised renewal processes(38), the quality of the repair $q$ is time-varying in response to outside factors: such as the force of selection from direct or inclusive fitness in biology(39, 40),  the future expectation of economic prices in vehicle systems(34, 41), or the accuracy and quality of replacement-part synthesis in both. In addition, preventative maintenance schedules – where old parts are replaced before they fail – are often assumed to be constant(36) but may vary widely and undergo time-varying rates of change. Increasing the intensity or accuracy of preventative maintenance per year or mile may, for example, lead to steadily falling mortality rates with age.

That is, it is not clear why it should be assumed that either repairable mechanical or biological systems must exhibit aging because of 'wear-and-tear'. The rate and accuracy of parts replacement and repair can vary substantially in response to extrinsic factors, such as the economic or emotional pressure to maintain a mechanical system, resource availability, or the evolutionary pressure to repair somatic cells or subcellular units(39, 42) – which declines with age under direct fitness but not inclusive fitness(39, 40) models. This flexibility of repair and maintenance processes may allow for anti-ageing patterns to arise, in either biological or mechanical systems, if pressure to improve the rate and accuracy of repair increases over time.

Indeed, economists have long suggested a very different picture of survival at the population level in machines. The average longevity of national vehicle fleets is seen as primarily determined by economic factors(34, 43) independent of fleet composition or quality, with around 80% of the variance in vehicle fleet longevity explained by the relative cost of parts replacement(34, 41). Our results suggest that socioeconomic factors may also modulate lifespan and ageing rates at the granular level of individual vehicles. The implication is that the capacity for repair is not constrained by systemic damage accumulation in complex organisms – such damage does not even cause irreversible aging in machines – but by the strength of selection for repair(39, 42).

All of this leaves a longstanding assumption – that biological damage accumulates irreversibly with age to cause accelerating mortality rates – on uncertain ground. The perceived resemblance of biological and mechanical survival curves(14) appears to be the result of incomplete observation, not the deep theoretical connection proposed by aging theorists(4, 8, 44). Machines behave in more complex ways than our models predict, and our picture of organisms as decaying machines may require a fundamental reconfiguration. Such reconfiguration would change the rules of mechanical aging, and help rewrite how we interpret organisms, and ourselves, as biological machines.

**Materials and Methods**

*Data and quality controls*

The UK Department for Transportation (DFT) and the Driver and Vehicle Standards Association (DVSA) provided quality-controlled data for all sixty-five million privately registered vehicles in Great Britain that underwent an 'MOT' vehicle inspection between 1 January 2005 to 31 December 2021 inclusive. The MOT roadworthy inspections are carried out from one year of age onwards in motorcycles and mopeds and three years of age onwards in other vehicles. Repair data from 397 million MOT inspections was linked with these vehicles, with each inspection having a unique date, repairer (mechanic) code, location identifier, and comprehensive testing results encoded by over 7,000 unique repair codes that detail the type of repair and location on the vehicle. For example, one unique repair code exists for replacing the headlight bulb, another for cambelt replacement, another for repairing a puncture in rear left tyre (as opposed to a front right tyre).

Two independent identifiers accompanying each vehicle: the vehicle identification number (VIN) attached to the vehicle engine and chassis, and the number plate of the vehicle, which was traced longitudinally by the DFT to account for legacy transfers, changes, and reissues. The VIN is an internationally regulated number, governed by a legal standard(45, 46), unique to every vehicle. In all global vehicles manufactured since 1981, the first eleven digits of the VIN also encode the year and country of manufacture, the factory floor on which the vehicle was made, the engine type (for each capacity, make, and fuel type), the trim level, and vehicle manufacturer via the World Manufacturer's Index(47).

Private vehicles were legally classified into several categories during MOT testing, including the major categories for motorcycles or mopeds below (Class 1) or above (Class 2) a 200cc engine size, three wheeled vehicles up to 450kg gross weight (Class 3; N = 32,555), cars and vans below 3000kg gross weight (Class 4), ambulances and vehicles with 14 or more seats (Class 5; N = 95,477), and vans of 3000-3500kg (Class 7; Table S1).

Vehicle records were subject to ongoing quality controls by the DFT, including cross-referencing of two independent vehicle identification systems – the number plate and the VIN – both recorded at each successive roadworthy test. The DVSA also provided two independent chronological ages of each vehicle, the date of first registration and the date of first use. Years of model manufacture, rather than the date of first use, were also encoded in the vehicle identification number(46) and the model year manually reported during each MOT inspection, allowing cross-referencing and cleaning of mismatches in reported age.

We implemented further strict quality controls, given the large population. Vehicles failed quality control if their age was ever reported as negative, if their odometer fell by any amount between successive MOTs, if the dates of registration or first MOT preceded the date of first use by more than 90 days, if the odometer reading exceeded a million miles or kilometres (depending on the reporting units), if the make of vehicle was not recorded, if any anomalous or incorrect dates were reported during an MOT (*e.g.* 2100 instead of 2010), and (most often) if the dates of registration or first use were missing or unknown. As errors nonlinearly accumulate over time in survival processes(48), vehicles over 110 years of age were excluded from analysis as they were likely largely age-coding errors. Collectively these stringent quality controls resulted in the loss of 2.7% of records (N = 1,733,139) by excluding vehicles with typographic errors, vehicles with incorrect odometer readings, imported used vehicles that did not have a known age at first registration, or vehicles that were re-registered after receiving a statutory off-road notification (SORN) that resulted in uncertain or unknown ages.

To allow us to trace the end-of-life, or loss-to-follow-up for exported vehicles, of each vehicle in Great Britain the DFT and DVSA provided certification data for every exported, scrapped, or destroyed vehicle that was privately registered in Great Britain (Table S1). Vitally, the odometer reading and unit (miles or kilometres) were reported alongside the inspection date at every annual roadworthy test: allowing both an observation of accumulated mileage over time, and the calculation of per-mile mechanical failures or vehicle deaths from these data (Software S1).

*Mortality and mechanical failure rates*

Mile- and age-specific life tables were calculated for all common makes and models of vehicle in Great Britain (Dataset S1; Software S1). After quality controls, mortality was classified as either by date issued on a certificate of destruction and/or scrappage certificate (N = 9,302,767), or by the failure of a vehicle to undertake the mandatory annual roadworthy test for eighteen months (N = 23,808,349). Inferred mortality in the latter case was back-dated to the last known date at which the vehicle underwent a roadworthy test inspection, causing binning of mortality rates to approximately year-long intervals after the first MOT inspection. This has a very limited effect on mortality rates calculated for annual life tables but was nonetheless a meaningful study limitation that precluded finer time-scale estimates of mortality rates. As survival curves are generally left-skewed distributions, vehicles de-registered using a Statutory Off-Road Notification (SORN) after their MOT inspection may have roughly 3-6 months longer average lifespan than our life table estimates, a limitation that could be addressed in future by linking SORN certificate data to the MOT roadworthy test data.

Accurate age-specific mortality data and life tables generally require cohorts of over 1000 individuals(49). After quality control, fifty-nine million vehicles (90.9%), measured in 355 million roadworthy tests, were from a sufficiently common make-model-year combination to satisfy this threshold. Age- and mile-specific hazard rates and life tables were calculated using the 'fmsb' software package(50) for all 6,281 different make-model-year combinations with 1,000 or more vehicles after quality controls and excluding all exported vehicles as lost-to-follow-up (N = 6,922,292 exported vehicles; Dataset S1). In addition, identically QC-filtered life tables were calculated for all 9,622 unique 11-letter VIN combinations with over 1,000 vehicles. Each 11-letter VIN is used by manufacturers to indicate each underpinning chassis and engine made in the same factory and thus includes 'badge-engineered' vehicles like the 2011 Aston Martin Cygnet – which is a 2011 Toyota IQ with a new badge – under a shared code and therefore larger sample size. All raw, non-identifiable risk-exposure, mortality rate data, and life tables are available in the Dataset S1.

To handle the complexity of repair codes and vehicle types, we leveraged a legal requirement for mechanics to classify the degree of repairs required for a vehicle to meet legal roadworthiness standards into three tiers: major repairs, minor 'prs' repairs conducted without a follow-up inspection, and no repairs. As all repairs are mandatory for continued registration, we used these closely regulated legal standards to calculate both incident rates of major mechanical failures discovered at each inspection, and the rates of past repairs for all used vehicles in Great Britain (Dataset S1). Data on rates of major and minor failures were aggregated by age in years and by 10,000-mile increment (rounded down) for all used vehicles. Tables showing major mechanical failure rates per inspection, for every common make-model-year (N=6,281) and for every mileage increment observed at 100 or more vehicle inspections are provided in Dataset S1.

*Theoretical Fit with Weibull Distribution*

The fit between observed data and the Weibull distribution was approximated using a quantile-quantile regression. This allowed for assessment of the nonparametric fit between theoretical quantiles of the two-parameter Weibull distribution and the observed quantiles of mortality, independent of overall mortality rates, longevity, or mortality compression. Theoretical quantiles were generated for the cumulative distribution function of the Weibull, normal, and lognormal distribution and calculating Pearson's correlation coefficient $R^2$ between these theoretical quantiles and the observed quantiles of cohort survival estimated from each life table (Fig 2; Software S1). Makes and models of vehicles observed for fewer than ten complete years were excluded as, given the annual binning of mortality rates, this provided an insufficient number of survival quantile for accurate regression (Software S1). To gauge the 'best-case scenario' possible for the upper bound for the Weibull distribution fit, the quantile-quantile regression was re-calculated using the best-fit two-parameter Weibull distribution for the observed data instead of a Weibull distribution with fixed parameters (Software S1). In line with established practice the minimum standard for accepting a 'good' fit between a theoretical and observed distribution was considered an $R^2$ of 0.9.

*Mechanical, social, and physical environment data*

To clarify the role of mechanical and non-mechanical determinants of vehicle longevity, vehicles were matched to contemporary socioeconomic and spatial data for the postcode region in which they were tested. It was assumed that vehicle owners were most likely to obtain a roadworthy inspection at a mechanic in a nearby postcode region and, therefore, that the postcode at the inspection point is broadly representative of vehicle owners' regional socioeconomic status. This assumption was necessary as vehicle owners' exact home addresses were not available for the study, which remains an important limitation.

Data from geographic regions at different scales, such as the Middle Layer Super Output areas (MSOA), were linked to postcode regions using the National Statistics postcode lookup tables(51). A targeted set of small-area social(52), population(52), and economic data(53) was constructed from 2011 census data(53) for each postcode region of Great Britain, as the census fell in the middle of our 2004-2021 sampling period and measured the fine geographic scales typical of British postcodes, which have a mean of only 99 square kilometres (*e.g.* roughly 10x10km or 6x6 miles).

Social variables were selected to capture major components of socioeconomic status and diversity (Table S2). Measures were aggregated up to the 2474 postcode regions containing an MOT inspection point, with only a few postcode regions in the Scottish highlands and outer islands outside this group, and included total population size, total households, the average fraction of the population in poverty before and after deduction of housing costs, average net income before and after deduction of housing costs, and the median Indices of Multiple Deprivation for areas contained in the postcode region(51, 54). The diversity of urban-rural environments was captured by eighteen discrete type classifications used for British postcodes(51): for example, a postcode within a postcode district may be classified as "Accessible Rural Area", "Remote Rural Area", or "Urban Major Conurbation" (Software S1). The presence of each type of classification was coded as a binary vector within each postcode region, and the total number of classifications counted to reflect the within-region diversity of urban settings. In addition, geographic variables were calculated for median altitude, median distance to the closest train station, and median distance to the sea within each postcode region, and binary variables used to encode presence within each of the London districts, or within a national park (Software S1).

Regional vote shares for the 2010 UK general election(55) were downloaded from the Election Commission(55) (30 Aug 2023; now archived) and linked to postcode regions via the National Statistics Postcode Look-up tables(51). Major party vote shares were calculated by region, with all minor parties earning less than 0.5% of the vote were pooled as 'other' vote shares. In addition, the frequency of each make-model-year combination at baseline was included as a predictor, as rare vehicles are associated with being less mechanically reliable, and socially and economically valued more highly (Software S1). In total 613 training variables were available for model construction (described in Software S1 and Table S3).

All used vehicles tested for an MOT roadworthy inspection during 2011, which passed the quality control thresholds, were captured. Some 176,063 vehicles were removed for having 10 or fewer vehicles of an identical make-model-year on the road, the vast majority these vehicles being typographic input errors (e.g. "Forb Focus 2008"). Of these, 434,571 vehicles were removed as lost-to follow-up when exported, 4,731 for having negative survival times, 143 for having test dates after 2021, and 249,576 vehicles were removed because their odometers went down between any two tests (caused by data entry errors, vehicles exceeding a million miles or kilometres, odometers reporting both kilometres and miles, or undetected 'wound back' odometer fraud). A final 439,370 vehicles were removed for failing the quality controls on vehicle names and engines introduced for testing emissions described in Newman *et al.* 2024(19), leaving 15,263,682 vehicles of all classes for constructing individual-vehicle survival models.

Large, targeted subsets were selected for model construction and training from this data. A million-vehicle population was randomly sampled from the general test class containing cars – which included a wide array of vehicles but excluded minibuses, ambulances, motorbikes, mopeds, and 3-3.5T light goods vehicles – hereafter called the million-vehicle sample or MVS. The MVS was balanced such that exactly half of the sampled vehicles died during follow-up (Software S1). Two populations were sampled to get a more detailed picture of nonlinear late-life changes in mortality risk: a balanced sample of chronologically-older vehicles (N = 150,000 class-4 vehicles aged 18+ years at baseline; sampled so 75,000 died during follow-up), and an unbalanced population oversampled for high-mileage vehicles (N = 50,000 class-4 vehicles with $\geq$ 200,000 miles on the odometer at baseline; Software S1). To extend this framework to other vehicle types, a randomly sampled population balanced by vehicle type was constructed with: 250,000 random motorbikes and mopeds (test class 1 and 2), 250,000 vans or light goods vehicles from 3-3.5T (class 7 vehicles), and 250,000 cars or light vans (class 4 vehicles), for a total 750,000 vehicles and 445,687 deaths in the training set (Table S3). These training datasets were subjected to imputation using a fast random forest imputation with an average terminal node size of 50 and broken into 50 subsets by random sampling under a random splitting rule, and default mtry and ntree parameters(56) (Software S1).

*Cox Proportional Hazards and Survival Random Forest Models*

Cox proportional hazards models were fit in a nested framework to each common vehicle makes and models (N>10,000 vehicles per make-model-year; 8.8 million vehicles in total) and stratified by make-model-year to predict survival to a census date of 1/6/2019. This census date was used to exclude observational biases that may have been introduced by delayed MOT inspections after the onset of the COVID pandemic, and to match the census date for inferring de-registration mortality of vehicles issued with a SORN used when calculating life table data. Model structure was established to provide the most straightforward set of predictors of longevity from a mechanical basis and included the predictor variables of the engine capacity in cubic centimetres, the fuel type, the make (*e.g.* "Honda"), the latitude and longitude of the test, the Julian days after the start of the year at the test, and the number of previous roadworthy tests conducted and failed (Software S1). However, all Cox proportional hazards models were omitted from results as they uniformly failed the z-test for proportionality(57), which was perhaps unsurprising given the irregular, diverse, nonlinear mortality patterns observed across cohorts. Expansion of this set of predictors to include social factors like area-level incomes, deprivation, or voting patterns (as a proxy for social diversity) also resulted in failed tests for proportionality (Software S1).

To address the failure of Cox models to satisfy the requirement of proportional hazards, random survival forests(58) were implemented on the MVS, the two vehicle populations oversampled for high years and mileage, and the sample of 750,000 motorbikes, cars, and vans using the 'randomForestSRC' package(26, 56) version 3.2.3. Survival forests were tuned across a grid-based search of training parameters, using a preliminary survival random forest restricted to a random sample of 5,000 vehicles at each tuning step and trained to a node depth of four to allow rapid parameter tuning (Software S1).

To allow generalizability to a larger number of vehicles node size was evaluated from 100 to 200 by increments of 20, and the mtry parameter was run at the default grid search settings.

Brier scores – an indicator of the probability of a misclassification error, relative to a random guess(59) – were computed for survival forests in predictive models across each survival random forest model (Fig S6a; Software S1). Variable importance scores were estimated using the approaches detailed in Ishawaran(30), built-in to the 'randomForestSRC' package(26, 56). The marginal survival effect of accumulated mileage, accumulated age, past repairs, and diverse social variables included in the million-vehicle survival random forest were evaluated in individual marginal plots using random forest corrected mortality rates(56) (Fig 4; Fig S3; Supplementary Information).

All analysis was conducted using R version(60) 4.0.5 (Software S1), with code and population-level data freely available in the Supplementary Information, and with individual-level data available upon application from the DVSA.

## Acknowledgments

The author would like to thank the DVSA and the Department for Transport for their substantial help and support, Dr Doug Leasure and Prof Charles Rahal for their kind encouragement, Prof Felix Tropf for his feedback, Dr Elena Racheva for her longstanding tolerance of vehicle statistics, and Mr Grant Thunder for his excellent help and advice, without all of whom this project would not be possible.

**Figures and Tables**

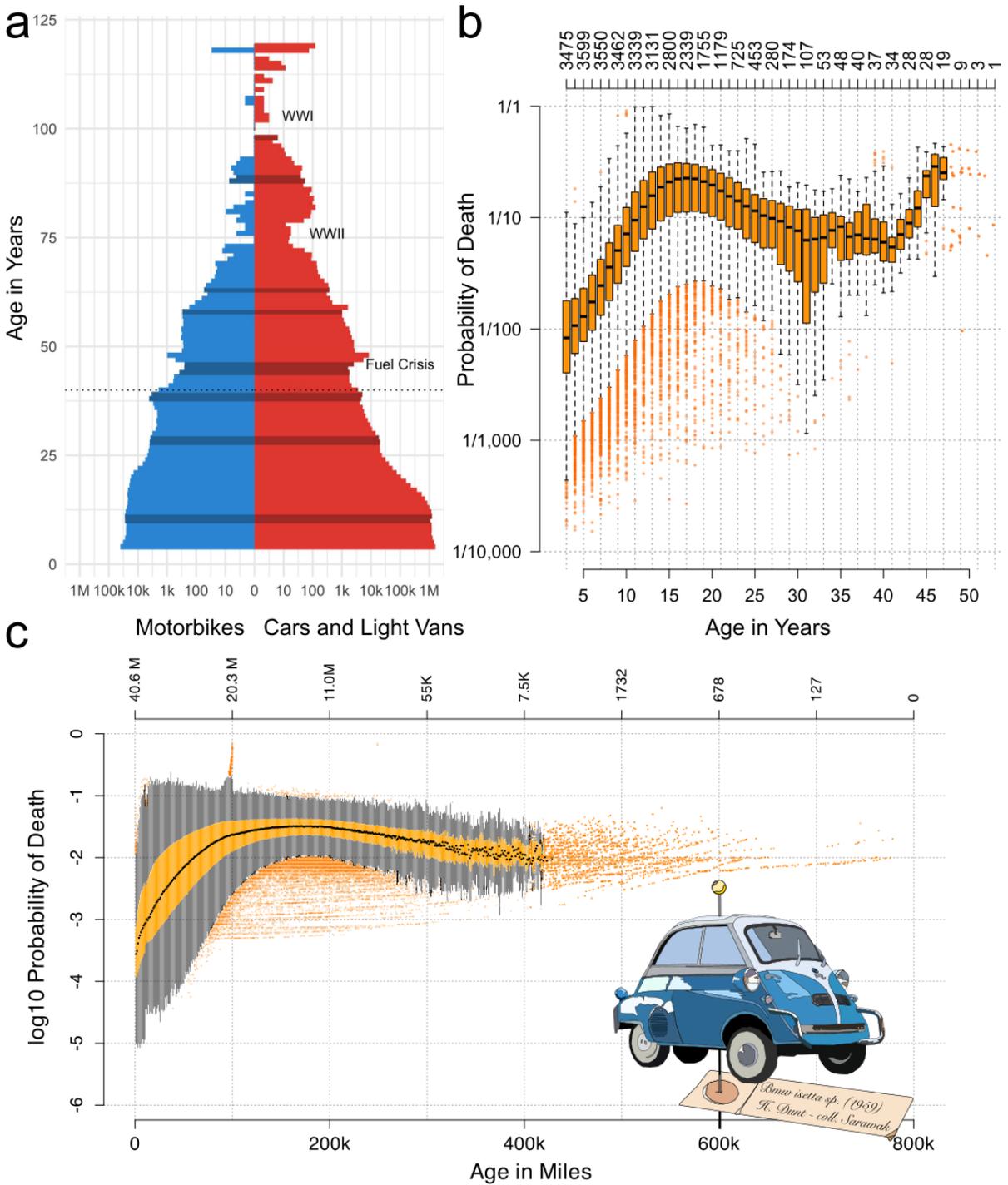

**Figure 1. The Life, Death, and Aging of Machines.** For cars and light vans (**a**, red), light goods vehicles, and motorbikes (blue), population pyramids (**a**) reveal changes in population structure corresponding to recessions (dark bands) and wars (labelled). Striking patterns are also observed in age-specific mortality rates **b,** which generally undergo an initial acceleration with age, before decelerating from around age 15 and falling exponentially or 'ageing in reverse' from around age 18, then finally changing nonlinearly beyond ages 30-35. In contrast, mile-specific survival rates – measured per 10,000 miles – capture responses to physical wear rather than chronological age. These mile-specific mortality rates **(c)** also initially accelerate, before falling progressively as miles, and wear-and-tear, accumulate. Thus, both age (**b**) and mile-specific (**c**) survival patterns disrupt theoretical expectations that machine survival rates will progressively degrade over time. Points in **b, c** represents each unique make-model-year of car containing N$\geq$100 surviving members; supplementary (top) x-axes show number of unique makes and models surviving in **b** and the number of total vehicles passing quality control observed at each age in **c**, age-specific mortality for motorcycles and vans are shown in Fig S1a-b.

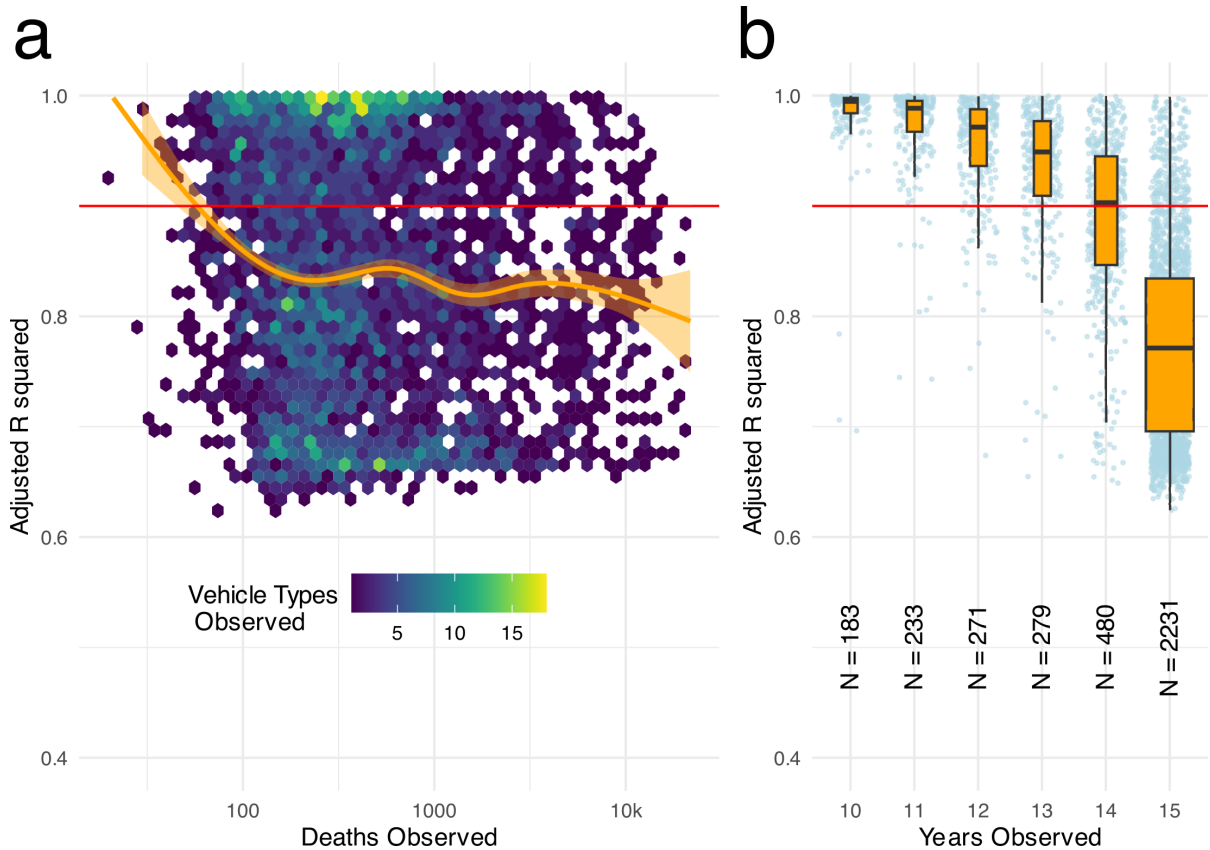

**Figure 2. The degrading fit of the Weibull distribution with increasing sample sizes and observation periods.** The fit between the theoretical quantiles of a best-fit Weibull distribution and the observed mortality pattern of each common and unique make, model, and year of vehicle (vehicle type; with *e.g.* "2002 Ford Focus" as one unique vehicle type) was generally below the minimum $R^2$ = 0.9 threshold required to support a Weibull distribution (N = 2342 or 63.4% of sample had adjusted $R^2$ < 0.9). In addition, the fit of the Weibull distribution degrades with an increasing number of observed vehicle deaths (**a**) or observation times (**b**; measured in whole years). That is, even when using a best-fit estimate, the model fit of the two-parameter Weibull distribution is poor and becomes worse with an increasing amount of data and longer observation times. Sample sizes in (**b**) indicate number of make-model-year combinations observed; orange trendline in (**a**) shows a locally weighted smoothed spline with 95% confidence intervals;

boxplots in (**b**) show median and interquartile ranges overlaid on jittered data (blue transparent points).

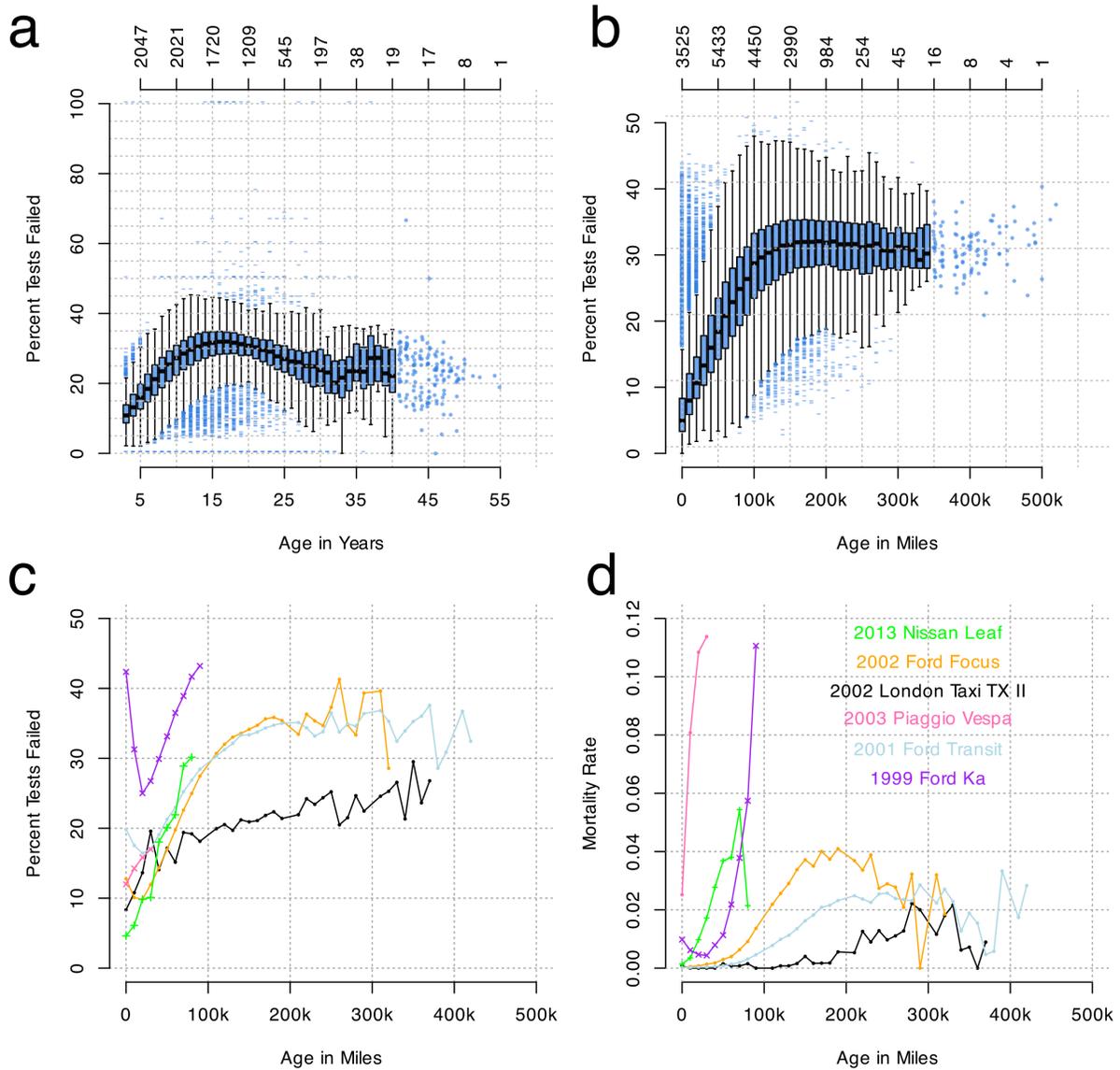

**Figure 3. Mechanical reliability and survival.** Vehicle reliability and rates of wear-and-tear, shown here as a function of age (**a**) and mileage (**b**), are measured by rates of major mechanical failures and over seven thousand unique mechanical failure codes captured during 397 million roadworthiness inspections. Despite the broad resemblance of failure rate curves (**a-b**) to mortality rates in Fig 1b-c, mechanical reliability has a flexible and nonlinear relationship with age-specific survival rates across vehicle types (N=6,262 make-model-year combinations). The flexibility of this relationship can be seen when exploring the age-specific repair rates (**c**) and mortality rates (**d**) of common or noteworthy makes and models. For example, the Piaggio Vespa (pink; the most common moped) has a low mechanical failure rate (**c**) but a high mile-specific mortality rate (excluding crashes; **d**); a pattern also seen in the 2013 Nissan Leaf (green; the most common electric car) at low miles. In contrast, the 2002 Ford Focus (orange; most common car) has a similar or lower mechanical failure rate (**c**) than the 2001 Ford Transit (blue; most common van), but a higher non-crash mortality rate (**d**).

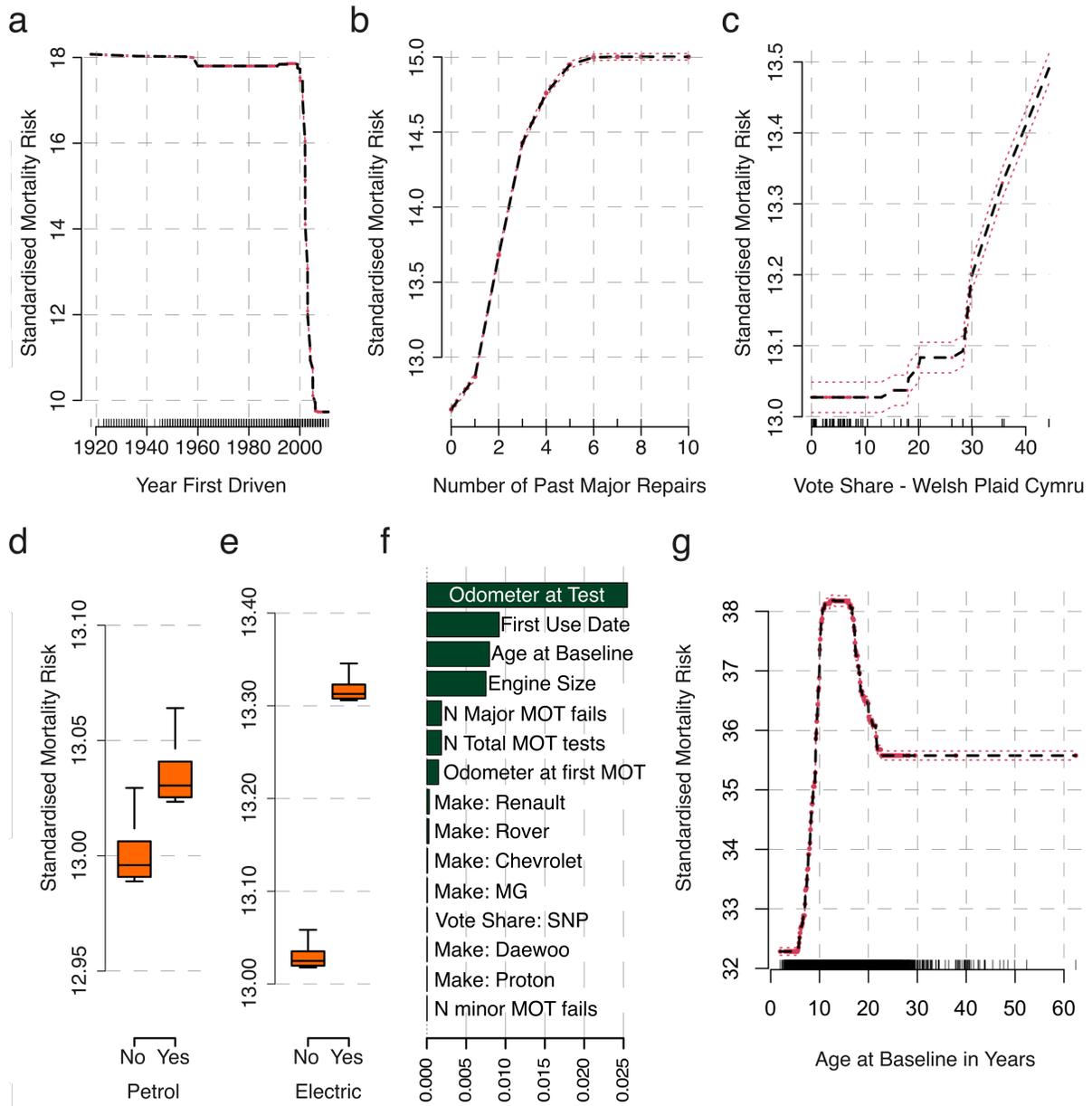

**Figure 4. Nonlinear patterns and social correlates detected by survival forests.**
Predicting mortality on a balanced random sample of a million class-4 vehicles (primarily cars) subjected to an MOT roadworthy inspection in 2010, with the sample balanced so that half a million vehicles died during follow-up, led to accurate predictions of mortality risk and the nonparametric estimate of mortality risk associated with diverse factors. Older vehicle cohorts (**a**) had a near-constant mortality risk, which reduced marginally in vehicles built from 1960-1991, and fell substantially for young vehicles. Mortality risk rose for increasing numbers of past major repairs (**b**) before a threshold of roughly five or six past major repairs, beyond which mortality risk was constant. Mortality risk also corresponded to some basic social predictors such as vote share (**c**) – which acts as a proxy of social and geographic divides – such as the Welsh national party Plaid Cymru vote share shown here. Binary comparisons are also possible, such as the marginally elevated mortality risk associated with petrol vehicles (**d**) and more substantially elevated risk associated with early electric vehicles (the Nissan Leaf and Tesla 3; **e**). These comparisons inform the direction and magnitude of variable importance scores assigned during random forest construction (**f**) which otherwise only indicate approximate predictive utility. More nuanced approaches were possible, for example shown in (**g**) by reconstructing the same model after oversampling for high-mileage vehicles: the post-20-year survival patterns of these higher mileage vehicles revealed a sizable fall in late-life mortality risk.

**Supporting Information for**

British Automobiles, Aging Theory, and the Death of Complex Machines

**This PDF file includes:**

       Figures S1 to S6

       Tables S1 to S3

       Legends for Datasets S1

       Legends for Software S1

**Other supporting materials for this manuscript include the following:**

       Datasets S1

       Software S1

**Supplementary Materials – note for the arxiv version**

Pages 60 to 272 are the first two hundred entries in the first volume of the Drive Dictionary, one of seven total volumes, containing data on all 6281 common makes and models of vehicle in Britain.

Please take a look. The full version will be made available shortly at DriveDictionary.com

The remaining pages are Software S1 in plain-text R code – from pdf page 272-476 inclusive.

**Figure S1. Diverse mortality patterns in motorbikes, mopeds, and light-medium vans.** The roadworthy inspection process in the UK differentiates different vehicle types, such as motorcycles and mopeds (denoted by test

classes 1 and 2) and light vans – that is, light goods vehicles – weighing between 3-3.5 metric tons (test class 7). The mortality profiles of motorcycles and mopeds (**a**) are strikingly different from other vehicle classes, in that they have high (non-crash) mortality rates that increase at a minimal, but non-zero, rate with age. This pattern continues before mortality rates rapidly fall at age 19, below the mortality rates observed at any earlier age. In contrast, vans or light goods vehicles (**b**) undergo a more 'car-like' mortality pattern with a gradual and near log-linear increase in mortality rates until flattening and inverting around age 15-19, followed by a longer-term decline in mortality rates. Mortality rates are only shown for common makes and models (N>1000 observed vehicles overall) and for years where at least 100 unique instances of each vehicle type are observed.

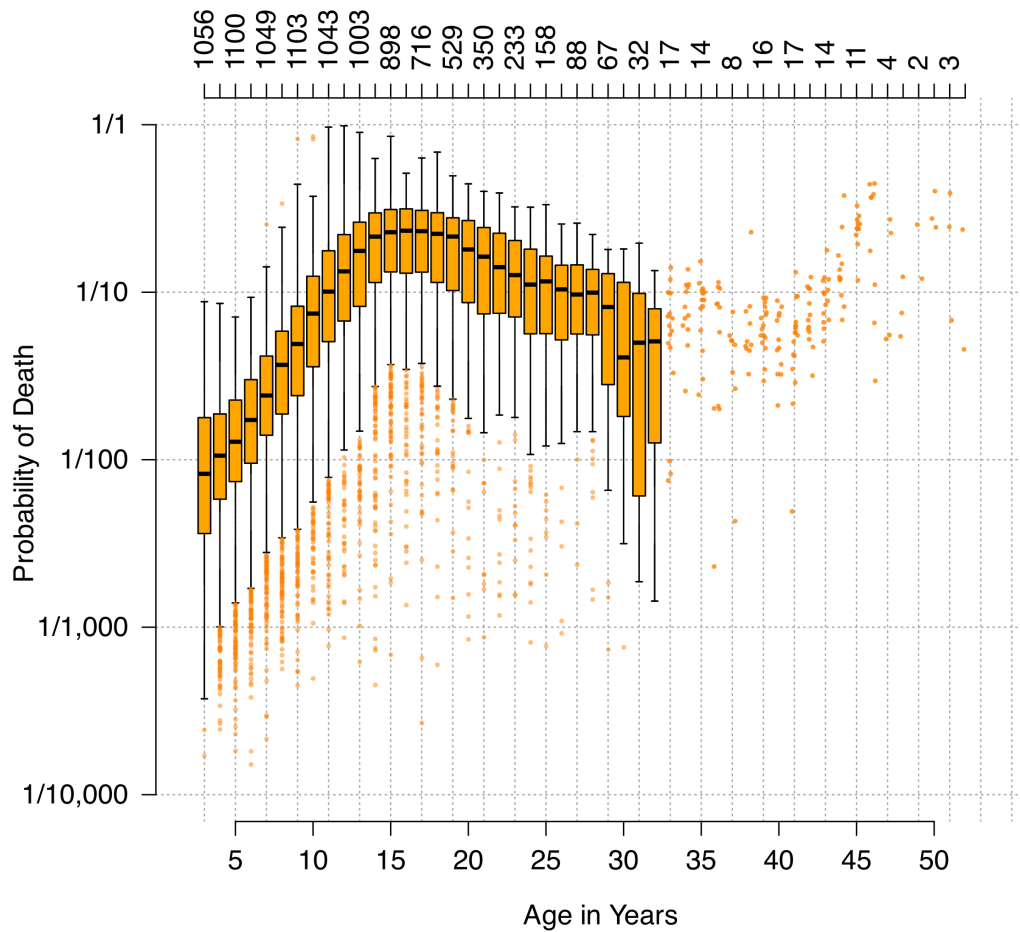

**Figure S2. Conserved mortality patterns when censoring for vehicles dying before 2011.** Artefactual distortions of mortality rates can be caused by under-reporting or delayed reporting of recent deaths. Added to this possibility, the last years of these data were distorted by the COVID pandemic, such as exclusions to regular testing introduced from March 2020 to allow certain vehicles to avoid roadworthy tests. To test the robustness of mortality patterns to these effects, all life tables were reconstructed using a filter to exclude any delayed reporting of deaths by only measuring mortality rates before January 1, 2011. Even after censoring vehicle deaths to allow for a 10-year delay in reporting, the nonlinear reverse-aging type patterns in Fig 1 were conserved, showing reductions in mortality risk after age 16-17 and sharply reducing the possibility these patterns are reporting artefacts. Total number of makes and models observed is shown on the supplementary (top) x-axis; full distributions are shown for ages with under twenty observed make-model combinations.

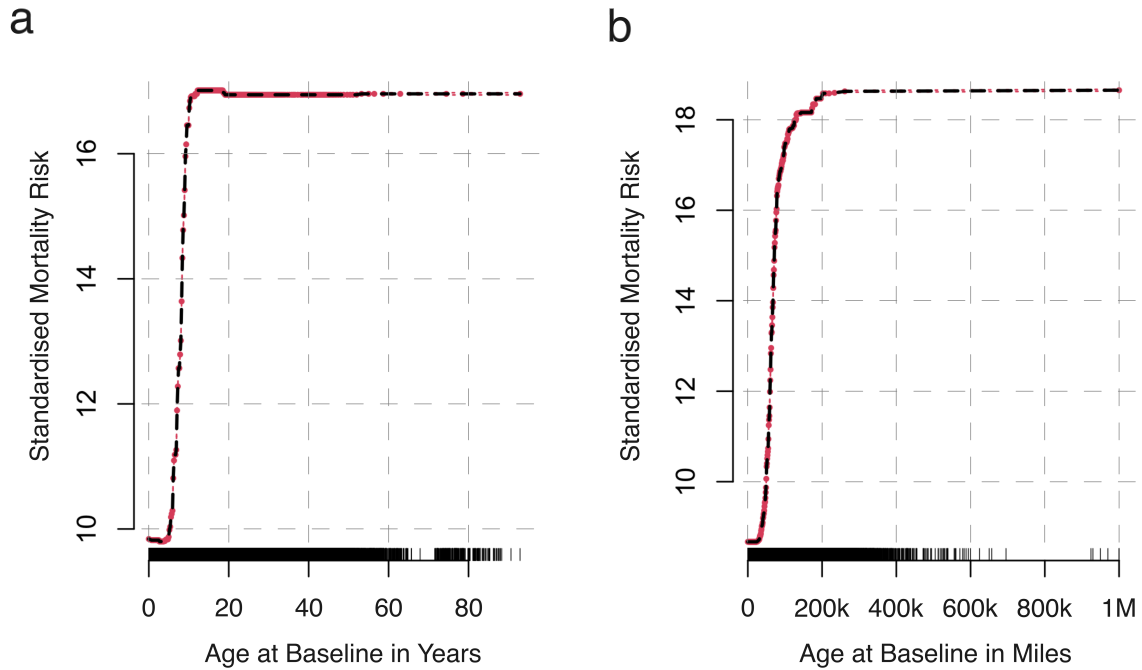

a

b

**Figure S3. The nonlinear marginal effects of linear increases in 'wear and tear'.** Training survival random forests – a basic machine-learning algorithm – to predict mortality risk allows non-parametric estimation of the marginal effects of variables on vehicle mortality. After correcting for a broad array of mechanical, geographic, and social other variables, in a million-vehicle cohort subjected to an MOT inspection during 2010, increasing chronological age is still associated with nonlinear mortality effects (**a**): marginal mortality risks initially fall slightly over the first few years' of vehicle use, rise rapidly from five to 17 years of age, fall very slightly again between 17 and 22, and then remain largely unaffected by subsequent increases in chronological age. A similar nonlinear response occurs in response to a linear increase in miles driven (**b**): mortality rates are near constant until warranties begin to expire, then increase rapidly, before mortality rates cease responding to increasing 'wear and tear' to remain constant above 270,000 miles. Neither trend supports the idea that accumulated 'wear and tear' will drive increasingly rapid, accelerated failures with age. Note the change in y-

axis scale and the standard errors, represented as barely-visible dashed red lines drawn around the adjusted mortality risk estimates.

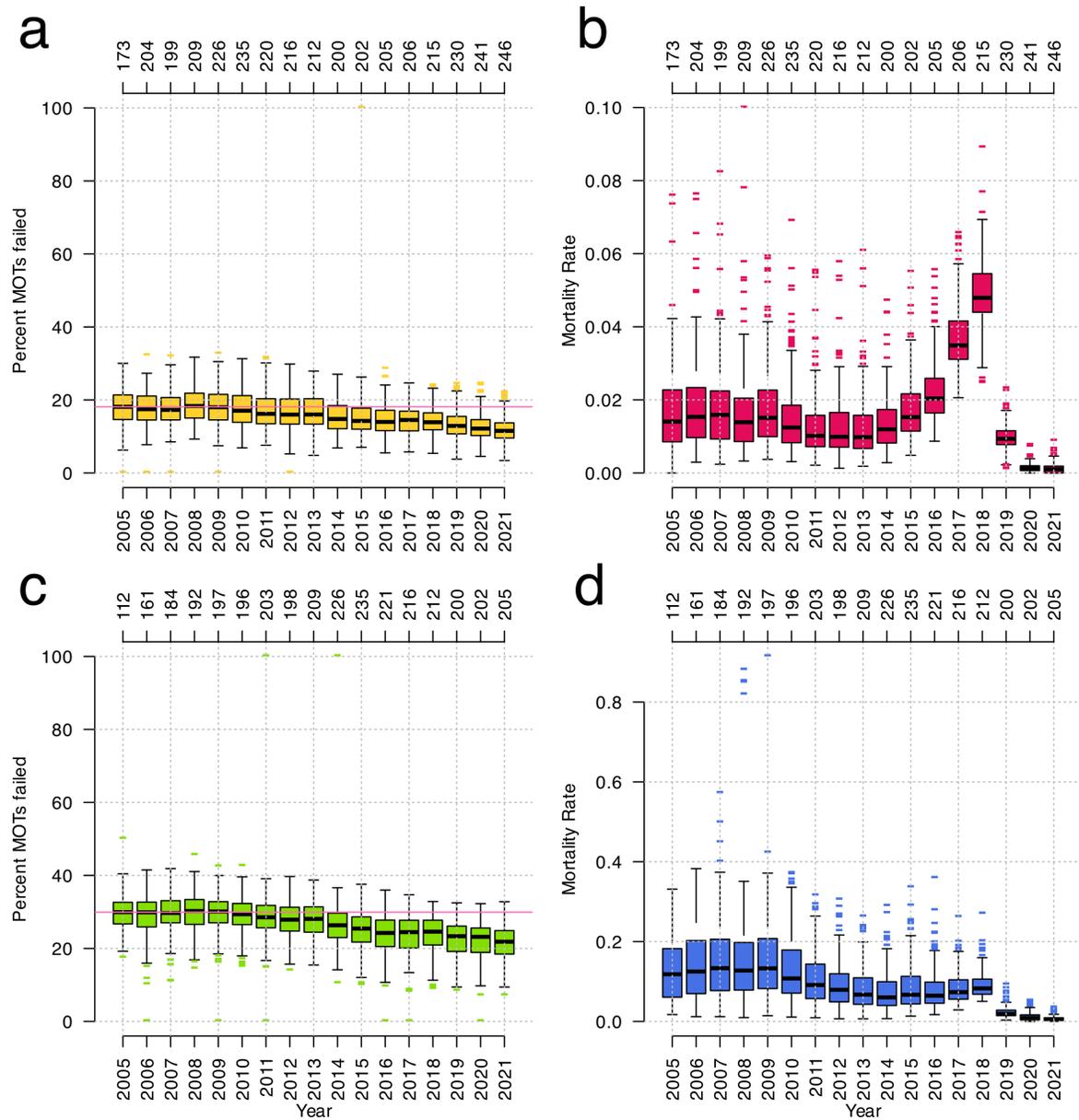

**Figure S4. Recent shifts in the mechanical failure rates and mortality rates for vehicles of a uniform age.** Mechanical reliability and mortality rates of vehicles that are five (**a-b**) and ten years old (**c-d**) show markedly different patterns across the study period. In 2005, five-year-old vehicles (**a**) had a 'major' failure in the roadworthiness test 18.1% of the time (horizontal pink line). This figure declined steadily toward a median failure rate per test of 11.5% by 2021,

despite longitudinal increases in the stringency of roadworthiness testing. In contrast, mortality rates in the same cohorts of five-year-old cars (**b**) did not track improvements in mechanical reliability, and instead fell sporadically until 2013, rose rapidly until 2021, and then fell sharply again - especially during the distortions of the pandemic when distances driven per year were minimal. As in five-year-old vehicles, ten-year-old vehicles (**c-d**) also revealed a pattern of steadily falling mechanical failure rates (**c**) with the median rate of major mechanical failures falling from 29.9% per test in 2005 (pink line) to 21.8% in 2021. Again, however, mortality rates in these 10-year-old vehicles (**d**) were uncoupled from this gradual improvement in mechanical failure rates and was also markedly different from the mortality patterns in five-year-old vehicles in (**b**). Points and numbers on the supplementary (top) x-axes show the number of observed makes and models of vehicle; note the large shift in y-axis scales between **b** and **d**.

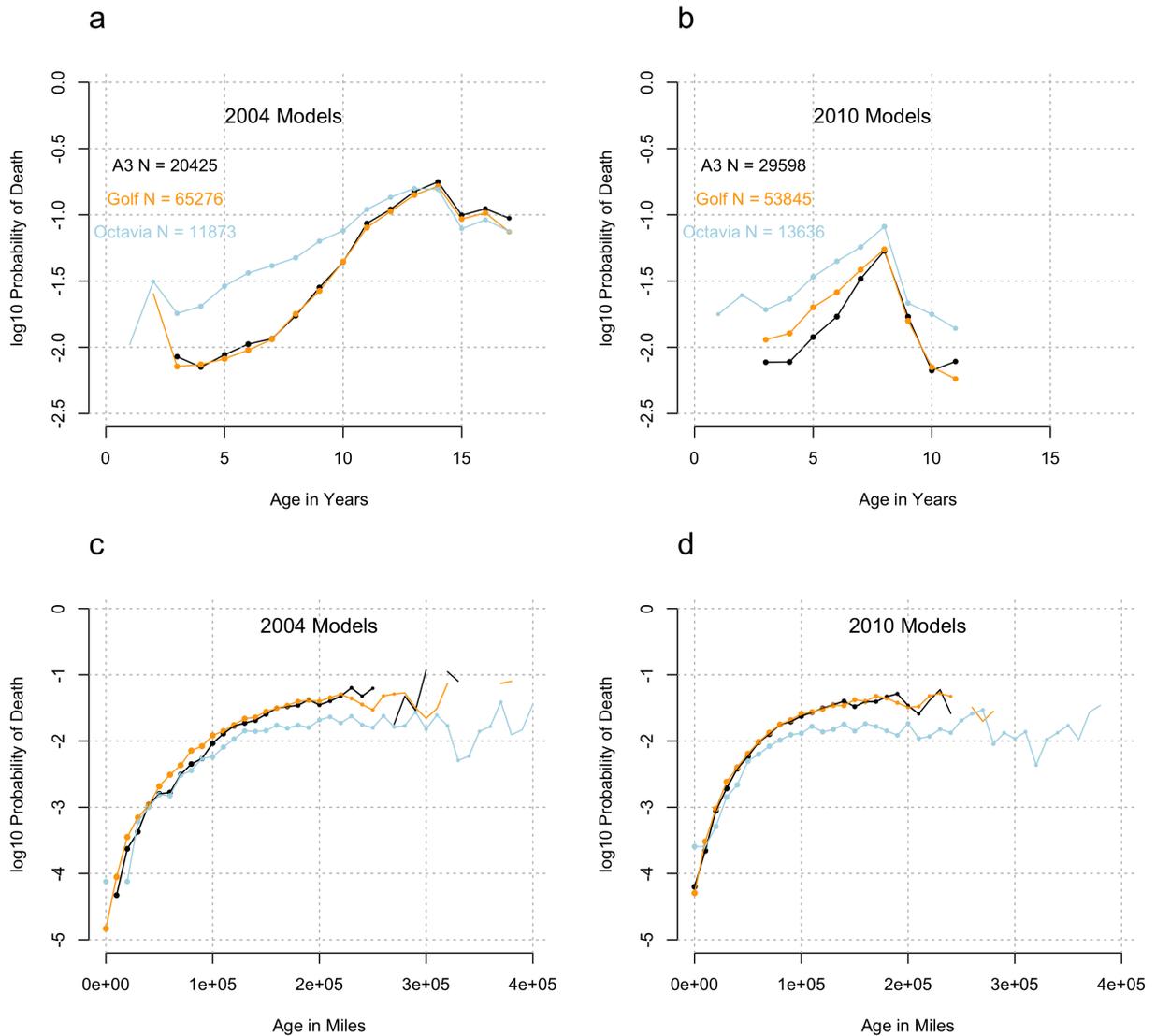

**Figure S5. Survival patterns of mechanically undifferentiated machines.**
Vehicles are often produced to identical manufacturing specifications in shared
factories, then sold under different badges with cosmetic differences in styling.
Plotting survival patterns of these socially differentiated but otherwise-identical
vehicles revealed surprising differences in survival patterns. Here, the
Volkswagen Golf (black) – also sold as the Audi A3 (orange) and Skoda Octavia
(light blue) – displays markedly different survival curves. Measuring the
probability of death by year in either 2004 cohorts (**a**) or 2010 cohorts (**b**) reveals
sizeable differences in the age-specific probability of death between models. The Octavia has a higher probability of death per year, but a lower probability of death at advanced ages in the older 2004-era vehicles. These probabilities of death are changed when measured per driven mile, as shown in the 2004 (**c**) and 2010 (**d**) cohorts. The diversity in these survival curves persists if survival rates are corrected for driving patterns, geography, and repair rates: such gaps may instead be determined by social differences between vehicle owners or, in the final three years of (**a-b**), by lower mileages driven during the COVID pandemic. Also of note is the increase in year- and mile-specific mortality rates in the A3 and Golf between 2004 and 2010, indicating the shorter lifespan of newer models, and the opposite trend in the per-year mortality rate of the Octavia.

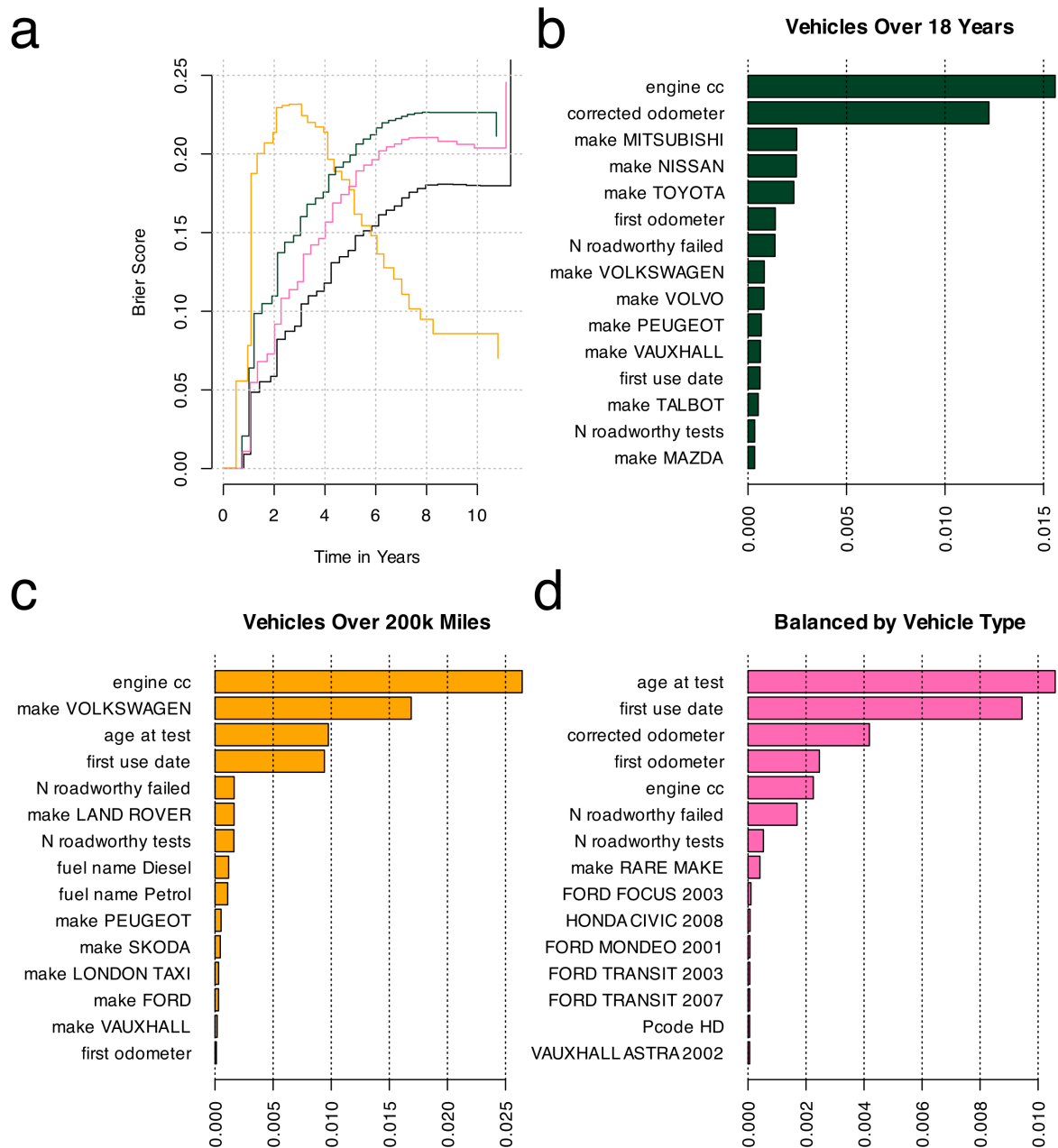

**Figure S6. Brier scores and variable importance for survival forests of vehicle survival.** The Brier score of survival forest models (**a**) showing the time-dependent accuracy of mortality models relative to a random guess for: the million-vehicle sample balanced to contain half a million deaths (black), a random sample of 150,000 vehicles over 18 years at baseline (racing green), a random

sample of 50,000 vehicles with over 200,000 miles at baseline (orange), and a population balanced by vehicle type to contain 250,000 random mopeds and motorbikes, 250,000 random light goods vehicles or vans, and 250,000 random cars or light vans (pink). These models degraded in predictive accuracy over time, but maintained relatively high concordance, apart from the population of high-mileage vehicles (orange). Predictive accuracy in this subset initially degraded, shown by the steep initial upward slope (orange), but then became more accurate over longer period.  There was also a relatively high concordance in variable importance scores across models (**b-d**). For example, engine size is a proxy for both broad (*e.g.* mopeds versus vans) and narrower vehicle types (*e.g.* city cars versus sports cars) and was ranked as the most important predictor of survival in vehicles over 18 years (**b**) or 200,000 miles at baseline (**c**), and the fifth most-important variable in the balanced subset (**d**) out of 612 predictor variables.

**Table S1. Descriptive data on British Vehicle Populations 2004-2021.**

| Variable | All Vehicle Classes 1-5 & 7 | Class 1 Mopeds and Motorbikes < 200cc | Class 2 Mopeds and Motorbikes ≥200cc | Class 4 Cars, light vans, and assorted vehicles | Class 7 3-3.5T Light Goods Vehicles (Vans) |
|---|---|---|---|---|---|
| Total | 64,910,416 | 904,868 | 1,760,655 | 60,465,934 | 1,615,431 |
| Exported | 1,405,204 | 20,244 | 40,089 | 1,308,669 | 32,154 |
| Certified Destroyed + Scrapped | 9,302,767 | 136,140 | 276,032 | 8,660,235 | 206,503 |
| Survived | 30,394,096 | 399,230 | 749,808 | 28,336,432 | 835,695 |
| Engine Capacity cc – mean (25-75%) | 1705 (1360 – 1995) | 103 (49-125) | 786 (599-998) | 1739 (1386 - 1995) | 2318 (2151 – 2463) |
| Petrol | 40,086,346 | 903,229 | 1,759,664 | 37,348,080 | 13,744 |
| Diesel | 24,213,316 | 251 | 655 | 22,510,973 | 1,599,745 |
| Hybrid | 436,076 | < 10 | < 10 | 436,043 | 29 |
| Electric | 95,874 | 1,362 | 309 | 93,971 | 159 |
| Median date of first use | 5 Nov 2005 | 17 Jan 2005 | 17 Sep 2004 | 14 Nov 2005 | 4 March 2007 |

**Table S2. Summary of mechanical and temporal variables included in survival forests.**

| Variable(s) | Type | Notes |
| --- | --- | --- |
| Odometer at baseline inspection | Numeric | Odometer reading, corrected to miles |
| Initial Odometer reading at first inspection | Numeric | Odometer reading, corrected to miles |
| Age at test | Numeric | |
| Total MOT/ roadworthy tests at baseline | Integer | |
| Total major (repaired) failures in an MOT test at baseline | Integer | |
| Total minor (repaired) failures in an MOT test | Integer | |
| Engine size in cubic centimetres | Numeric | Electric vehicles coded as zero |
| Base_freq | Numeric | Total surviving number of this vehicle type at baseline |
| "fuel_name_" coded variables | Binary | Six binary-coded variables indicating fuel type as Diesel, Petrol, Electric, Hybrid Electric, LPG, Other |
| "test_class_" coded variables | Binary | Six binary-coded variables indicating roadworthy test categories 1-5 and 7. |
| "make_" coded variables | Binary | 86 binary variables coded to indicate common makes (e.g. "make_HONDA"; with a binary 0/1 coding to indicate if Honda was the primary manufacturer) |
| make_RARE_MAKE | Binary | Additional binary variable coded to indicate if the manufacturer was a rare (N<10,000) make and model - for example, the Aston Martin Cygnus |
| Common make-model-year coded variables e.g. "FORD__MONDEO__1997" | Binary | 325 binary-coded variables indicating each common (N>10,000) vehicle make, model, and year |

**Table S2 continued. Summary of social, temporal, and spatial variables included in survival forests.**

| Variable(s) | Type | Notes |
|---|---|---|
| JDAT | Numeric | Julian Days at Test from the start of the year - corrects seasonal trends in testing |
| First_use_date | Date as numeric | The date of first registration and use - corrects for 'unsold time' spent on *e.g.* the car lot |
| "Pcode_" postcode-region coded variables | Binary | 123 binary-coded variables indicating location of vehicle at testing |
| approxLat | Numeric | Approximated (mean of postcodes) latitude of test centre |
| approxLong | Numeric | Approximated (mean of postcodes) longitude of test centre |
| Altitude | Numeric | |
| Households | Numeric | Number of households in region |
| Population | Numeric | |
| IMD | Numeric | Indices of Multiple Deprivation |
| Distance_Station | Numeric | Distance to the closest train station, indicating public transport accessibility |
| Distance_sea | Numeric | Distance to the sea |
| "vote_" coded variables | Numeric | Indicates percentage votes for each major party in the 2010 general election, including a category "vote_OTHER" showing the share of votes allocated to minor parties. |
| NatParks | Binary | Is there a national park in the postcode |
| "London_zone_" coded variables | Binary | Seven binary-coded variables indicating residence in London postcode zones |
| Average_income | Numeric | Average annual household incomes |
| Average_wk_income | Numeric | Average weekly incomes |
| Pct_poverty_AH | Numeric | Percent households in poverty after housing costs |
| Pct_poverty_BH | Numeric | Percent households in poverty before housing costs |
| NIAH | Numeric | Normalised Income After Housing Costs |
| NIBH | Numeric | Normalised Income Before Housing Costs |
| maxQuality | Numeric | Indicators of data quality for social estimates |
| minQuality | Numeric | Indicators of data quality for social estimates |

**Table S3. Model training parameters and description.**

|  | Million-Vehicle Sample | High Years (>17 years) | High Mileage (>200,000 miles) | Balanced Vehicle Types |
|---|---|---|---|---|
| N | 1000000 | 150000 | 50000 | 750000 |
| Of Which: |  |  |  |  |
| Cars and vans - Class 4 | 1000000 | 150000 | 50000 | 250000 |
| Motorbikes & Mopeds – Class 1 & 2 | - | - | - | 250000 |
| Vans and LGVs – Class 6 | - | - | - | 250000 |
| Deaths: | 500000 | 75000 | 44853 | 445687 |
| Number of trees: | 100 | 100 | 100 | 100 |
| Forest terminal node size: | 160 | 120 | 180 | 120 |
| Average no. of terminal nodes: | 47.7 | 34.3 | 36.2 | 48.2 |
| mtry parameter: | 381 | 148 | 159 | 247 |
| Training variables: | 613 | 613 | 613 | 613 |
| (OOB) CRPS: | 0.14 | 0.18 | 0.14 | 0.16 |
| (OOB) Requested performance error: | 0.25 | 0.35 | 0.38 | 0.32 |
| Maximum Brier Score | 0.27 | 0.23 | 0.23 | 0.25 |

# Dr Newman's Drive Dictionary

Dr Saul Justin Newman, University College London and University of Oxford

03/04/2025

## Volume 1 - Abarth to Citroen

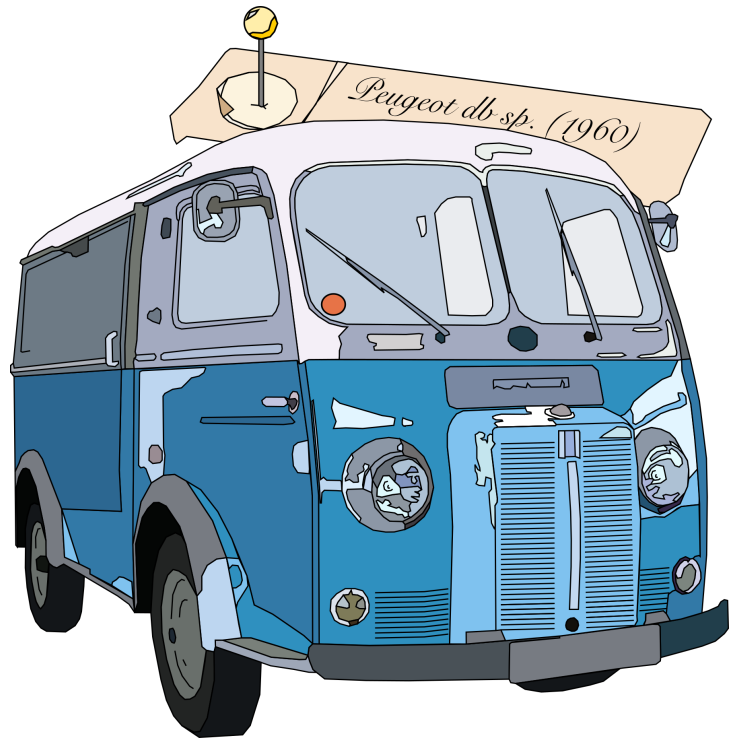

## Understanding the Dictionary

Welcome to the Drive Dictionary.

This guide exists to fill a rather surprising shortfall in public knowledge. Society pays enormous amounts of money for vehicles and transport but, despite our governments collecting the data, nobody tells us which vehicles are most likely to fail or die (become de-registered or scrapped).

My guide tries to fix that problem.

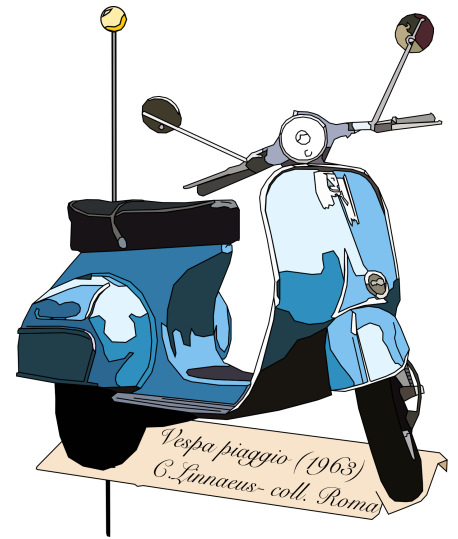

Vehicles are the second-most expensive asset most people will ever buy. Yet, everyone is left to guess what the longest-lived and most-reliable vehicles are, based on perception and reputation alone. Most of these guesses are unfortunately wrong, often because car manufacturers fight so hard and pay so much to change our perception of what is reliable.

It does not have to be this way. Data on vehicle reliability exists, and is made without the interference of car manufacturers or private interests. Every year, mandatory government roadworthy tests and registration data *directly measure* the reliability of vehicles in the real world, by subjecting vehicles to uniform, legally-standardized testing by trained mechanics.

Here is that data, in a simple dictionary: shareable, easy-to-use, and hopefully useful.

The drive dictionary provides you with a measure of the longest-lived and most mechanically reliable (used) vehicles in the United Kingdom. Just look a vehicle up, much like a regular dictionary: every common vehicle in the UK is in here, first sorted alphabetically by make and model, and then sorted chronologically by year.

If you look a vehicle up, you can find out its mortality rate at different ages, and compare that to similar vehicles made in the same year. Even better, you can find out how often the vehicle fails a UK roadworthy test, and compare this to the failure rate of its peers. This data on reliability is even broken down by mileage: for example, looking up a 2010 model Abarth 500 (the first vehicle in this book) with 20,000 miles on the odometer, and you find that they fail the mechanical roadworthy test 16% of the time, and have the tenth-highest inspection failure rate of any car built in 2010.

If all of this seems confusing, don't worry. A user guide is provided in the first few pages. So is a 'best-of' league table. But please, be cautious. Vehicle survival behaves in unexpected ways.

The past survival of a vehicle is not necessarily a good guide to future survival. Remarkably, vehicle survival and breakdown rates seem to depend more on people than on parts: a modern vehicle is so reliable that longevity appears to be mostly determined by a vehicle owners' behavior, not the nuts-and-bolts in the car.

Why is this important? Obviously buying a green Mini Cooper will not make you drive like Mr Bean, in the same way that buying a Bentley does not make you as rich as a typical Bentley owner (quite the opposite). Therefore, buying a long-lived vehicle will not necessarily make **your** vehicle live for a long time - you cannot buy the behavioral patterns of the owners themselves, and these patterns are a substantial cause of vehicle reliability.

A better guide may be the observed rate at which vehicles failed a standard roadworthy inspection, which I term the mechanical reliability, but again this is not fool-proof. Vehicle maintenance and repairs depend on patterns of vehicle use, and the money available for repairs, all of which also seem to depend on the owner (or the economy) as well as the vehicle.

Keep all this in mind when flipping through the guide.

Please also keep in mind that this document is not financial advice, it is a scientific document recording the survival of machines, made readable for a curious public. Please only enjoy it as a resource, a way to settle arguments, or as a pleasant curiosity. Finally, if you are buying a vehicle, I would consider that most modern cars are mechanically very reliable. You are better off considering the best vehicle for the environment[1], which are almost always far cheaper to run and maintain, than the best vehicle at the mechanics.

---

[1]Or consider if you can use a bicycle - to borrow a meme, cars run on money and make you fat, bikes run on fat and save you money. Cycling is also, on average, a far faster commute in the UK. If you know anyone you know who cycle-commutes, ask them for tips.

(the table of contents is clickable and hyperlinked)

# Contents

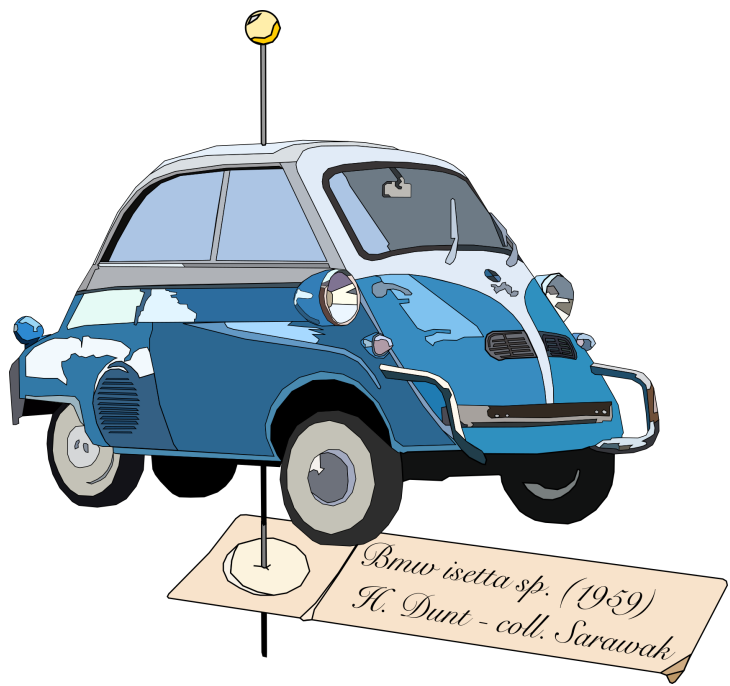

*Bmw isetta sp. (1959)*
*K. Dunt - coll. Sarawak*

## The League Tables

The most common thing I have been asked since starting this project is: what is the 'best' vehicle, in terms of reliability?

As such, I have constructed a league table showing the most and least reliable vehicles observed in the UK government data - with reliability judged by the fraction failing a standardized roadworthy inspection due to a major failure.

This does not necessarily make these vehicles the 'best' because, as I said, different vehicles do different things and are maintained by different drivers. Bentley Continentals, for example, may have low breakdown rates because they have rich drivers, not necessarily because they are reliable vehicles.

But for the curious, I have arranged the tables to rank failure rates at moderate mileages for each of the major British government vehicle classes used for inspections: mopeds and motorcycles below 200cc (Table 1), mopeds and motorcycles over 200cc (Table 2), cars and light vans (Table 3), and vans or light goods vehicles (Table 4). The failure rates are measured, respectively, at 10,000 miles for motorbikes and 100,000 miles for cars and vans.

Inspection Pass-Fail Rates - Top 10 Mopeds and Motorcycles < 200cc
For all inspections conducted at 10,000 miles

| Make-Model-Year | N Tests | Percent Passed | Percent Major Failures |
|---|---|---|---|
| Honda Nsc 2017 | 708 | 83.3 | 7.63 |
| Piaggio Vespa 2015 | 695 | 84.5 | 9.35 |
| Honda Ww 2016 | 704 | 83.5 | 10.20 |
| Piaggio Vespa 2014 | 959 | 83.8 | 10.40 |
| Piaggio Vespa 2013 | 1211 | 82.0 | 10.70 |
| Piaggio Vespa 2016 | 393 | 82.7 | 10.90 |
| Honda Nsc 2016 | 737 | 79.8 | 11.00 |
| Piaggio Vespa 2012 | 1063 | 80.4 | 11.10 |
| Honda Ww 2017 | 526 | 82.3 | 11.20 |
| Honda Ww 2018 | 293 | 83.3 | 11.30 |

Inspection Pass-Fail Rates - Top 10 Mopeds and Motorcycles 200cc and over
For all inspections conducted at 10,000 miles

| Make-Model-Year | N Tests | Percent Passed | Percent Major Failures |
|---|---|---|---|
| Bmw R1200 2018 | 664 | 94.9 | 1.36 |
| Bmw R 1200 2016 | 1749 | 95.3 | 1.94 |
| Bmw R Series 2012 | 1929 | 94.7 | 2.13 |
| Honda Crf 2017 | 532 | 94.4 | 2.26 |
| Bmw R 1200 2015 | 3031 | 95.4 | 2.28 |
| Bmw R1200 2010 | 1596 | 95.1 | 2.38 |
| Bmw R Series 2011 | 1764 | 94.8 | 2.49 |
| Bmw R Series 2013 | 2531 | 94.1 | 2.65 |
| Bmw R Series 2009 | 1496 | 93.2 | 2.74 |
| Bmw R1200 2017 | 1455 | 94.7 | 2.75 |

### Inspection Pass-Fail Rates - Top 20 Vans and LGVs
### For all inspections conducted at 100,000 miles

| Make-Model-Year | N Tests | Percent Passed | Percent Major Failures |
|---|---|---|---|
| Mercedes-Benz Sprinter 2018 | 832 | 81.2 | 13.9 |
| Iveco Daily 2016 | 494 | 81.2 | 14.0 |
| Iveco Daily 2017 | 189 | 76.7 | 14.8 |
| Mercedes-Benz Sprinter 2017 | 1856 | 78.7 | 15.7 |
| Iveco Daily 2015 | 717 | 78.5 | 16.0 |
| Mercedes-Benz Sprinter 2016 | 2496 | 78.6 | 16.1 |
| Mercedes-Benz Sprinter 2015 | 3954 | 77.5 | 16.7 |
| Mercedes-Benz Sprinter 2014 | 4651 | 76.4 | 17.8 |
| Fiat Ducato 2017 | 233 | 70.8 | 18.5 |
| Nissan Nv400 2015 | 134 | 76.1 | 18.7 |
| Ford Transit 2018 | 672 | 73.8 | 18.8 |
| Volkswagen Crafter 2017 | 325 | 75.4 | 18.8 |
| Iveco Daily 2013 | 725 | 75.0 | 19.7 |
| Mercedes-Benz Sprinter 2011 | 4686 | 72.6 | 19.9 |
| Mercedes-Benz Sprinter 2013 | 4770 | 73.9 | 19.9 |
| Renault Master 2018 | 166 | 72.9 | 20.5 |
| Fiat Ducato 2015 | 449 | 72.4 | 20.5 |
| Mercedes-Benz Sprinter 2012 | 4817 | 73.2 | 20.6 |
| Peugeot Boxer 2017 | 595 | 72.8 | 20.7 |
| Vauxhall Movano 2014 | 809 | 68.9 | 20.9 |

Inspection Pass-Fail Rates - Top 20 Cars and Light Vans
For all inspections conducted at 100,000 miles

| Make-Model-Year | N Tests | Percent Passed | Percent Major Failures |
|---|---|---|---|
| Mercedes-Benz E 2018 | 140 | 89.3 | 4.29 |
| Audi A8 2012 | 216 | 91.2 | 5.09 |
| Toyota Prius 2018 | 195 | 90.8 | 5.13 |
| Toyota Prius 2016 | 1118 | 87.0 | 6.17 |
| Ford Galaxy 2018 | 128 | 76.6 | 6.25 |
| Volkswagen Sharan 2016 | 204 | 88.2 | 6.37 |
| Ford Galaxy 2017 | 246 | 83.3 | 6.50 |
| Toyota Prius 2014 | 1504 | 78.6 | 6.52 |
| Toyota Prius 2015 | 1578 | 81.1 | 6.53 |
| Bmw X1 2016 | 120 | 86.7 | 6.67 |
| Audi A8 2014 | 116 | 92.2 | 6.90 |
| Lexus Rx450h 2012 | 127 | 85.0 | 7.09 |
| Toyota Prius 2013 | 1847 | 79.7 | 7.09 |
| Mercedes-Benz Ml 2014 | 167 | 88.0 | 7.19 |
| Mercedes-Benz E 2017 | 594 | 88.9 | 7.24 |
| Bmw 430 2015 | 108 | 92.6 | 7.41 |
| Mercedes-Benz S 2014 | 221 | 89.1 | 7.69 |
| Toyota Prius 2017 | 595 | 86.9 | 7.73 |
| Bmw X1 2015 | 102 | 86.3 | 7.84 |
| Seat Alhambra 2017 | 102 | 85.3 | 7.84 |

Please note that virtually all these top-ranked cars are also commonly used as taxis. This may be because they are chosen as taxis because they are reliable or, perversely, they may instead be more likely to pass inspections because they are used as taxis. That is, it is possible that the average taxi company does a bettter job of looking after a car than the average owner (by keeping up on reglar servicing, oil changes, maintenance etc.), therefore making the car more likely to pass an inspection. It's not clear if one or (I suspect) both of these things is ensuring most top-ranked cars are taxis.

## The One-Page Guides

Each page in the dictionary contains a unique make, model, and year of vehicle. I try and stick to basic, robust numbers and comparisons.

A vehicle must be old enough to undergo a mandatory roadworthy test, and be reasonably common to be included in the dictionary. If your vehicle is younger than a 2019 model or is very rare, I can only congratulate you on owning something too new or funky for the dictionary, and apologize that I cannot robustly measure rare or new vehicles. As they are not subjected to a standard roadworthy, heavy goods vehicles are also sadly missing.

Mortality rates are calculated by observing the apparent de-registration rate, including vehicles that are certified destroyed or scrapped, but excluding vehicles that are exported. Each mortality rate is equivalent to the odds of mortality at a given age —- with 1 being certain death and 0 occurring when no deaths were observed —- such that a mortality rate of 0.01 is equal to a 1% mortality risk.

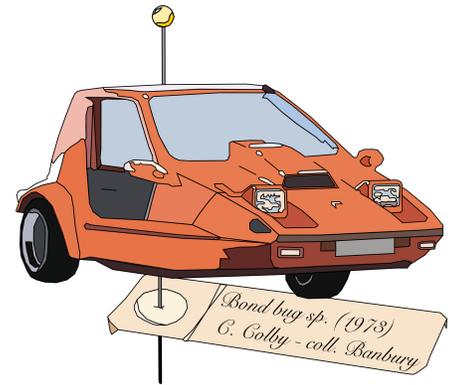

The mechanical reliability rates are calculated by observing the number of *major* failures per MOT inspection. The MOT is the mandatory annual British motor inspection, carried out in accordance with strict legal and reporting guidelines by a registered mechanic, conducted on all cars over 3 years of age and all motorbikes and mopeds over 1 year of age in Great Britain. These are calculated both per year and, as we know the odometer reading at each MOT inspection, per 1000-mile interval.

Mortality rates include vehicles that were crashed, destroyed, or scrapped while still working - that is, vehicles that died while still mechanically sound. This means that mortality rates are strongly affected by owner behaviors, and may not reflect mechanical reliability at all.

The next three pages give you a visual guide to reading the report - if you are not very comfortable with graphs or numbers, this can help. Generally, there are two figures (top and bottom) and two tables (left and right). The first (top) figure shows the mortality rate of each vehicle, relative to vehicles of the same type and age (*e.g.* > 200cc-engine motorbikes built in 2008 will only be compared to other >200cc-engine motorbikes built in 2008 not, say, vans built in 2012). The black and white boxes and crosses in both figures show the survival rates of these comparable 'same type and age' vehicles: they are explained on the next page.

The mortality (top figure) or failure rates (bottom figure) of the vehicle that we want to know about, the one written at the top of the page, is shown by the colored dots and lines.

Mortality rates per year are plotted in the top figure, are also written in the bottom-left table (the one labeled 'mortality rates'). Multiply them by 100 to get an idea of the percentage dying in each year.

However, a better guide to vehicle reliability may be the failure rates of vehicles per mile, shown in the bottom figure, and the right-hand table (the one labeled 'Mechanical Reliability Rates'). These data show how often a vehicle fails an MOT roadworthy inspection, for all vehicles tested at a given mileage. That is, if the 'mileage at test' column is 20,000, and the 'Pct failed' column is 10.2, that means 10.2% of vehicles tested with 20,000 miles on their odometer failed a mechanical roadworthy test. Mileage is binned to the nearest 10,000 miles.

To be included, a make-model has to have at least 1000 registered vehicles tested in any one year. Below this number, the reliability of our estimates degrades because of random sampling effects. This threshold, and our quality control, has the effect of removing rare makes and models. Collectively, all rare makes and models account for only ~3% of all vehicles on the British roads.

**Reading Boxplots**

A boxplot is a way to summarise all other vehicles, of the same age and type, and show how the target vehicle performs. Simply, the higher up the colourful dot is on the chart, the worse the vehicle is relative to other comparable vehicles.

A 'comparable' vehicle is, remember, a vehicle of the same type (*e.g.* a car or a van or a motorbike) made in the same year. If there *are* no comparable vehicles, we just plot the vehicle by itself and write 'too unique' on the chart.

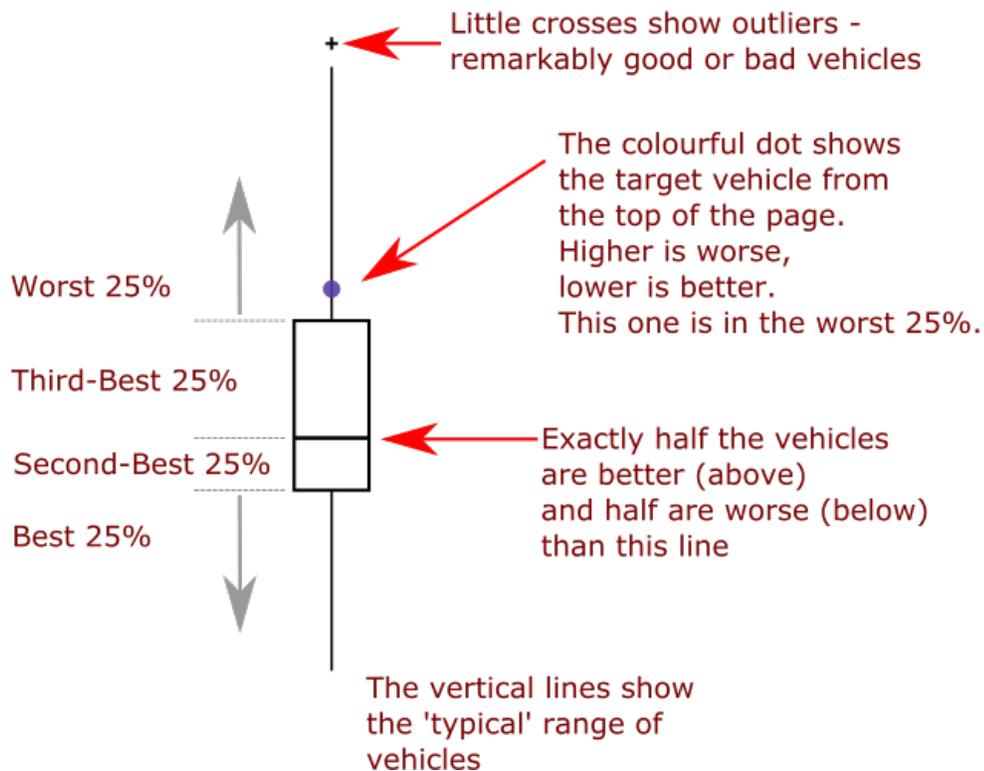

If the colourful dot is above the line, it's worse than at least 50% of comparable vehicles. If it is above the box, it is worse than 75% of vehicles. And if it is so high that it overlaps a tiny little cross, then it is a remarkable outlier for either high mortality rates or high inspection test failure rates.

Another way to picture it is that exactly half of comparable vehicles are inside the box, a quarter are above the box (worse) and a quarter are below the box (better). The boxplots therefore cuts the population into quarters: the worst-performing quarter is above the box, the second-worst is the top part of the box above the line, the second-best is the bottom part of the box below the line, and the best quarter is below the box.

Where your little coloured dot falls will tell you how your vehicle ranks for each age or mileage. I hope that helps.

**Reading the Mortality Rate Charts**

This figure shows mortality rates for the vehicle written at the top of each page.

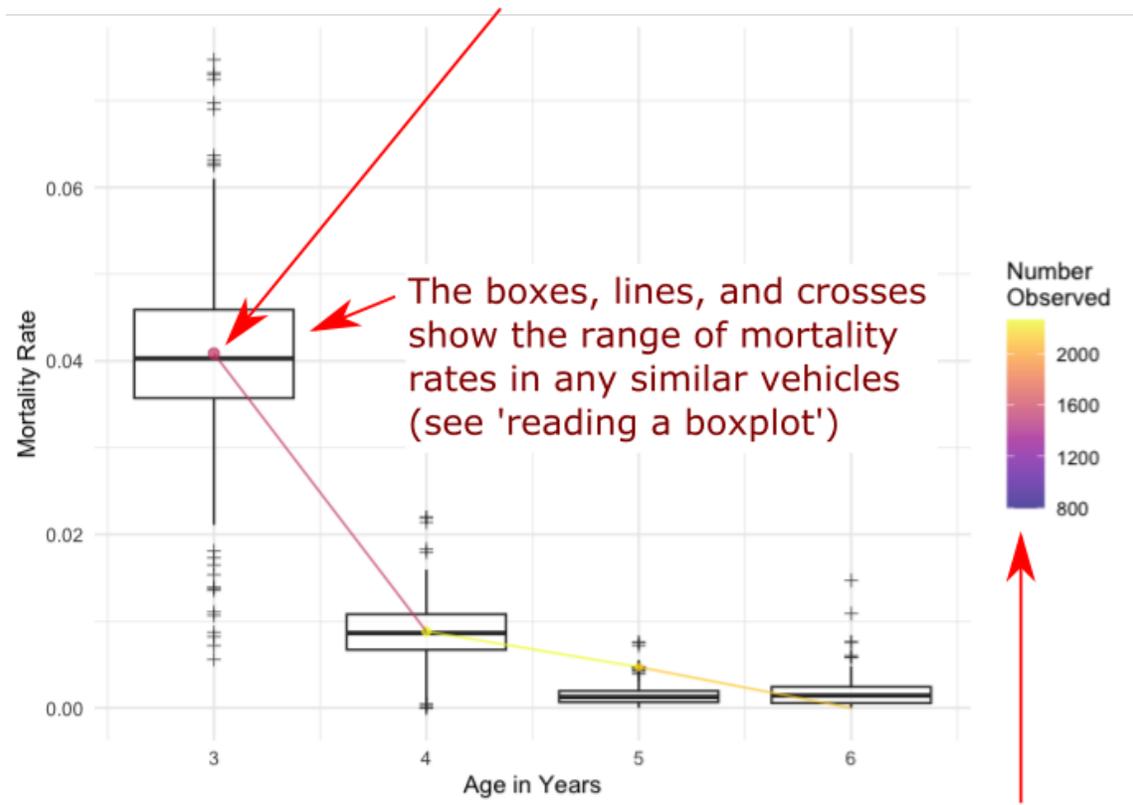

Each colourful dot shows the 'target' vehicle, written at the top of each page, at a given age. Ages are always rounded down.

This dot shows a mortality rate of about 4% (follow the horizontal line across to '0.04') at age 3 years (follow the vertical line down)

The boxes, lines, and crosses show the range of mortality rates in any similar vehicles (see 'reading a boxplot')

The colour of the dots shows how many vehicles were observed each year: more is better

Here, for the first two years (at 3 and 4 years of age), this vehicle is right about average. The colourful dot is sitting on the '50%' line, the median line, meaning half of the population had a higher mortality rate, and half had a lower mortality rate. The mortality rate then drifts around: at age 5 the mortality rate was way above average (in the worst quarter above the box) and for year six the mortality rate was virtually zero (below the box in the best quarter of vehicles). Mortality rates often wander around like this - don't read too much into it.

**Reading the Reliability Charts**

Each colourful dot shows the 'target' vehicle, written at the top of each page, at a given mileage.

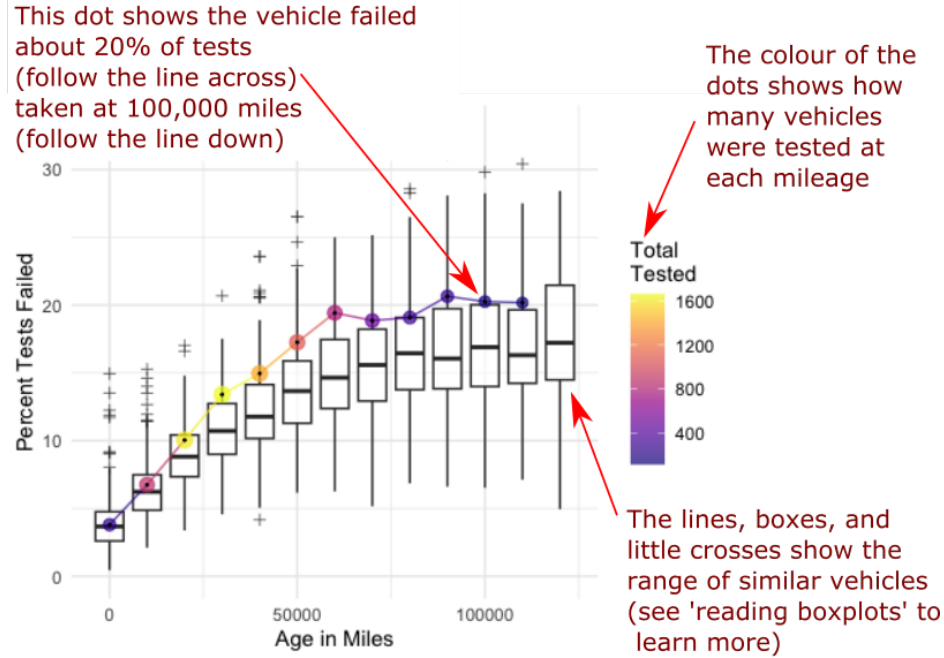

This dot shows the vehicle failed about 20% of tests (follow the line across) taken at 100,000 miles (follow the line down)

The colour of the dots shows how many vehicles were tested at each mileage

The lines, boxes, and little crosses show the range of similar vehicles (see 'reading boxplots' to learn more)

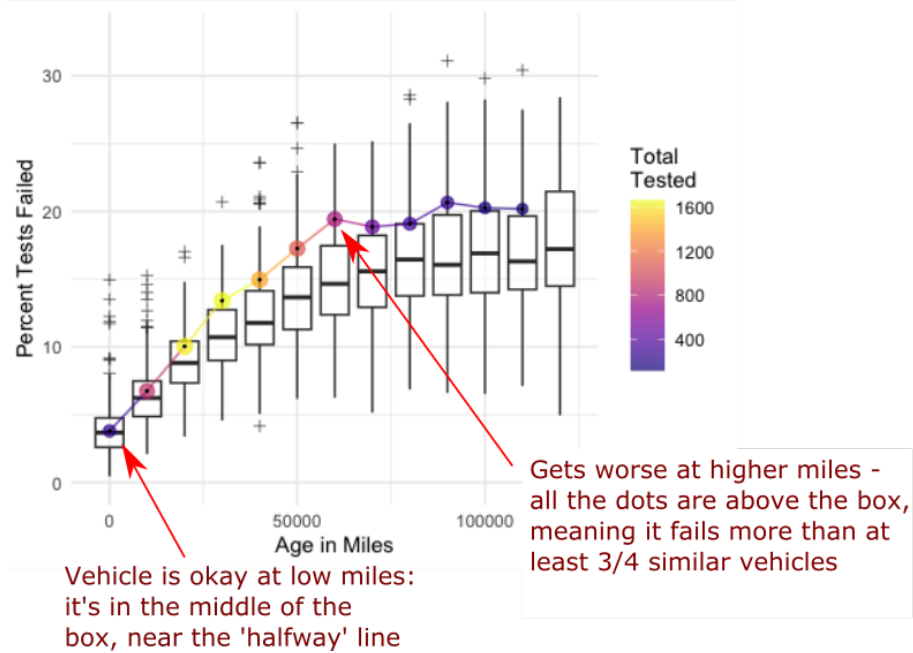

Vehicle is okay at low miles: it's in the middle of the box, near the 'halfway' line

Gets worse at higher miles - all the dots are above the box, meaning it fails more than at least 3/4 similar vehicles

Mileages are pooled to show all inspections/tests taken at 10,000 mile increments, starting at 0-9,999 miles, and always rounded down. Vehicle reliability or test-failure rates may have little connection to vehicle survival, and we only measure 'major' failures - not minor failures that can are fixed without re-testing such as replacing lightbulbs. There are usually far more tests than vehicles.

**A Note on the Mortality Patterns**

You may notice that the last two years' worth of mortality data is often much lower than the previous year, which is itself often lower than the years before that. This is a real effect on the observed mortality rates caused by incomplete reporting - a pattern I was careful to exclude from my scientific paper - and by the far larger effect of the COVID pandemic.

So, why did the mortality rates (per year) drop so dramatically?

Obviously a huge factor was the collapse in the number of miles driven per car, the disappearance of most traffic on the road (and the reduced number of crashes), and the long periods of lockdown in which many vehicles accumulated zero miles.

Added to these substantial protective effects on vehicles were a few less-obvious or less measurable effects that likely contributed to lower vehicle deaths. The lack of vehicle traffic during COVID allowed for unchanged or even accelerated road repair and maintenance works, improving the driving conditions during and long after lockdowns. Traffic rates also remained persistently lower, with people driving less long after lockdowns were lifted. Lower traffic rates also led many people to take up cycling, and many roads had cycle lanes added during pandemic, accelerating a longer-term trend towards lower rates of city driving. Finally, testing regimes changed. Vehicle deaths are measured here at a cut-off date, by looking for vehicles that leave the population and are not exported. However, this method was thrown off by the COVID pandemic, because from March 2020 the UK government allowed people to skip roadworthy tests for six months. This means that death rates are under-estimated, to a completely uniform degree across vehicle classes, during 2020. This effect bled back into 2019 estimates because some death were not visible until late 2020, when the vehicles failed to turn up for their (delayed and legally required) roadworthiness test. That is why mortality rates in the third-last year is often lower.

This is all to say that mortality rates - per year - were lower because people drove less. Again, I excluded all of this from my scintific paper.

So, why include these effects in my charts now?

Well first, because the effect of any testing changes are uniformly applied to all vehicles within a test class (*e.g.* all vans are affected in the same way at the same time) comparisons of the *relative* mortality rates are still fair.

Second, reporting of the rates of exported, scrapped, and crashed vehicles are unchanged.

Third, many vehicles are only observed during this period and their exclusion from mortality statistics would remove a large fraction of relevant data, and the most recent data, from the dictionary.

Fourth, you can simply ignore the last three years' observations if you prefer.

It is also important to note that the reliability estimates are unaffected by all this, as they are based on the percentage of tested vehicles that fail a roadworthy inspection. This is arguably a more useful statistic for many motorists anyway - as it is not affected by crashes, SORN-ing, or scrappage rates - and because it reflects the mechanical reliability of the vehicle in a way that includes less serious mechanical failures than a complete write-off.

I hope that all helps.

**Abarth**

**Abarth 500 2010** At 5 years of age, the mortality rate of a Abarth 500 2010 (manufactured as a Car or Light Van) ranked number 93 out of 206 vehicles of the same age and type (any Car or Light Van constructed in 2010). One is the lowest (or best) and 206 the highest mortality rate. For vehicles reaching 20000 miles, its unreliability score (rate of failing an inspection) ranked 10 out of 201 vehicles of the same age, type, and mileage. One is the highest (or worst) and 201 the lowest rate of failing an inspection.

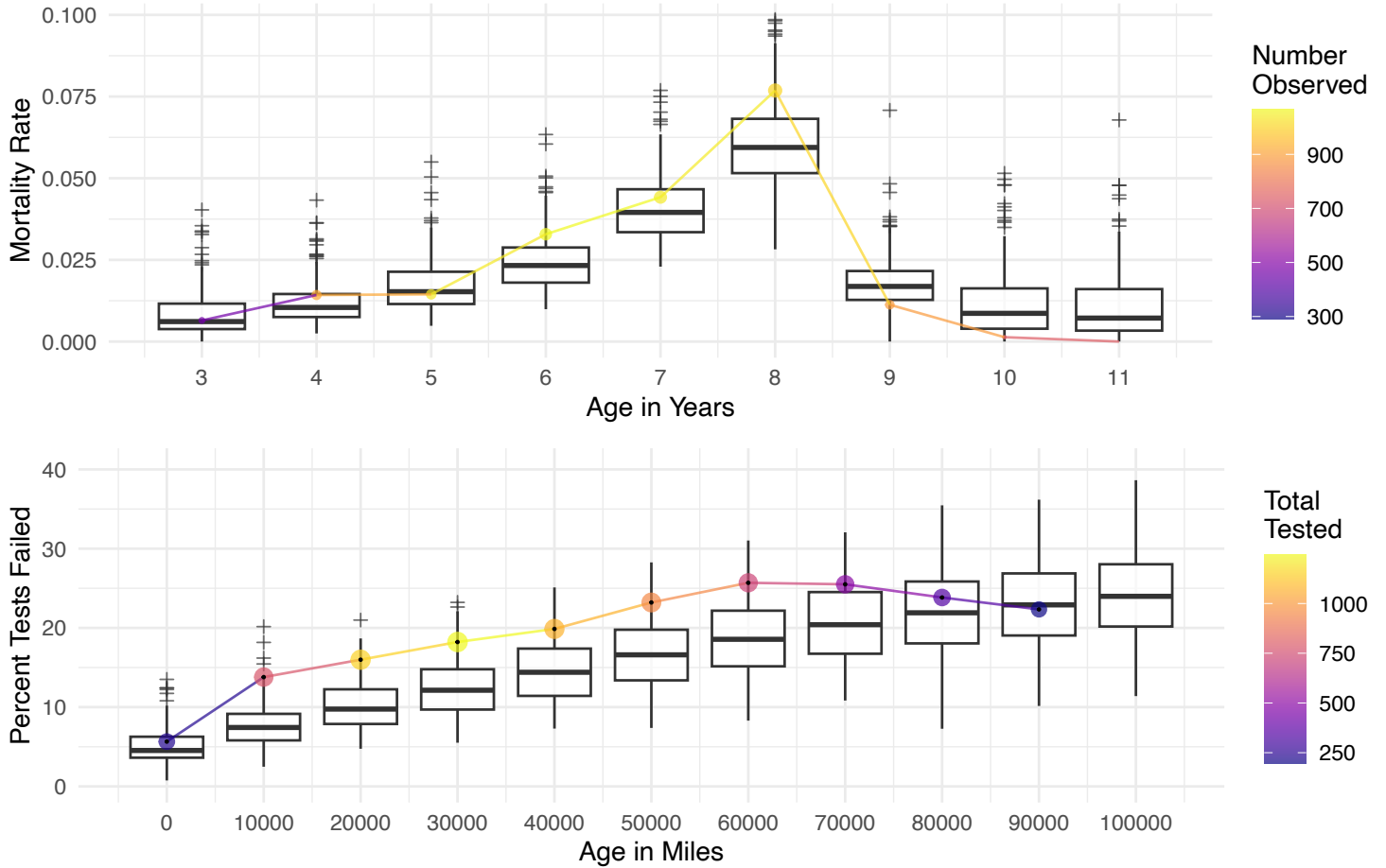

| Mortality rates | | | |
|:---:|:---:|:---:|:---:|
| Age in Years | Observed | Died | Mortality Rate |
| 3 | 464 | 3 | 0.00647 |
| 4 | 912 | 13 | 0.01430 |
| 5 | 1037 | 15 | 0.01450 |
| 6 | 1065 | 35 | 0.03290 |
| 7 | 1041 | 46 | 0.04420 |
| 8 | 1002 | 77 | 0.07680 |
| 9 | 887 | 10 | 0.01130 |
| 10 | 742 | 1 | 0.00135 |
| 11 | 293 | 0 | 0.00000 |

| Mechanical Reliability Rates | | |
|:---:|:---:|:---:|
| Mileage at test | N tested | Pct failed |
| 0 | 230 | 5.65 |
| 10000 | 783 | 13.80 |
| 20000 | 1151 | 16.00 |
| 30000 | 1246 | 18.20 |
| 40000 | 1072 | 19.90 |
| 50000 | 948 | 23.20 |
| 60000 | 724 | 25.70 |
| 70000 | 498 | 25.50 |
| 80000 | 344 | 23.80 |
| 90000 | 197 | 22.30 |

**Abarth 500 2011**

At 5 years of age, the mortality rate of a Abarth 500 2011 (manufactured as a Car or Light Van) ranked number 78 out of 211 vehicles of the same age and type (any Car or Light Van constructed in 2011). One is the lowest (or best) and 211 the highest mortality rate. For vehicles reaching 20000 miles, its unreliability score (rate of failing an inspection) ranked 7 out of 205 vehicles of the same age, type, and mileage. One is the highest (or worst) and 205 the lowest rate of failing an inspection.

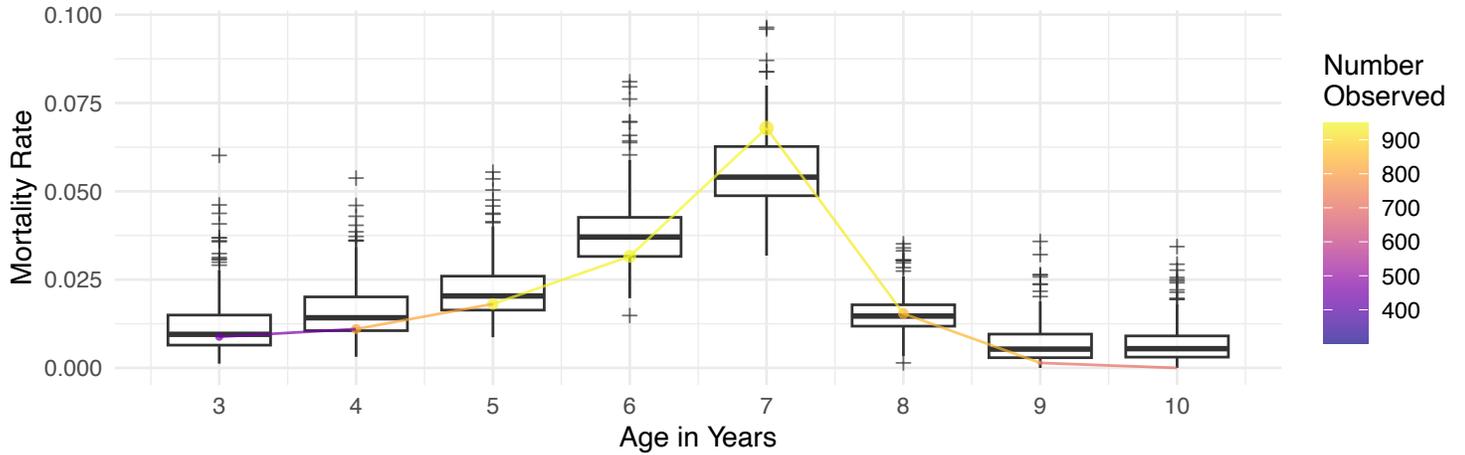

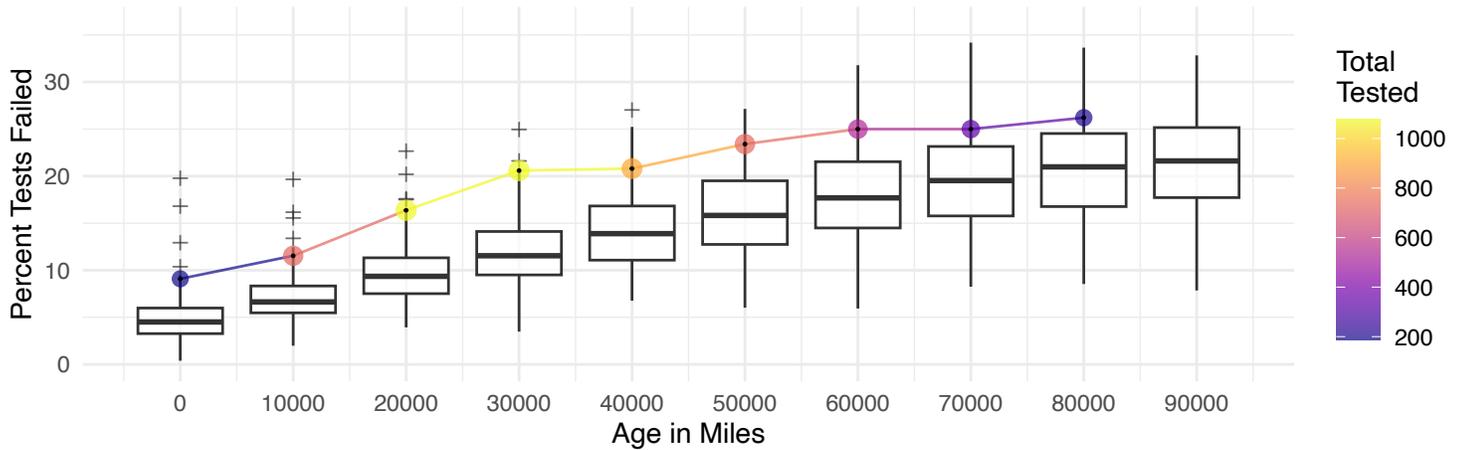

Mortality rates

| Age in Years | Observed | Died | Mortality Rate |
|---|---|---|---|
| 3 | 454 | 4 | 0.00881 |
| 4 | 817 | 9 | 0.01100 |
| 5 | 939 | 17 | 0.01810 |
| 6 | 950 | 30 | 0.03160 |
| 7 | 928 | 63 | 0.06790 |
| 8 | 845 | 13 | 0.01540 |
| 9 | 704 | 1 | 0.00142 |
| 10 | 301 | 0 | 0.00000 |

Mechanical Reliability Rates

| Mileage at test | N tested | Pct failed |
|---|---|---|
| 0 | 187 | 9.09 |
| 10000 | 728 | 11.50 |
| 20000 | 1075 | 16.40 |
| 30000 | 1078 | 20.60 |
| 40000 | 899 | 20.80 |
| 50000 | 739 | 23.40 |
| 60000 | 532 | 25.00 |
| 70000 | 332 | 25.00 |
| 80000 | 206 | 26.20 |

# Abarth 595 2015

At 5 years of age, the mortality rate of a Abarth 595 2015 (manufactured as a Car or Light Van) ranked number 168 out of 247 vehicles of the same age and type (any Car or Light Van constructed in 2015). One is the lowest (or best) and 247 the highest mortality rate. For vehicles reaching 20000 miles, its unreliability score (rate of failing an inspection) ranked 15 out of 241 vehicles of the same age, type, and mileage. One is the highest (or worst) and 241 the lowest rate of failing an inspection.

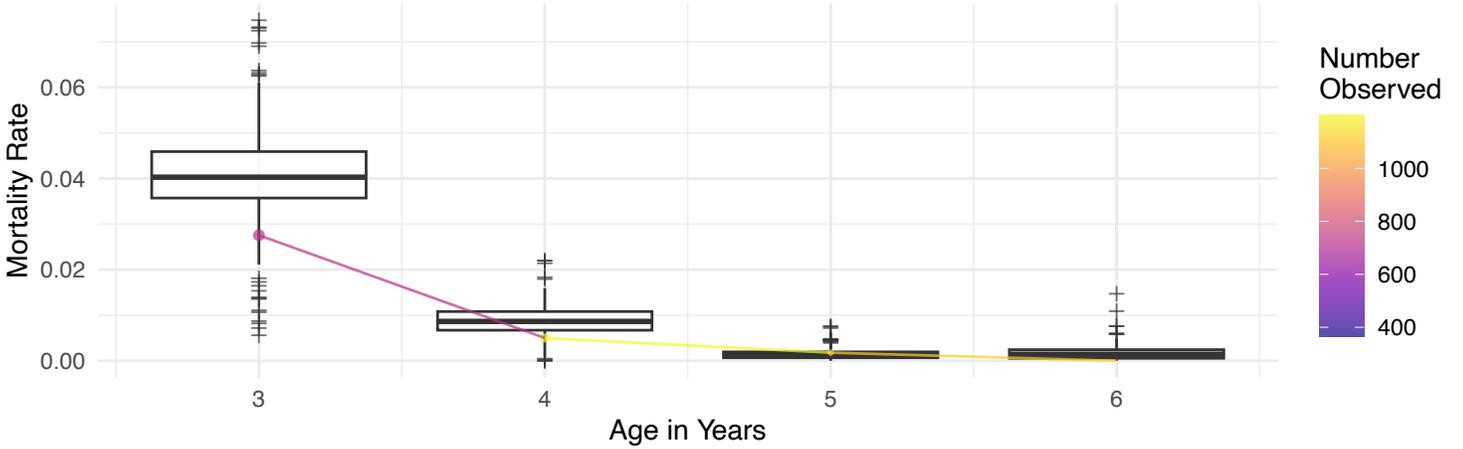

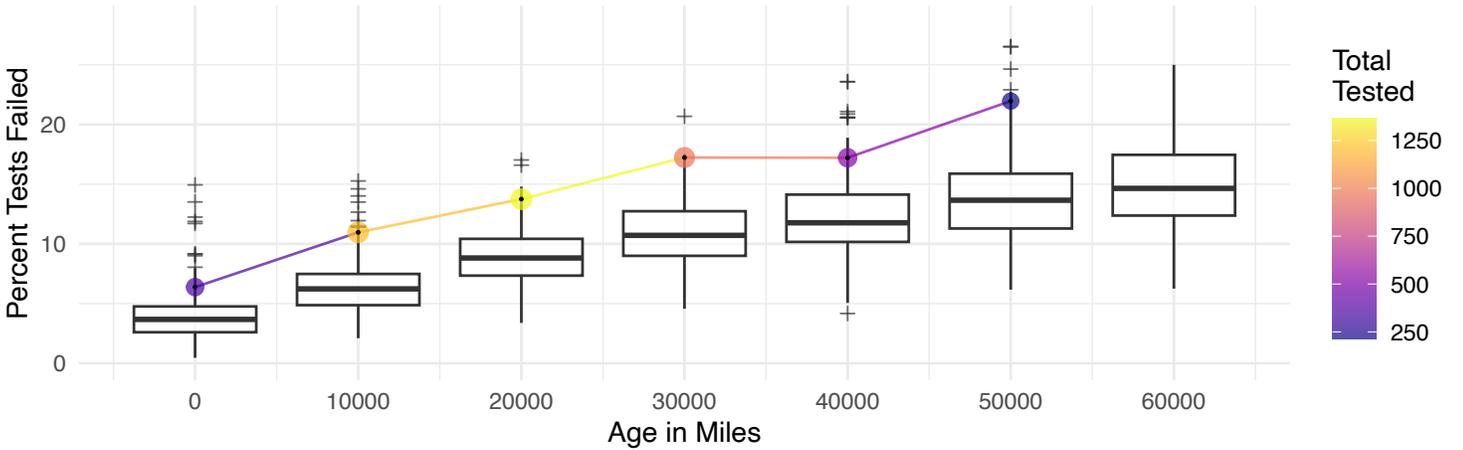

Mortality rates

| Age in Years | Observed | Died | Mortality Rate |
|---|---|---|---|
| 3 | 726 | 20 | 0.02750 |
| 4 | 1199 | 6 | 0.00500 |
| 5 | 1116 | 2 | 0.00179 |
| 6 | 366 | 0 | 0.00000 |

Mechanical Reliability Rates

| Mileage at test | N tested | Pct failed |
|---|---|---|
| 0 | 345 | 6.38 |
| 10000 | 1204 | 11.00 |
| 20000 | 1367 | 13.80 |
| 30000 | 975 | 17.20 |
| 40000 | 517 | 17.20 |
| 50000 | 214 | 22.00 |

**Abarth 595 2016**

At 5 years of age, the mortality rate of a Abarth 595 2016 (manufactured as a Car or Light Van) ranked number 154 out of 252 vehicles of the same age and type (any Car or Light Van constructed in 2016). One is the lowest (or best) and 252 the highest mortality rate. For vehicles reaching 20000 miles, its unreliability score (rate of failing an inspection) ranked 21 out of 246 vehicles of the same age, type, and mileage. One is the highest (or worst) and 246 the lowest rate of failing an inspection.

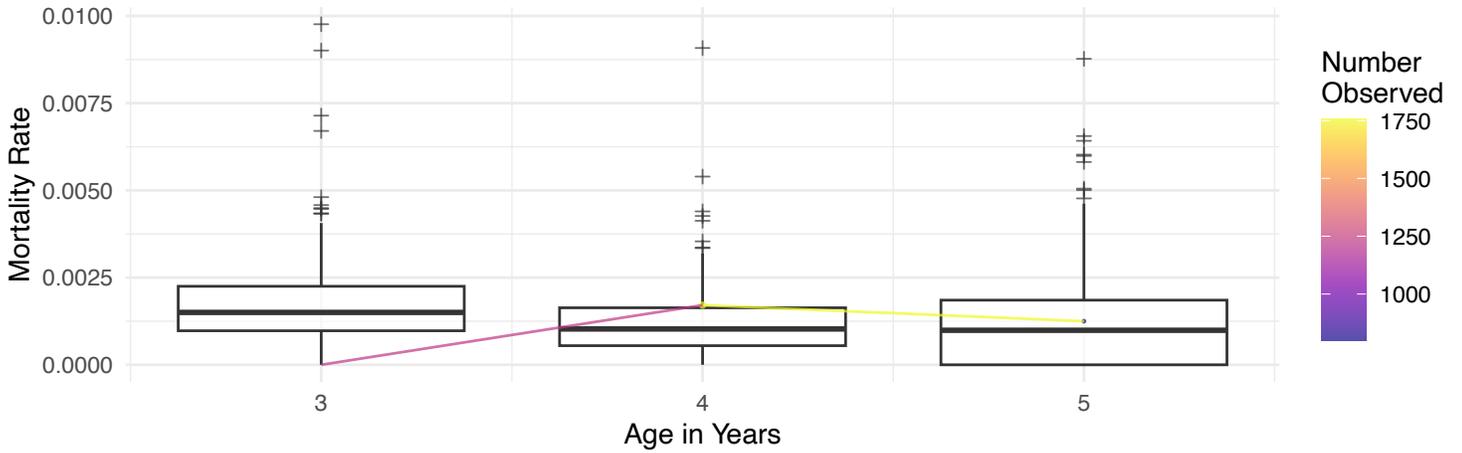

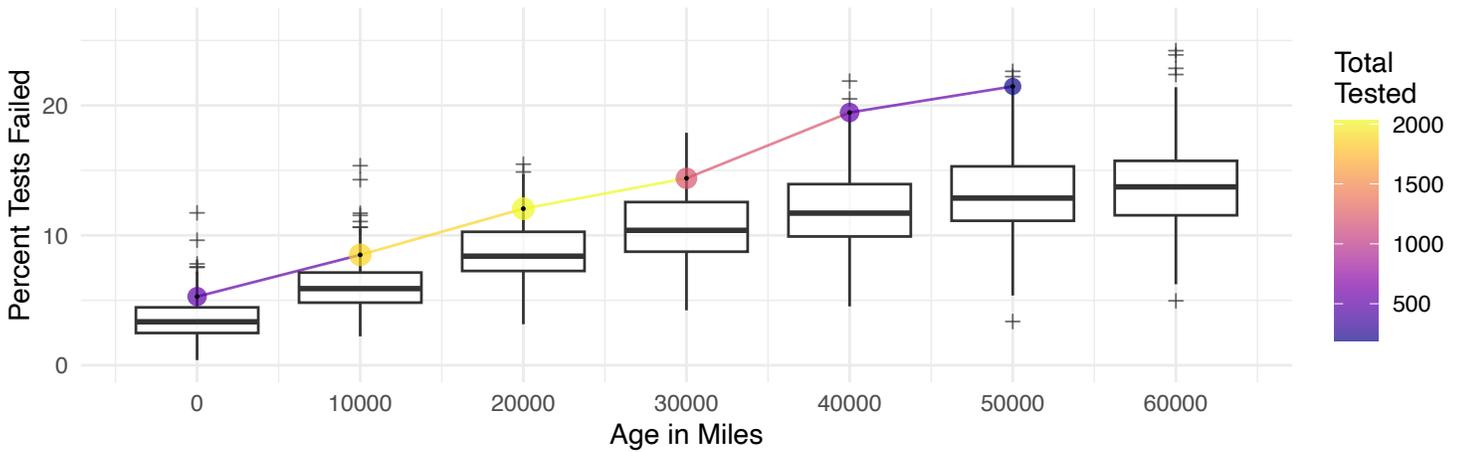

<table>
<tr><th colspan="4">Mortality rates</th></tr>
<tr><th>Age in Years</th><th>Observed</th><th>Died</th><th>Mortality Rate</th></tr>
<tr><td>3</td><td>1224</td><td>0</td><td>0.00000</td></tr>
<tr><td>4</td><td>1754</td><td>3</td><td>0.00171</td></tr>
<tr><td>5</td><td>800</td><td>1</td><td>0.00125</td></tr>
</table>

<table>
<tr><th colspan="3">Mechanical Reliability Rates</th></tr>
<tr><th>Mileage at test</th><th>N tested</th><th>Pct failed</th></tr>
<tr><td>0</td><td>492</td><td>5.28</td></tr>
<tr><td>10000</td><td>1883</td><td>8.50</td></tr>
<tr><td>20000</td><td>2033</td><td>12.10</td></tr>
<tr><td>30000</td><td>1209</td><td>14.40</td></tr>
<tr><td>40000</td><td>514</td><td>19.50</td></tr>
<tr><td>50000</td><td>191</td><td>21.50</td></tr>
</table>

**Abarth 595 2017**

At 3 years of age, the mortality rate of a Abarth 595 2017 (manufactured as a Car or Light Van) ranked number 196 out of 247 vehicles of the same age and type (any Car or Light Van constructed in 2017). One is the lowest (or best) and 247 the highest mortality rate. For vehicles reaching 20000 miles, its unreliability score (rate of failing an inspection) ranked 21 out of 240 vehicles of the same age, type, and mileage. One is the highest (or worst) and 240 the lowest rate of failing an inspection.

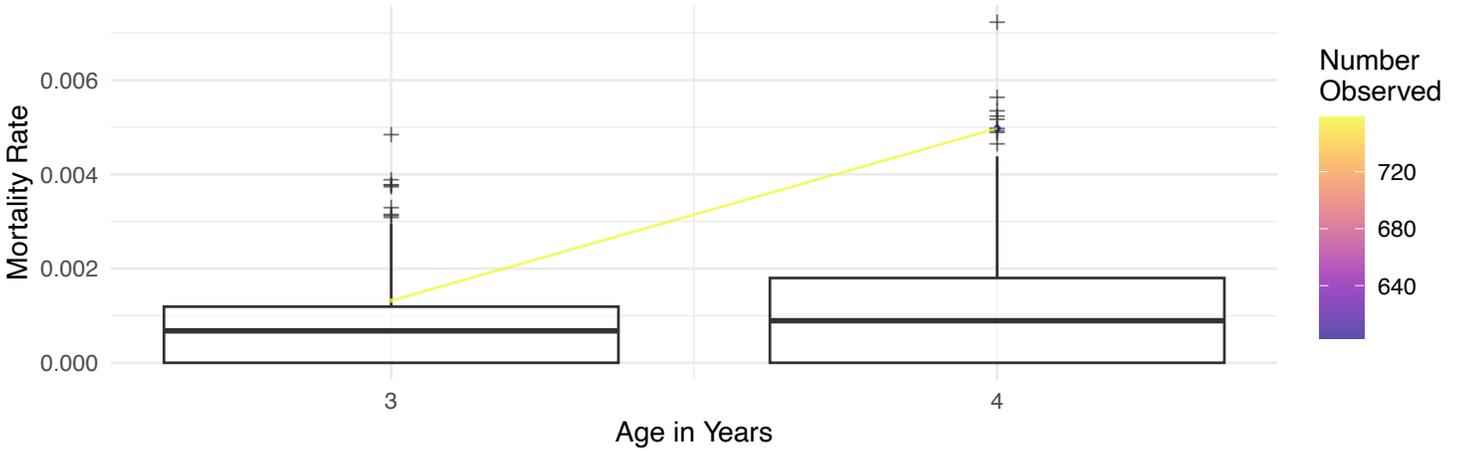

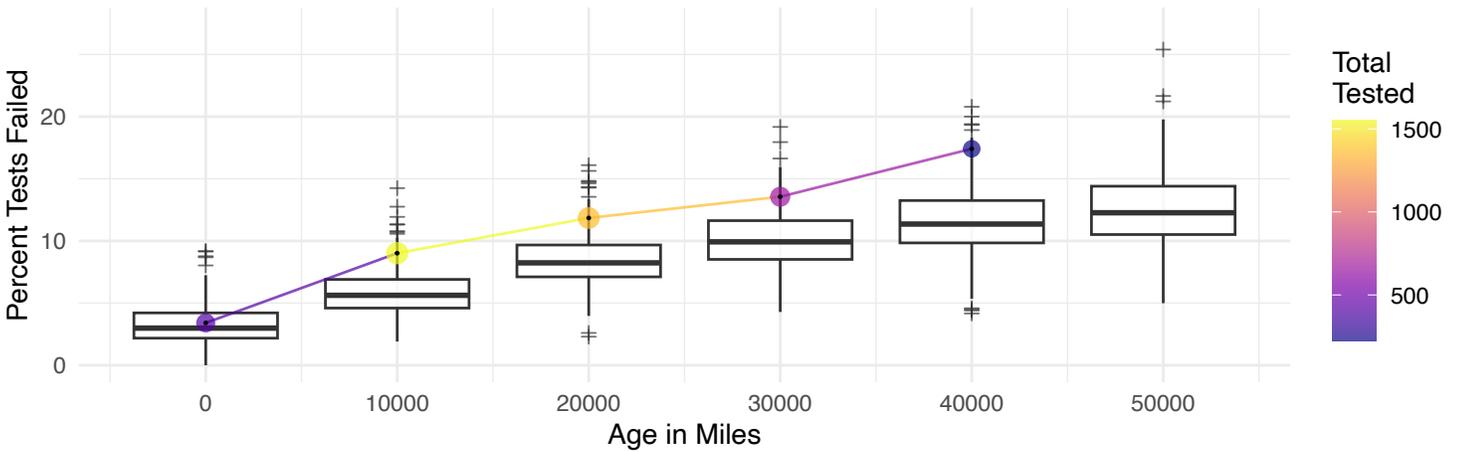

Mortality rates

| Age in Years | Observed | Died | Mortality Rate |
|---|---|---|---|
| 3 | 758 | 1 | 0.00132 |
| 4 | 603 | 3 | 0.00498 |

Mechanical Reliability Rates

| Mileage at test | N tested | Pct failed |
|---|---|---|
| 0 | 411 | 3.41 |
| 10000 | 1553 | 9.01 |
| 20000 | 1359 | 11.80 |
| 30000 | 664 | 13.60 |
| 40000 | 224 | 17.40 |

**Abarth 595 2018**

At 3 years of age, the mortality rate of a Abarth 595 2018 (manufactured as a Car or Light Van) ranked number 1 out of 222 vehicles of the same age and type (any Car or Light Van constructed in 2018). One is the lowest (or best) and 222 the highest mortality rate. For vehicles reaching 20000 miles, its unreliability score (rate of failing an inspection) ranked 16 out of 215 vehicles of the same age, type, and mileage. One is the highest (or worst) and 215 the lowest rate of failing an inspection.

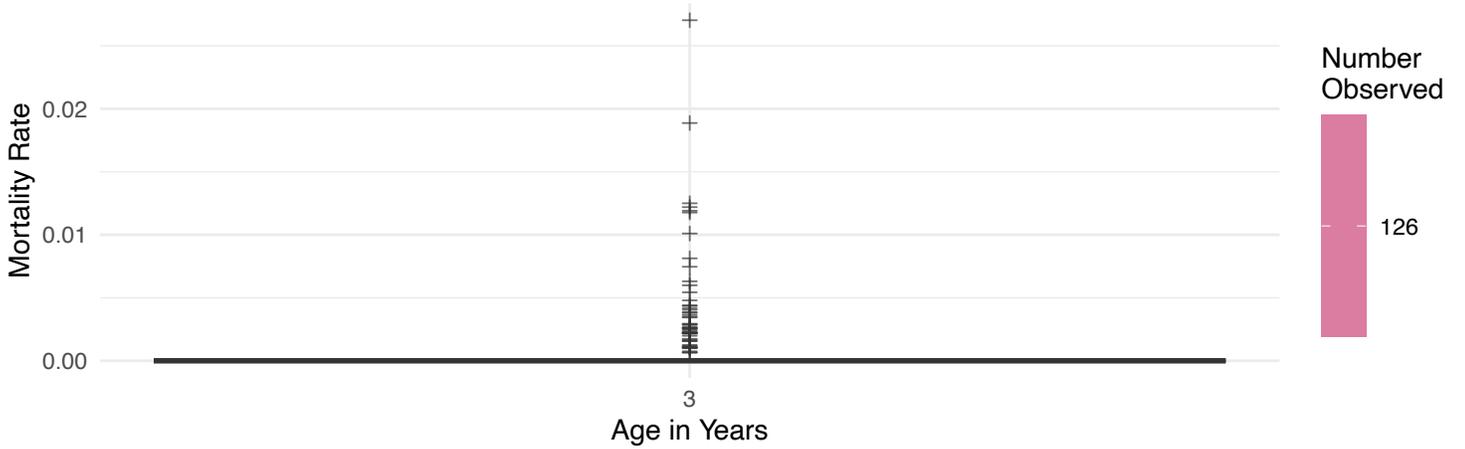

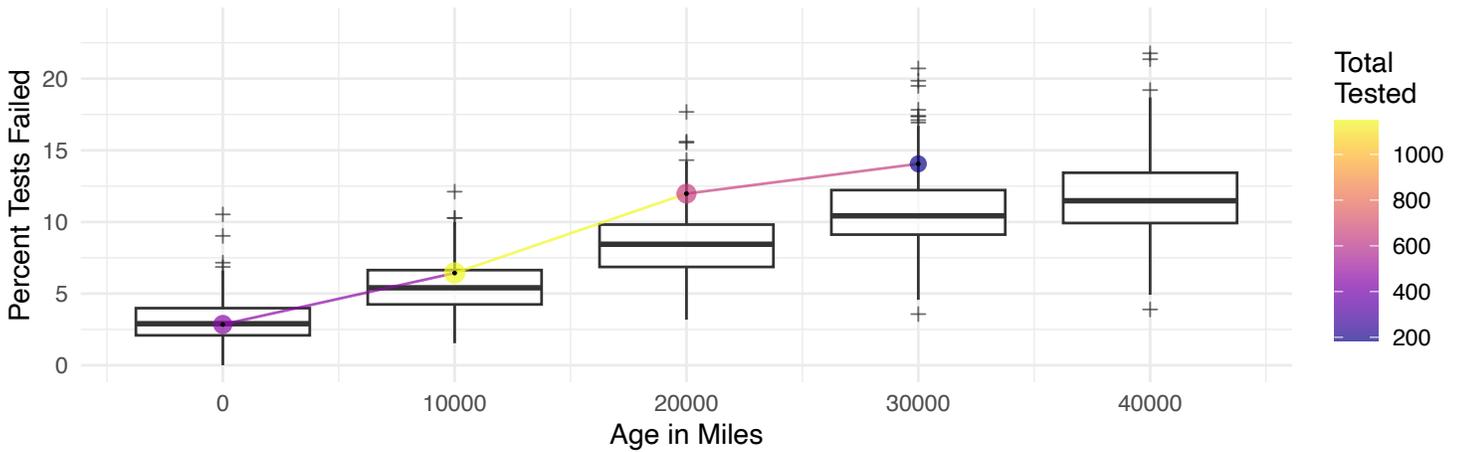

Mortality rates

| Age in Years | Observed | Died | Mortality Rate |
|---|---|---|---|
| 3 | 126 | 0 | 0 |

Mechanical Reliability Rates

| Mileage at test | N tested | Pct failed |
|---|---|---|
| 0 | 457 | 2.84 |
| 10000 | 1150 | 6.43 |
| 20000 | 643 | 12.00 |
| 30000 | 185 | 14.10 |

## Alfa Romeo

**Alfa Romeo 146 1996** At 10 years of age, the mortality rate of a Alfa Romeo 146 1996 (manufactured as a Car or Light Van) ranked number 156 out of 162 vehicles of the same age and type (any Car or Light Van constructed in 1996). One is the lowest (or best) and 162 the highest mortality rate. For vehicles reaching 120000 miles, its unreliability score (rate of failing an inspection) ranked 12 out of 147 vehicles of the same age, type, and mileage. One is the highest (or worst) and 147 the lowest rate of failing an inspection.

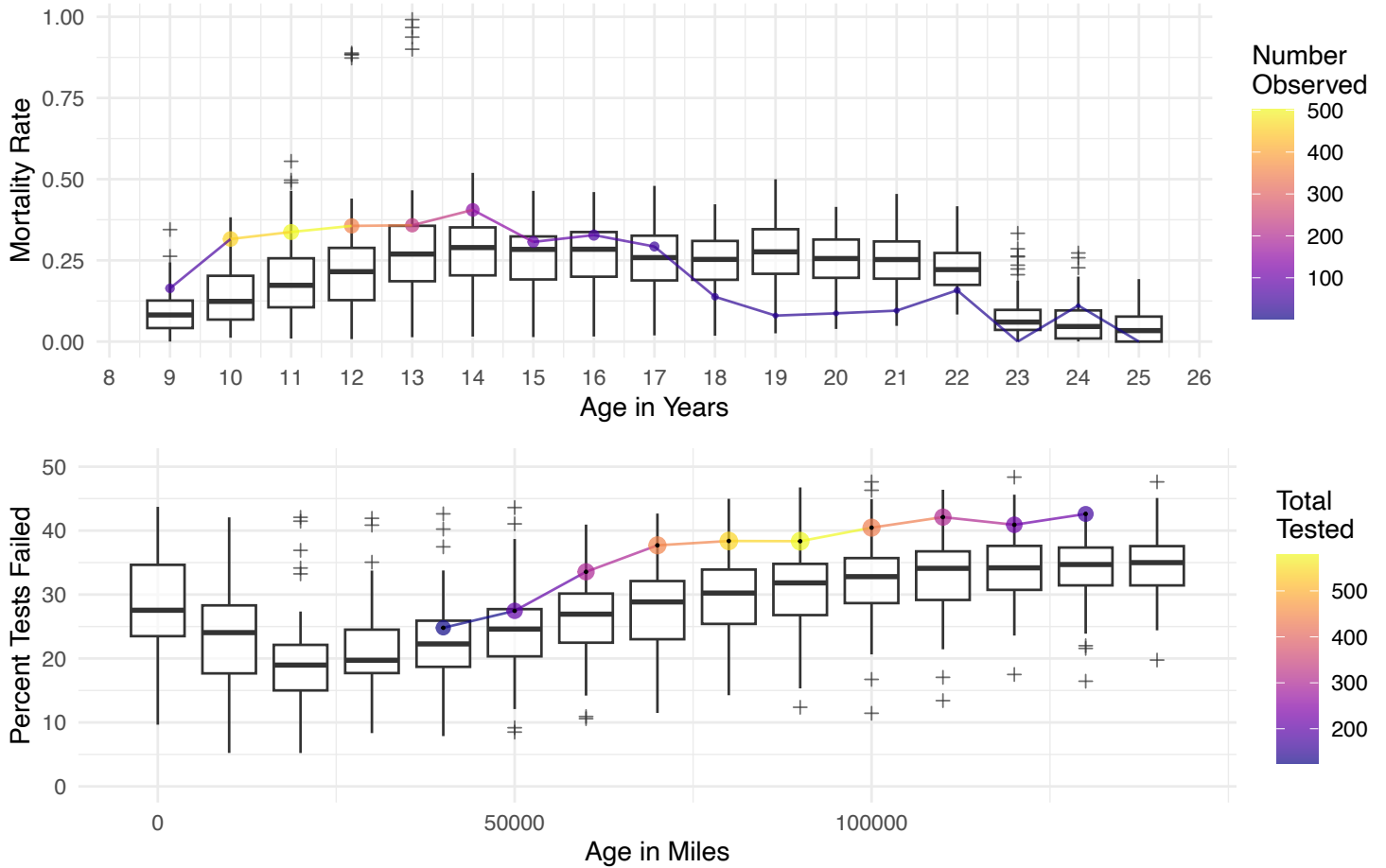

Mortality rates

| Age in Years | Observed | Died | Mortality Rate |
|---|---|---|---|
| 9 | 61 | 10 | 0.1640 |
| 10 | 456 | 144 | 0.3160 |
| 11 | 501 | 169 | 0.3370 |
| 12 | 351 | 125 | 0.3560 |
| 13 | 229 | 82 | 0.3580 |
| 14 | 148 | 60 | 0.4050 |
| 15 | 88 | 27 | 0.3070 |
| 16 | 61 | 20 | 0.3280 |
| 17 | 41 | 12 | 0.2930 |
| 18 | 29 | 4 | 0.1380 |
| 19 | 25 | 2 | 0.0800 |
| 20 | 23 | 2 | 0.0870 |
| 21 | 21 | 2 | 0.0952 |
| 22 | 19 | 3 | 0.1580 |
| 23 | 14 | 0 | 0.0000 |

Mechanical Reliability Rates

| Mileage at test | N tested | Pct failed |
|---|---|---|
| 40000 | 125 | 24.8 |
| 50000 | 204 | 27.5 |
| 60000 | 298 | 33.6 |
| 70000 | 443 | 37.7 |
| 80000 | 542 | 38.4 |
| 90000 | 579 | 38.3 |
| 100000 | 435 | 40.5 |
| 110000 | 304 | 42.1 |
| 120000 | 220 | 40.9 |
| 130000 | 169 | 42.6 |

## Alfa Romeo 146 1997

At 10 years of age, the mortality rate of a Alfa Romeo 146 1997 (manufactured as a Car or Light Van) ranked number 165 out of 187 vehicles of the same age and type (any Car or Light Van constructed in 1997). One is the lowest (or best) and 187 the highest mortality rate. For vehicles reaching 120000 miles, its unreliability score (rate of failing an inspection) ranked 4 out of 167 vehicles of the same age, type, and mileage. One is the highest (or worst) and 167 the lowest rate of failing an inspection.

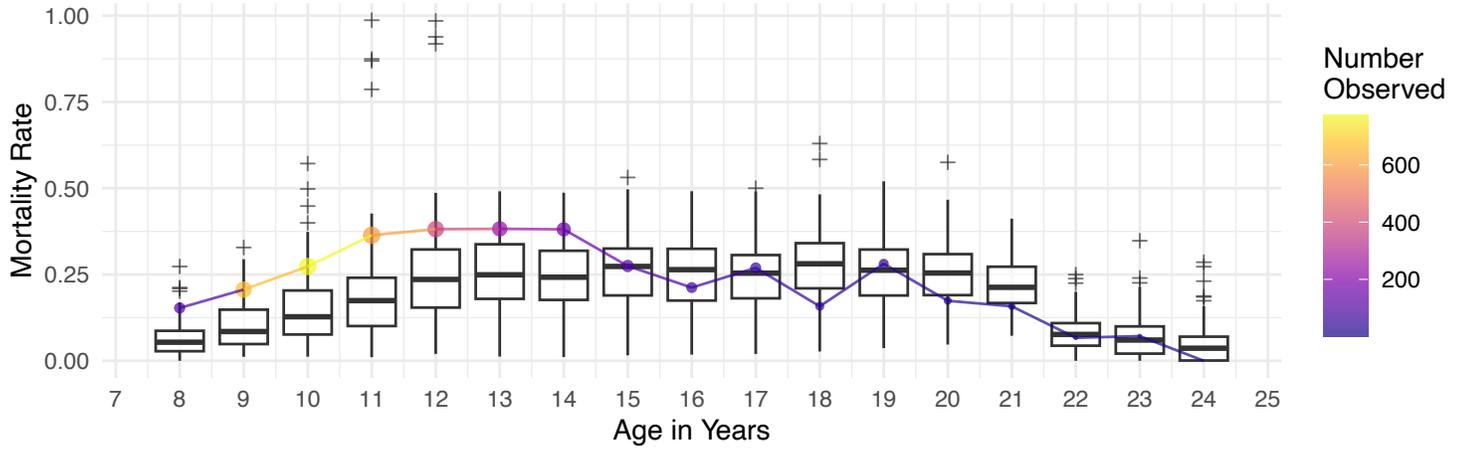

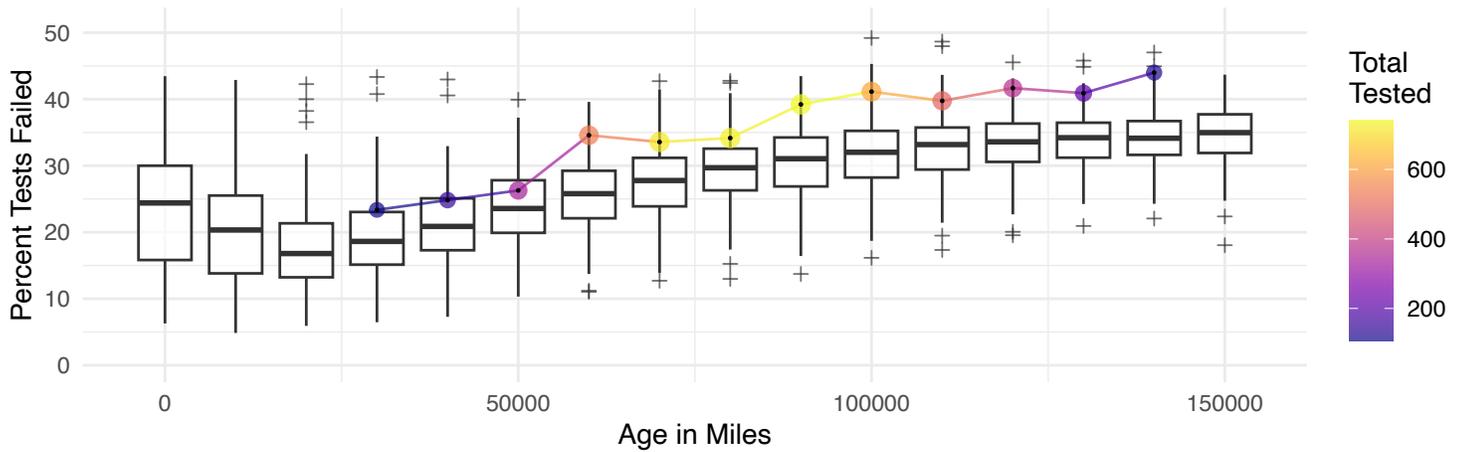

### Mortality rates

| Age in Years | Observed | Died | Mortality Rate |
|:---:|:---:|:---:|:---:|
| 8 | 98 | 15 | 0.1530 |
| 9 | 659 | 136 | 0.2060 |
| 10 | 772 | 211 | 0.2730 |
| 11 | 591 | 215 | 0.3640 |
| 12 | 380 | 145 | 0.3820 |
| 13 | 238 | 91 | 0.3820 |
| 14 | 147 | 56 | 0.3810 |
| 15 | 91 | 25 | 0.2750 |
| 16 | 66 | 14 | 0.2120 |
| 17 | 52 | 14 | 0.2690 |
| 18 | 38 | 6 | 0.1580 |
| 19 | 32 | 9 | 0.2810 |
| 20 | 23 | 4 | 0.1740 |
| 21 | 19 | 3 | 0.1580 |
| 22 | 15 | 1 | 0.0667 |
| 23 | 14 | 1 | 0.0714 |

### Mechanical Reliability Rates

| Mileage at test | N tested | Pct failed |
|:---:|:---:|:---:|
| 30000 | 107 | 23.4 |
| 40000 | 145 | 24.8 |
| 50000 | 331 | 26.3 |
| 60000 | 535 | 34.6 |
| 70000 | 718 | 33.6 |
| 80000 | 729 | 34.2 |
| 90000 | 742 | 39.2 |
| 100000 | 615 | 41.1 |
| 110000 | 488 | 39.8 |
| 120000 | 372 | 41.7 |
| 130000 | 198 | 40.9 |
| 140000 | 125 | 44.0 |

**Alfa Romeo 146 1998**

At 10 years of age, the mortality rate of a Alfa Romeo 146 1998 (manufactured as a Car or Light Van) ranked number 187 out of 196 vehicles of the same age and type (any Car or Light Van constructed in 1998). One is the lowest (or best) and 196 the highest mortality rate. For vehicles reaching 120000 miles, its unreliability score (rate of failing an inspection) ranked 4 out of 172 vehicles of the same age, type, and mileage. One is the highest (or worst) and 172 the lowest rate of failing an inspection.

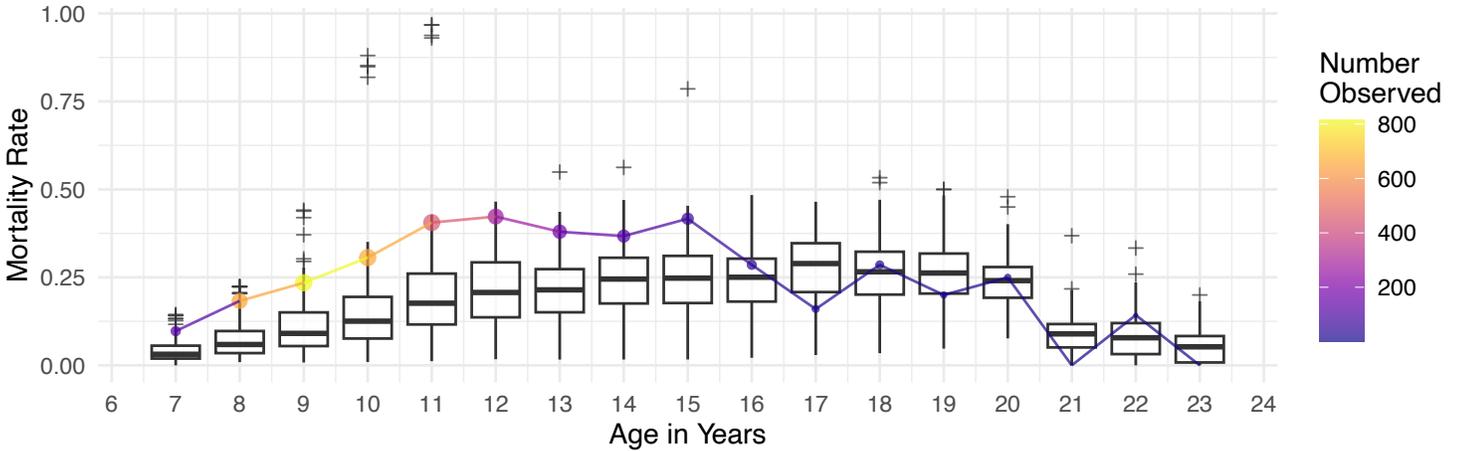

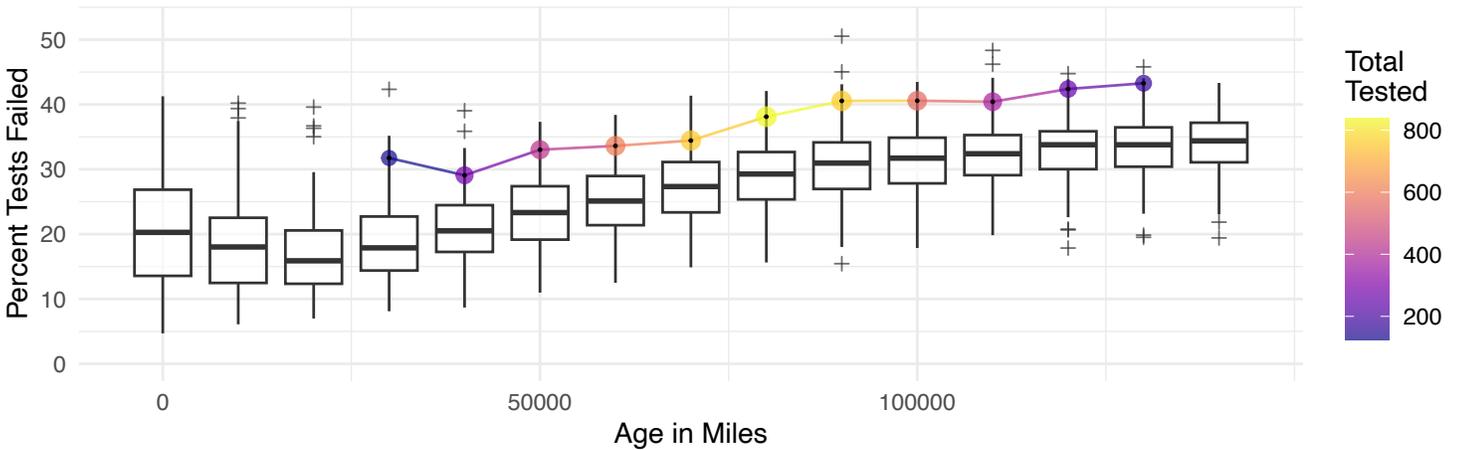

Mortality rates

| Age in Years | Observed | Died | Mortality Rate |
|---|---|---|---|
| 7 | 113 | 11 | 0.0973 |
| 8 | 644 | 118 | 0.1830 |
| 9 | 814 | 191 | 0.2350 |
| 10 | 653 | 200 | 0.3060 |
| 11 | 456 | 185 | 0.4060 |
| 12 | 272 | 115 | 0.4230 |
| 13 | 158 | 60 | 0.3800 |
| 14 | 98 | 36 | 0.3670 |
| 15 | 60 | 25 | 0.4170 |
| 16 | 35 | 10 | 0.2860 |
| 17 | 25 | 4 | 0.1600 |
| 18 | 21 | 6 | 0.2860 |
| 19 | 15 | 3 | 0.2000 |
| 20 | 12 | 3 | 0.2500 |

Mechanical Reliability Rates

| Mileage at test | N tested | Pct failed |
|---|---|---|
| 30000 | 126 | 31.7 |
| 40000 | 282 | 29.1 |
| 50000 | 427 | 33.0 |
| 60000 | 601 | 33.6 |
| 70000 | 755 | 34.4 |
| 80000 | 837 | 38.1 |
| 90000 | 762 | 40.6 |
| 100000 | 557 | 40.6 |
| 110000 | 381 | 40.4 |
| 120000 | 243 | 42.4 |
| 130000 | 171 | 43.3 |

## Alfa Romeo 147 2001

At 5 years of age, the mortality rate of a Alfa Romeo 147 2001 (manufactured as a Car or Light Van) ranked number 170 out of 205 vehicles of the same age and type (any Car or Light Van constructed in 2001). One is the lowest (or best) and 205 the highest mortality rate. For vehicles reaching 120000 miles, its unreliability score (rate of failing an inspection) ranked 33 out of 194 vehicles of the same age, type, and mileage. One is the highest (or worst) and 194 the lowest rate of failing an inspection.

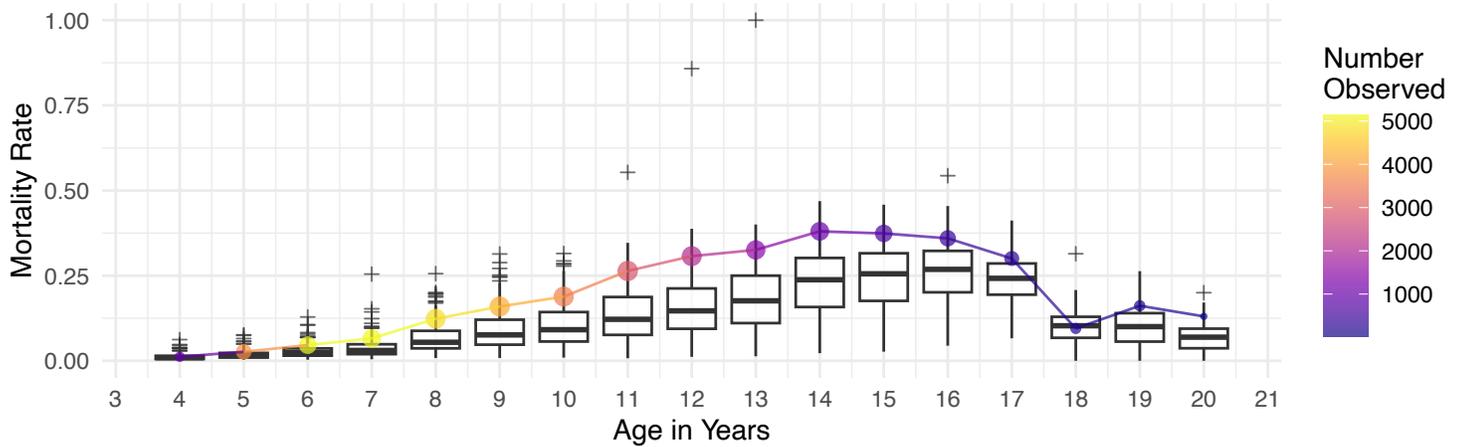

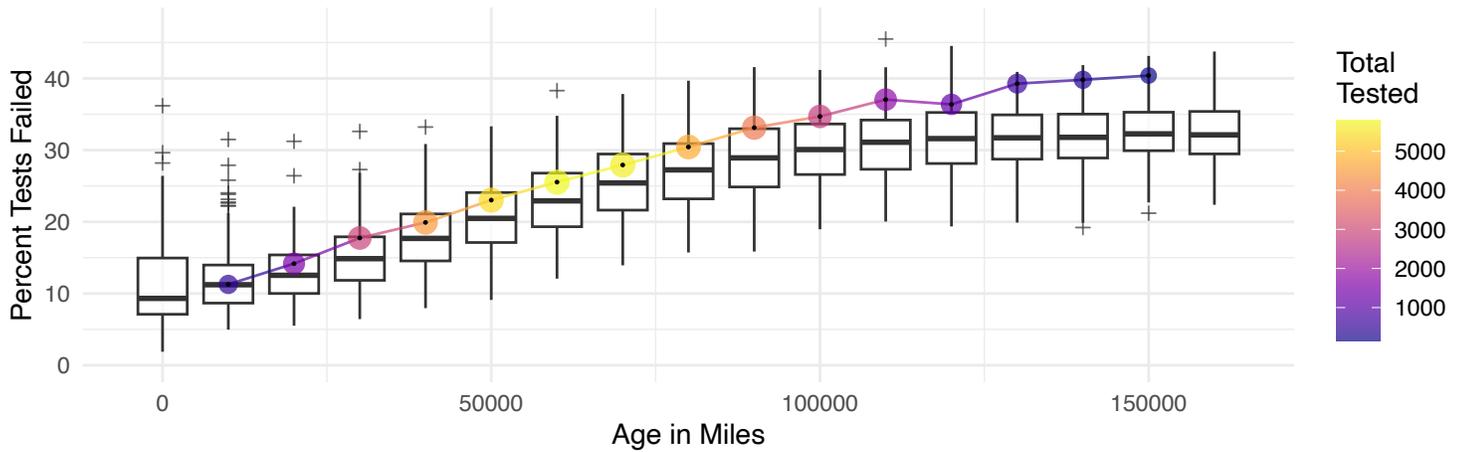

| Mortality rates | | | |
|---|---|---|---|
| Age in Years | Observed | Died | Mortality Rate |
| 4 | 930 | 11 | 0.0118 |
| 5 | 3788 | 100 | 0.0264 |
| 6 | 5122 | 236 | 0.0461 |
| 7 | 5126 | 337 | 0.0657 |
| 8 | 4789 | 591 | 0.1230 |
| 9 | 4187 | 667 | 0.1590 |
| 10 | 3517 | 663 | 0.1890 |
| 11 | 2847 | 751 | 0.2640 |
| 12 | 2099 | 645 | 0.3070 |
| 13 | 1453 | 473 | 0.3260 |
| 14 | 981 | 373 | 0.3800 |
| 15 | 607 | 227 | 0.3740 |
| 16 | 381 | 137 | 0.3600 |
| 17 | 243 | 73 | 0.3000 |
| 18 | 158 | 15 | 0.0949 |
| 19 | 99 | 16 | 0.1620 |
| 20 | 23 | 3 | 0.1300 |

| Mechanical Reliability Rates | | |
|---|---|---|
| Mileage at test | N tested | Pct failed |
| 10000 | 479 | 11.3 |
| 20000 | 1503 | 14.2 |
| 30000 | 2930 | 17.7 |
| 40000 | 4469 | 19.9 |
| 50000 | 5406 | 23.0 |
| 60000 | 5795 | 25.5 |
| 70000 | 5676 | 27.9 |
| 80000 | 4731 | 30.4 |
| 90000 | 3939 | 33.1 |
| 100000 | 2810 | 34.7 |
| 110000 | 1822 | 37.0 |
| 120000 | 1138 | 36.4 |
| 130000 | 629 | 39.3 |
| 140000 | 309 | 39.8 |
| 150000 | 151 | 40.4 |

## Alfa Romeo 147 2002

At 5 years of age, the mortality rate of a Alfa Romeo 147 2002 (manufactured as a Car or Light Van) ranked number 163 out of 202 vehicles of the same age and type (any Car or Light Van constructed in 2002). One is the lowest (or best) and 202 the highest mortality rate. For vehicles reaching 120000 miles, its unreliability score (rate of failing an inspection) ranked 29 out of 193 vehicles of the same age, type, and mileage. One is the highest (or worst) and 193 the lowest rate of failing an inspection.

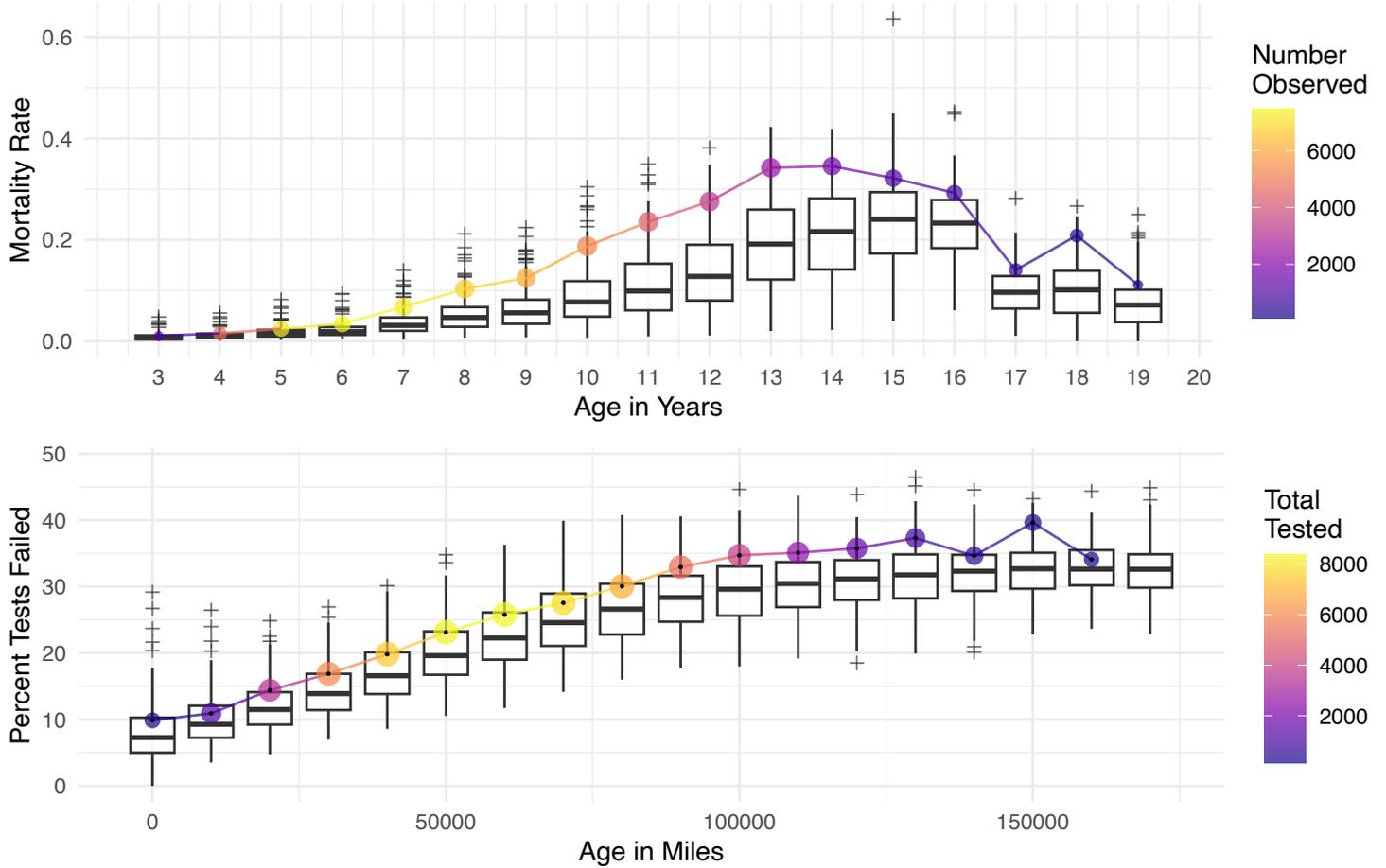

| Mortality rates | | | |
|---|---|---|---|
| Age in Years | Observed | Died | Mortality Rate |
| 3 | 839 | 9 | 0.0107 |
| 4 | 4716 | 72 | 0.0153 |
| 5 | 7181 | 177 | 0.0246 |
| 6 | 7456 | 255 | 0.0342 |
| 7 | 7214 | 488 | 0.0676 |
| 8 | 6723 | 693 | 0.1030 |
| 9 | 6024 | 750 | 0.1250 |
| 10 | 5265 | 988 | 0.1880 |
| 11 | 4267 | 1005 | 0.2360 |
| 12 | 3261 | 899 | 0.2760 |
| 13 | 2359 | 807 | 0.3420 |
| 14 | 1546 | 534 | 0.3450 |
| 15 | 1013 | 326 | 0.3220 |
| 16 | 687 | 201 | 0.2930 |
| 17 | 456 | 64 | 0.1400 |
| 18 | 307 | 64 | 0.2080 |
| 19 | 108 | 12 | 0.1110 |

| Mechanical Reliability Rates | | |
|---|---|---|
| Mileage at test | N tested | Pct failed |
| 0 | 162 | 9.88 |
| 10000 | 1209 | 10.90 |
| 20000 | 3451 | 14.40 |
| 30000 | 5984 | 16.90 |
| 40000 | 7448 | 19.80 |
| 50000 | 8286 | 23.10 |
| 60000 | 8401 | 25.80 |
| 70000 | 7856 | 27.50 |
| 80000 | 6841 | 30.00 |
| 90000 | 5342 | 32.90 |
| 100000 | 3927 | 34.70 |
| 110000 | 2490 | 35.10 |
| 120000 | 1636 | 35.80 |
| 130000 | 930 | 37.30 |
| 140000 | 514 | 34.60 |
| 150000 | 270 | 39.60 |
| 160000 | 132 | 34.10 |

**Alfa Romeo 147 2003**

At 5 years of age, the mortality rate of a Alfa Romeo 147 2003 (manufactured as a Car or Light Van) ranked number 155 out of 213 vehicles of the same age and type (any Car or Light Van constructed in 2003). One is the lowest (or best) and 213 the highest mortality rate. For vehicles reaching 20000 miles, its unreliability score (rate of failing an inspection) ranked 76 out of 209 vehicles of the same age, type, and mileage. One is the highest (or worst) and 209 the lowest rate of failing an inspection.

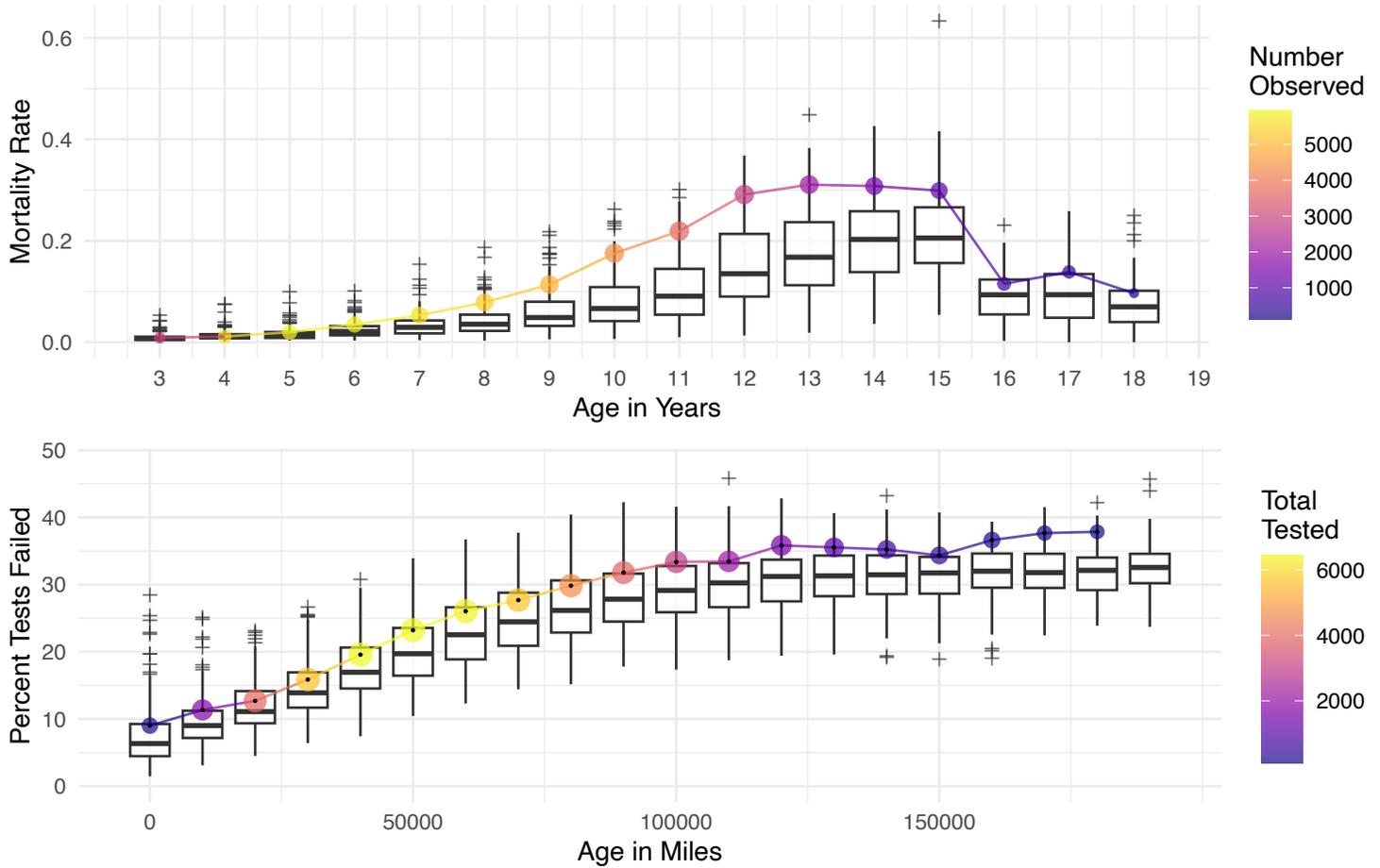

| Mortality rates | | | |
|---|---|---|---|
| Age in Years | Observed | Died | Mortality Rate |
| 3 | 2905 | 26 | 0.00895 |
| 4 | 5550 | 63 | 0.01140 |
| 5 | 5935 | 118 | 0.01990 |
| 6 | 5837 | 205 | 0.03510 |
| 7 | 5628 | 299 | 0.05310 |
| 8 | 5323 | 418 | 0.07850 |
| 9 | 4902 | 557 | 0.11400 |
| 10 | 4339 | 761 | 0.17500 |
| 11 | 3578 | 784 | 0.21900 |
| 12 | 2783 | 810 | 0.29100 |
| 13 | 1970 | 612 | 0.31100 |
| 14 | 1358 | 418 | 0.30800 |
| 15 | 940 | 281 | 0.29900 |
| 16 | 608 | 70 | 0.11500 |
| 17 | 404 | 56 | 0.13900 |
| 18 | 124 | 12 | 0.09680 |

| Mechanical Reliability Rates | | |
|---|---|---|
| Mileage at test | N tested | Pct failed |
| 0 | 211 | 9.0 |
| 10000 | 1745 | 11.3 |
| 20000 | 4015 | 12.7 |
| 30000 | 5658 | 15.9 |
| 40000 | 6445 | 19.6 |
| 50000 | 6456 | 23.2 |
| 60000 | 6366 | 26.0 |
| 70000 | 5662 | 27.7 |
| 80000 | 4729 | 29.8 |
| 90000 | 3882 | 31.8 |
| 100000 | 2966 | 33.4 |
| 110000 | 2137 | 33.5 |
| 120000 | 1448 | 35.8 |
| 130000 | 937 | 35.5 |
| 140000 | 613 | 35.2 |
| 150000 | 428 | 34.3 |
| 160000 | 273 | 36.6 |

**Alfa Romeo 147 2004**

At 5 years of age, the mortality rate of a Alfa Romeo 147 2004 (manufactured as a Car or Light Van) ranked number 171 out of 229 vehicles of the same age and type (any Car or Light Van constructed in 2004). One is the lowest (or best) and 229 the highest mortality rate. For vehicles reaching 40000 miles, its unreliability score (rate of failing an inspection) ranked 53 out of 227 vehicles of the same age, type, and mileage. One is the highest (or worst) and 227 the lowest rate of failing an inspection.

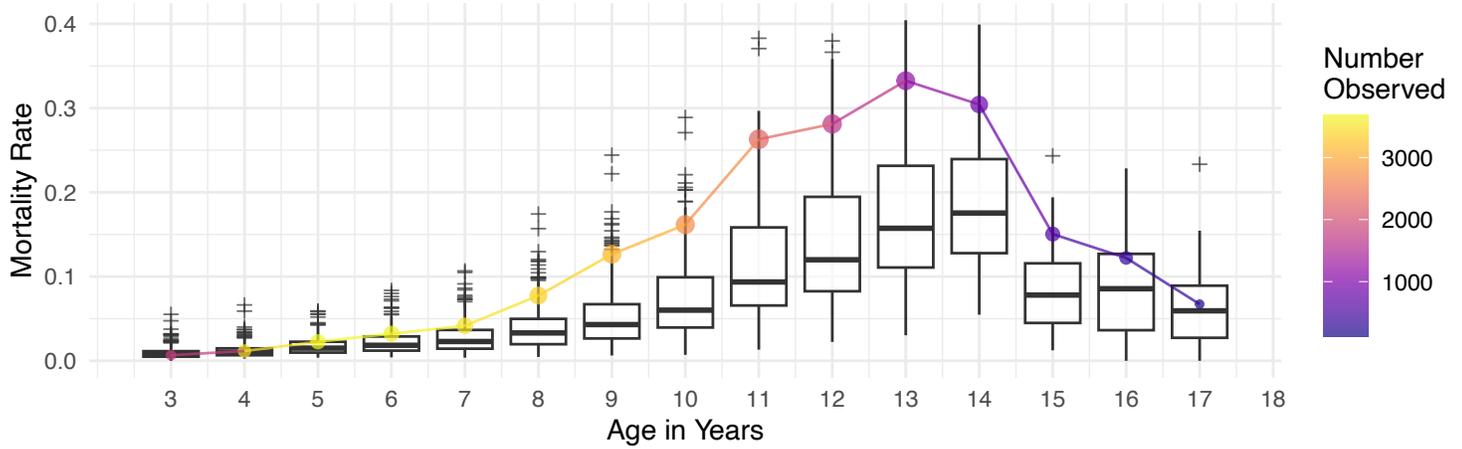

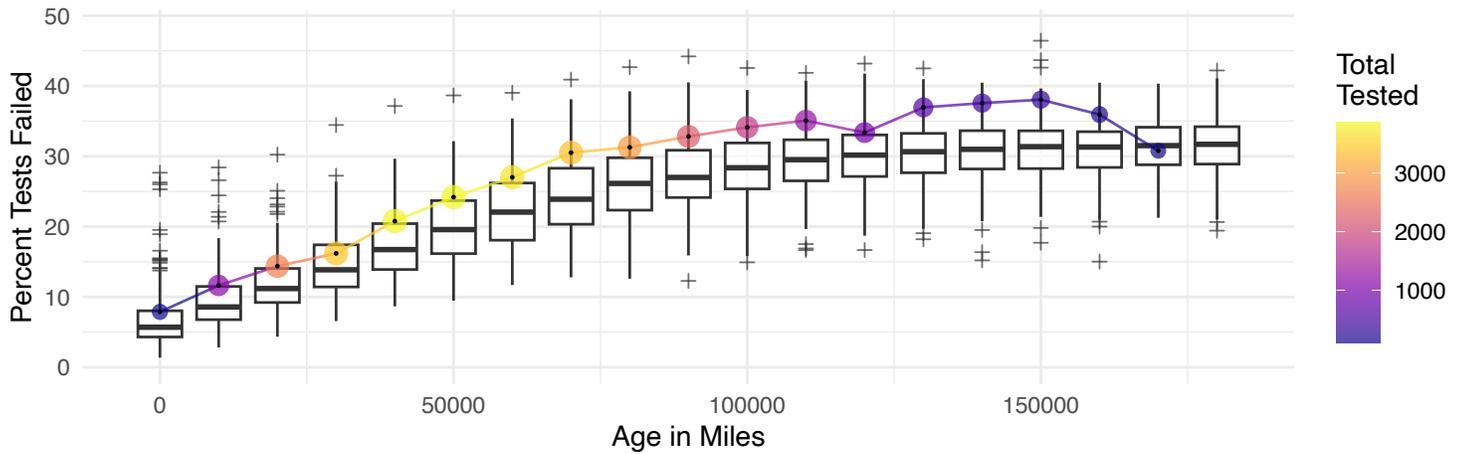

<table>
<tr><td colspan="4" align="center">Mortality rates</td></tr>
<tr><td>Age in Years</td><td>Observed</td><td>Died</td><td>Mortality Rate</td></tr>
<tr><td>3</td><td>1799</td><td>12</td><td>0.00667</td></tr>
<tr><td>4</td><td>3446</td><td>40</td><td>0.01160</td></tr>
<tr><td>5</td><td>3675</td><td>83</td><td>0.02260</td></tr>
<tr><td>6</td><td>3607</td><td>116</td><td>0.03220</td></tr>
<tr><td>7</td><td>3492</td><td>145</td><td>0.04150</td></tr>
<tr><td>8</td><td>3345</td><td>259</td><td>0.07740</td></tr>
<tr><td>9</td><td>3082</td><td>390</td><td>0.12700</td></tr>
<tr><td>10</td><td>2691</td><td>435</td><td>0.16200</td></tr>
<tr><td>11</td><td>2251</td><td>592</td><td>0.26300</td></tr>
<tr><td>12</td><td>1657</td><td>466</td><td>0.28100</td></tr>
<tr><td>13</td><td>1188</td><td>395</td><td>0.33200</td></tr>
<tr><td>14</td><td>792</td><td>241</td><td>0.30400</td></tr>
<tr><td>15</td><td>525</td><td>79</td><td>0.15000</td></tr>
<tr><td>16</td><td>344</td><td>42</td><td>0.12200</td></tr>
<tr><td>17</td><td>119</td><td>8</td><td>0.06720</td></tr>
</table>

| | Mechanical Reliability Rates | |
| --- | --- | --- |
| Mileage at test | N tested | Pct failed |
| 0 | 127 | 7.87 |
| 10000 | 1126 | 11.60 |
| 20000 | 2693 | 14.40 |
| 30000 | 3410 | 16.20 |
| 40000 | 3827 | 20.80 |
| 50000 | 3869 | 24.20 |
| 60000 | 3683 | 27.00 |
| 70000 | 3353 | 30.50 |
| 80000 | 2977 | 31.30 |
| 90000 | 2252 | 32.80 |
| 100000 | 1782 | 34.10 |
| 110000 | 1269 | 35.10 |
| 120000 | 914 | 33.40 |
| 130000 | 655 | 36.90 |
| 140000 | 442 | 37.60 |
| 150000 | 297 | 38.00 |
| 160000 | 167 | 35.90 |

## Alfa Romeo 147 2005

At 5 years of age, the mortality rate of a Alfa Romeo 147 2005 (manufactured as a Car or Light Van) ranked number 196 out of 240 vehicles of the same age and type (any Car or Light Van constructed in 2005). One is the lowest (or best) and 240 the highest mortality rate. For vehicles reaching 20000 miles, its unreliability score (rate of failing an inspection) ranked 91 out of 235 vehicles of the same age, type, and mileage. One is the highest (or worst) and 235 the lowest rate of failing an inspection.

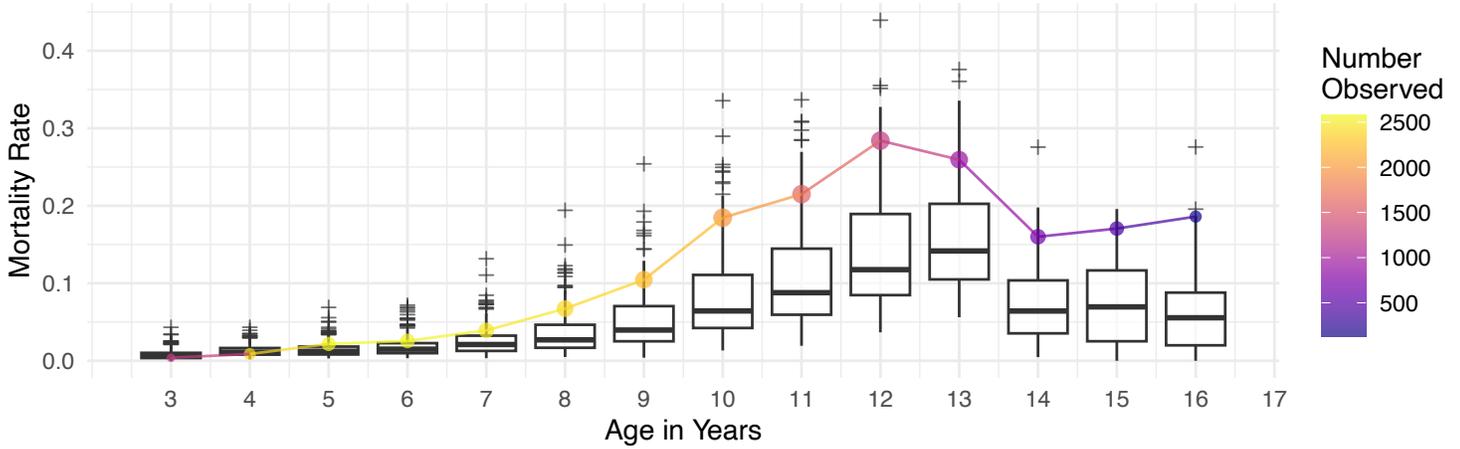

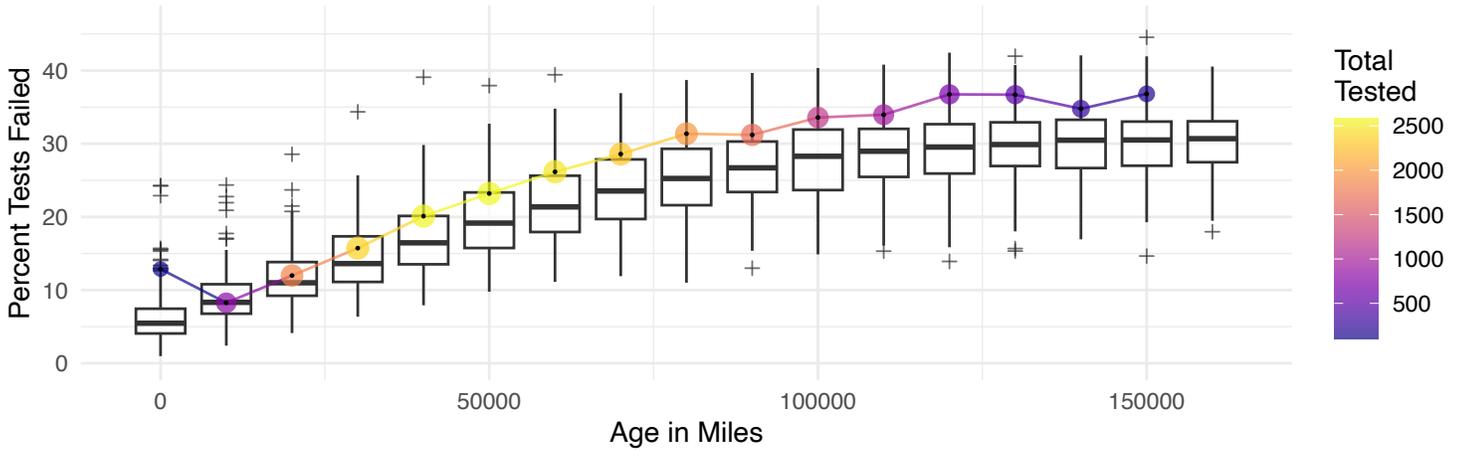

<table>
<tr><th colspan="4">Mortality rates</th></tr>
<tr><th>Age in Years</th><th>Observed</th><th>Died</th><th>Mortality Rate</th></tr>
<tr><td>3</td><td>1268</td><td>5</td><td>0.00394</td></tr>
<tr><td>4</td><td>2389</td><td>21</td><td>0.00879</td></tr>
<tr><td>5</td><td>2572</td><td>56</td><td>0.02180</td></tr>
<tr><td>6</td><td>2537</td><td>65</td><td>0.02560</td></tr>
<tr><td>7</td><td>2474</td><td>97</td><td>0.03920</td></tr>
<tr><td>8</td><td>2377</td><td>160</td><td>0.06730</td></tr>
<tr><td>9</td><td>2218</td><td>232</td><td>0.10500</td></tr>
<tr><td>10</td><td>1983</td><td>366</td><td>0.18500</td></tr>
<tr><td>11</td><td>1614</td><td>347</td><td>0.21500</td></tr>
<tr><td>12</td><td>1264</td><td>359</td><td>0.28400</td></tr>
<tr><td>13</td><td>902</td><td>234</td><td>0.25900</td></tr>
<tr><td>14</td><td>631</td><td>101</td><td>0.16000</td></tr>
<tr><td>15</td><td>381</td><td>65</td><td>0.17100</td></tr>
<tr><td>16</td><td>129</td><td>24</td><td>0.18600</td></tr>
</table>

| Mechanical Reliability Rates | | |
|---|---|---|
| Mileage at test | N tested | Pct failed |
| 0 | 101 | 12.90 |
| 10000 | 811 | 8.26 |
| 20000 | 1844 | 12.00 |
| 30000 | 2365 | 15.70 |
| 40000 | 2567 | 20.10 |
| 50000 | 2587 | 23.20 |
| 60000 | 2481 | 26.20 |
| 70000 | 2302 | 28.60 |
| 80000 | 1980 | 31.40 |
| 90000 | 1676 | 31.20 |
| 100000 | 1209 | 33.60 |
| 110000 | 960 | 34.00 |
| 120000 | 694 | 36.70 |
| 130000 | 466 | 36.70 |
| 140000 | 256 | 34.80 |
| 150000 | 163 | 36.80 |

# Alfa Romeo 147 2006

At 5 years of age, the mortality rate of a Alfa Romeo 147 2006 (manufactured as a Car or Light Van) ranked number 118 out of 225 vehicles of the same age and type (any Car or Light Van constructed in 2006). One is the lowest (or best) and 225 the highest mortality rate. For vehicles reaching 40000 miles, its unreliability score (rate of failing an inspection) ranked 38 out of 220 vehicles of the same age, type, and mileage. One is the highest (or worst) and 220 the lowest rate of failing an inspection.

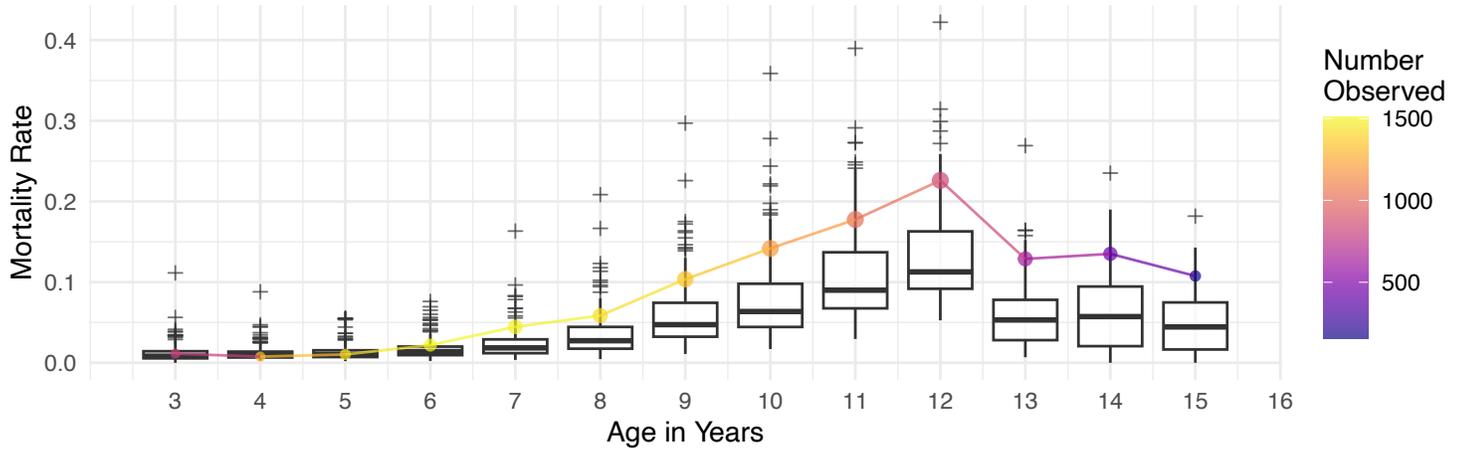

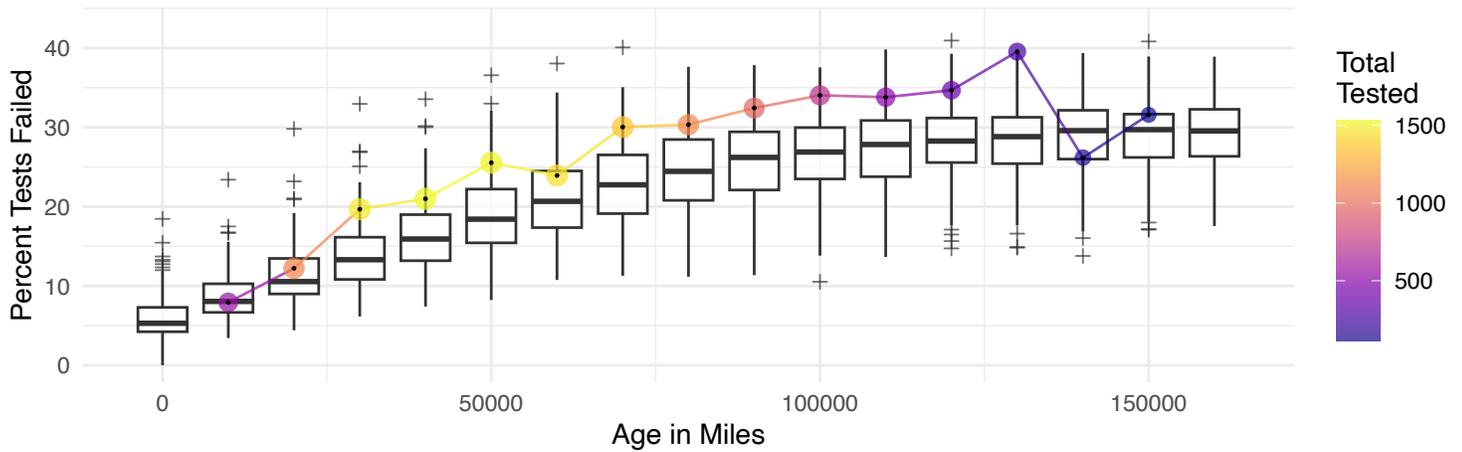

<table>
<tr><td colspan="4" align="center">Mortality rates</td></tr>
</table>

| Age in Years | Observed | Died | Mortality Rate |
|---|---|---|---|
| 3 | 796 | 9 | 0.01130 |
| 4 | 1306 | 10 | 0.00766 |
| 5 | 1442 | 15 | 0.01040 |
| 6 | 1505 | 33 | 0.02190 |
| 7 | 1481 | 66 | 0.04460 |
| 8 | 1418 | 83 | 0.05850 |
| 9 | 1332 | 138 | 0.10400 |
| 10 | 1192 | 169 | 0.14200 |
| 11 | 1019 | 181 | 0.17800 |
| 12 | 836 | 189 | 0.22600 |
| 13 | 629 | 81 | 0.12900 |
| 14 | 422 | 57 | 0.13500 |
| 15 | 158 | 17 | 0.10800 |

Mechanical Reliability Rates

| Mileage at test | N tested | Pct failed |
|---|---|---|
| 10000 | 555 | 7.93 |
| 20000 | 1113 | 12.20 |
| 30000 | 1484 | 19.70 |
| 40000 | 1534 | 21.00 |
| 50000 | 1523 | 25.50 |
| 60000 | 1446 | 23.90 |
| 70000 | 1355 | 30.00 |
| 80000 | 1124 | 30.30 |
| 90000 | 965 | 32.40 |
| 100000 | 711 | 34.00 |
| 110000 | 497 | 33.80 |
| 120000 | 372 | 34.70 |
| 130000 | 263 | 39.50 |
| 140000 | 149 | 26.20 |
| 150000 | 114 | 31.60 |

## Alfa Romeo 147 2007

At 5 years of age, the mortality rate of a Alfa Romeo 147 2007 (manufactured as a Car or Light Van) ranked number 134 out of 219 vehicles of the same age and type (any Car or Light Van constructed in 2007). One is the lowest (or best) and 219 the highest mortality rate. For vehicles reaching 20000 miles, its unreliability score (rate of failing an inspection) ranked 63 out of 214 vehicles of the same age, type, and mileage. One is the highest (or worst) and 214 the lowest rate of failing an inspection.

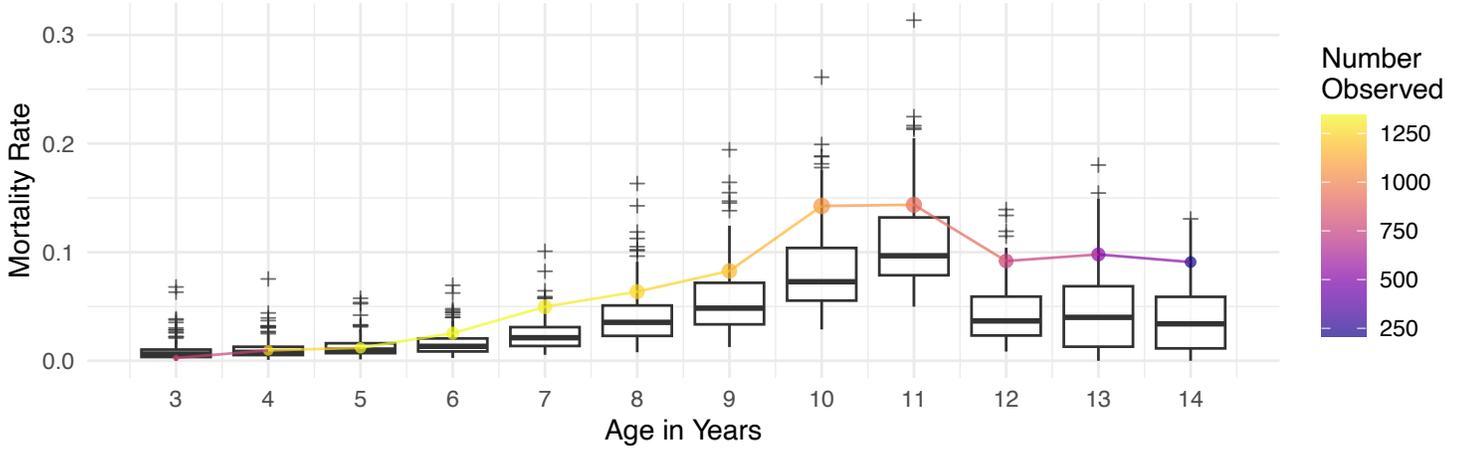

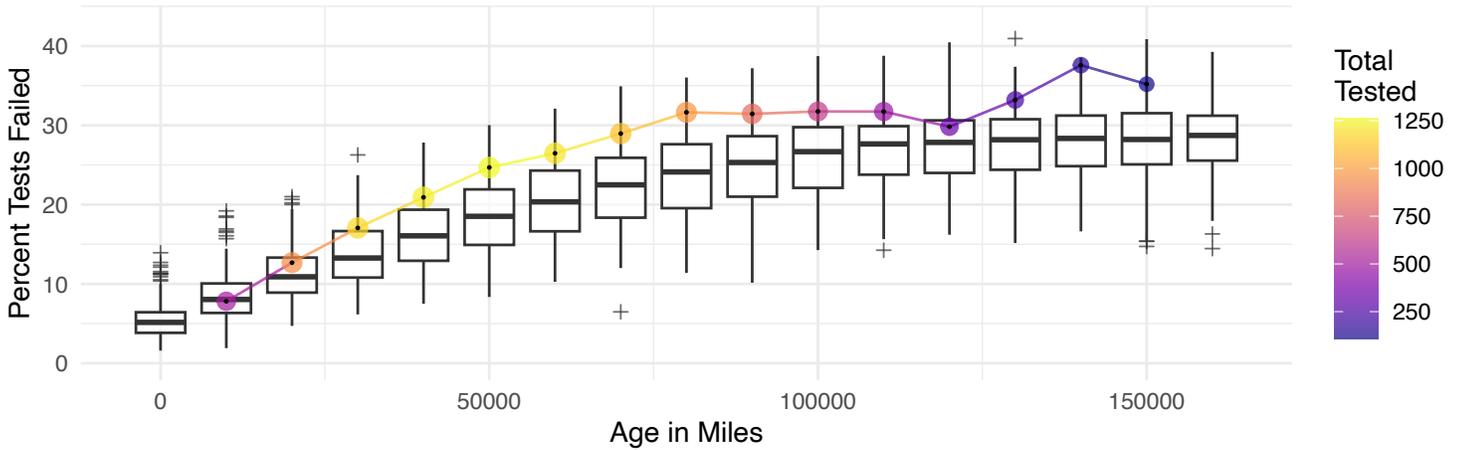

<table>
<tr><td colspan="4" align="center">Mortality rates</td></tr>
<tr><th>Age in Years</th><th>Observed</th><th>Died</th><th>Mortality Rate</th></tr>
<tr><td>3</td><td>769</td><td>2</td><td>0.0026</td></tr>
<tr><td>4</td><td>1250</td><td>12</td><td>0.0096</td></tr>
<tr><td>5</td><td>1341</td><td>16</td><td>0.0119</td></tr>
<tr><td>6</td><td>1338</td><td>34</td><td>0.0254</td></tr>
<tr><td>7</td><td>1306</td><td>65</td><td>0.0498</td></tr>
<tr><td>8</td><td>1241</td><td>79</td><td>0.0637</td></tr>
<tr><td>9</td><td>1160</td><td>96</td><td>0.0828</td></tr>
<tr><td>10</td><td>1059</td><td>151</td><td>0.1430</td></tr>
<tr><td>11</td><td>905</td><td>130</td><td>0.1440</td></tr>
<tr><td>12</td><td>740</td><td>68</td><td>0.0919</td></tr>
<tr><td>13</td><td>531</td><td>52</td><td>0.0979</td></tr>
<tr><td>14</td><td>209</td><td>19</td><td>0.0909</td></tr>
</table>

| Mechanical Reliability Rates | | |
|---|---|---|
| Mileage at test | N tested | Pct failed |
| 10000 | 512 | 7.81 |
| 20000 | 954 | 12.70 |
| 30000 | 1172 | 17.10 |
| 40000 | 1238 | 20.90 |
| 50000 | 1264 | 24.70 |
| 60000 | 1209 | 26.50 |
| 70000 | 1105 | 29.00 |
| 80000 | 983 | 31.60 |
| 90000 | 830 | 31.40 |
| 100000 | 630 | 31.70 |
| 110000 | 479 | 31.70 |
| 120000 | 332 | 29.80 |
| 130000 | 241 | 33.20 |
| 140000 | 149 | 37.60 |
| 150000 | 108 | 35.20 |

## Alfa Romeo 147 2008

At 5 years of age, the mortality rate of a Alfa Romeo 147 2008 (manufactured as a Car or Light Van) ranked number 166 out of 218 vehicles of the same age and type (any Car or Light Van constructed in 2008). One is the lowest (or best) and 218 the highest mortality rate. For vehicles reaching 20000 miles, its unreliability score (rate of failing an inspection) ranked 70 out of 212 vehicles of the same age, type, and mileage. One is the highest (or worst) and 212 the lowest rate of failing an inspection.

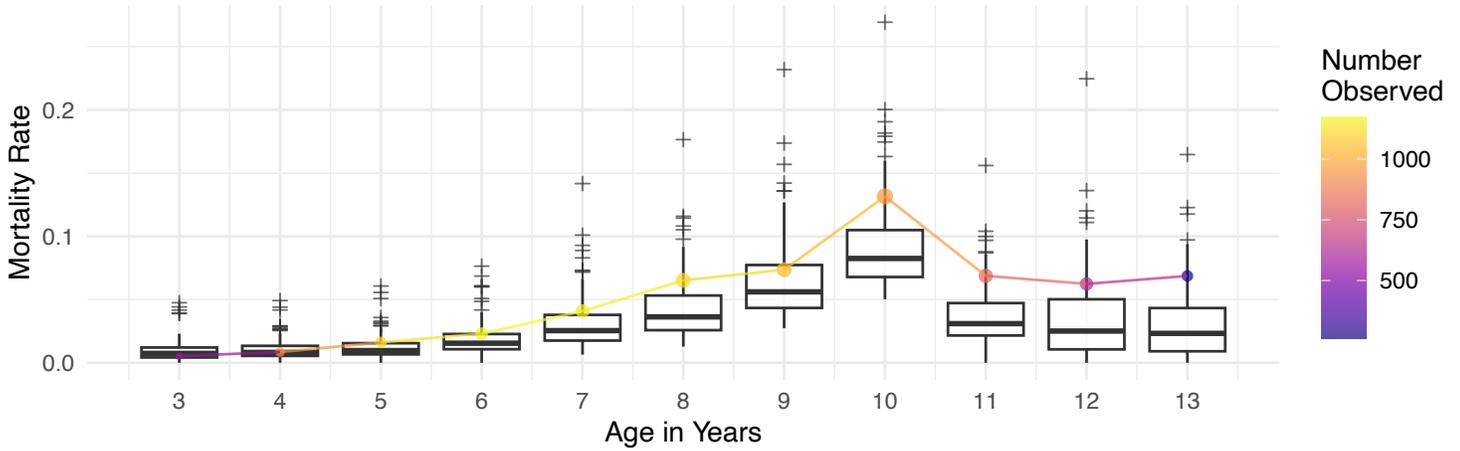

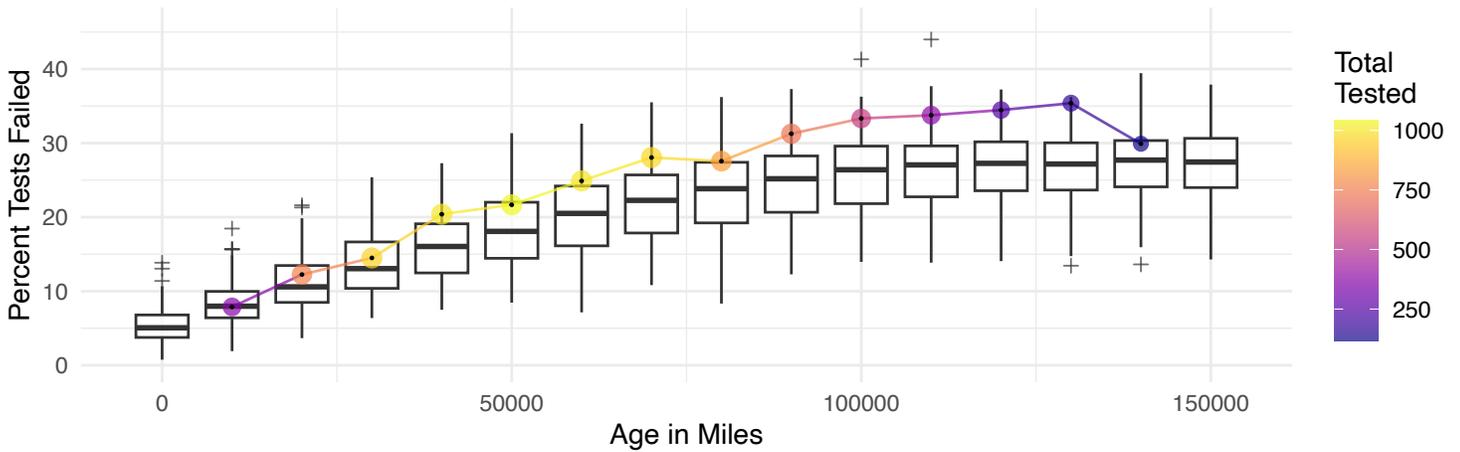

<table>
<tr><td colspan="4" align="center">Mortality rates</td></tr>
<tr><td>Age in Years</td><td>Observed</td><td>Died</td><td>Mortality Rate</td></tr>
<tr><td>3</td><td>562</td><td>3</td><td>0.00534</td></tr>
<tr><td>4</td><td>936</td><td>8</td><td>0.00855</td></tr>
<tr><td>5</td><td>1116</td><td>18</td><td>0.01610</td></tr>
<tr><td>6</td><td>1171</td><td>27</td><td>0.02310</td></tr>
<tr><td>7</td><td>1150</td><td>47</td><td>0.04090</td></tr>
<tr><td>8</td><td>1104</td><td>72</td><td>0.06520</td></tr>
<tr><td>9</td><td>1032</td><td>76</td><td>0.07360</td></tr>
<tr><td>10</td><td>957</td><td>126</td><td>0.13200</td></tr>
<tr><td>11</td><td>814</td><td>56</td><td>0.06880</td></tr>
<tr><td>12</td><td>641</td><td>40</td><td>0.06240</td></tr>
<tr><td>13</td><td>262</td><td>18</td><td>0.06870</td></tr>
</table>

| | Mechanical Reliability Rates | |
| Mileage at test | N tested | Pct failed |
| --- | --- | --- |
| 10000 | 368 | 7.88 |
| 20000 | 759 | 12.30 |
| 30000 | 959 | 14.50 |
| 40000 | 1005 | 20.40 |
| 50000 | 1043 | 21.70 |
| 60000 | 1008 | 24.90 |
| 70000 | 973 | 28.10 |
| 80000 | 838 | 27.60 |
| 90000 | 726 | 31.30 |
| 100000 | 549 | 33.30 |
| 110000 | 382 | 33.80 |
| 120000 | 235 | 34.50 |
| 130000 | 178 | 35.40 |
| 140000 | 117 | 29.90 |

**Alfa Romeo 156 1998**

At 10 years of age, the mortality rate of a Alfa Romeo 156 1998 (manufactured as a Car or Light Van) ranked number 175 out of 196 vehicles of the same age and type (any Car or Light Van constructed in 1998). One is the lowest (or best) and 196 the highest mortality rate. For vehicles reaching 120000 miles, its unreliability score (rate of failing an inspection) ranked 15 out of 172 vehicles of the same age, type, and mileage. One is the highest (or worst) and 172 the lowest rate of failing an inspection.

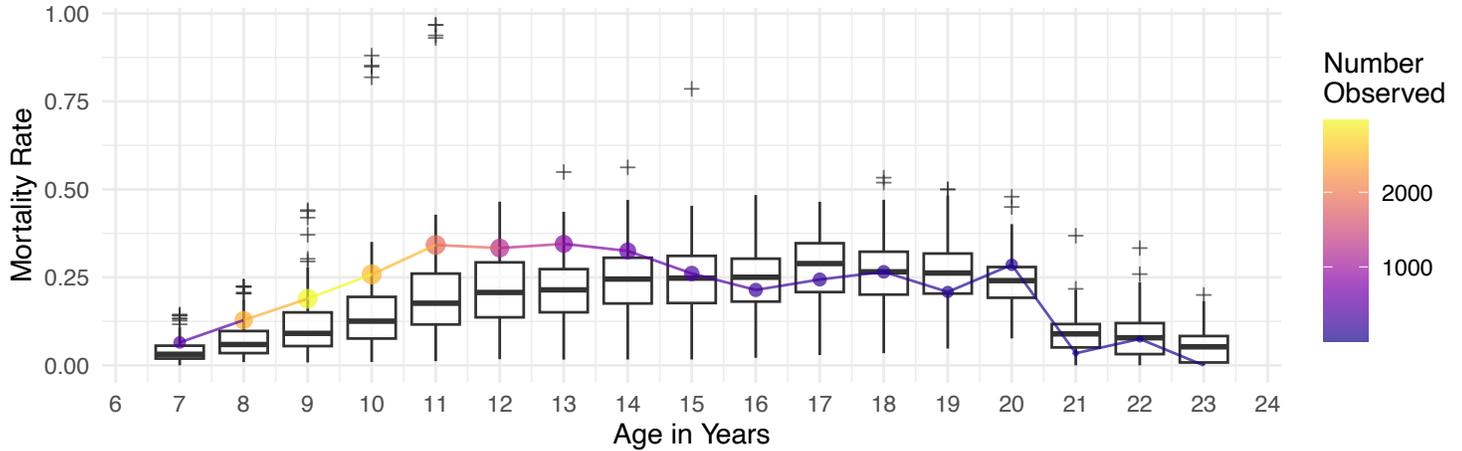

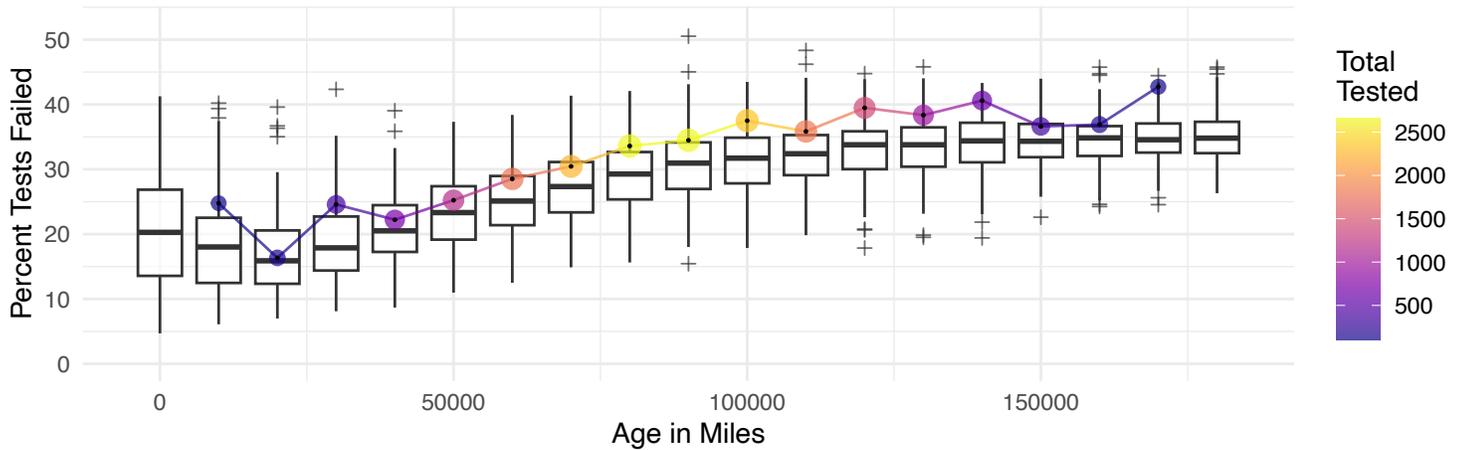

Mortality rates

| Age in Years | Observed | Died | Mortality Rate |
|---|---|---|---|
| 7 | 429 | 28 | 0.0653 |
| 8 | 2451 | 315 | 0.1290 |
| 9 | 2964 | 561 | 0.1890 |
| 10 | 2499 | 646 | 0.2590 |
| 11 | 1856 | 635 | 0.3420 |
| 12 | 1221 | 407 | 0.3330 |
| 13 | 811 | 280 | 0.3450 |
| 14 | 535 | 174 | 0.3250 |
| 15 | 360 | 94 | 0.2610 |
| 16 | 266 | 57 | 0.2140 |
| 17 | 209 | 51 | 0.2440 |
| 18 | 158 | 42 | 0.2660 |
| 19 | 115 | 24 | 0.2090 |
| 20 | 91 | 26 | 0.2860 |
| 21 | 58 | 2 | 0.0345 |
| 22 | 40 | 3 | 0.0750 |

Mechanical Reliability Rates

| Mileage at test | N tested | Pct failed |
|---|---|---|
| 10000 | 109 | 24.8 |
| 20000 | 153 | 16.3 |
| 30000 | 362 | 24.6 |
| 40000 | 711 | 22.2 |
| 50000 | 1181 | 25.2 |
| 60000 | 1788 | 28.5 |
| 70000 | 2249 | 30.5 |
| 80000 | 2655 | 33.6 |
| 90000 | 2654 | 34.5 |
| 100000 | 2347 | 37.5 |
| 110000 | 1819 | 35.8 |
| 120000 | 1421 | 39.5 |
| 130000 | 949 | 38.4 |
| 140000 | 606 | 40.6 |
| 150000 | 355 | 36.6 |
| 160000 | 187 | 36.9 |
| 170000 | 117 | 42.7 |

## Alfa Romeo 156 1999

At 10 years of age, the mortality rate of a Alfa Romeo 156 1999 (manufactured as a Car or Light Van) ranked number 189 out of 201 vehicles of the same age and type (any Car or Light Van constructed in 1999). One is the lowest (or best) and 201 the highest mortality rate. For vehicles reaching 120000 miles, its unreliability score (rate of failing an inspection) ranked 21 out of 181 vehicles of the same age, type, and mileage. One is the highest (or worst) and 181 the lowest rate of failing an inspection.

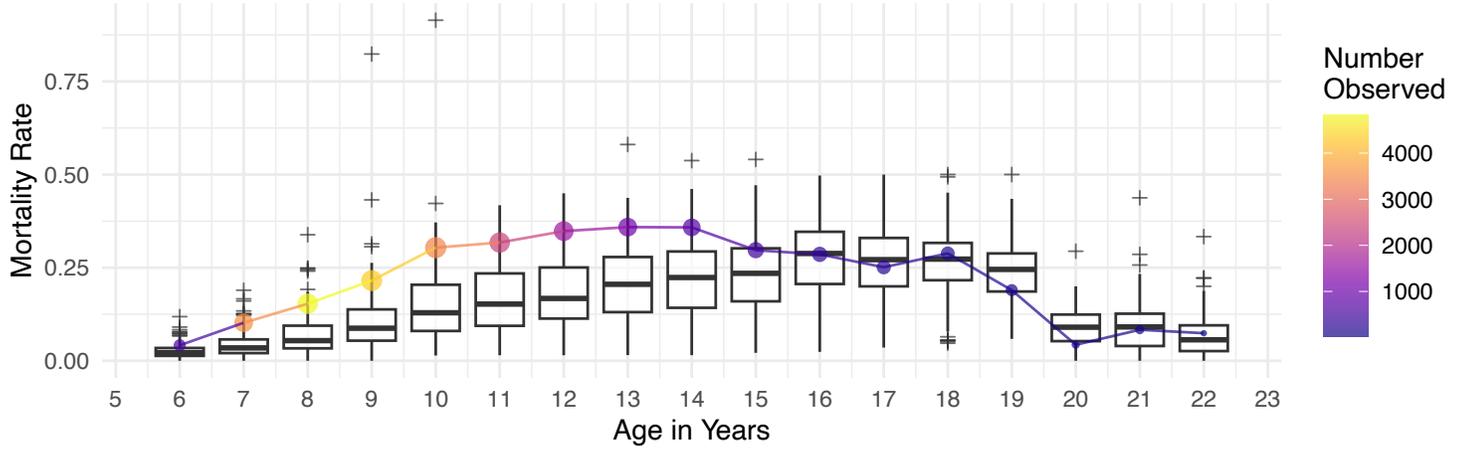

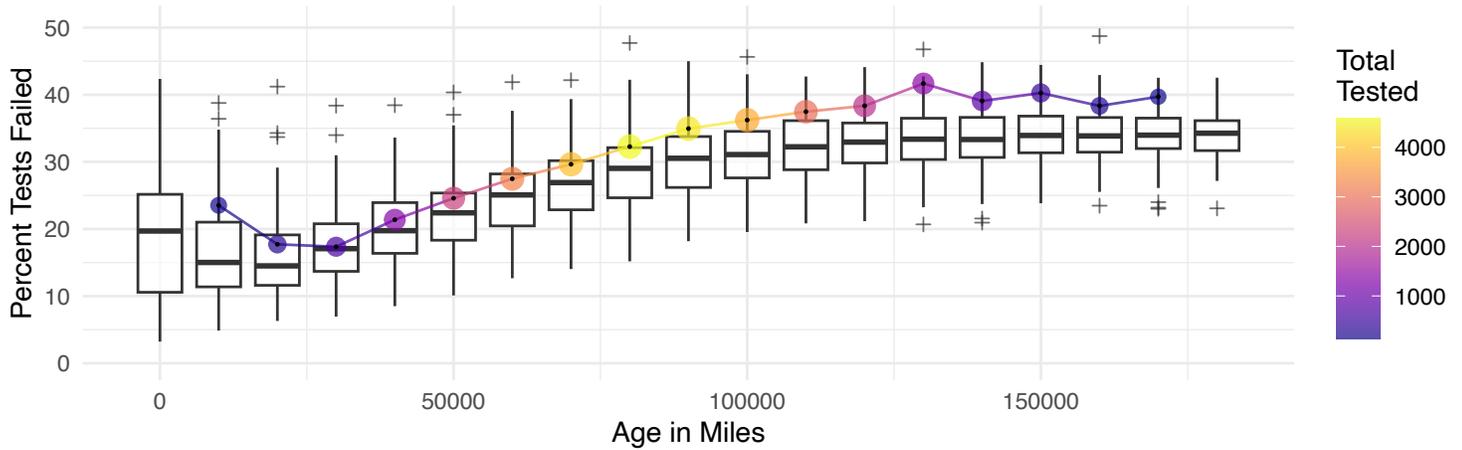

| Mortality rates | | | |
|---|---|---|---|
| Age in Years | Observed | Died | Mortality Rate |
| 6 | 552 | 23 | 0.0417 |
| 7 | 3526 | 360 | 0.1020 |
| 8 | 4813 | 737 | 0.1530 |
| 9 | 4301 | 926 | 0.2150 |
| 10 | 3381 | 1027 | 0.3040 |
| 11 | 2353 | 747 | 0.3170 |
| 12 | 1603 | 558 | 0.3480 |
| 13 | 1045 | 375 | 0.3590 |
| 14 | 670 | 240 | 0.3580 |
| 15 | 431 | 128 | 0.2970 |
| 16 | 301 | 86 | 0.2860 |
| 17 | 215 | 54 | 0.2510 |
| 18 | 163 | 47 | 0.2880 |
| 19 | 116 | 22 | 0.1900 |
| 20 | 93 | 4 | 0.0430 |
| 21 | 72 | 6 | 0.0833 |
| 22 | 27 | 2 | 0.0741 |

| Mechanical Reliability Rates | | |
|---|---|---|
| Mileage at test | N tested | Pct failed |
| 10000 | 170 | 23.5 |
| 20000 | 316 | 17.7 |
| 30000 | 651 | 17.4 |
| 40000 | 1356 | 21.4 |
| 50000 | 2282 | 24.6 |
| 60000 | 3260 | 27.5 |
| 70000 | 3977 | 29.6 |
| 80000 | 4589 | 32.3 |
| 90000 | 4408 | 35.0 |
| 100000 | 3746 | 36.2 |
| 110000 | 2946 | 37.5 |
| 120000 | 2026 | 38.4 |
| 130000 | 1399 | 41.7 |
| 140000 | 796 | 39.1 |
| 150000 | 452 | 40.3 |
| 160000 | 253 | 38.3 |
| 170000 | 136 | 39.7 |

## Alfa Romeo 156 2000

At 5 years of age, the mortality rate of a Alfa Romeo 156 2000 (manufactured as a Car or Light Van) ranked number 188 out of 198 vehicles of the same age and type (any Car or Light Van constructed in 2000). One is the lowest (or best) and 198 the highest mortality rate. For vehicles reaching 120000 miles, its unreliability score (rate of failing an inspection) ranked 46 out of 184 vehicles of the same age, type, and mileage. One is the highest (or worst) and 184 the lowest rate of failing an inspection.

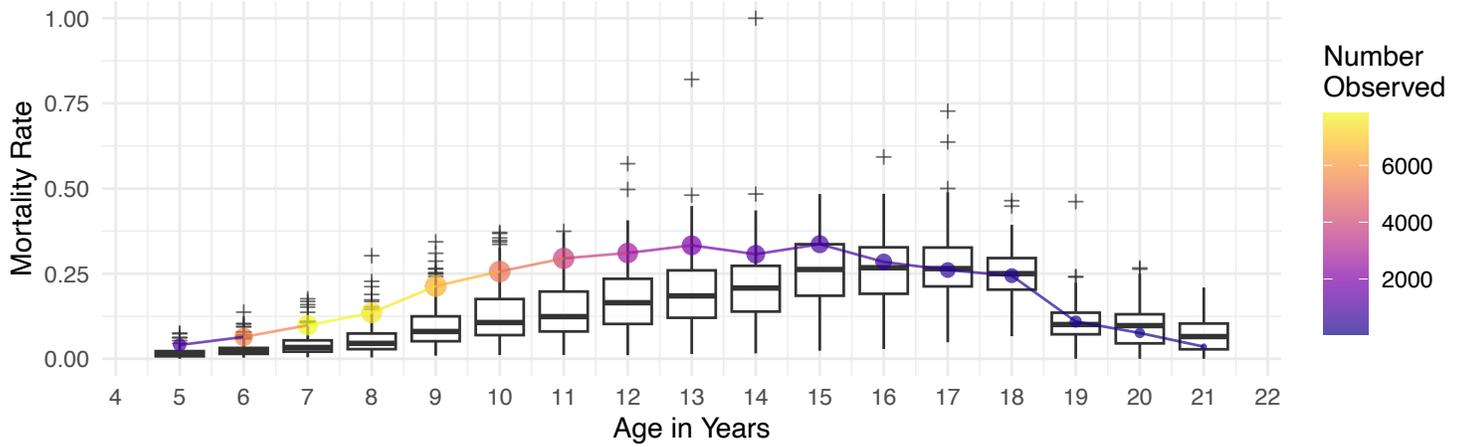

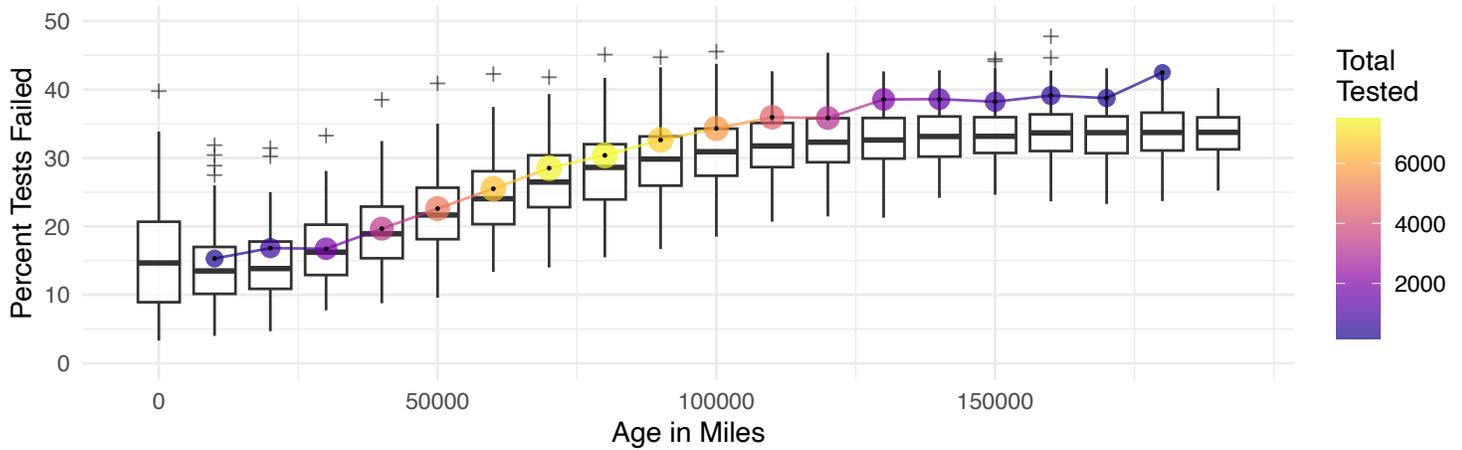

| Mortality rates | | | |
|---|---|---|---|
| Age in Years | Observed | Died | Mortality Rate |
| 5 | 935 | 38 | 0.0406 |
| 6 | 5483 | 352 | 0.0642 |
| 7 | 7818 | 770 | 0.0985 |
| 8 | 7461 | 1010 | 0.1350 |
| 9 | 6467 | 1381 | 0.2140 |
| 10 | 5079 | 1302 | 0.2560 |
| 11 | 3777 | 1114 | 0.2950 |
| 12 | 2662 | 827 | 0.3110 |
| 13 | 1834 | 611 | 0.3330 |
| 14 | 1221 | 375 | 0.3070 |
| 15 | 844 | 284 | 0.3360 |
| 16 | 559 | 159 | 0.2840 |
| 17 | 399 | 104 | 0.2610 |
| 18 | 295 | 72 | 0.2440 |
| 19 | 210 | 23 | 0.1100 |
| 20 | 145 | 11 | 0.0759 |
| 21 | 56 | 2 | 0.0357 |

| Mechanical Reliability Rates | | |
|---|---|---|
| Mileage at test | N tested | Pct failed |
| 10000 | 268 | 15.3 |
| 20000 | 755 | 16.8 |
| 30000 | 1805 | 16.7 |
| 40000 | 3377 | 19.7 |
| 50000 | 5011 | 22.6 |
| 60000 | 6494 | 25.5 |
| 70000 | 7383 | 28.5 |
| 80000 | 7503 | 30.4 |
| 100000 | 5715 | 34.3 |
| 110000 | 4418 | 35.9 |
| 120000 | 3167 | 35.8 |
| 130000 | 1982 | 38.5 |
| 140000 | 1319 | 38.6 |
| 150000 | 785 | 38.2 |
| 160000 | 455 | 39.1 |
| 170000 | 302 | 38.7 |
| 180000 | 160 | 42.5 |

# Alfa Romeo 156 2001

At 5 years of age, the mortality rate of a Alfa Romeo 156 2001 (manufactured as a Car or Light Van) ranked number 196 out of 205 vehicles of the same age and type (any Car or Light Van constructed in 2001). One is the lowest (or best) and 205 the highest mortality rate. For vehicles reaching 120000 miles, its unreliability score (rate of failing an inspection) ranked 53 out of 194 vehicles of the same age, type, and mileage. One is the highest (or worst) and 194 the lowest rate of failing an inspection.

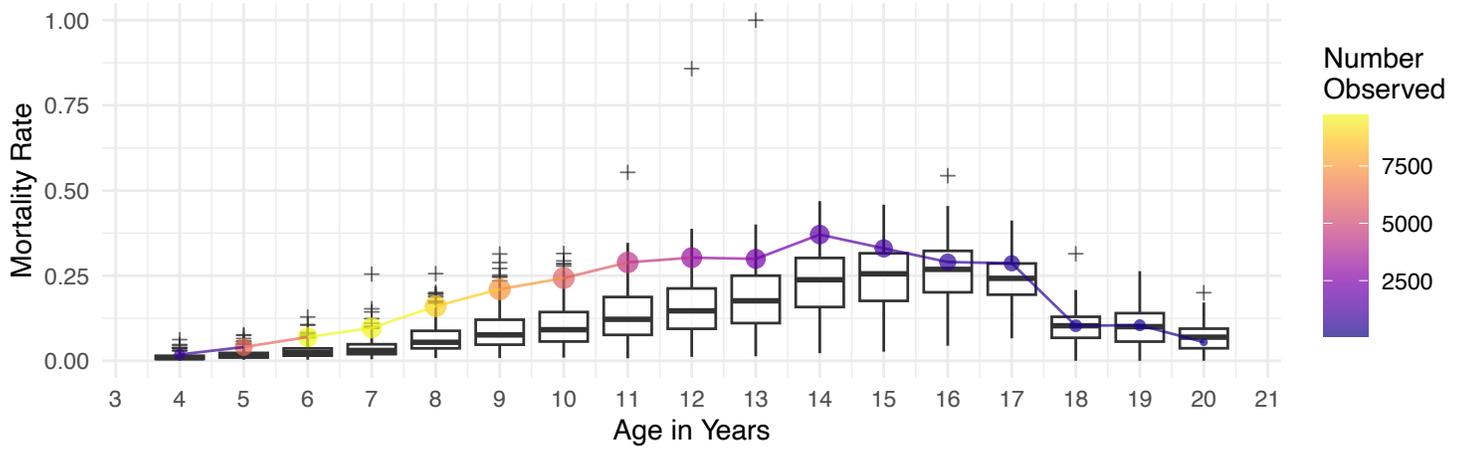

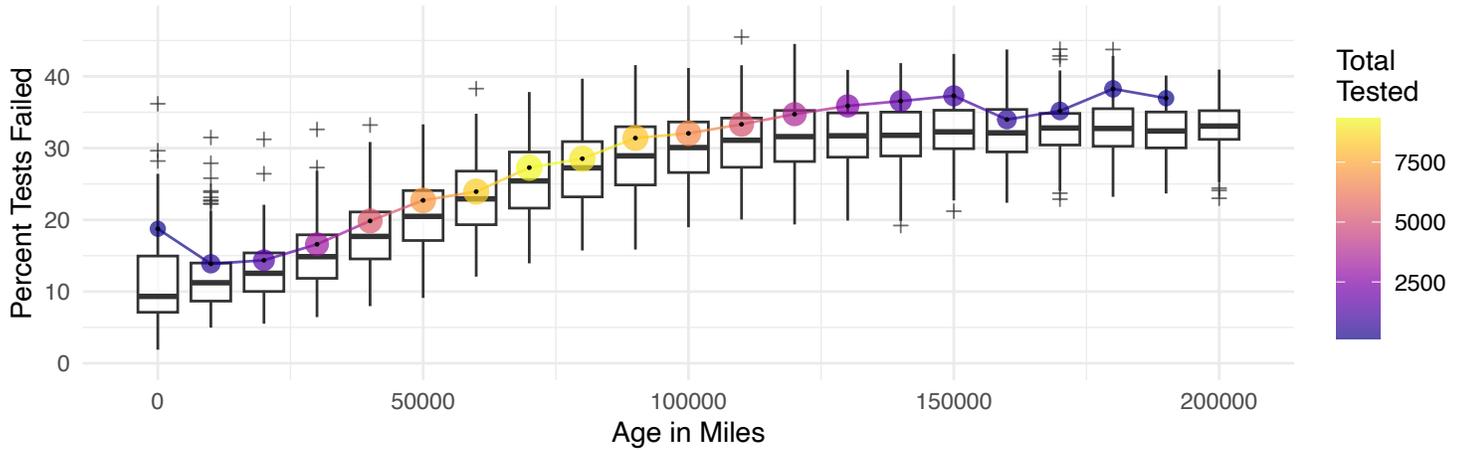

| Mortality rates | | | |
|---|---|---|---|
| Age in Years | Observed | Died | Mortality Rate |
| 4 | 859 | 16 | 0.0186 |
| 5 | 6091 | 246 | 0.0404 |
| 6 | 9690 | 675 | 0.0697 |
| 7 | 9614 | 927 | 0.0964 |
| 8 | 8709 | 1396 | 0.1600 |
| 9 | 7303 | 1535 | 0.2100 |
| 10 | 5747 | 1396 | 0.2430 |
| 11 | 4344 | 1255 | 0.2890 |
| 12 | 3084 | 934 | 0.3030 |
| 13 | 2145 | 642 | 0.2990 |
| 14 | 1500 | 556 | 0.3710 |
| 15 | 942 | 311 | 0.3300 |
| 16 | 631 | 183 | 0.2900 |
| 17 | 447 | 128 | 0.2860 |
| 18 | 301 | 31 | 0.1030 |
| 19 | 201 | 21 | 0.1040 |
| 20 | 73 | 4 | 0.0548 |

| Mechanical Reliability Rates | | |
|---|---|---|
| Mileage at test | N tested | Pct failed |
| 0 | 128 | 18.8 |
| 10000 | 476 | 13.9 |
| 20000 | 1379 | 14.4 |
| 30000 | 3194 | 16.6 |
| 40000 | 5280 | 19.8 |
| 50000 | 7126 | 22.7 |
| 60000 | 8403 | 23.9 |
| 70000 | 9341 | 27.3 |
| 80000 | 8967 | 28.5 |
| 90000 | 8215 | 31.4 |
| 100000 | 6713 | 32.1 |
| 110000 | 5167 | 33.3 |
| 120000 | 3685 | 34.7 |
| 130000 | 2522 | 35.9 |
| 140000 | 1529 | 36.6 |
| 150000 | 1019 | 37.3 |
| 160000 | 562 | 34.0 |

## Alfa Romeo 156 2002

At 5 years of age, the mortality rate of a Alfa Romeo 156 2002 (manufactured as a Car or Light Van) ranked number 189 out of 202 vehicles of the same age and type (any Car or Light Van constructed in 2002). One is the lowest (or best) and 202 the highest mortality rate. For vehicles reaching 120000 miles, its unreliability score (rate of failing an inspection) ranked 59 out of 193 vehicles of the same age, type, and mileage. One is the highest (or worst) and 193 the lowest rate of failing an inspection.

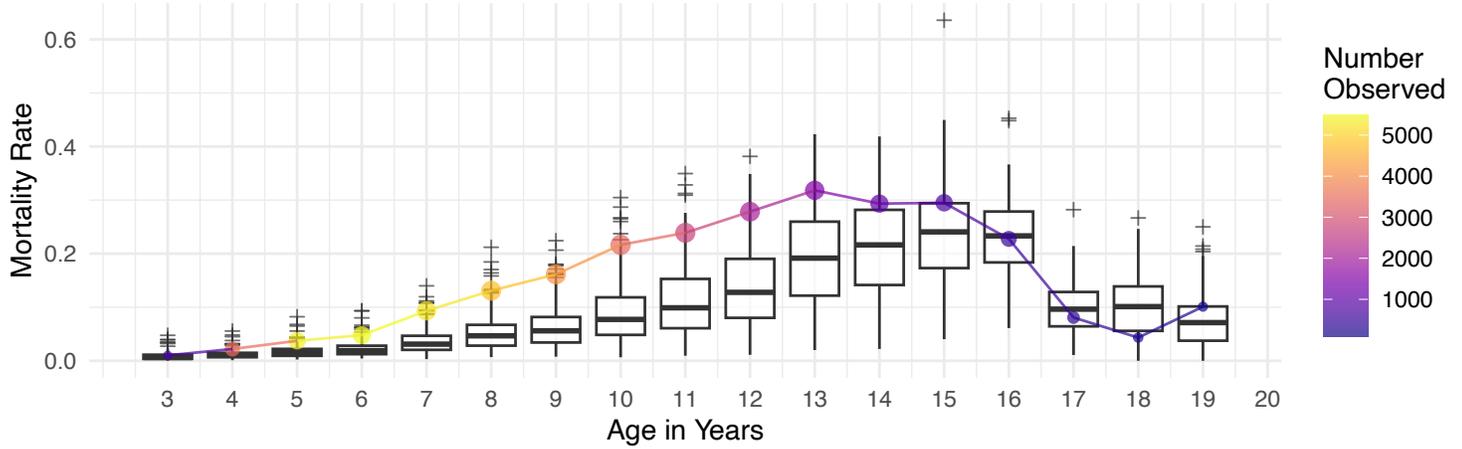

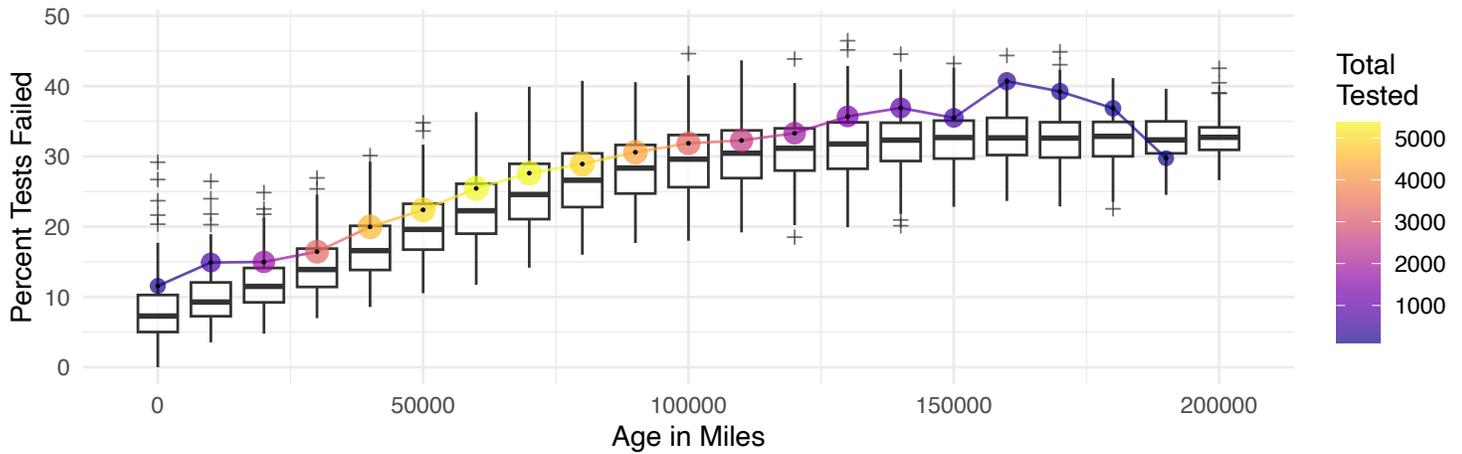

| Mortality rates | | | |
|---|---|---|---|
| Age in Years | Observed | Died | Mortality Rate |
| 3 | 689 | 7 | 0.0102 |
| 4 | 3494 | 77 | 0.0220 |
| 5 | 5354 | 200 | 0.0374 |
| 6 | 5475 | 262 | 0.0479 |
| 7 | 5236 | 489 | 0.0934 |
| 8 | 4746 | 621 | 0.1310 |
| 9 | 4119 | 665 | 0.1610 |
| 10 | 3451 | 747 | 0.2160 |
| 11 | 2704 | 646 | 0.2390 |
| 12 | 2055 | 572 | 0.2780 |
| 13 | 1480 | 471 | 0.3180 |
| 14 | 1009 | 296 | 0.2930 |
| 15 | 712 | 210 | 0.2950 |
| 16 | 501 | 114 | 0.2280 |
| 17 | 359 | 29 | 0.0808 |
| 18 | 254 | 11 | 0.0433 |
| 19 | 99 | 10 | 0.1010 |

| Mechanical Reliability Rates | | |
|---|---|---|
| Mileage at test | N tested | Pct failed |
| 0 | 121 | 11.6 |
| 10000 | 624 | 14.9 |
| 20000 | 1789 | 15.0 |
| 30000 | 3309 | 16.4 |
| 40000 | 4511 | 20.0 |
| 50000 | 5041 | 22.4 |
| 60000 | 5376 | 25.4 |
| 70000 | 5320 | 27.6 |
| 80000 | 4914 | 28.9 |
| 90000 | 4240 | 30.6 |
| 100000 | 3298 | 31.9 |
| 110000 | 2498 | 32.3 |
| 120000 | 1856 | 33.3 |
| 130000 | 1331 | 35.7 |
| 140000 | 889 | 36.9 |
| 150000 | 575 | 35.5 |
| 160000 | 317 | 40.7 |

## Alfa Romeo 156 2003

At 5 years of age, the mortality rate of a Alfa Romeo 156 2003 (manufactured as a Car or Light Van) ranked number 197 out of 213 vehicles of the same age and type (any Car or Light Van constructed in 2003). One is the lowest (or best) and 213 the highest mortality rate. For vehicles reaching 20000 miles, its unreliability score (rate of failing an inspection) ranked 26 out of 209 vehicles of the same age, type, and mileage. One is the highest (or worst) and 209 the lowest rate of failing an inspection.

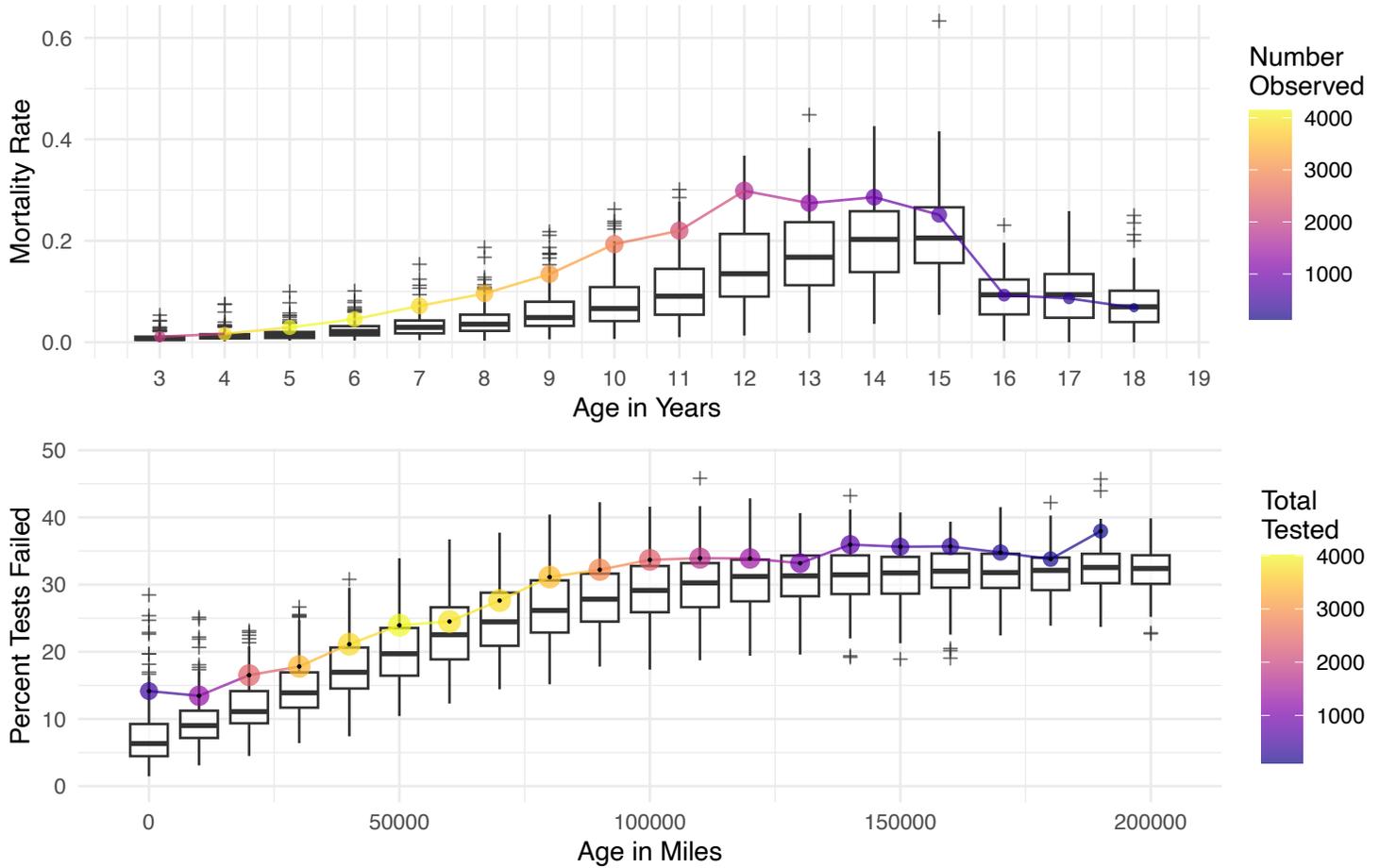

### Mortality rates

| Age in Years | Observed | Died | Mortality Rate |
|---|---|---|---|
| 3 | 2008 | 22 | 0.0110 |
| 4 | 3872 | 65 | 0.0168 |
| 5 | 4130 | 121 | 0.0293 |
| 6 | 4018 | 183 | 0.0455 |
| 7 | 3829 | 275 | 0.0718 |
| 8 | 3551 | 341 | 0.0960 |
| 9 | 3201 | 430 | 0.1340 |
| 10 | 2766 | 535 | 0.1930 |
| 11 | 2223 | 489 | 0.2200 |
| 12 | 1732 | 517 | 0.2980 |
| 13 | 1211 | 332 | 0.2740 |
| 14 | 878 | 251 | 0.2860 |
| 15 | 625 | 157 | 0.2510 |
| 16 | 452 | 42 | 0.0929 |
| 17 | 312 | 27 | 0.0865 |
| 18 | 132 | 9 | 0.0682 |

### Mechanical Reliability Rates

| Mileage at test | N tested | Pct failed |
|---|---|---|
| 0 | 424 | 14.2 |
| 10000 | 1286 | 13.5 |
| 20000 | 2397 | 16.5 |
| 30000 | 3374 | 17.8 |
| 40000 | 3730 | 21.1 |
| 50000 | 4014 | 23.9 |
| 60000 | 3819 | 24.5 |
| 70000 | 3761 | 27.6 |
| 80000 | 3463 | 31.1 |
| 90000 | 2868 | 32.2 |
| 100000 | 2357 | 33.7 |
| 110000 | 1759 | 33.9 |
| 120000 | 1375 | 33.9 |
| 130000 | 949 | 33.2 |
| 140000 | 698 | 36.0 |
| 160000 | 325 | 35.7 |
| 190000 | 108 | 38.0 |

## Alfa Romeo 156 2004

At 5 years of age, the mortality rate of a Alfa Romeo 156 2004 (manufactured as a Car or Light Van) ranked number 186 out of 229 vehicles of the same age and type (any Car or Light Van constructed in 2004). One is the lowest (or best) and 229 the highest mortality rate. For vehicles reaching 20000 miles, its unreliability score (rate of failing an inspection) ranked 28 out of 225 vehicles of the same age, type, and mileage. One is the highest (or worst) and 225 the lowest rate of failing an inspection.

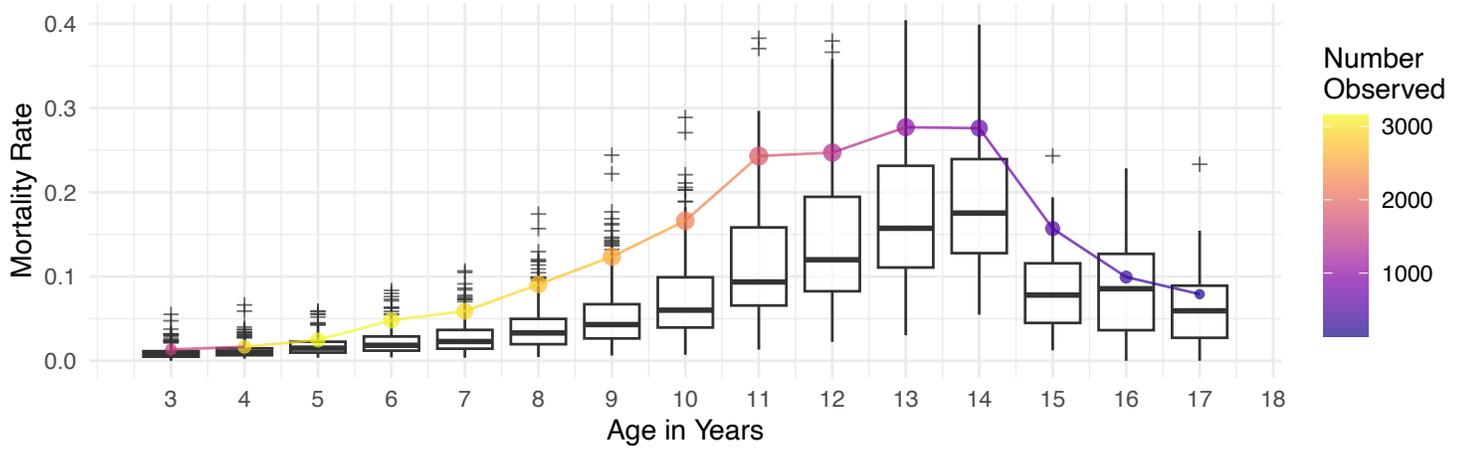

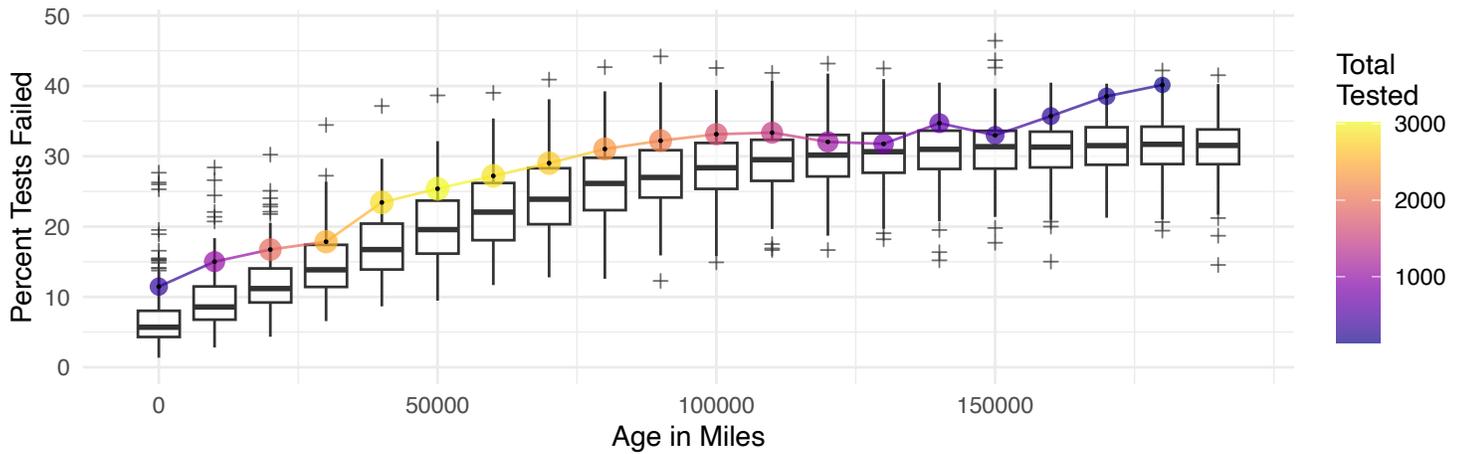

<table>
<tr><td colspan="4" align="center">Mortality rates</td></tr>
<tr><th>Age in Years</th><th>Observed</th><th>Died</th><th>Mortality Rate</th></tr>
<tr><td>3</td><td>1560</td><td>21</td><td>0.0135</td></tr>
<tr><td>4</td><td>2949</td><td>49</td><td>0.0166</td></tr>
<tr><td>5</td><td>3146</td><td>78</td><td>0.0248</td></tr>
<tr><td>6</td><td>3076</td><td>148</td><td>0.0481</td></tr>
<tr><td>7</td><td>2923</td><td>172</td><td>0.0588</td></tr>
<tr><td>8</td><td>2747</td><td>249</td><td>0.0906</td></tr>
<tr><td>9</td><td>2492</td><td>308</td><td>0.1240</td></tr>
<tr><td>10</td><td>2177</td><td>362</td><td>0.1660</td></tr>
<tr><td>11</td><td>1806</td><td>439</td><td>0.2430</td></tr>
<tr><td>12</td><td>1355</td><td>335</td><td>0.2470</td></tr>
<tr><td>13</td><td>1017</td><td>282</td><td>0.2770</td></tr>
<tr><td>14</td><td>735</td><td>203</td><td>0.2760</td></tr>
<tr><td>15</td><td>516</td><td>81</td><td>0.1570</td></tr>
<tr><td>16</td><td>342</td><td>34</td><td>0.0994</td></tr>
<tr><td>17</td><td>139</td><td>11</td><td>0.0791</td></tr>
</table>

<table>
<tr><td colspan="3" align="center">Mechanical Reliability Rates</td></tr>
<tr><th>Mileage at test</th><th>N tested</th><th>Pct failed</th></tr>
<tr><td>0</td><td>314</td><td>11.5</td></tr>
<tr><td>10000</td><td>1019</td><td>15.0</td></tr>
<tr><td>20000</td><td>1936</td><td>16.7</td></tr>
<tr><td>30000</td><td>2526</td><td>17.9</td></tr>
<tr><td>40000</td><td>2846</td><td>23.4</td></tr>
<tr><td>50000</td><td>3021</td><td>25.4</td></tr>
<tr><td>60000</td><td>2874</td><td>27.2</td></tr>
<tr><td>70000</td><td>2708</td><td>29.0</td></tr>
<tr><td>80000</td><td>2307</td><td>31.0</td></tr>
<tr><td>90000</td><td>2114</td><td>32.2</td></tr>
<tr><td>100000</td><td>1693</td><td>33.1</td></tr>
<tr><td>110000</td><td>1454</td><td>33.4</td></tr>
<tr><td>120000</td><td>1089</td><td>32.0</td></tr>
<tr><td>130000</td><td>765</td><td>31.8</td></tr>
<tr><td>140000</td><td>605</td><td>34.7</td></tr>
<tr><td>150000</td><td>379</td><td>33.0</td></tr>
<tr><td>170000</td><td>218</td><td>38.5</td></tr>
</table>

# Alfa Romeo 156 2005

At 5 years of age, the mortality rate of a Alfa Romeo 156 2005 (manufactured as a Car or Light Van) ranked number 218 out of 240 vehicles of the same age and type (any Car or Light Van constructed in 2005). One is the lowest (or best) and 240 the highest mortality rate. For vehicles reaching 100000 miles, its unreliability score (rate of failing an inspection) ranked 32 out of 234 vehicles of the same age, type, and mileage. One is the highest (or worst) and 234 the lowest rate of failing an inspection.

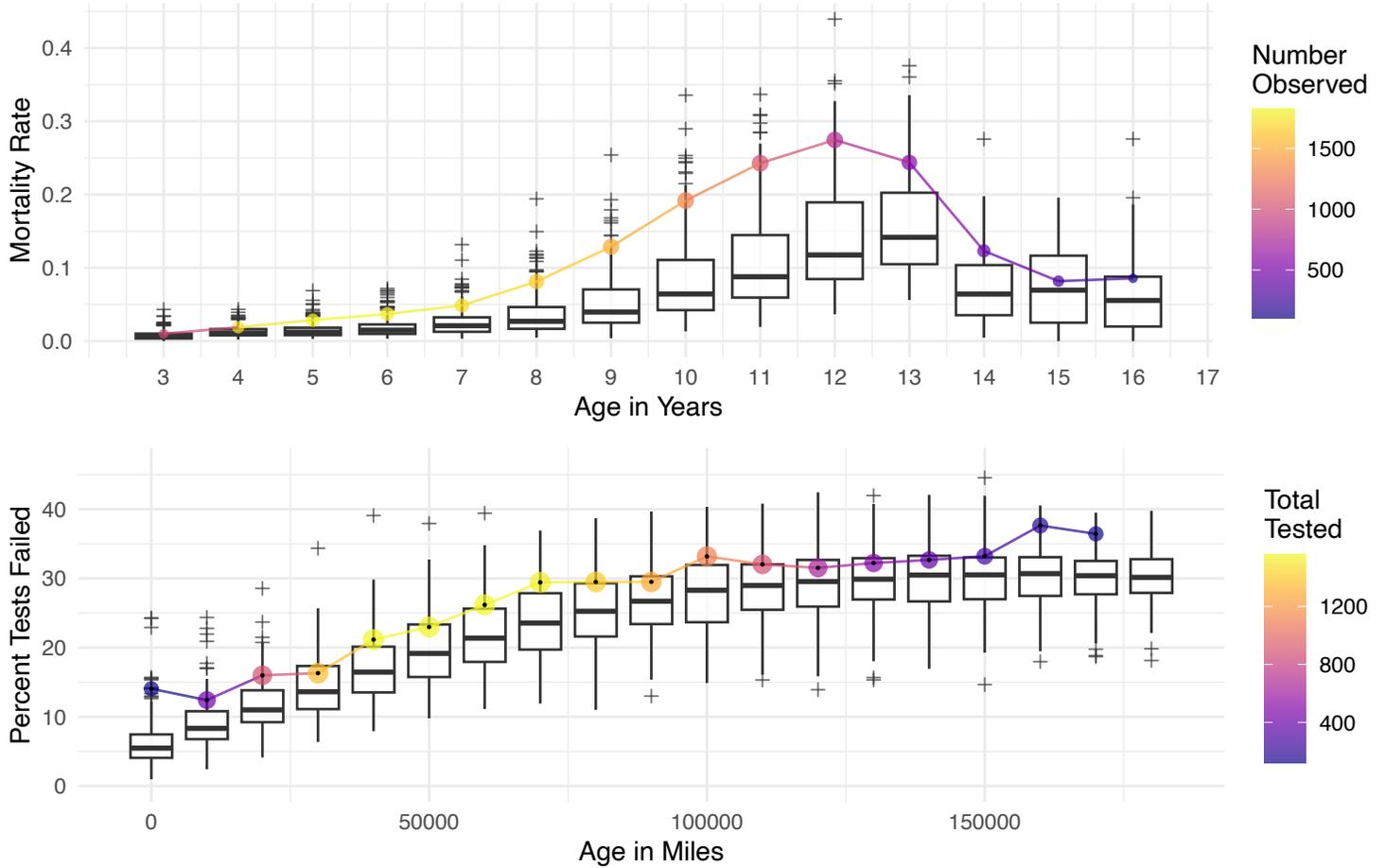

<table>
<tr><td colspan="4" align="center">Mortality rates</td></tr>
<tr><th>Age in Years</th><th>Observed</th><th>Died</th><th>Mortality Rate</th></tr>
<tr><td>3</td><td>984</td><td>10</td><td>0.0102</td></tr>
<tr><td>4</td><td>1739</td><td>33</td><td>0.0190</td></tr>
<tr><td>5</td><td>1822</td><td>52</td><td>0.0285</td></tr>
<tr><td>6</td><td>1778</td><td>66</td><td>0.0371</td></tr>
<tr><td>7</td><td>1706</td><td>83</td><td>0.0487</td></tr>
<tr><td>8</td><td>1623</td><td>132</td><td>0.0813</td></tr>
<tr><td>9</td><td>1490</td><td>192</td><td>0.1290</td></tr>
<tr><td>10</td><td>1293</td><td>248</td><td>0.1920</td></tr>
<tr><td>11</td><td>1038</td><td>252</td><td>0.2430</td></tr>
<tr><td>12</td><td>783</td><td>215</td><td>0.2750</td></tr>
<tr><td>13</td><td>566</td><td>138</td><td>0.2440</td></tr>
<tr><td>14</td><td>406</td><td>50</td><td>0.1230</td></tr>
<tr><td>15</td><td>281</td><td>23</td><td>0.0819</td></tr>
<tr><td>16</td><td>105</td><td>9</td><td>0.0857</td></tr>
</table>

| | Mechanical Reliability Rates | |
|---|---|---|
| Mileage at test | N tested | Pct failed |
| 0 | 121 | 14.0 |
| 10000 | 411 | 12.4 |
| 20000 | 888 | 16.0 |
| 30000 | 1349 | 16.3 |
| 40000 | 1560 | 21.2 |
| 50000 | 1550 | 23.0 |
| 60000 | 1555 | 26.2 |
| 70000 | 1526 | 29.4 |
| 80000 | 1404 | 29.5 |
| 90000 | 1275 | 29.5 |
| 100000 | 1106 | 33.2 |
| 110000 | 890 | 32.0 |
| 120000 | 644 | 31.5 |
| 130000 | 509 | 32.2 |
| 140000 | 349 | 32.7 |
| 150000 | 256 | 33.2 |
| 160000 | 178 | 37.6 |

## Alfa Romeo 159 2006

At 5 years of age, the mortality rate of a Alfa Romeo 159 2006 (manufactured as a Car or Light Van) ranked number 142 out of 225 vehicles of the same age and type (any Car or Light Van constructed in 2006). One is the lowest (or best) and 225 the highest mortality rate. For vehicles reaching 120000 miles, its unreliability score (rate of failing an inspection) ranked 45 out of 207 vehicles of the same age, type, and mileage. One is the highest (or worst) and 207 the lowest rate of failing an inspection.

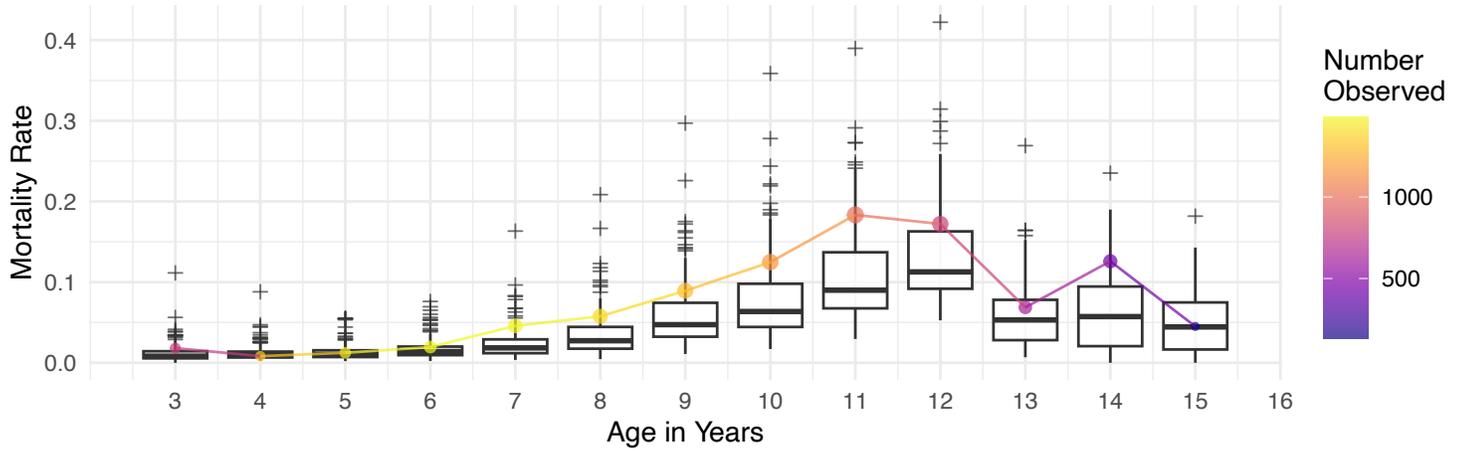

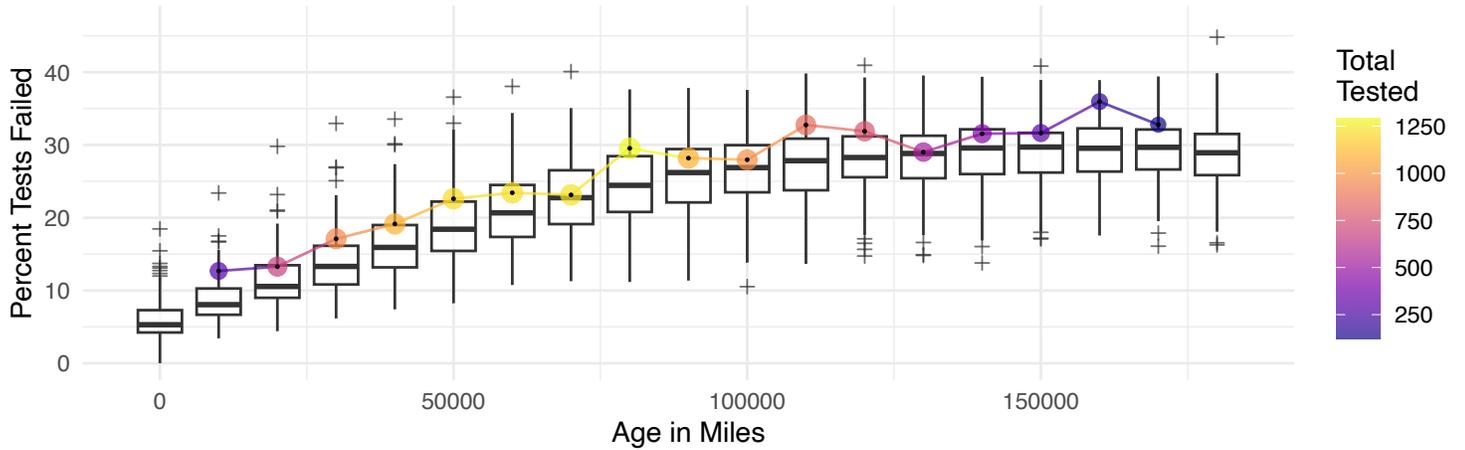

Mortality rates

| Age in Years | Observed | Died | Mortality Rate |
|---|---|---|---|
| 3 | 770 | 14 | 0.01820 |
| 4 | 1328 | 11 | 0.00828 |
| 5 | 1455 | 18 | 0.01240 |
| 6 | 1487 | 29 | 0.01950 |
| 7 | 1462 | 67 | 0.04580 |
| 8 | 1388 | 80 | 0.05760 |
| 9 | 1288 | 115 | 0.08930 |
| 10 | 1154 | 144 | 0.12500 |
| 11 | 998 | 183 | 0.18300 |
| 12 | 802 | 138 | 0.17200 |
| 13 | 626 | 43 | 0.06870 |
| 14 | 429 | 54 | 0.12600 |
| 15 | 133 | 6 | 0.04510 |

Mechanical Reliability Rates

| Mileage at test | N tested | Pct failed |
|---|---|---|
| 10000 | 276 | 12.7 |
| 20000 | 678 | 13.3 |
| 30000 | 959 | 17.1 |
| 40000 | 1097 | 19.1 |
| 50000 | 1212 | 22.6 |
| 60000 | 1225 | 23.4 |
| 70000 | 1232 | 23.1 |
| 80000 | 1293 | 29.5 |
| 90000 | 1089 | 28.2 |
| 100000 | 987 | 28.0 |
| 110000 | 858 | 32.8 |
| 120000 | 756 | 31.9 |
| 130000 | 541 | 29.0 |
| 140000 | 428 | 31.5 |
| 150000 | 294 | 31.6 |
| 160000 | 178 | 36.0 |
| 170000 | 122 | 32.8 |

## Alfa Romeo 159 2007

At 5 years of age, the mortality rate of a Alfa Romeo 159 2007 (manufactured as a Car or Light Van) ranked number 174 out of 219 vehicles of the same age and type (any Car or Light Van constructed in 2007). One is the lowest (or best) and 219 the highest mortality rate. For vehicles reaching 20000 miles, its unreliability score (rate of failing an inspection) ranked 35 out of 214 vehicles of the same age, type, and mileage. One is the highest (or worst) and 214 the lowest rate of failing an inspection.

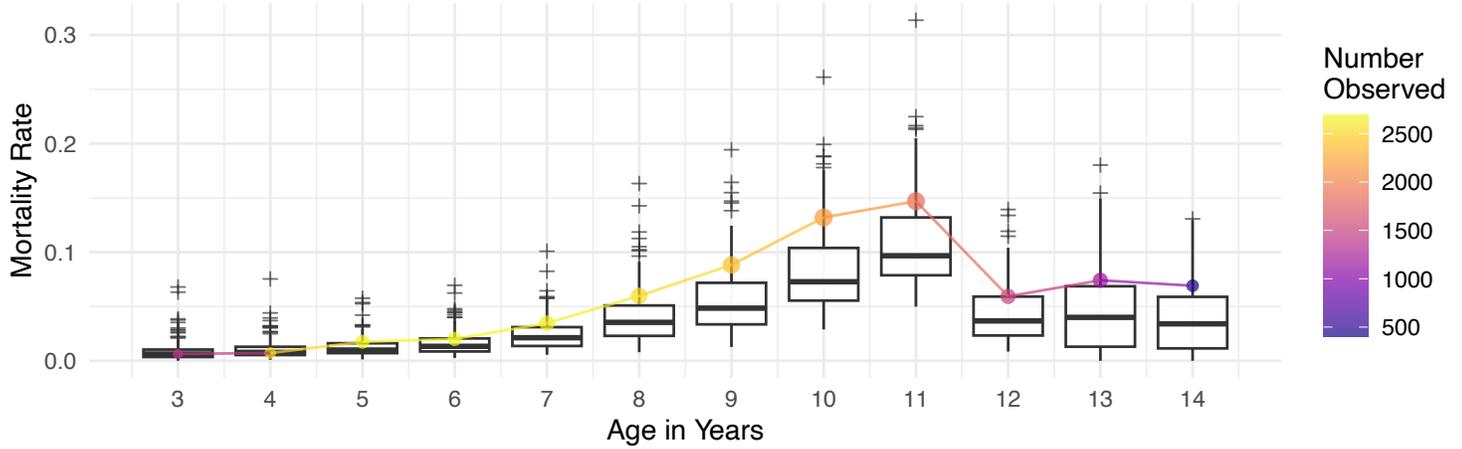

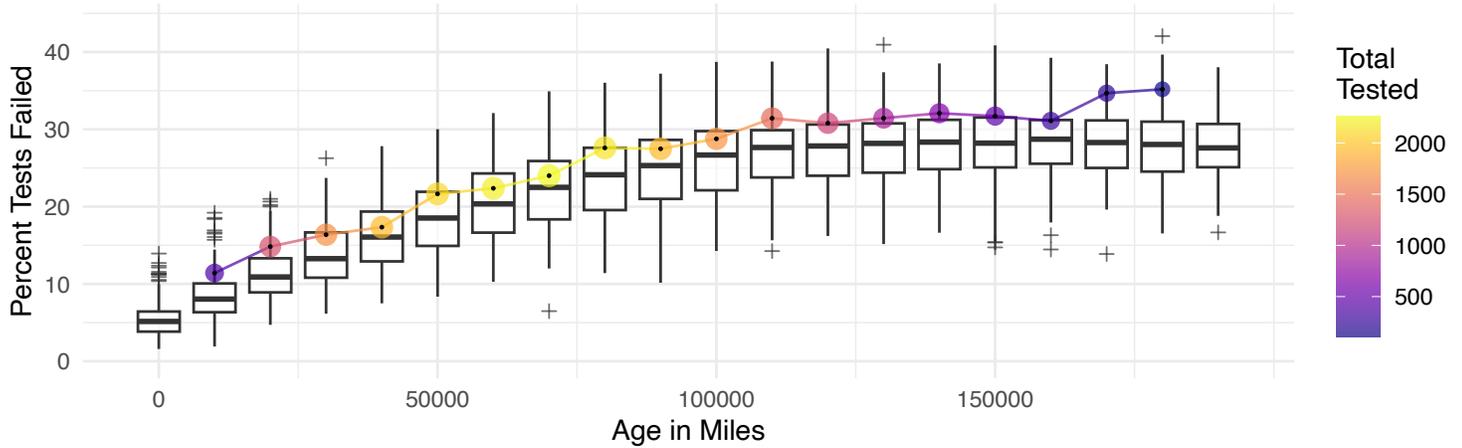

Mortality rates

| Age in Years | Observed | Died | Mortality Rate |
|---|---|---|---|
| 3 | 1421 | 9 | 0.00633 |
| 4 | 2516 | 18 | 0.00715 |
| 5 | 2686 | 47 | 0.01750 |
| 6 | 2683 | 54 | 0.02010 |
| 7 | 2630 | 91 | 0.03460 |
| 8 | 2527 | 151 | 0.05980 |
| 9 | 2355 | 208 | 0.08830 |
| 10 | 2129 | 281 | 0.13200 |
| 11 | 1837 | 270 | 0.14700 |
| 12 | 1483 | 88 | 0.05930 |
| 13 | 1133 | 84 | 0.07410 |
| 14 | 405 | 28 | 0.06910 |

Mechanical Reliability Rates

| Mileage at test | N tested | Pct failed |
|---|---|---|
| 10000 | 368 | 11.4 |
| 20000 | 1274 | 14.8 |
| 30000 | 1704 | 16.4 |
| 40000 | 1943 | 17.3 |
| 50000 | 2097 | 21.6 |
| 60000 | 2208 | 22.4 |
| 70000 | 2262 | 24.0 |
| 80000 | 2167 | 27.6 |
| 90000 | 1932 | 27.5 |
| 100000 | 1724 | 28.8 |
| 110000 | 1454 | 31.4 |
| 120000 | 1182 | 30.8 |
| 130000 | 881 | 31.4 |
| 140000 | 664 | 32.1 |
| 150000 | 473 | 31.7 |
| 160000 | 283 | 31.1 |
| 170000 | 199 | 34.7 |

## Alfa Romeo 159 2008

At 5 years of age, the mortality rate of a Alfa Romeo 159 2008 (manufactured as a Car or Light Van) ranked number 141 out of 218 vehicles of the same age and type (any Car or Light Van constructed in 2008). One is the lowest (or best) and 218 the highest mortality rate. For vehicles reaching 100000 miles, its unreliability score (rate of failing an inspection) ranked 49 out of 206 vehicles of the same age, type, and mileage. One is the highest (or worst) and 206 the lowest rate of failing an inspection.

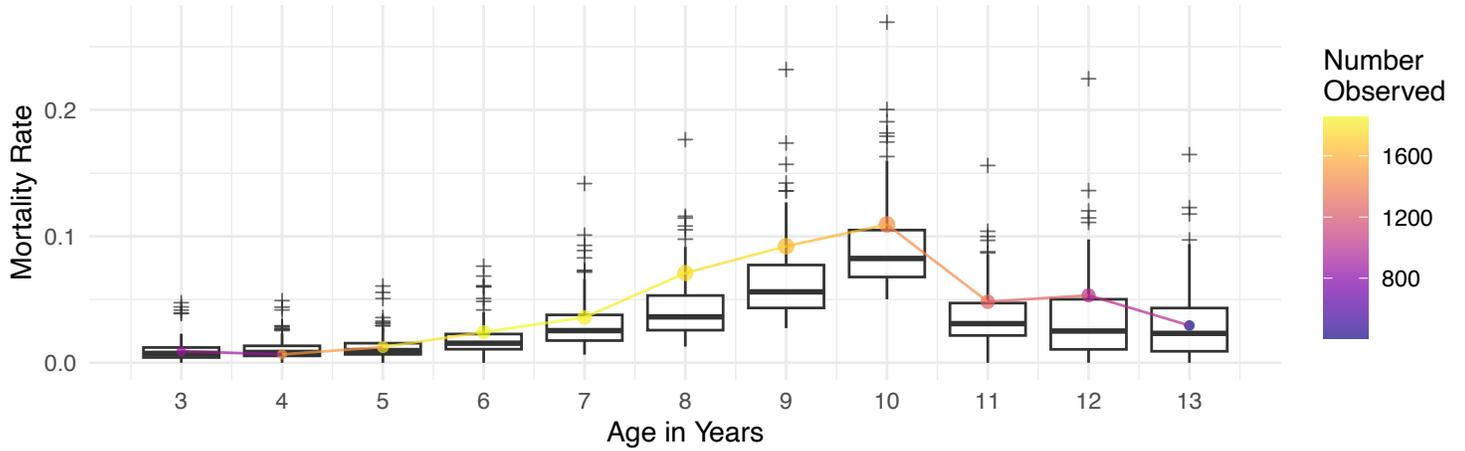

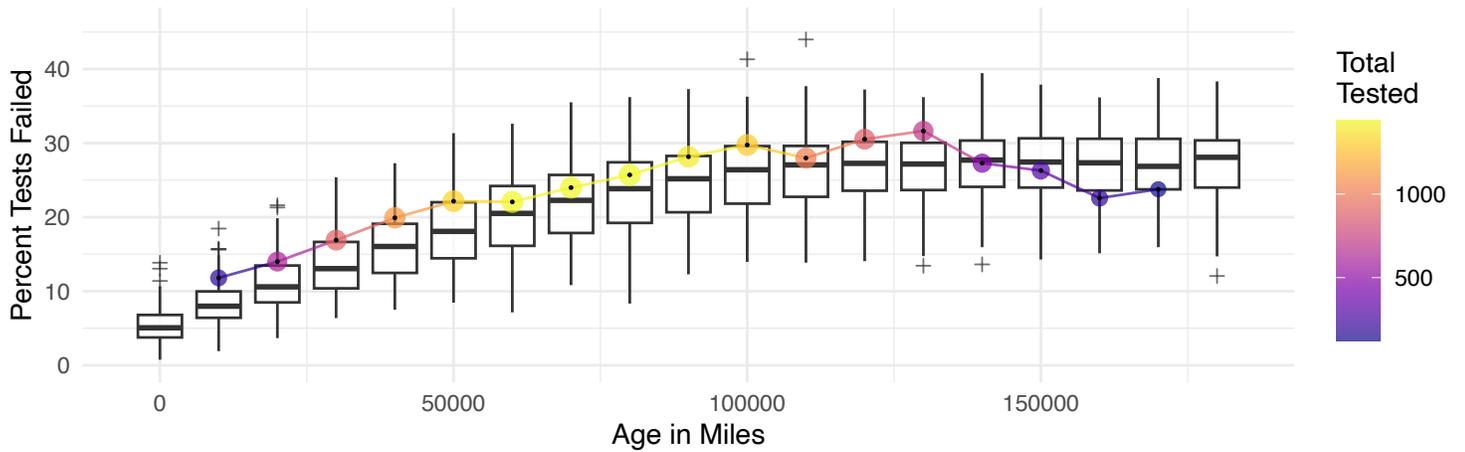

Mortality rates

| Age in Years | Observed | Died | Mortality Rate |
|---|---|---|---|
| 3 | 880 | 8 | 0.00909 |
| 4 | 1505 | 10 | 0.00664 |
| 5 | 1769 | 22 | 0.01240 |
| 6 | 1850 | 45 | 0.02430 |
| 7 | 1811 | 65 | 0.03590 |
| 8 | 1747 | 124 | 0.07100 |
| 9 | 1616 | 149 | 0.09220 |
| 10 | 1454 | 159 | 0.10900 |
| 11 | 1263 | 61 | 0.04830 |
| 12 | 993 | 53 | 0.05340 |
| 13 | 407 | 12 | 0.02950 |

Mechanical Reliability Rates

| Mileage at test | N tested | Pct failed |
|---|---|---|
| 10000 | 178 | 11.8 |
| 20000 | 607 | 14.0 |
| 30000 | 911 | 16.9 |
| 40000 | 1120 | 19.9 |
| 50000 | 1327 | 22.2 |
| 60000 | 1437 | 22.1 |
| 70000 | 1434 | 24.0 |
| 80000 | 1443 | 25.7 |
| 90000 | 1399 | 28.2 |
| 100000 | 1294 | 29.8 |
| 110000 | 1022 | 28.0 |
| 120000 | 881 | 30.5 |
| 130000 | 708 | 31.6 |
| 140000 | 447 | 27.3 |
| 150000 | 289 | 26.3 |
| 160000 | 186 | 22.6 |
| 170000 | 122 | 23.8 |

**Alfa Romeo 159 2009**

At 5 years of age, the mortality rate of a Alfa Romeo 159 2009 (manufactured as a Car or Light Van) ranked number 138 out of 205 vehicles of the same age and type (any Car or Light Van constructed in 2009). One is the lowest (or best) and 205 the highest mortality rate. For vehicles reaching 20000 miles, its unreliability score (rate of failing an inspection) ranked 18 out of 200 vehicles of the same age, type, and mileage. One is the highest (or worst) and 200 the lowest rate of failing an inspection.

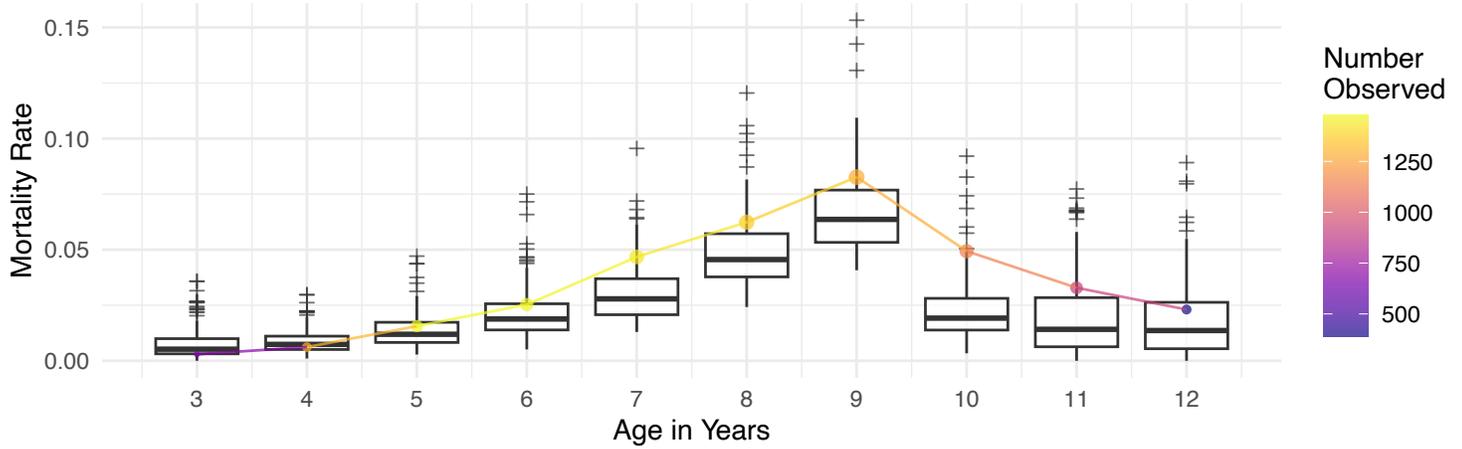

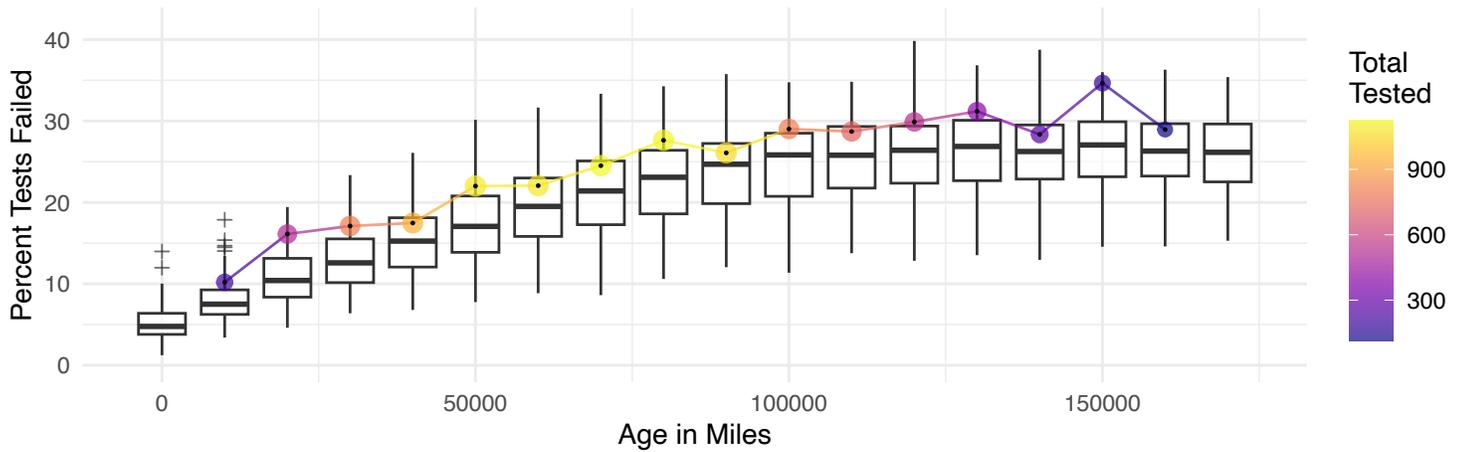

Mortality rates

| Age in Years | Observed | Died | Mortality Rate |
|---|---|---|---|
| 3 | 681 | 2 | 0.00294 |
| 4 | 1296 | 8 | 0.00617 |
| 5 | 1475 | 23 | 0.01560 |
| 6 | 1463 | 37 | 0.02530 |
| 7 | 1433 | 67 | 0.04680 |
| 8 | 1363 | 85 | 0.06240 |
| 9 | 1270 | 105 | 0.08270 |
| 10 | 1135 | 56 | 0.04930 |
| 11 | 913 | 30 | 0.03290 |
| 12 | 390 | 9 | 0.02310 |

Mechanical Reliability Rates

| Mileage at test | N tested | Pct failed |
|---|---|---|
| 10000 | 206 | 10.2 |
| 20000 | 515 | 16.1 |
| 30000 | 801 | 17.1 |
| 40000 | 950 | 17.5 |
| 50000 | 1095 | 22.0 |
| 60000 | 1088 | 22.1 |
| 70000 | 1126 | 24.5 |
| 80000 | 1122 | 27.6 |
| 90000 | 1058 | 26.1 |
| 100000 | 796 | 29.0 |
| 110000 | 721 | 28.7 |
| 120000 | 532 | 29.9 |
| 130000 | 388 | 31.2 |
| 140000 | 261 | 28.4 |
| 150000 | 176 | 34.7 |
| 160000 | 114 | 28.9 |

**Alfa Romeo 159 2010**

At 5 years of age, the mortality rate of a Alfa Romeo 159 2010 (manufactured as a Car or Light Van) ranked number 152 out of 206 vehicles of the same age and type (any Car or Light Van constructed in 2010). One is the lowest (or best) and 206 the highest mortality rate. For vehicles reaching 20000 miles, its unreliability score (rate of failing an inspection) ranked 43 out of 201 vehicles of the same age, type, and mileage. One is the highest (or worst) and 201 the lowest rate of failing an inspection.

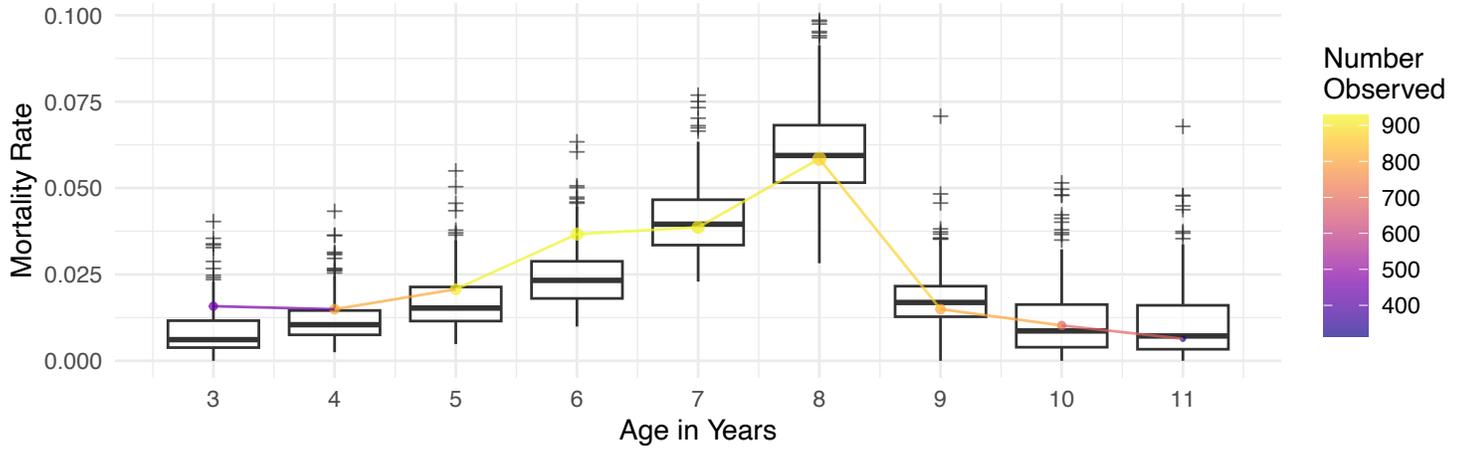

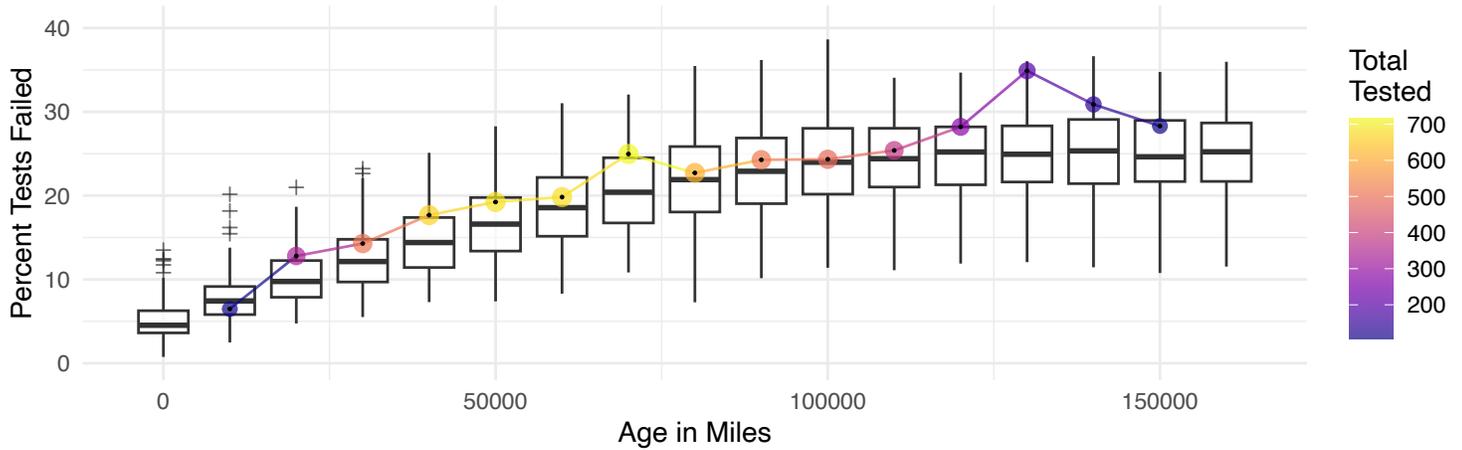

Mortality rates

| Age in Years | Observed | Died | Mortality Rate |
|---|---|---|---|
| 3 | 443 | 7 | 0.01580 |
| 4 | 804 | 12 | 0.01490 |
| 5 | 917 | 19 | 0.02070 |
| 6 | 927 | 34 | 0.03670 |
| 7 | 906 | 35 | 0.03860 |
| 8 | 872 | 51 | 0.05850 |
| 9 | 803 | 12 | 0.01490 |
| 10 | 685 | 7 | 0.01020 |
| 11 | 314 | 2 | 0.00637 |

Mechanical Reliability Rates

| Mileage at test | N tested | Pct failed |
|---|---|---|
| 10000 | 108 | 6.48 |
| 20000 | 336 | 12.80 |
| 30000 | 511 | 14.30 |
| 40000 | 662 | 17.70 |
| 50000 | 686 | 19.20 |
| 60000 | 681 | 19.80 |
| 70000 | 717 | 25.00 |
| 80000 | 612 | 22.70 |
| 90000 | 515 | 24.30 |
| 100000 | 489 | 24.30 |
| 110000 | 382 | 25.40 |
| 120000 | 273 | 28.20 |
| 130000 | 192 | 34.90 |
| 140000 | 136 | 30.90 |
| 150000 | 106 | 28.30 |

**Alfa Romeo Giulia 2017**

At 3 years of age, the mortality rate of a Alfa Romeo Giulia 2017 (manufactured as a Car or Light Van) ranked number 1 out of 247 vehicles of the same age and type (any Car or Light Van constructed in 2017). One is the lowest (or best) and 247 the highest mortality rate. For vehicles reaching 20000 miles, its unreliability score (rate of failing an inspection) ranked 13 out of 240 vehicles of the same age, type, and mileage. One is the highest (or worst) and 240 the lowest rate of failing an inspection.

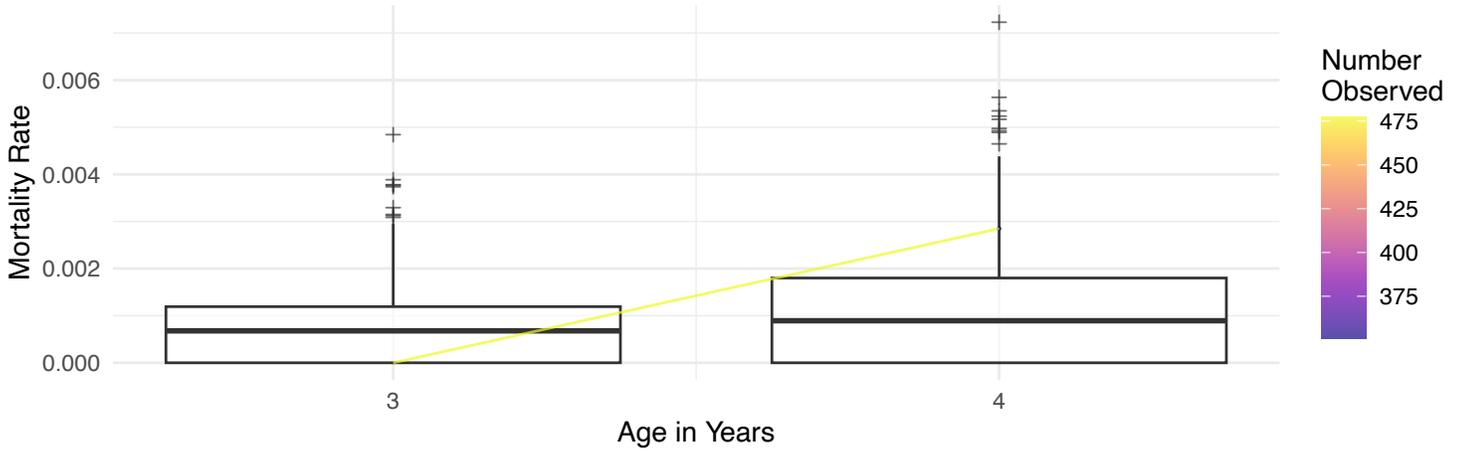

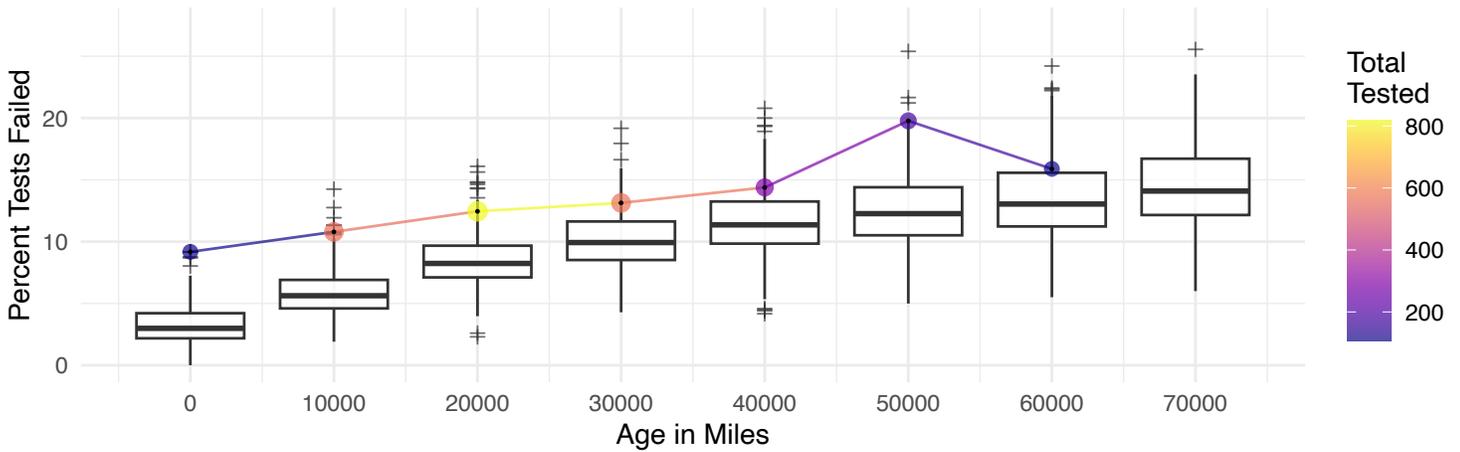

Mortality rates

| Age in Years | Observed | Died | Mortality Rate |
|---|---|---|---|
| 3 | 477 | 0 | 0.00000 |
| 4 | 351 | 1 | 0.00285 |

Mechanical Reliability Rates

| Mileage at test | N tested | Pct failed |
|---|---|---|
| 0 | 109 | 9.17 |
| 10000 | 556 | 10.80 |
| 20000 | 819 | 12.50 |
| 30000 | 571 | 13.10 |
| 40000 | 285 | 14.40 |
| 50000 | 177 | 19.80 |
| 60000 | 107 | 15.90 |

## Alfa Romeo Giulietta 2010

At 5 years of age, the mortality rate of a Alfa Romeo Giulietta 2010 (manufactured as a Car or Light Van) ranked number 103 out of 206 vehicles of the same age and type (any Car or Light Van constructed in 2010). One is the lowest (or best) and 206 the highest mortality rate. For vehicles reaching 120000 miles, its unreliability score (rate of failing an inspection) ranked 108 out of 177 vehicles of the same age, type, and mileage. One is the highest (or worst) and 177 the lowest rate of failing an inspection.

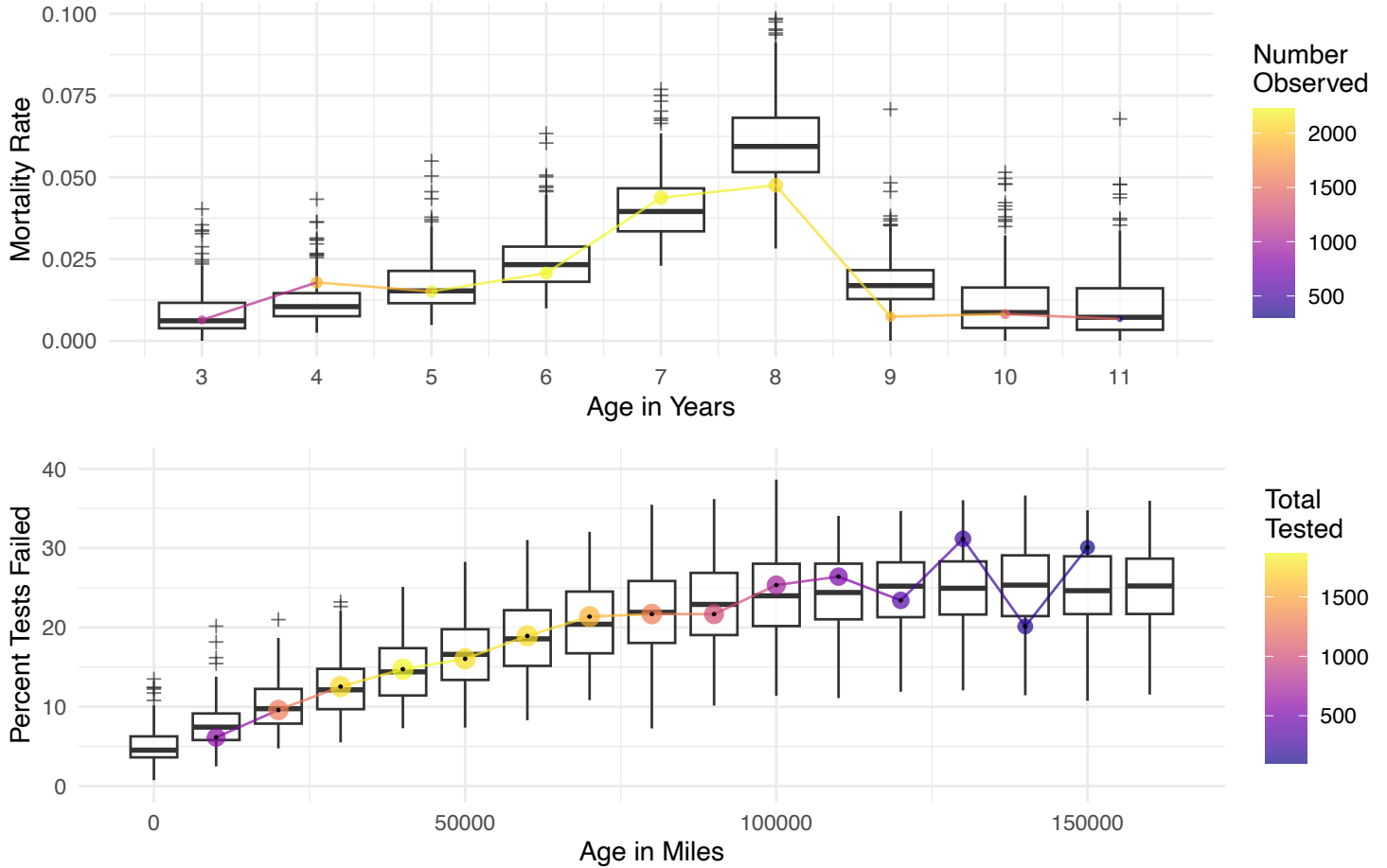

### Mortality rates

| Age in Years | Observed | Died | Mortality Rate |
|---|---|---|---|
| 3 | 1096 | 7 | 0.00639 |
| 4 | 1905 | 34 | 0.01780 |
| 5 | 2187 | 33 | 0.01510 |
| 6 | 2224 | 46 | 0.02070 |
| 7 | 2195 | 96 | 0.04370 |
| 8 | 2102 | 100 | 0.04760 |
| 9 | 1896 | 14 | 0.00738 |
| 10 | 1467 | 12 | 0.00818 |
| 11 | 301 | 2 | 0.00664 |

### Mechanical Reliability Rates

| Mileage at test | N tested | Pct failed |
|---|---|---|
| 10000 | 602 | 6.15 |
| 20000 | 1272 | 9.59 |
| 30000 | 1738 | 12.50 |
| 40000 | 1860 | 14.70 |
| 50000 | 1745 | 16.00 |
| 60000 | 1743 | 18.90 |
| 70000 | 1554 | 21.40 |
| 80000 | 1282 | 21.70 |
| 90000 | 1015 | 21.70 |
| 100000 | 730 | 25.30 |
| 110000 | 515 | 26.40 |
| 120000 | 346 | 23.40 |
| 130000 | 231 | 31.20 |
| 140000 | 159 | 20.10 |
| 150000 | 103 | 30.10 |

## Alfa Romeo Giulietta 2011

At 5 years of age, the mortality rate of a Alfa Romeo Giulietta 2011 (manufactured as a Car or Light Van) ranked number 129 out of 211 vehicles of the same age and type (any Car or Light Van constructed in 2011). One is the lowest (or best) and 211 the highest mortality rate. For vehicles reaching 20000 miles, its unreliability score (rate of failing an inspection) ranked 78 out of 205 vehicles of the same age, type, and mileage. One is the highest (or worst) and 205 the lowest rate of failing an inspection.

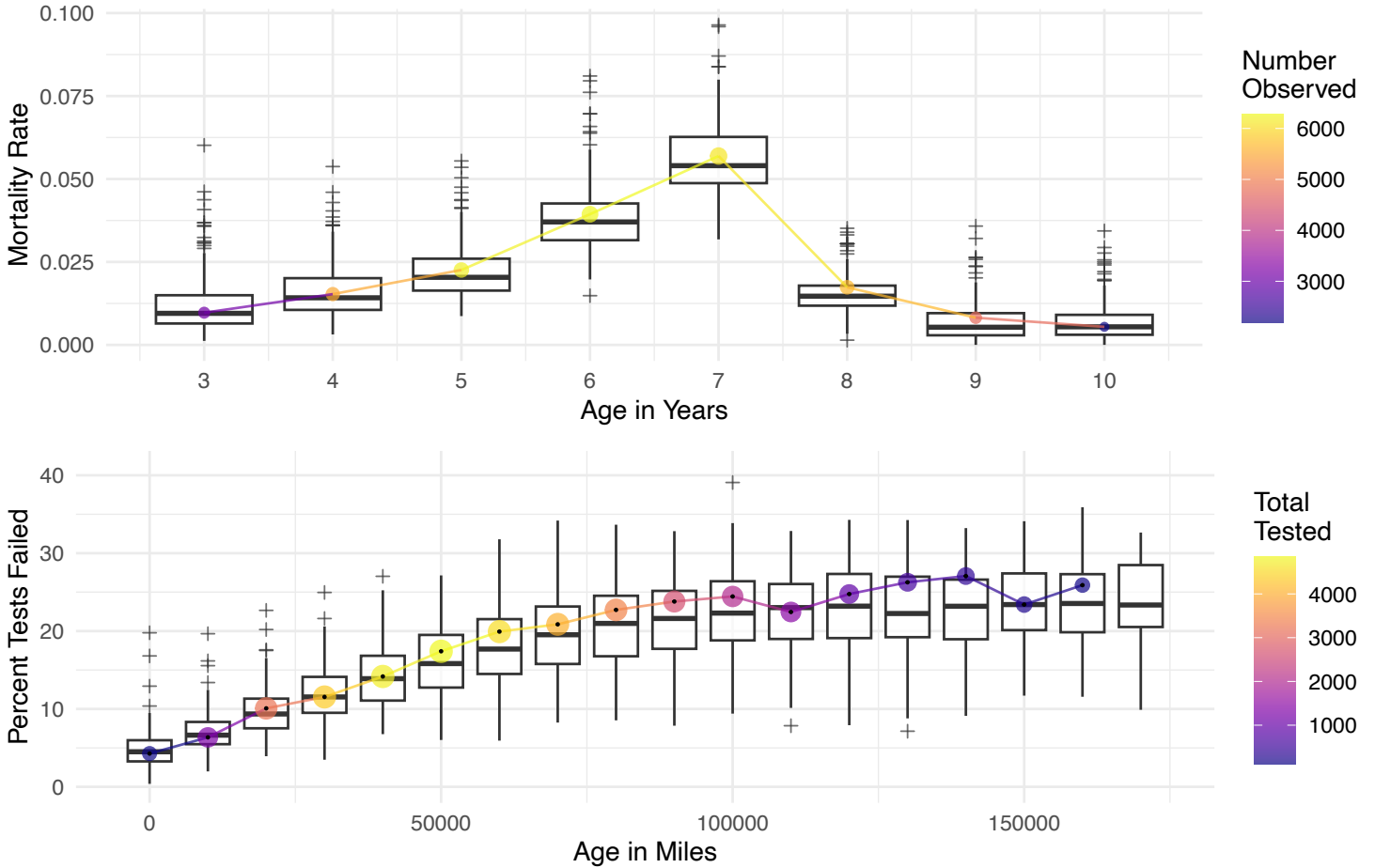

### Mortality rates

| Age in Years | Observed | Died | Mortality Rate |
|---|---|---|---|
| 3 | 3092 | 30 | 0.00970 |
| 4 | 5487 | 84 | 0.01530 |
| 5 | 6201 | 140 | 0.02260 |
| 6 | 6272 | 247 | 0.03940 |
| 7 | 6097 | 347 | 0.05690 |
| 8 | 5642 | 98 | 0.01740 |
| 9 | 4756 | 39 | 0.00820 |
| 10 | 2203 | 12 | 0.00545 |

### Mechanical Reliability Rates

| Mileage at test | N tested | Pct failed |
|---|---|---|
| 0 | 117 | 4.27 |
| 10000 | 1306 | 6.36 |
| 20000 | 3225 | 10.10 |
| 30000 | 4349 | 11.50 |
| 40000 | 4756 | 14.20 |
| 50000 | 4854 | 17.40 |
| 60000 | 4560 | 19.90 |
| 70000 | 4035 | 20.90 |
| 80000 | 3446 | 22.70 |
| 90000 | 2639 | 23.80 |
| 100000 | 2025 | 24.40 |
| 110000 | 1452 | 22.50 |
| 120000 | 929 | 24.80 |
| 130000 | 632 | 26.30 |
| 140000 | 399 | 27.10 |
| 150000 | 235 | 23.40 |
| 160000 | 139 | 25.90 |

**Alfa Romeo Giulietta 2012**

At 5 years of age, the mortality rate of a Alfa Romeo Giulietta 2012 (manufactured as a Car or Light Van) ranked number 40 out of 212 vehicles of the same age and type (any Car or Light Van constructed in 2012). One is the lowest (or best) and 212 the highest mortality rate. For vehicles reaching 20000 miles, its unreliability score (rate of failing an inspection) ranked 92 out of 206 vehicles of the same age, type, and mileage. One is the highest (or worst) and 206 the lowest rate of failing an inspection.

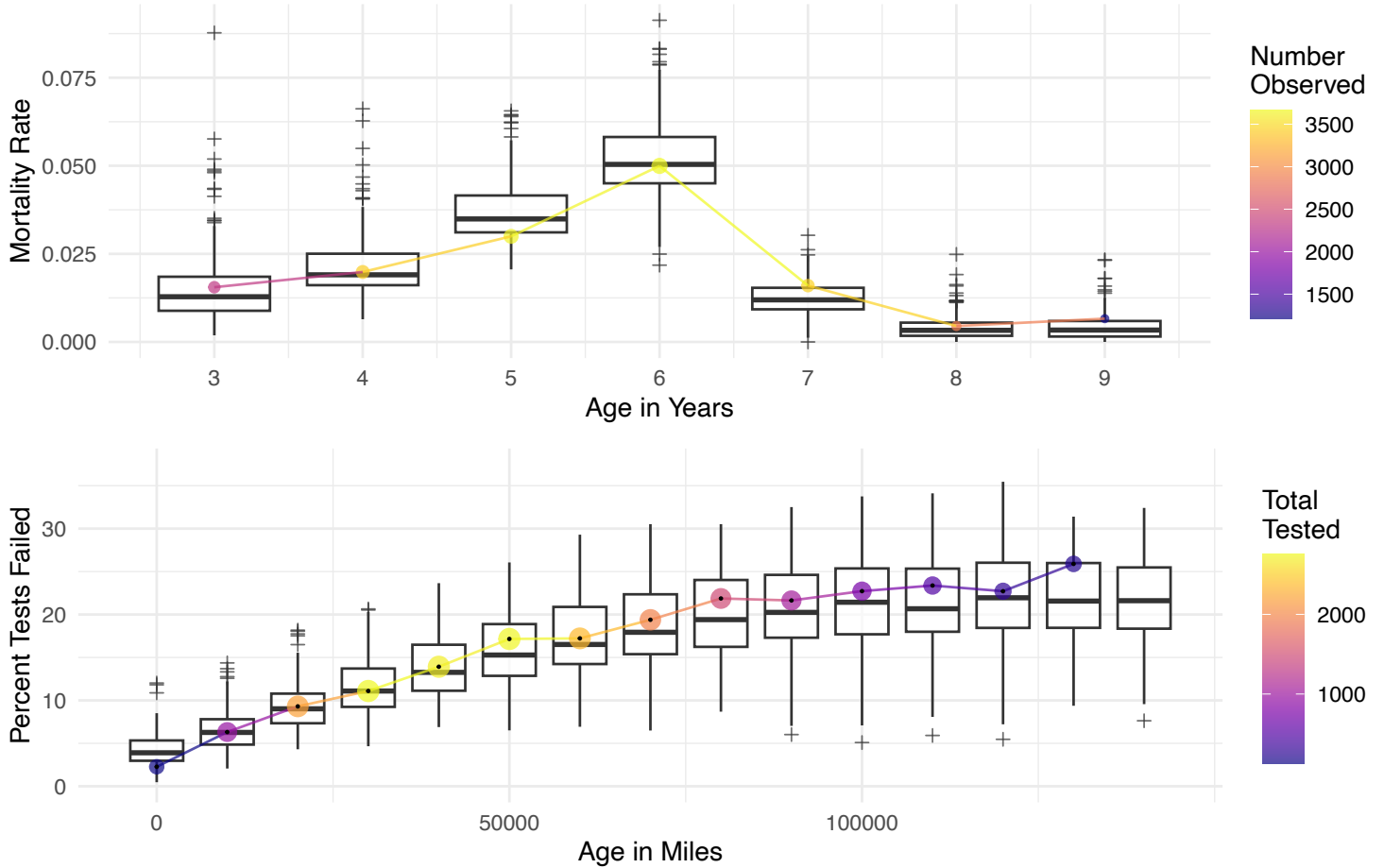

Mortality rates

| Age in Years | Observed | Died | Mortality Rate |
|---|---|---|---|
| 3 | 2318 | 36 | 0.01550 |
| 4 | 3424 | 68 | 0.01990 |
| 5 | 3637 | 109 | 0.03000 |
| 6 | 3663 | 183 | 0.05000 |
| 7 | 3441 | 55 | 0.01600 |
| 8 | 2874 | 13 | 0.00452 |
| 9 | 1210 | 8 | 0.00661 |

Mechanical Reliability Rates

| Mileage at test | N tested | Pct failed |
|---|---|---|
| 0 | 132 | 2.27 |
| 10000 | 966 | 6.31 |
| 20000 | 2182 | 9.30 |
| 30000 | 2749 | 11.10 |
| 40000 | 2694 | 13.90 |
| 50000 | 2734 | 17.20 |
| 60000 | 2346 | 17.20 |
| 70000 | 1955 | 19.40 |
| 80000 | 1477 | 21.90 |
| 90000 | 1063 | 21.60 |
| 100000 | 735 | 22.70 |
| 110000 | 492 | 23.40 |
| 120000 | 295 | 22.70 |
| 130000 | 224 | 25.90 |

## Alfa Romeo Giulietta 2013

At 5 years of age, the mortality rate of a Alfa Romeo Giulietta 2013 (manufactured as a Car or Light Van) ranked number 56 out of 221 vehicles of the same age and type (any Car or Light Van constructed in 2013). One is the lowest (or best) and 221 the highest mortality rate. For vehicles reaching 20000 miles, its unreliability score (rate of failing an inspection) ranked 134 out of 215 vehicles of the same age, type, and mileage. One is the highest (or worst) and 215 the lowest rate of failing an inspection.

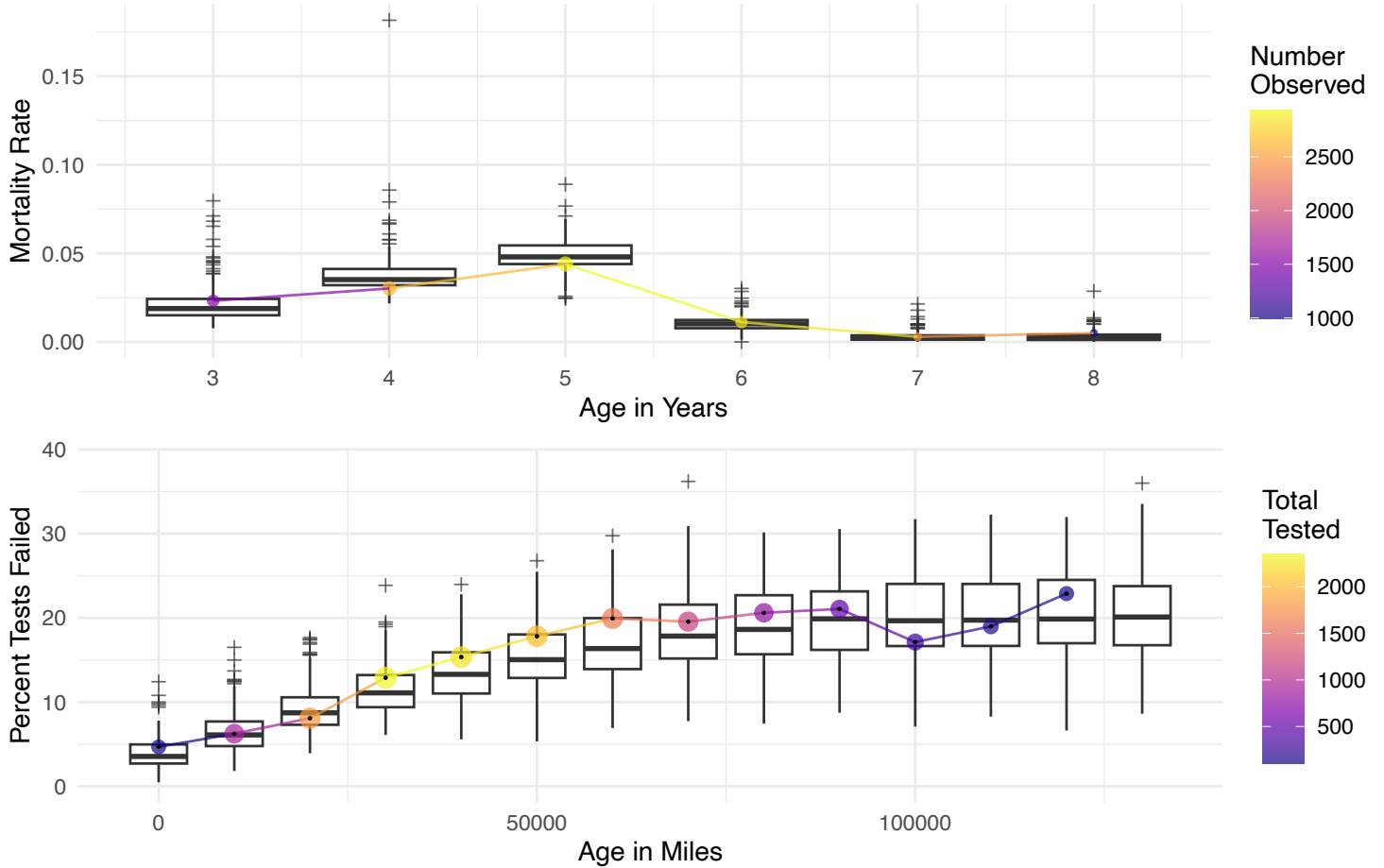

<table>
<tr><td colspan="4" align="center">Mortality rates</td></tr>
</table>

| Age in Years | Observed | Died | Mortality Rate |
|---|---|---|---|
| 3 | 1506 | 35 | 0.02320 |
| 4 | 2615 | 79 | 0.03020 |
| 5 | 2933 | 129 | 0.04400 |
| 6 | 2853 | 32 | 0.01120 |
| 7 | 2425 | 7 | 0.00289 |
| 8 | 992 | 5 | 0.00504 |

Mechanical Reliability Rates

| Mileage at test | N tested | Pct failed |
|---|---|---|
| 0 | 107 | 4.67 |
| 10000 | 912 | 6.25 |
| 20000 | 1819 | 8.08 |
| 30000 | 2343 | 12.90 |
| 40000 | 2241 | 15.40 |
| 50000 | 2015 | 17.80 |
| 60000 | 1595 | 19.90 |
| 70000 | 1166 | 19.60 |
| 80000 | 796 | 20.60 |
| 90000 | 508 | 21.10 |
| 100000 | 292 | 17.10 |
| 110000 | 158 | 19.00 |
| 120000 | 118 | 22.90 |

## Alfa Romeo Giulietta 2014

At 5 years of age, the mortality rate of a Alfa Romeo Giulietta 2014 (manufactured as a Car or Light Van) ranked number 37 out of 236 vehicles of the same age and type (any Car or Light Van constructed in 2014). One is the lowest (or best) and 236 the highest mortality rate. For vehicles reaching 20000 miles, its unreliability score (rate of failing an inspection) ranked 111 out of 230 vehicles of the same age, type, and mileage. One is the highest (or worst) and 230 the lowest rate of failing an inspection.

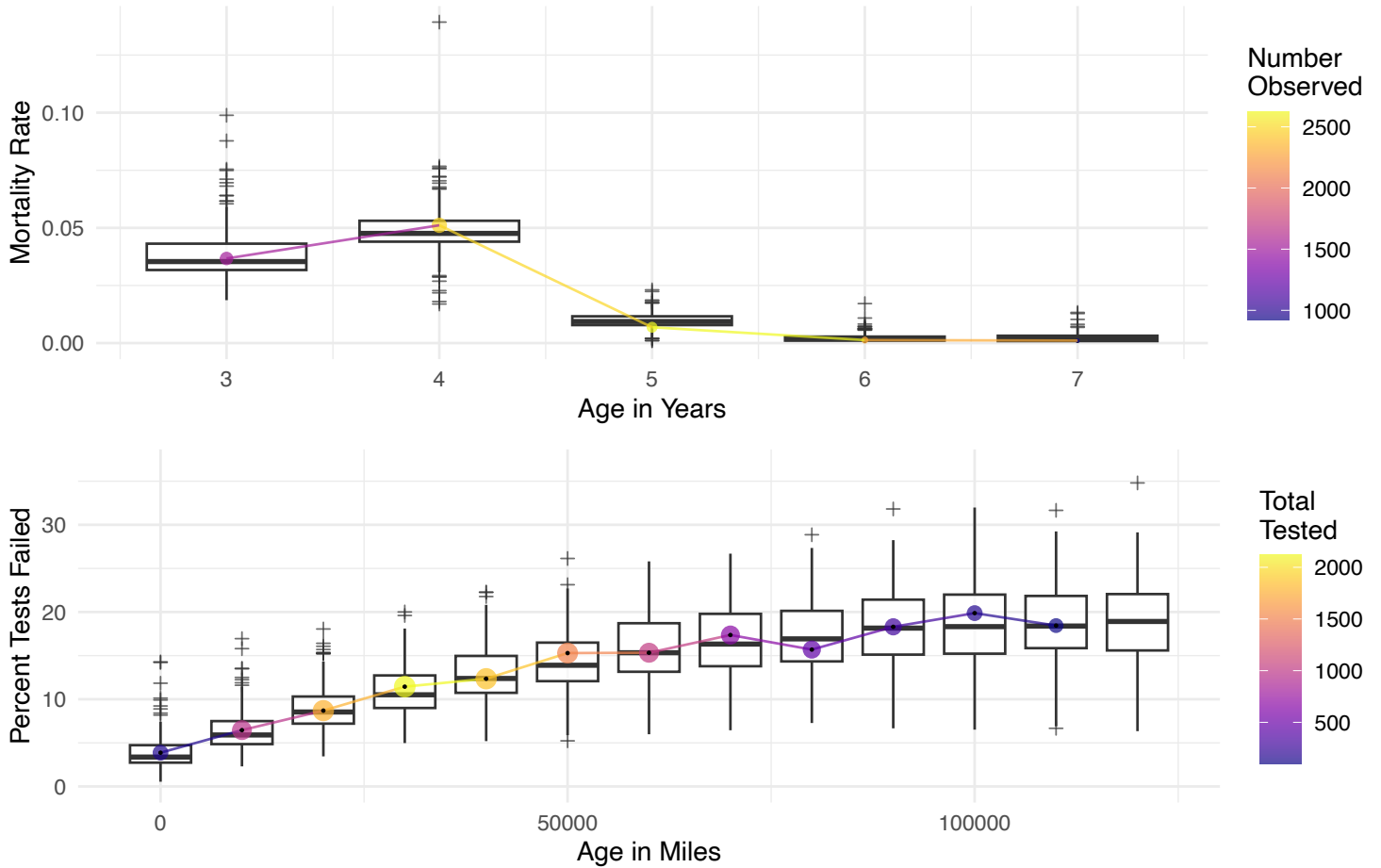

### Mortality rates

| Age in Years | Observed | Died | Mortality Rate |
|---|---|---|---|
| 3 | 1525 | 56 | 0.03670 |
| 4 | 2484 | 127 | 0.05110 |
| 5 | 2621 | 18 | 0.00687 |
| 6 | 2299 | 3 | 0.00130 |
| 7 | 921 | 1 | 0.00109 |

### Mechanical Reliability Rates

| Mileage at test | N tested | Pct failed |
|---|---|---|
| 0 | 155 | 3.87 |
| 10000 | 960 | 6.46 |
| 20000 | 1771 | 8.70 |
| 30000 | 2115 | 11.40 |
| 40000 | 1862 | 12.40 |
| 50000 | 1504 | 15.30 |
| 60000 | 1024 | 15.30 |
| 70000 | 668 | 17.40 |
| 80000 | 446 | 15.70 |
| 90000 | 295 | 18.30 |
| 100000 | 161 | 19.90 |
| 110000 | 103 | 18.40 |

## Alfa Romeo Giulietta 2015

At 5 years of age, the mortality rate of a Alfa Romeo Giulietta 2015 (manufactured as a Car or Light Van) ranked number 186 out of 247 vehicles of the same age and type (any Car or Light Van constructed in 2015). One is the lowest (or best) and 247 the highest mortality rate. For vehicles reaching 20000 miles, its unreliability score (rate of failing an inspection) ranked 91 out of 241 vehicles of the same age, type, and mileage. One is the highest (or worst) and 241 the lowest rate of failing an inspection.

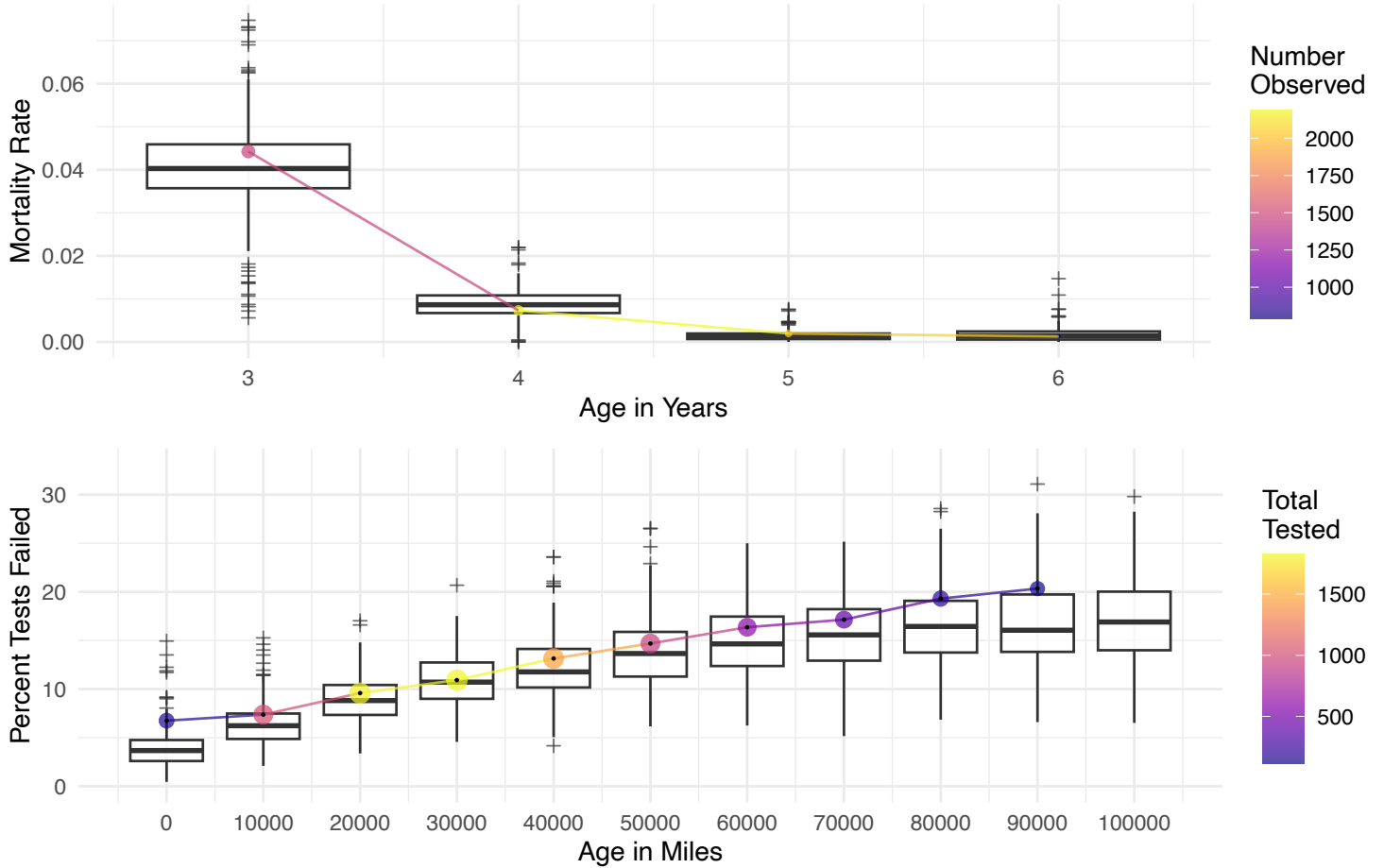

Mortality rates

| Age in Years | Observed | Died | Mortality Rate |
|---|---|---|---|
| 3 | 1469 | 65 | 0.04420 |
| 4 | 2187 | 16 | 0.00732 |
| 5 | 2043 | 4 | 0.00196 |
| 6 | 788 | 1 | 0.00127 |

Mechanical Reliability Rates

| Mileage at test | N tested | Pct failed |
|---|---|---|
| 0 | 163 | 6.75 |
| 10000 | 1004 | 7.37 |
| 20000 | 1794 | 9.59 |
| 30000 | 1822 | 10.90 |
| 40000 | 1453 | 13.10 |
| 50000 | 926 | 14.70 |
| 60000 | 599 | 16.40 |
| 70000 | 385 | 17.10 |
| 80000 | 202 | 19.30 |
| 90000 | 118 | 20.30 |

## Alfa Romeo Giulietta 2016

At 5 years of age, the mortality rate of a Alfa Romeo Giulietta 2016 (manufactured as a Car or Light Van) ranked number 181 out of 252 vehicles of the same age and type (any Car or Light Van constructed in 2016). One is the lowest (or best) and 252 the highest mortality rate. For vehicles reaching 20000 miles, its unreliability score (rate of failing an inspection) ranked 83 out of 246 vehicles of the same age, type, and mileage. One is the highest (or worst) and 246 the lowest rate of failing an inspection.

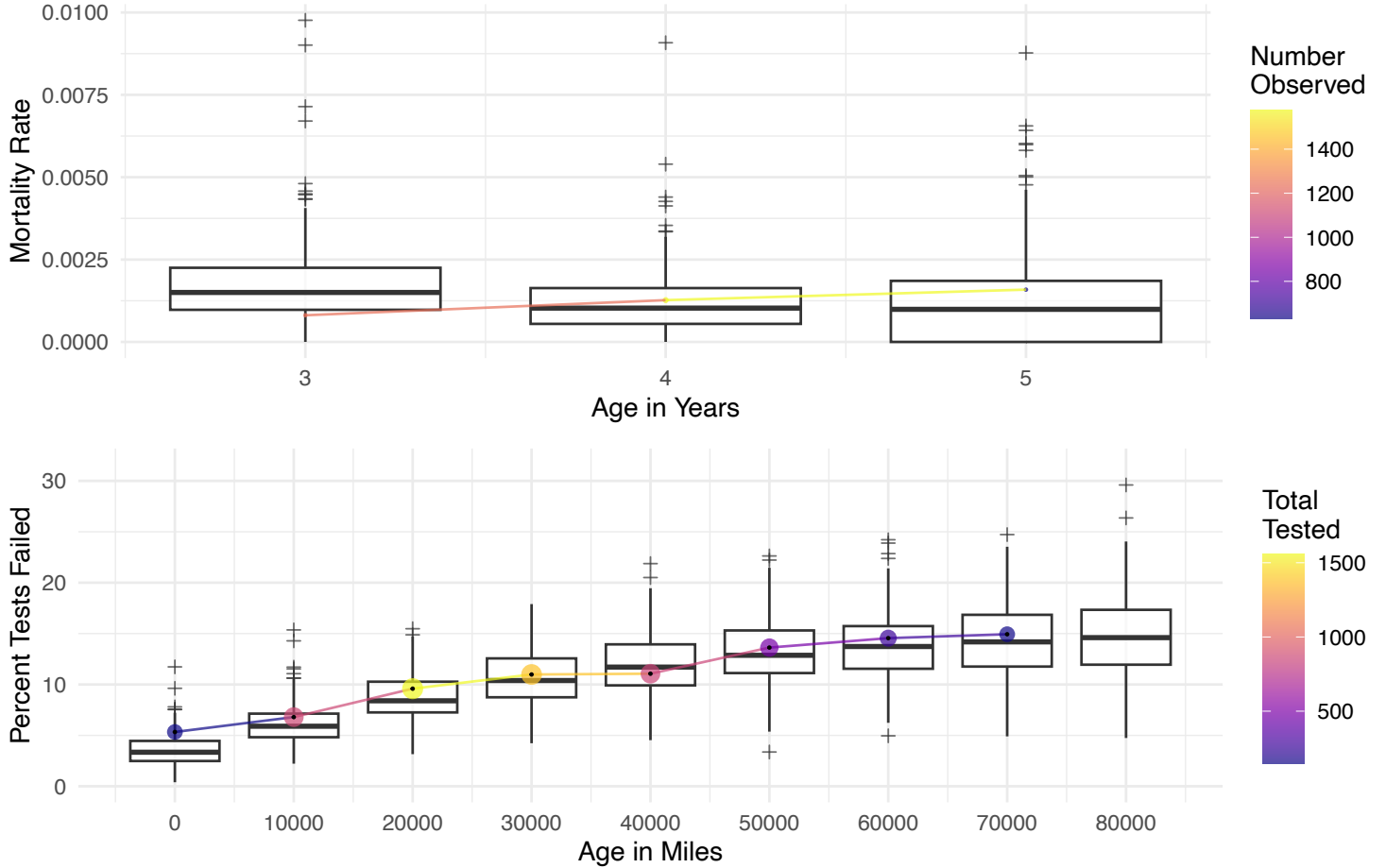

### Mortality rates

| Age in Years | Observed | Died | Mortality Rate |
|---|---|---|---|
| 3 | 1235 | 1 | 0.00081 |
| 4 | 1576 | 2 | 0.00127 |
| 5 | 630 | 1 | 0.00159 |

### Mechanical Reliability Rates

| Mileage at test | N tested | Pct failed |
|---|---|---|
| 0 | 150 | 5.33 |
| 10000 | 868 | 6.80 |
| 20000 | 1554 | 9.59 |
| 30000 | 1375 | 11.00 |
| 40000 | 850 | 11.10 |
| 50000 | 477 | 13.60 |
| 60000 | 268 | 14.60 |
| 70000 | 154 | 14.90 |

**Alfa Romeo Giulietta 2017**

At 3 years of age, the mortality rate of a Alfa Romeo Giulietta 2017 (manufactured as a Car or Light Van) ranked number 2 out of 247 vehicles of the same age and type (any Car or Light Van constructed in 2017). One is the lowest (or best) and 247 the highest mortality rate. For vehicles reaching 20000 miles, its unreliability score (rate of failing an inspection) ranked 47 out of 240 vehicles of the same age, type, and mileage. One is the highest (or worst) and 240 the lowest rate of failing an inspection.

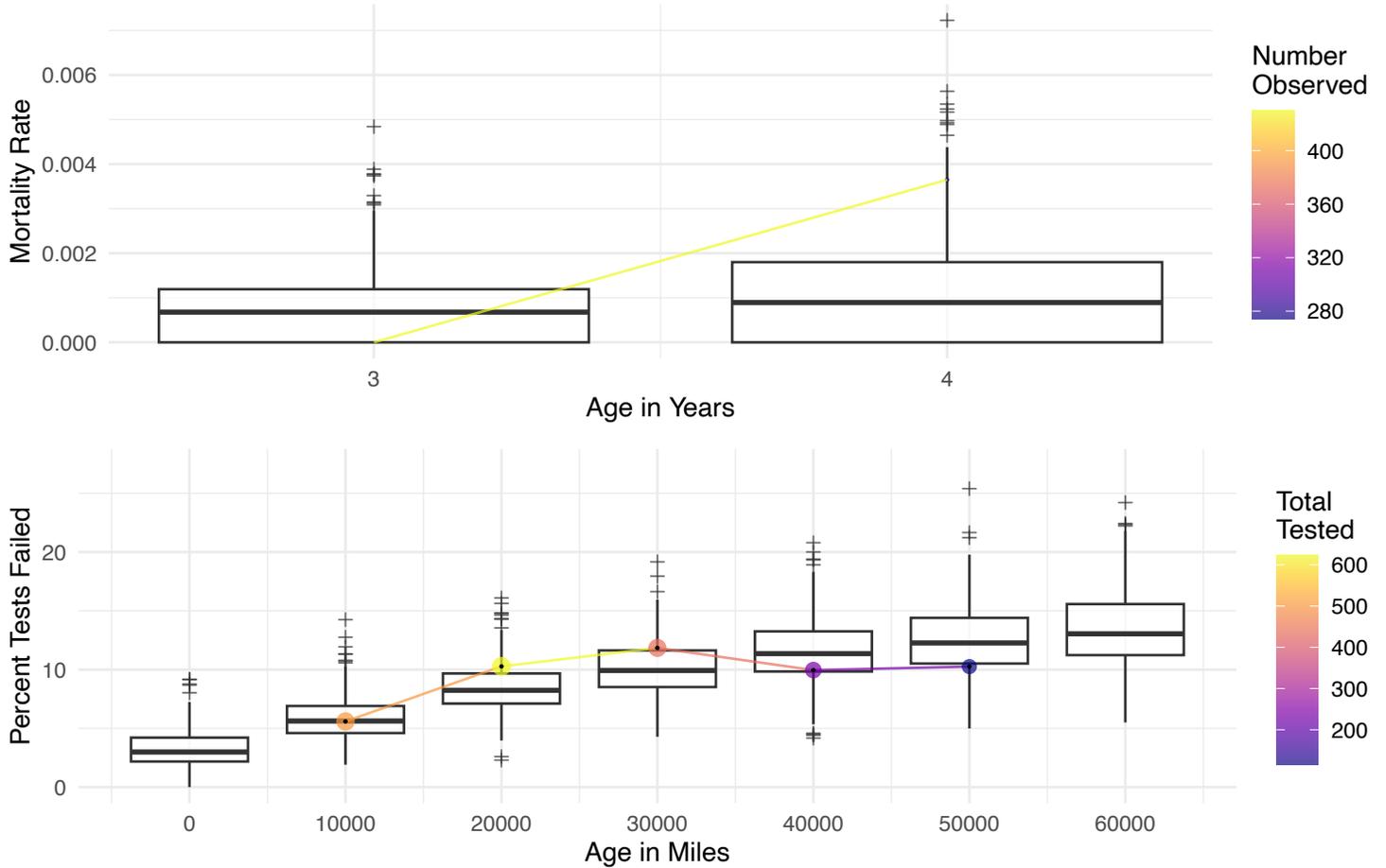

Mortality rates

| Age in Years | Observed | Died | Mortality Rate |
|---|---|---|---|
| 3 | 430 | 0 | 0.00000 |
| 4 | 274 | 1 | 0.00365 |

Mechanical Reliability Rates

| Mileage at test | N tested | Pct failed |
|---|---|---|
| 10000 | 501 | 5.59 |
| 20000 | 623 | 10.30 |
| 30000 | 439 | 11.80 |
| 40000 | 231 | 9.96 |
| 50000 | 117 | 10.30 |

# Alfa Romeo Gt 2004

At 5 years of age, the mortality rate of a Alfa Romeo Gt 2004 (manufactured as a Car or Light Van) ranked number 156 out of 229 vehicles of the same age and type (any Car or Light Van constructed in 2004). One is the lowest (or best) and 229 the highest mortality rate. For vehicles reaching 20000 miles, its unreliability score (rate of failing an inspection) ranked 68 out of 225 vehicles of the same age, type, and mileage. One is the highest (or worst) and 225 the lowest rate of failing an inspection.

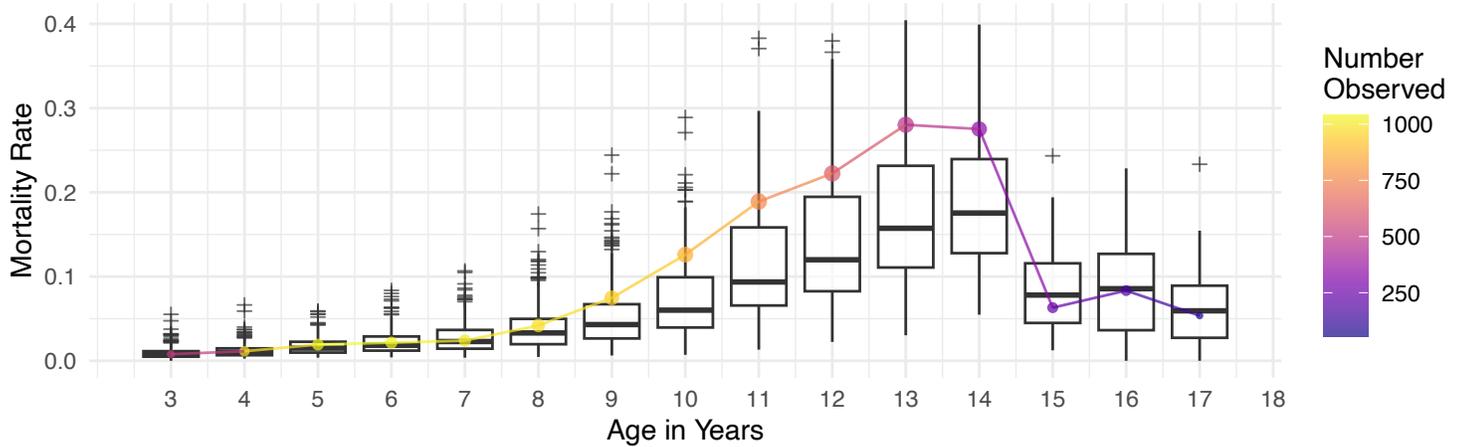

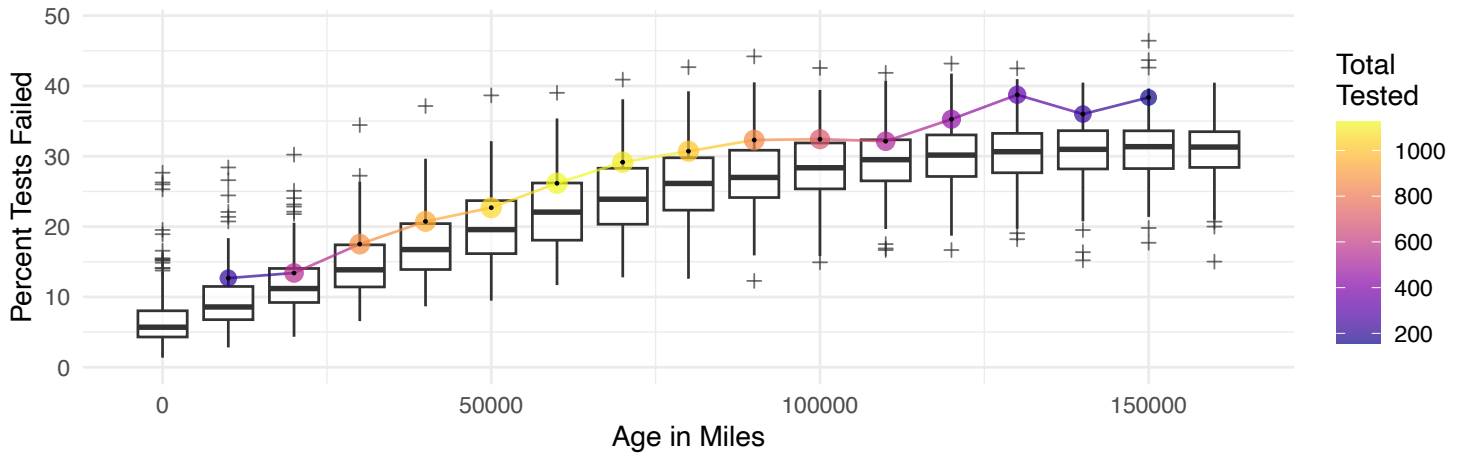

| Mortality rates | | | |
|---|---|---|---|
| Age in Years | Observed | Died | Mortality Rate |
| 3 | 509 | 4 | 0.00786 |
| 4 | 972 | 11 | 0.01130 |
| 5 | 1038 | 20 | 0.01930 |
| 6 | 1022 | 22 | 0.02150 |
| 7 | 1001 | 24 | 0.02400 |
| 8 | 978 | 41 | 0.04190 |
| 9 | 937 | 70 | 0.07470 |
| 10 | 865 | 109 | 0.12600 |
| 11 | 751 | 142 | 0.18900 |
| 12 | 607 | 135 | 0.22200 |
| 13 | 471 | 132 | 0.28000 |
| 14 | 338 | 93 | 0.27500 |
| 15 | 222 | 14 | 0.06310 |
| 16 | 156 | 13 | 0.08330 |
| 17 | 56 | 3 | 0.05360 |

| Mechanical Reliability Rates | | |
|---|---|---|
| Mileage at test | N tested | Pct failed |
| 10000 | 213 | 12.7 |
| 20000 | 552 | 13.4 |
| 30000 | 868 | 17.5 |
| 40000 | 945 | 20.7 |
| 50000 | 1049 | 22.7 |
| 60000 | 1124 | 26.2 |
| 70000 | 1094 | 29.2 |
| 80000 | 1006 | 30.7 |
| 90000 | 851 | 32.3 |
| 100000 | 700 | 32.4 |
| 110000 | 529 | 32.1 |
| 120000 | 448 | 35.3 |
| 130000 | 302 | 38.7 |
| 140000 | 225 | 36.0 |
| 150000 | 159 | 38.4 |

# Alfa Romeo Gt 2005

At 5 years of age, the mortality rate of a Alfa Romeo Gt 2005 (manufactured as a Car or Light Van) ranked number 156 out of 240 vehicles of the same age and type (any Car or Light Van constructed in 2005). One is the lowest (or best) and 240 the highest mortality rate. For vehicles reaching 20000 miles, its unreliability score (rate of failing an inspection) ranked 55 out of 235 vehicles of the same age, type, and mileage. One is the highest (or worst) and 235 the lowest rate of failing an inspection.

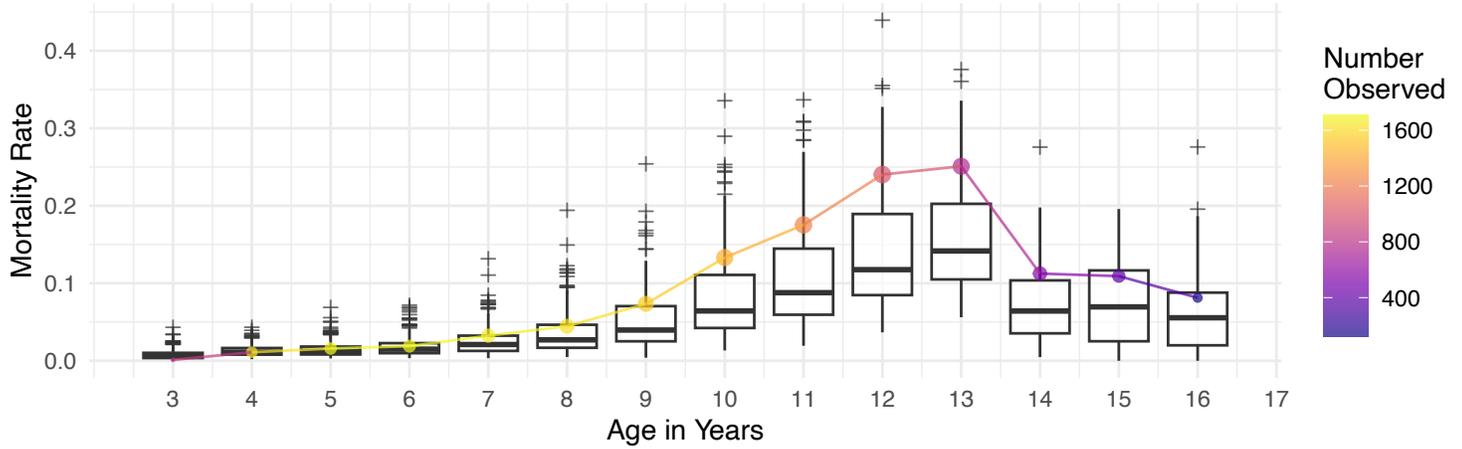

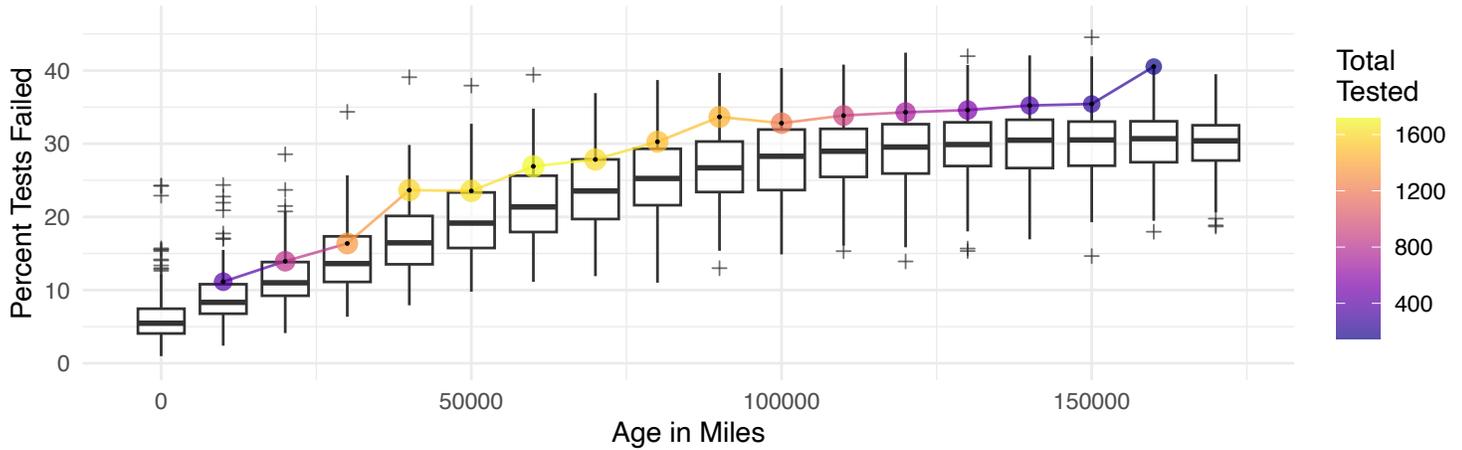

| | Mortality rates | | |
|---|---|---|---|
| Age in Years | Observed | Died | Mortality Rate |
| 3 | 854 | 1 | 0.00117 |
| 4 | 1625 | 18 | 0.01110 |
| 5 | 1704 | 27 | 0.01580 |
| 6 | 1695 | 32 | 0.01890 |
| 7 | 1667 | 54 | 0.03240 |
| 8 | 1614 | 72 | 0.04460 |
| 9 | 1538 | 113 | 0.07350 |
| 10 | 1419 | 189 | 0.13300 |
| 11 | 1226 | 215 | 0.17500 |
| 12 | 1003 | 241 | 0.24000 |
| 13 | 761 | 191 | 0.25100 |
| 14 | 542 | 61 | 0.11300 |
| 15 | 357 | 39 | 0.10900 |
| 16 | 123 | 10 | 0.08130 |

| Mechanical Reliability Rates | | |
|---|---|---|
| Mileage at test | N tested | Pct failed |
| 10000 | 359 | 11.1 |
| 20000 | 825 | 13.9 |
| 30000 | 1338 | 16.4 |
| 40000 | 1581 | 23.7 |
| 50000 | 1631 | 23.5 |
| 60000 | 1717 | 26.9 |
| 70000 | 1609 | 27.8 |
| 80000 | 1500 | 30.3 |
| 90000 | 1450 | 33.7 |
| 100000 | 1201 | 32.8 |
| 110000 | 904 | 33.8 |
| 120000 | 697 | 34.3 |
| 130000 | 526 | 34.6 |
| 140000 | 335 | 35.2 |
| 150000 | 223 | 35.4 |
| 160000 | 148 | 40.5 |

## Alfa Romeo Gt 2006

At 5 years of age, the mortality rate of a Alfa Romeo Gt 2006 (manufactured as a Car or Light Van) ranked number 149 out of 225 vehicles of the same age and type (any Car or Light Van constructed in 2006). One is the lowest (or best) and 225 the highest mortality rate. For vehicles reaching 20000 miles, its unreliability score (rate of failing an inspection) ranked 30 out of 220 vehicles of the same age, type, and mileage. One is the highest (or worst) and 220 the lowest rate of failing an inspection.

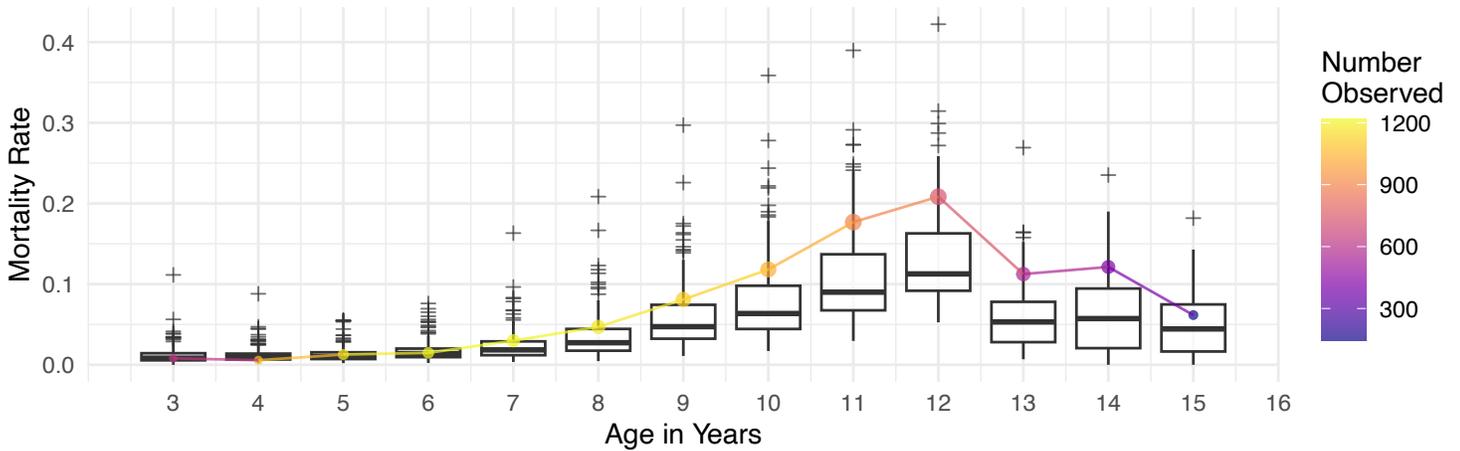

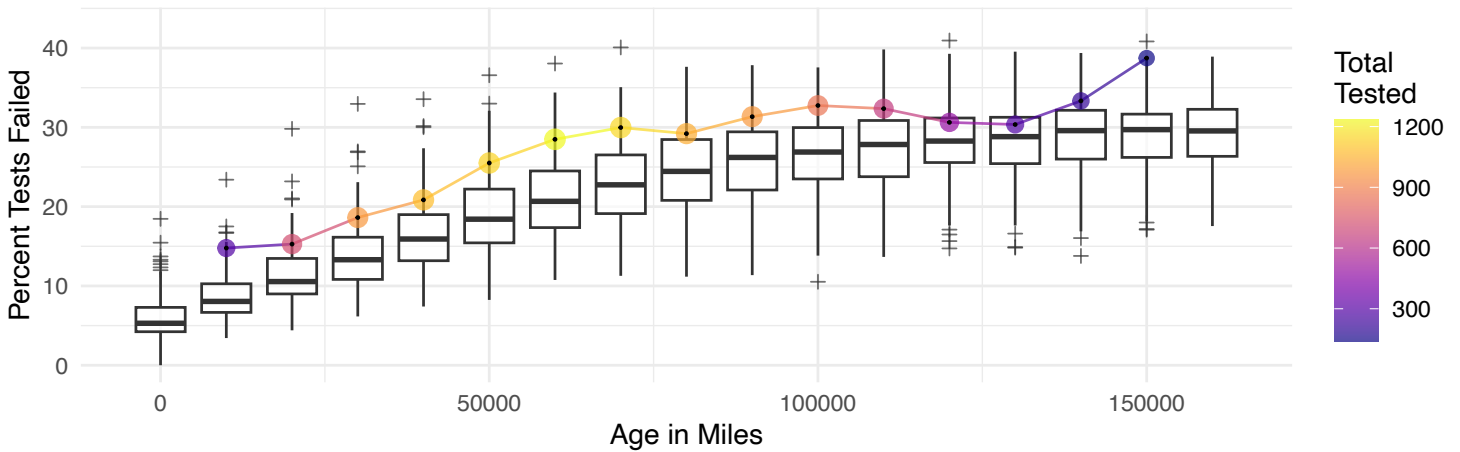

<table>
<tr><td colspan="4" align="center">Mortality rates</td></tr>
<tr><th>Age in Years</th><th>Observed</th><th>Died</th><th>Mortality Rate</th></tr>
<tr><td>3</td><td>622</td><td>5</td><td>0.00804</td></tr>
<tr><td>4</td><td>1054</td><td>6</td><td>0.00569</td></tr>
<tr><td>5</td><td>1180</td><td>15</td><td>0.01270</td></tr>
<tr><td>6</td><td>1215</td><td>18</td><td>0.01480</td></tr>
<tr><td>7</td><td>1205</td><td>36</td><td>0.02990</td></tr>
<tr><td>8</td><td>1170</td><td>55</td><td>0.04700</td></tr>
<tr><td>9</td><td>1112</td><td>90</td><td>0.08090</td></tr>
<tr><td>10</td><td>1024</td><td>121</td><td>0.11800</td></tr>
<tr><td>11</td><td>899</td><td>159</td><td>0.17700</td></tr>
<tr><td>12</td><td>739</td><td>154</td><td>0.20800</td></tr>
<tr><td>13</td><td>560</td><td>63</td><td>0.11200</td></tr>
<tr><td>14</td><td>404</td><td>49</td><td>0.12100</td></tr>
<tr><td>15</td><td>146</td><td>9</td><td>0.06160</td></tr>
</table>

| | Mechanical Reliability Rates | |
|---|---|---|
| Mileage at test | N tested | Pct failed |
| 10000 | 291 | 14.8 |
| 20000 | 720 | 15.3 |
| 30000 | 972 | 18.6 |
| 40000 | 1079 | 20.9 |
| 50000 | 1141 | 25.5 |
| 60000 | 1232 | 28.5 |
| 70000 | 1151 | 30.0 |
| 80000 | 1034 | 29.2 |
| 90000 | 973 | 31.3 |
| 100000 | 861 | 32.8 |
| 110000 | 649 | 32.4 |
| 120000 | 480 | 30.6 |
| 130000 | 280 | 30.4 |
| 140000 | 213 | 33.3 |
| 150000 | 142 | 38.7 |

# Alfa Romeo Gt 2007

At 5 years of age, the mortality rate of a Alfa Romeo Gt 2007 (manufactured as a Car or Light Van) ranked number 105 out of 219 vehicles of the same age and type (any Car or Light Van constructed in 2007). One is the lowest (or best) and 219 the highest mortality rate. For vehicles reaching 40000 miles, its unreliability score (rate of failing an inspection) ranked 44 out of 214 vehicles of the same age, type, and mileage. One is the highest (or worst) and 214 the lowest rate of failing an inspection.

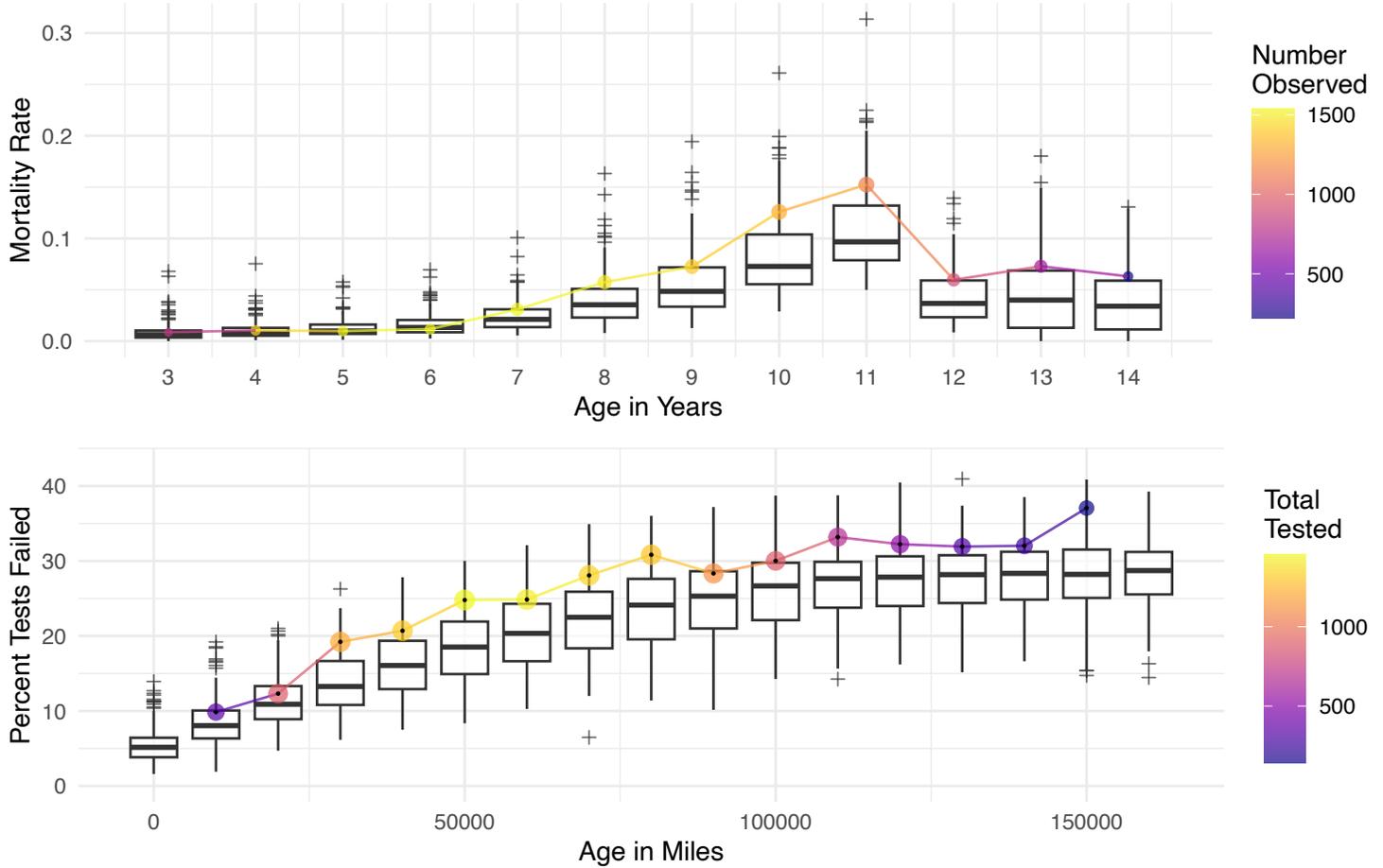

### Mortality rates

| Age in Years | Observed | Died | Mortality Rate |
|---|---|---|---|
| 3 | 807 | 7 | 0.00867 |
| 4 | 1430 | 15 | 0.01050 |
| 5 | 1532 | 15 | 0.00979 |
| 6 | 1535 | 18 | 0.01170 |
| 7 | 1521 | 47 | 0.03090 |
| 8 | 1480 | 85 | 0.05740 |
| 9 | 1393 | 101 | 0.07250 |
| 10 | 1287 | 162 | 0.12600 |
| 11 | 1122 | 171 | 0.15200 |
| 12 | 904 | 54 | 0.05970 |
| 13 | 672 | 49 | 0.07290 |
| 14 | 222 | 14 | 0.06310 |

### Mechanical Reliability Rates

| Mileage at test | N tested | Pct failed |
|---|---|---|
| 10000 | 335 | 9.85 |
| 20000 | 885 | 12.30 |
| 30000 | 1207 | 19.20 |
| 40000 | 1319 | 20.70 |
| 50000 | 1456 | 24.80 |
| 60000 | 1420 | 24.90 |
| 70000 | 1346 | 28.10 |
| 80000 | 1294 | 30.80 |
| 90000 | 1115 | 28.30 |
| 100000 | 896 | 30.00 |
| 110000 | 681 | 33.20 |
| 120000 | 518 | 32.20 |
| 130000 | 307 | 31.90 |
| 140000 | 256 | 32.00 |
| 150000 | 143 | 37.10 |

## Alfa Romeo Gt 2008

At 5 years of age, the mortality rate of a Alfa Romeo Gt 2008 (manufactured as a Car or Light Van) ranked number 58 out of 218 vehicles of the same age and type (any Car or Light Van constructed in 2008). One is the lowest (or best) and 218 the highest mortality rate. For vehicles reaching 20000 miles, its unreliability score (rate of failing an inspection) ranked 51 out of 212 vehicles of the same age, type, and mileage. One is the highest (or worst) and 212 the lowest rate of failing an inspection.

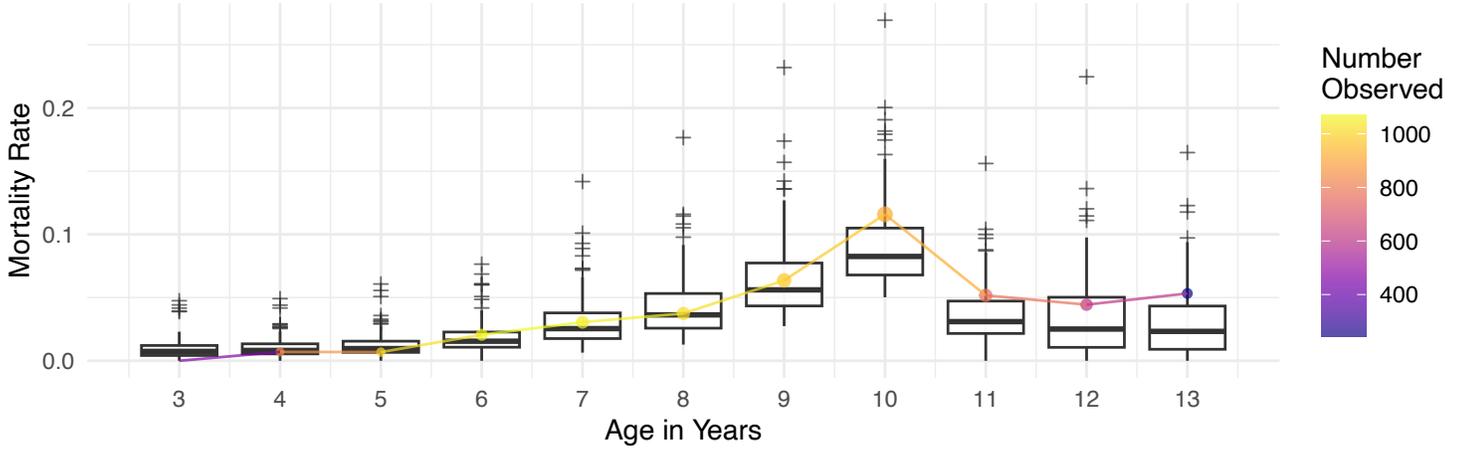

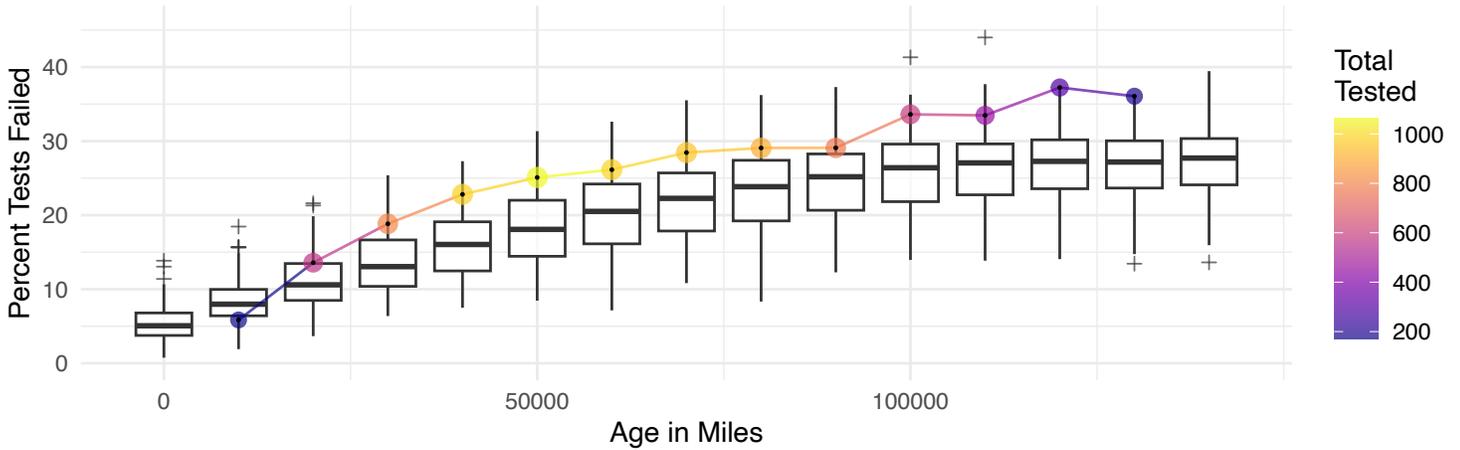

<table>
<tr><td colspan="4" align="center">Mortality rates</td></tr>
<tr><th>Age in Years</th><th>Observed</th><th>Died</th><th>Mortality Rate</th></tr>
<tr><td>3</td><td>461</td><td>0</td><td>0.00000</td></tr>
<tr><td>4</td><td>852</td><td>6</td><td>0.00704</td></tr>
<tr><td>5</td><td>1015</td><td>7</td><td>0.00690</td></tr>
<tr><td>6</td><td>1068</td><td>22</td><td>0.02060</td></tr>
<tr><td>7</td><td>1048</td><td>32</td><td>0.03050</td></tr>
<tr><td>8</td><td>1015</td><td>38</td><td>0.03740</td></tr>
<tr><td>9</td><td>975</td><td>62</td><td>0.06360</td></tr>
<tr><td>10</td><td>915</td><td>106</td><td>0.11600</td></tr>
<tr><td>11</td><td>773</td><td>40</td><td>0.05170</td></tr>
<tr><td>12</td><td>608</td><td>27</td><td>0.04440</td></tr>
<tr><td>13</td><td>244</td><td>13</td><td>0.05330</td></tr>
</table>

Mortality rates

| Age in Years | Observed | Died | Mortality Rate |
|---|---|---|---|
| 3 | 461 | 0 | 0.00000 |
| 4 | 852 | 6 | 0.00704 |
| 5 | 1015 | 7 | 0.00690 |
| 6 | 1068 | 22 | 0.02060 |
| 7 | 1048 | 32 | 0.03050 |
| 8 | 1015 | 38 | 0.03740 |
| 9 | 975 | 62 | 0.06360 |
| 10 | 915 | 106 | 0.11600 |
| 11 | 773 | 40 | 0.05170 |
| 12 | 608 | 27 | 0.04440 |
| 13 | 244 | 13 | 0.05330 |

Mechanical Reliability Rates

| Mileage at test | N tested | Pct failed |
|---|---|---|
| 10000 | 171 | 5.85 |
| 20000 | 567 | 13.60 |
| 30000 | 807 | 18.80 |
| 40000 | 982 | 22.80 |
| 50000 | 1064 | 25.10 |
| 60000 | 987 | 26.10 |
| 70000 | 956 | 28.50 |
| 80000 | 901 | 29.10 |
| 90000 | 770 | 29.10 |
| 100000 | 592 | 33.60 |
| 110000 | 454 | 33.50 |
| 120000 | 290 | 37.20 |
| 130000 | 183 | 36.10 |

**Alfa Romeo Gtv 1997**

At 10 years of age, the mortality rate of a Alfa Romeo Gtv 1997 (manufactured as a Car or Light Van) ranked number 102 out of 187 vehicles of the same age and type (any Car or Light Van constructed in 1997). One is the lowest (or best) and 187 the highest mortality rate. For vehicles reaching 120000 miles, its unreliability score (rate of failing an inspection) ranked 55 out of 167 vehicles of the same age, type, and mileage. One is the highest (or worst) and 167 the lowest rate of failing an inspection.

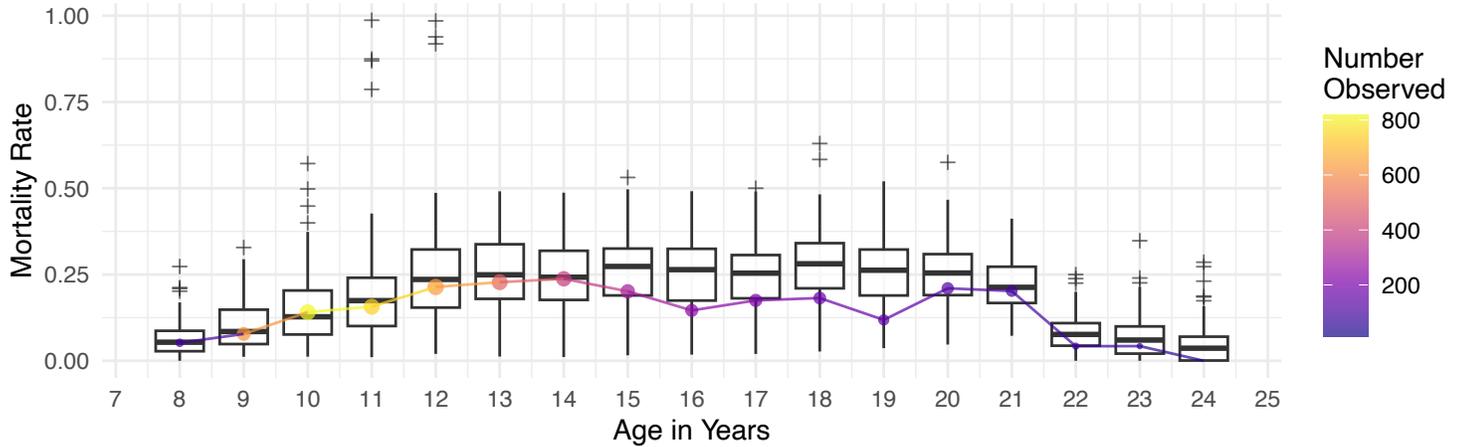

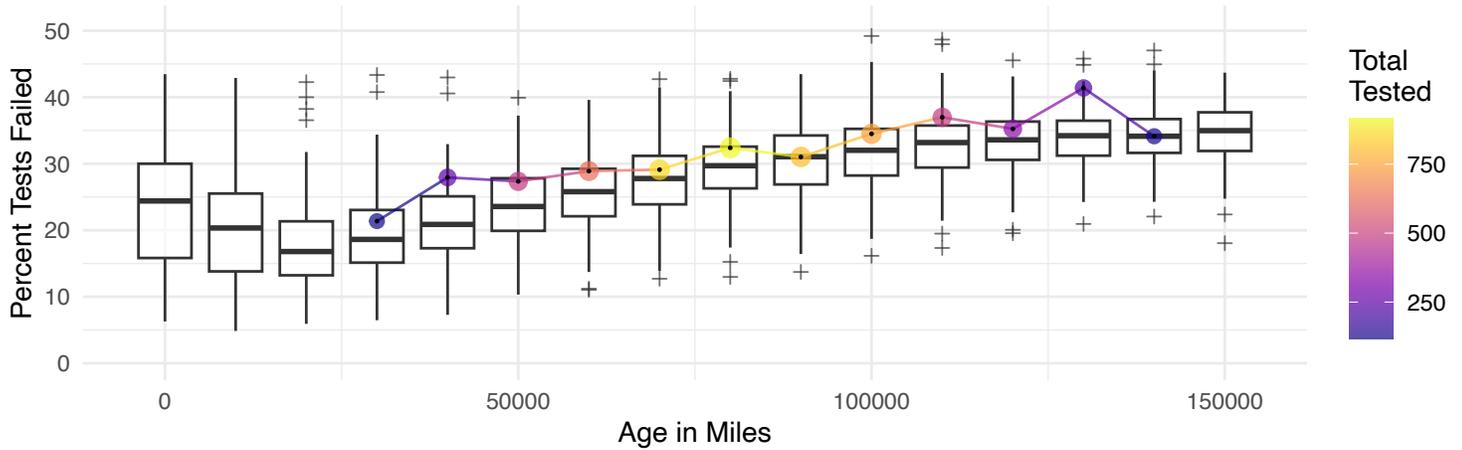

| | Mortality rates | | |
|---|---|---|---|
| Age in Years | Observed | Died | Mortality Rate |
| 8 | 96 | 5 | 0.0521 |
| 9 | 617 | 48 | 0.0778 |
| 10 | 816 | 115 | 0.1410 |
| 11 | 740 | 116 | 0.1570 |
| 12 | 631 | 135 | 0.2140 |
| 13 | 496 | 113 | 0.2280 |
| 14 | 383 | 91 | 0.2380 |
| 15 | 293 | 59 | 0.2010 |
| 16 | 233 | 34 | 0.1460 |
| 17 | 200 | 35 | 0.1750 |
| 18 | 165 | 30 | 0.1820 |
| 19 | 135 | 16 | 0.1190 |
| 20 | 119 | 25 | 0.2100 |
| 21 | 94 | 19 | 0.2020 |
| 22 | 71 | 3 | 0.0423 |
| 23 | 47 | 2 | 0.0426 |
| 24 | 13 | 0 | 0.0000 |

| Mechanical Reliability Rates | | |
|---|---|---|
| Mileage at test | N tested | Pct failed |
| 30000 | 117 | 21.4 |
| 40000 | 247 | 27.9 |
| 50000 | 479 | 27.3 |
| 60000 | 623 | 28.9 |
| 70000 | 859 | 29.1 |
| 80000 | 917 | 32.4 |
| 90000 | 806 | 31.0 |
| 100000 | 745 | 34.5 |
| 110000 | 484 | 37.0 |
| 120000 | 349 | 35.2 |
| 130000 | 215 | 41.4 |
| 140000 | 126 | 34.1 |

## Alfa Romeo Gtv 1998

At 10 years of age, the mortality rate of a Alfa Romeo Gtv 1998 (manufactured as a Car or Light Van) ranked number 102 out of 196 vehicles of the same age and type (any Car or Light Van constructed in 1998). One is the lowest (or best) and 196 the highest mortality rate. For vehicles reaching 120000 miles, its unreliability score (rate of failing an inspection) ranked 51 out of 172 vehicles of the same age, type, and mileage. One is the highest (or worst) and 172 the lowest rate of failing an inspection.

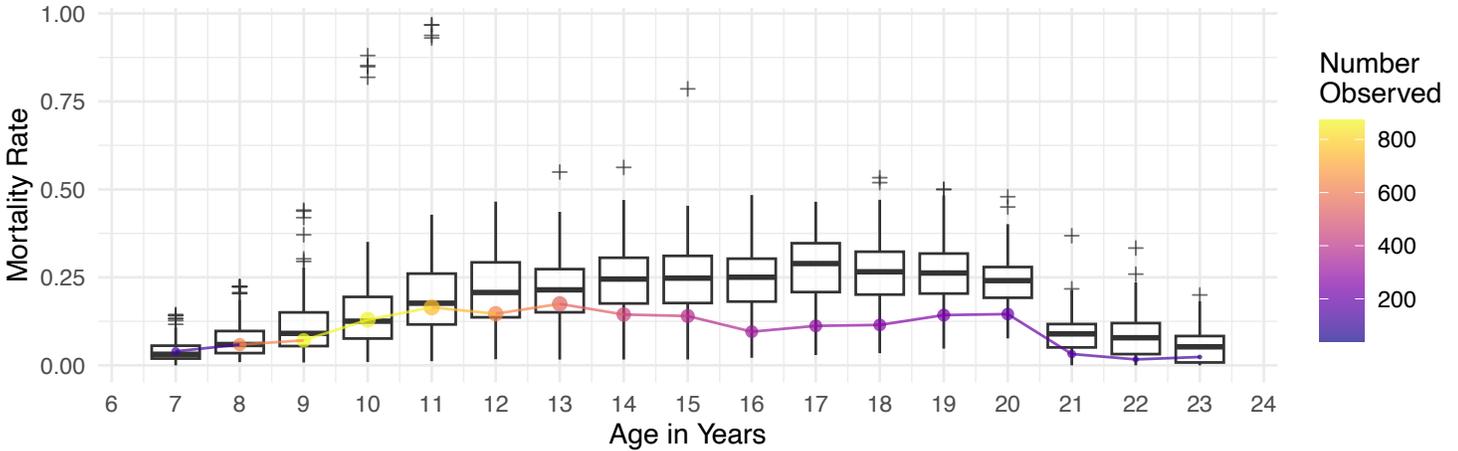

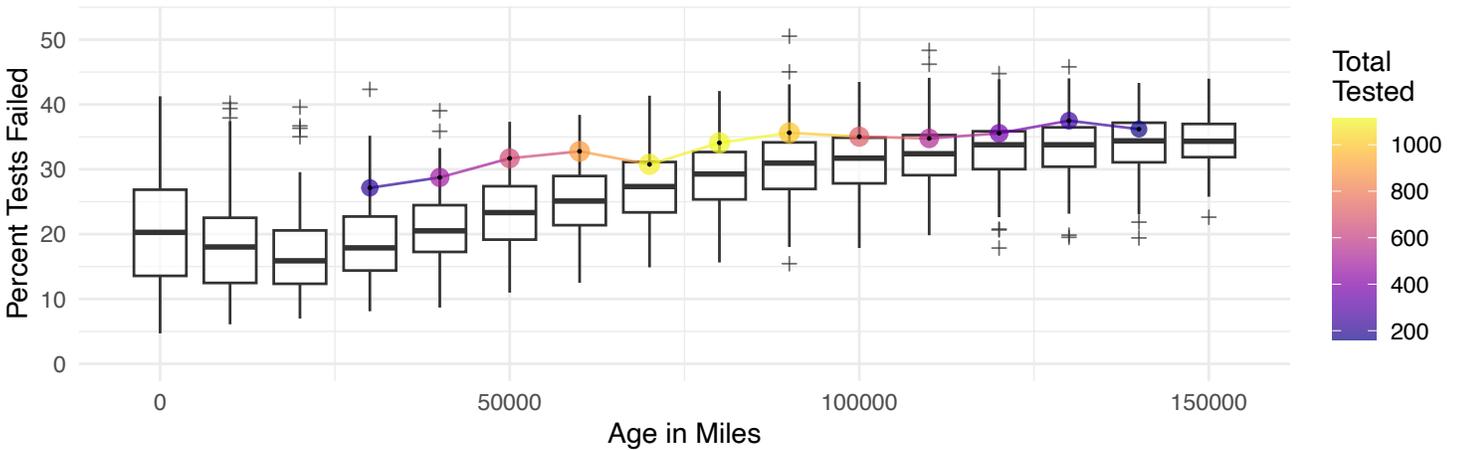

Mortality rates

| Age in Years | Observed | Died | Mortality Rate |
|---|---|---|---|
| 7 | 125 | 5 | 0.0400 |
| 8 | 627 | 37 | 0.0590 |
| 9 | 872 | 62 | 0.0711 |
| 10 | 858 | 111 | 0.1290 |
| 11 | 752 | 124 | 0.1650 |
| 12 | 627 | 92 | 0.1470 |
| 13 | 538 | 94 | 0.1750 |
| 14 | 443 | 64 | 0.1440 |
| 15 | 378 | 53 | 0.1400 |
| 16 | 324 | 31 | 0.0957 |
| 17 | 294 | 33 | 0.1120 |
| 18 | 261 | 30 | 0.1150 |
| 19 | 231 | 33 | 0.1430 |
| 20 | 199 | 29 | 0.1460 |
| 21 | 155 | 5 | 0.0323 |
| 22 | 118 | 2 | 0.0169 |
| 23 | 42 | 1 | 0.0238 |

Mechanical Reliability Rates

| Mileage at test | N tested | Pct failed |
|---|---|---|
| 30000 | 210 | 27.1 |
| 40000 | 487 | 28.7 |
| 50000 | 650 | 31.7 |
| 60000 | 891 | 32.8 |
| 70000 | 1089 | 30.8 |
| 80000 | 1112 | 34.1 |
| 90000 | 999 | 35.6 |
| 100000 | 705 | 35.0 |
| 110000 | 535 | 34.8 |
| 120000 | 360 | 35.6 |
| 130000 | 240 | 37.5 |
| 140000 | 163 | 36.2 |

## Alfa Romeo Gtv 1999

At 10 years of age, the mortality rate of a Alfa Romeo Gtv 1999 (manufactured as a Car or Light Van) ranked number 94 out of 201 vehicles of the same age and type (any Car or Light Van constructed in 1999). One is the lowest (or best) and 201 the highest mortality rate. For vehicles reaching 120000 miles, its unreliability score (rate of failing an inspection) ranked 28 out of 181 vehicles of the same age, type, and mileage. One is the highest (or worst) and 181 the lowest rate of failing an inspection.

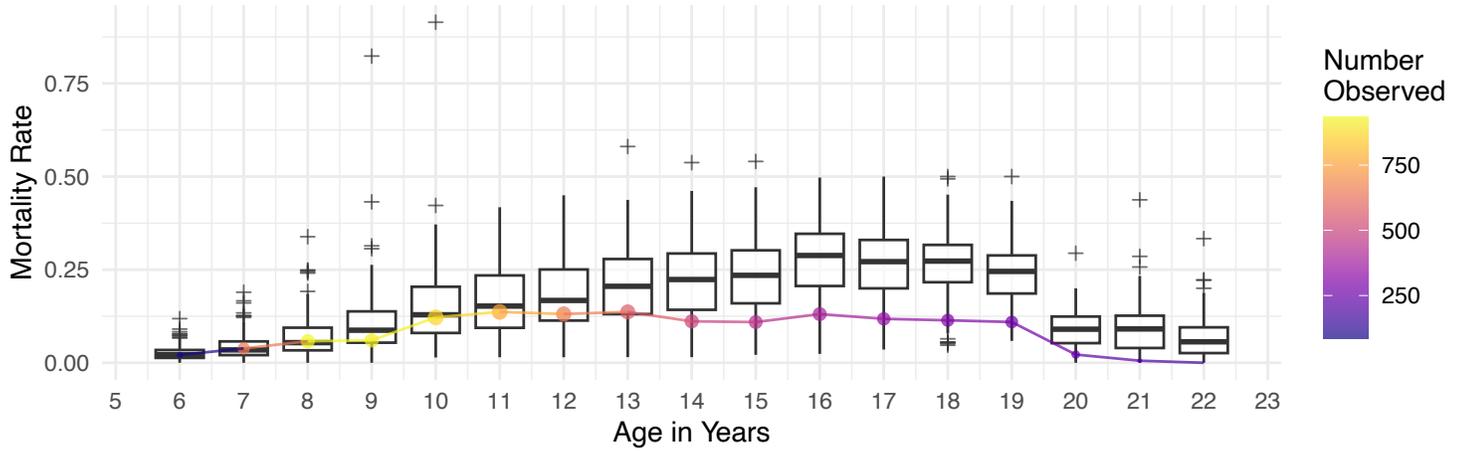

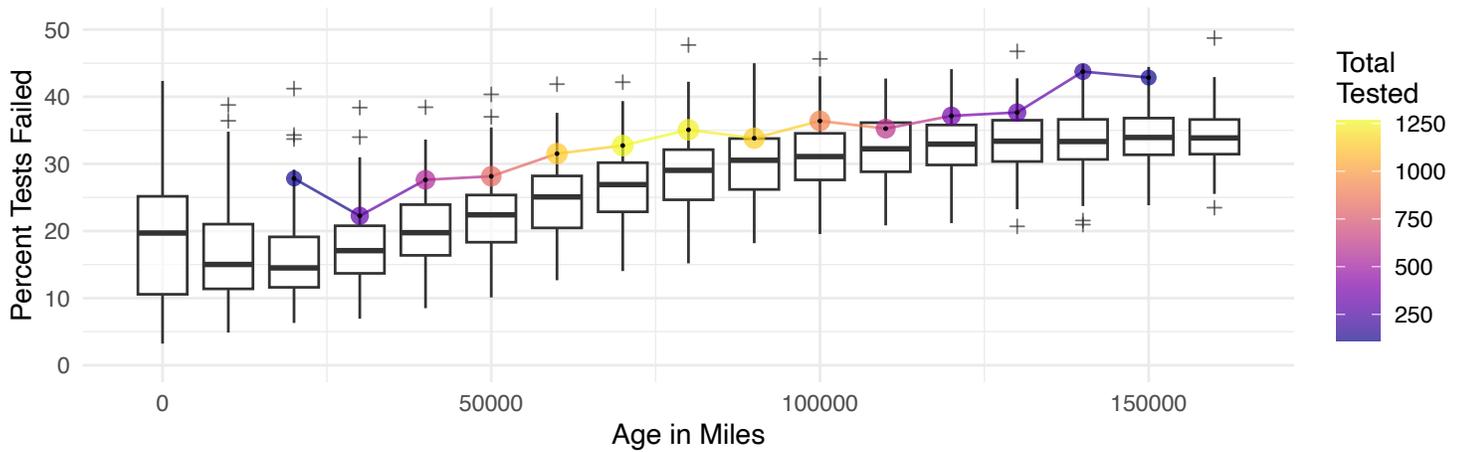

| Mortality rates | | | |
|---|---|---|---|
| Age in Years | Observed | Died | Mortality Rate |
| 6 | 95 | 2 | 0.02110 |
| 7 | 648 | 25 | 0.03860 |
| 8 | 932 | 54 | 0.05790 |
| 9 | 912 | 55 | 0.06030 |
| 10 | 862 | 105 | 0.12200 |
| 11 | 761 | 104 | 0.13700 |
| 12 | 657 | 86 | 0.13100 |
| 13 | 571 | 78 | 0.13700 |
| 14 | 494 | 55 | 0.11100 |
| 15 | 439 | 48 | 0.10900 |
| 16 | 390 | 51 | 0.13100 |
| 17 | 339 | 40 | 0.11800 |
| 18 | 298 | 34 | 0.11400 |
| 19 | 265 | 29 | 0.10900 |
| 20 | 226 | 5 | 0.02210 |
| 21 | 179 | 1 | 0.00559 |
| 22 | 86 | 0 | 0.00000 |

| Mechanical Reliability Rates | | |
|---|---|---|
| Mileage at test | N tested | Pct failed |
| 20000 | 115 | 27.8 |
| 30000 | 310 | 22.3 |
| 40000 | 561 | 27.6 |
| 50000 | 817 | 28.2 |
| 60000 | 1142 | 31.5 |
| 70000 | 1268 | 32.7 |
| 80000 | 1243 | 35.1 |
| 90000 | 1153 | 33.8 |
| 100000 | 915 | 36.4 |
| 110000 | 607 | 35.3 |
| 120000 | 358 | 37.2 |
| 130000 | 308 | 37.7 |
| 140000 | 176 | 43.8 |
| 150000 | 112 | 42.9 |

## Alfa Romeo Mito 2009

At 5 years of age, the mortality rate of a Alfa Romeo Mito 2009 (manufactured as a Car or Light Van) ranked number 76 out of 205 vehicles of the same age and type (any Car or Light Van constructed in 2009). One is the lowest (or best) and 205 the highest mortality rate. For vehicles reaching 20000 miles, its unreliability score (rate of failing an inspection) ranked 15 out of 200 vehicles of the same age, type, and mileage. One is the highest (or worst) and 200 the lowest rate of failing an inspection.

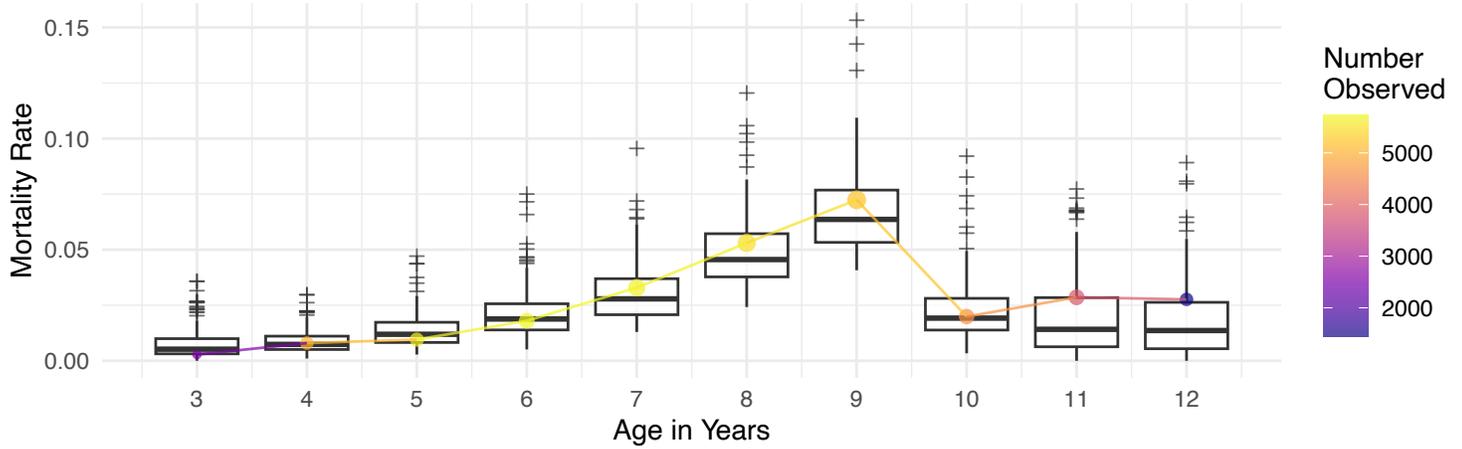

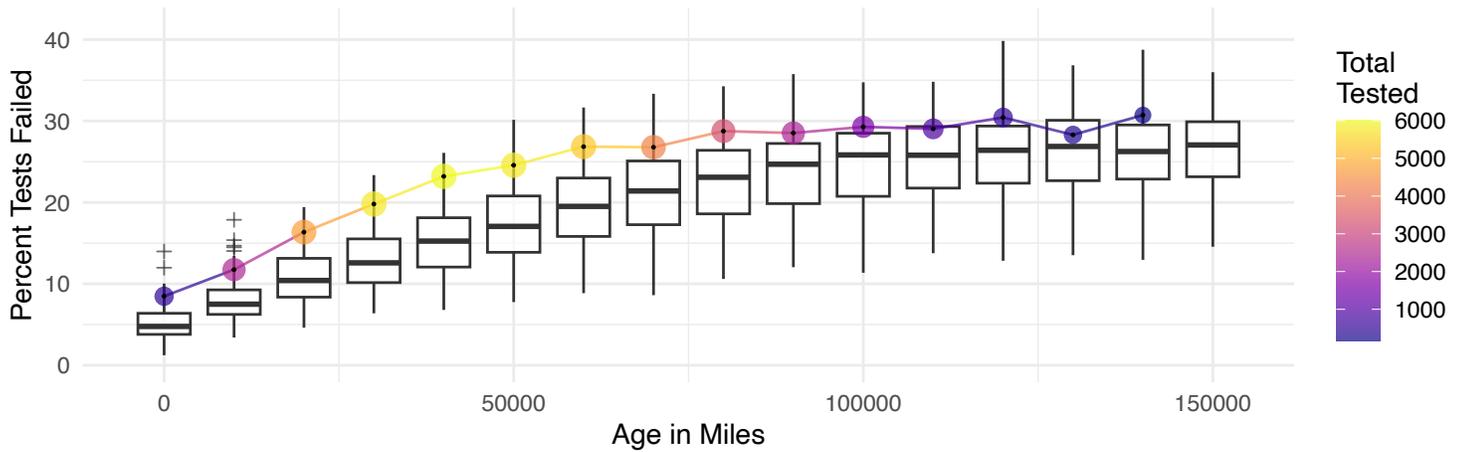

Mortality rates

| Age in Years | Observed | Died | Mortality Rate |
|---|---|---|---|
| 3 | 2513 | 7 | 0.00279 |
| 4 | 5017 | 40 | 0.00797 |
| 5 | 5655 | 54 | 0.00955 |
| 6 | 5722 | 103 | 0.01800 |
| 7 | 5652 | 186 | 0.03290 |
| 8 | 5467 | 290 | 0.05300 |
| 9 | 5178 | 375 | 0.07240 |
| 10 | 4623 | 92 | 0.01990 |
| 11 | 3815 | 109 | 0.02860 |
| 12 | 1448 | 40 | 0.02760 |

Mechanical Reliability Rates

| Mileage at test | N tested | Pct failed |
|---|---|---|
| 0 | 472 | 8.47 |
| 10000 | 2461 | 11.70 |
| 20000 | 4722 | 16.30 |
| 30000 | 5803 | 19.80 |
| 40000 | 6013 | 23.20 |
| 50000 | 5829 | 24.60 |
| 60000 | 5349 | 26.80 |
| 70000 | 4261 | 26.80 |
| 80000 | 3290 | 28.80 |
| 90000 | 2412 | 28.50 |
| 100000 | 1571 | 29.30 |
| 110000 | 892 | 29.00 |
| 120000 | 529 | 30.40 |
| 130000 | 311 | 28.30 |
| 140000 | 166 | 30.70 |

# Alfa Romeo Mito 2010

At 5 years of age, the mortality rate of a Alfa Romeo Mito 2010 (manufactured as a Car or Light Van) ranked number 99 out of 206 vehicles of the same age and type (any Car or Light Van constructed in 2010). One is the lowest (or best) and 206 the highest mortality rate. For vehicles reaching 20000 miles, its unreliability score (rate of failing an inspection) ranked 20 out of 201 vehicles of the same age, type, and mileage. One is the highest (or worst) and 201 the lowest rate of failing an inspection.

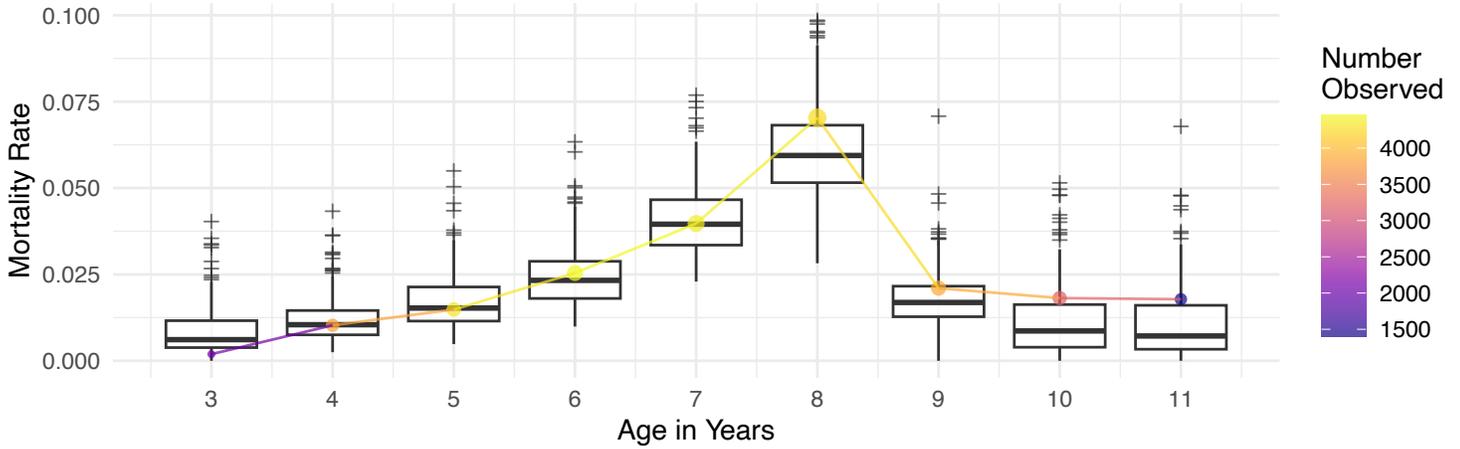

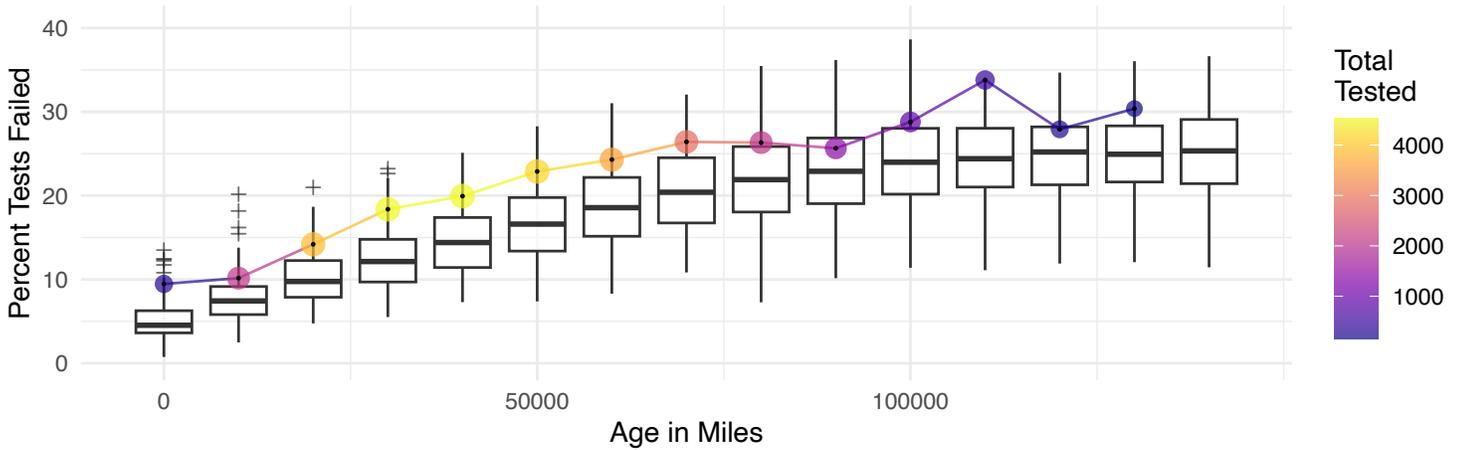

### Mortality rates

| Age in Years | Observed | Died | Mortality Rate |
|---|---|---|---|
| 3 | 2083 | 4 | 0.00192 |
| 4 | 3793 | 39 | 0.01030 |
| 5 | 4314 | 64 | 0.01480 |
| 6 | 4448 | 113 | 0.02540 |
| 7 | 4379 | 174 | 0.03970 |
| 8 | 4235 | 298 | 0.07040 |
| 9 | 3853 | 81 | 0.02100 |
| 10 | 3195 | 58 | 0.01820 |
| 11 | 1403 | 25 | 0.01780 |

### Mechanical Reliability Rates

| Mileage at test | N tested | Pct failed |
|---|---|---|
| 0 | 317 | 9.46 |
| 10000 | 2087 | 10.20 |
| 20000 | 3897 | 14.20 |
| 30000 | 4533 | 18.40 |
| 40000 | 4534 | 19.90 |
| 50000 | 4191 | 22.90 |
| 60000 | 3622 | 24.30 |
| 70000 | 2862 | 26.40 |
| 80000 | 2070 | 26.30 |
| 90000 | 1388 | 25.60 |
| 100000 | 858 | 28.80 |
| 110000 | 503 | 33.80 |
| 120000 | 258 | 27.90 |
| 130000 | 158 | 30.40 |

## Alfa Romeo Mito 2011

At 5 years of age, the mortality rate of a Alfa Romeo Mito 2011 (manufactured as a Car or Light Van) ranked number 120 out of 211 vehicles of the same age and type (any Car or Light Van constructed in 2011). One is the lowest (or best) and 211 the highest mortality rate. For vehicles reaching 20000 miles, its unreliability score (rate of failing an inspection) ranked 31 out of 205 vehicles of the same age, type, and mileage. One is the highest (or worst) and 205 the lowest rate of failing an inspection.

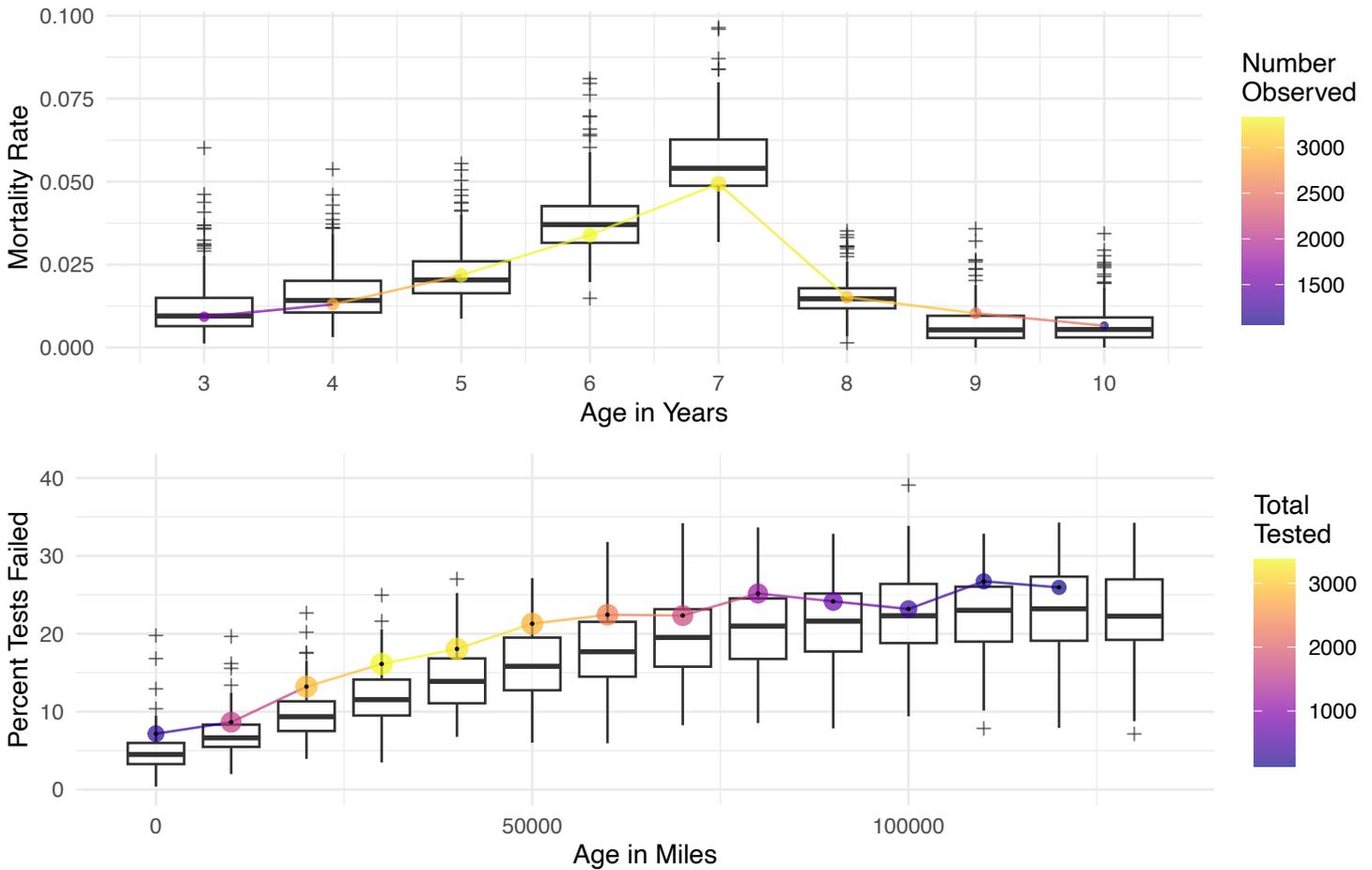

### Mortality rates

| Age in Years | Observed | Died | Mortality Rate |
|---|---|---|---|
| 3 | 1619 | 15 | 0.00926 |
| 4 | 2924 | 38 | 0.01300 |
| 5 | 3302 | 72 | 0.02180 |
| 6 | 3332 | 113 | 0.03390 |
| 7 | 3261 | 161 | 0.04940 |
| 8 | 3032 | 46 | 0.01520 |
| 9 | 2516 | 26 | 0.01030 |
| 10 | 1063 | 7 | 0.00659 |

### Mechanical Reliability Rates

| Mileage at test | N tested | Pct failed |
|---|---|---|
| 0 | 279 | 7.17 |
| 10000 | 1596 | 8.65 |
| 20000 | 2891 | 13.20 |
| 30000 | 3375 | 16.10 |
| 40000 | 3201 | 18.10 |
| 50000 | 2799 | 21.30 |
| 60000 | 2317 | 22.40 |
| 70000 | 1719 | 22.30 |
| 80000 | 1176 | 25.20 |
| 90000 | 712 | 24.20 |
| 100000 | 384 | 23.20 |
| 110000 | 202 | 26.70 |
| 120000 | 131 | 26.00 |

## Alfa Romeo Mito 2012

At 5 years of age, the mortality rate of a Alfa Romeo Mito 2012 (manufactured as a Car or Light Van) ranked number 125 out of 212 vehicles of the same age and type (any Car or Light Van constructed in 2012). One is the lowest (or best) and 212 the highest mortality rate. For vehicles reaching 20000 miles, its unreliability score (rate of failing an inspection) ranked 18 out of 206 vehicles of the same age, type, and mileage. One is the highest (or worst) and 206 the lowest rate of failing an inspection.

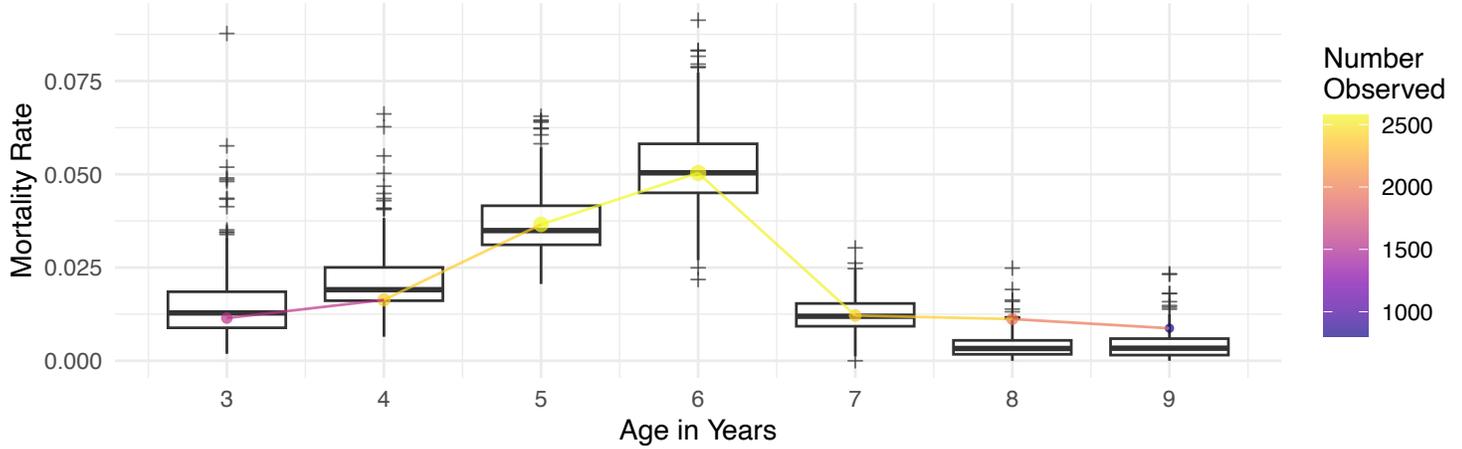

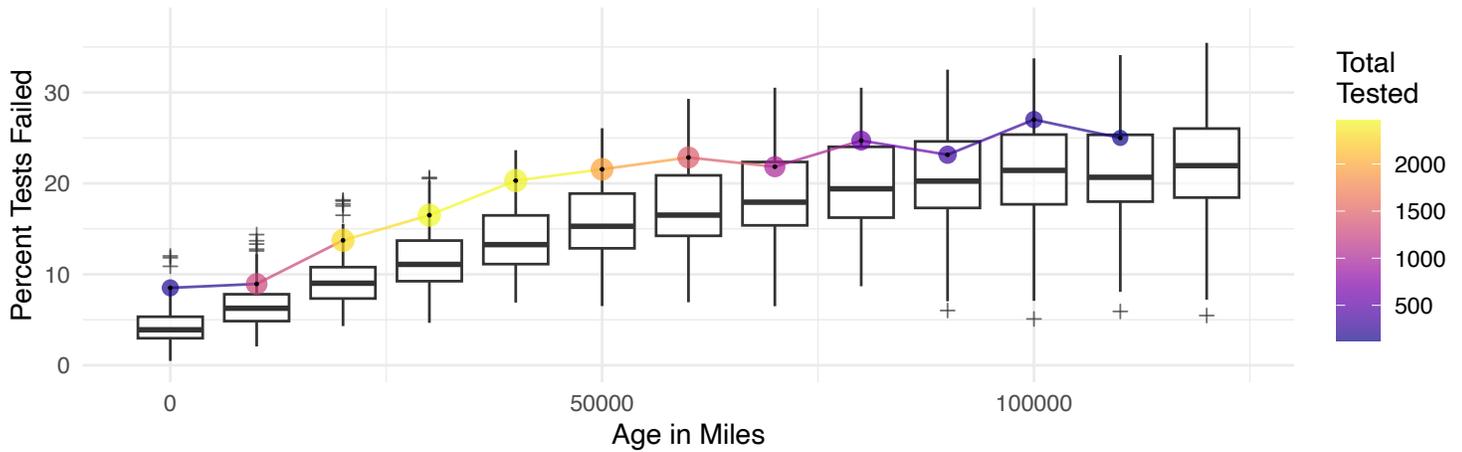

Mortality rates

| Age in Years | Observed | Died | Mortality Rate |
|---|---|---|---|
| 3 | 1572 | 18 | 0.01150 |
| 4 | 2388 | 39 | 0.01630 |
| 5 | 2572 | 94 | 0.03650 |
| 6 | 2542 | 128 | 0.05040 |
| 7 | 2388 | 29 | 0.01210 |
| 8 | 1966 | 22 | 0.01120 |
| 9 | 803 | 7 | 0.00872 |

Mechanical Reliability Rates

| Mileage at test | N tested | Pct failed |
|---|---|---|
| 0 | 188 | 8.51 |
| 10000 | 1275 | 8.94 |
| 20000 | 2300 | 13.70 |
| 30000 | 2466 | 16.50 |
| 40000 | 2433 | 20.30 |
| 50000 | 1958 | 21.60 |
| 60000 | 1453 | 22.80 |
| 70000 | 1003 | 21.80 |
| 80000 | 591 | 24.70 |
| 90000 | 311 | 23.20 |
| 100000 | 211 | 27.00 |
| 110000 | 124 | 25.00 |

# Alfa Romeo Mito 2013

At 5 years of age, the mortality rate of a Alfa Romeo Mito 2013 (manufactured as a Car or Light Van) ranked number 59 out of 221 vehicles of the same age and type (any Car or Light Van constructed in 2013). One is the lowest (or best) and 221 the highest mortality rate. For vehicles reaching 20000 miles, its unreliability score (rate of failing an inspection) ranked 12 out of 215 vehicles of the same age, type, and mileage. One is the highest (or worst) and 215 the lowest rate of failing an inspection.

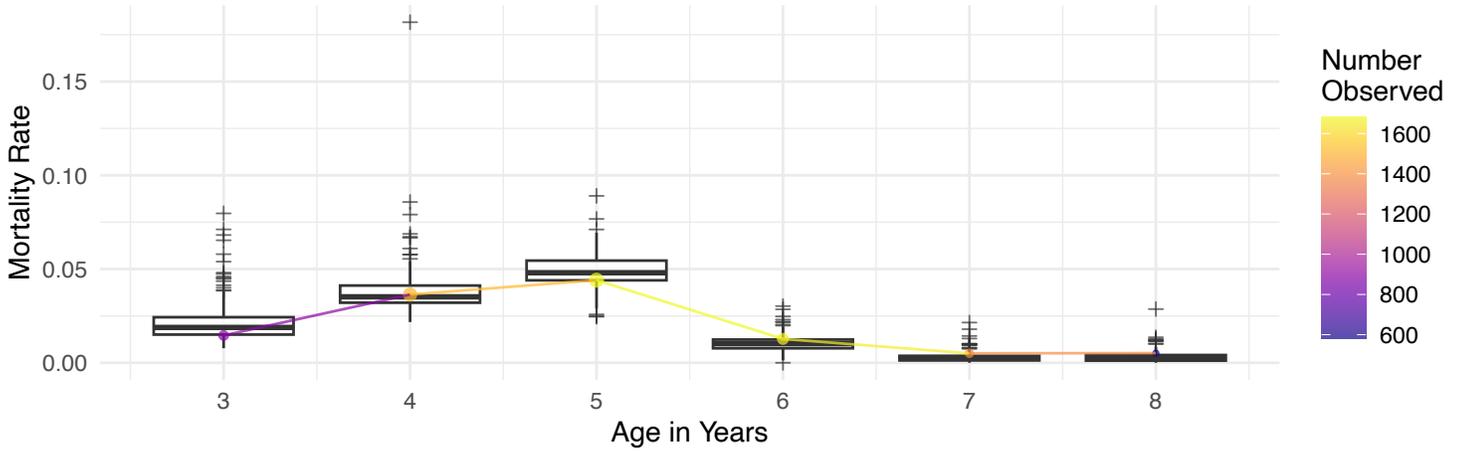

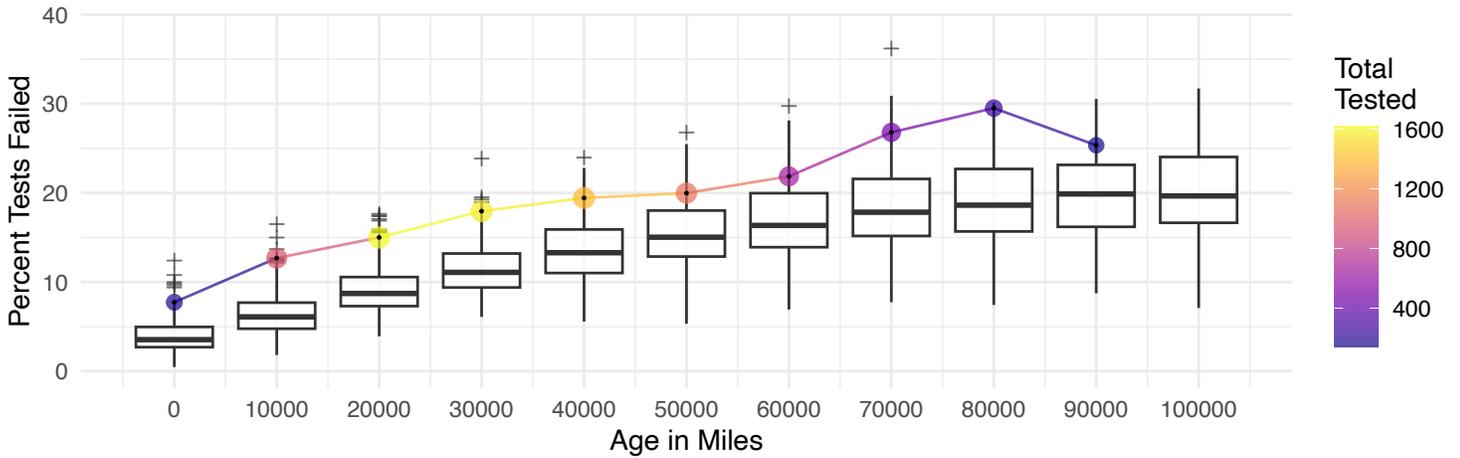

Mortality rates

| Age in Years | Observed | Died | Mortality Rate |
|---|---|---|---|
| 3 | 891 | 13 | 0.01460 |
| 4 | 1511 | 55 | 0.03640 |
| 5 | 1679 | 74 | 0.04410 |
| 6 | 1649 | 21 | 0.01270 |
| 7 | 1385 | 7 | 0.00505 |
| 8 | 582 | 3 | 0.00515 |

Mechanical Reliability Rates

| Mileage at test | N tested | Pct failed |
|---|---|---|
| 0 | 168 | 7.74 |
| 10000 | 953 | 12.70 |
| 20000 | 1619 | 15.00 |
| 30000 | 1576 | 18.00 |
| 40000 | 1389 | 19.40 |
| 50000 | 1086 | 20.00 |
| 60000 | 654 | 21.90 |
| 70000 | 459 | 26.80 |
| 80000 | 227 | 29.50 |
| 90000 | 142 | 25.40 |

# Alfa Romeo Mito 2014

At 5 years of age, the mortality rate of a Alfa Romeo Mito 2014 (manufactured as a Car or Light Van) ranked number 200 out of 236 vehicles of the same age and type (any Car or Light Van constructed in 2014). One is the lowest (or best) and 236 the highest mortality rate. For vehicles reaching 20000 miles, its unreliability score (rate of failing an inspection) ranked 6 out of 230 vehicles of the same age, type, and mileage. One is the highest (or worst) and 230 the lowest rate of failing an inspection.

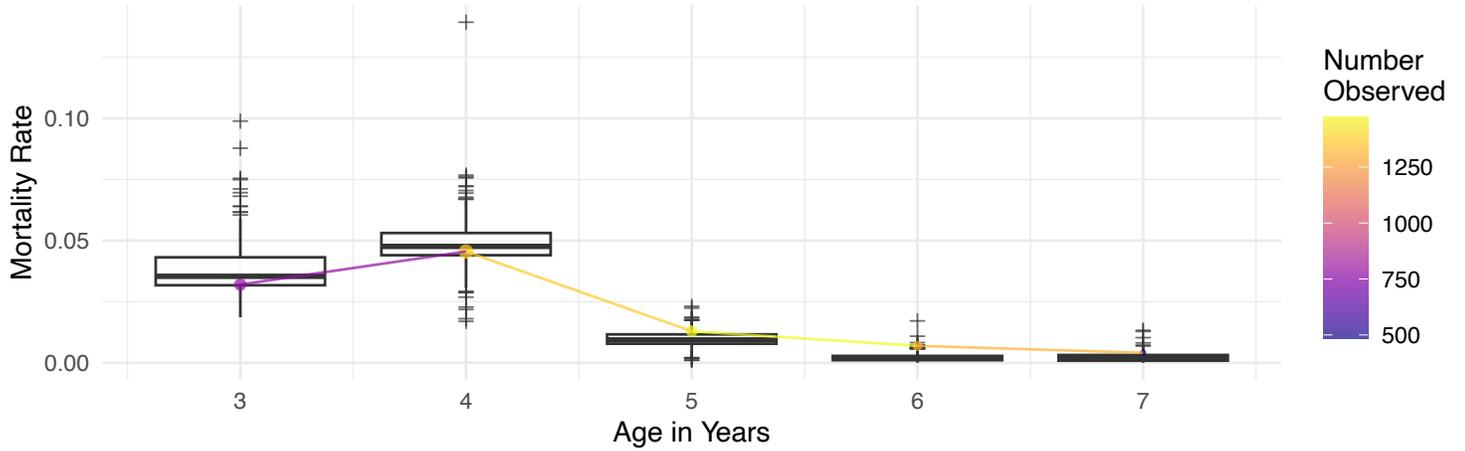

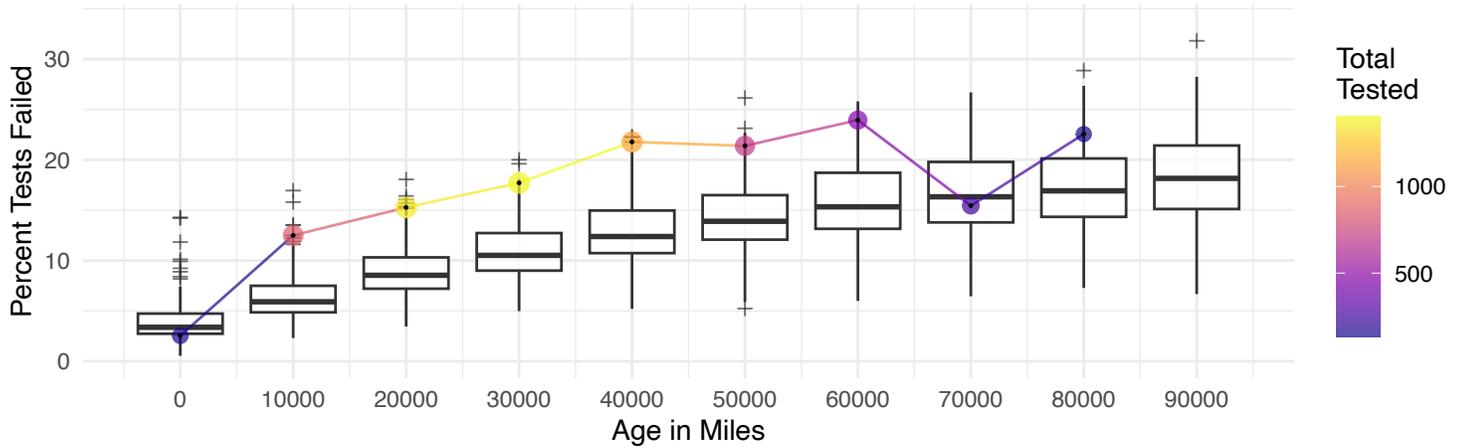

Mortality rates

| Age in Years | Observed | Died | Mortality Rate |
|---|---|---|---|
| 3 | 781 | 25 | 0.03200 |
| 4 | 1362 | 62 | 0.04550 |
| 5 | 1471 | 19 | 0.01290 |
| 6 | 1299 | 9 | 0.00693 |
| 7 | 485 | 2 | 0.00412 |

Mechanical Reliability Rates

| Mileage at test | N tested | Pct failed |
|---|---|---|
| 0 | 158 | 2.53 |
| 10000 | 864 | 12.50 |
| 20000 | 1362 | 15.30 |
| 30000 | 1406 | 17.70 |
| 40000 | 1148 | 21.80 |
| 50000 | 706 | 21.40 |
| 60000 | 451 | 23.90 |
| 70000 | 246 | 15.40 |
| 80000 | 133 | 22.60 |

## Alfa Romeo Mito 2015

At 5 years of age, the mortality rate of a Alfa Romeo Mito 2015 (manufactured as a Car or Light Van) ranked number 241 out of 247 vehicles of the same age and type (any Car or Light Van constructed in 2015). One is the lowest (or best) and 247 the highest mortality rate. For vehicles reaching 20000 miles, its unreliability score (rate of failing an inspection) ranked 12 out of 241 vehicles of the same age, type, and mileage. One is the highest (or worst) and 241 the lowest rate of failing an inspection.

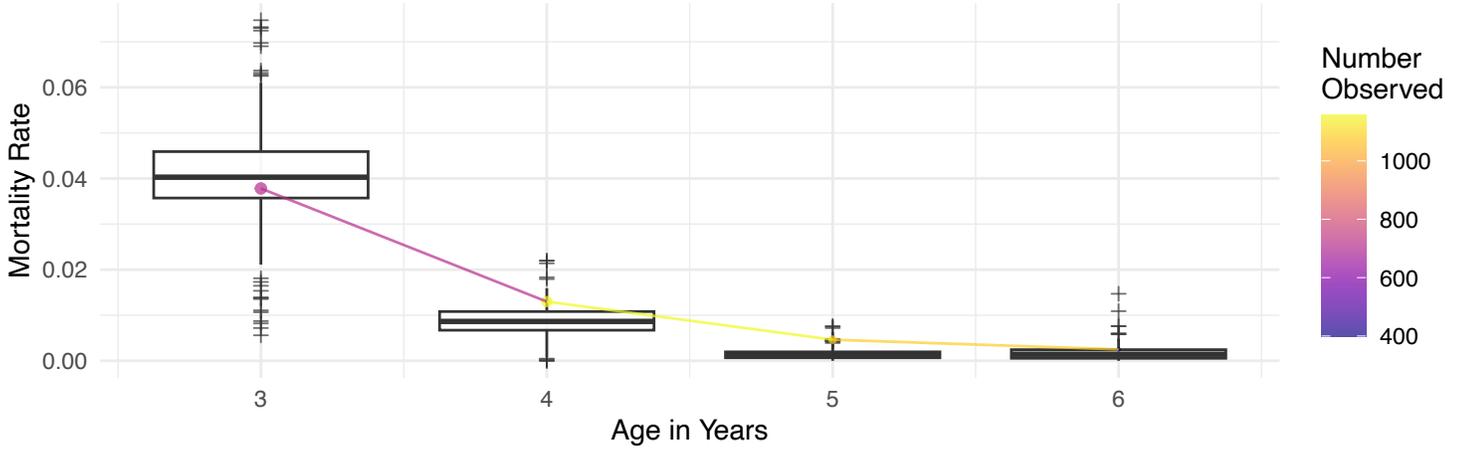

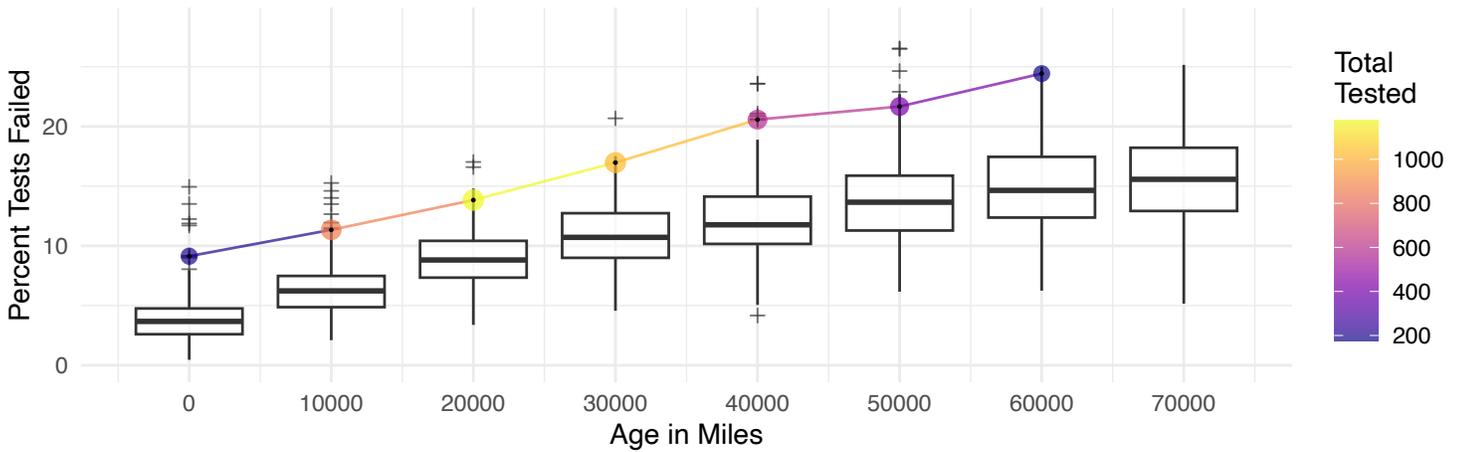

Mortality rates

| Age in Years | Observed | Died | Mortality Rate |
|---|---|---|---|
| 3 | 714 | 27 | 0.03780 |
| 4 | 1155 | 15 | 0.01300 |
| 5 | 1080 | 5 | 0.00463 |
| 6 | 399 | 1 | 0.00251 |

Mechanical Reliability Rates

| Mileage at test | N tested | Pct failed |
|---|---|---|
| 0 | 197 | 9.14 |
| 10000 | 855 | 11.30 |
| 20000 | 1178 | 13.80 |
| 30000 | 1025 | 17.00 |
| 40000 | 588 | 20.60 |
| 50000 | 406 | 21.70 |
| 60000 | 176 | 24.40 |

**Alfa Romeo Mito 2016**

At 5 years of age, the mortality rate of a Alfa Romeo Mito 2016 (manufactured as a Car or Light Van) ranked number 218 out of 252 vehicles of the same age and type (any Car or Light Van constructed in 2016). One is the lowest (or best) and 252 the highest mortality rate. For vehicles reaching 20000 miles, its unreliability score (rate of failing an inspection) ranked 62 out of 246 vehicles of the same age, type, and mileage. One is the highest (or worst) and 246 the lowest rate of failing an inspection.

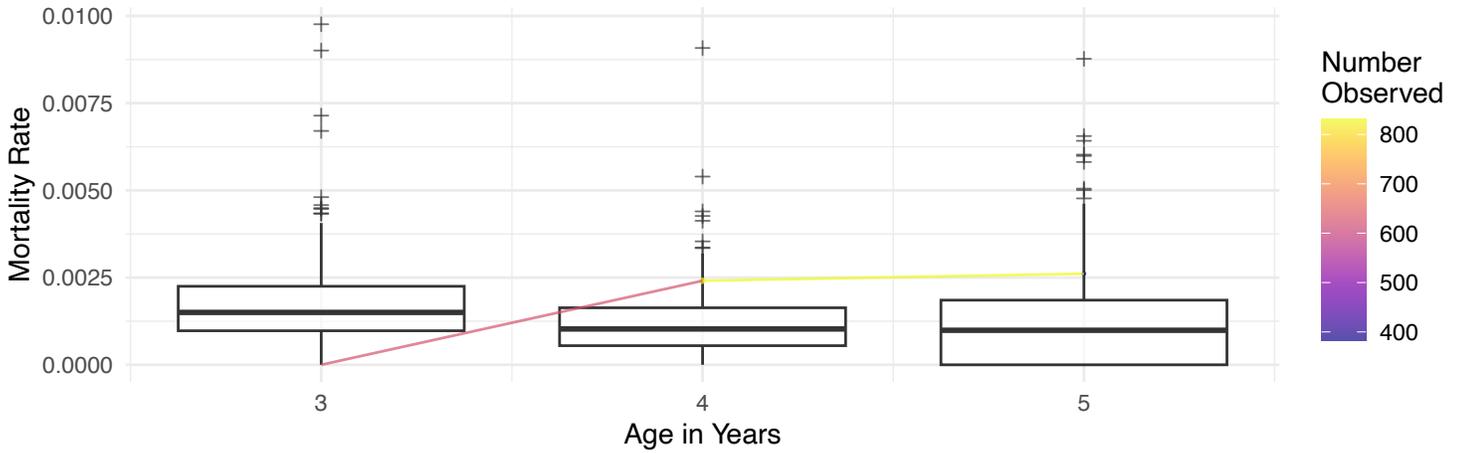

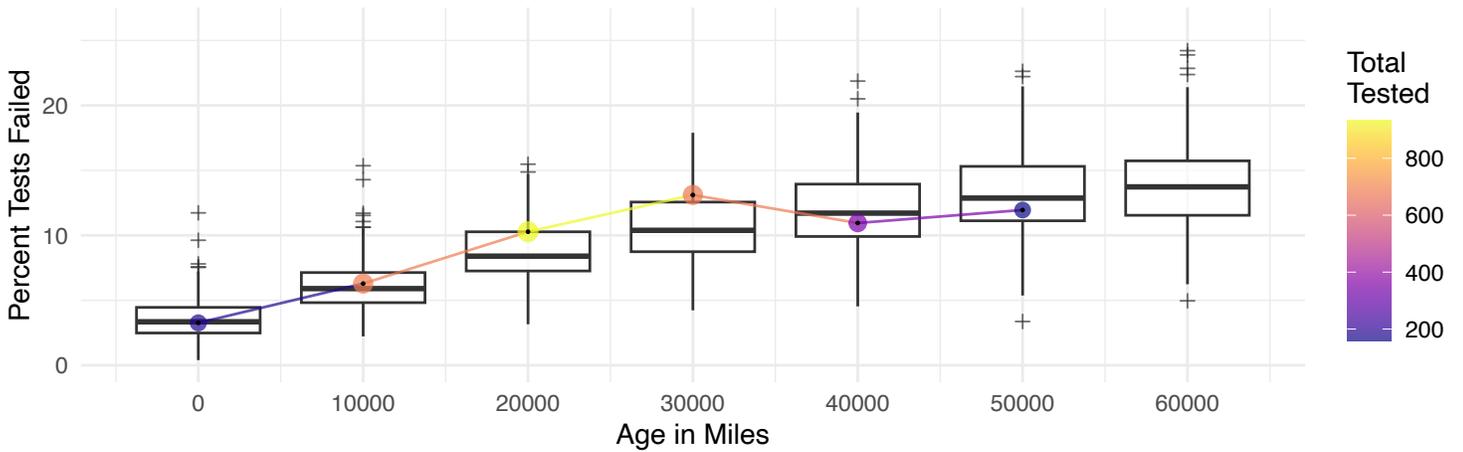

<table>
<tr><td colspan="4" align="center">Mortality rates</td></tr>
</table>

| Age in Years | Observed | Died | Mortality Rate |
|:---:|:---:|:---:|:---:|
| 3 | 627 | 0 | 0.00000 |
| 4 | 830 | 2 | 0.00241 |
| 5 | 383 | 1 | 0.00261 |

<table>
<tr><td colspan="3" align="center">Mechanical Reliability Rates</td></tr>
</table>

| Mileage at test | N tested | Pct failed |
|:---:|:---:|:---:|
| 0 | 184 | 3.26 |
| 10000 | 684 | 6.29 |
| 20000 | 934 | 10.30 |
| 30000 | 679 | 13.10 |
| 40000 | 347 | 11.00 |
| 50000 | 159 | 11.90 |

## Aprilia

**Aprilia Rsv 2002** At 5 years of age, the mortality rate of a Aprilia Rsv 2002 (manufactured as a Motorbike of 200cc or over) ranked number 13 out of 17 vehicles of the same age and type (any Motorbike of 200cc or over constructed in 2002). One is the lowest (or best) and 17 the highest mortality rate. For vehicles reaching 20000 miles, its unreliability score (rate of failing an inspection) ranked 5 out of 17 vehicles of the same age, type, and mileage. One is the highest (or worst) and 17 the lowest rate of failing an inspection.

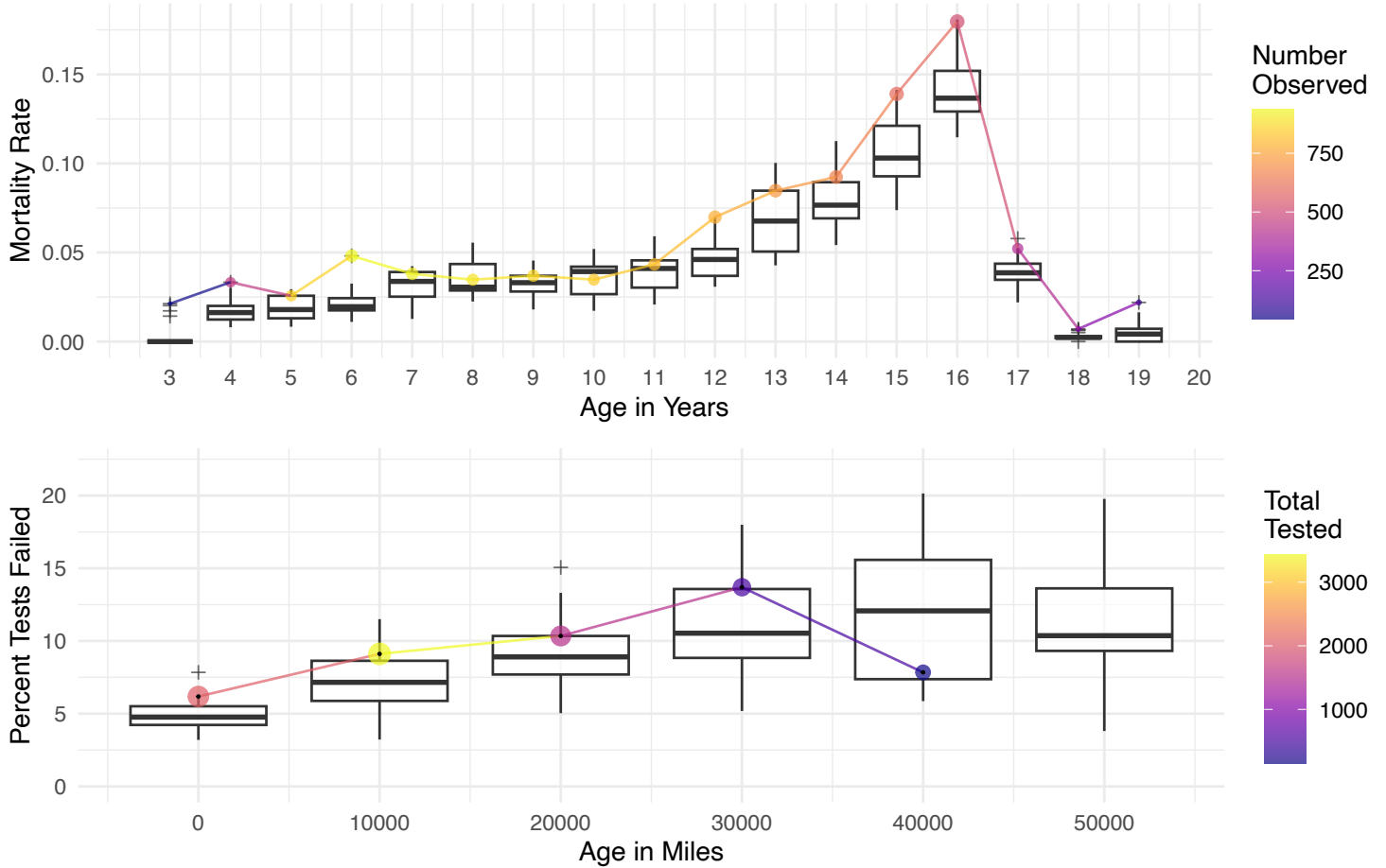

### Mortality rates

| Age in Years | Observed | Died | Mortality Rate |
|---|---|---|---|
| 3 | 47 | 1 | 0.02130 |
| 4 | 450 | 15 | 0.03330 |
| 5 | 855 | 22 | 0.02570 |
| 6 | 935 | 45 | 0.04810 |
| 7 | 920 | 35 | 0.03800 |
| 8 | 892 | 31 | 0.03480 |
| 9 | 866 | 32 | 0.03700 |
| 10 | 834 | 29 | 0.03480 |
| 11 | 807 | 35 | 0.04340 |
| 12 | 773 | 54 | 0.06990 |
| 13 | 720 | 61 | 0.08470 |
| 14 | 660 | 61 | 0.09240 |
| 15 | 597 | 83 | 0.13900 |
| 16 | 512 | 92 | 0.18000 |
| 17 | 402 | 21 | 0.05220 |
| 18 | 283 | 2 | 0.00707 |
| 19 | 91 | 2 | 0.02200 |

### Mechanical Reliability Rates

| Mileage at test | N tested | Pct failed |
|---|---|---|
| 0 | 2054 | 6.18 |
| 10000 | 3435 | 9.11 |
| 20000 | 1527 | 10.30 |
| 30000 | 511 | 13.70 |
| 40000 | 153 | 7.84 |

## Aprilia Rsv 2003

At 5 years of age, the mortality rate of a Aprilia Rsv 2003 (manufactured as a Motorbike of 200cc or over) ranked number 13 out of 15 vehicles of the same age and type (any Motorbike of 200cc or over constructed in 2003). One is the lowest (or best) and 15 the highest mortality rate. For vehicles reaching 20000 miles, its unreliability score (rate of failing an inspection) ranked 9 out of 15 vehicles of the same age, type, and mileage. One is the highest (or worst) and 15 the lowest rate of failing an inspection.

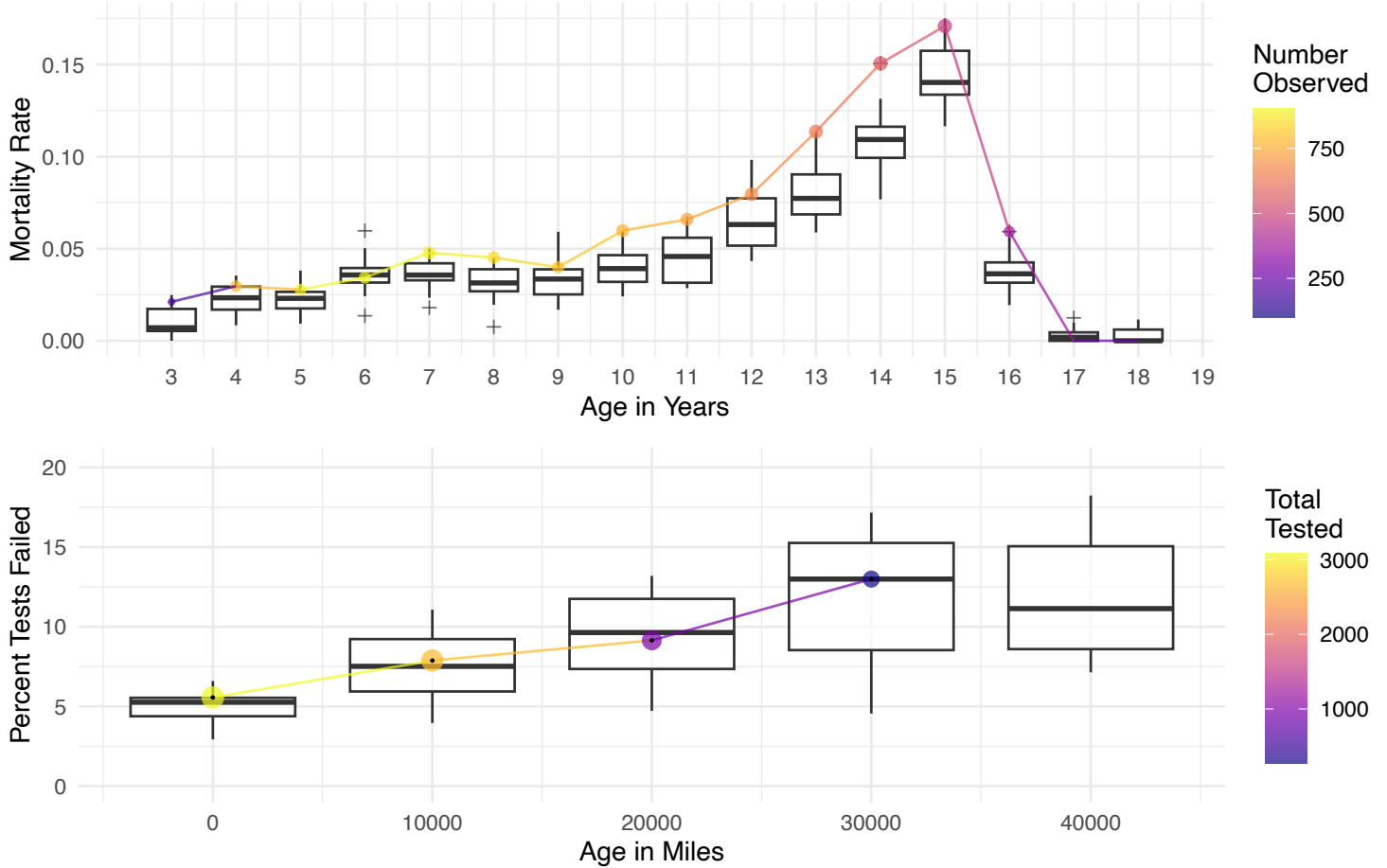

<table>
<tr><td colspan="4" align="center">Mortality rates</td></tr>
</table>

| Age in Years | Observed | Died | Mortality Rate |
|:---:|:---:|:---:|:---:|
| 3 | 189 | 4 | 0.0212 |
| 4 | 776 | 23 | 0.0296 |
| 5 | 900 | 25 | 0.0278 |
| 6 | 903 | 31 | 0.0343 |
| 7 | 879 | 42 | 0.0478 |
| 8 | 841 | 38 | 0.0452 |
| 9 | 801 | 32 | 0.0400 |
| 10 | 769 | 46 | 0.0598 |
| 11 | 728 | 48 | 0.0659 |
| 12 | 679 | 54 | 0.0795 |
| 13 | 625 | 71 | 0.1140 |
| 14 | 551 | 83 | 0.1510 |
| 15 | 468 | 80 | 0.1710 |
| 16 | 371 | 22 | 0.0593 |
| 17 | 269 | 0 | 0.0000 |
| 18 | 99 | 0 | 0.0000 |

Mechanical Reliability Rates

| Mileage at test | N tested | Pct failed |
|:---:|:---:|:---:|
| 0 | 3075 | 5.56 |
| 10000 | 2665 | 7.88 |
| 20000 | 962 | 9.15 |
| 30000 | 277 | 13.00 |

## Aprilia Rsv 2004

At 5 years of age, the mortality rate of a Aprilia Rsv 2004 (manufactured as a Motorbike of 200cc or over) ranked number 8 out of 11 vehicles of the same age and type (any Motorbike of 200cc or over constructed in 2004). One is the lowest (or best) and 11 the highest mortality rate. For vehicles reaching 20000 miles, its unreliability score (rate of failing an inspection) ranked 10 out of 11 vehicles of the same age, type, and mileage. One is the highest (or worst) and 11 the lowest rate of failing an inspection.

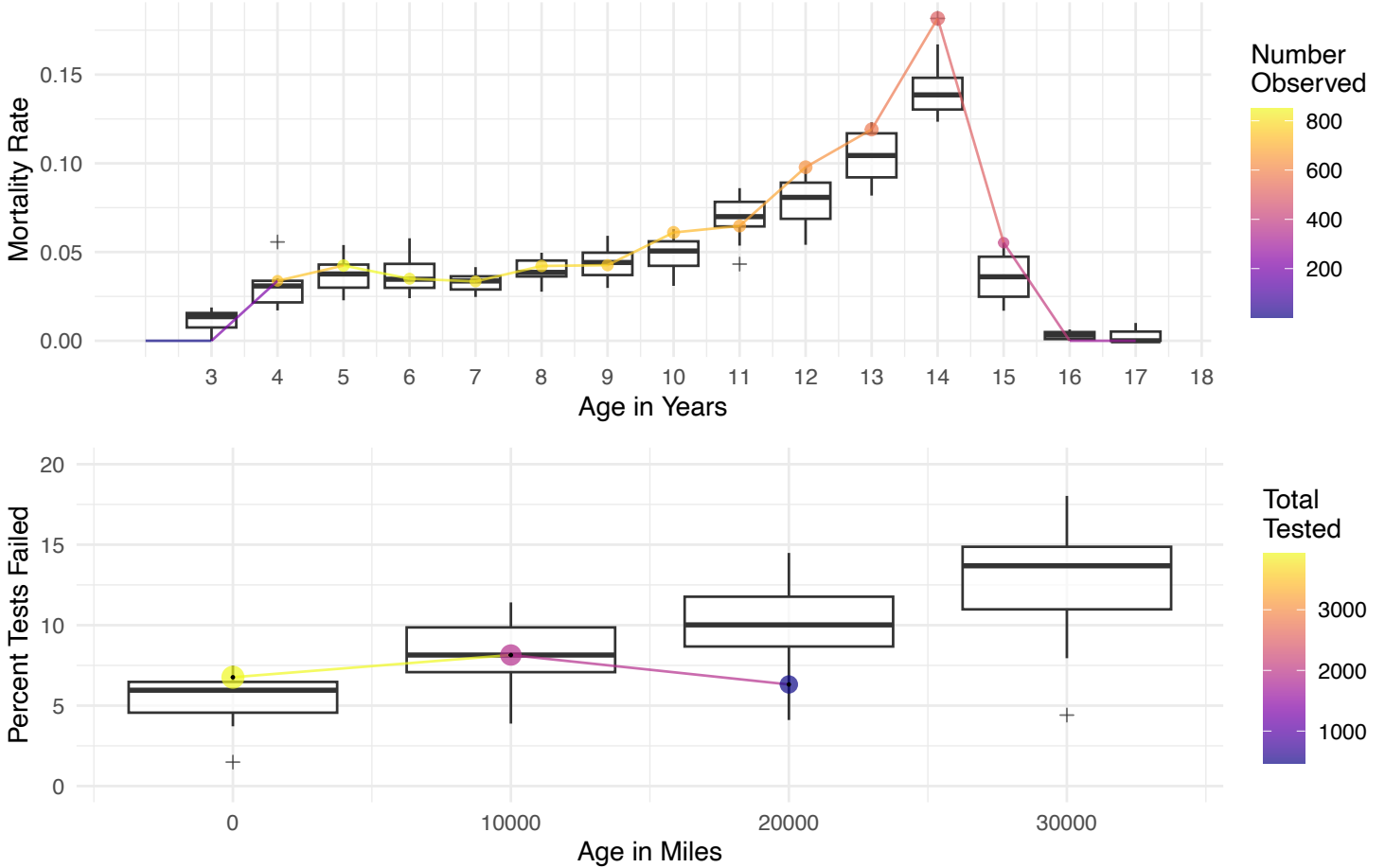

Mortality rates

| Age in Years | Observed | Died | Mortality Rate |
|---|---|---|---|
| 3 | 211 | 0 | 0.0000 |
| 4 | 740 | 25 | 0.0338 |
| 5 | 849 | 36 | 0.0424 |
| 6 | 829 | 29 | 0.0350 |
| 7 | 807 | 27 | 0.0335 |
| 8 | 783 | 33 | 0.0421 |
| 9 | 751 | 32 | 0.0426 |
| 10 | 721 | 44 | 0.0610 |
| 11 | 680 | 44 | 0.0647 |
| 12 | 634 | 62 | 0.0978 |
| 13 | 571 | 68 | 0.1190 |
| 14 | 501 | 91 | 0.1820 |
| 15 | 398 | 22 | 0.0553 |
| 16 | 292 | 0 | 0.0000 |
| 17 | 146 | 0 | 0.0000 |

Mechanical Reliability Rates

| Mileage at test | N tested | Pct failed |
|---|---|---|
| 0 | 3918 | 6.76 |
| 10000 | 1891 | 8.14 |
| 20000 | 475 | 6.32 |

## Aston Martin

**Aston Martin Db9 2005** At 5 years of age, the mortality rate of a Aston Martin Db9 2005 (manufactured as a Car or Light Van) ranked number 7 out of 240 vehicles of the same age and type (any Car or Light Van constructed in 2005). One is the lowest (or best) and 240 the highest mortality rate. For vehicles reaching 20000 miles, its unreliability score (rate of failing an inspection) ranked 199 out of 235 vehicles of the same age, type, and mileage. One is the highest (or worst) and 235 the lowest rate of failing an inspection.

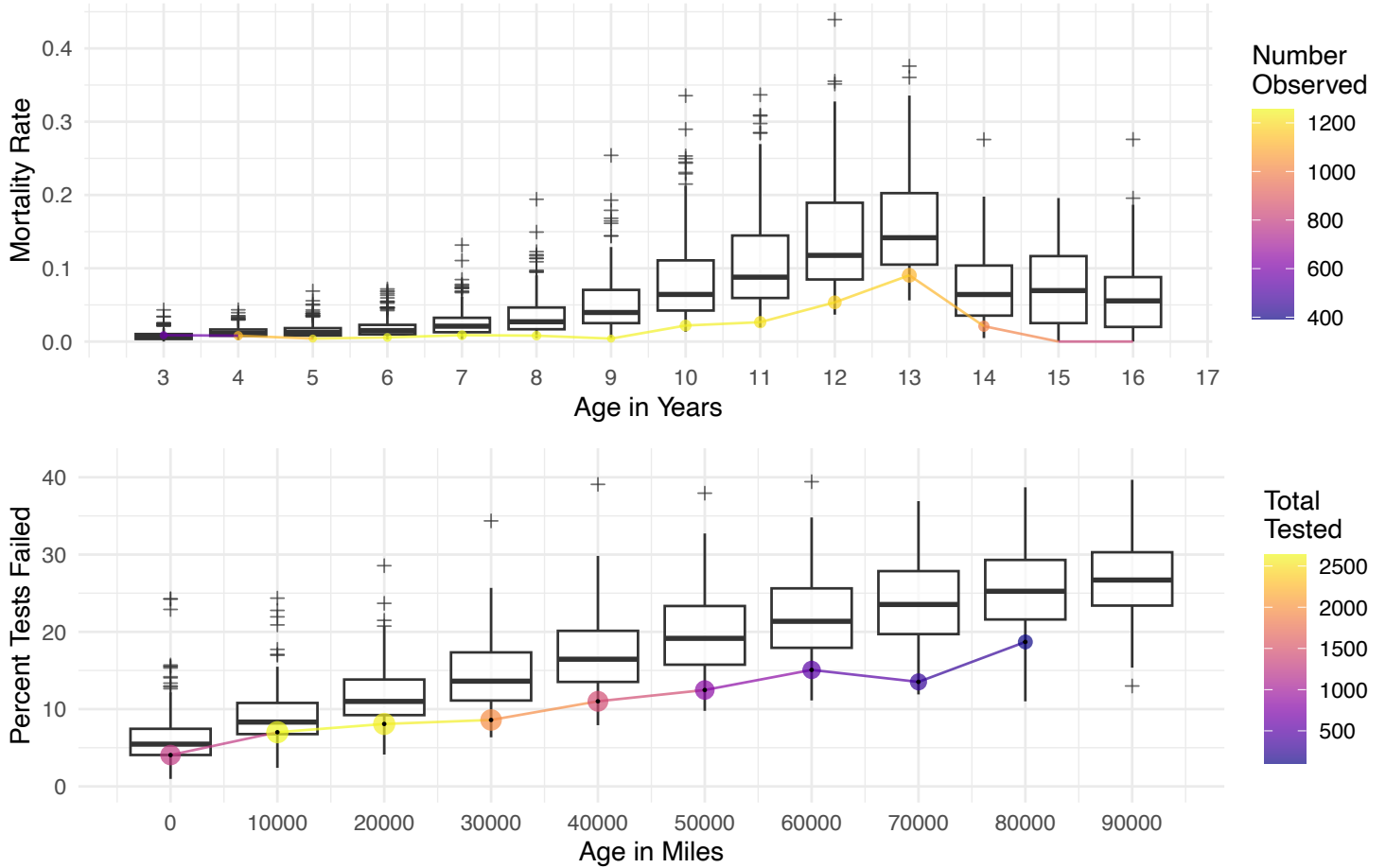

| Mortality rates | | | |
|---|---|---|---|
| Age in Years | Observed | Died | Mortality Rate |
| 3 | 572 | 5 | 0.00874 |
| 4 | 1127 | 9 | 0.00799 |
| 5 | 1228 | 5 | 0.00407 |
| 6 | 1254 | 7 | 0.00558 |
| 7 | 1253 | 11 | 0.00878 |
| 8 | 1245 | 10 | 0.00803 |
| 9 | 1238 | 5 | 0.00404 |
| 10 | 1234 | 27 | 0.02190 |
| 11 | 1209 | 32 | 0.02650 |
| 12 | 1177 | 63 | 0.05350 |
| 13 | 1117 | 101 | 0.09040 |
| 14 | 1002 | 21 | 0.02100 |
| 15 | 820 | 0 | 0.00000 |
| 16 | 395 | 0 | 0.00000 |

| Mechanical Reliability Rates | | |
|---|---|---|
| Mileage at test | N tested | Pct failed |
| 0 | 1257 | 4.06 |
| 10000 | 2639 | 7.01 |
| 20000 | 2575 | 8.08 |
| 30000 | 1954 | 8.60 |
| 40000 | 1426 | 11.00 |
| 50000 | 834 | 12.50 |
| 60000 | 491 | 15.10 |
| 70000 | 281 | 13.50 |
| 80000 | 107 | 18.70 |

**Aston Martin V8 Vantage 2006**

At 5 years of age, the mortality rate of a Aston Martin V8 Vantage 2006 (manufactured as a Car or Light Van) ranked number 5 out of 225 vehicles of the same age and type (any Car or Light Van constructed in 2006). One is the lowest (or best) and 225 the highest mortality rate. For vehicles reaching 20000 miles, its unreliability score (rate of failing an inspection) ranked 209 out of 220 vehicles of the same age, type, and mileage. One is the highest (or worst) and 220 the lowest rate of failing an inspection.

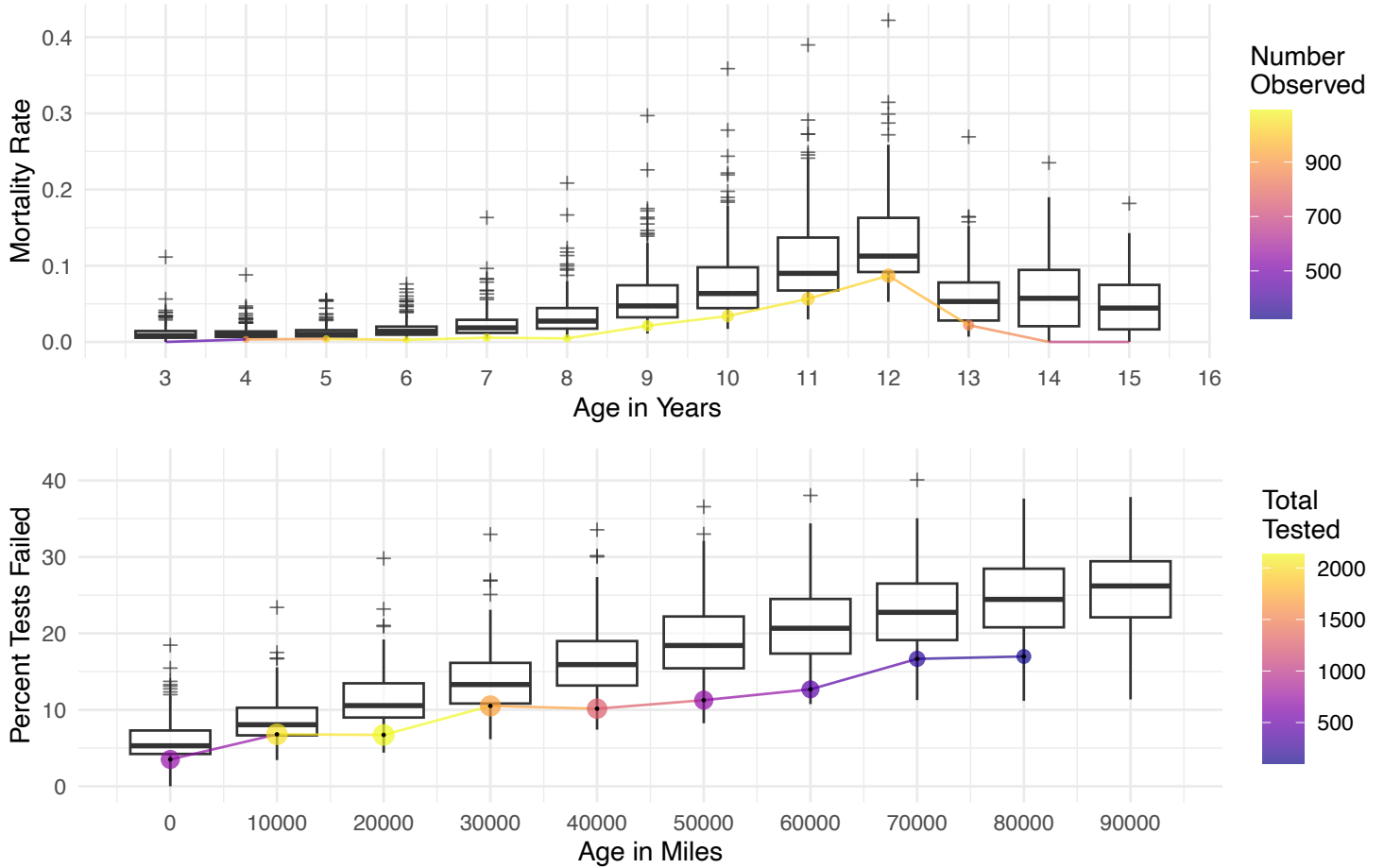

<table>
<tr><td colspan="4" align="center">Mortality rates</td></tr>
</table>

| Age in Years | Observed | Died | Mortality Rate |
|---|---|---|---|
| 3 | 435 | 0 | 0.00000 |
| 4 | 902 | 3 | 0.00333 |
| 5 | 1032 | 4 | 0.00388 |
| 6 | 1084 | 3 | 0.00277 |
| 7 | 1090 | 6 | 0.00550 |
| 8 | 1086 | 5 | 0.00460 |
| 9 | 1083 | 23 | 0.02120 |
| 10 | 1059 | 36 | 0.03400 |
| 11 | 1025 | 58 | 0.05660 |
| 12 | 965 | 84 | 0.08700 |
| 13 | 861 | 19 | 0.02210 |
| 14 | 685 | 0 | 0.00000 |
| 15 | 324 | 0 | 0.00000 |

Mechanical Reliability Rates

| Mileage at test | N tested | Pct failed |
|---|---|---|
| 0 | 712 | 3.51 |
| 10000 | 2030 | 6.80 |
| 20000 | 2128 | 6.72 |
| 30000 | 1684 | 10.50 |
| 40000 | 1251 | 10.20 |
| 50000 | 719 | 11.30 |
| 60000 | 418 | 12.70 |
| 70000 | 198 | 16.70 |
| 80000 | 106 | 17.00 |

## Audi

**Audi 100 1992** At 15 years of age, the mortality rate of a Audi 100 1992 (manufactured as a Car or Light Van) ranked number 59 out of 90 vehicles of the same age and type (any Car or Light Van constructed in 1992). One is the lowest (or best) and 90 the highest mortality rate. For vehicles reaching 120000 miles, its unreliability score (rate of failing an inspection) ranked 33 out of 75 vehicles of the same age, type, and mileage. One is the highest (or worst) and 75 the lowest rate of failing an inspection.

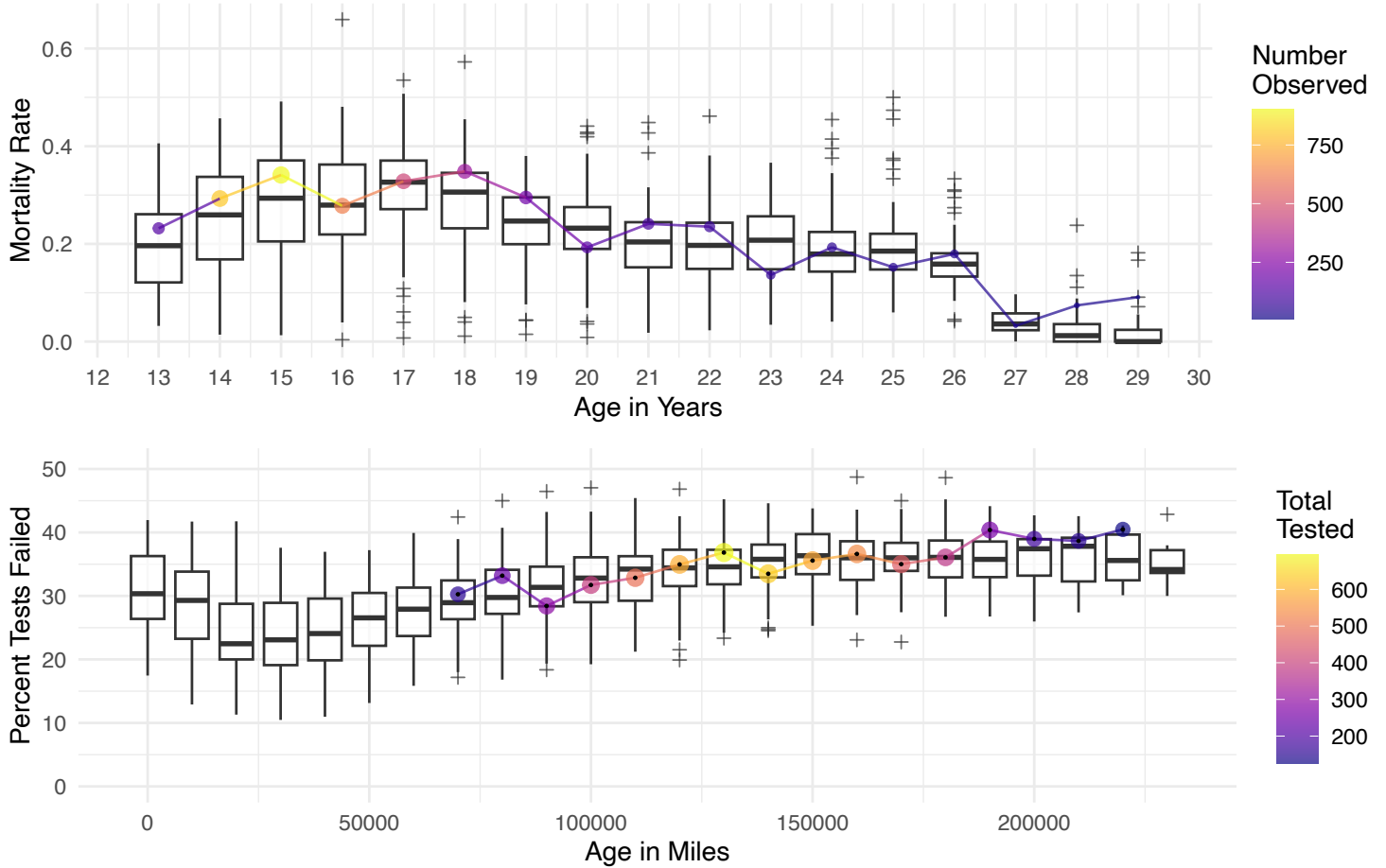

|  | Mortality rates | | |
|---|---|---|---|
| Age in Years | Observed | Died | Mortality Rate |
| 13 | 138 | 32 | 0.2320 |
| 14 | 799 | 234 | 0.2930 |
| 15 | 900 | 307 | 0.3410 |
| 16 | 626 | 174 | 0.2780 |
| 17 | 457 | 150 | 0.3280 |
| 18 | 307 | 107 | 0.3490 |
| 19 | 200 | 59 | 0.2950 |
| 20 | 140 | 27 | 0.1930 |
| 21 | 112 | 27 | 0.2410 |
| 22 | 85 | 20 | 0.2350 |
| 23 | 66 | 9 | 0.1360 |
| 24 | 57 | 11 | 0.1930 |
| 25 | 46 | 7 | 0.1520 |
| 26 | 39 | 7 | 0.1790 |
| 27 | 31 | 1 | 0.0323 |
| 28 | 27 | 2 | 0.0741 |
| 29 | 11 | 1 | 0.0909 |

|  | Mechanical Reliability Rates | |
|---|---|---|
| Mileage at test | N tested | Pct failed |
| 70000 | 152 | 30.3 |
| 80000 | 235 | 33.2 |
| 90000 | 292 | 28.4 |
| 100000 | 394 | 31.7 |
| 110000 | 493 | 32.9 |
| 120000 | 592 | 35.0 |
| 130000 | 695 | 36.8 |
| 140000 | 636 | 33.5 |
| 150000 | 582 | 35.6 |
| 160000 | 538 | 36.6 |
| 170000 | 457 | 35.0 |
| 180000 | 377 | 36.1 |
| 190000 | 260 | 40.4 |
| 200000 | 177 | 39.0 |
| 210000 | 150 | 38.7 |
| 220000 | 126 | 40.5 |

**Audi 100 1993**

At 15 years of age, the mortality rate of a Audi 100 1993 (manufactured as a Car or Light Van) ranked number 61 out of 115 vehicles of the same age and type (any Car or Light Van constructed in 1993). One is the lowest (or best) and 115 the highest mortality rate. For vehicles reaching 120000 miles, its unreliability score (rate of failing an inspection) ranked 75 out of 98 vehicles of the same age, type, and mileage. One is the highest (or worst) and 98 the lowest rate of failing an inspection.

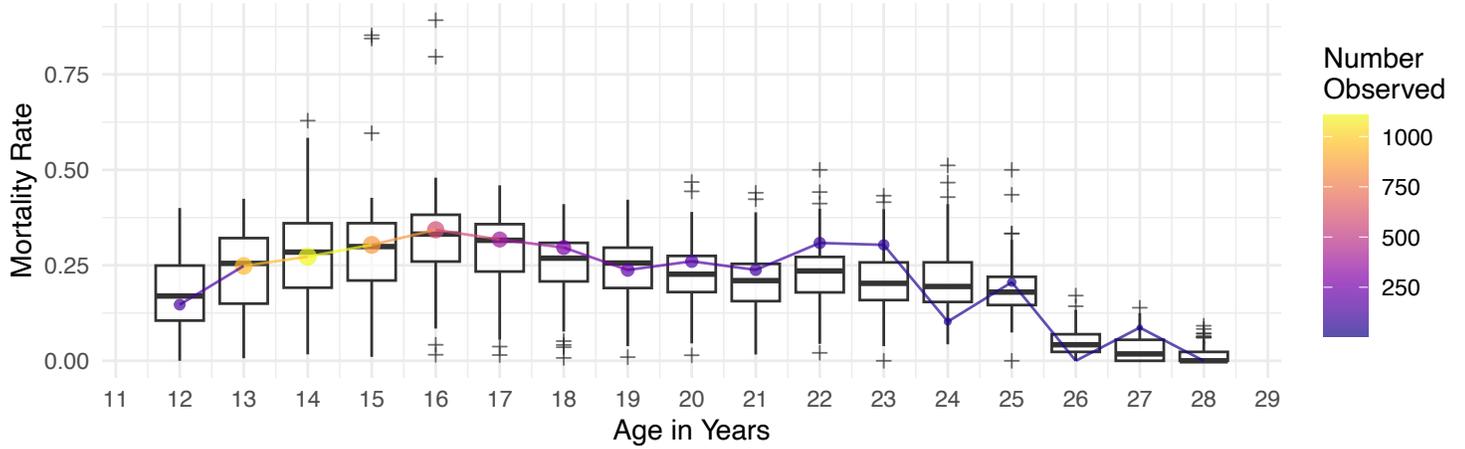

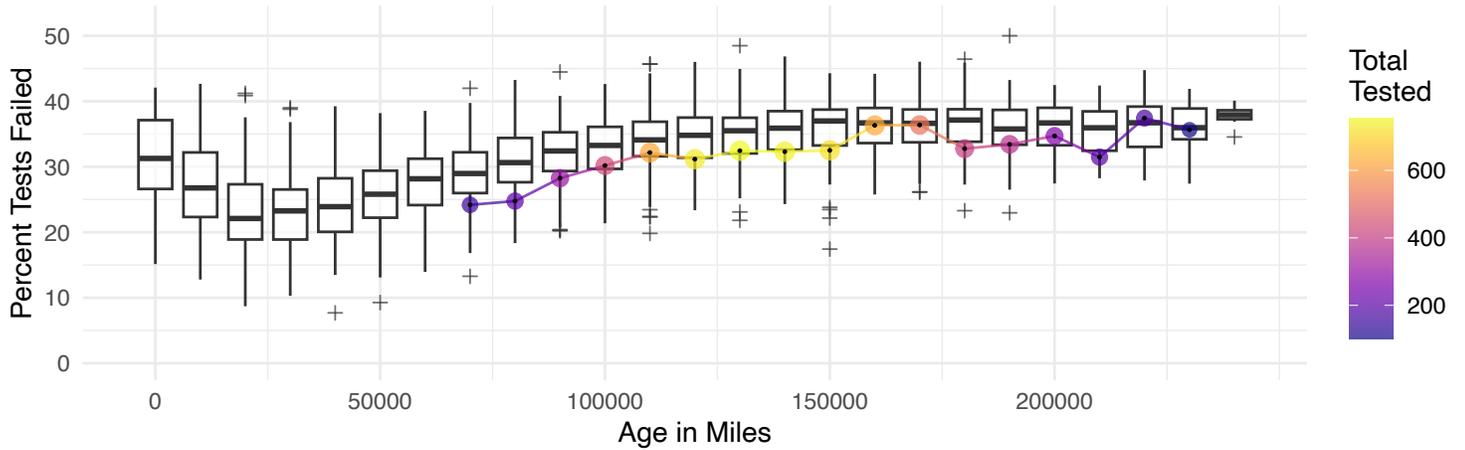

| Mortality rates | | | |
|---|---|---|---|
| Age in Years | Observed | Died | Mortality Rate |
| 12 | 136 | 20 | 0.147 |
| 13 | 942 | 234 | 0.248 |
| 14 | 1104 | 301 | 0.273 |
| 15 | 840 | 255 | 0.304 |
| 16 | 583 | 200 | 0.343 |
| 17 | 381 | 121 | 0.318 |
| 18 | 263 | 78 | 0.297 |
| 19 | 185 | 44 | 0.238 |
| 20 | 142 | 37 | 0.261 |
| 21 | 105 | 25 | 0.238 |
| 22 | 81 | 25 | 0.309 |
| 23 | 56 | 17 | 0.304 |
| 24 | 39 | 4 | 0.103 |
| 25 | 34 | 7 | 0.206 |
| 26 | 25 | 0 | 0.000 |
| 27 | 23 | 2 | 0.087 |

| Mechanical Reliability Rates | | |
|---|---|---|
| Mileage at test | N tested | Pct failed |
| 70000 | 153 | 24.2 |
| 80000 | 206 | 24.8 |
| 90000 | 315 | 28.3 |
| 100000 | 434 | 30.2 |
| 110000 | 628 | 32.2 |
| 120000 | 735 | 31.2 |
| 130000 | 755 | 32.5 |
| 140000 | 743 | 32.3 |
| 150000 | 720 | 32.5 |
| 160000 | 625 | 36.3 |
| 170000 | 525 | 36.4 |
| 180000 | 409 | 32.8 |
| 190000 | 374 | 33.4 |
| 200000 | 268 | 34.7 |
| 210000 | 197 | 31.5 |
| 220000 | 179 | 37.4 |
| 230000 | 101 | 35.6 |

**Audi 100 1994**

At 15 years of age, the mortality rate of a Audi 100 1994 (manufactured as a Car or Light Van) ranked number 53 out of 120 vehicles of the same age and type (any Car or Light Van constructed in 1994). One is the lowest (or best) and 120 the highest mortality rate. For vehicles reaching 120000 miles, its unreliability score (rate of failing an inspection) ranked 74 out of 112 vehicles of the same age, type, and mileage. One is the highest (or worst) and 112 the lowest rate of failing an inspection.

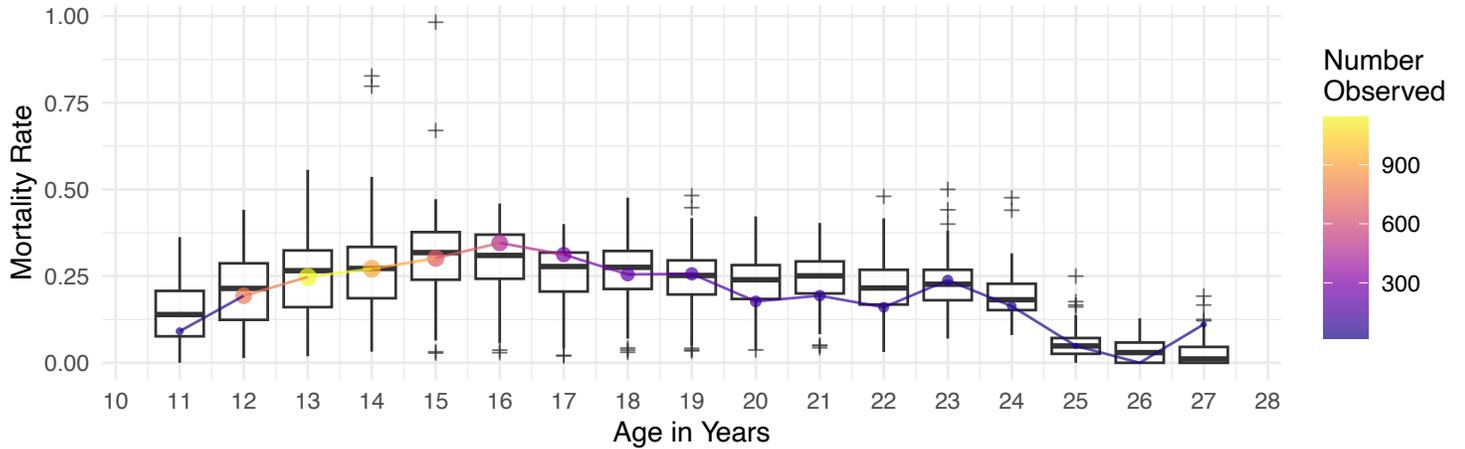

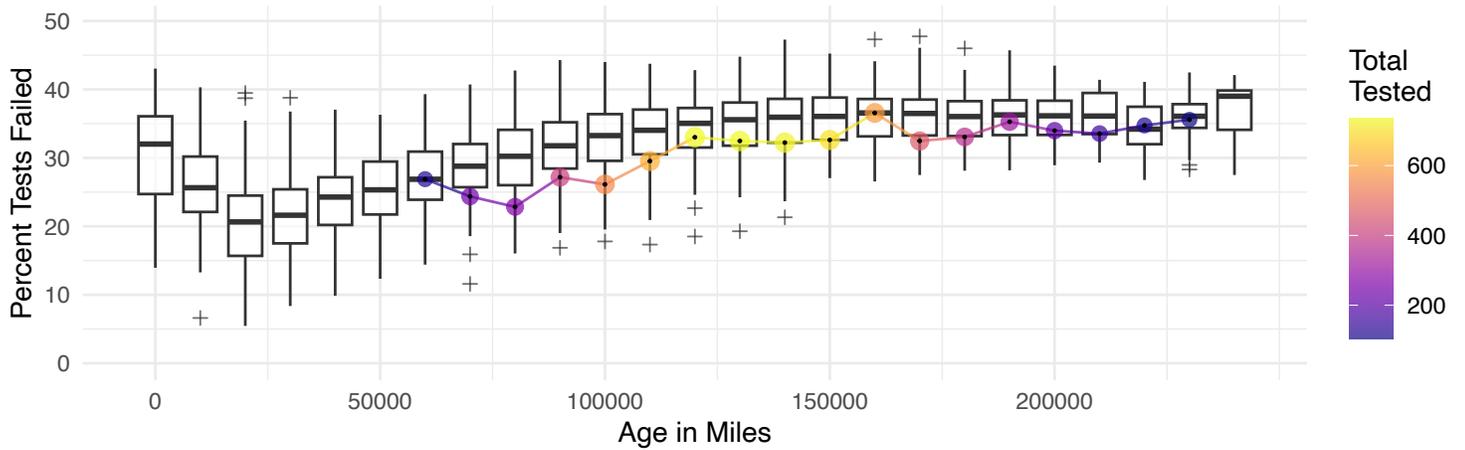

Mortality rates

| Age in Years | Observed | Died | Mortality Rate |
|---|---|---|---|
| 11 | 44 | 4 | 0.0909 |
| 12 | 771 | 149 | 0.1930 |
| 13 | 1139 | 282 | 0.2480 |
| 14 | 911 | 247 | 0.2710 |
| 15 | 663 | 200 | 0.3020 |
| 16 | 460 | 159 | 0.3460 |
| 17 | 298 | 93 | 0.3120 |
| 18 | 204 | 52 | 0.2550 |
| 19 | 152 | 39 | 0.2570 |
| 20 | 113 | 20 | 0.1770 |
| 21 | 93 | 18 | 0.1940 |
| 22 | 75 | 12 | 0.1600 |
| 23 | 63 | 15 | 0.2380 |
| 24 | 49 | 8 | 0.1630 |
| 25 | 41 | 2 | 0.0488 |
| 26 | 32 | 0 | 0.0000 |
| 27 | 18 | 2 | 0.1110 |

Mechanical Reliability Rates

| Mileage at test | N tested | Pct failed |
|---|---|---|
| 60000 | 119 | 26.9 |
| 70000 | 238 | 24.4 |
| 80000 | 267 | 22.8 |
| 100000 | 555 | 26.1 |
| 110000 | 623 | 29.5 |
| 120000 | 736 | 33.0 |
| 130000 | 736 | 32.5 |
| 140000 | 726 | 32.2 |
| 150000 | 702 | 32.6 |
| 160000 | 588 | 36.6 |
| 170000 | 462 | 32.5 |
| 180000 | 369 | 33.1 |
| 190000 | 309 | 35.3 |
| 200000 | 206 | 34.0 |
| 210000 | 155 | 33.5 |
| 220000 | 118 | 34.7 |
| 230000 | 104 | 35.6 |

## Audi 80 1988

At 20 years of age, the mortality rate of a Audi 80 1988 (manufactured as a Car or Light Van) ranked number 34 out of 40 vehicles of the same age and type (any Car or Light Van constructed in 1988). One is the lowest (or best) and 40 the highest mortality rate. For vehicles reaching 120000 miles, its unreliability score (rate of failing an inspection) ranked 17 out of 28 vehicles of the same age, type, and mileage. One is the highest (or worst) and 28 the lowest rate of failing an inspection.

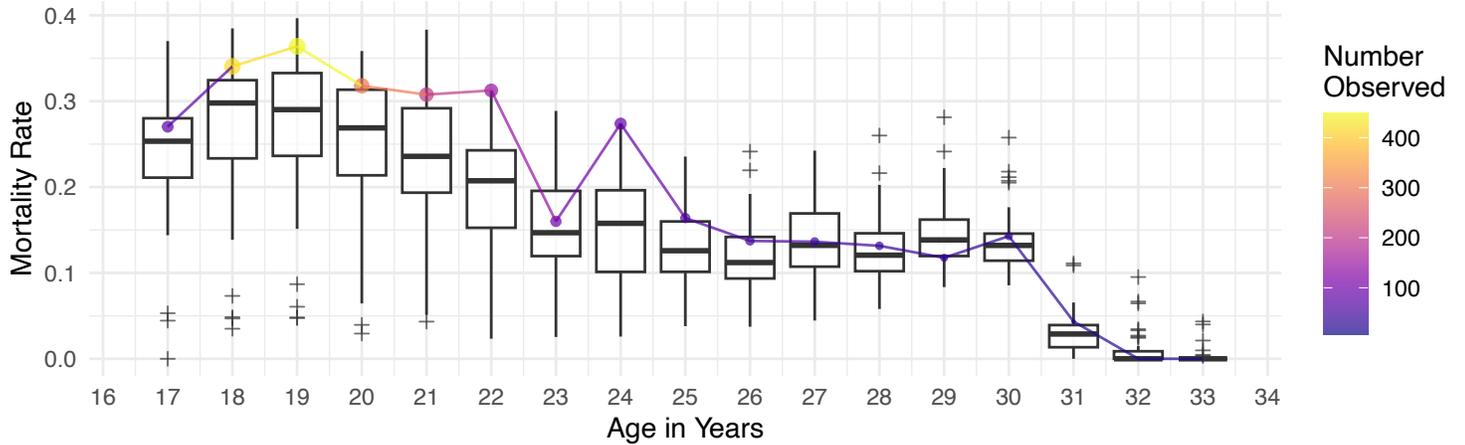

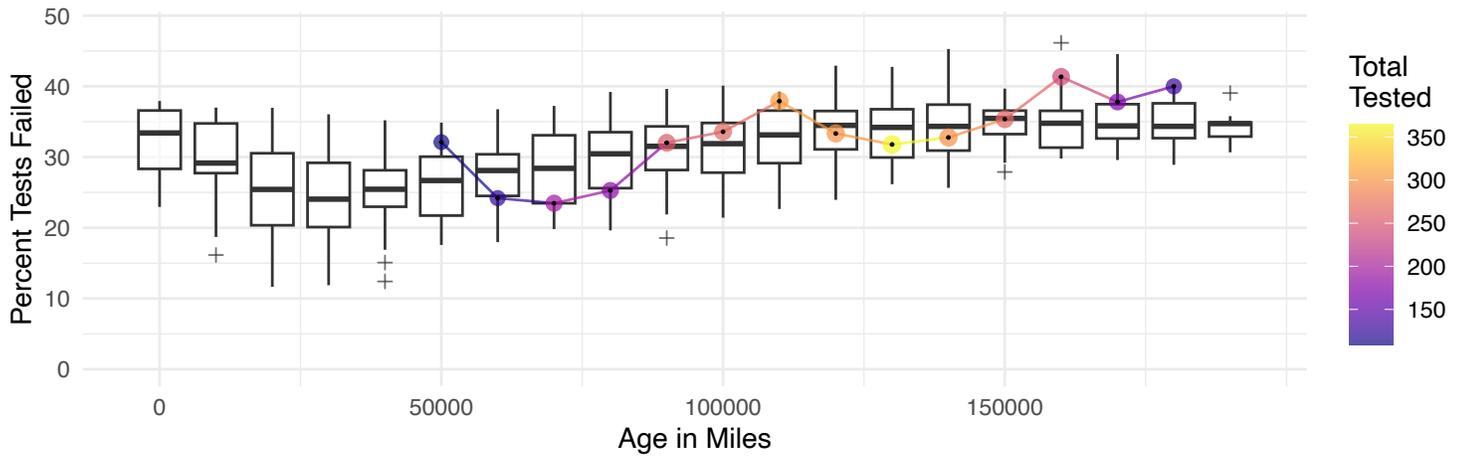

| Mortality rates | | | |
|---|---|---|---|
| Age in Years | Observed | Died | Mortality Rate |
| 17 | 74 | 20 | 0.2700 |
| 18 | 420 | 143 | 0.3400 |
| 19 | 448 | 163 | 0.3640 |
| 20 | 302 | 96 | 0.3180 |
| 21 | 208 | 64 | 0.3080 |
| 22 | 144 | 45 | 0.3120 |
| 23 | 100 | 16 | 0.1600 |
| 24 | 84 | 23 | 0.2740 |
| 25 | 61 | 10 | 0.1640 |
| 26 | 51 | 7 | 0.1370 |
| 27 | 44 | 6 | 0.1360 |
| 28 | 38 | 5 | 0.1320 |
| 29 | 34 | 4 | 0.1180 |
| 30 | 28 | 4 | 0.1430 |
| 31 | 23 | 1 | 0.0435 |
| 32 | 17 | 0 | 0.0000 |

| Mechanical Reliability Rates | | |
|---|---|---|
| Mileage at test | N tested | Pct failed |
| 50000 | 109 | 32.1 |
| 60000 | 124 | 24.2 |
| 70000 | 196 | 23.5 |
| 80000 | 182 | 25.3 |
| 90000 | 253 | 32.0 |
| 100000 | 265 | 33.6 |
| 110000 | 306 | 37.9 |
| 120000 | 300 | 33.3 |
| 130000 | 365 | 31.8 |
| 140000 | 296 | 32.8 |
| 150000 | 263 | 35.4 |
| 160000 | 237 | 41.4 |
| 170000 | 172 | 37.8 |
| 180000 | 130 | 40.0 |

## Audi 80 1989

At 20 years of age, the mortality rate of a Audi 80 1989 (manufactured as a Car or Light Van) ranked number 37 out of 61 vehicles of the same age and type (any Car or Light Van constructed in 1989). One is the lowest (or best) and 61 the highest mortality rate. For vehicles reaching 120000 miles, its unreliability score (rate of failing an inspection) ranked 26 out of 49 vehicles of the same age, type, and mileage. One is the highest (or worst) and 49 the lowest rate of failing an inspection.

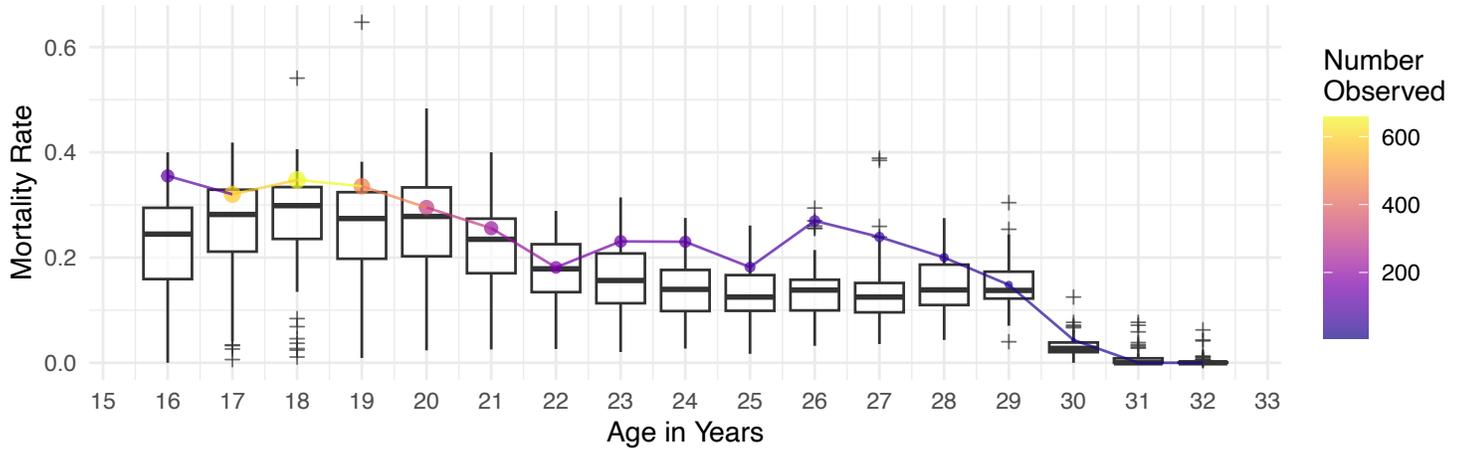

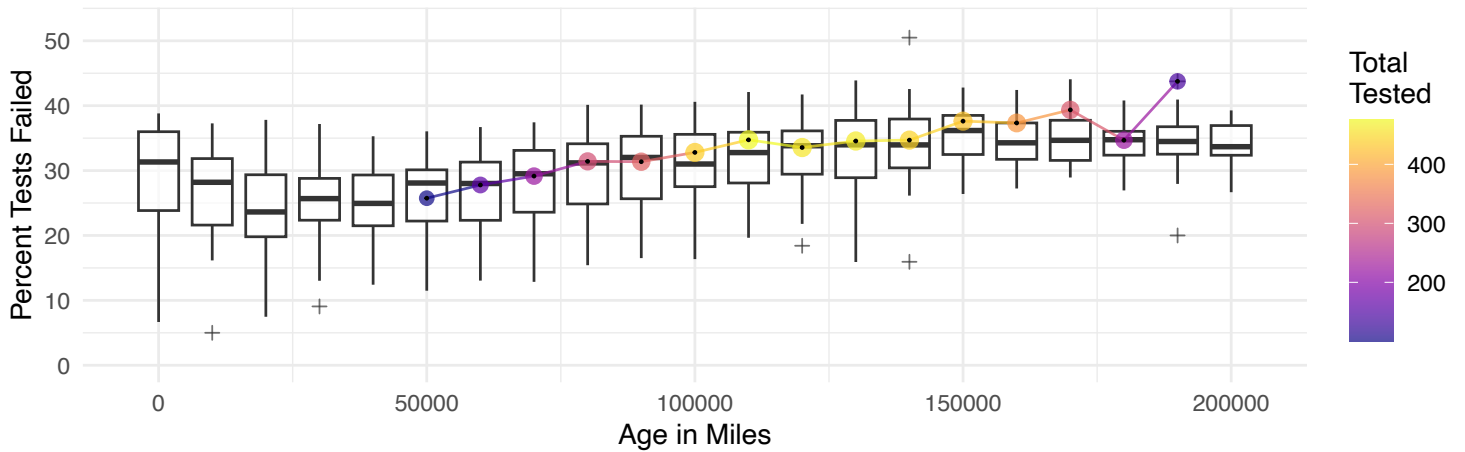

Mortality rates

| Age in Years | Observed | Died | Mortality Rate |
|---|---|---|---|
| 16 | 107 | 38 | 0.3550 |
| 17 | 584 | 187 | 0.3200 |
| 18 | 656 | 228 | 0.3480 |
| 19 | 456 | 153 | 0.3360 |
| 20 | 305 | 90 | 0.2950 |
| 21 | 215 | 55 | 0.2560 |
| 22 | 160 | 29 | 0.1810 |
| 23 | 130 | 30 | 0.2310 |
| 24 | 100 | 23 | 0.2300 |
| 25 | 77 | 14 | 0.1820 |
| 26 | 63 | 17 | 0.2700 |
| 27 | 46 | 11 | 0.2390 |
| 28 | 35 | 7 | 0.2000 |
| 29 | 27 | 4 | 0.1480 |
| 30 | 23 | 1 | 0.0435 |
| 31 | 18 | 0 | 0.0000 |

Mechanical Reliability Rates

| Mileage at test | N tested | Pct failed |
|---|---|---|
| 50000 | 101 | 25.7 |
| 60000 | 162 | 27.8 |
| 70000 | 223 | 29.1 |
| 80000 | 296 | 31.4 |
| 90000 | 322 | 31.4 |
| 100000 | 436 | 32.8 |
| 110000 | 475 | 34.7 |
| 120000 | 459 | 33.6 |
| 130000 | 463 | 34.6 |
| 140000 | 441 | 34.7 |
| 150000 | 428 | 37.6 |
| 160000 | 383 | 37.3 |
| 170000 | 310 | 39.4 |
| 180000 | 222 | 34.7 |
| 190000 | 144 | 43.8 |

**Audi 80 1990**

At 15 years of age, the mortality rate of a Audi 80 1990 (manufactured as a Car or Light Van) ranked number 60 out of 77 vehicles of the same age and type (any Car or Light Van constructed in 1990). One is the lowest (or best) and 77 the highest mortality rate. For vehicles reaching 120000 miles, its unreliability score (rate of failing an inspection) ranked 26 out of 60 vehicles of the same age, type, and mileage. One is the highest (or worst) and 60 the lowest rate of failing an inspection.

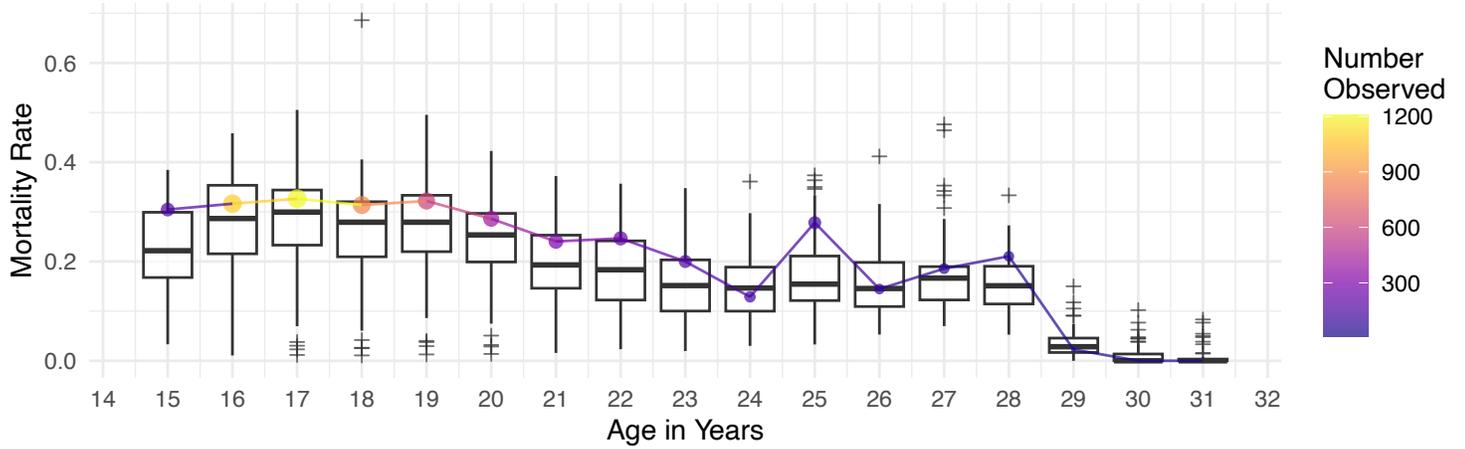

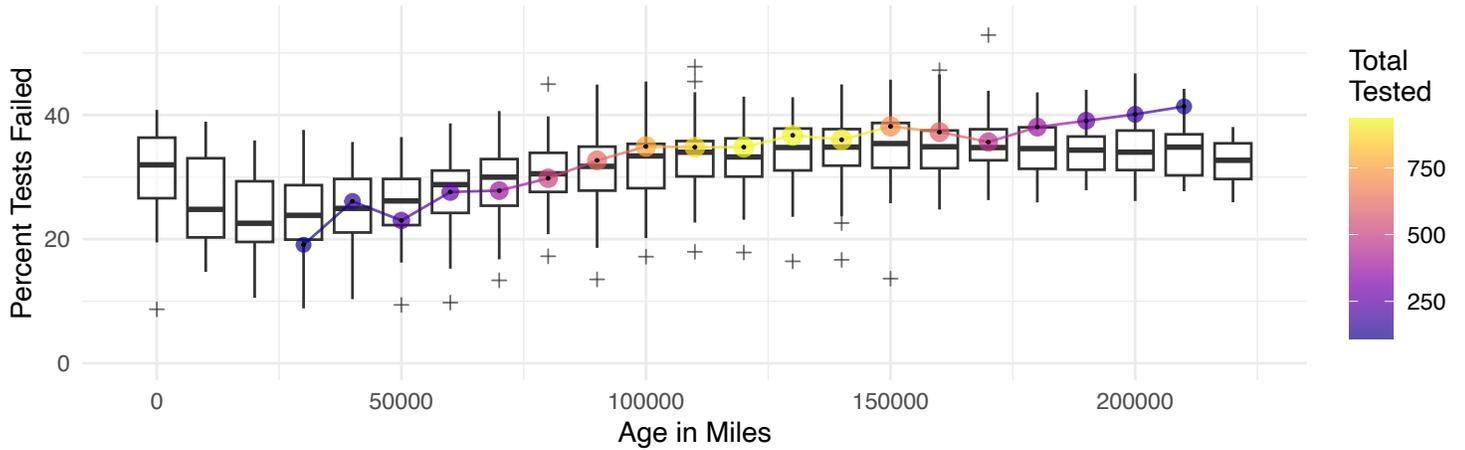

| Mortality rates | | | |
|---|---|---|---|
| Age in Years | Observed | Died | Mortality Rate |
| 15 | 151 | 46 | 0.3050 |
| 16 | 1084 | 343 | 0.3160 |
| 17 | 1204 | 393 | 0.3260 |
| 18 | 854 | 268 | 0.3140 |
| 19 | 587 | 189 | 0.3220 |
| 20 | 402 | 115 | 0.2860 |
| 21 | 287 | 69 | 0.2400 |
| 22 | 219 | 54 | 0.2470 |
| 23 | 165 | 33 | 0.2000 |
| 24 | 132 | 17 | 0.1290 |
| 25 | 115 | 32 | 0.2780 |
| 26 | 83 | 12 | 0.1450 |
| 27 | 70 | 13 | 0.1860 |
| 28 | 57 | 12 | 0.2110 |
| 29 | 44 | 1 | 0.0227 |
| 30 | 31 | 0 | 0.0000 |
| 31 | 11 | 0 | 0.0000 |

| Mechanical Reliability Rates | | |
|---|---|---|
| Mileage at test | N tested | Pct failed |
| 30000 | 110 | 19.1 |
| 40000 | 161 | 26.1 |
| 50000 | 213 | 23.0 |
| 60000 | 250 | 27.6 |
| 70000 | 381 | 27.8 |
| 100000 | 773 | 34.9 |
| 110000 | 873 | 34.8 |
| 120000 | 936 | 34.8 |
| 130000 | 931 | 36.7 |
| 140000 | 908 | 36.0 |
| 150000 | 710 | 38.2 |
| 160000 | 588 | 37.2 |
| 170000 | 460 | 35.7 |
| 180000 | 381 | 38.1 |
| 190000 | 261 | 39.1 |
| 200000 | 177 | 40.1 |
| 210000 | 116 | 41.4 |

**Audi 80 1991**

At 15 years of age, the mortality rate of a Audi 80 1991 (manufactured as a Car or Light Van) ranked number 48 out of 76 vehicles of the same age and type (any Car or Light Van constructed in 1991). One is the lowest (or best) and 76 the highest mortality rate. For vehicles reaching 120000 miles, its unreliability score (rate of failing an inspection) ranked 12 out of 61 vehicles of the same age, type, and mileage. One is the highest (or worst) and 61 the lowest rate of failing an inspection.

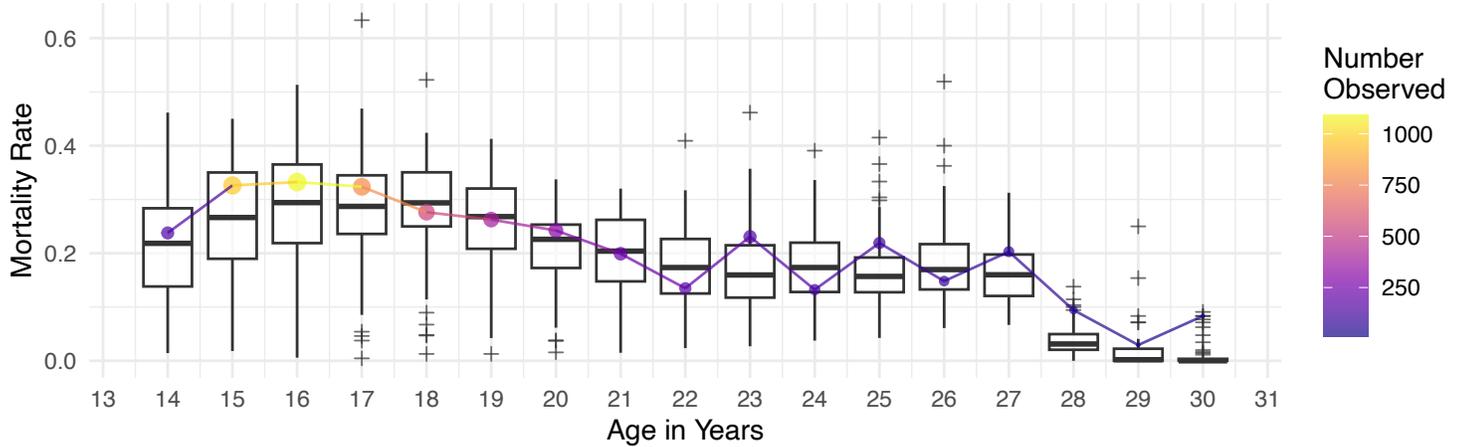

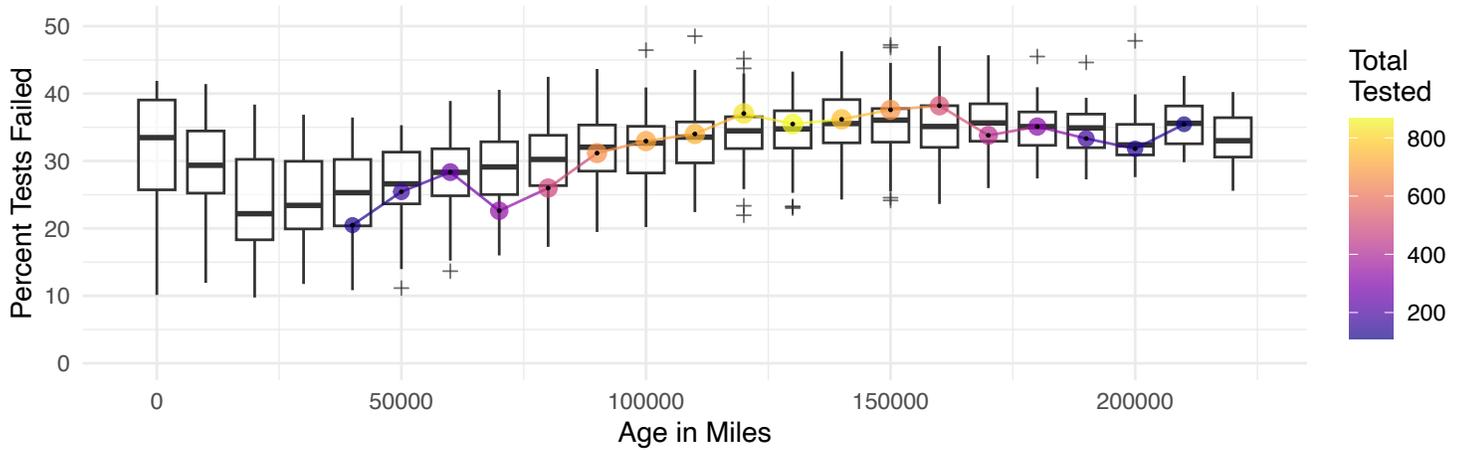

<table>
<tr><td colspan="4" align="center">Mortality rates</td></tr>
<tr><th>Age in Years</th><th>Observed</th><th>Died</th><th>Mortality Rate</th></tr>
<tr><td>14</td><td>143</td><td>34</td><td>0.2380</td></tr>
<tr><td>15</td><td>969</td><td>316</td><td>0.3260</td></tr>
<tr><td>16</td><td>1089</td><td>362</td><td>0.3320</td></tr>
<tr><td>17</td><td>791</td><td>256</td><td>0.3240</td></tr>
<tr><td>18</td><td>543</td><td>150</td><td>0.2760</td></tr>
<tr><td>19</td><td>396</td><td>104</td><td>0.2630</td></tr>
<tr><td>20</td><td>293</td><td>71</td><td>0.2420</td></tr>
<tr><td>21</td><td>221</td><td>44</td><td>0.1990</td></tr>
<tr><td>22</td><td>178</td><td>24</td><td>0.1350</td></tr>
<tr><td>23</td><td>156</td><td>36</td><td>0.2310</td></tr>
<tr><td>24</td><td>121</td><td>16</td><td>0.1320</td></tr>
<tr><td>25</td><td>105</td><td>23</td><td>0.2190</td></tr>
<tr><td>26</td><td>81</td><td>12</td><td>0.1480</td></tr>
<tr><td>27</td><td>69</td><td>14</td><td>0.2030</td></tr>
<tr><td>28</td><td>53</td><td>5</td><td>0.0943</td></tr>
<tr><td>29</td><td>34</td><td>1</td><td>0.0294</td></tr>
<tr><td>30</td><td>12</td><td>1</td><td>0.0833</td></tr>
</table>

<table>
<tr><td colspan="3" align="center">Mechanical Reliability Rates</td></tr>
<tr><th>Mileage at test</th><th>N tested</th><th>Pct failed</th></tr>
<tr><td>40000</td><td>122</td><td>20.5</td></tr>
<tr><td>50000</td><td>169</td><td>25.4</td></tr>
<tr><td>60000</td><td>261</td><td>28.4</td></tr>
<tr><td>70000</td><td>336</td><td>22.6</td></tr>
<tr><td>80000</td><td>481</td><td>26.0</td></tr>
<tr><td>100000</td><td>704</td><td>33.0</td></tr>
<tr><td>110000</td><td>735</td><td>34.0</td></tr>
<tr><td>120000</td><td>834</td><td>37.1</td></tr>
<tr><td>130000</td><td>868</td><td>35.5</td></tr>
<tr><td>140000</td><td>760</td><td>36.2</td></tr>
<tr><td>150000</td><td>673</td><td>37.6</td></tr>
<tr><td>160000</td><td>513</td><td>38.2</td></tr>
<tr><td>170000</td><td>426</td><td>33.8</td></tr>
<tr><td>180000</td><td>319</td><td>35.1</td></tr>
<tr><td>190000</td><td>174</td><td>33.3</td></tr>
<tr><td>200000</td><td>132</td><td>31.8</td></tr>
<tr><td>210000</td><td>110</td><td>35.5</td></tr>
</table>

## Audi 80 1992

At 15 years of age, the mortality rate of a Audi 80 1992 (manufactured as a Car or Light Van) ranked number 51 out of 90 vehicles of the same age and type (any Car or Light Van constructed in 1992). One is the lowest (or best) and 90 the highest mortality rate. For vehicles reaching 120000 miles, its unreliability score (rate of failing an inspection) ranked 23 out of 75 vehicles of the same age, type, and mileage. One is the highest (or worst) and 75 the lowest rate of failing an inspection.

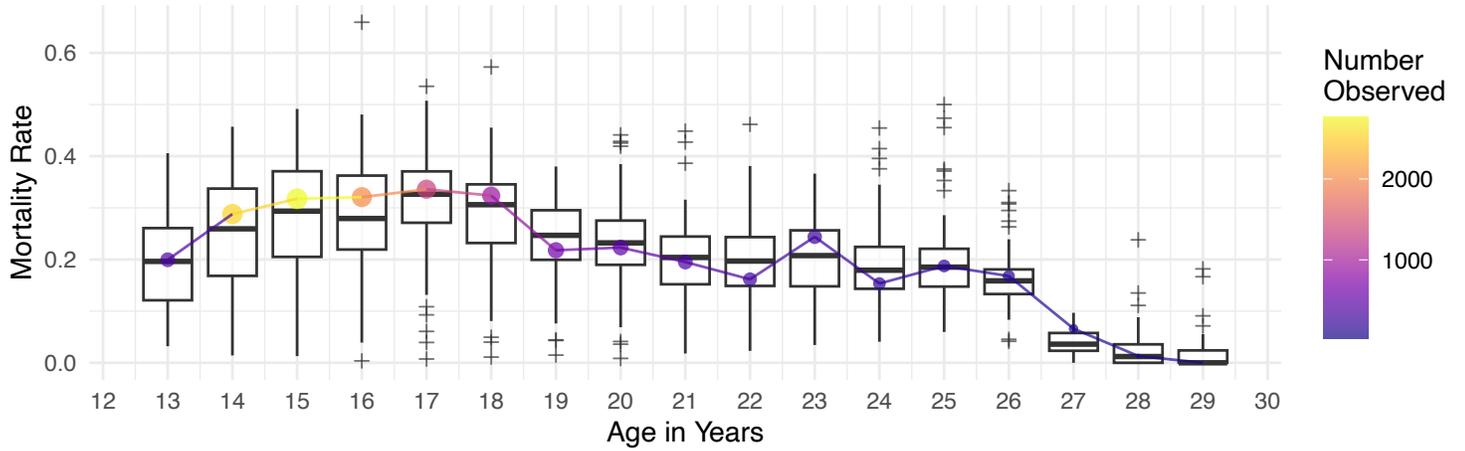

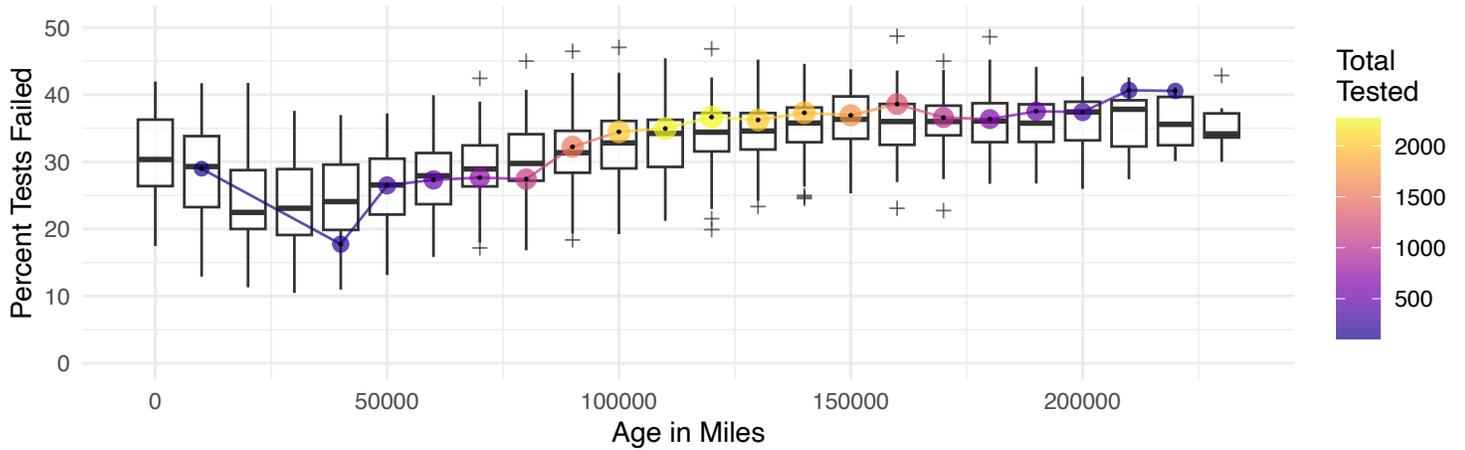

| Mortality rates | | | |
|---|---|---|---|
| Age in Years | Observed | Died | Mortality Rate |
| 13 | 356 | 71 | 0.1990 |
| 14 | 2525 | 727 | 0.2880 |
| 15 | 2758 | 876 | 0.3180 |
| 16 | 1983 | 636 | 0.3210 |
| 17 | 1351 | 454 | 0.3360 |
| 18 | 896 | 290 | 0.3240 |
| 19 | 606 | 132 | 0.2180 |
| 20 | 475 | 106 | 0.2230 |
| 21 | 369 | 72 | 0.1950 |
| 22 | 297 | 48 | 0.1620 |
| 23 | 250 | 61 | 0.2440 |
| 24 | 189 | 29 | 0.1530 |
| 25 | 160 | 30 | 0.1880 |
| 26 | 131 | 22 | 0.1680 |
| 27 | 106 | 7 | 0.0660 |
| 28 | 76 | 1 | 0.0132 |
| 29 | 30 | 0 | 0.0000 |

| Mechanical Reliability Rates | | |
|---|---|---|
| Mileage at test | N tested | Pct failed |
| 10000 | 107 | 29.0 |
| 40000 | 203 | 17.7 |
| 50000 | 302 | 26.5 |
| 60000 | 498 | 27.3 |
| 100000 | 2022 | 34.5 |
| 110000 | 2260 | 35.0 |
| 120000 | 2272 | 36.7 |
| 130000 | 2034 | 36.2 |
| 140000 | 1949 | 37.3 |
| 150000 | 1698 | 36.9 |
| 160000 | 1298 | 38.6 |
| 170000 | 1000 | 36.6 |
| 180000 | 660 | 36.4 |
| 190000 | 493 | 37.5 |
| 200000 | 318 | 37.4 |
| 210000 | 182 | 40.7 |
| 220000 | 143 | 40.6 |

## Audi 80 1993

At 15 years of age, the mortality rate of a Audi 80 1993 (manufactured as a Car or Light Van) ranked number 51 out of 115 vehicles of the same age and type (any Car or Light Van constructed in 1993). One is the lowest (or best) and 115 the highest mortality rate. For vehicles reaching 120000 miles, its unreliability score (rate of failing an inspection) ranked 63 out of 98 vehicles of the same age, type, and mileage. One is the highest (or worst) and 98 the lowest rate of failing an inspection.

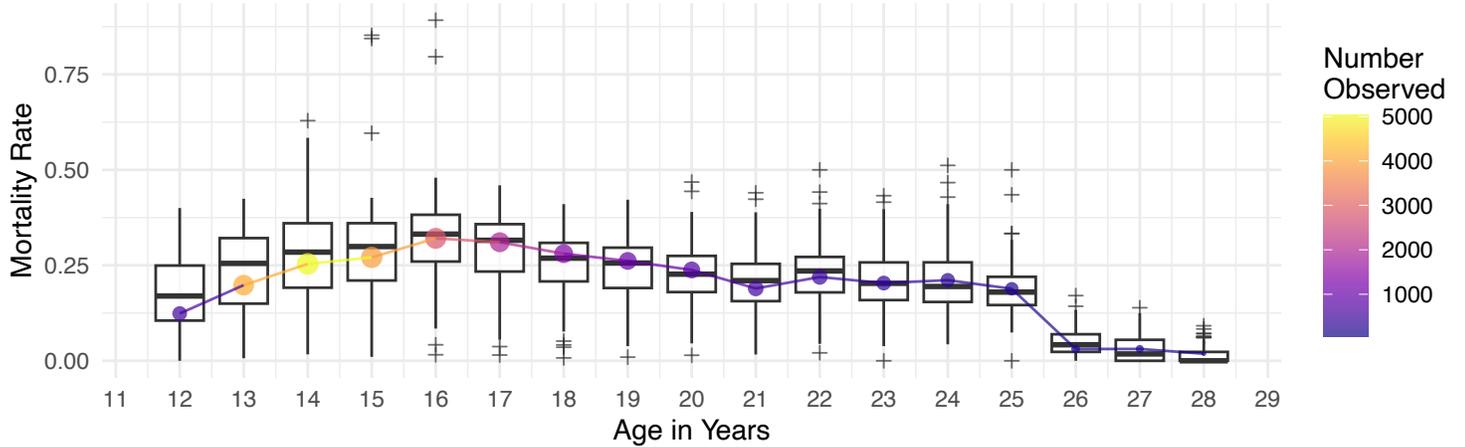

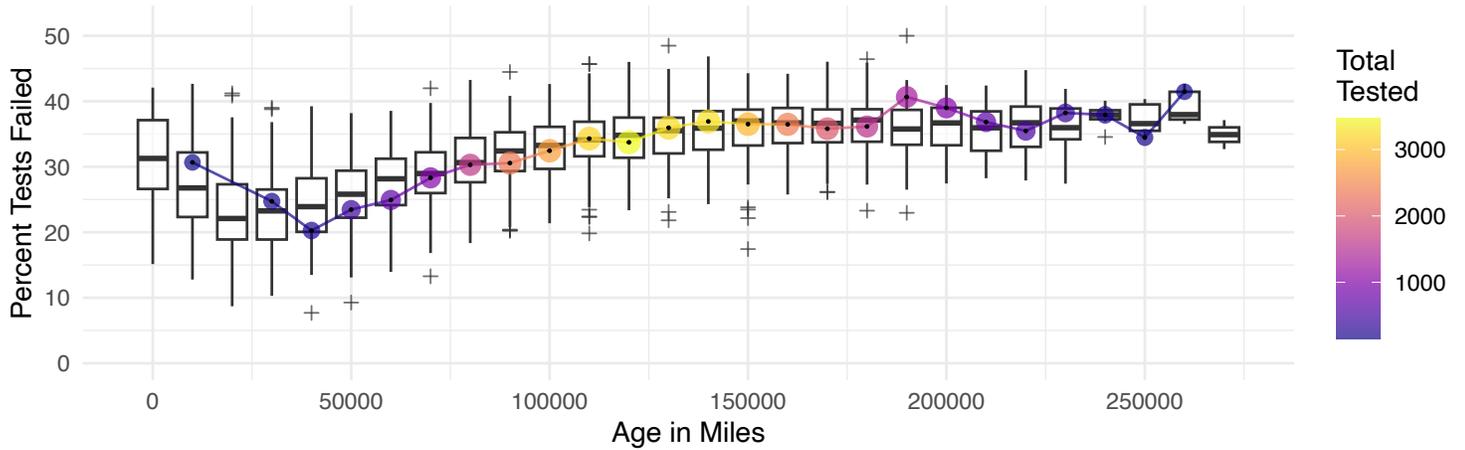

| Mortality rates | | | |
|---|---|---|---|
| Age in Years | Observed | Died | Mortality Rate |
| 12 | 541 | 67 | 0.1240 |
| 13 | 4077 | 808 | 0.1980 |
| 14 | 5018 | 1273 | 0.2540 |
| 15 | 3910 | 1059 | 0.2710 |
| 16 | 2847 | 913 | 0.3210 |
| 17 | 1929 | 599 | 0.3110 |
| 18 | 1326 | 372 | 0.2810 |
| 19 | 952 | 249 | 0.2620 |
| 20 | 702 | 167 | 0.2380 |
| 21 | 533 | 101 | 0.1890 |
| 22 | 433 | 95 | 0.2190 |
| 23 | 339 | 69 | 0.2040 |
| 24 | 270 | 57 | 0.2110 |
| 25 | 212 | 40 | 0.1890 |
| 26 | 162 | 5 | 0.0309 |
| 27 | 129 | 4 | 0.0310 |
| 28 | 54 | 1 | 0.0185 |

| Mechanical Reliability Rates | | |
|---|---|---|
| Mileage at test | N tested | Pct failed |
| 10000 | 150 | 30.7 |
| 100000 | 2737 | 32.4 |
| 110000 | 3184 | 34.3 |
| 120000 | 3474 | 33.7 |
| 130000 | 3246 | 36.0 |
| 140000 | 3291 | 36.9 |
| 150000 | 2953 | 36.5 |
| 160000 | 2399 | 36.5 |
| 170000 | 2005 | 35.8 |
| 180000 | 1541 | 36.1 |
| 190000 | 1301 | 40.7 |
| 210000 | 687 | 36.8 |
| 220000 | 530 | 35.5 |
| 230000 | 382 | 38.2 |
| 240000 | 232 | 37.9 |
| 250000 | 177 | 34.5 |
| 260000 | 152 | 41.4 |

## Audi 80 1994

At 15 years of age, the mortality rate of a Audi 80 1994 (manufactured as a Car or Light Van) ranked number 42 out of 120 vehicles of the same age and type (any Car or Light Van constructed in 1994). One is the lowest (or best) and 120 the highest mortality rate. For vehicles reaching 120000 miles, its unreliability score (rate of failing an inspection) ranked 58 out of 112 vehicles of the same age, type, and mileage. One is the highest (or worst) and 112 the lowest rate of failing an inspection.

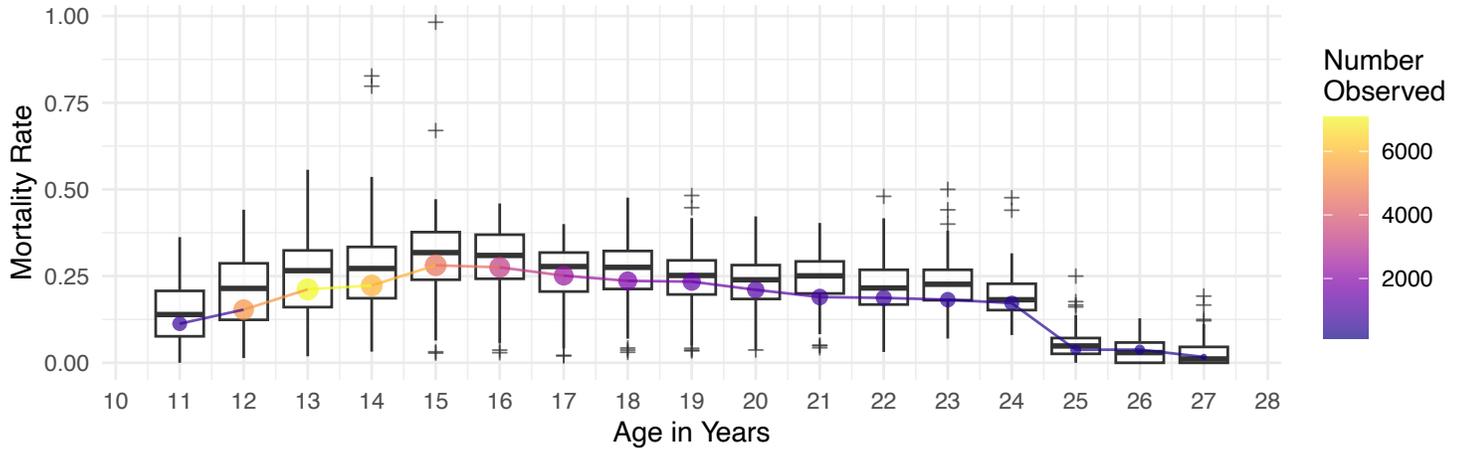

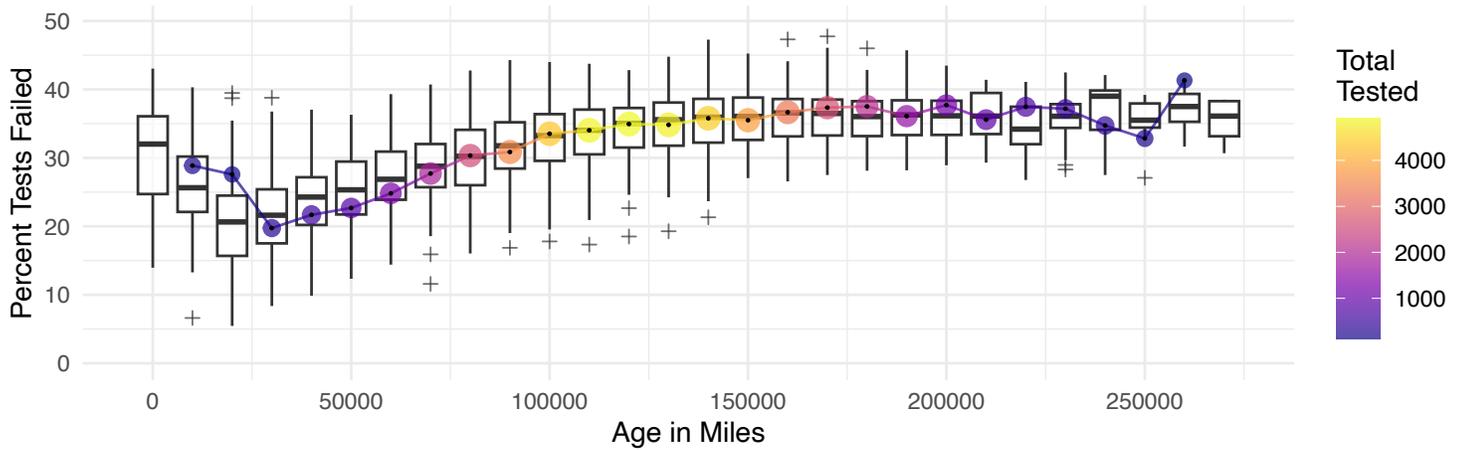

| Mortality rates | | | |
|---|---|---|---|
| Age in Years | Observed | Died | Mortality Rate |
| 11 | 621 | 70 | 0.1130 |
| 12 | 5286 | 811 | 0.1530 |
| 13 | 7058 | 1495 | 0.2120 |
| 14 | 5864 | 1307 | 0.2230 |
| 15 | 4548 | 1279 | 0.2810 |
| 16 | 3266 | 899 | 0.2750 |
| 17 | 2359 | 593 | 0.2510 |
| 18 | 1765 | 417 | 0.2360 |
| 19 | 1346 | 315 | 0.2340 |
| 20 | 1031 | 217 | 0.2100 |
| 21 | 812 | 154 | 0.1900 |
| 22 | 657 | 123 | 0.1870 |
| 23 | 533 | 97 | 0.1820 |
| 24 | 436 | 75 | 0.1720 |
| 25 | 347 | 13 | 0.0375 |
| 26 | 263 | 10 | 0.0380 |
| 27 | 119 | 2 | 0.0168 |

| Mechanical Reliability Rates | | |
|---|---|---|
| Mileage at test | N tested | Pct failed |
| 10000 | 187 | 28.9 |
| 20000 | 145 | 27.6 |
| 100000 | 4343 | 33.5 |
| 110000 | 4860 | 34.0 |
| 120000 | 4919 | 34.9 |
| 130000 | 4798 | 34.8 |
| 140000 | 4423 | 35.8 |
| 150000 | 3840 | 35.5 |
| 160000 | 3308 | 36.7 |
| 170000 | 2678 | 37.3 |
| 180000 | 2122 | 37.5 |
| 190000 | 1604 | 36.1 |
| 210000 | 888 | 35.6 |
| 220000 | 662 | 37.5 |
| 230000 | 468 | 37.2 |
| 240000 | 357 | 34.7 |
| 250000 | 204 | 32.8 |

**Audi 80 1995**

At 10 years of age, the mortality rate of a Audi 80 1995 (manufactured as a Car or Light Van) ranked number 34 out of 148 vehicles of the same age and type (any Car or Light Van constructed in 1995). One is the lowest (or best) and 148 the highest mortality rate. For vehicles reaching 120000 miles, its unreliability score (rate of failing an inspection) ranked 80 out of 135 vehicles of the same age, type, and mileage. One is the highest (or worst) and 135 the lowest rate of failing an inspection.

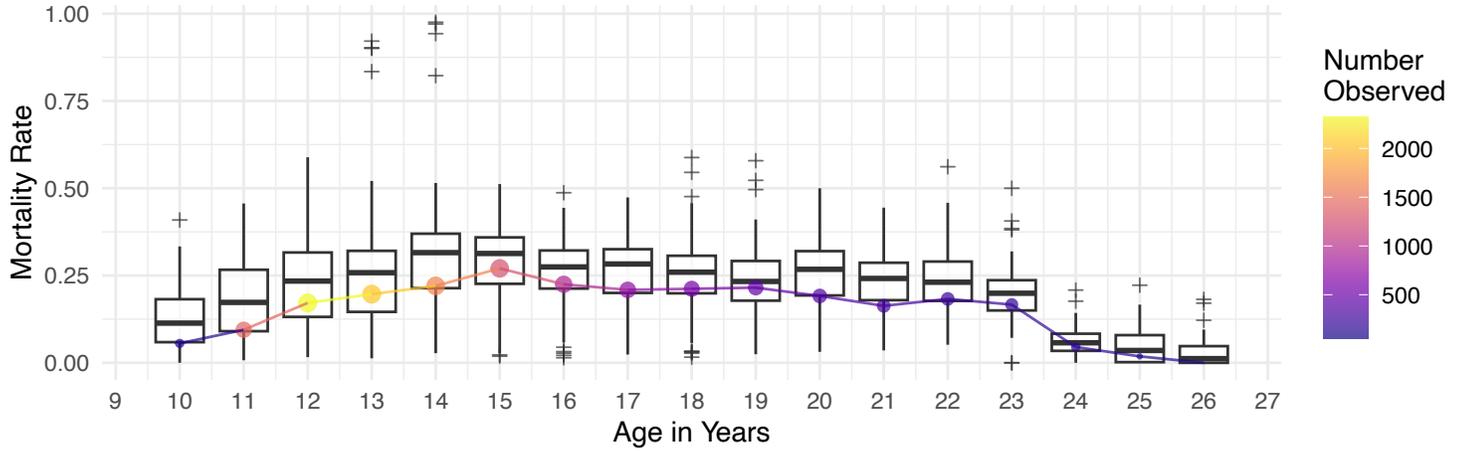

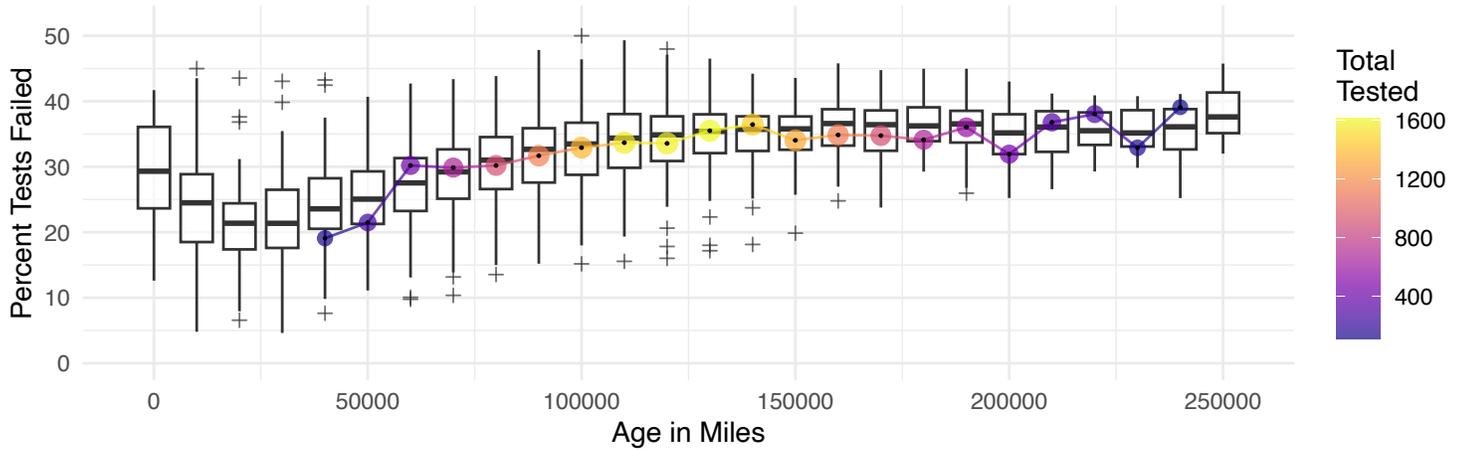

| Mortality rates | | | |
|---|---|---|---|
| Age in Years | Observed | Died | Mortality Rate |
| 10 | 126 | 7 | 0.0556 |
| 11 | 1428 | 135 | 0.0945 |
| 12 | 2316 | 397 | 0.1710 |
| 13 | 2039 | 401 | 0.1970 |
| 14 | 1634 | 360 | 0.2200 |
| 15 | 1272 | 344 | 0.2700 |
| 16 | 925 | 208 | 0.2250 |
| 17 | 718 | 150 | 0.2090 |
| 18 | 566 | 120 | 0.2120 |
| 19 | 446 | 96 | 0.2150 |
| 20 | 350 | 67 | 0.1910 |
| 21 | 282 | 46 | 0.1630 |
| 22 | 235 | 43 | 0.1830 |
| 23 | 192 | 32 | 0.1670 |
| 24 | 155 | 7 | 0.0452 |
| 25 | 109 | 2 | 0.0183 |
| 26 | 53 | 0 | 0.0000 |

| Mechanical Reliability Rates | | |
|---|---|---|
| Mileage at test | N tested | Pct failed |
| 40000 | 131 | 19.1 |
| 50000 | 242 | 21.5 |
| 60000 | 414 | 30.2 |
| 100000 | 1367 | 32.9 |
| 110000 | 1565 | 33.7 |
| 120000 | 1588 | 33.6 |
| 130000 | 1614 | 35.5 |
| 140000 | 1457 | 36.4 |
| 150000 | 1322 | 34.0 |
| 160000 | 1133 | 34.9 |
| 170000 | 892 | 34.8 |
| 180000 | 677 | 34.1 |
| 190000 | 583 | 36.0 |
| 200000 | 426 | 31.9 |
| 210000 | 318 | 36.8 |
| 220000 | 276 | 38.0 |
| 230000 | 152 | 32.9 |

## Audi A1 2010

At 5 years of age, the mortality rate of a Audi A1 2010 (manufactured as a Car or Light Van) ranked number 24 out of 206 vehicles of the same age and type (any Car or Light Van constructed in 2010). One is the lowest (or best) and 206 the highest mortality rate. For vehicles reaching 20000 miles, its unreliability score (rate of failing an inspection) ranked 180 out of 201 vehicles of the same age, type, and mileage. One is the highest (or worst) and 201 the lowest rate of failing an inspection.

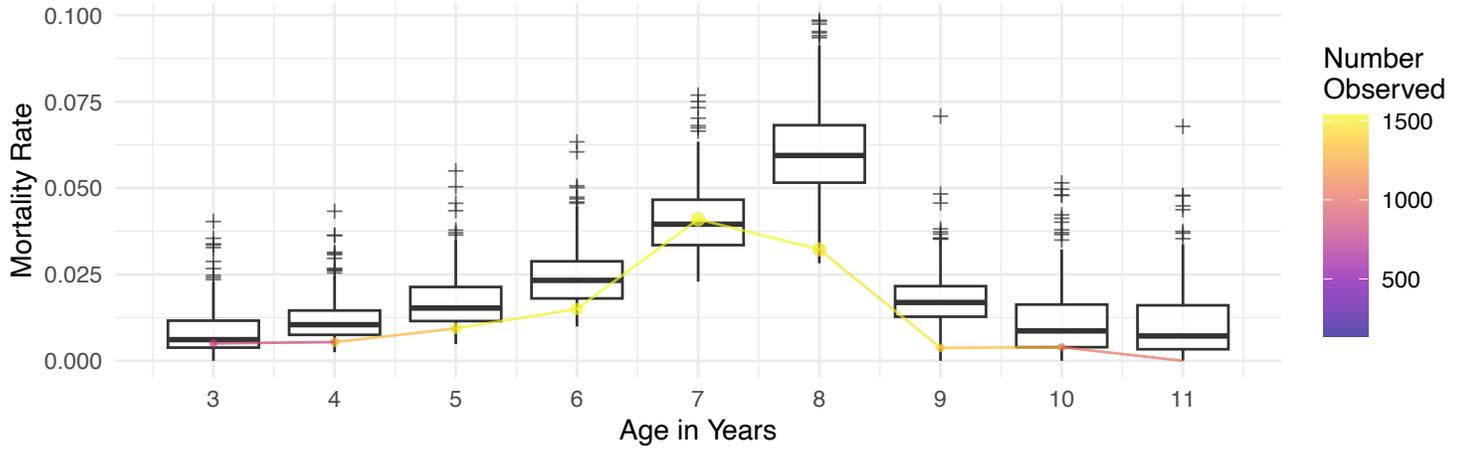

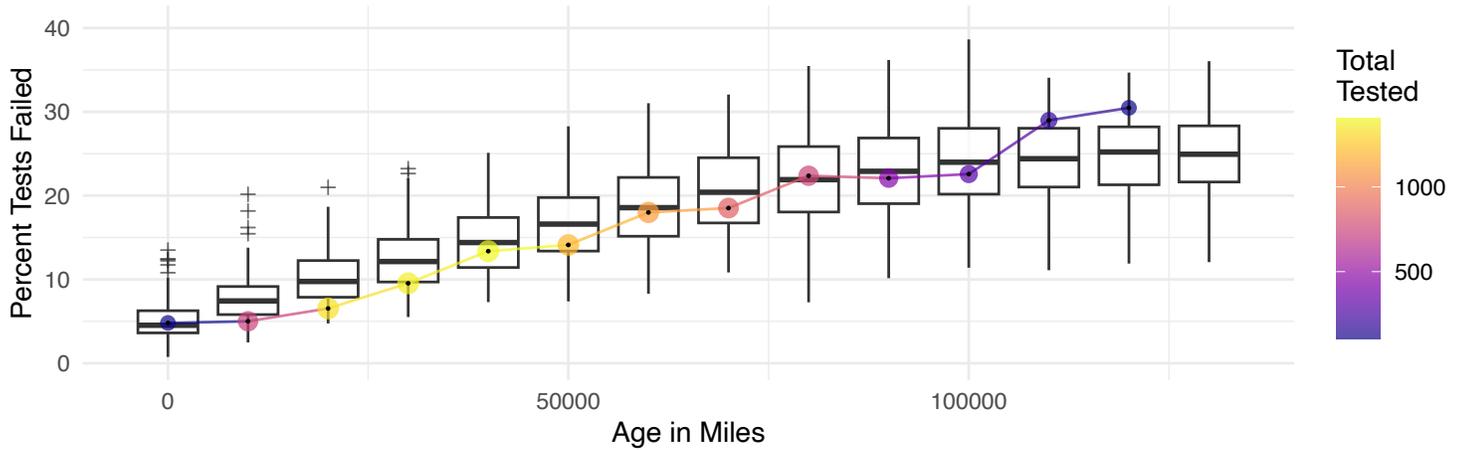

Mortality rates

| Age in Years | Observed | Died | Mortality Rate |
|---|---|---|---|
| 3 | 793 | 4 | 0.00504 |
| 4 | 1290 | 7 | 0.00543 |
| 5 | 1495 | 14 | 0.00936 |
| 6 | 1532 | 23 | 0.01500 |
| 7 | 1514 | 62 | 0.04100 |
| 8 | 1457 | 47 | 0.03230 |
| 9 | 1335 | 5 | 0.00375 |
| 10 | 1019 | 4 | 0.00393 |
| 11 | 137 | 0 | 0.00000 |

Mechanical Reliability Rates

| Mileage at test | N tested | Pct failed |
|---|---|---|
| 0 | 104 | 4.81 |
| 10000 | 739 | 5.01 |
| 20000 | 1331 | 6.54 |
| 30000 | 1373 | 9.54 |
| 40000 | 1406 | 13.40 |
| 50000 | 1204 | 14.10 |
| 60000 | 1117 | 18.00 |
| 70000 | 864 | 18.50 |
| 80000 | 751 | 22.40 |
| 90000 | 453 | 22.10 |
| 100000 | 288 | 22.60 |
| 110000 | 176 | 29.00 |
| 120000 | 105 | 30.50 |

**Audi A1 2011**

At 5 years of age, the mortality rate of a Audi A1 2011 (manufactured as a Car or Light Van) ranked number 62 out of 211 vehicles of the same age and type (any Car or Light Van constructed in 2011). One is the lowest (or best) and 211 the highest mortality rate. For vehicles reaching 20000 miles, its unreliability score (rate of failing an inspection) ranked 180 out of 205 vehicles of the same age, type, and mileage. One is the highest (or worst) and 205 the lowest rate of failing an inspection.

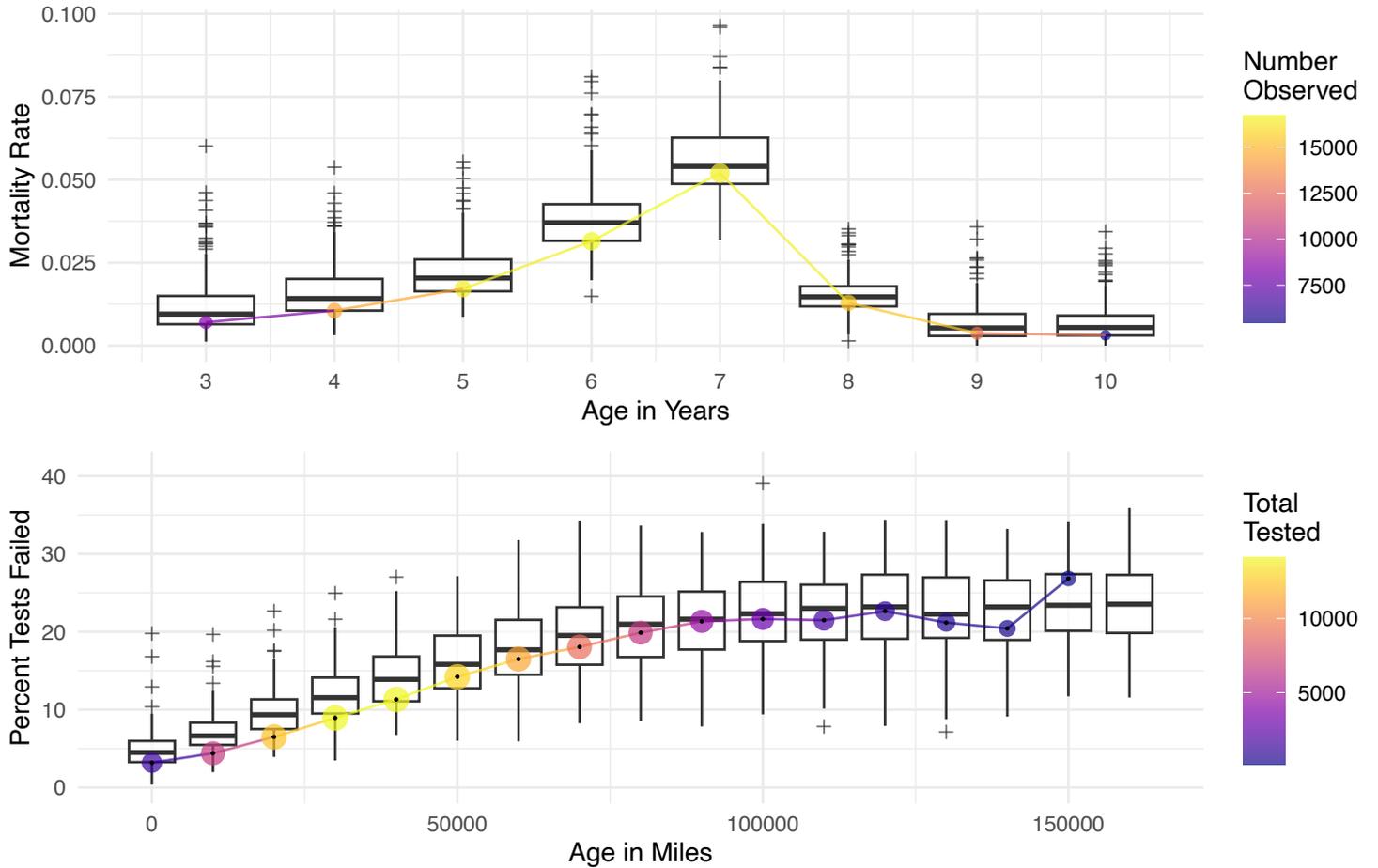

Mortality rates

| Age in Years | Observed | Died | Mortality Rate |
|---|---|---|---|
| 3 | 8379 | 59 | 0.00704 |
| 4 | 14513 | 153 | 0.01050 |
| 5 | 16427 | 281 | 0.01710 |
| 6 | 16725 | 526 | 0.03140 |
| 7 | 16417 | 854 | 0.05200 |
| 8 | 15346 | 198 | 0.01290 |
| 9 | 12842 | 48 | 0.00374 |
| 10 | 5462 | 17 | 0.00311 |

Mechanical Reliability Rates

| Mileage at test | N tested | Pct failed |
|---|---|---|
| 0 | 1254 | 3.19 |
| 10000 | 6604 | 4.41 |
| 20000 | 12367 | 6.50 |
| 30000 | 14133 | 8.95 |
| 40000 | 14001 | 11.30 |
| 50000 | 12985 | 14.20 |
| 60000 | 11330 | 16.50 |
| 70000 | 8994 | 18.10 |
| 80000 | 6547 | 19.90 |
| 90000 | 4452 | 21.30 |
| 100000 | 2781 | 21.60 |
| 110000 | 1666 | 21.50 |
| 120000 | 945 | 22.60 |
| 130000 | 524 | 21.20 |
| 140000 | 279 | 20.40 |
| 150000 | 149 | 26.80 |

## Audi A1 2012

At 5 years of age, the mortality rate of a Audi A1 2012 (manufactured as a Car or Light Van) ranked number 72 out of 212 vehicles of the same age and type (any Car or Light Van constructed in 2012). One is the lowest (or best) and 212 the highest mortality rate. For vehicles reaching 20000 miles, its unreliability score (rate of failing an inspection) ranked 142 out of 206 vehicles of the same age, type, and mileage. One is the highest (or worst) and 206 the lowest rate of failing an inspection.

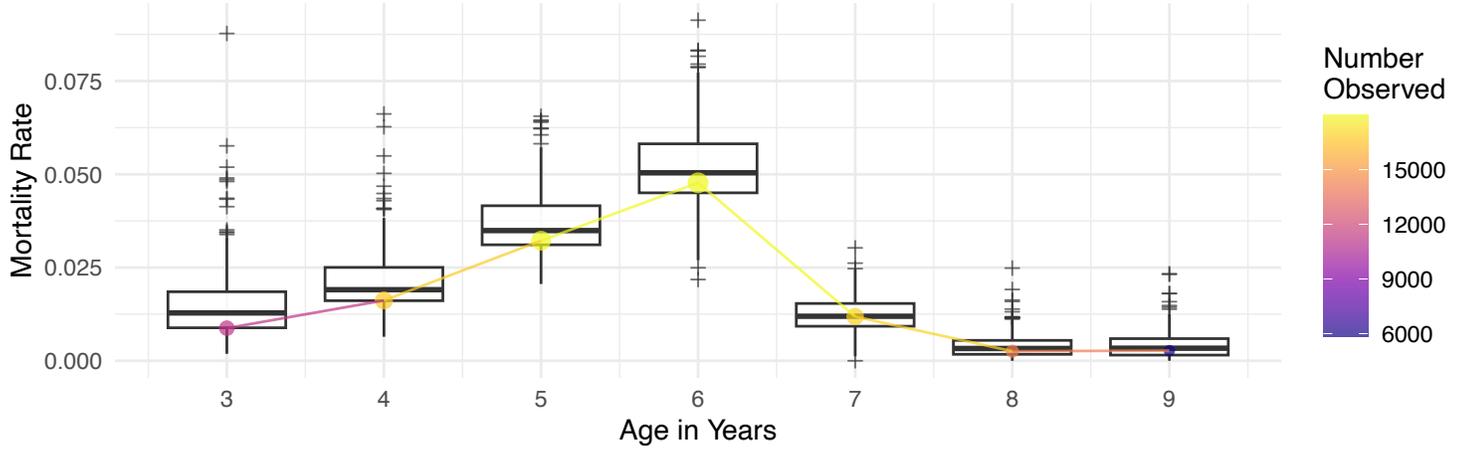

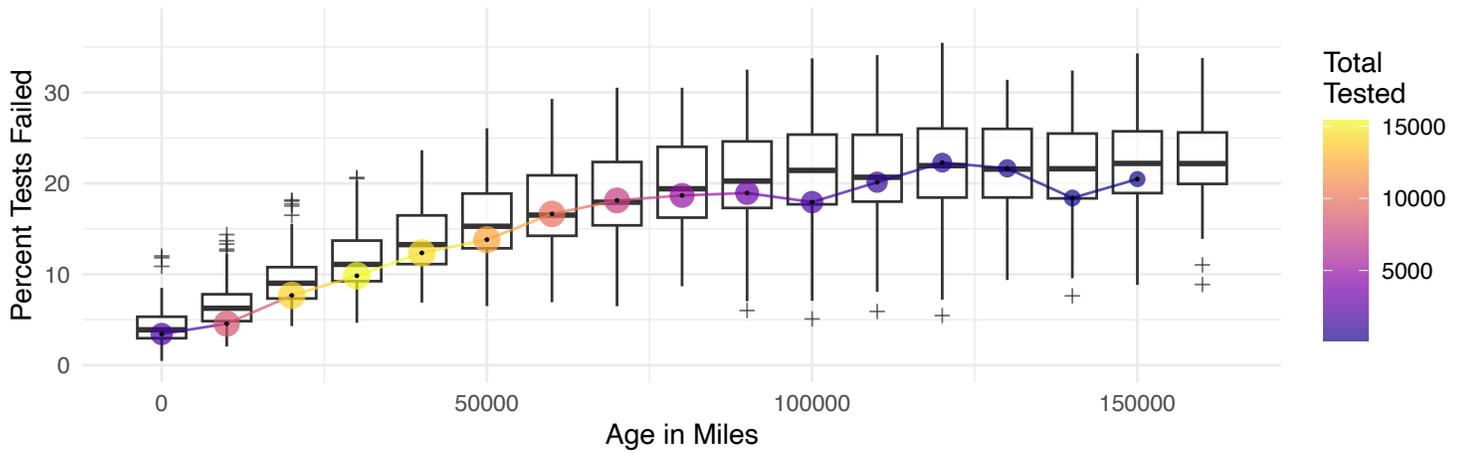

### Mortality rates

| Age in Years | Observed | Died | Mortality Rate |
|---|---|---|---|
| 3 | 11206 | 98 | 0.00875 |
| 4 | 16629 | 269 | 0.01620 |
| 5 | 17926 | 577 | 0.03220 |
| 6 | 17961 | 857 | 0.04770 |
| 7 | 16879 | 200 | 0.01180 |
| 8 | 14081 | 36 | 0.00256 |
| 9 | 5853 | 16 | 0.00273 |

### Mechanical Reliability Rates

| Mileage at test | N tested | Pct failed |
|---|---|---|
| 0 | 1717 | 3.44 |
| 10000 | 8376 | 4.57 |
| 20000 | 14183 | 7.68 |
| 30000 | 15429 | 9.84 |
| 40000 | 14481 | 12.40 |
| 50000 | 12296 | 13.80 |
| 60000 | 9777 | 16.60 |
| 70000 | 7178 | 18.10 |
| 80000 | 4959 | 18.70 |
| 90000 | 3094 | 18.90 |
| 100000 | 1739 | 17.90 |
| 110000 | 1068 | 20.10 |
| 120000 | 588 | 22.30 |
| 130000 | 314 | 21.70 |
| 140000 | 174 | 18.40 |
| 150000 | 127 | 20.50 |

**Audi A1 2013**

At 5 years of age, the mortality rate of a Audi A1 2013 (manufactured as a Car or Light Van) ranked number 117 out of 221 vehicles of the same age and type (any Car or Light Van constructed in 2013). One is the lowest (or best) and 221 the highest mortality rate. For vehicles reaching 20000 miles, its unreliability score (rate of failing an inspection) ranked 181 out of 215 vehicles of the same age, type, and mileage. One is the highest (or worst) and 215 the lowest rate of failing an inspection.

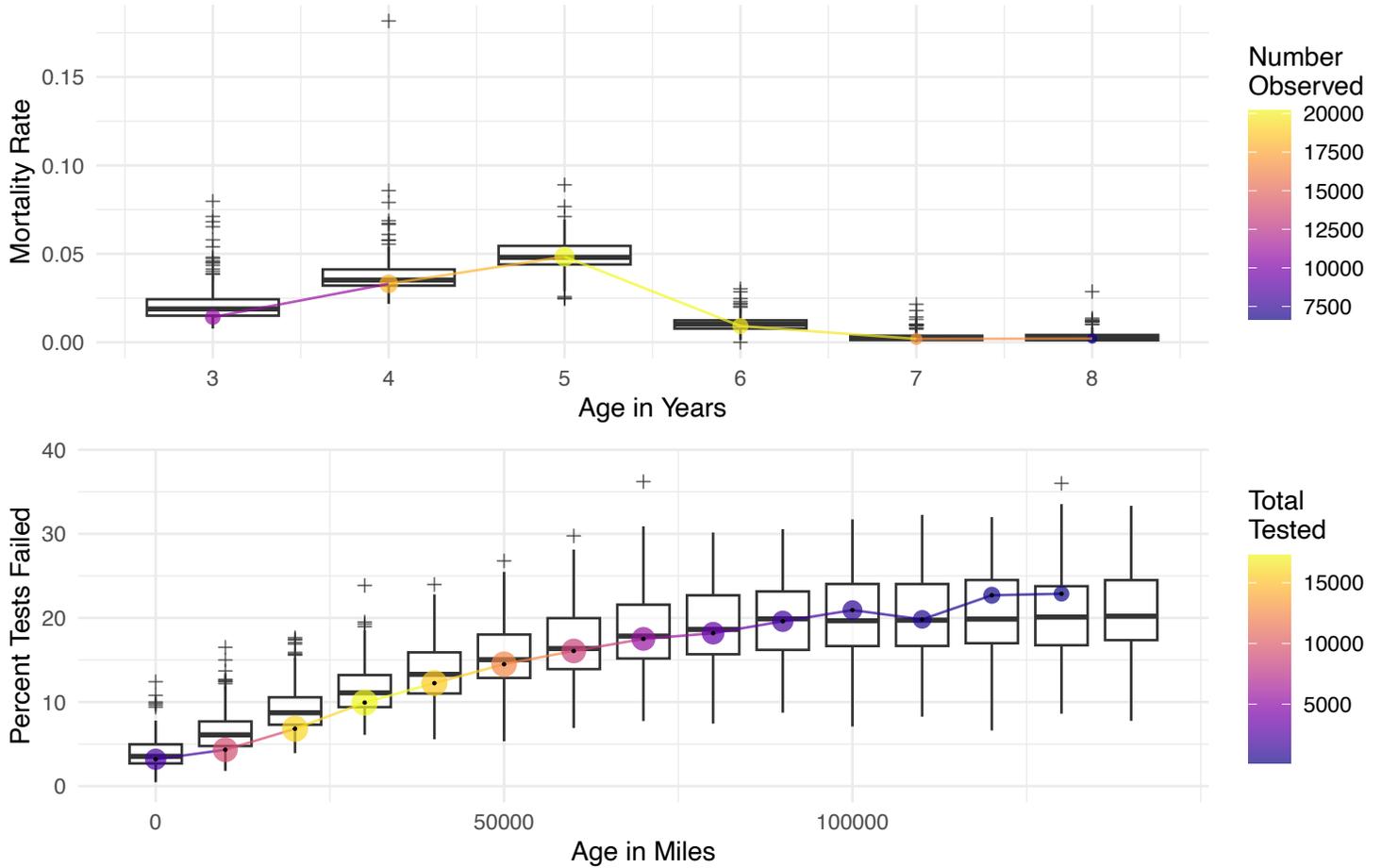

Mortality rates

| Age in Years | Observed | Died | Mortality Rate |
|---|---|---|---|
| 3 | 10938 | 156 | 0.01430 |
| 4 | 18025 | 596 | 0.03310 |
| 5 | 20162 | 976 | 0.04840 |
| 6 | 19658 | 181 | 0.00921 |
| 7 | 16634 | 32 | 0.00192 |
| 8 | 6676 | 14 | 0.00210 |

Mechanical Reliability Rates

| Mileage at test | N tested | Pct failed |
|---|---|---|
| 0 | 1908 | 3.20 |
| 10000 | 9219 | 4.33 |
| 20000 | 15945 | 6.82 |
| 30000 | 17301 | 9.94 |
| 40000 | 15457 | 12.20 |
| 50000 | 12331 | 14.50 |
| 60000 | 8516 | 16.10 |
| 70000 | 5626 | 17.50 |
| 80000 | 3349 | 18.20 |
| 90000 | 1898 | 19.60 |
| 100000 | 946 | 20.90 |
| 110000 | 585 | 19.80 |
| 120000 | 282 | 22.70 |
| 130000 | 153 | 22.90 |

**Audi A1 2014**

At 5 years of age, the mortality rate of a Audi A1 2014 (manufactured as a Car or Light Van) ranked number 154 out of 236 vehicles of the same age and type (any Car or Light Van constructed in 2014). One is the lowest (or best) and 236 the highest mortality rate. For vehicles reaching 20000 miles, its unreliability score (rate of failing an inspection) ranked 185 out of 230 vehicles of the same age, type, and mileage. One is the highest (or worst) and 230 the lowest rate of failing an inspection.

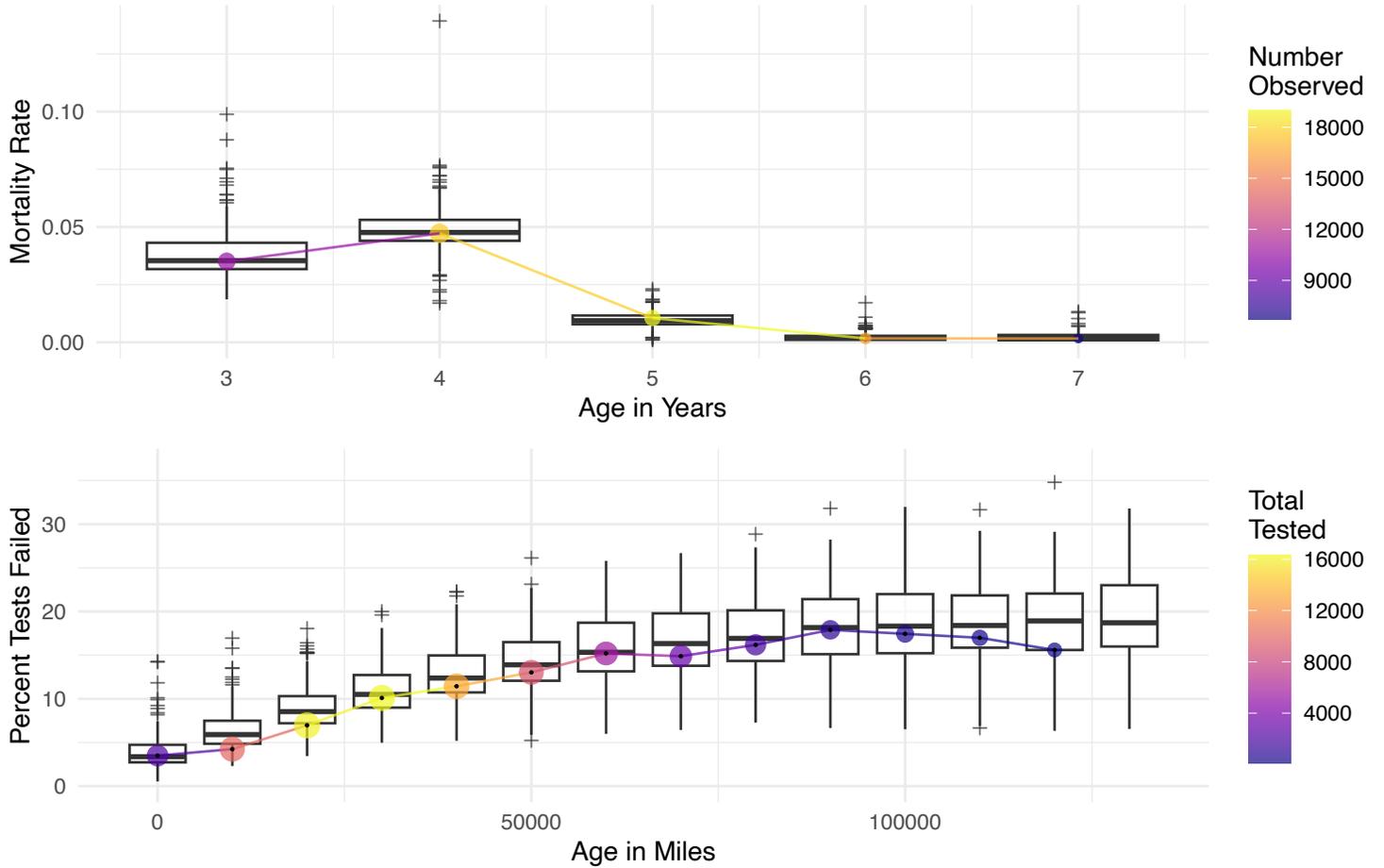

Mortality rates

| Age in Years | Observed | Died | Mortality Rate |
|---|---|---|---|
| 3 | 10578 | 371 | 0.03510 |
| 4 | 17697 | 836 | 0.04720 |
| 5 | 18951 | 199 | 0.01050 |
| 6 | 16488 | 28 | 0.00170 |
| 7 | 6724 | 11 | 0.00164 |

Mechanical Reliability Rates

| Mileage at test | N tested | Pct failed |
|---|---|---|
| 0 | 2173 | 3.50 |
| 10000 | 9979 | 4.24 |
| 20000 | 16068 | 6.97 |
| 30000 | 16342 | 10.10 |
| 40000 | 13215 | 11.40 |
| 50000 | 8978 | 13.00 |
| 60000 | 5639 | 15.20 |
| 70000 | 3151 | 14.90 |
| 80000 | 1755 | 16.20 |
| 90000 | 838 | 17.90 |
| 100000 | 407 | 17.40 |
| 110000 | 206 | 17.00 |
| 120000 | 109 | 15.60 |

**Audi A1 2015**

At 5 years of age, the mortality rate of a Audi A1 2015 (manufactured as a Car or Light Van) ranked number 95 out of 247 vehicles of the same age and type (any Car or Light Van constructed in 2015). One is the lowest (or best) and 247 the highest mortality rate. For vehicles reaching 20000 miles, its unreliability score (rate of failing an inspection) ranked 192 out of 241 vehicles of the same age, type, and mileage. One is the highest (or worst) and 241 the lowest rate of failing an inspection.

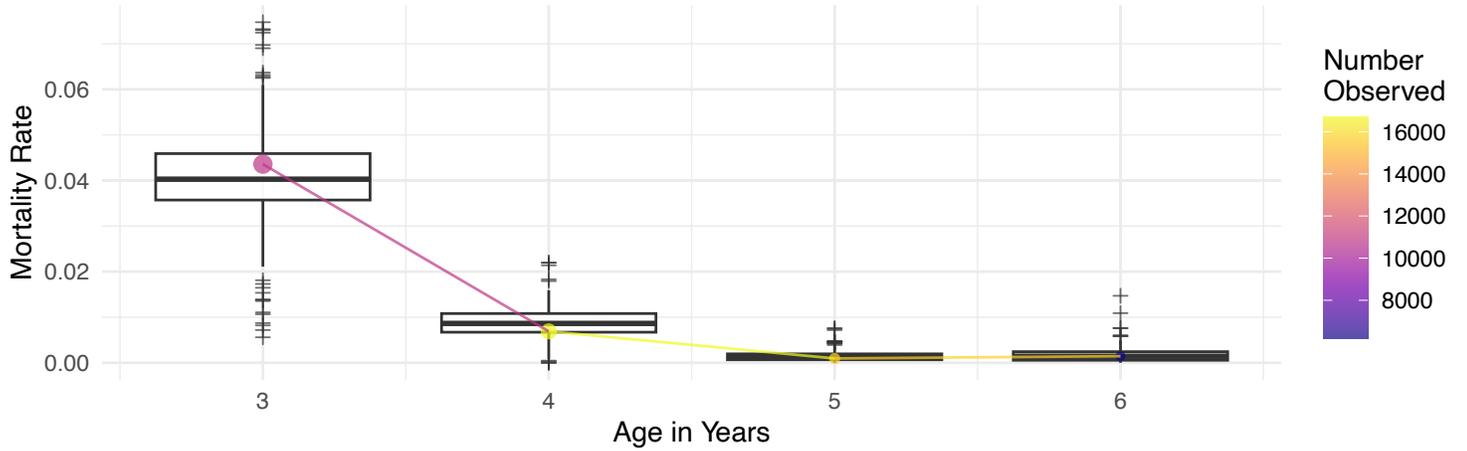

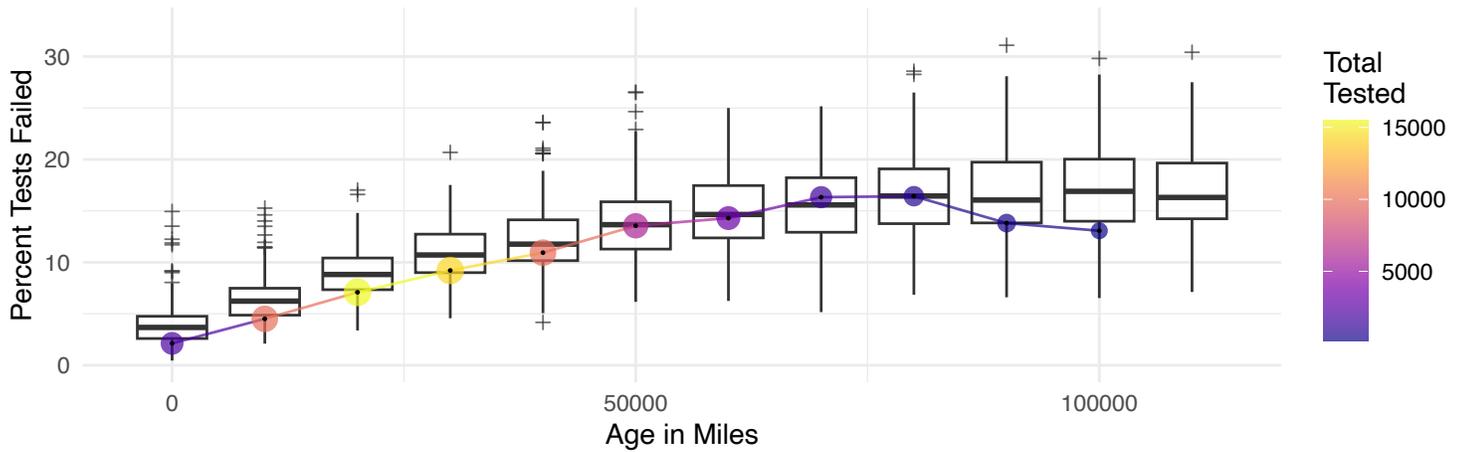

Mortality rates

| Age in Years | Observed | Died | Mortality Rate |
|---|---|---|---|
| 3 | 10829 | 472 | 0.043600 |
| 4 | 16668 | 115 | 0.006900 |
| 5 | 15689 | 15 | 0.000956 |
| 6 | 6196 | 9 | 0.001450 |

Mechanical Reliability Rates

| Mileage at test | N tested | Pct failed |
|---|---|---|
| 0 | 2069 | 2.13 |
| 10000 | 9912 | 4.51 |
| 20000 | 15504 | 7.09 |
| 30000 | 14292 | 9.21 |
| 40000 | 9874 | 10.90 |
| 50000 | 6043 | 13.60 |
| 60000 | 3252 | 14.30 |
| 70000 | 1586 | 16.30 |
| 80000 | 761 | 16.40 |
| 90000 | 362 | 13.80 |
| 100000 | 176 | 13.10 |

**Audi A1 2016**

At 5 years of age, the mortality rate of a Audi A1 2016 (manufactured as a Car or Light Van) ranked number 128 out of 252 vehicles of the same age and type (any Car or Light Van constructed in 2016). One is the lowest (or best) and 252 the highest mortality rate. For vehicles reaching 40000 miles, its unreliability score (rate of failing an inspection) ranked 128 out of 243 vehicles of the same age, type, and mileage. One is the highest (or worst) and 243 the lowest rate of failing an inspection.

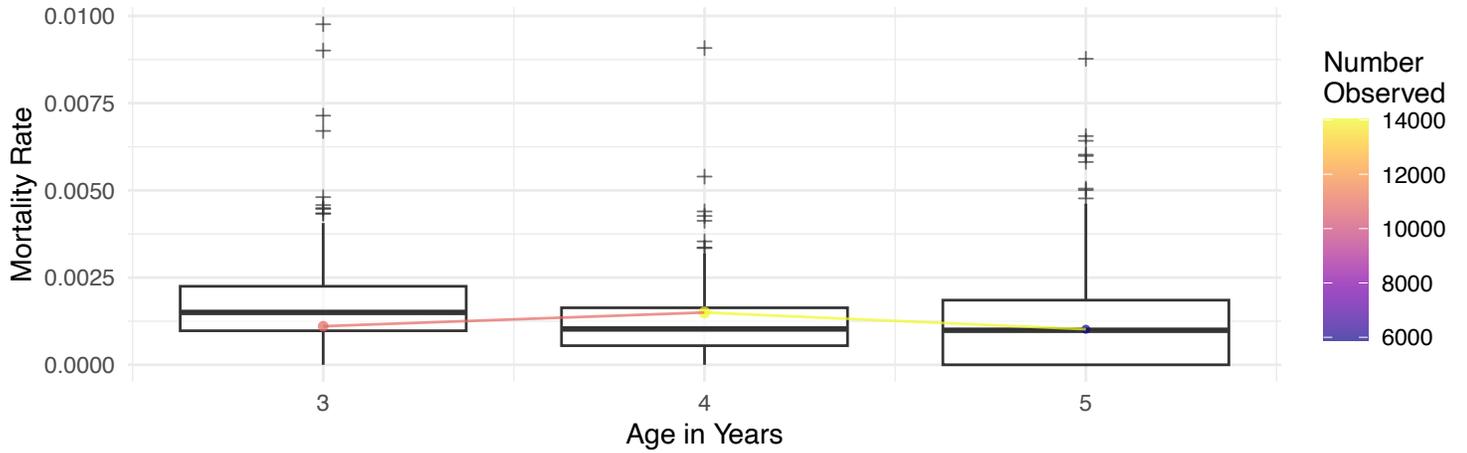

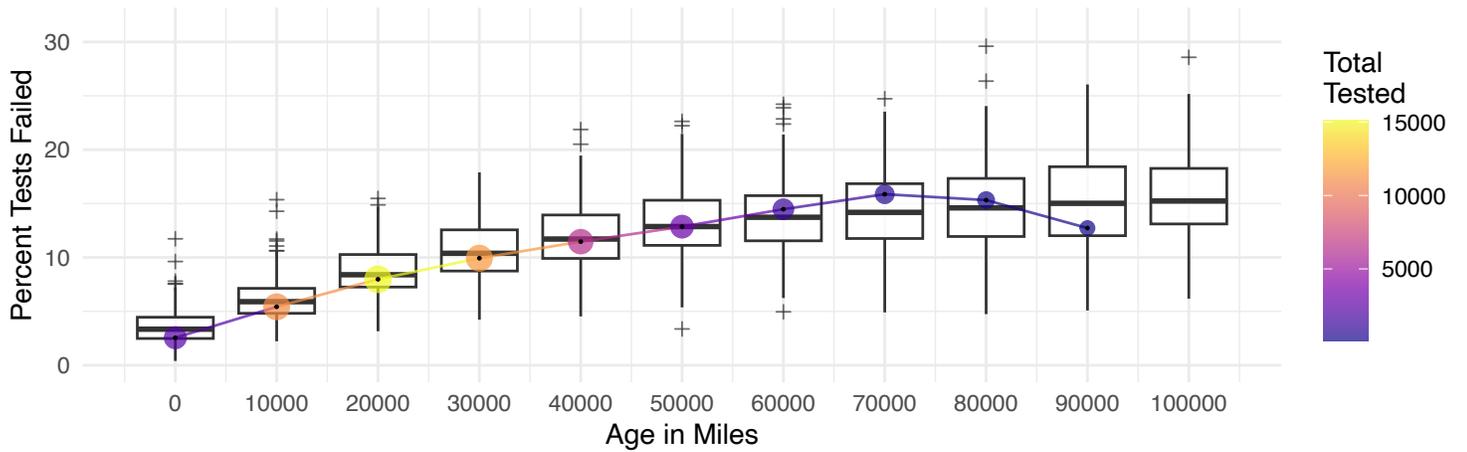

Mortality rates

| Age in Years | Observed | Died | Mortality Rate |
|---|---|---|---|
| 3 | 10843 | 12 | 0.00111 |
| 4 | 14010 | 21 | 0.00150 |
| 5 | 5889 | 6 | 0.00102 |

Mechanical Reliability Rates

| Mileage at test | N tested | Pct failed |
|---|---|---|
| 0 | 2235 | 2.55 |
| 10000 | 10992 | 5.42 |
| 20000 | 15130 | 7.98 |
| 30000 | 11485 | 9.93 |
| 40000 | 6323 | 11.50 |
| 50000 | 3041 | 12.90 |
| 60000 | 1271 | 14.50 |
| 70000 | 586 | 15.90 |
| 80000 | 248 | 15.30 |
| 90000 | 110 | 12.70 |

**Audi A1 2017**

At 3 years of age, the mortality rate of a Audi A1 2017 (manufactured as a Car or Light Van) ranked number 174 out of 247 vehicles of the same age and type (any Car or Light Van constructed in 2017). One is the lowest (or best) and 247 the highest mortality rate. For vehicles reaching 20000 miles, its unreliability score (rate of failing an inspection) ranked 117 out of 240 vehicles of the same age, type, and mileage. One is the highest (or worst) and 240 the lowest rate of failing an inspection.

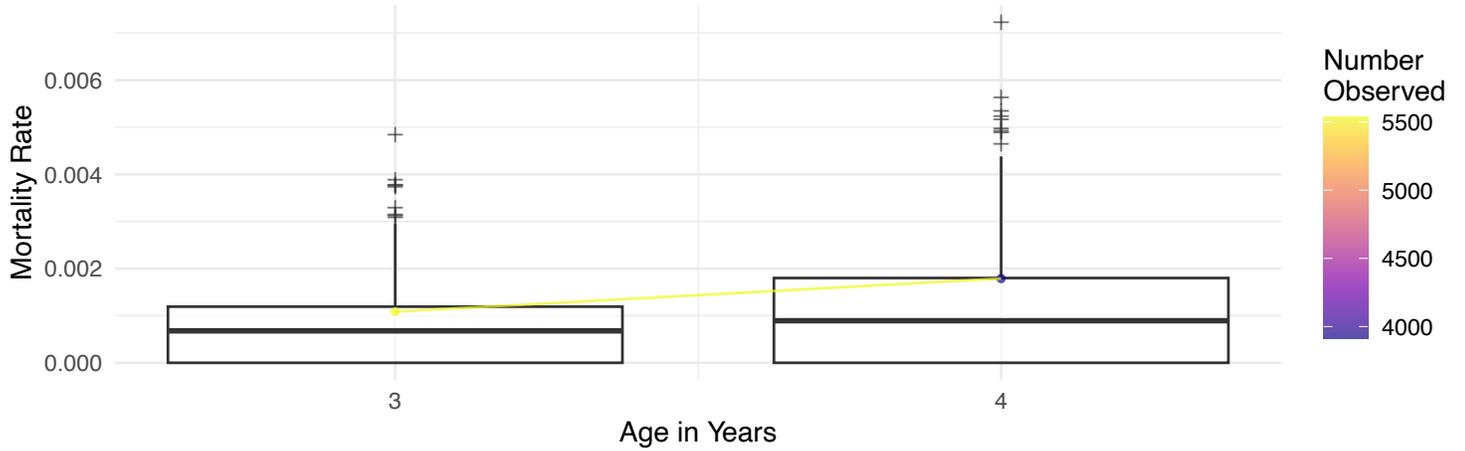

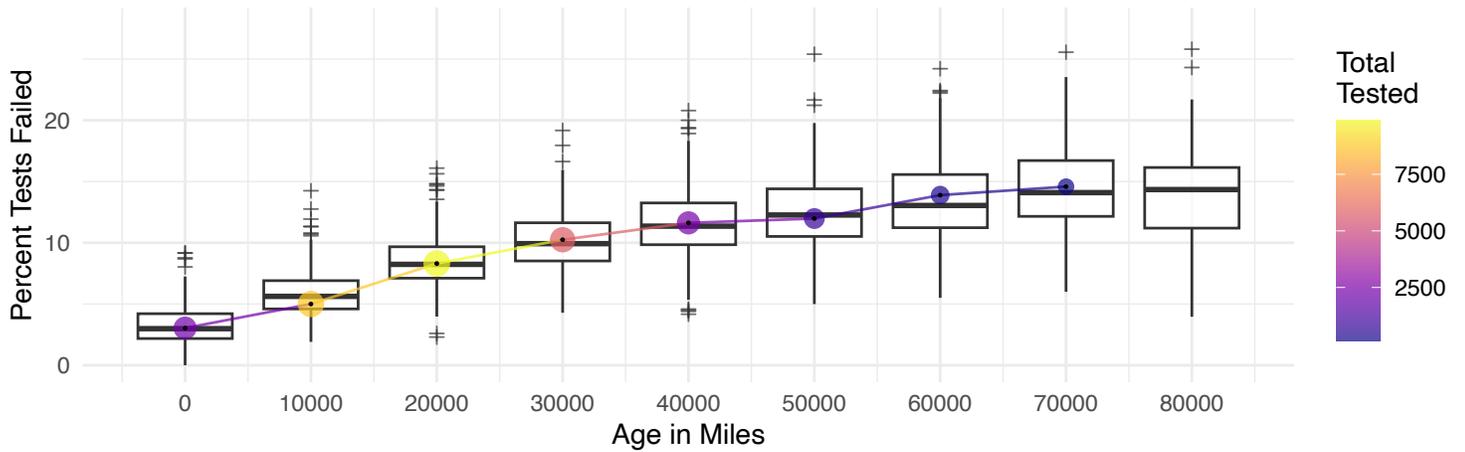

Mortality rates

| Age in Years | Observed | Died | Mortality Rate |
|---|---|---|---|
| 3 | 5532 | 6 | 0.00108 |
| 4 | 3914 | 7 | 0.00179 |

Mechanical Reliability Rates

| Mileage at test | N tested | Pct failed |
|---|---|---|
| 0 | 2240 | 3.04 |
| 10000 | 8780 | 5.00 |
| 20000 | 9896 | 8.31 |
| 30000 | 5826 | 10.20 |
| 40000 | 2374 | 11.60 |
| 50000 | 885 | 12.00 |
| 60000 | 360 | 13.90 |
| 70000 | 144 | 14.60 |

**Audi A1 2018**

At 3 years of age, the mortality rate of a Audi A1 2018 (manufactured as a Car or Light Van) ranked number 182 out of 222 vehicles of the same age and type (any Car or Light Van constructed in 2018). One is the lowest (or best) and 222 the highest mortality rate. For vehicles reaching 20000 miles, its unreliability score (rate of failing an inspection) ranked 84 out of 215 vehicles of the same age, type, and mileage. One is the highest (or worst) and 215 the lowest rate of failing an inspection.

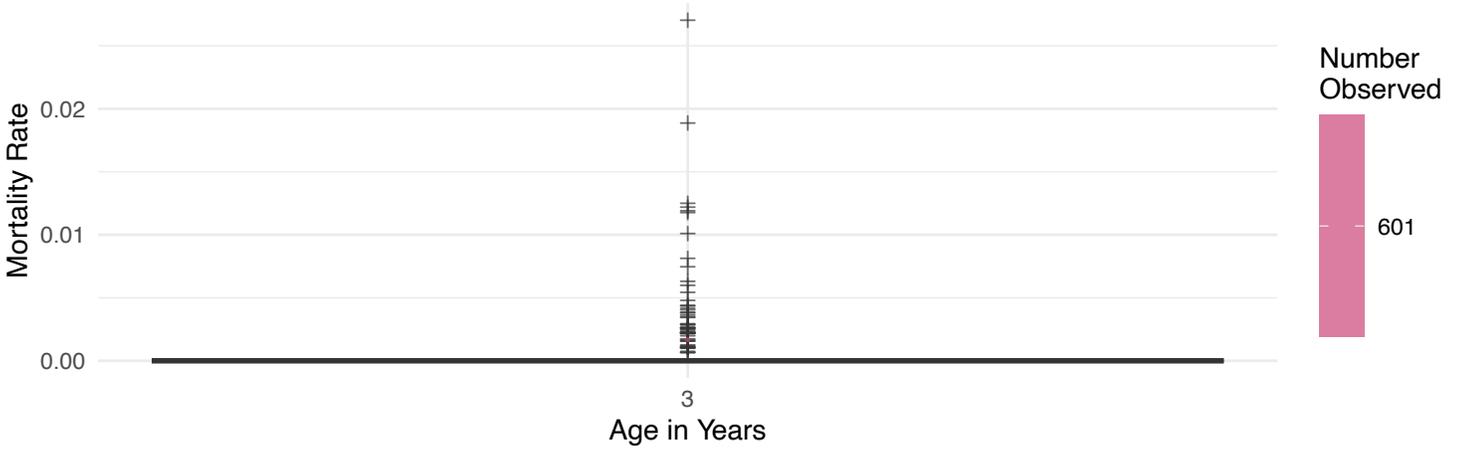

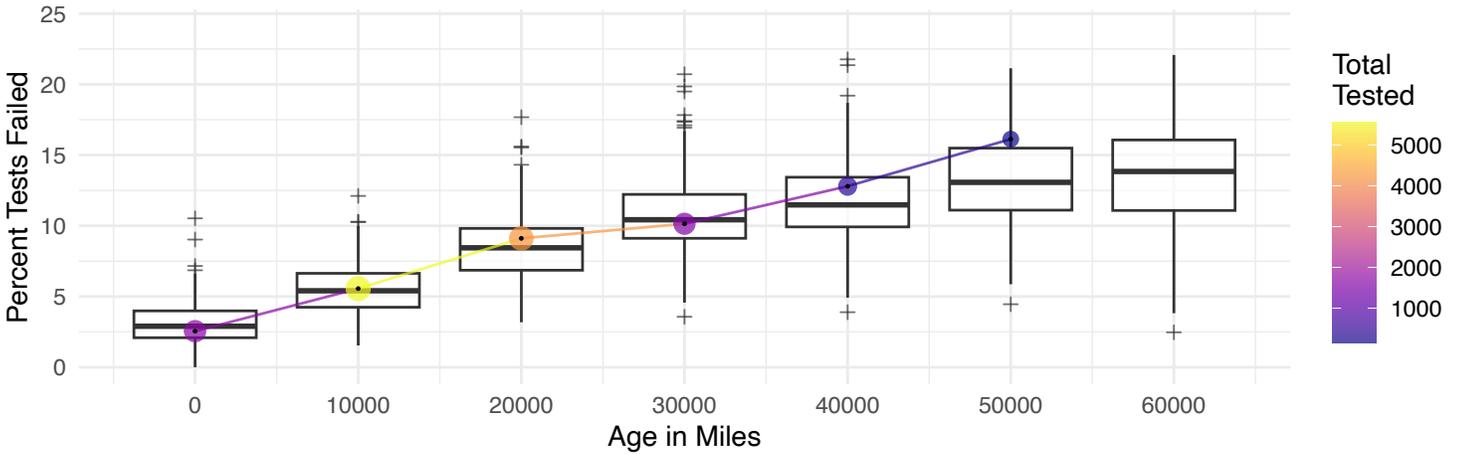

Mortality rates

| Age in Years | Observed | Died | Mortality Rate |
|---|---|---|---|
| 3 | 601 | 1 | 0.00166 |

Mechanical Reliability Rates

| Mileage at test | N tested | Pct failed |
|---|---|---|
| 0 | 1687 | 2.55 |
| 10000 | 5557 | 5.56 |
| 20000 | 4215 | 9.11 |
| 30000 | 1558 | 10.10 |
| 40000 | 422 | 12.80 |
| 50000 | 155 | 16.10 |

## Audi A2 2001

At 5 years of age, the mortality rate of a Audi A2 2001 (manufactured as a Car or Light Van) ranked number 26 out of 205 vehicles of the same age and type (any Car or Light Van constructed in 2001). One is the lowest (or best) and 205 the highest mortality rate. For vehicles reaching 120000 miles, its unreliability score (rate of failing an inspection) ranked 133 out of 194 vehicles of the same age, type, and mileage. One is the highest (or worst) and 194 the lowest rate of failing an inspection.

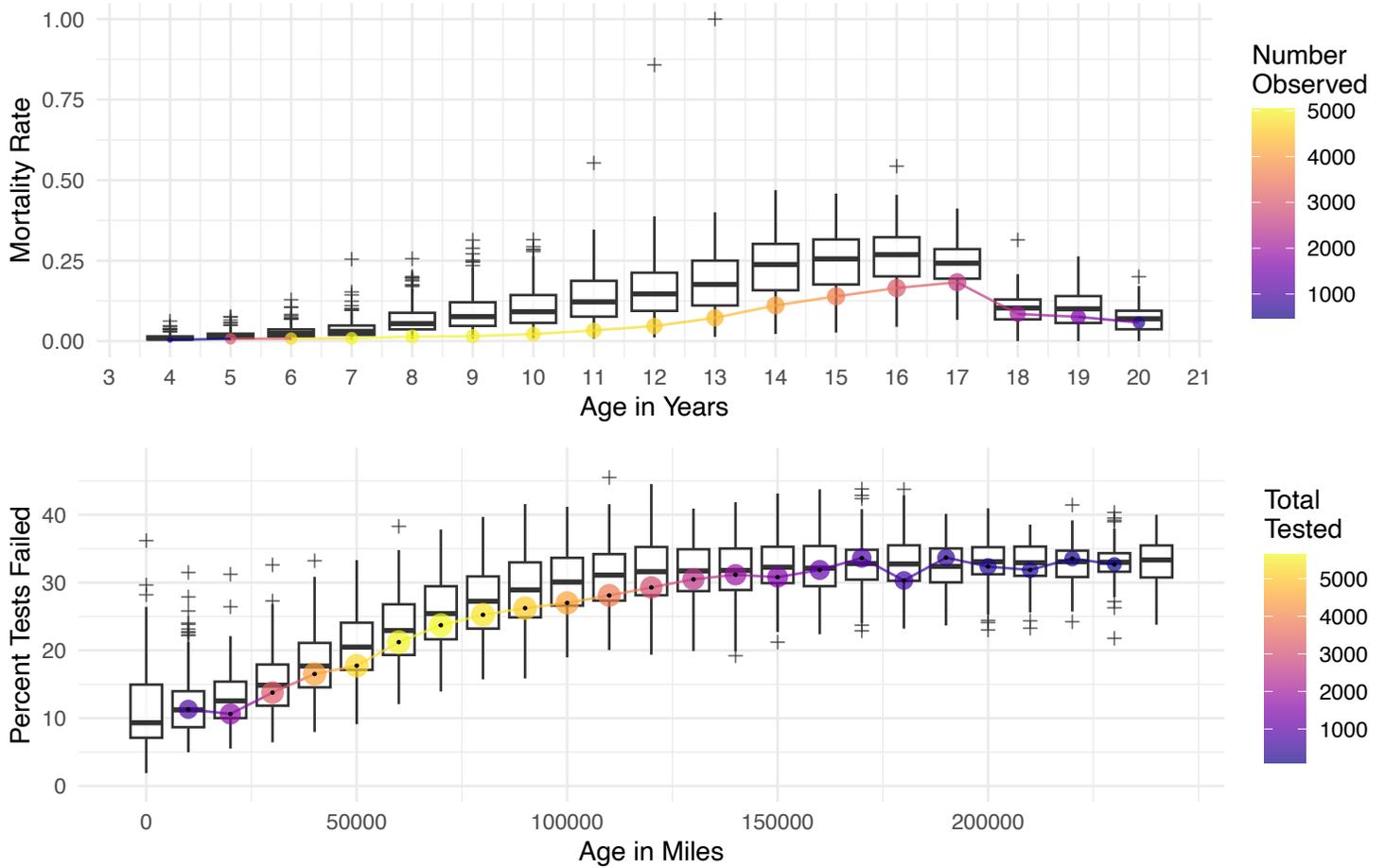

| Mortality rates |||||
| Age in Years | Observed | Died | Mortality Rate |
| --- | --- | --- | --- |
| 4 | 477 | 2 | 0.00419 |
| 5 | 3051 | 22 | 0.00721 |
| 6 | 4758 | 37 | 0.00778 |
| 7 | 5029 | 43 | 0.00855 |
| 8 | 5001 | 71 | 0.01420 |
| 9 | 4928 | 76 | 0.01540 |
| 10 | 4850 | 106 | 0.02190 |
| 11 | 4737 | 158 | 0.03340 |
| 12 | 4567 | 214 | 0.04690 |
| 13 | 4346 | 315 | 0.07250 |
| 14 | 4020 | 446 | 0.11100 |
| 15 | 3555 | 495 | 0.13900 |
| 16 | 3054 | 504 | 0.16500 |
| 17 | 2543 | 465 | 0.18300 |
| 18 | 2002 | 169 | 0.08440 |
| 19 | 1499 | 113 | 0.07540 |
| 20 | 590 | 34 | 0.05760 |

| Mechanical Reliability Rates |||
| Mileage at test | N tested | Pct failed |
| --- | --- | --- |
| 10000 | 734 | 11.3 |
| 20000 | 1723 | 10.6 |
| 30000 | 3282 | 13.8 |
| 40000 | 4462 | 16.5 |
| 50000 | 5209 | 17.7 |
| 60000 | 5573 | 21.2 |
| 70000 | 5652 | 23.7 |
| 80000 | 5396 | 25.2 |
| 90000 | 5029 | 26.2 |
| 100000 | 4467 | 27.0 |
| 110000 | 3804 | 28.1 |
| 120000 | 3026 | 29.3 |
| 130000 | 2503 | 30.5 |
| 140000 | 1910 | 31.2 |
| 150000 | 1484 | 30.8 |
| 160000 | 1086 | 31.9 |
| 170000 | 812 | 33.6 |

**Audi A2 2002**

At 5 years of age, the mortality rate of a Audi A2 2002 (manufactured as a Car or Light Van) ranked number 8 out of 202 vehicles of the same age and type (any Car or Light Van constructed in 2002). One is the lowest (or best) and 202 the highest mortality rate. For vehicles reaching 120000 miles, its unreliability score (rate of failing an inspection) ranked 111 out of 193 vehicles of the same age, type, and mileage. One is the highest (or worst) and 193 the lowest rate of failing an inspection.

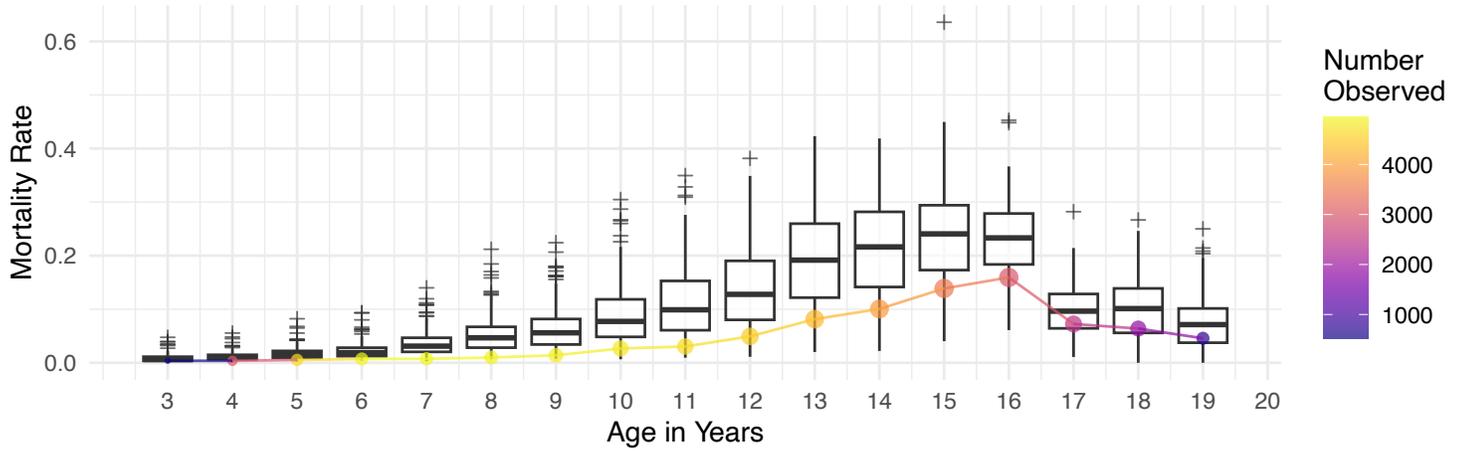

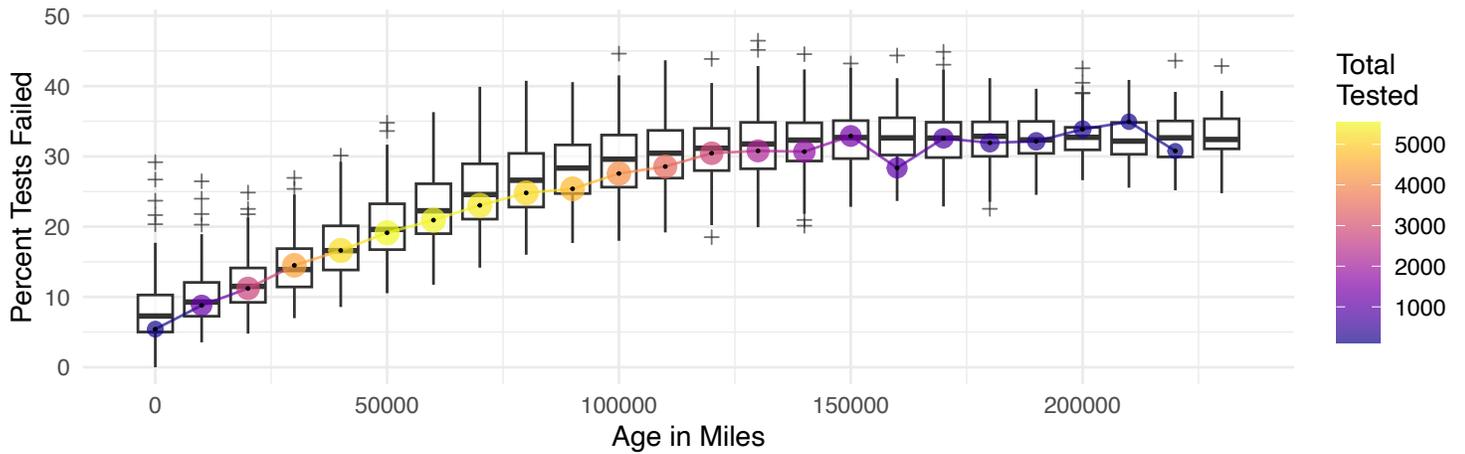

| Mortality rates | | | |
|---|---|---|---|
| Age in Years | Observed | Died | Mortality Rate |
| 3 | 524 | 2 | 0.00382 |
| 4 | 3010 | 11 | 0.00365 |
| 5 | 4652 | 25 | 0.00537 |
| 6 | 4937 | 35 | 0.00709 |
| 7 | 4918 | 37 | 0.00752 |
| 8 | 4877 | 48 | 0.00984 |
| 9 | 4827 | 68 | 0.01410 |
| 10 | 4759 | 127 | 0.02670 |
| 11 | 4623 | 141 | 0.03050 |
| 12 | 4479 | 221 | 0.04930 |
| 13 | 4247 | 346 | 0.08150 |
| 14 | 3891 | 391 | 0.10000 |
| 15 | 3488 | 484 | 0.13900 |
| 16 | 2993 | 477 | 0.15900 |
| 17 | 2442 | 177 | 0.07250 |
| 18 | 1846 | 118 | 0.06390 |
| 19 | 743 | 34 | 0.04580 |

| Mechanical Reliability Rates | | |
|---|---|---|
| Mileage at test | N tested | Pct failed |
| 0 | 166 | 5.42 |
| 10000 | 1113 | 8.81 |
| 20000 | 2782 | 11.20 |
| 30000 | 4394 | 14.50 |
| 40000 | 5212 | 16.60 |
| 50000 | 5546 | 19.10 |
| 60000 | 5508 | 20.90 |
| 70000 | 5344 | 23.00 |
| 80000 | 5148 | 24.80 |
| 90000 | 4707 | 25.40 |
| 100000 | 4082 | 27.60 |
| 110000 | 3470 | 28.60 |
| 120000 | 2831 | 30.40 |
| 130000 | 2209 | 30.80 |
| 140000 | 1747 | 30.70 |
| 170000 | 789 | 32.60 |
| 210000 | 146 | 34.90 |

## Audi A2 2003

At 5 years of age, the mortality rate of a Audi A2 2003 (manufactured as a Car or Light Van) ranked number 12 out of 213 vehicles of the same age and type (any Car or Light Van constructed in 2003). One is the lowest (or best) and 213 the highest mortality rate. For vehicles reaching 100000 miles, its unreliability score (rate of failing an inspection) ranked 141 out of 208 vehicles of the same age, type, and mileage. One is the highest (or worst) and 208 the lowest rate of failing an inspection.

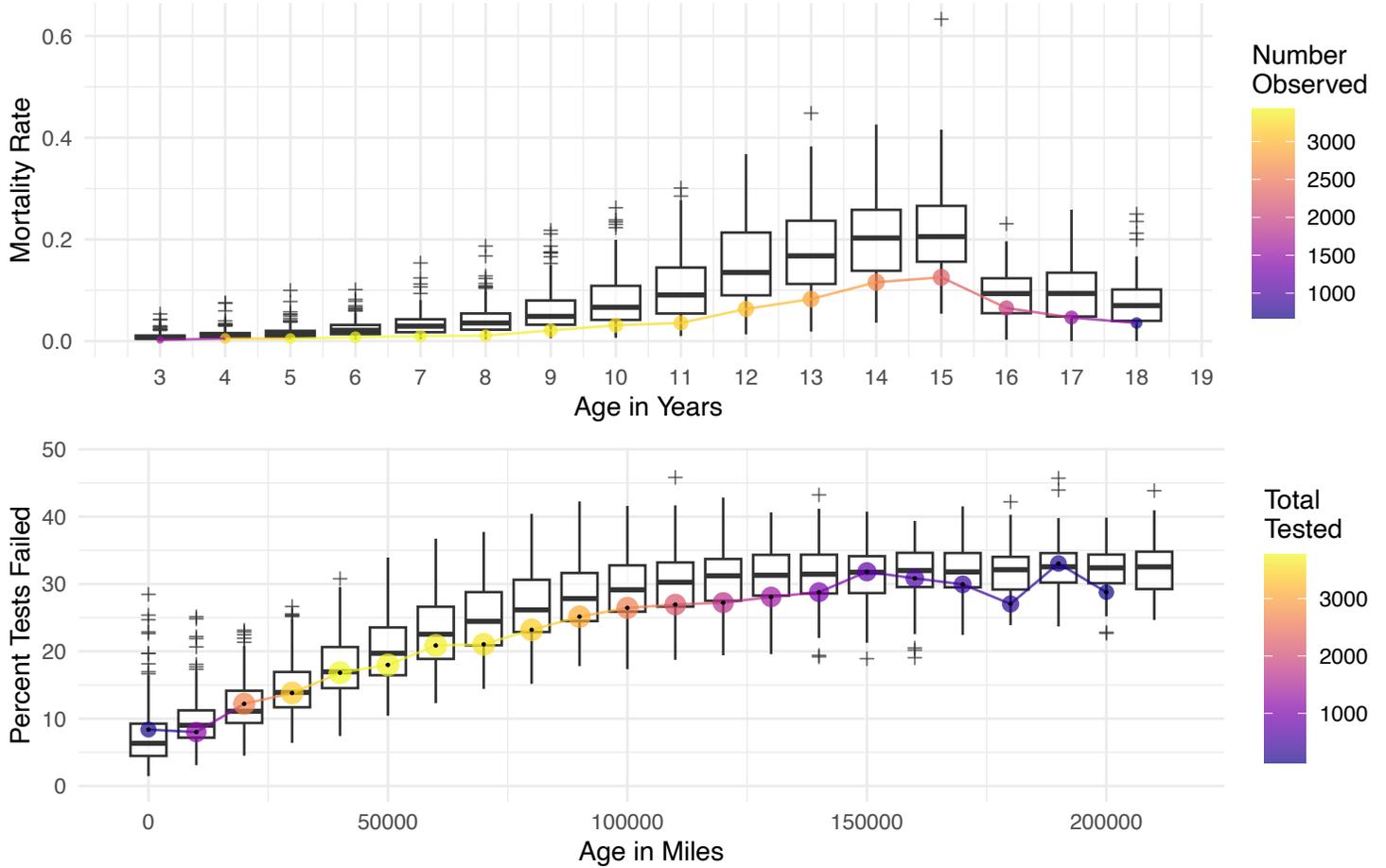

### Mortality rates

| Age in Years | Observed | Died | Mortality Rate |
|---|---|---|---|
| 3 | 1617 | 4 | 0.00247 |
| 4 | 3151 | 16 | 0.00508 |
| 5 | 3423 | 17 | 0.00497 |
| 6 | 3417 | 26 | 0.00761 |
| 7 | 3387 | 34 | 0.01000 |
| 8 | 3350 | 36 | 0.01070 |
| 9 | 3312 | 70 | 0.02110 |
| 10 | 3238 | 101 | 0.03120 |
| 11 | 3135 | 112 | 0.03570 |
| 12 | 3017 | 191 | 0.06330 |
| 13 | 2817 | 233 | 0.08270 |
| 14 | 2579 | 299 | 0.11600 |
| 15 | 2279 | 286 | 0.12500 |
| 16 | 1950 | 127 | 0.06510 |
| 17 | 1508 | 70 | 0.04640 |
| 18 | 673 | 24 | 0.03570 |

### Mechanical Reliability Rates

| Mileage at test | N tested | Pct failed |
|---|---|---|
| 0 | 179 | 8.38 |
| 10000 | 1200 | 8.00 |
| 20000 | 2679 | 12.20 |
| 30000 | 3395 | 13.80 |
| 40000 | 3774 | 16.80 |
| 50000 | 3721 | 18.00 |
| 60000 | 3695 | 20.90 |
| 70000 | 3603 | 21.00 |
| 80000 | 3420 | 23.20 |
| 90000 | 3086 | 25.10 |
| 100000 | 2654 | 26.50 |
| 110000 | 2149 | 26.90 |
| 120000 | 1660 | 27.20 |
| 130000 | 1364 | 28.10 |
| 140000 | 1018 | 28.80 |
| 150000 | 748 | 31.80 |
| 160000 | 561 | 30.80 |

# Audi A2 2004

At 5 years of age, the mortality rate of a Audi A2 2004 (manufactured as a Car or Light Van) ranked number 7 out of 229 vehicles of the same age and type (any Car or Light Van constructed in 2004). One is the lowest (or best) and 229 the highest mortality rate. For vehicles reaching 20000 miles, its unreliability score (rate of failing an inspection) ranked 129 out of 225 vehicles of the same age, type, and mileage. One is the highest (or worst) and 225 the lowest rate of failing an inspection.

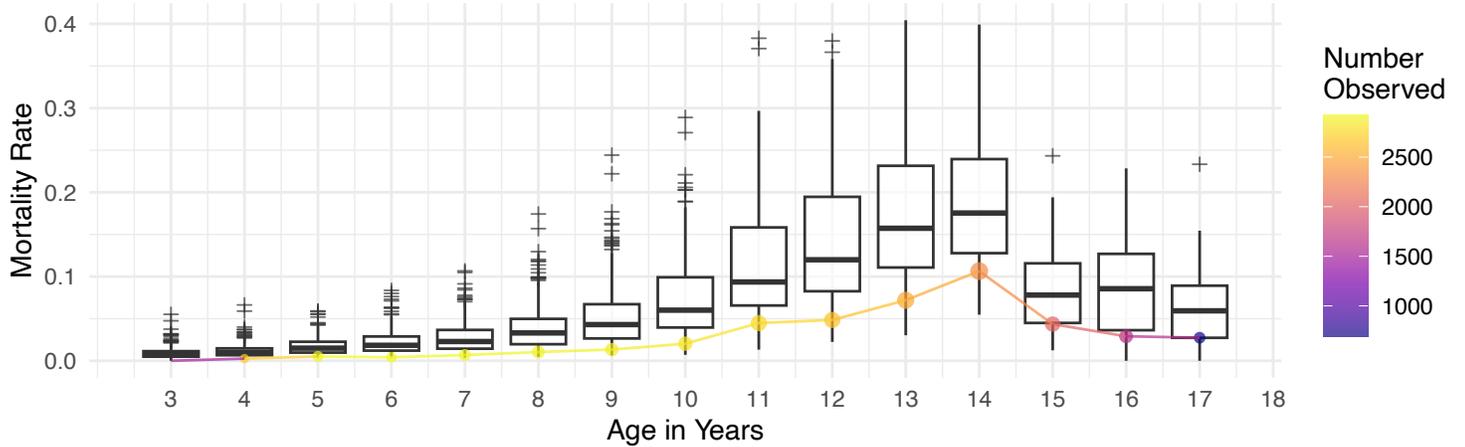

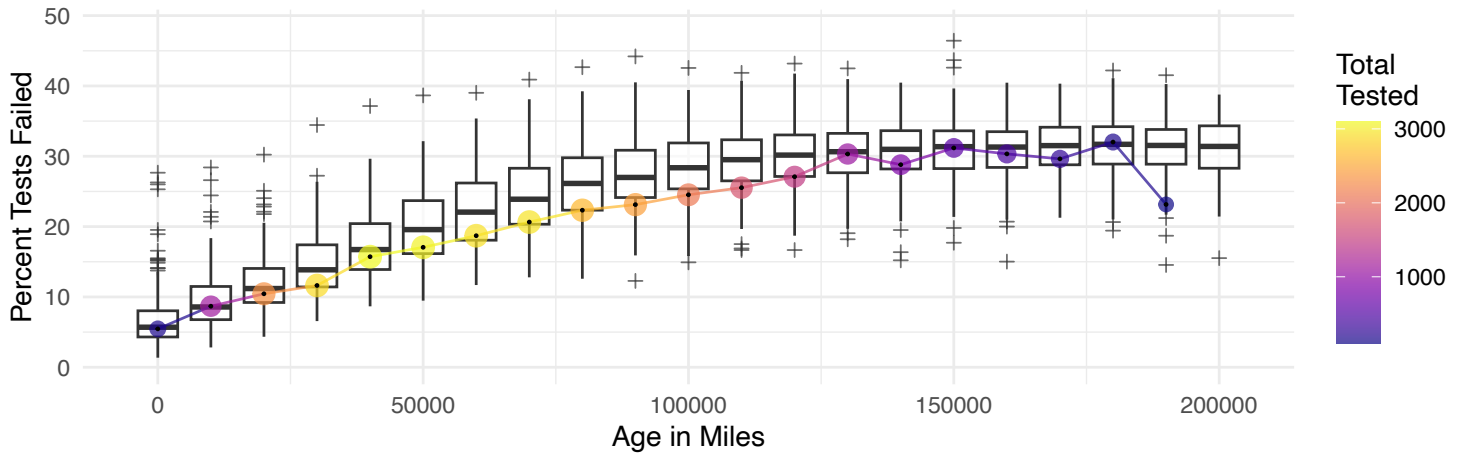

Mortality rates

| Age in Years | Observed | Died | Mortality Rate |
|---|---|---|---|
| 3 | 1501 | 0 | 0.00000 |
| 4 | 2696 | 7 | 0.00260 |
| 5 | 2918 | 15 | 0.00514 |
| 6 | 2914 | 12 | 0.00412 |
| 7 | 2904 | 20 | 0.00689 |
| 8 | 2881 | 30 | 0.01040 |
| 9 | 2849 | 38 | 0.01330 |
| 10 | 2806 | 57 | 0.02030 |
| 11 | 2745 | 123 | 0.04480 |
| 12 | 2620 | 127 | 0.04850 |
| 13 | 2489 | 179 | 0.07190 |
| 14 | 2311 | 246 | 0.10600 |
| 15 | 2020 | 88 | 0.04360 |
| 16 | 1615 | 47 | 0.02910 |
| 17 | 693 | 19 | 0.02740 |

Mechanical Reliability Rates

| Mileage at test | N tested | Pct failed |
|---|---|---|
| 0 | 165 | 5.45 |
| 10000 | 1129 | 8.68 |
| 20000 | 2317 | 10.40 |
| 30000 | 2881 | 11.60 |
| 40000 | 3092 | 15.70 |
| 50000 | 3043 | 17.10 |
| 60000 | 2930 | 18.70 |
| 70000 | 2945 | 20.60 |
| 80000 | 2629 | 22.30 |
| 90000 | 2439 | 23.10 |
| 100000 | 2100 | 24.50 |
| 110000 | 1727 | 25.50 |
| 120000 | 1469 | 27.10 |
| 130000 | 1085 | 30.30 |
| 140000 | 854 | 28.80 |
| 150000 | 574 | 31.20 |
| 160000 | 445 | 30.30 |

## Audi A2 2005

At 5 years of age, the mortality rate of a Audi A2 2005 (manufactured as a Car or Light Van) ranked number 4 out of 240 vehicles of the same age and type (any Car or Light Van constructed in 2005). One is the lowest (or best) and 240 the highest mortality rate. For vehicles reaching 20000 miles, its unreliability score (rate of failing an inspection) ranked 121 out of 235 vehicles of the same age, type, and mileage. One is the highest (or worst) and 235 the lowest rate of failing an inspection.

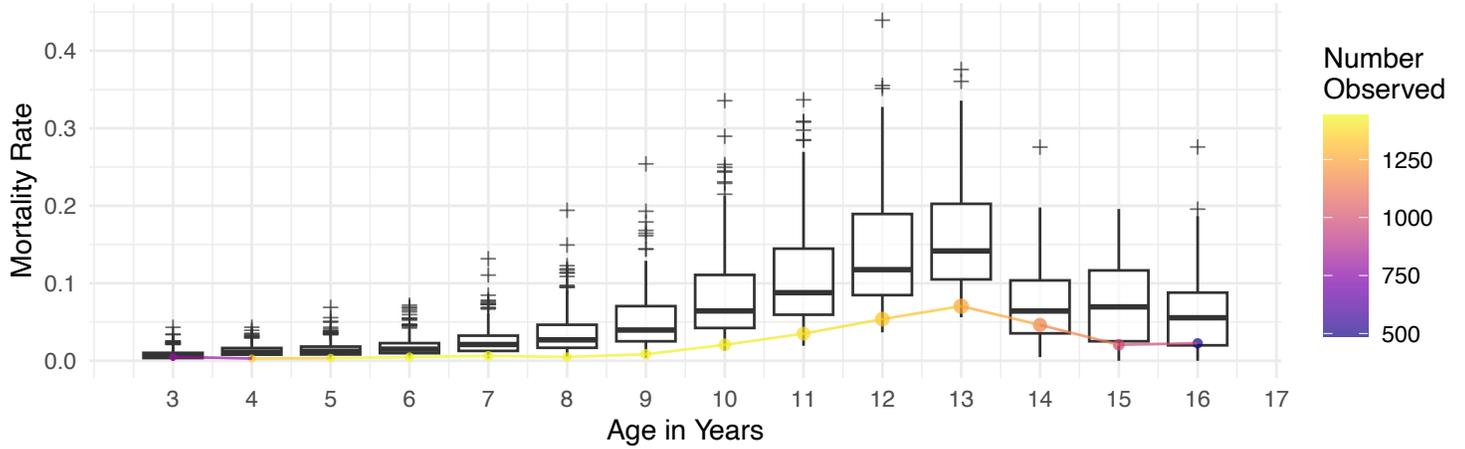

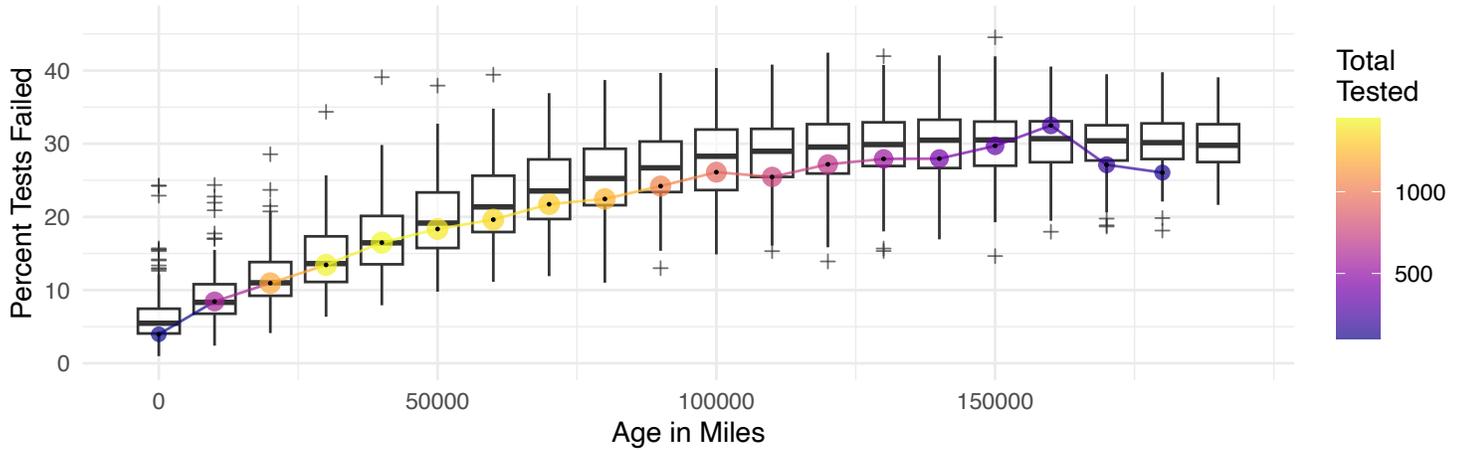

| Mortality rates | | | |
|---|---|---|---|
| Age in Years | Observed | Died | Mortality Rate |
| 3 | 788 | 4 | 0.00508 |
| 4 | 1331 | 4 | 0.00301 |
| 5 | 1430 | 5 | 0.00350 |
| 6 | 1439 | 7 | 0.00486 |
| 7 | 1433 | 9 | 0.00628 |
| 8 | 1425 | 7 | 0.00491 |
| 9 | 1418 | 12 | 0.00846 |
| 10 | 1404 | 29 | 0.02070 |
| 11 | 1370 | 48 | 0.03500 |
| 12 | 1320 | 71 | 0.05380 |
| 13 | 1248 | 88 | 0.07050 |
| 14 | 1145 | 53 | 0.04630 |
| 15 | 965 | 20 | 0.02070 |
| 16 | 488 | 11 | 0.02250 |

| Mechanical Reliability Rates | | |
|---|---|---|
| Mileage at test | N tested | Pct failed |
| 0 | 101 | 3.96 |
| 10000 | 604 | 8.44 |
| 20000 | 1178 | 11.00 |
| 30000 | 1429 | 13.40 |
| 40000 | 1450 | 16.50 |
| 50000 | 1401 | 18.30 |
| 60000 | 1350 | 19.60 |
| 70000 | 1307 | 21.70 |
| 80000 | 1225 | 22.40 |
| 90000 | 1041 | 24.20 |
| 100000 | 923 | 26.10 |
| 110000 | 758 | 25.50 |
| 120000 | 688 | 27.20 |
| 130000 | 548 | 27.90 |
| 140000 | 397 | 28.00 |
| 150000 | 340 | 29.70 |
| 160000 | 240 | 32.50 |

# Audi A3 1997

At 10 years of age, the mortality rate of a Audi A3 1997 (manufactured as a Car or Light Van) ranked number 37 out of 187 vehicles of the same age and type (any Car or Light Van constructed in 1997). One is the lowest (or best) and 187 the highest mortality rate. For vehicles reaching 120000 miles, its unreliability score (rate of failing an inspection) ranked 92 out of 167 vehicles of the same age, type, and mileage. One is the highest (or worst) and 167 the lowest rate of failing an inspection.

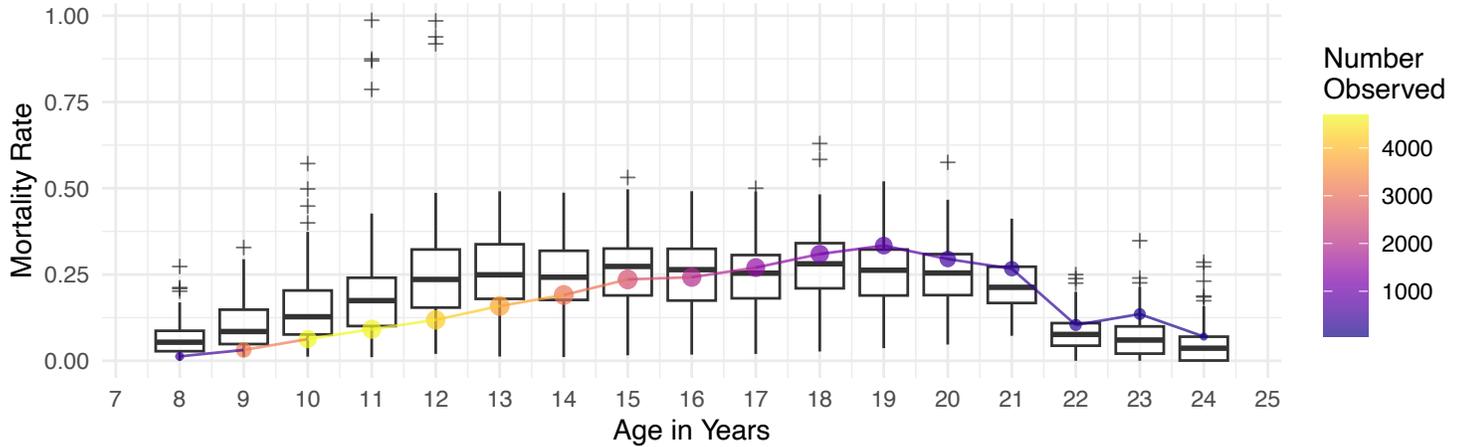

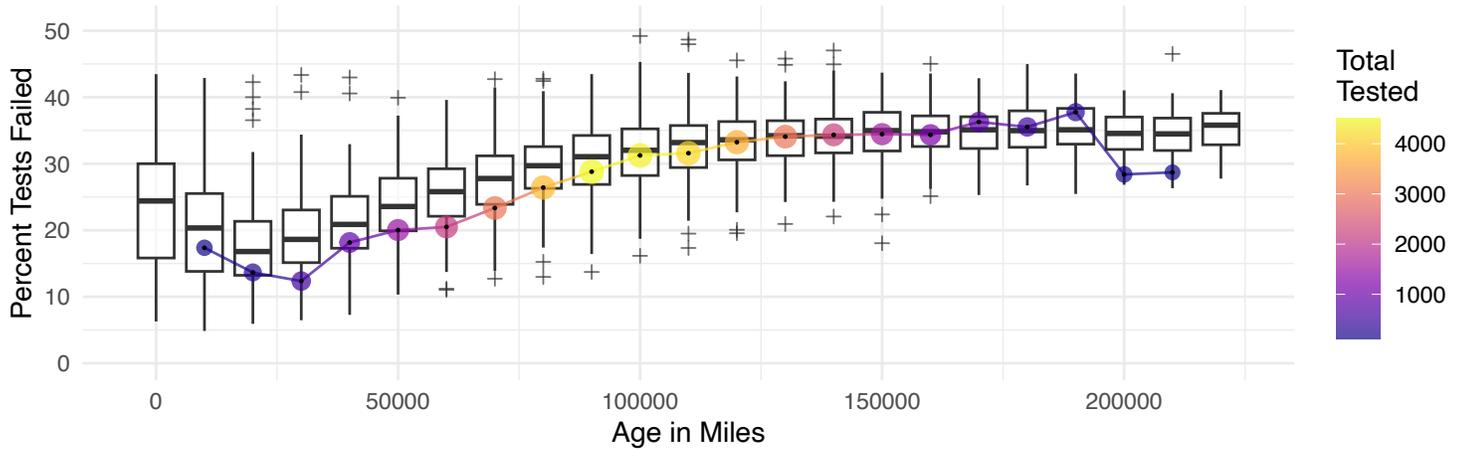

| Mortality rates | | | |
|---|---|---|---|
| Age in Years | Observed | Died | Mortality Rate |
| 8 | 466 | 6 | 0.0129 |
| 9 | 3188 | 100 | 0.0314 |
| 10 | 4677 | 294 | 0.0629 |
| 11 | 4627 | 423 | 0.0914 |
| 12 | 4214 | 500 | 0.1190 |
| 13 | 3702 | 588 | 0.1590 |
| 14 | 3105 | 592 | 0.1910 |
| 15 | 2510 | 592 | 0.2360 |
| 16 | 1920 | 465 | 0.2420 |
| 17 | 1451 | 392 | 0.2700 |
| 18 | 1057 | 327 | 0.3090 |
| 19 | 728 | 243 | 0.3340 |
| 20 | 485 | 143 | 0.2950 |
| 21 | 341 | 91 | 0.2670 |
| 22 | 231 | 24 | 0.1040 |
| 23 | 155 | 21 | 0.1350 |
| 24 | 57 | 4 | 0.0702 |

| Mechanical Reliability Rates | | |
|---|---|---|
| Mileage at test | N tested | Pct failed |
| 10000 | 144 | 17.4 |
| 20000 | 271 | 13.7 |
| 30000 | 518 | 12.4 |
| 40000 | 887 | 18.2 |
| 50000 | 1508 | 20.0 |
| 60000 | 2171 | 20.5 |
| 100000 | 4436 | 31.2 |
| 110000 | 4170 | 31.6 |
| 120000 | 3725 | 33.2 |
| 130000 | 2988 | 34.1 |
| 140000 | 2219 | 34.3 |
| 150000 | 1626 | 34.4 |
| 160000 | 1162 | 34.3 |
| 170000 | 769 | 36.3 |
| 180000 | 501 | 35.5 |
| 200000 | 162 | 28.4 |
| 210000 | 115 | 28.7 |

**Audi A3 1998**

At 10 years of age, the mortality rate of a Audi A3 1998 (manufactured as a Car or Light Van) ranked number 32 out of 196 vehicles of the same age and type (any Car or Light Van constructed in 1998). One is the lowest (or best) and 196 the highest mortality rate. For vehicles reaching 120000 miles, its unreliability score (rate of failing an inspection) ranked 115 out of 172 vehicles of the same age, type, and mileage. One is the highest (or worst) and 172 the lowest rate of failing an inspection.

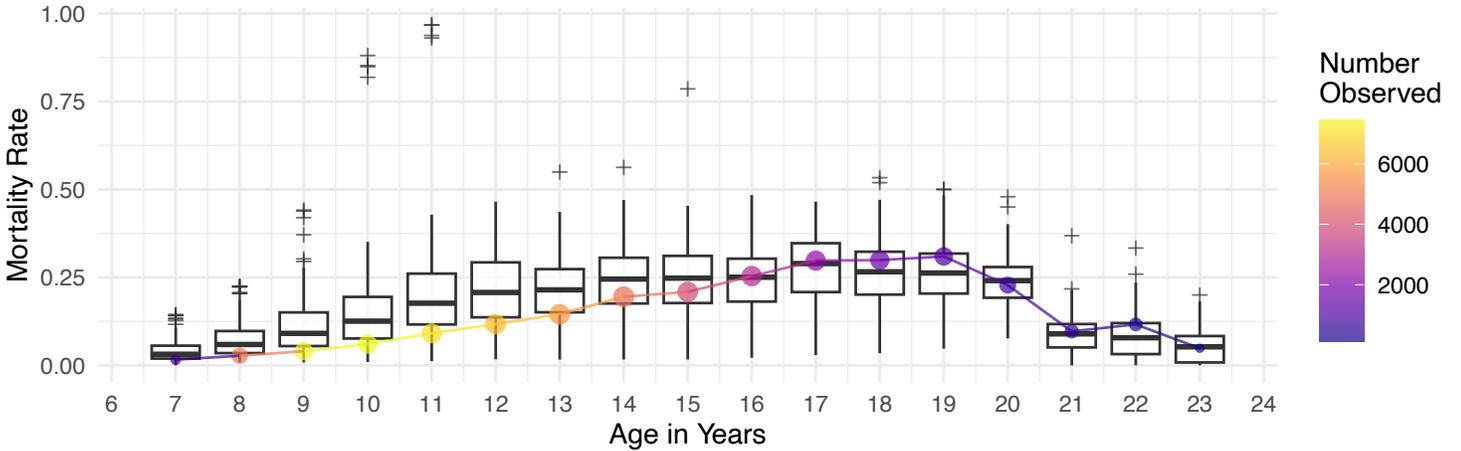

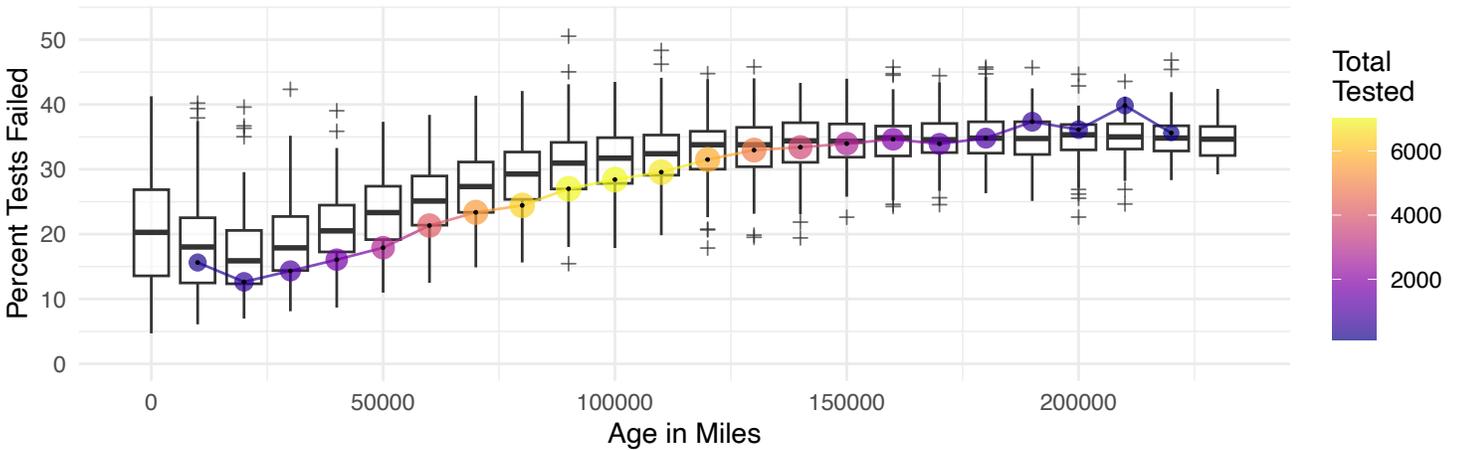

| Mortality rates | | | |
|---|---|---|---|
| Age in Years | Observed | Died | Mortality Rate |
| 7 | 735 | 12 | 0.0163 |
| 8 | 4986 | 136 | 0.0273 |
| 9 | 7349 | 297 | 0.0404 |
| 10 | 7437 | 450 | 0.0605 |
| 11 | 7002 | 637 | 0.0910 |
| 12 | 6363 | 746 | 0.1170 |
| 13 | 5606 | 815 | 0.1450 |
| 14 | 4776 | 931 | 0.1950 |
| 15 | 3835 | 801 | 0.2090 |
| 16 | 3027 | 768 | 0.2540 |
| 17 | 2248 | 669 | 0.2980 |
| 18 | 1578 | 472 | 0.2990 |
| 19 | 1105 | 342 | 0.3100 |
| 20 | 763 | 174 | 0.2280 |
| 21 | 559 | 54 | 0.0966 |
| 22 | 378 | 44 | 0.1160 |
| 23 | 144 | 7 | 0.0486 |

| Mechanical Reliability Rates | | |
|---|---|---|
| Mileage at test | N tested | Pct failed |
| 10000 | 288 | 15.6 |
| 20000 | 538 | 12.6 |
| 30000 | 998 | 14.3 |
| 40000 | 1743 | 16.1 |
| 50000 | 2831 | 17.9 |
| 60000 | 4131 | 21.3 |
| 70000 | 5440 | 23.4 |
| 100000 | 7004 | 28.4 |
| 110000 | 6752 | 29.6 |
| 120000 | 5736 | 31.5 |
| 130000 | 4881 | 33.0 |
| 140000 | 3652 | 33.4 |
| 150000 | 2783 | 34.0 |
| 160000 | 1943 | 34.6 |
| 170000 | 1341 | 33.9 |
| 180000 | 928 | 34.8 |
| 190000 | 600 | 37.3 |

## Audi A3 1999

At 10 years of age, the mortality rate of a Audi A3 1999 (manufactured as a Car or Light Van) ranked number 28 out of 201 vehicles of the same age and type (any Car or Light Van constructed in 1999). One is the lowest (or best) and 201 the highest mortality rate. For vehicles reaching 120000 miles, its unreliability score (rate of failing an inspection) ranked 129 out of 181 vehicles of the same age, type, and mileage. One is the highest (or worst) and 181 the lowest rate of failing an inspection.

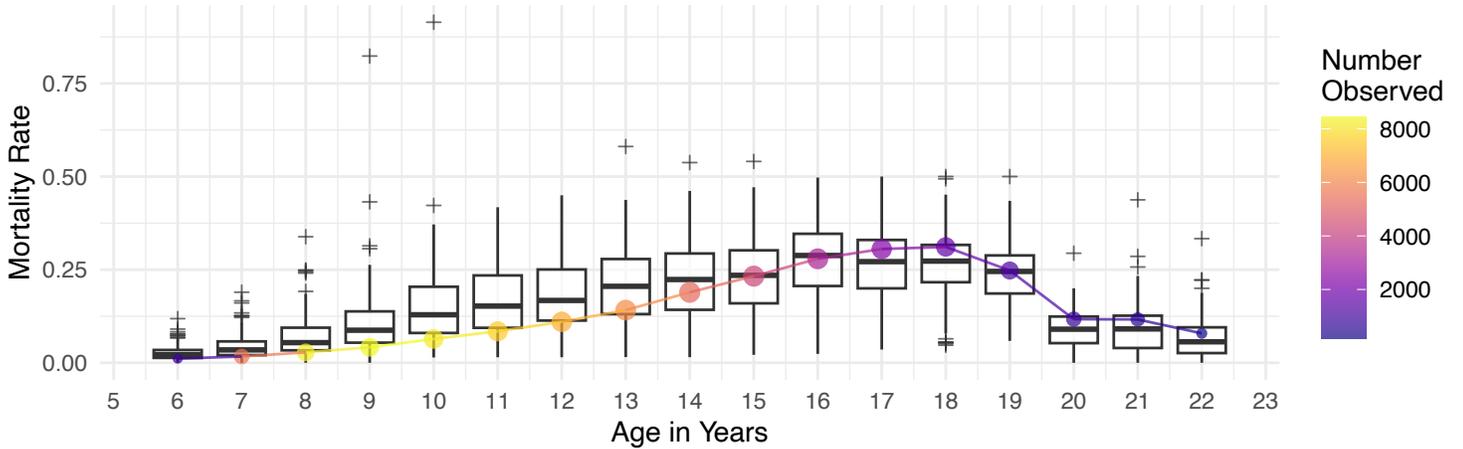

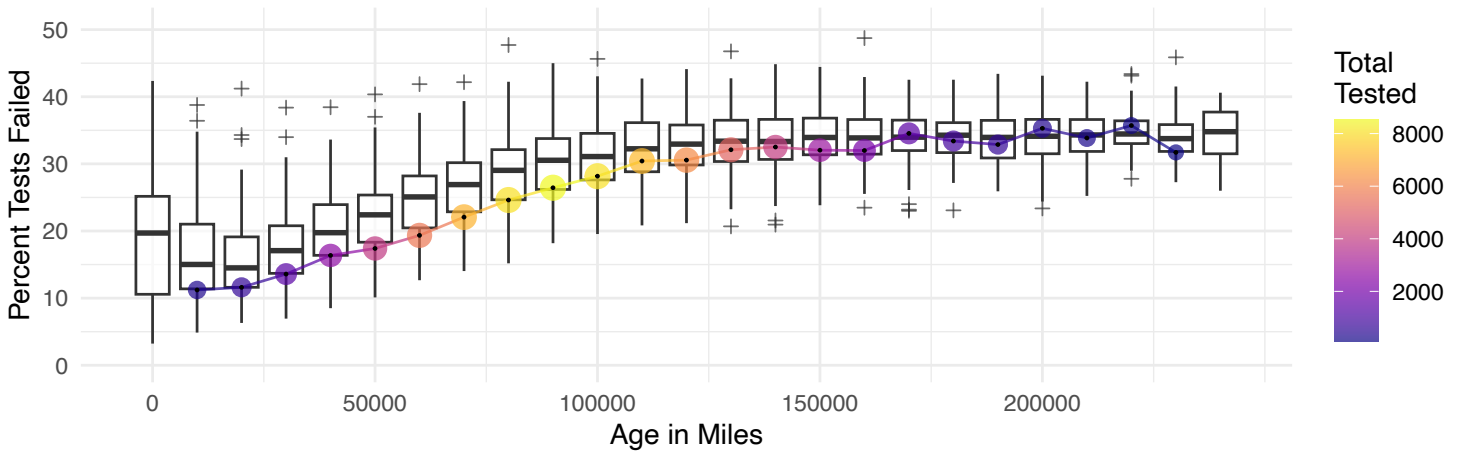

| Mortality rates | | | |
|---|---|---|---|
| Age in Years | Observed | Died | Mortality Rate |
| 6 | 890 | 10 | 0.0112 |
| 7 | 5608 | 95 | 0.0169 |
| 8 | 8204 | 230 | 0.0280 |
| 9 | 8417 | 350 | 0.0416 |
| 10 | 8088 | 518 | 0.0640 |
| 11 | 7557 | 639 | 0.0846 |
| 12 | 6910 | 760 | 0.1100 |
| 13 | 6130 | 867 | 0.1410 |
| 14 | 5261 | 996 | 0.1890 |
| 15 | 4251 | 989 | 0.2330 |
| 16 | 3249 | 907 | 0.2790 |
| 17 | 2334 | 713 | 0.3050 |
| 18 | 1620 | 504 | 0.3110 |
| 19 | 1114 | 276 | 0.2480 |
| 20 | 777 | 91 | 0.1170 |
| 21 | 507 | 59 | 0.1160 |
| 22 | 177 | 14 | 0.0791 |

| Mechanical Reliability Rates | | |
|---|---|---|
| Mileage at test | N tested | Pct failed |
| 10000 | 321 | 11.2 |
| 20000 | 646 | 11.6 |
| 30000 | 1384 | 13.6 |
| 40000 | 2585 | 16.4 |
| 50000 | 3955 | 17.4 |
| 60000 | 5670 | 19.3 |
| 70000 | 7115 | 22.1 |
| 80000 | 8039 | 24.6 |
| 90000 | 8513 | 26.4 |
| 100000 | 8063 | 28.2 |
| 110000 | 7267 | 30.4 |
| 120000 | 6188 | 30.6 |
| 130000 | 5152 | 32.1 |
| 140000 | 4011 | 32.5 |
| 150000 | 2884 | 32.0 |
| 160000 | 2085 | 32.0 |
| 170000 | 1459 | 34.5 |

**Audi A3 2000**

At 5 years of age, the mortality rate of a Audi A3 2000 (manufactured as a Car or Light Van) ranked number 39 out of 198 vehicles of the same age and type (any Car or Light Van constructed in 2000). One is the lowest (or best) and 198 the highest mortality rate. For vehicles reaching 120000 miles, its unreliability score (rate of failing an inspection) ranked 115 out of 184 vehicles of the same age, type, and mileage. One is the highest (or worst) and 184 the lowest rate of failing an inspection.

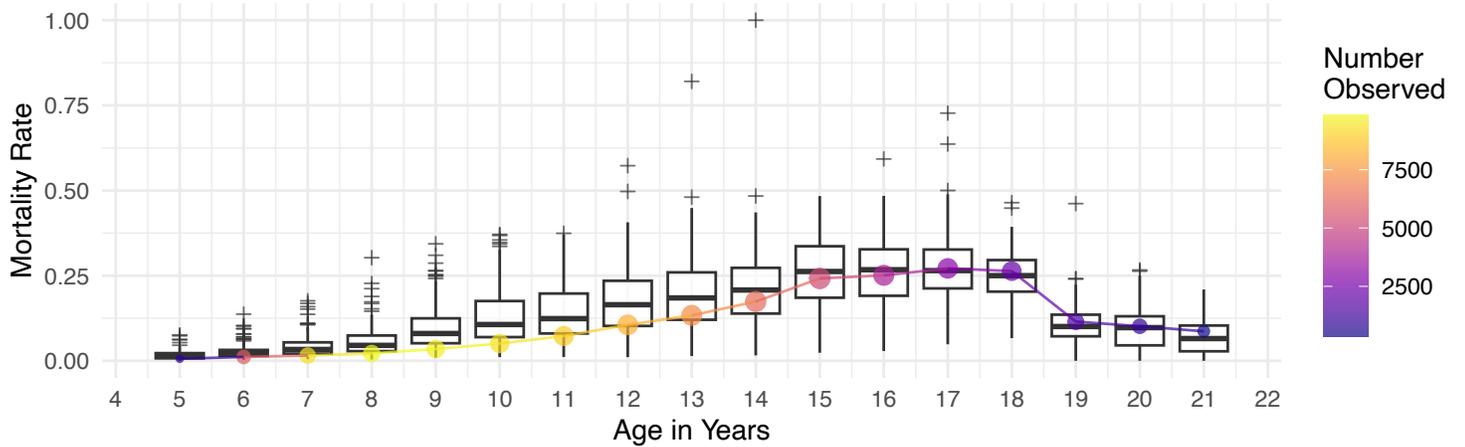

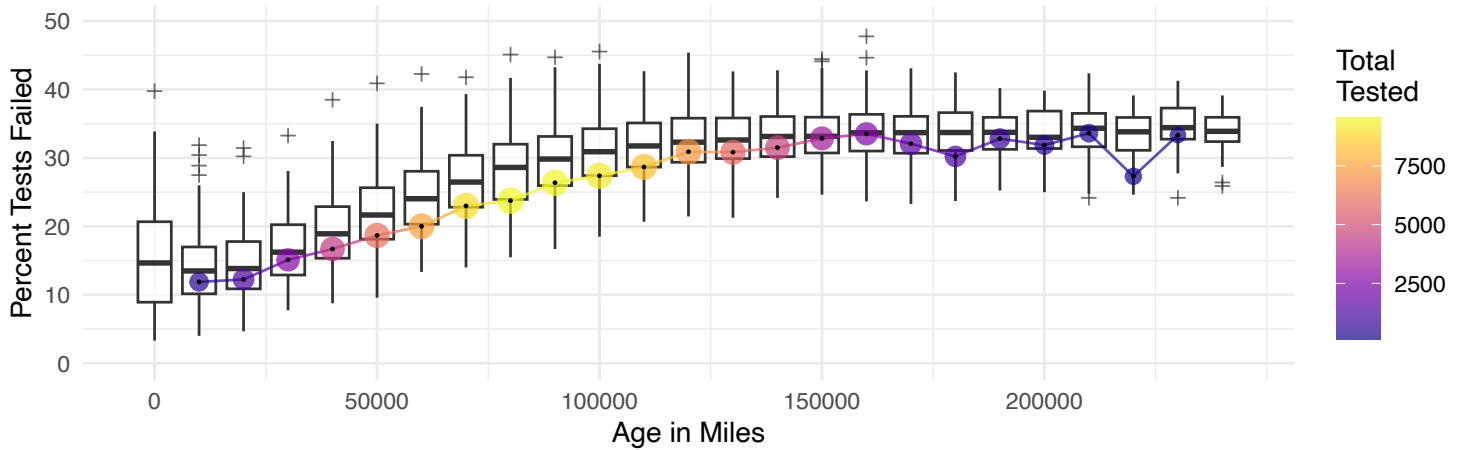

| Mortality rates | | | |
|---|---|---|---|
| Age in Years | Observed | Died | Mortality Rate |
| 5 | 833 | 5 | 0.0060 |
| 6 | 5845 | 66 | 0.0113 |
| 7 | 9334 | 144 | 0.0154 |
| 8 | 9818 | 211 | 0.0215 |
| 9 | 9613 | 333 | 0.0346 |
| 10 | 9271 | 470 | 0.0507 |
| 11 | 8771 | 638 | 0.0727 |
| 12 | 8111 | 856 | 0.1060 |
| 13 | 7237 | 965 | 0.1330 |
| 14 | 6254 | 1088 | 0.1740 |
| 15 | 5143 | 1244 | 0.2420 |
| 16 | 3877 | 973 | 0.2510 |
| 17 | 2897 | 786 | 0.2710 |
| 18 | 2107 | 555 | 0.2630 |
| 19 | 1469 | 169 | 0.1150 |
| 20 | 976 | 99 | 0.1010 |
| 21 | 358 | 31 | 0.0866 |

| Mechanical Reliability Rates | | |
|---|---|---|
| Mileage at test | N tested | Pct failed |
| 10000 | 530 | 11.9 |
| 20000 | 1250 | 12.2 |
| 30000 | 2668 | 15.1 |
| 40000 | 4365 | 16.7 |
| 50000 | 6111 | 18.7 |
| 60000 | 7583 | 20.0 |
| 70000 | 8939 | 23.0 |
| 80000 | 9529 | 23.8 |
| 90000 | 9490 | 26.4 |
| 100000 | 9192 | 27.4 |
| 110000 | 8249 | 28.7 |
| 120000 | 7230 | 30.9 |
| 130000 | 5834 | 30.8 |
| 140000 | 4650 | 31.5 |
| 150000 | 3400 | 32.9 |
| 160000 | 2574 | 33.5 |
| 170000 | 1699 | 32.1 |

## Audi A3 2001

At 5 years of age, the mortality rate of a Audi A3 2001 (manufactured as a Car or Light Van) ranked number 48 out of 205 vehicles of the same age and type (any Car or Light Van constructed in 2001). One is the lowest (or best) and 205 the highest mortality rate. For vehicles reaching 120000 miles, its unreliability score (rate of failing an inspection) ranked 131 out of 194 vehicles of the same age, type, and mileage. One is the highest (or worst) and 194 the lowest rate of failing an inspection.

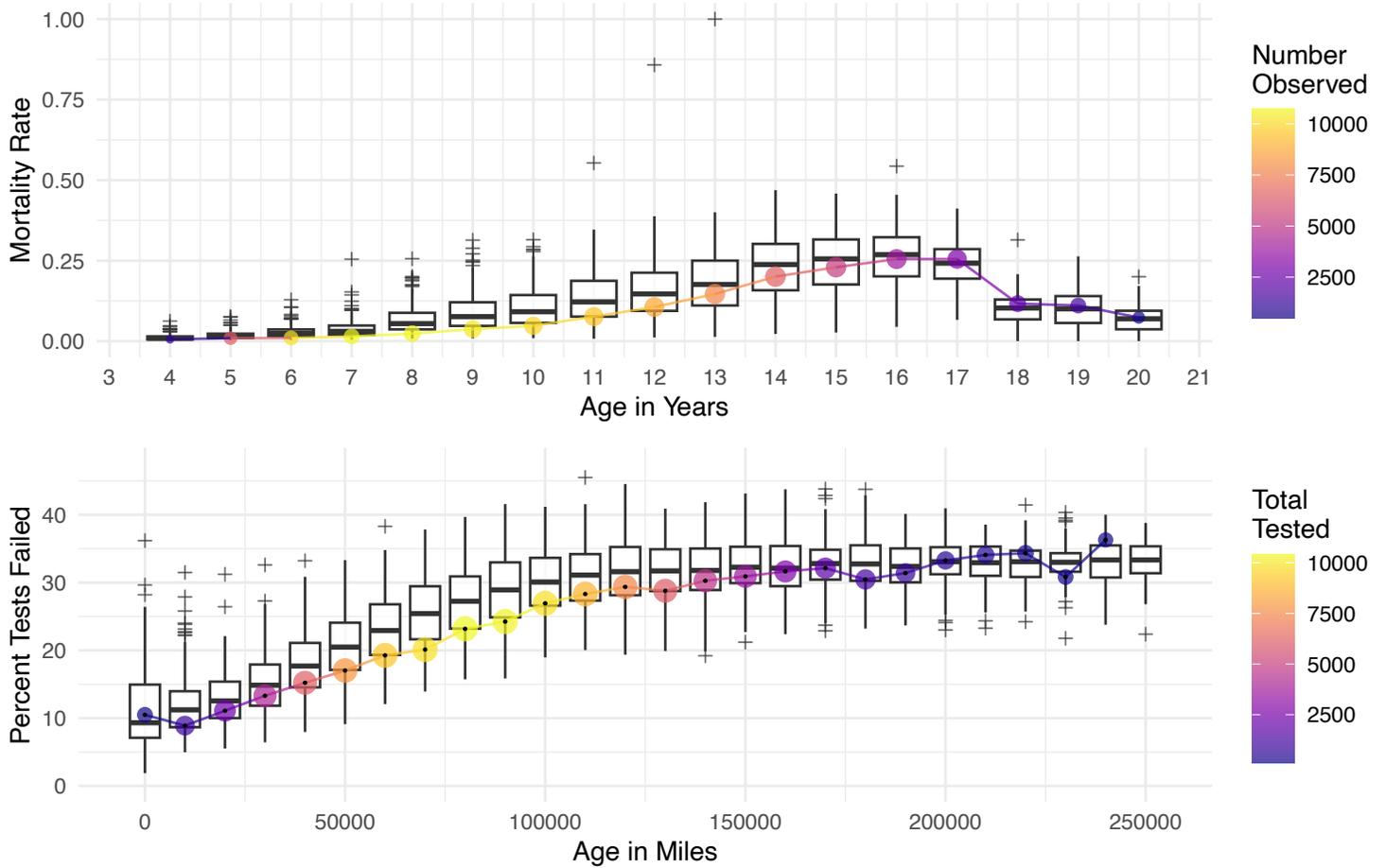

| Mortality rates | | | |
|---|---|---|---|
| Age in Years | Observed | Died | Mortality Rate |
| 4 | 1070 | 6 | 0.00561 |
| 5 | 6391 | 60 | 0.00939 |
| 6 | 10146 | 101 | 0.00995 |
| 7 | 10710 | 155 | 0.01450 |
| 8 | 10582 | 246 | 0.02320 |
| 9 | 10315 | 384 | 0.03720 |
| 10 | 9907 | 472 | 0.04760 |
| 11 | 9409 | 716 | 0.07610 |
| 12 | 8670 | 919 | 0.10600 |
| 13 | 7722 | 1130 | 0.14600 |
| 14 | 6551 | 1313 | 0.20000 |
| 15 | 5207 | 1195 | 0.22900 |
| 16 | 3990 | 1017 | 0.25500 |
| 17 | 2961 | 755 | 0.25500 |
| 18 | 2103 | 245 | 0.11700 |
| 19 | 1428 | 158 | 0.11100 |
| 20 | 503 | 37 | 0.07360 |

| Mechanical Reliability Rates | | |
|---|---|---|
| Mileage at test | N tested | Pct failed |
| 0 | 162 | 10.5 |
| 10000 | 921 | 8.9 |
| 20000 | 2208 | 11.1 |
| 30000 | 4044 | 13.3 |
| 40000 | 6131 | 15.2 |
| 50000 | 7882 | 17.0 |
| 60000 | 9199 | 19.3 |
| 70000 | 9850 | 20.1 |
| 80000 | 10423 | 23.2 |
| 100000 | 9816 | 26.9 |
| 110000 | 8690 | 28.3 |
| 120000 | 7431 | 29.4 |
| 130000 | 6054 | 28.8 |
| 140000 | 4559 | 30.3 |
| 150000 | 3504 | 30.9 |
| 160000 | 2642 | 31.6 |
| 170000 | 1740 | 32.1 |

## Audi A3 2002

At 5 years of age, the mortality rate of a Audi A3 2002 (manufactured as a Car or Light Van) ranked number 32 out of 202 vehicles of the same age and type (any Car or Light Van constructed in 2002). One is the lowest (or best) and 202 the highest mortality rate. For vehicles reaching 120000 miles, its unreliability score (rate of failing an inspection) ranked 134 out of 193 vehicles of the same age, type, and mileage. One is the highest (or worst) and 193 the lowest rate of failing an inspection.

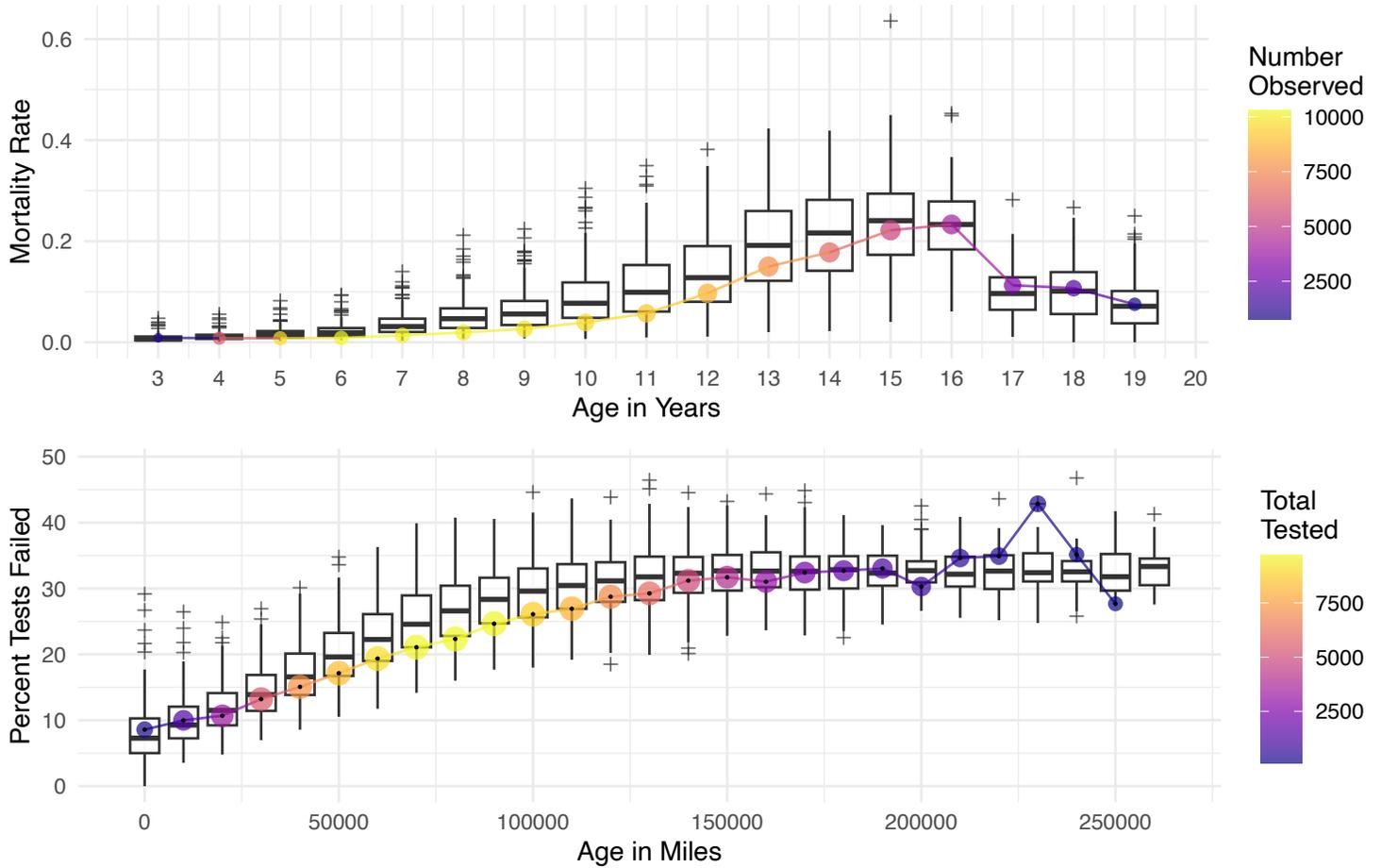

| Mortality rates | | | |
|---|---|---|---|
| Age in Years | Observed | Died | Mortality Rate |
| 3 | 1106 | 10 | 0.00904 |
| 4 | 6207 | 50 | 0.00806 |
| 5 | 9679 | 77 | 0.00796 |
| 6 | 10263 | 89 | 0.00867 |
| 7 | 10195 | 139 | 0.01360 |
| 8 | 10034 | 195 | 0.01940 |
| 9 | 9831 | 260 | 0.02640 |
| 10 | 9545 | 374 | 0.03920 |
| 11 | 9147 | 524 | 0.05730 |
| 12 | 8578 | 831 | 0.09690 |
| 13 | 7706 | 1155 | 0.15000 |
| 14 | 6516 | 1159 | 0.17800 |
| 15 | 5323 | 1180 | 0.22200 |
| 16 | 4139 | 965 | 0.23300 |
| 17 | 3041 | 343 | 0.11300 |
| 18 | 2102 | 225 | 0.10700 |
| 19 | 772 | 58 | 0.07510 |

| Mechanical Reliability Rates | | |
|---|---|---|
| Mileage at test | N tested | Pct failed |
| 0 | 233 | 8.58 |
| 10000 | 1401 | 9.99 |
| 20000 | 3245 | 10.70 |
| 30000 | 5485 | 13.20 |
| 40000 | 7246 | 15.10 |
| 50000 | 8347 | 17.10 |
| 60000 | 9141 | 19.40 |
| 70000 | 9658 | 21.10 |
| 80000 | 9723 | 22.30 |
| 90000 | 9630 | 24.70 |
| 100000 | 8735 | 26.10 |
| 110000 | 8062 | 26.90 |
| 120000 | 6997 | 28.80 |
| 130000 | 5987 | 29.30 |
| 140000 | 4882 | 31.20 |
| 150000 | 3980 | 31.70 |
| 160000 | 3045 | 31.00 |

# Audi A3 2003

At 5 years of age, the mortality rate of a Audi A3 2003 (manufactured as a Car or Light Van) ranked number 46 out of 213 vehicles of the same age and type (any Car or Light Van constructed in 2003). One is the lowest (or best) and 213 the highest mortality rate. For vehicles reaching 100000 miles, its unreliability score (rate of failing an inspection) ranked 160 out of 208 vehicles of the same age, type, and mileage. One is the highest (or worst) and 208 the lowest rate of failing an inspection.

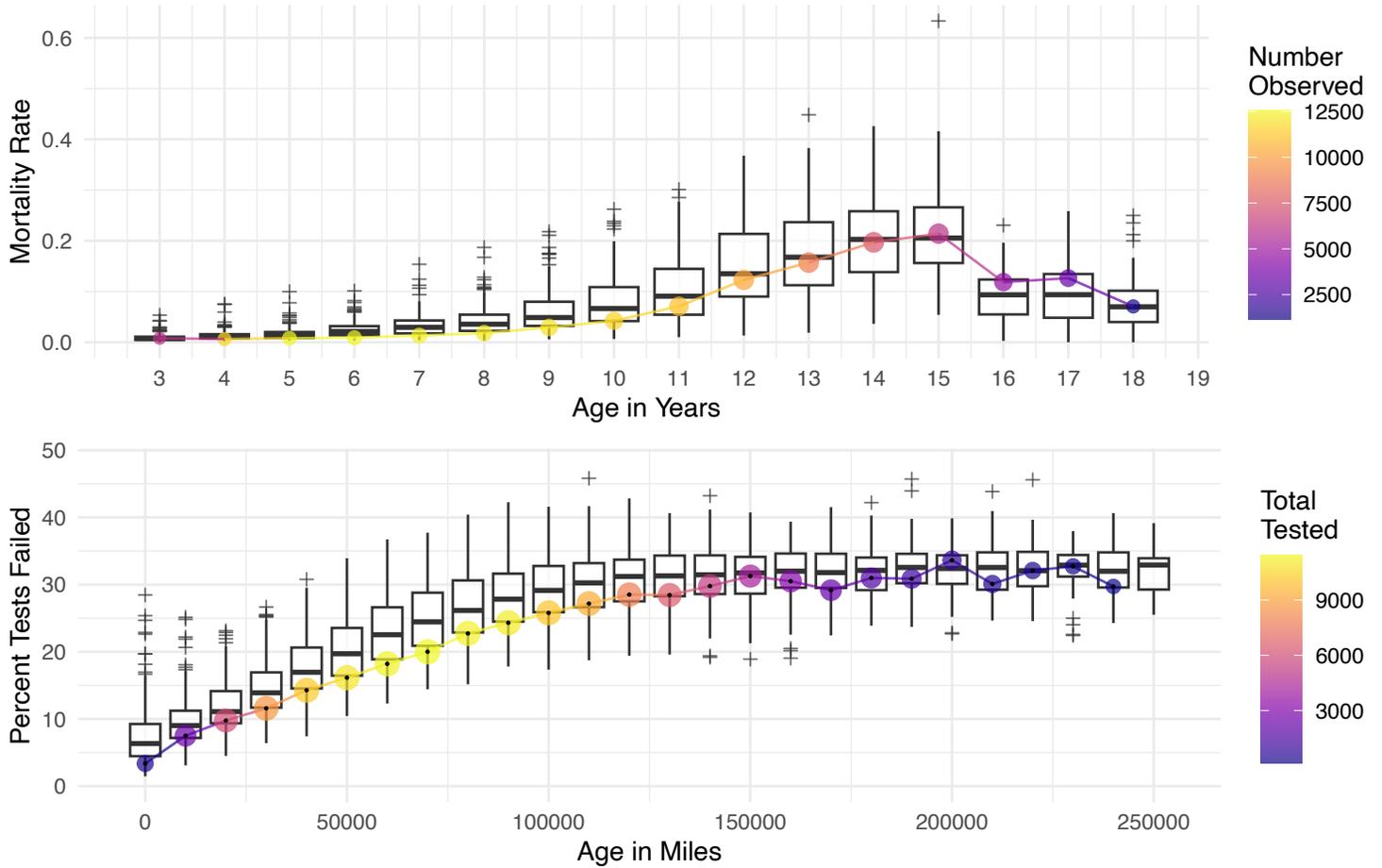

### Mortality rates

| Age in Years | Observed | Died | Mortality Rate |
|---|---|---|---|
| 3 | 6037 | 48 | 0.00795 |
| 4 | 11577 | 74 | 0.00639 |
| 5 | 12526 | 101 | 0.00806 |
| 6 | 12449 | 112 | 0.00900 |
| 7 | 12314 | 163 | 0.01320 |
| 8 | 12152 | 220 | 0.01810 |
| 9 | 11909 | 348 | 0.02920 |
| 10 | 11552 | 491 | 0.04250 |
| 11 | 11018 | 781 | 0.07090 |
| 12 | 10188 | 1253 | 0.12300 |
| 13 | 8878 | 1394 | 0.15700 |
| 14 | 7453 | 1468 | 0.19700 |
| 15 | 5975 | 1279 | 0.21400 |
| 16 | 4496 | 533 | 0.11900 |
| 17 | 3118 | 394 | 0.12600 |
| 18 | 1160 | 82 | 0.07070 |

### Mechanical Reliability Rates

| Mileage at test | N tested | Pct failed |
|---|---|---|
| 0 | 297 | 3.37 |
| 10000 | 2346 | 7.50 |
| 20000 | 6008 | 9.77 |
| 30000 | 8646 | 11.60 |
| 40000 | 10234 | 14.30 |
| 50000 | 10963 | 16.10 |
| 60000 | 11300 | 18.20 |
| 70000 | 11453 | 20.00 |
| 80000 | 11282 | 22.70 |
| 90000 | 11127 | 24.30 |
| 100000 | 10100 | 25.80 |
| 110000 | 9351 | 27.20 |
| 120000 | 7857 | 28.50 |
| 130000 | 6594 | 28.50 |
| 140000 | 5372 | 29.80 |
| 150000 | 4221 | 31.30 |
| 160000 | 3153 | 30.50 |

## Audi A3 2004

At 5 years of age, the mortality rate of a Audi A3 2004 (manufactured as a Car or Light Van) ranked number 43 out of 229 vehicles of the same age and type (any Car or Light Van constructed in 2004). One is the lowest (or best) and 229 the highest mortality rate. For vehicles reaching 120000 miles, its unreliability score (rate of failing an inspection) ranked 165 out of 220 vehicles of the same age, type, and mileage. One is the highest (or worst) and 220 the lowest rate of failing an inspection.

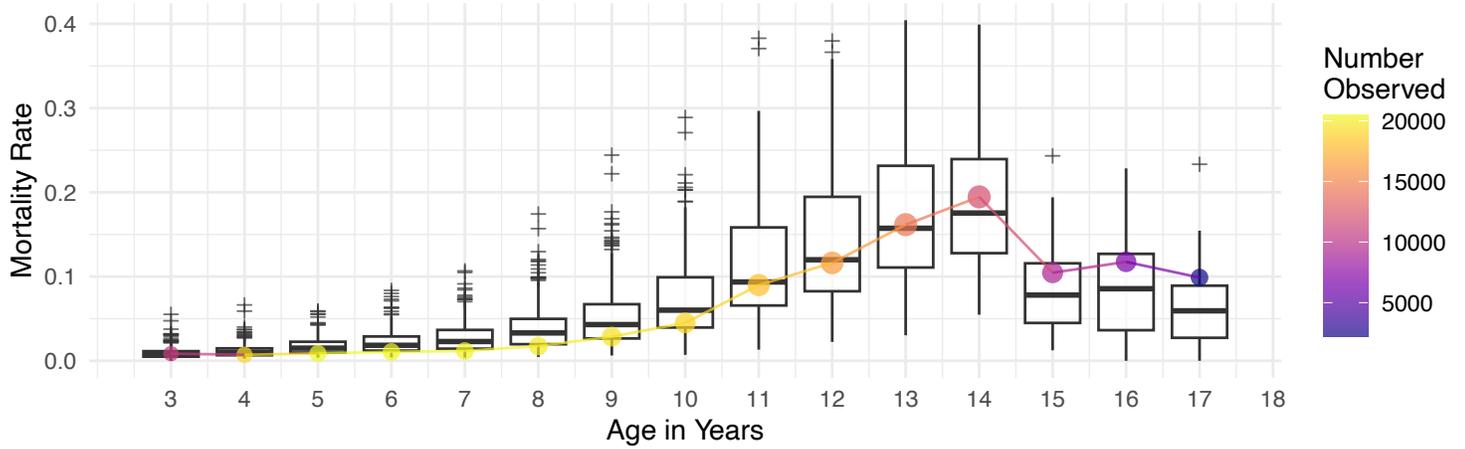

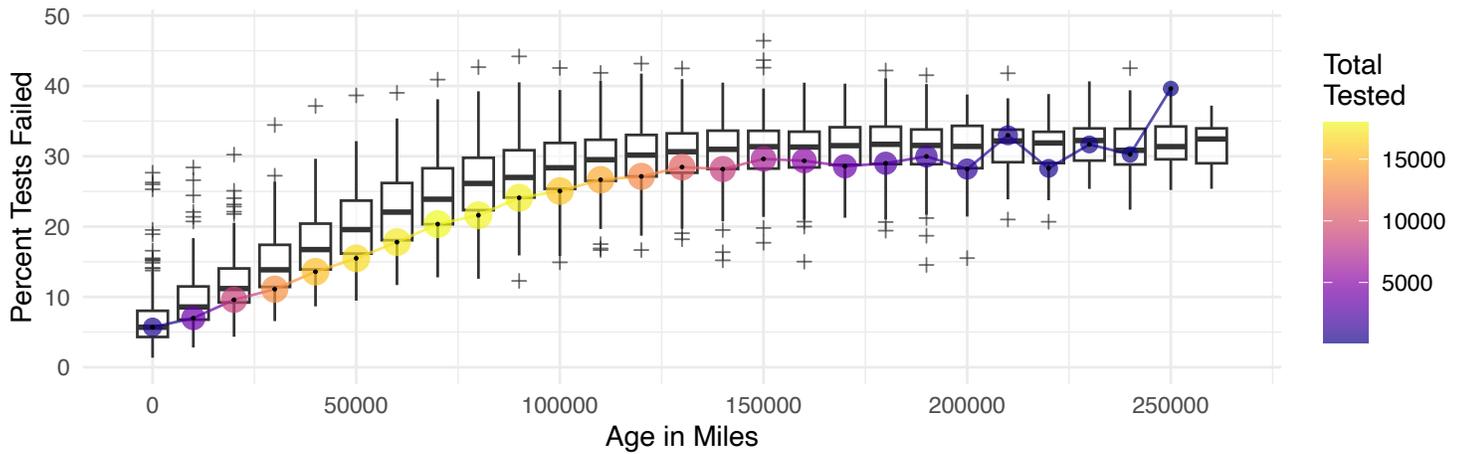

<table>
<tr><td colspan="4" align="center">Mortality rates</td></tr>
<tr><th>Age in Years</th><th>Observed</th><th>Died</th><th>Mortality Rate</th></tr>
<tr><td>3</td><td>10771</td><td>92</td><td>0.00854</td></tr>
<tr><td>4</td><td>18954</td><td>135</td><td>0.00712</td></tr>
<tr><td>5</td><td>20425</td><td>180</td><td>0.00881</td></tr>
<tr><td>6</td><td>20321</td><td>216</td><td>0.01060</td></tr>
<tr><td>7</td><td>20116</td><td>235</td><td>0.01170</td></tr>
<tr><td>8</td><td>19869</td><td>347</td><td>0.01750</td></tr>
<tr><td>9</td><td>19510</td><td>560</td><td>0.02870</td></tr>
<tr><td>10</td><td>18898</td><td>850</td><td>0.04500</td></tr>
<tr><td>11</td><td>17931</td><td>1613</td><td>0.09000</td></tr>
<tr><td>12</td><td>16210</td><td>1884</td><td>0.11600</td></tr>
<tr><td>13</td><td>14231</td><td>2300</td><td>0.16200</td></tr>
<tr><td>14</td><td>11876</td><td>2311</td><td>0.19500</td></tr>
<tr><td>15</td><td>9181</td><td>960</td><td>0.10500</td></tr>
<tr><td>16</td><td>6555</td><td>770</td><td>0.11700</td></tr>
<tr><td>17</td><td>2225</td><td>220</td><td>0.09890</td></tr>
</table>

<table>
<tr><td colspan="3" align="center">Mechanical Reliability Rates</td></tr>
<tr><th>Mileage at test</th><th>N tested</th><th>Pct failed</th></tr>
<tr><td>0</td><td>474</td><td>5.70</td></tr>
<tr><td>10000</td><td>3578</td><td>6.99</td></tr>
<tr><td>20000</td><td>8788</td><td>9.59</td></tr>
<tr><td>30000</td><td>13297</td><td>11.10</td></tr>
<tr><td>40000</td><td>15624</td><td>13.60</td></tr>
<tr><td>50000</td><td>16821</td><td>15.50</td></tr>
<tr><td>60000</td><td>17381</td><td>17.80</td></tr>
<tr><td>70000</td><td>18015</td><td>20.40</td></tr>
<tr><td>80000</td><td>17815</td><td>21.60</td></tr>
<tr><td>90000</td><td>17623</td><td>24.10</td></tr>
<tr><td>100000</td><td>15843</td><td>25.10</td></tr>
<tr><td>110000</td><td>14606</td><td>26.70</td></tr>
<tr><td>120000</td><td>12908</td><td>27.10</td></tr>
<tr><td>130000</td><td>10737</td><td>28.50</td></tr>
<tr><td>140000</td><td>8633</td><td>28.20</td></tr>
<tr><td>160000</td><td>5229</td><td>29.40</td></tr>
<tr><td>230000</td><td>262</td><td>31.70</td></tr>
</table>

## Audi A3 2005

At 5 years of age, the mortality rate of a Audi A3 2005 (manufactured as a Car or Light Van) ranked number 16 out of 240 vehicles of the same age and type (any Car or Light Van constructed in 2005). One is the lowest (or best) and 240 the highest mortality rate. For vehicles reaching 120000 miles, its unreliability score (rate of failing an inspection) ranked 165 out of 226 vehicles of the same age, type, and mileage. One is the highest (or worst) and 226 the lowest rate of failing an inspection.

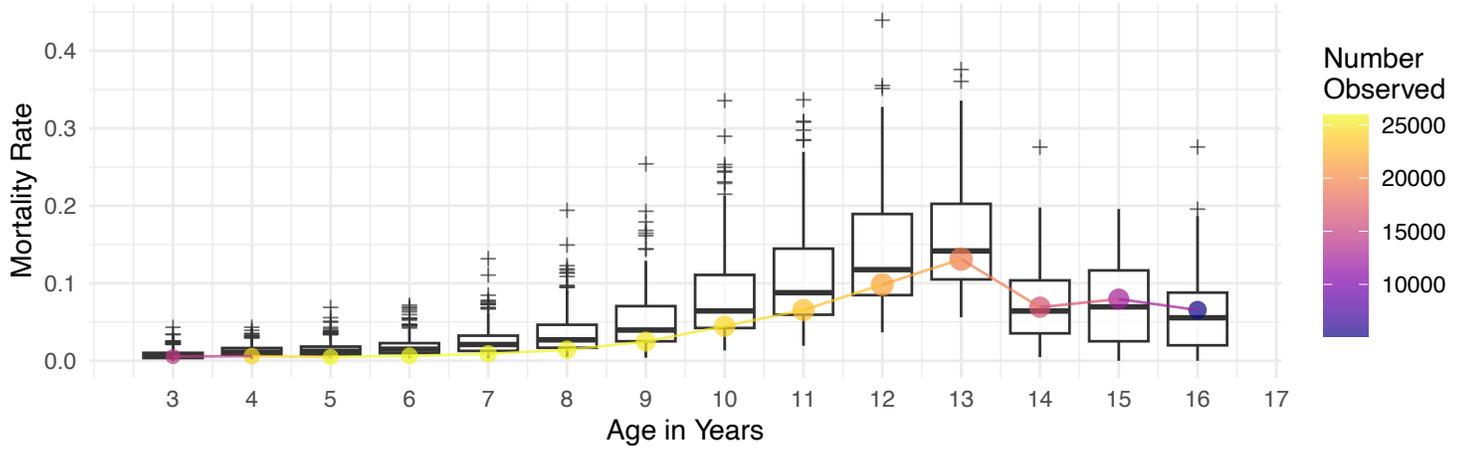

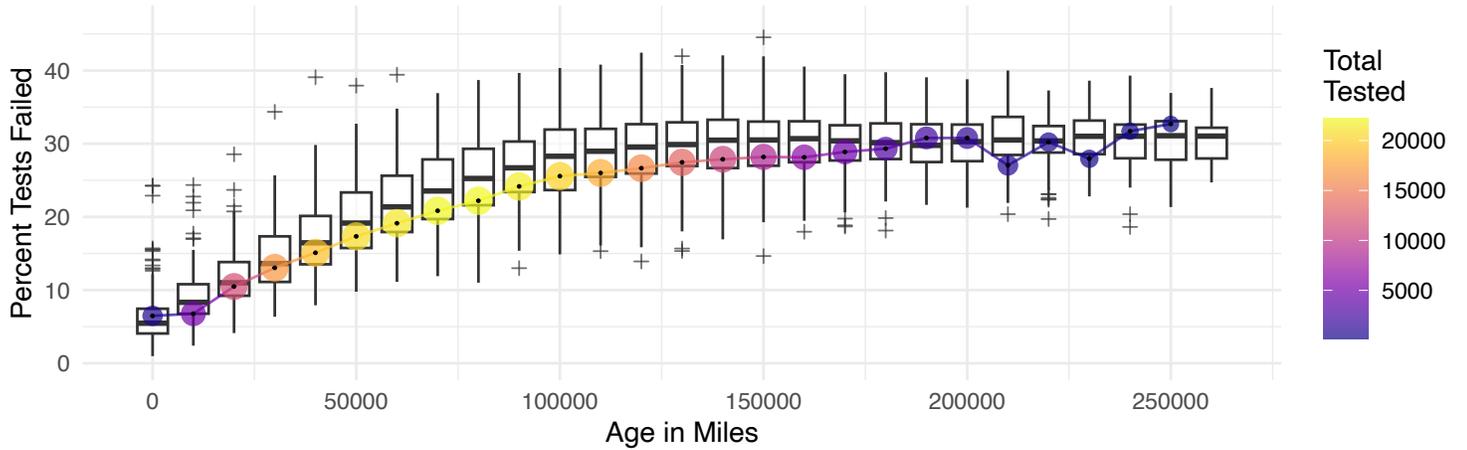

| Mortality rates | | | |
|---|---|---|---|
| Age in Years | Observed | Died | Mortality Rate |
| 3 | 14470 | 80 | 0.00553 |
| 4 | 24183 | 146 | 0.00604 |
| 5 | 25782 | 126 | 0.00489 |
| 6 | 25878 | 165 | 0.00638 |
| 7 | 25729 | 231 | 0.00898 |
| 8 | 25484 | 362 | 0.01420 |
| 9 | 25073 | 620 | 0.02470 |
| 10 | 24341 | 1087 | 0.04470 |
| 11 | 23105 | 1514 | 0.06550 |
| 12 | 21443 | 2109 | 0.09840 |
| 13 | 19243 | 2523 | 0.13100 |
| 14 | 16236 | 1123 | 0.06920 |
| 15 | 12515 | 997 | 0.07970 |
| 16 | 5114 | 335 | 0.06550 |

| Mechanical Reliability Rates | | |
|---|---|---|
| Mileage at test | N tested | Pct failed |
| 0 | 741 | 6.48 |
| 10000 | 4905 | 6.77 |
| 20000 | 11688 | 10.50 |
| 30000 | 16817 | 13.00 |
| 40000 | 19347 | 15.10 |
| 50000 | 20750 | 17.30 |
| 60000 | 21396 | 19.10 |
| 70000 | 22113 | 20.80 |
| 80000 | 22242 | 22.20 |
| 90000 | 21622 | 24.20 |
| 100000 | 20266 | 25.60 |
| 110000 | 18207 | 26.00 |
| 120000 | 15968 | 26.70 |
| 130000 | 13479 | 27.50 |
| 140000 | 10914 | 27.90 |
| 150000 | 8695 | 28.20 |
| 160000 | 6502 | 28.10 |

## Audi A3 2006

At 5 years of age, the mortality rate of a Audi A3 2006 (manufactured as a Car or Light Van) ranked number 24 out of 225 vehicles of the same age and type (any Car or Light Van constructed in 2006). One is the lowest (or best) and 225 the highest mortality rate. For vehicles reaching 160000 miles, its unreliability score (rate of failing an inspection) ranked 107 out of 161 vehicles of the same age, type, and mileage. One is the highest (or worst) and 161 the lowest rate of failing an inspection.

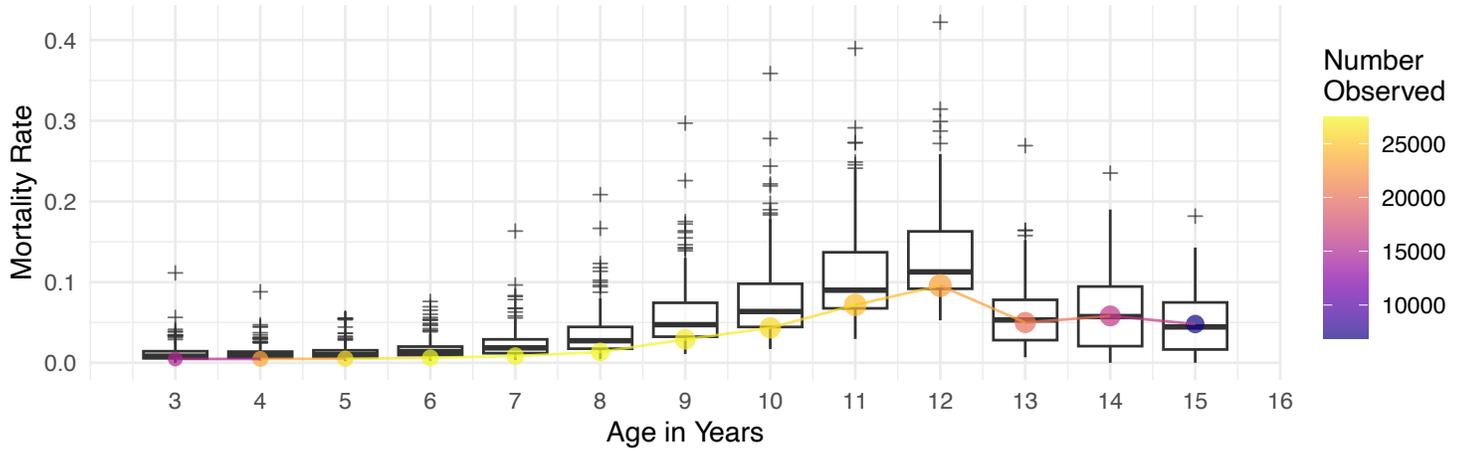

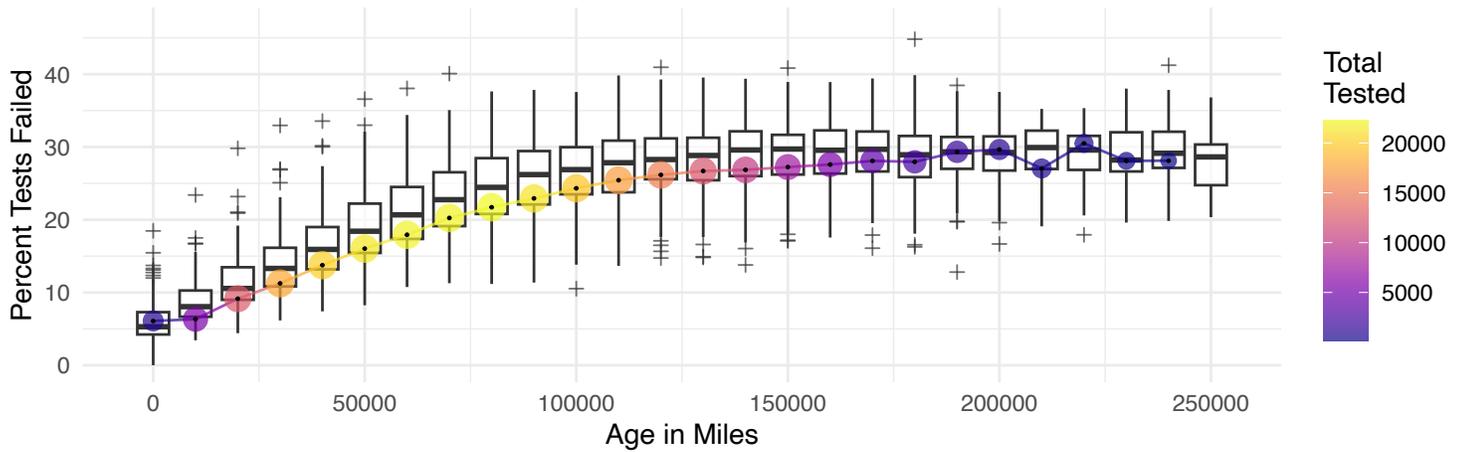

### Mortality rates

| Age in Years | Observed | Died | Mortality Rate |
|---|---|---|---|
| 3 | 14902 | 71 | 0.00476 |
| 4 | 23400 | 112 | 0.00479 |
| 5 | 26197 | 132 | 0.00504 |
| 6 | 27414 | 168 | 0.00613 |
| 7 | 27413 | 228 | 0.00832 |
| 8 | 27214 | 366 | 0.01340 |
| 9 | 26799 | 779 | 0.02910 |
| 10 | 25934 | 1124 | 0.04330 |
| 11 | 24685 | 1772 | 0.07180 |
| 12 | 22825 | 2184 | 0.09570 |
| 13 | 20081 | 1002 | 0.04990 |
| 14 | 16070 | 933 | 0.05810 |
| 15 | 6912 | 332 | 0.04800 |

### Mechanical Reliability Rates

| Mileage at test | N tested | Pct failed |
|---|---|---|
| 0 | 953 | 6.09 |
| 10000 | 5529 | 6.33 |
| 20000 | 12728 | 9.15 |
| 30000 | 17950 | 11.20 |
| 40000 | 20188 | 13.80 |
| 50000 | 21469 | 16.00 |
| 60000 | 21910 | 17.90 |
| 70000 | 22001 | 20.20 |
| 80000 | 22301 | 21.70 |
| 90000 | 21309 | 22.90 |
| 100000 | 19687 | 24.30 |
| 110000 | 17382 | 25.40 |
| 120000 | 15278 | 26.20 |
| 130000 | 12329 | 26.70 |
| 140000 | 10130 | 26.90 |
| 150000 | 7736 | 27.20 |
| 160000 | 5536 | 27.60 |

## Audi A3 2007

At 5 years of age, the mortality rate of a Audi A3 2007 (manufactured as a Car or Light Van) ranked number 20 out of 219 vehicles of the same age and type (any Car or Light Van constructed in 2007). One is the lowest (or best) and 219 the highest mortality rate. For vehicles reaching 40000 miles, its unreliability score (rate of failing an inspection) ranked 179 out of 214 vehicles of the same age, type, and mileage. One is the highest (or worst) and 214 the lowest rate of failing an inspection.

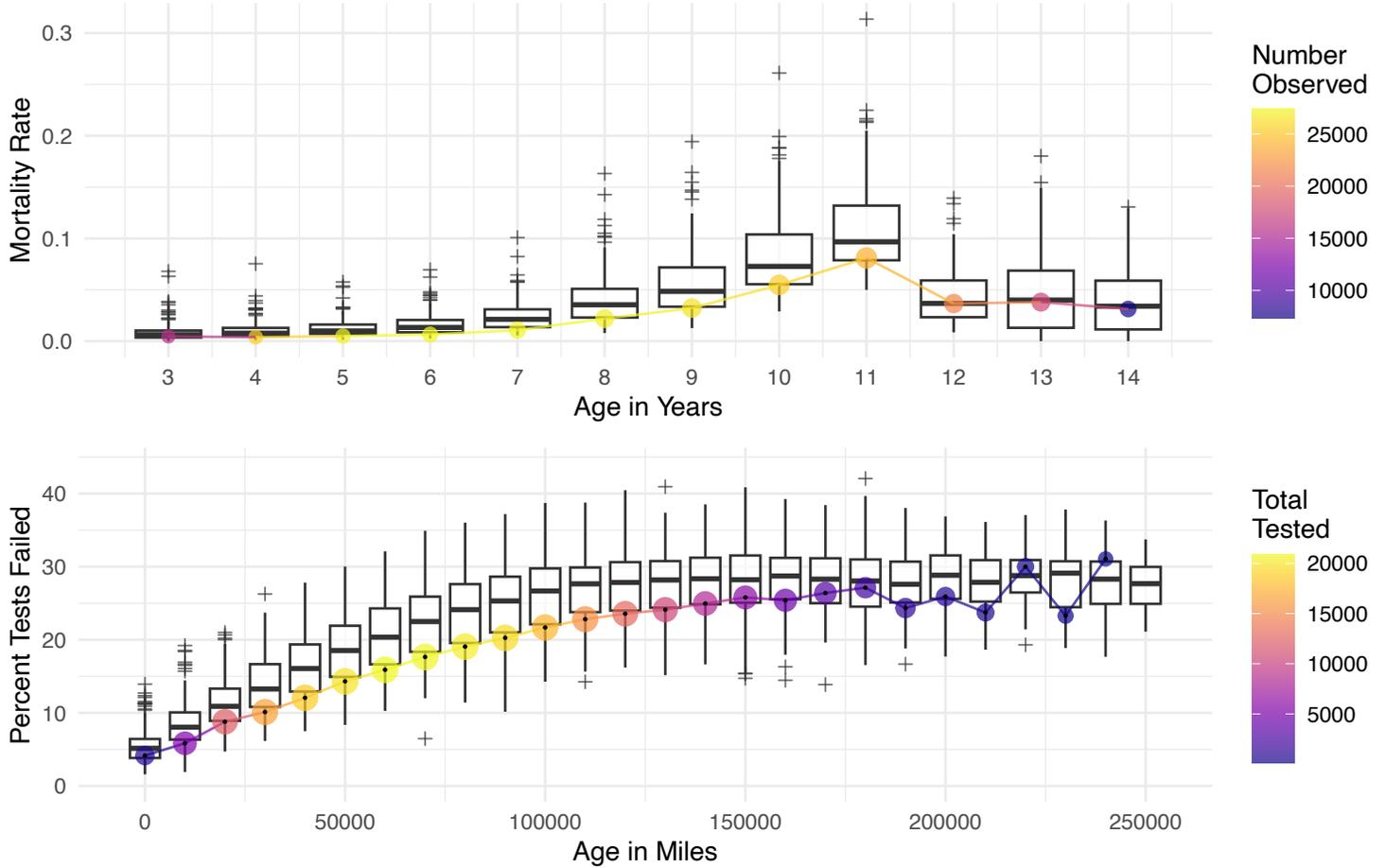

### Mortality rates

| Age in Years | Observed | Died | Mortality Rate |
|---|---|---|---|
| 3 | 15981 | 73 | 0.00457 |
| 4 | 25479 | 99 | 0.00389 |
| 5 | 27185 | 134 | 0.00493 |
| 6 | 27355 | 176 | 0.00643 |
| 7 | 27277 | 293 | 0.01070 |
| 8 | 26968 | 590 | 0.02190 |
| 9 | 26307 | 846 | 0.03220 |
| 10 | 25374 | 1389 | 0.05470 |
| 11 | 23910 | 1937 | 0.08100 |
| 12 | 21448 | 785 | 0.03660 |
| 13 | 17393 | 661 | 0.03800 |
| 14 | 7358 | 230 | 0.03130 |

### Mechanical Reliability Rates

| Mileage at test | N tested | Pct failed |
|---|---|---|
| 0 | 794 | 4.16 |
| 10000 | 5323 | 5.84 |
| 20000 | 12105 | 8.79 |
| 30000 | 16473 | 10.10 |
| 40000 | 18647 | 12.10 |
| 50000 | 19905 | 14.30 |
| 60000 | 20911 | 15.90 |
| 70000 | 20802 | 17.60 |
| 80000 | 20337 | 19.10 |
| 90000 | 19551 | 20.30 |
| 100000 | 17616 | 21.70 |
| 110000 | 15453 | 22.80 |
| 120000 | 12931 | 23.60 |
| 130000 | 10564 | 24.10 |
| 200000 | 791 | 25.90 |
| 210000 | 430 | 23.70 |
| 230000 | 189 | 23.30 |

## Audi A3 2008

At 5 years of age, the mortality rate of a Audi A3 2008 (manufactured as a Car or Light Van) ranked number 47 out of 218 vehicles of the same age and type (any Car or Light Van constructed in 2008). One is the lowest (or best) and 218 the highest mortality rate. For vehicles reaching 120000 miles, its unreliability score (rate of failing an inspection) ranked 155 out of 194 vehicles of the same age, type, and mileage. One is the highest (or worst) and 194 the lowest rate of failing an inspection.

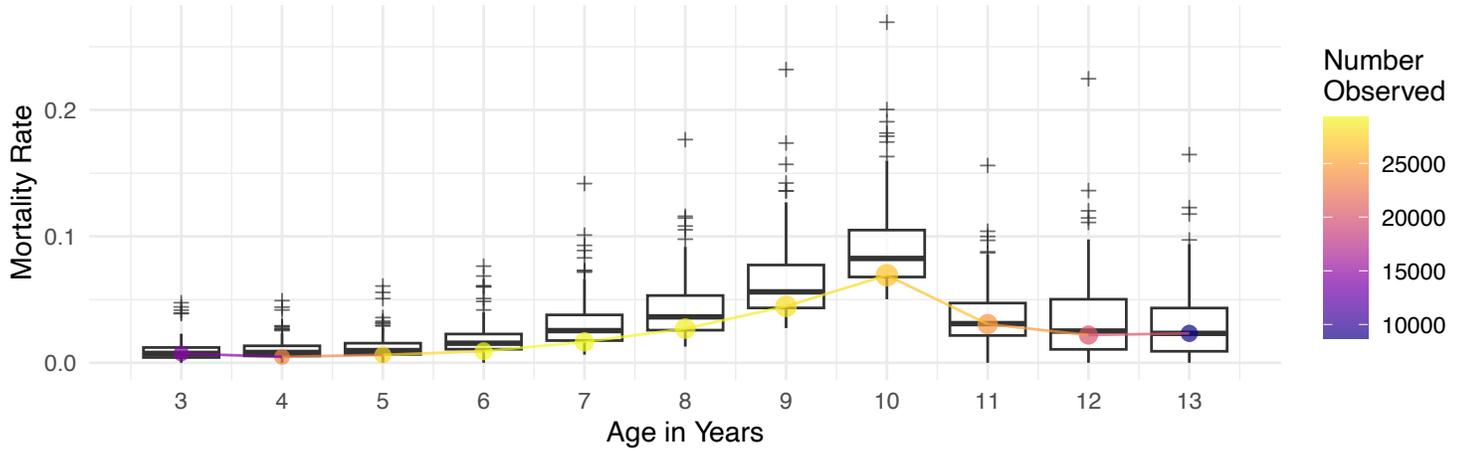

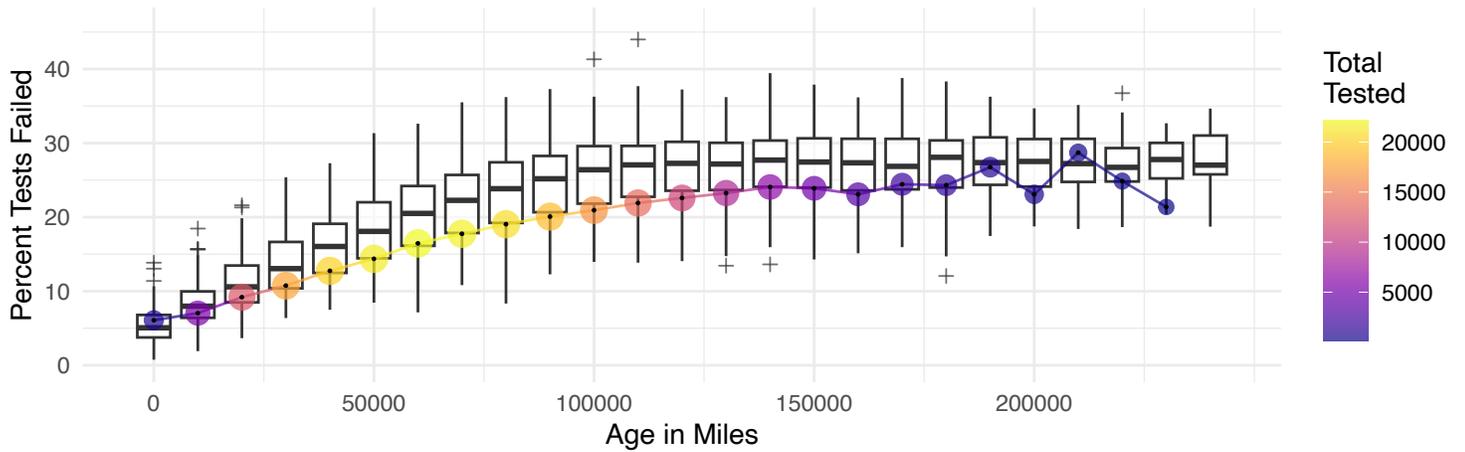

| Mechanical Reliability Rates | | |
| --- | --- | --- |
| Mileage at test | N tested | Pct failed |
| 0 | 675 | 6.07 |
| 10000 | 5444 | 7.04 |
| 20000 | 12627 | 9.19 |
| 30000 | 17713 | 10.80 |
| 40000 | 20186 | 12.80 |
| 50000 | 21522 | 14.40 |
| 60000 | 22191 | 16.50 |
| 70000 | 21778 | 17.80 |
| 80000 | 20841 | 19.00 |
| 90000 | 18974 | 20.10 |
| 100000 | 16565 | 20.90 |
| 110000 | 13946 | 21.90 |
| 120000 | 11150 | 22.60 |
| 130000 | 8828 | 23.30 |
| 140000 | 6580 | 24.10 |
| 150000 | 4788 | 23.90 |
| 160000 | 3205 | 23.10 |

| Mortality rates | | | |
| --- | --- | --- | --- |
| Age in Years | Observed | Died | Mortality Rate |
| 3 | 14280 | 105 | 0.00735 |
| 4 | 23147 | 109 | 0.00471 |
| 5 | 27612 | 173 | 0.00627 |
| 6 | 29259 | 267 | 0.00913 |
| 7 | 29215 | 483 | 0.01650 |
| 8 | 28795 | 778 | 0.02700 |
| 9 | 27961 | 1249 | 0.04470 |
| 10 | 26661 | 1846 | 0.06920 |
| 11 | 24260 | 742 | 0.03060 |
| 12 | 20035 | 440 | 0.02200 |
| 13 | 8715 | 203 | 0.02330 |

## Audi A3 2009

At 5 years of age, the mortality rate of a Audi A3 2009 (manufactured as a Car or Light Van) ranked number 65 out of 205 vehicles of the same age and type (any Car or Light Van constructed in 2009). One is the lowest (or best) and 205 the highest mortality rate. For vehicles reaching 20000 miles, its unreliability score (rate of failing an inspection) ranked 136 out of 200 vehicles of the same age, type, and mileage. One is the highest (or worst) and 200 the lowest rate of failing an inspection.

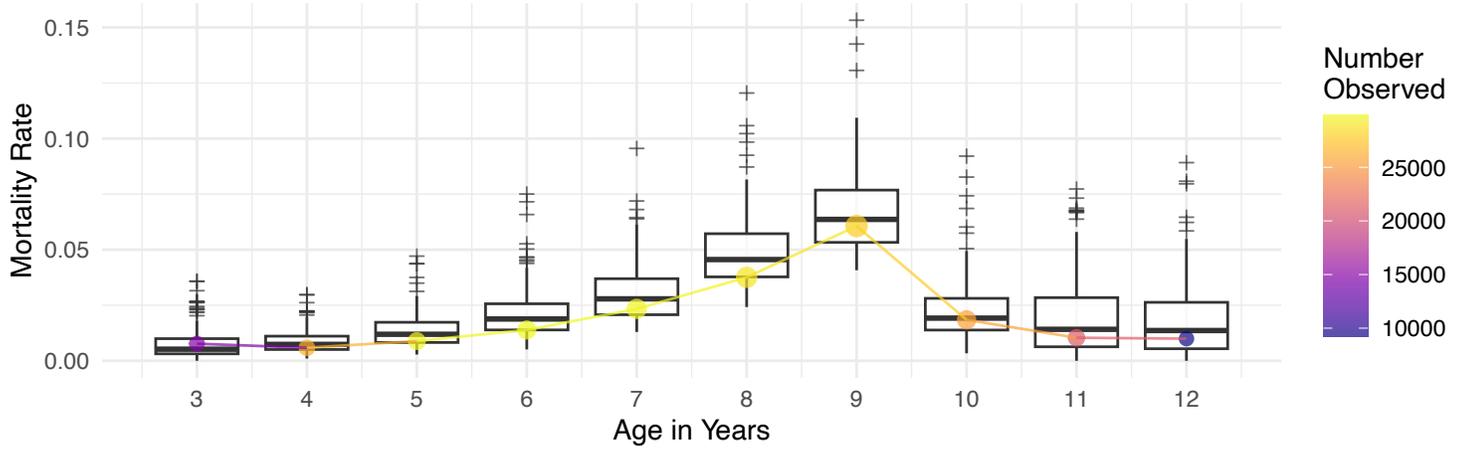

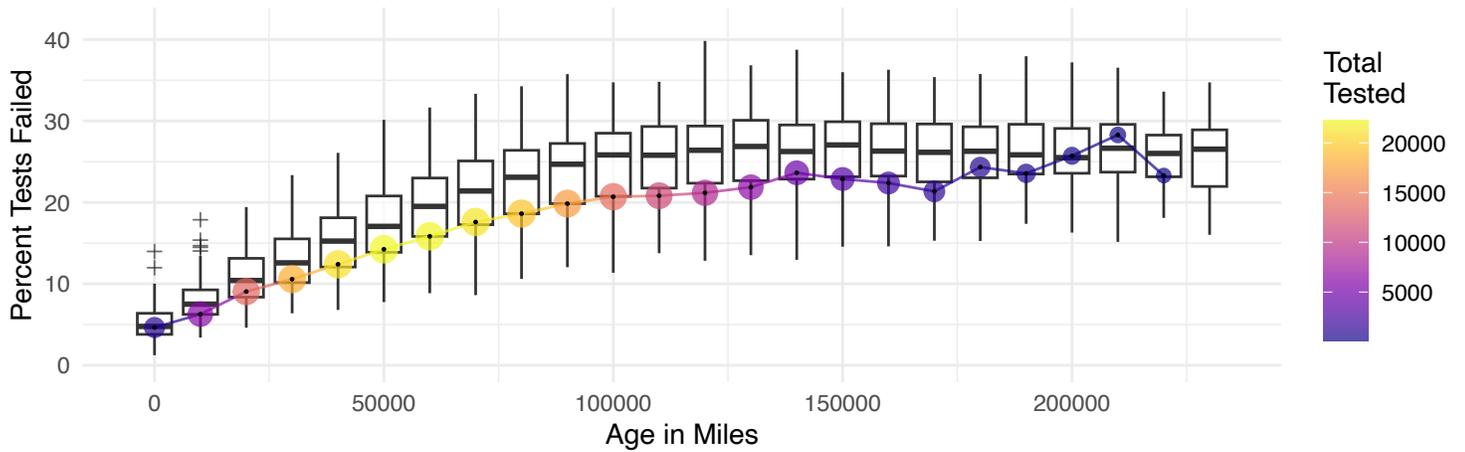

### Mortality rates

| Age in Years | Observed | Died | Mortality Rate |
|---|---|---|---|
| 3 | 14656 | 113 | 0.00771 |
| 4 | 26246 | 154 | 0.00587 |
| 5 | 29562 | 263 | 0.00890 |
| 6 | 29818 | 414 | 0.01390 |
| 7 | 29571 | 691 | 0.02340 |
| 8 | 28888 | 1085 | 0.03760 |
| 9 | 27768 | 1684 | 0.06060 |
| 10 | 25374 | 468 | 0.01840 |
| 11 | 21202 | 220 | 0.01040 |
| 12 | 9241 | 92 | 0.00996 |

### Mechanical Reliability Rates

| Mileage at test | N tested | Pct failed |
|---|---|---|
| 0 | 1014 | 4.64 |
| 10000 | 6312 | 6.27 |
| 20000 | 13626 | 9.05 |
| 30000 | 18344 | 10.60 |
| 40000 | 21072 | 12.40 |
| 50000 | 22281 | 14.20 |
| 60000 | 22075 | 15.80 |
| 70000 | 21249 | 17.60 |
| 80000 | 19474 | 18.60 |
| 90000 | 17044 | 19.90 |
| 100000 | 13968 | 20.70 |
| 110000 | 11322 | 20.80 |
| 120000 | 8896 | 21.20 |
| 130000 | 6498 | 21.90 |
| 140000 | 4684 | 23.60 |
| 150000 | 3187 | 22.90 |
| 160000 | 2073 | 22.40 |

## Audi A3 2010

At 5 years of age, the mortality rate of a Audi A3 2010 (manufactured as a Car or Light Van) ranked number 57 out of 206 vehicles of the same age and type (any Car or Light Van constructed in 2010). One is the lowest (or best) and 206 the highest mortality rate. For vehicles reaching 20000 miles, its unreliability score (rate of failing an inspection) ranked 141 out of 201 vehicles of the same age, type, and mileage. One is the highest (or worst) and 201 the lowest rate of failing an inspection.

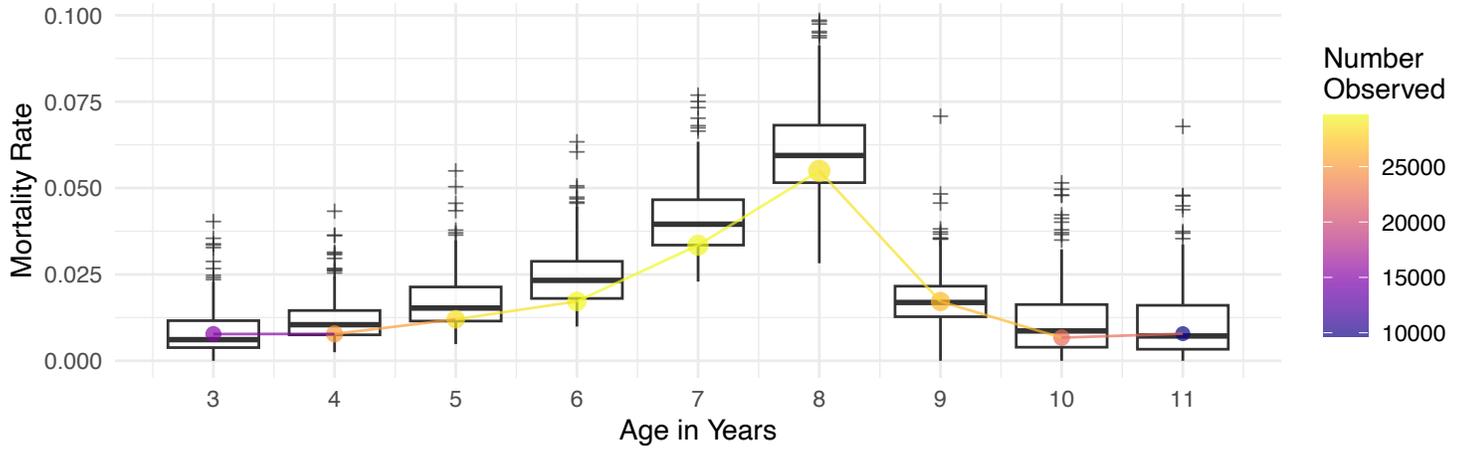

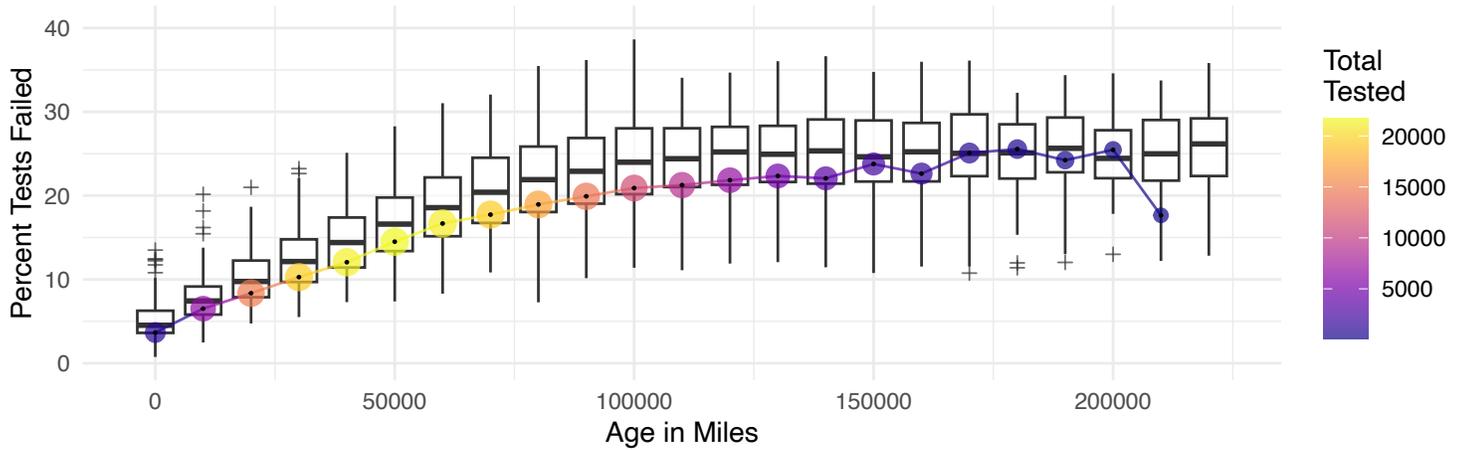

Mortality rates

| Age in Years | Observed | Died | Mortality Rate |
|---|---|---|---|
| 3 | 15103 | 117 | 0.00775 |
| 4 | 25055 | 195 | 0.00778 |
| 5 | 28535 | 343 | 0.01200 |
| 6 | 29598 | 509 | 0.01720 |
| 7 | 29471 | 986 | 0.03350 |
| 8 | 28594 | 1572 | 0.05500 |
| 9 | 26525 | 455 | 0.01720 |
| 10 | 22232 | 149 | 0.00670 |
| 11 | 9682 | 76 | 0.00785 |

Mechanical Reliability Rates

| Mileage at test | N tested | Pct failed |
|---|---|---|
| 0 | 928 | 3.66 |
| 10000 | 6647 | 6.51 |
| 20000 | 14784 | 8.36 |
| 30000 | 19649 | 10.30 |
| 40000 | 21358 | 12.10 |
| 50000 | 21753 | 14.50 |
| 60000 | 21165 | 16.70 |
| 70000 | 19585 | 17.70 |
| 80000 | 17140 | 18.90 |
| 90000 | 14239 | 19.90 |
| 100000 | 11268 | 20.90 |
| 110000 | 9033 | 21.30 |
| 120000 | 6818 | 21.80 |
| 130000 | 4762 | 22.30 |
| 140000 | 3377 | 22.10 |
| 150000 | 2246 | 23.80 |
| 160000 | 1485 | 22.60 |

## Audi A3 2011

At 5 years of age, the mortality rate of a Audi A3 2011 (manufactured as a Car or Light Van) ranked number 95 out of 211 vehicles of the same age and type (any Car or Light Van constructed in 2011). One is the lowest (or best) and 211 the highest mortality rate. For vehicles reaching 120000 miles, its unreliability score (rate of failing an inspection) ranked 127 out of 170 vehicles of the same age, type, and mileage. One is the highest (or worst) and 170 the lowest rate of failing an inspection.

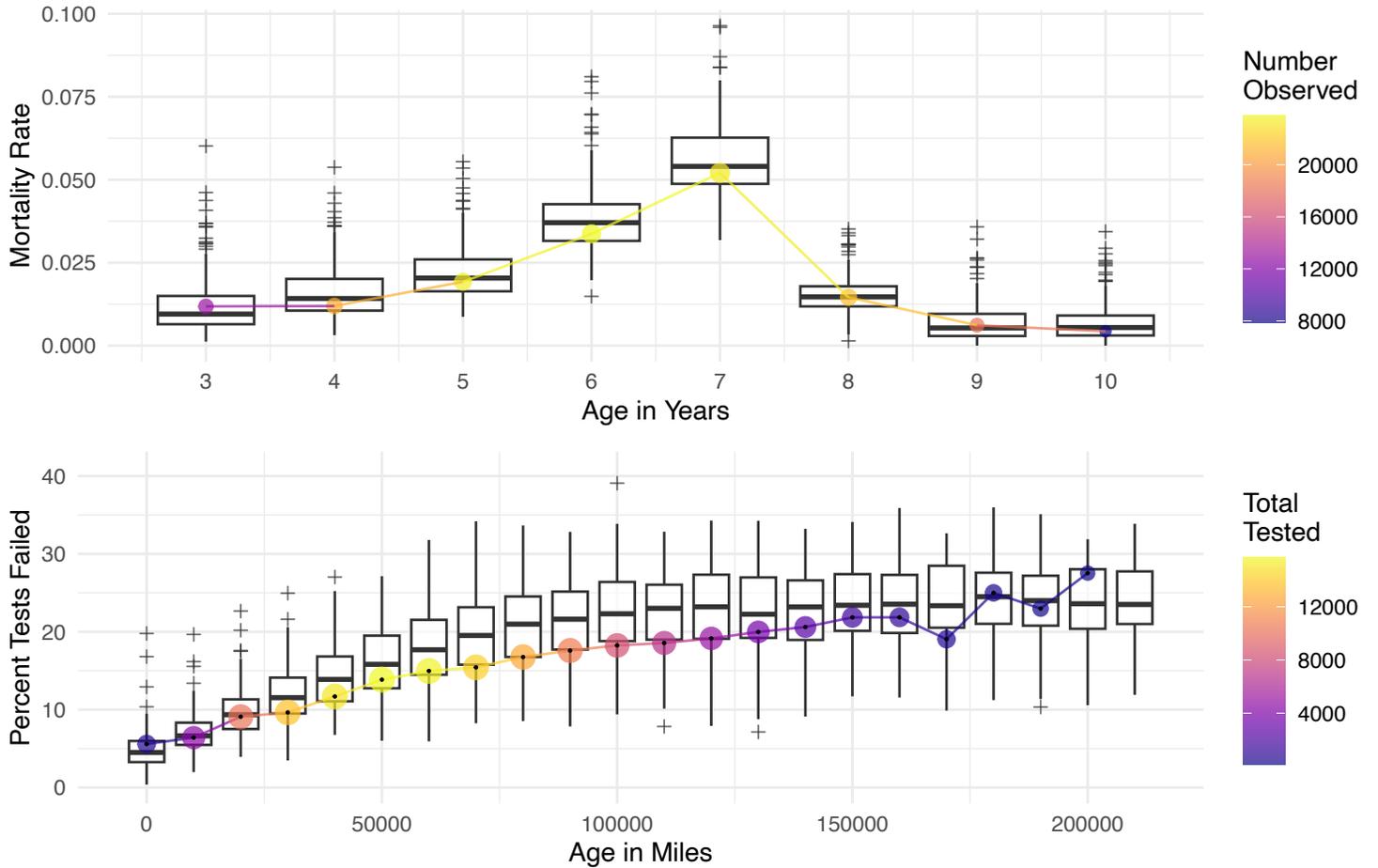

### Mortality rates

| Age in Years | Observed | Died | Mortality Rate |
|---|---|---|---|
| 3 | 13025 | 154 | 0.01180 |
| 4 | 20950 | 250 | 0.01190 |
| 5 | 23490 | 452 | 0.01920 |
| 6 | 23799 | 802 | 0.03370 |
| 7 | 23256 | 1211 | 0.05210 |
| 8 | 21642 | 315 | 0.01460 |
| 9 | 18194 | 112 | 0.00616 |
| 10 | 7870 | 34 | 0.00432 |

### Mechanical Reliability Rates

| Mileage at test | N tested | Pct failed |
|---|---|---|
| 0 | 608 | 5.59 |
| 10000 | 4524 | 6.41 |
| 20000 | 10280 | 9.10 |
| 30000 | 13669 | 9.65 |
| 40000 | 15071 | 11.70 |
| 50000 | 15724 | 13.90 |
| 60000 | 15451 | 15.00 |
| 70000 | 14248 | 15.40 |
| 80000 | 12632 | 16.80 |
| 90000 | 10658 | 17.60 |
| 100000 | 8466 | 18.20 |
| 110000 | 6782 | 18.60 |
| 120000 | 5117 | 19.20 |
| 130000 | 3618 | 20.00 |
| 140000 | 2479 | 20.60 |
| 150000 | 1524 | 21.90 |
| 160000 | 1002 | 21.90 |

**Audi A3 2012**

At 5 years of age, the mortality rate of a Audi A3 2012 (manufactured as a Car or Light Van) ranked number 66 out of 212 vehicles of the same age and type (any Car or Light Van constructed in 2012). One is the lowest (or best) and 212 the highest mortality rate. For vehicles reaching 20000 miles, its unreliability score (rate of failing an inspection) ranked 102 out of 206 vehicles of the same age, type, and mileage. One is the highest (or worst) and 206 the lowest rate of failing an inspection.

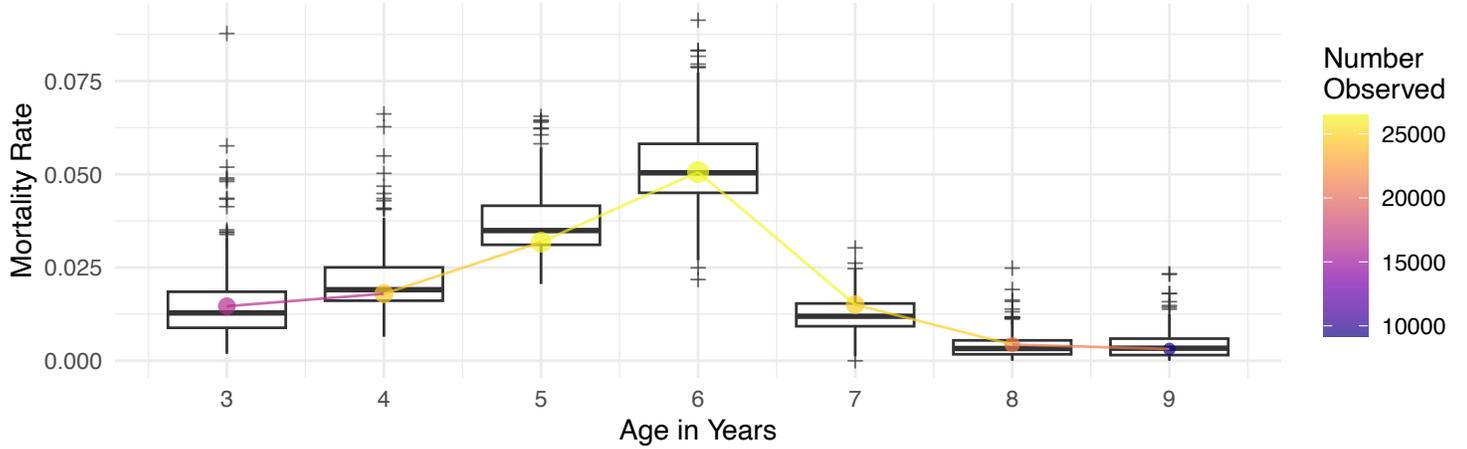

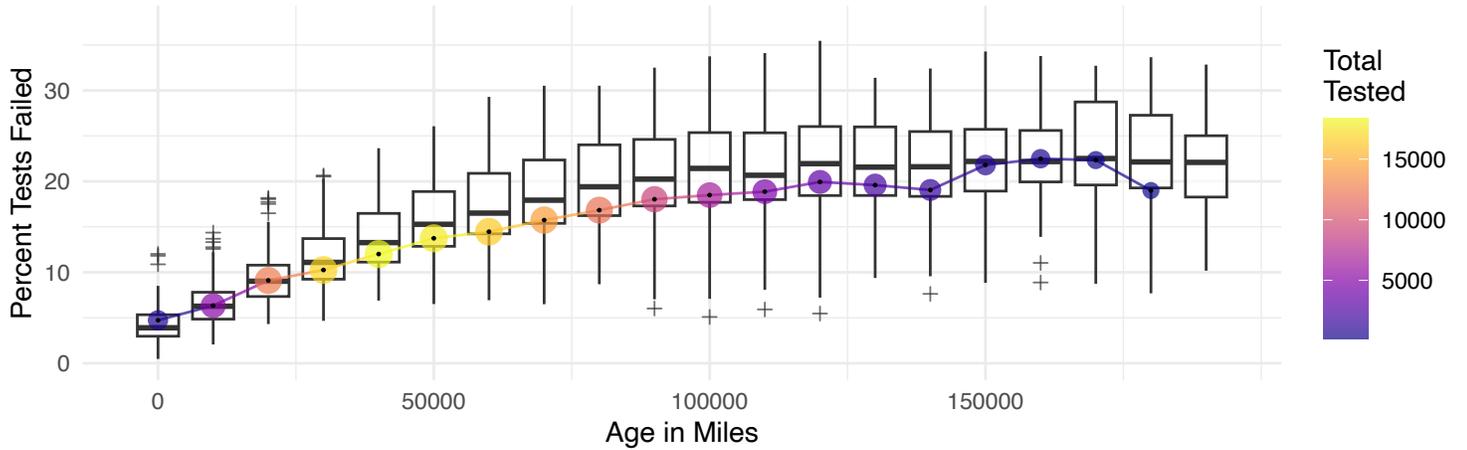

Mortality rates

| Age in Years | Observed | Died | Mortality Rate |
|---|---|---|---|
| 3 | 16330 | 239 | 0.01460 |
| 4 | 24462 | 440 | 0.01800 |
| 5 | 26344 | 838 | 0.03180 |
| 6 | 26433 | 1337 | 0.05060 |
| 7 | 24794 | 373 | 0.01500 |
| 8 | 20932 | 91 | 0.00435 |
| 9 | 9181 | 29 | 0.00316 |

Mechanical Reliability Rates

| Mileage at test | N tested | Pct failed |
|---|---|---|
| 0 | 698 | 4.73 |
| 10000 | 5160 | 6.34 |
| 20000 | 12563 | 9.11 |
| 30000 | 16878 | 10.30 |
| 40000 | 18392 | 12.00 |
| 50000 | 17806 | 13.70 |
| 60000 | 16354 | 14.50 |
| 70000 | 14310 | 15.70 |
| 80000 | 11879 | 16.80 |
| 90000 | 9010 | 18.00 |
| 100000 | 6716 | 18.50 |
| 110000 | 4674 | 18.90 |
| 120000 | 3280 | 19.90 |
| 130000 | 2205 | 19.60 |
| 140000 | 1417 | 19.10 |
| 150000 | 812 | 21.80 |
| 180000 | 179 | 19.00 |

## Audi A3 2013

At 5 years of age, the mortality rate of a Audi A3 2013 (manufactured as a Car or Light Van) ranked number 105 out of 221 vehicles of the same age and type (any Car or Light Van constructed in 2013). One is the lowest (or best) and 221 the highest mortality rate. For vehicles reaching 20000 miles, its unreliability score (rate of failing an inspection) ranked 185 out of 215 vehicles of the same age, type, and mileage. One is the highest (or worst) and 215 the lowest rate of failing an inspection.

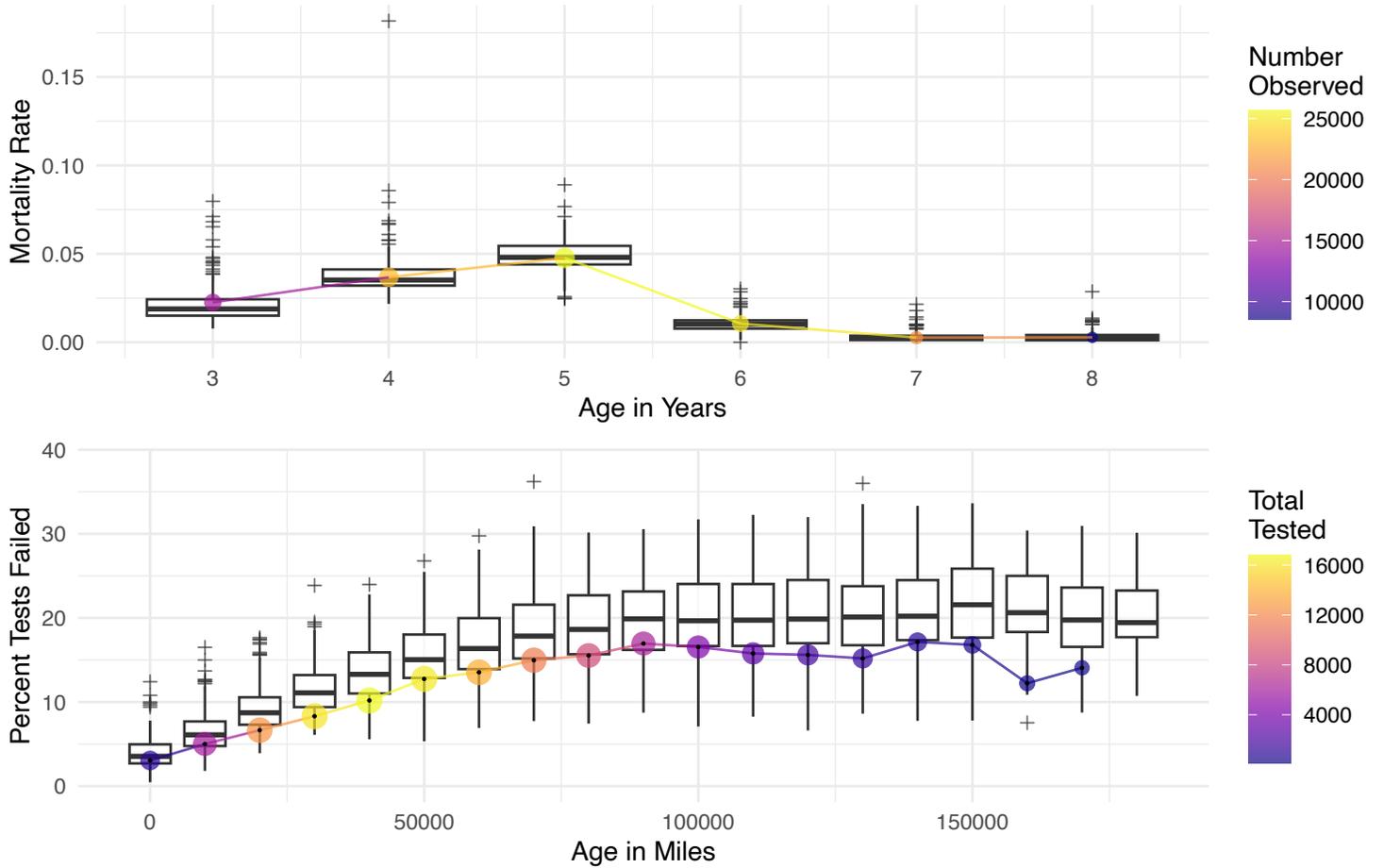

#### Mortality rates

| Age in Years | Observed | Died | Mortality Rate |
|---|---|---|---|
| 3 | 14505 | 327 | 0.02250 |
| 4 | 23096 | 849 | 0.03680 |
| 5 | 25622 | 1222 | 0.04770 |
| 6 | 24877 | 263 | 0.01060 |
| 7 | 20989 | 56 | 0.00267 |
| 8 | 8556 | 24 | 0.00281 |

#### Mechanical Reliability Rates

| Mileage at test | N tested | Pct failed |
|---|---|---|
| 0 | 885 | 3.05 |
| 10000 | 5947 | 5.01 |
| 20000 | 12567 | 6.65 |
| 30000 | 16070 | 8.31 |
| 40000 | 16819 | 10.20 |
| 50000 | 16277 | 12.70 |
| 60000 | 13882 | 13.50 |
| 70000 | 11413 | 15.00 |
| 80000 | 8645 | 15.50 |
| 90000 | 6257 | 17.00 |
| 100000 | 4268 | 16.50 |
| 110000 | 2839 | 15.80 |
| 120000 | 1775 | 15.60 |
| 130000 | 1159 | 15.20 |
| 140000 | 694 | 17.10 |
| 150000 | 387 | 16.80 |
| 160000 | 188 | 12.20 |

## Audi A3 2014

At 5 years of age, the mortality rate of a Audi A3 2014 (manufactured as a Car or Light Van) ranked number 153 out of 236 vehicles of the same age and type (any Car or Light Van constructed in 2014). One is the lowest (or best) and 236 the highest mortality rate. For vehicles reaching 20000 miles, its unreliability score (rate of failing an inspection) ranked 186 out of 230 vehicles of the same age, type, and mileage. One is the highest (or worst) and 230 the lowest rate of failing an inspection.

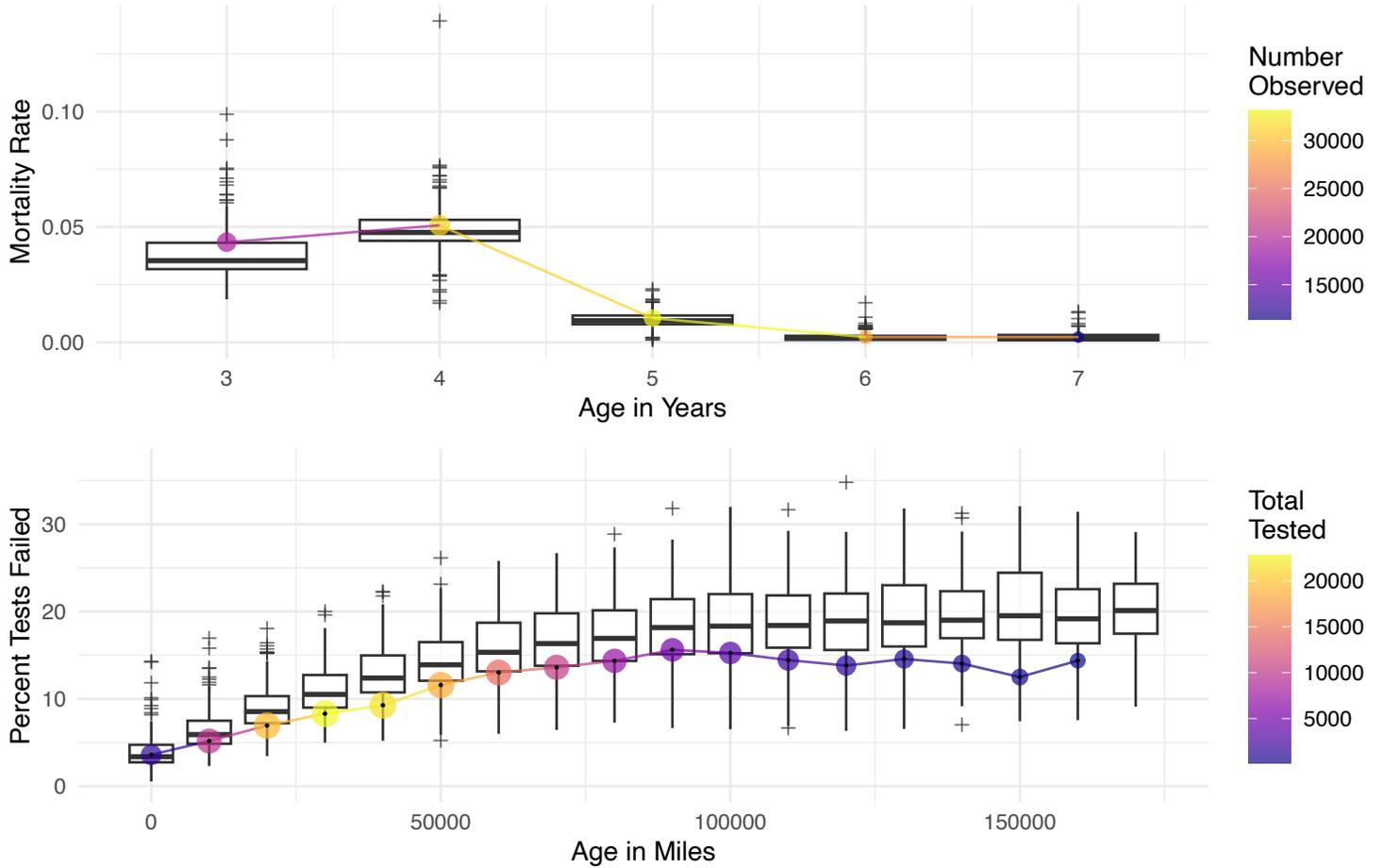

### Mortality rates

| Age in Years | Observed | Died | Mortality Rate |
|---|---|---|---|
| 3 | 19226 | 834 | 0.04340 |
| 4 | 31104 | 1577 | 0.05070 |
| 5 | 33050 | 346 | 0.01050 |
| 6 | 28593 | 65 | 0.00227 |
| 7 | 11412 | 26 | 0.00228 |

### Mechanical Reliability Rates

| Mileage at test | N tested | Pct failed |
|---|---|---|
| 0 | 1502 | 3.60 |
| 10000 | 9880 | 5.17 |
| 20000 | 19345 | 6.95 |
| 30000 | 22819 | 8.30 |
| 40000 | 21665 | 9.27 |
| 50000 | 18488 | 11.60 |
| 60000 | 14135 | 13.00 |
| 70000 | 10506 | 13.60 |
| 80000 | 7282 | 14.40 |
| 90000 | 4989 | 15.60 |
| 100000 | 3219 | 15.30 |
| 110000 | 1925 | 14.40 |
| 120000 | 1238 | 13.80 |
| 130000 | 776 | 14.60 |
| 140000 | 406 | 14.00 |
| 150000 | 256 | 12.50 |
| 160000 | 160 | 14.40 |

**Audi A3 2015**

At 5 years of age, the mortality rate of a Audi A3 2015 (manufactured as a Car or Light Van) ranked number 138 out of 247 vehicles of the same age and type (any Car or Light Van constructed in 2015). One is the lowest (or best) and 247 the highest mortality rate. For vehicles reaching 20000 miles, its unreliability score (rate of failing an inspection) ranked 189 out of 241 vehicles of the same age, type, and mileage. One is the highest (or worst) and 241 the lowest rate of failing an inspection.

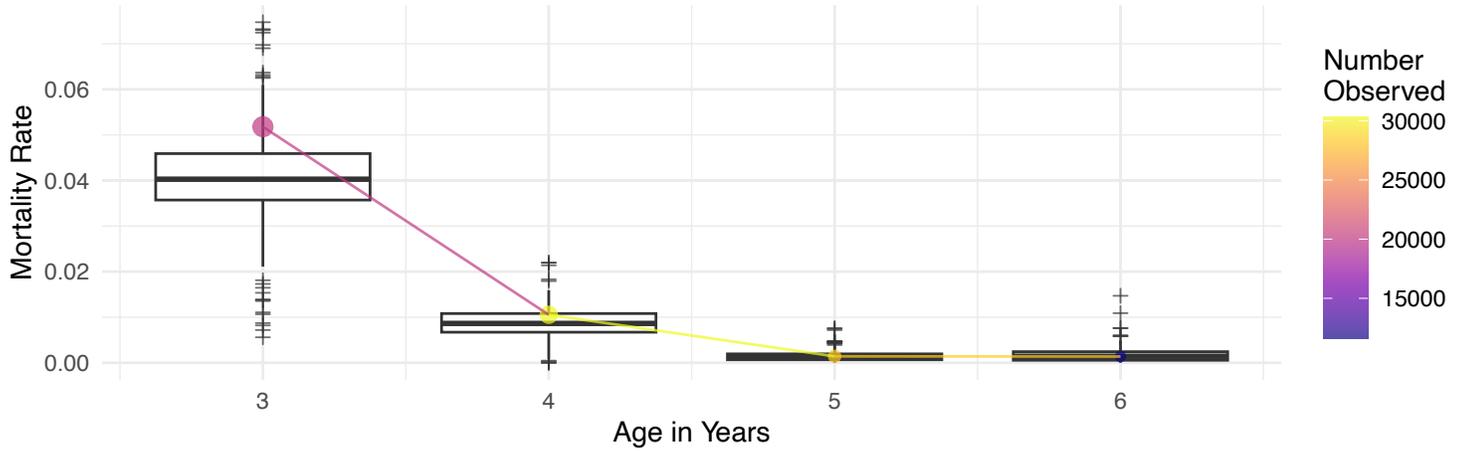

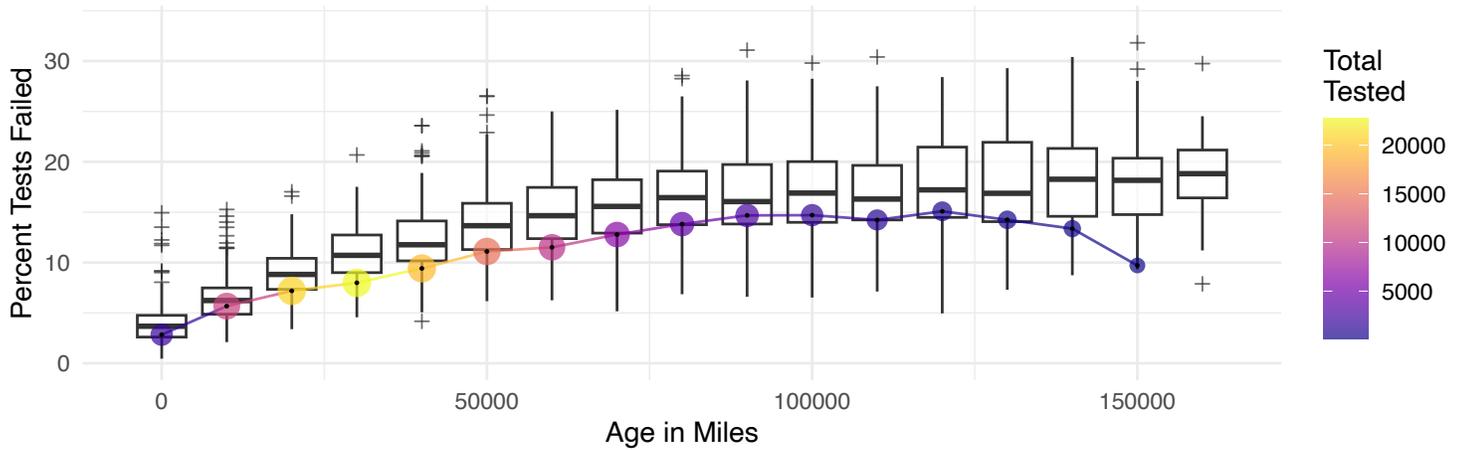

Mortality rates

| Age in Years | Observed | Died | Mortality Rate |
|---|---|---|---|
| 3 | 20060 | 1039 | 0.05180 |
| 4 | 30264 | 319 | 0.01050 |
| 5 | 28217 | 40 | 0.00142 |
| 6 | 11622 | 16 | 0.00138 |

Mechanical Reliability Rates

| Mileage at test | N tested | Pct failed |
|---|---|---|
| 0 | 1727 | 2.84 |
| 10000 | 11021 | 5.67 |
| 20000 | 20828 | 7.19 |
| 30000 | 22782 | 7.99 |
| 40000 | 19393 | 9.42 |
| 50000 | 14569 | 11.10 |
| 60000 | 9886 | 11.50 |
| 70000 | 6533 | 12.80 |
| 80000 | 4046 | 13.80 |
| 90000 | 2704 | 14.70 |
| 100000 | 1639 | 14.70 |
| 110000 | 1019 | 14.20 |
| 120000 | 570 | 15.10 |
| 130000 | 344 | 14.20 |
| 140000 | 202 | 13.40 |
| 150000 | 103 | 9.71 |

**Audi A3 2016**

At 5 years of age, the mortality rate of a Audi A3 2016 (manufactured as a Car or Light Van) ranked number 196 out of 252 vehicles of the same age and type (any Car or Light Van constructed in 2016). One is the lowest (or best) and 252 the highest mortality rate. For vehicles reaching 20000 miles, its unreliability score (rate of failing an inspection) ranked 176 out of 246 vehicles of the same age, type, and mileage. One is the highest (or worst) and 246 the lowest rate of failing an inspection.

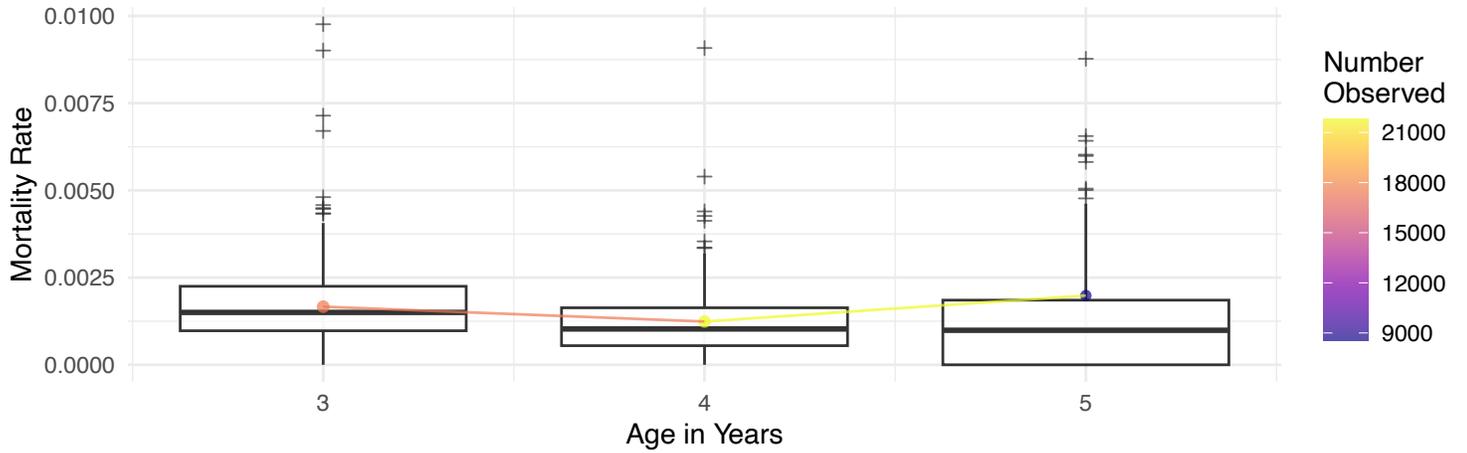

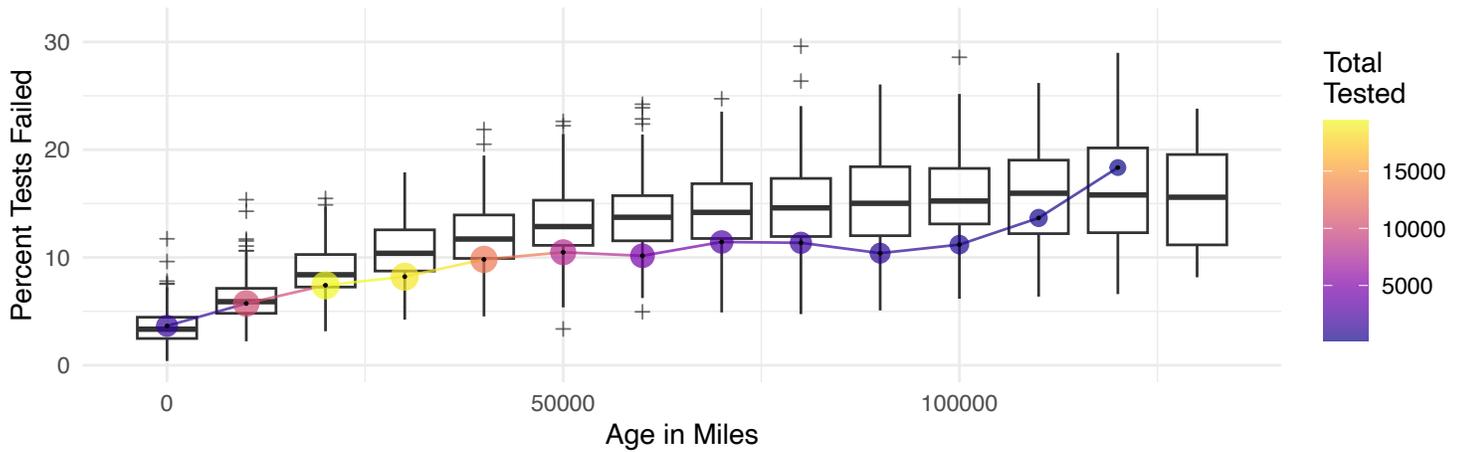

Mortality rates

| Age in Years | Observed | Died | Mortality Rate |
|--------------|----------|------|----------------|
| 3 | 17406 | 29 | 0.00167 |
| 4 | 21767 | 27 | 0.00124 |
| 5 | 8560 | 17 | 0.00199 |

Mechanical Reliability Rates

| Mileage at test | N tested | Pct failed |
|-----------------|----------|------------|
| 0 | 1452 | 3.65 |
| 10000 | 10297 | 5.73 |
| 20000 | 19460 | 7.42 |
| 30000 | 18622 | 8.22 |
| 40000 | 12930 | 9.82 |
| 50000 | 7986 | 10.50 |
| 60000 | 4529 | 10.20 |
| 70000 | 2710 | 11.40 |
| 80000 | 1611 | 11.40 |
| 90000 | 952 | 10.40 |
| 100000 | 554 | 11.20 |
| 110000 | 300 | 13.70 |
| 120000 | 169 | 18.30 |

**Audi A3 2017**

At 3 years of age, the mortality rate of a Audi A3 2017 (manufactured as a Car or Light Van) ranked number 159 out of 247 vehicles of the same age and type (any Car or Light Van constructed in 2017). One is the lowest (or best) and 247 the highest mortality rate. For vehicles reaching 20000 miles, its unreliability score (rate of failing an inspection) ranked 137 out of 240 vehicles of the same age, type, and mileage. One is the highest (or worst) and 240 the lowest rate of failing an inspection.

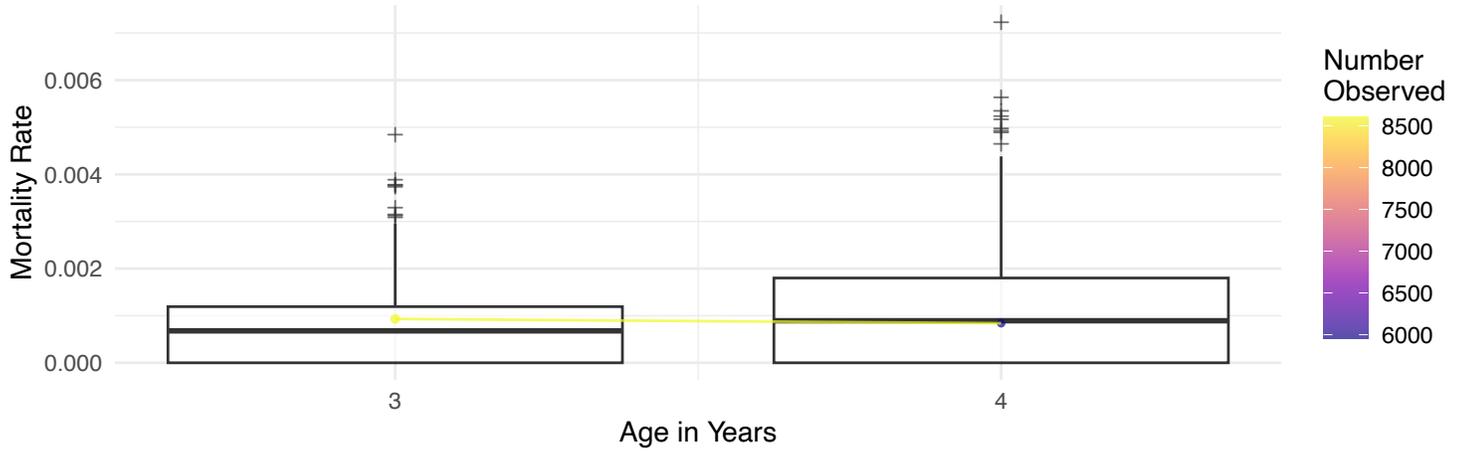

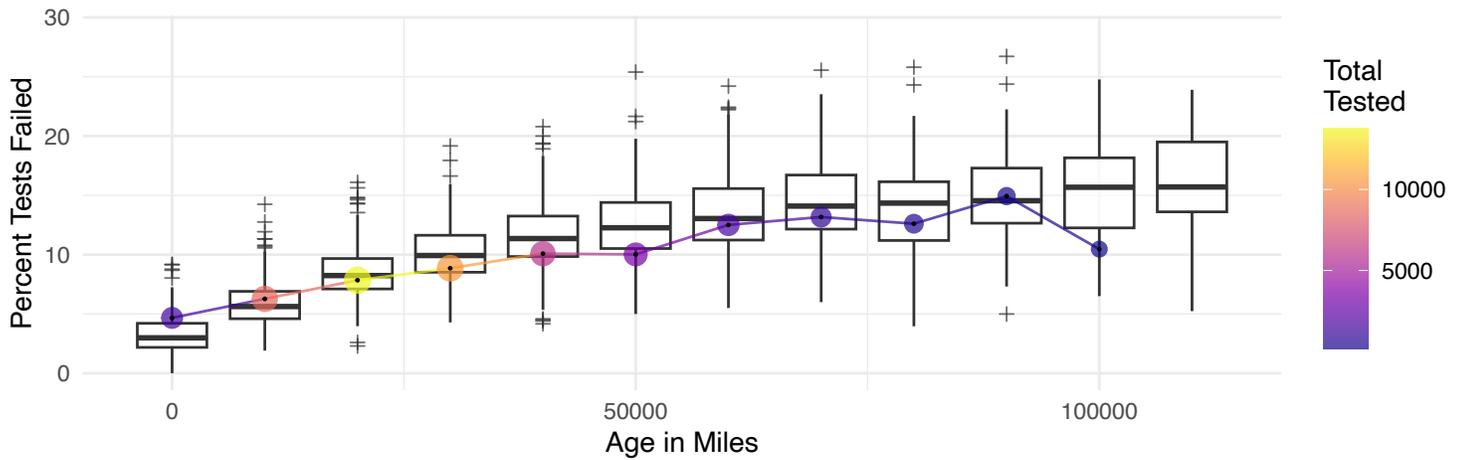

Mortality rates

| Age in Years | Observed | Died | Mortality Rate |
|---|---|---|---|
| 3 | 8597 | 8 | 0.000931 |
| 4 | 5957 | 5 | 0.000839 |

Mechanical Reliability Rates

| Mileage at test | N tested | Pct failed |
|---|---|---|
| 0 | 1435 | 4.67 |
| 10000 | 8485 | 6.27 |
| 20000 | 13782 | 7.84 |
| 30000 | 10526 | 8.85 |
| 40000 | 5969 | 10.10 |
| 50000 | 3313 | 10.00 |
| 60000 | 1711 | 12.50 |
| 70000 | 964 | 13.20 |
| 80000 | 579 | 12.60 |
| 90000 | 308 | 14.90 |
| 100000 | 172 | 10.50 |

**Audi A3 2018**

At 3 years of age, the mortality rate of a Audi A3 2018 (manufactured as a Car or Light Van) ranked number 2 out of 222 vehicles of the same age and type (any Car or Light Van constructed in 2018). One is the lowest (or best) and 222 the highest mortality rate. For vehicles reaching 20000 miles, its unreliability score (rate of failing an inspection) ranked 81 out of 215 vehicles of the same age, type, and mileage. One is the highest (or worst) and 215 the lowest rate of failing an inspection.

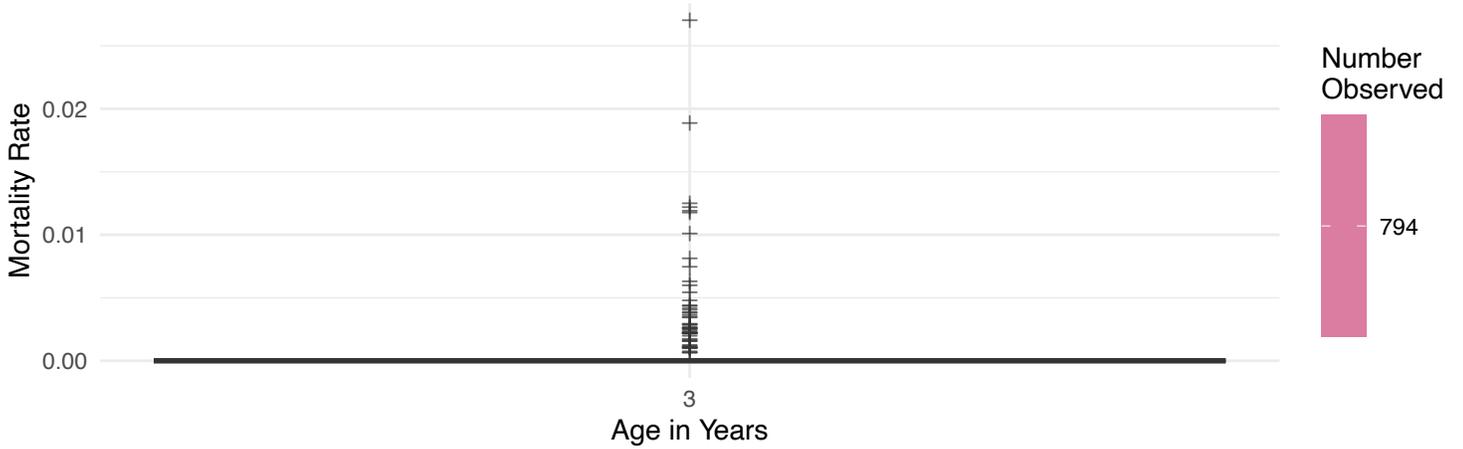

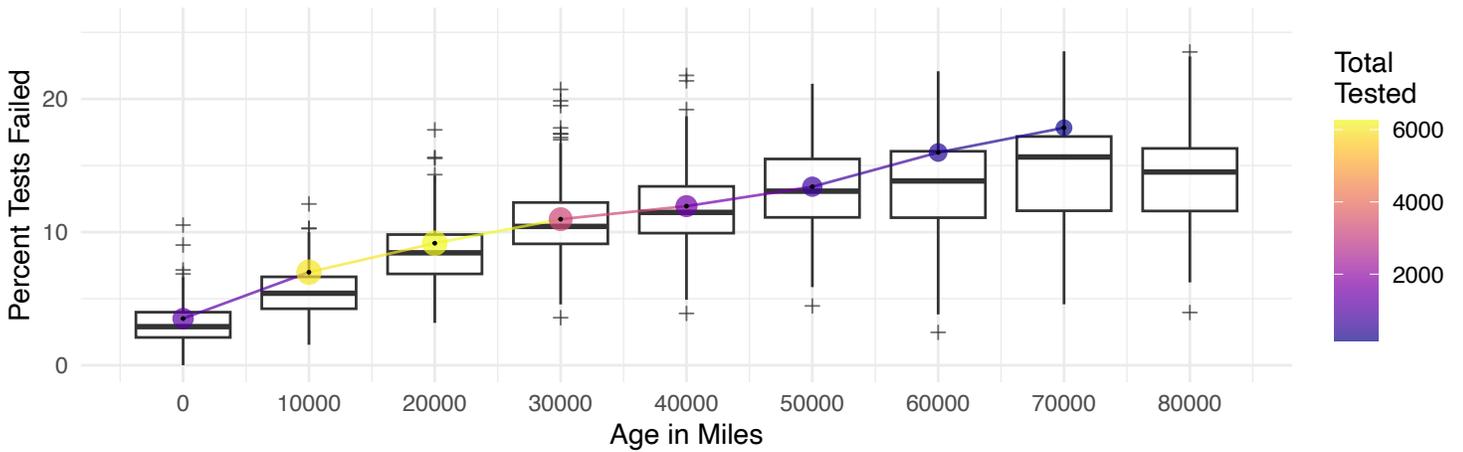

Mortality rates

| Age in Years | Observed | Died | Mortality Rate |
|---|---|---|---|
| 3 | 794 | 0 | 0 |

Mechanical Reliability Rates

| Mileage at test | N tested | Pct failed |
|---|---|---|
| 0 | 1112 | 3.51 |
| 10000 | 5980 | 6.99 |
| 20000 | 6264 | 9.16 |
| 30000 | 3218 | 11.00 |
| 40000 | 1473 | 11.90 |
| 50000 | 716 | 13.40 |
| 60000 | 363 | 16.00 |
| 70000 | 157 | 17.80 |

## Audi A4 1995

At 10 years of age, the mortality rate of a Audi A4 1995 (manufactured as a Car or Light Van) ranked number 56 out of 148 vehicles of the same age and type (any Car or Light Van constructed in 1995). One is the lowest (or best) and 148 the highest mortality rate. For vehicles reaching 120000 miles, its unreliability score (rate of failing an inspection) ranked 52 out of 135 vehicles of the same age, type, and mileage. One is the highest (or worst) and 135 the lowest rate of failing an inspection.

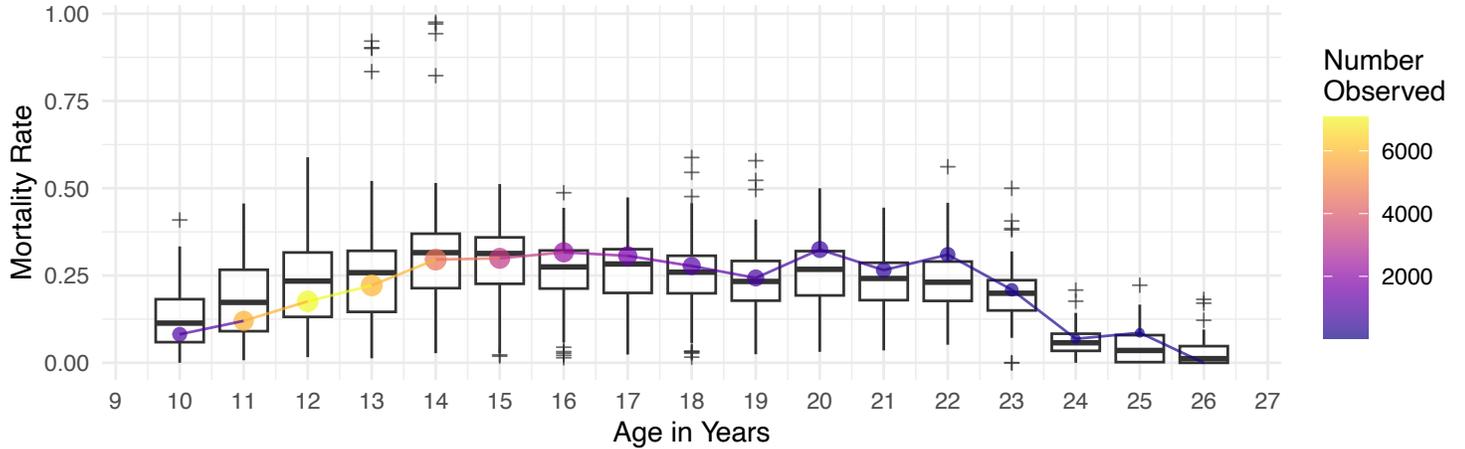

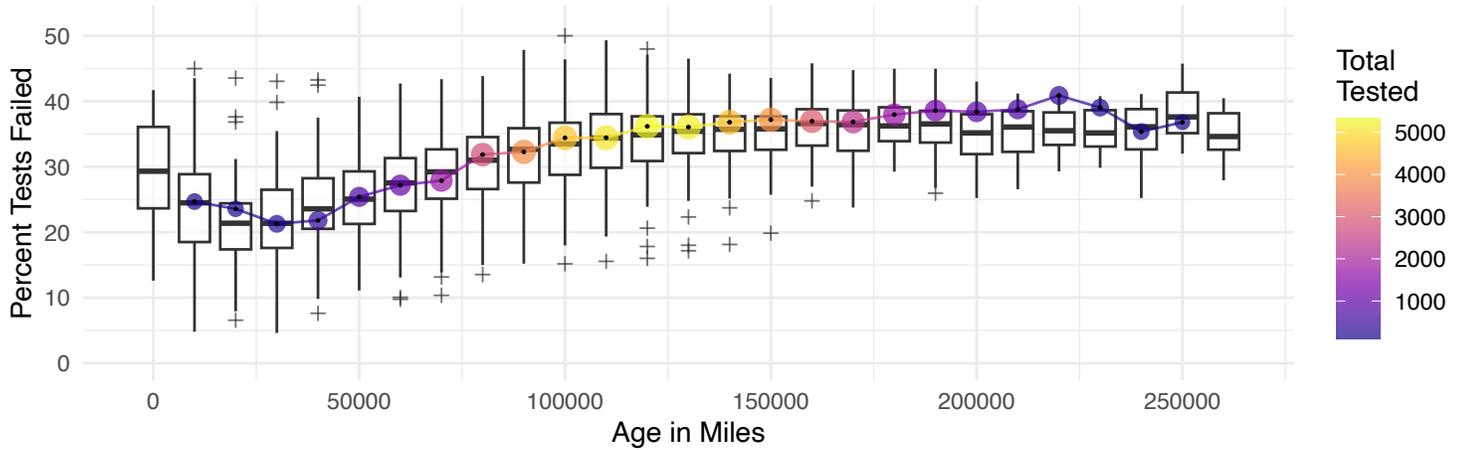

| Mortality rates | | | |
|---|---|---|---|
| Age in Years | Observed | Died | Mortality Rate |
| 10 | 919 | 75 | 0.0816 |
| 11 | 5844 | 702 | 0.1200 |
| 12 | 7058 | 1242 | 0.1760 |
| 13 | 5910 | 1310 | 0.2220 |
| 14 | 4569 | 1351 | 0.2960 |
| 15 | 3191 | 955 | 0.2990 |
| 16 | 2221 | 703 | 0.3170 |
| 17 | 1511 | 463 | 0.3060 |
| 18 | 1046 | 290 | 0.2770 |
| 19 | 757 | 184 | 0.2430 |
| 20 | 573 | 186 | 0.3250 |
| 21 | 388 | 103 | 0.2650 |
| 22 | 284 | 88 | 0.3100 |
| 23 | 196 | 41 | 0.2090 |
| 24 | 131 | 9 | 0.0687 |
| 25 | 93 | 8 | 0.0860 |
| 26 | 25 | 0 | 0.0000 |

| Mechanical Reliability Rates | | |
|---|---|---|
| Mileage at test | N tested | Pct failed |
| 10000 | 158 | 24.7 |
| 20000 | 140 | 23.6 |
| 100000 | 4690 | 34.4 |
| 110000 | 5106 | 34.4 |
| 120000 | 5336 | 36.2 |
| 130000 | 5212 | 36.1 |
| 140000 | 4557 | 36.8 |
| 150000 | 3975 | 37.2 |
| 160000 | 3074 | 37.0 |
| 170000 | 2382 | 36.9 |
| 180000 | 1720 | 38.0 |
| 190000 | 1309 | 38.6 |
| 210000 | 632 | 38.8 |
| 220000 | 406 | 40.9 |
| 230000 | 246 | 39.0 |
| 240000 | 164 | 35.4 |
| 250000 | 106 | 36.8 |

## Audi A4 1996

At 10 years of age, the mortality rate of a Audi A4 1996 (manufactured as a Car or Light Van) ranked number 45 out of 162 vehicles of the same age and type (any Car or Light Van constructed in 1996). One is the lowest (or best) and 162 the highest mortality rate. For vehicles reaching 120000 miles, its unreliability score (rate of failing an inspection) ranked 66 out of 147 vehicles of the same age, type, and mileage. One is the highest (or worst) and 147 the lowest rate of failing an inspection.

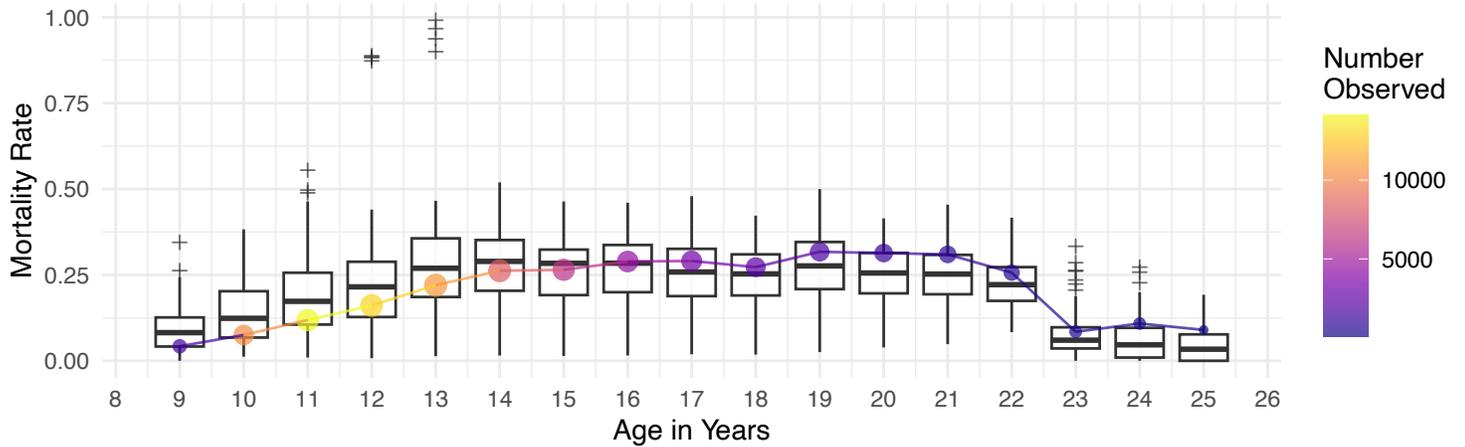

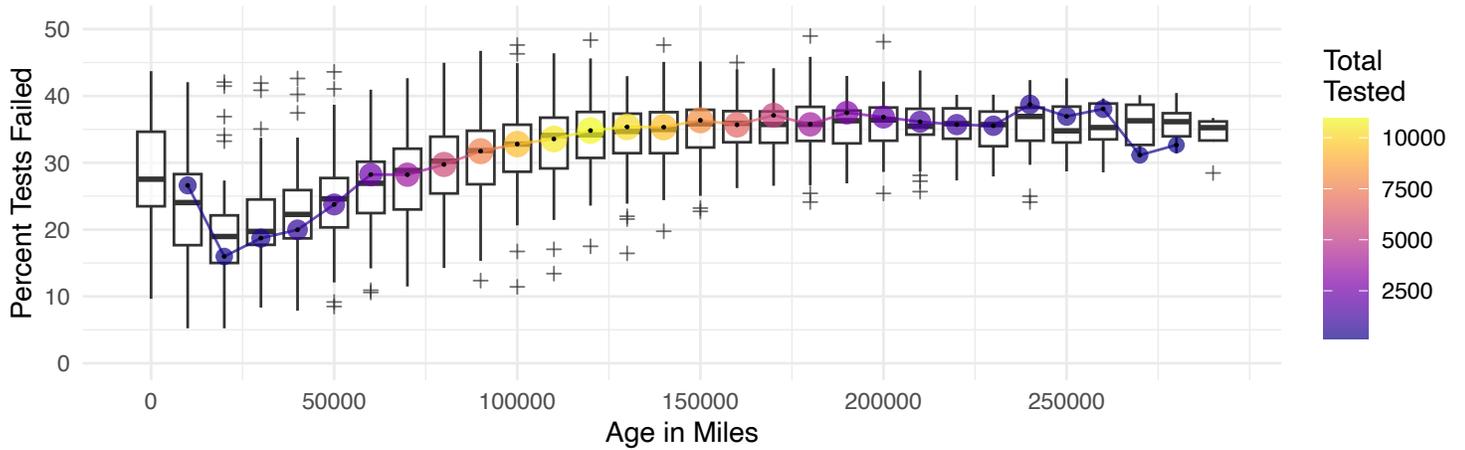

| Mortality rates | | | |
|---|---|---|---|
| Age in Years | Observed | Died | Mortality Rate |
| 9 | 1442 | 61 | 0.0423 |
| 10 | 10296 | 774 | 0.0752 |
| 11 | 14091 | 1672 | 0.1190 |
| 12 | 12860 | 2072 | 0.1610 |
| 13 | 10723 | 2360 | 0.2200 |
| 14 | 8296 | 2175 | 0.2620 |
| 15 | 6090 | 1614 | 0.2650 |
| 16 | 4463 | 1290 | 0.2890 |
| 17 | 3169 | 921 | 0.2910 |
| 18 | 2249 | 613 | 0.2730 |
| 19 | 1632 | 518 | 0.3170 |
| 20 | 1116 | 350 | 0.3140 |
| 21 | 765 | 237 | 0.3100 |
| 22 | 526 | 135 | 0.2570 |
| 23 | 367 | 31 | 0.0845 |
| 24 | 258 | 28 | 0.1090 |
| 25 | 89 | 8 | 0.0899 |

| Mechanical Reliability Rates | | |
|---|---|---|
| Mileage at test | N tested | Pct failed |
| 10000 | 278 | 26.6 |
| 20000 | 219 | 16.0 |
| 100000 | 9430 | 32.8 |
| 110000 | 10583 | 33.6 |
| 120000 | 10963 | 34.8 |
| 130000 | 10259 | 35.4 |
| 140000 | 9479 | 35.3 |
| 150000 | 8091 | 36.4 |
| 160000 | 6694 | 35.7 |
| 170000 | 5304 | 37.1 |
| 180000 | 3969 | 35.8 |
| 190000 | 2945 | 37.5 |
| 210000 | 1450 | 36.1 |
| 220000 | 1078 | 35.7 |
| 230000 | 754 | 35.5 |
| 240000 | 563 | 38.7 |
| 250000 | 406 | 36.9 |

## Audi A4 1997

At 10 years of age, the mortality rate of a Audi A4 1997 (manufactured as a Car or Light Van) ranked number 58 out of 187 vehicles of the same age and type (any Car or Light Van constructed in 1997). One is the lowest (or best) and 187 the highest mortality rate. For vehicles reaching 120000 miles, its unreliability score (rate of failing an inspection) ranked 79 out of 167 vehicles of the same age, type, and mileage. One is the highest (or worst) and 167 the lowest rate of failing an inspection.

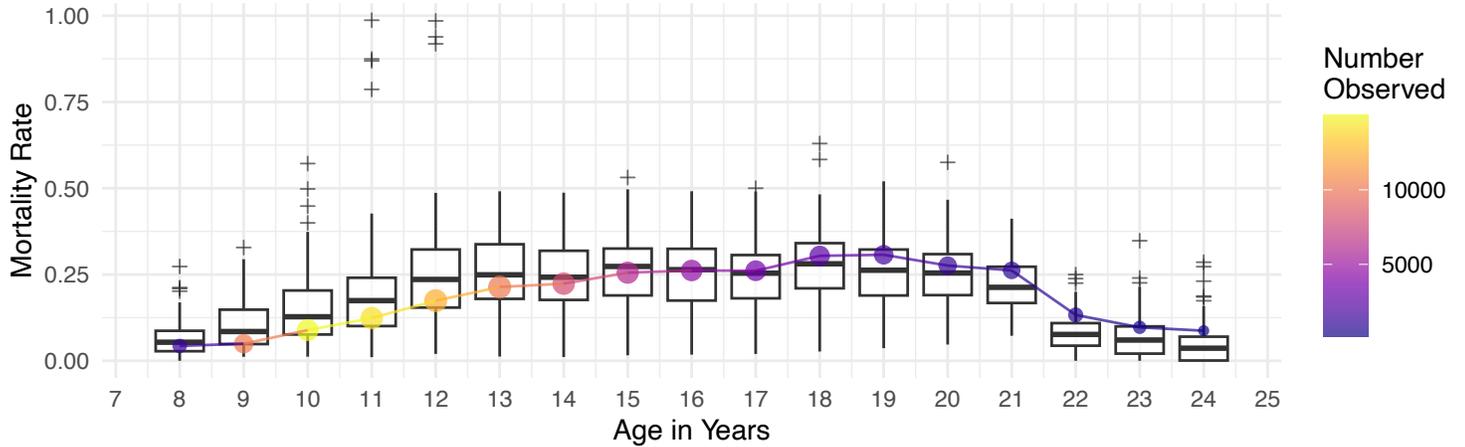

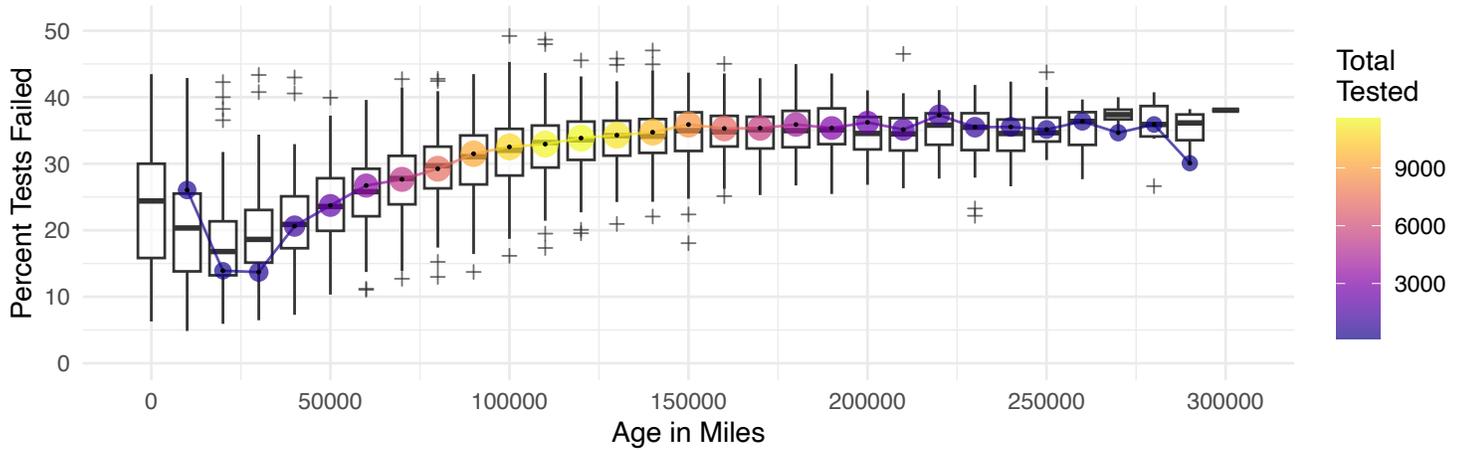

| Mortality rates | | | |
|---|---|---|---|
| Age in Years | Observed | Died | Mortality Rate |
| 8 | 1395 | 60 | 0.0430 |
| 9 | 10198 | 505 | 0.0495 |
| 10 | 14979 | 1332 | 0.0889 |
| 11 | 14174 | 1752 | 0.1240 |
| 12 | 12350 | 2151 | 0.1740 |
| 13 | 10124 | 2164 | 0.2140 |
| 14 | 7912 | 1769 | 0.2240 |
| 15 | 6121 | 1563 | 0.2550 |
| 16 | 4537 | 1187 | 0.2620 |
| 17 | 3342 | 872 | 0.2610 |
| 18 | 2463 | 749 | 0.3040 |
| 19 | 1712 | 526 | 0.3070 |
| 20 | 1185 | 327 | 0.2760 |
| 21 | 855 | 224 | 0.2620 |
| 22 | 602 | 80 | 0.1330 |
| 23 | 392 | 38 | 0.0969 |
| 24 | 172 | 15 | 0.0872 |

| Mechanical Reliability Rates | | |
|---|---|---|
| Mileage at test | N tested | Pct failed |
| 10000 | 334 | 26.0 |
| 20000 | 223 | 13.9 |
| 100000 | 10683 | 32.5 |
| 110000 | 11500 | 32.9 |
| 120000 | 11583 | 33.9 |
| 130000 | 11046 | 34.3 |
| 140000 | 9961 | 34.7 |
| 150000 | 8679 | 35.9 |
| 160000 | 7054 | 35.3 |
| 170000 | 5463 | 35.3 |
| 180000 | 4185 | 35.9 |
| 190000 | 3102 | 35.4 |
| 210000 | 1722 | 35.1 |
| 220000 | 1239 | 37.3 |
| 230000 | 837 | 35.5 |
| 240000 | 591 | 35.5 |
| 250000 | 387 | 35.1 |

## Audi A4 1998

At 10 years of age, the mortality rate of a Audi A4 1998 (manufactured as a Car or Light Van) ranked number 65 out of 196 vehicles of the same age and type (any Car or Light Van constructed in 1998). One is the lowest (or best) and 196 the highest mortality rate. For vehicles reaching 120000 miles, its unreliability score (rate of failing an inspection) ranked 108 out of 172 vehicles of the same age, type, and mileage. One is the highest (or worst) and 172 the lowest rate of failing an inspection.

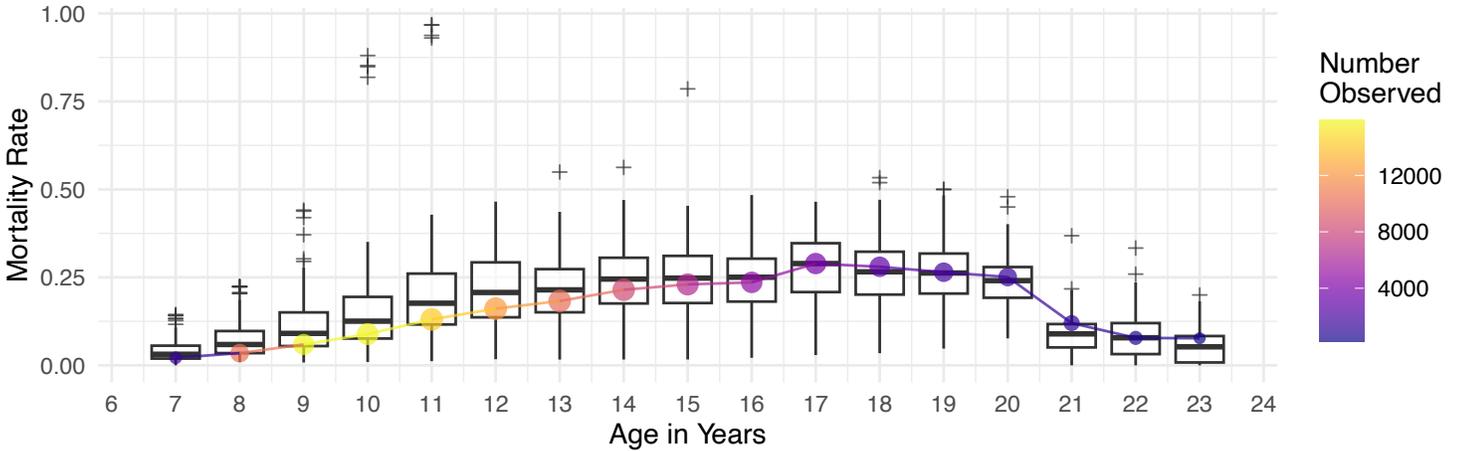

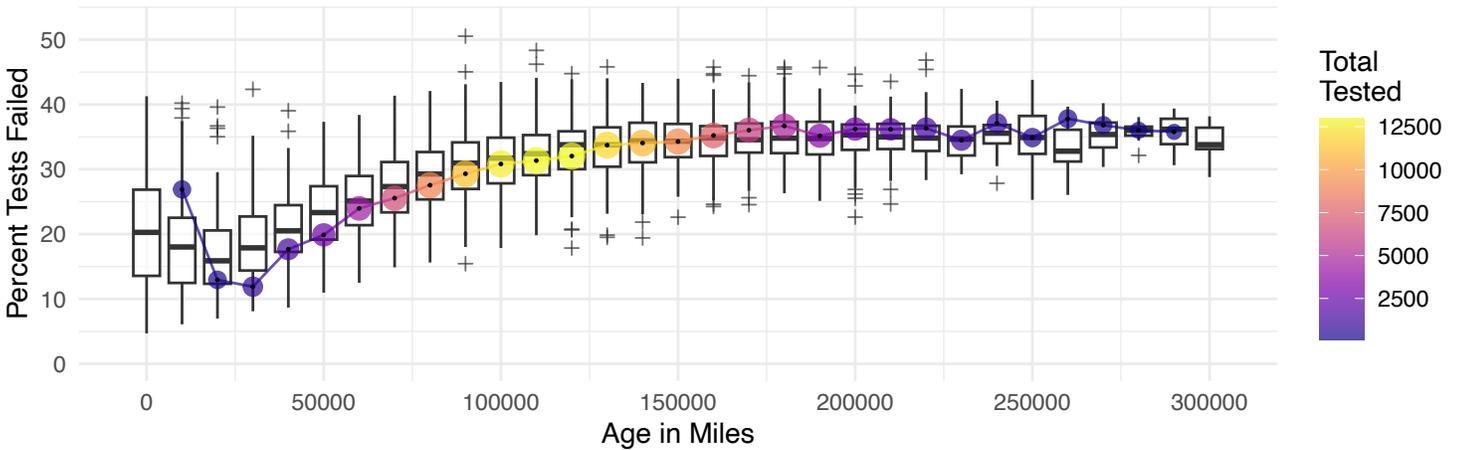

| Mortality rates | | | |
|---|---|---|---|
| Age in Years | Observed | Died | Mortality Rate |
| 7 | 1343 | 29 | 0.0216 |
| 8 | 10388 | 361 | 0.0348 |
| 9 | 15909 | 945 | 0.0594 |
| 10 | 15702 | 1396 | 0.0889 |
| 11 | 14244 | 1860 | 0.1310 |
| 12 | 12299 | 1981 | 0.1610 |
| 13 | 10262 | 1877 | 0.1830 |
| 14 | 8352 | 1794 | 0.2150 |
| 15 | 6533 | 1503 | 0.2300 |
| 16 | 5004 | 1180 | 0.2360 |
| 17 | 3803 | 1100 | 0.2890 |
| 18 | 2699 | 756 | 0.2800 |
| 19 | 1938 | 513 | 0.2650 |
| 20 | 1422 | 357 | 0.2510 |
| 21 | 1016 | 122 | 0.1200 |
| 22 | 679 | 53 | 0.0781 |
| 23 | 273 | 21 | 0.0769 |

| Mechanical Reliability Rates | | |
|---|---|---|
| Mileage at test | N tested | Pct failed |
| 10000 | 316 | 26.9 |
| 20000 | 363 | 12.9 |
| 100000 | 12629 | 30.8 |
| 110000 | 12958 | 31.3 |
| 120000 | 12966 | 32.0 |
| 130000 | 12043 | 33.7 |
| 140000 | 10931 | 34.0 |
| 150000 | 9248 | 34.3 |
| 160000 | 7592 | 35.2 |
| 170000 | 6110 | 36.0 |
| 180000 | 4810 | 36.7 |
| 190000 | 3531 | 35.2 |
| 210000 | 1844 | 36.2 |
| 220000 | 1363 | 36.3 |
| 230000 | 1044 | 34.5 |
| 240000 | 736 | 37.1 |
| 250000 | 513 | 34.9 |

## Audi A4 1999

At 10 years of age, the mortality rate of a Audi A4 1999 (manufactured as a Car or Light Van) ranked number 51 out of 201 vehicles of the same age and type (any Car or Light Van constructed in 1999). One is the lowest (or best) and 201 the highest mortality rate. For vehicles reaching 120000 miles, its unreliability score (rate of failing an inspection) ranked 101 out of 181 vehicles of the same age, type, and mileage. One is the highest (or worst) and 181 the lowest rate of failing an inspection.

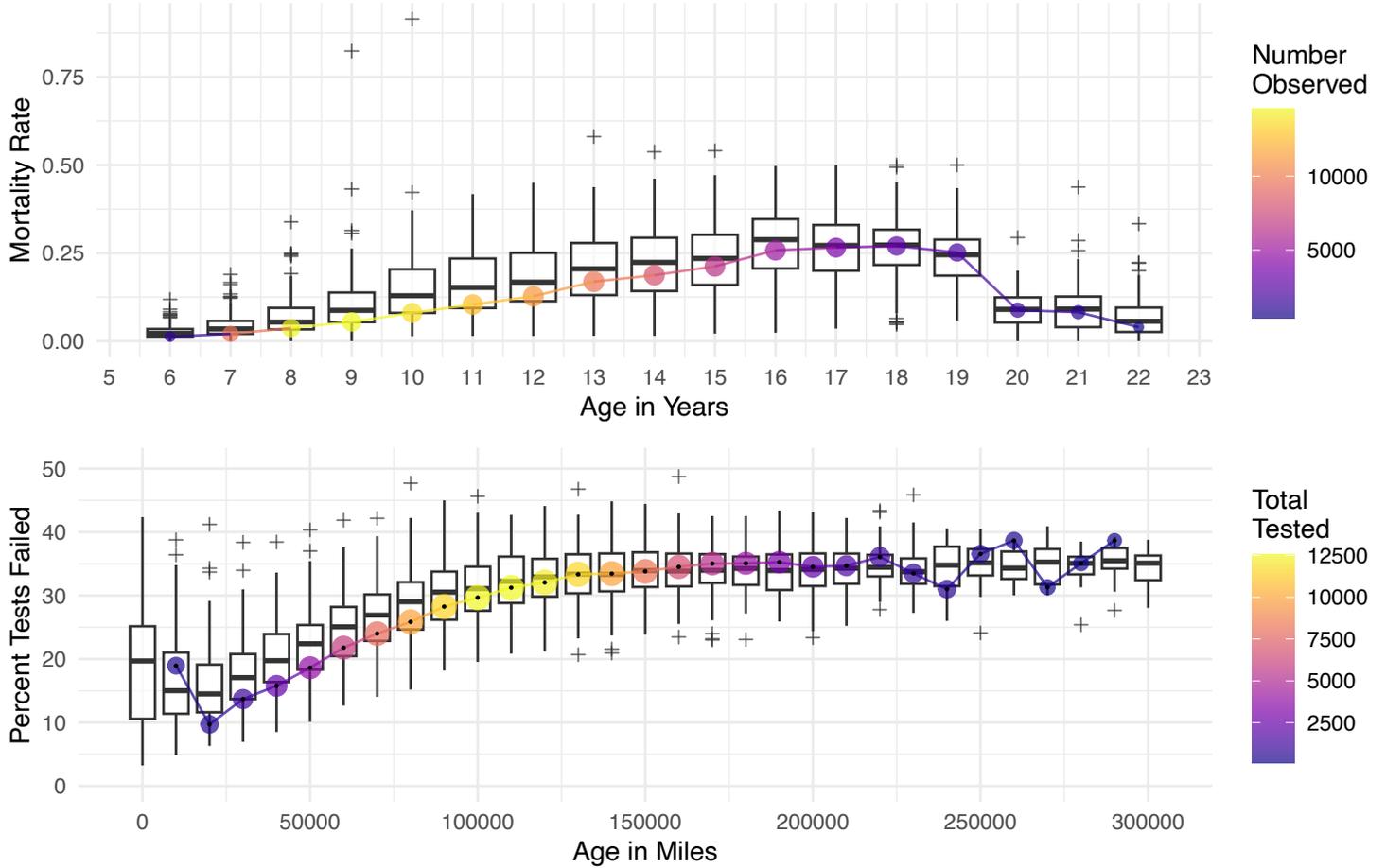

| Mortality rates | | | |
|---|---|---|---|
| Age in Years | Observed | Died | Mortality Rate |
| 6 | 1321 | 17 | 0.0129 |
| 7 | 9571 | 197 | 0.0206 |
| 8 | 14366 | 536 | 0.0373 |
| 9 | 14518 | 793 | 0.0546 |
| 10 | 13690 | 1096 | 0.0801 |
| 11 | 12500 | 1299 | 0.1040 |
| 12 | 11122 | 1416 | 0.1270 |
| 13 | 9631 | 1627 | 0.1690 |
| 14 | 7961 | 1491 | 0.1870 |
| 15 | 6434 | 1366 | 0.2120 |
| 16 | 5028 | 1297 | 0.2580 |
| 17 | 3703 | 983 | 0.2650 |
| 18 | 2713 | 733 | 0.2700 |
| 19 | 1974 | 497 | 0.2520 |
| 20 | 1386 | 122 | 0.0880 |
| 21 | 986 | 81 | 0.0822 |
| 22 | 403 | 16 | 0.0397 |

| Mechanical Reliability Rates | | |
|---|---|---|
| Mileage at test | N tested | Pct failed |
| 10000 | 401 | 19.0 |
| 20000 | 536 | 9.7 |
| 100000 | 12515 | 29.7 |
| 110000 | 12567 | 31.2 |
| 120000 | 12282 | 32.1 |
| 130000 | 11564 | 33.4 |
| 140000 | 9945 | 33.5 |
| 150000 | 8327 | 33.8 |
| 160000 | 6759 | 34.5 |
| 170000 | 5302 | 35.0 |
| 180000 | 4073 | 35.1 |
| 190000 | 3142 | 35.3 |
| 210000 | 1631 | 34.7 |
| 220000 | 1164 | 36.1 |
| 230000 | 842 | 33.5 |
| 240000 | 603 | 31.0 |
| 250000 | 454 | 36.6 |

## Audi A4 2000

At 5 years of age, the mortality rate of a Audi A4 2000 (manufactured as a Car or Light Van) ranked number 63 out of 198 vehicles of the same age and type (any Car or Light Van constructed in 2000). One is the lowest (or best) and 198 the highest mortality rate. For vehicles reaching 120000 miles, its unreliability score (rate of failing an inspection) ranked 120 out of 184 vehicles of the same age, type, and mileage. One is the highest (or worst) and 184 the lowest rate of failing an inspection.

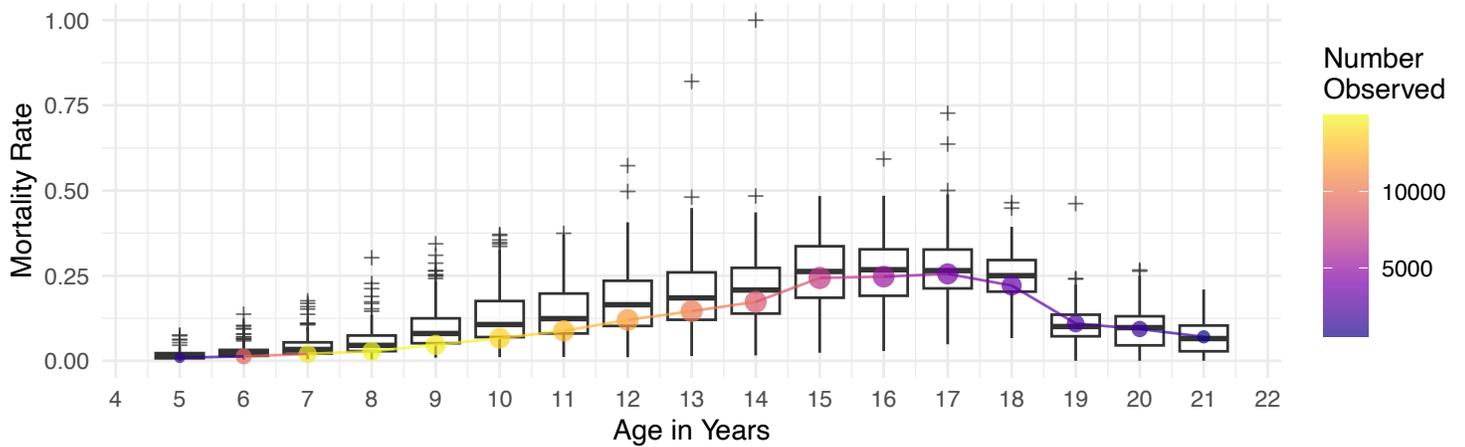

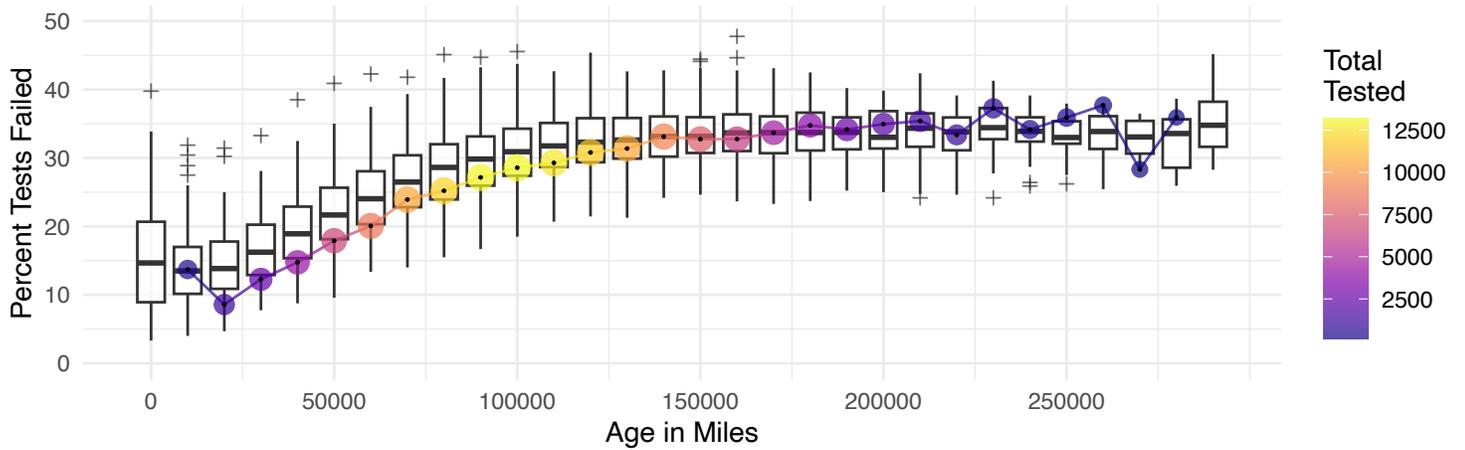

| Mortality rates | | | |
|---|---|---|---|
| Age in Years | Observed | Died | Mortality Rate |
| 5 | 1441 | 13 | 0.00902 |
| 6 | 9386 | 122 | 0.01300 |
| 7 | 14464 | 293 | 0.02030 |
| 8 | 14962 | 417 | 0.02790 |
| 9 | 14524 | 679 | 0.04680 |
| 10 | 13748 | 912 | 0.06630 |
| 11 | 12761 | 1121 | 0.08780 |
| 12 | 11552 | 1388 | 0.12000 |
| 13 | 10111 | 1474 | 0.14600 |
| 14 | 8588 | 1486 | 0.17300 |
| 15 | 7037 | 1712 | 0.24300 |
| 16 | 5293 | 1308 | 0.24700 |
| 17 | 3961 | 1011 | 0.25500 |
| 18 | 2945 | 651 | 0.22100 |
| 19 | 2183 | 239 | 0.10900 |
| 20 | 1499 | 140 | 0.09340 |
| 21 | 526 | 37 | 0.07030 |

| Mechanical Reliability Rates | | |
|---|---|---|
| Mileage at test | N tested | Pct failed |
| 10000 | 511 | 13.70 |
| 20000 | 1085 | 8.57 |
| 30000 | 2348 | 12.30 |
| 40000 | 4239 | 14.70 |
| 50000 | 6598 | 17.90 |
| 60000 | 8676 | 20.10 |
| 100000 | 13212 | 28.60 |
| 110000 | 12746 | 29.30 |
| 120000 | 11899 | 30.80 |
| 130000 | 10616 | 31.40 |
| 140000 | 9222 | 33.10 |
| 150000 | 7606 | 32.70 |
| 160000 | 6163 | 32.80 |
| 170000 | 4634 | 33.70 |
| 180000 | 3770 | 34.70 |
| 190000 | 2746 | 34.20 |
| 210000 | 1438 | 35.40 |

# Audi A4 2001

At 5 years of age, the mortality rate of a Audi A4 2001 (manufactured as a Car or Light Van) ranked number 97 out of 205 vehicles of the same age and type (any Car or Light Van constructed in 2001). One is the lowest (or best) and 205 the highest mortality rate. For vehicles reaching 120000 miles, its unreliability score (rate of failing an inspection) ranked 169 out of 194 vehicles of the same age, type, and mileage. One is the highest (or worst) and 194 the lowest rate of failing an inspection.

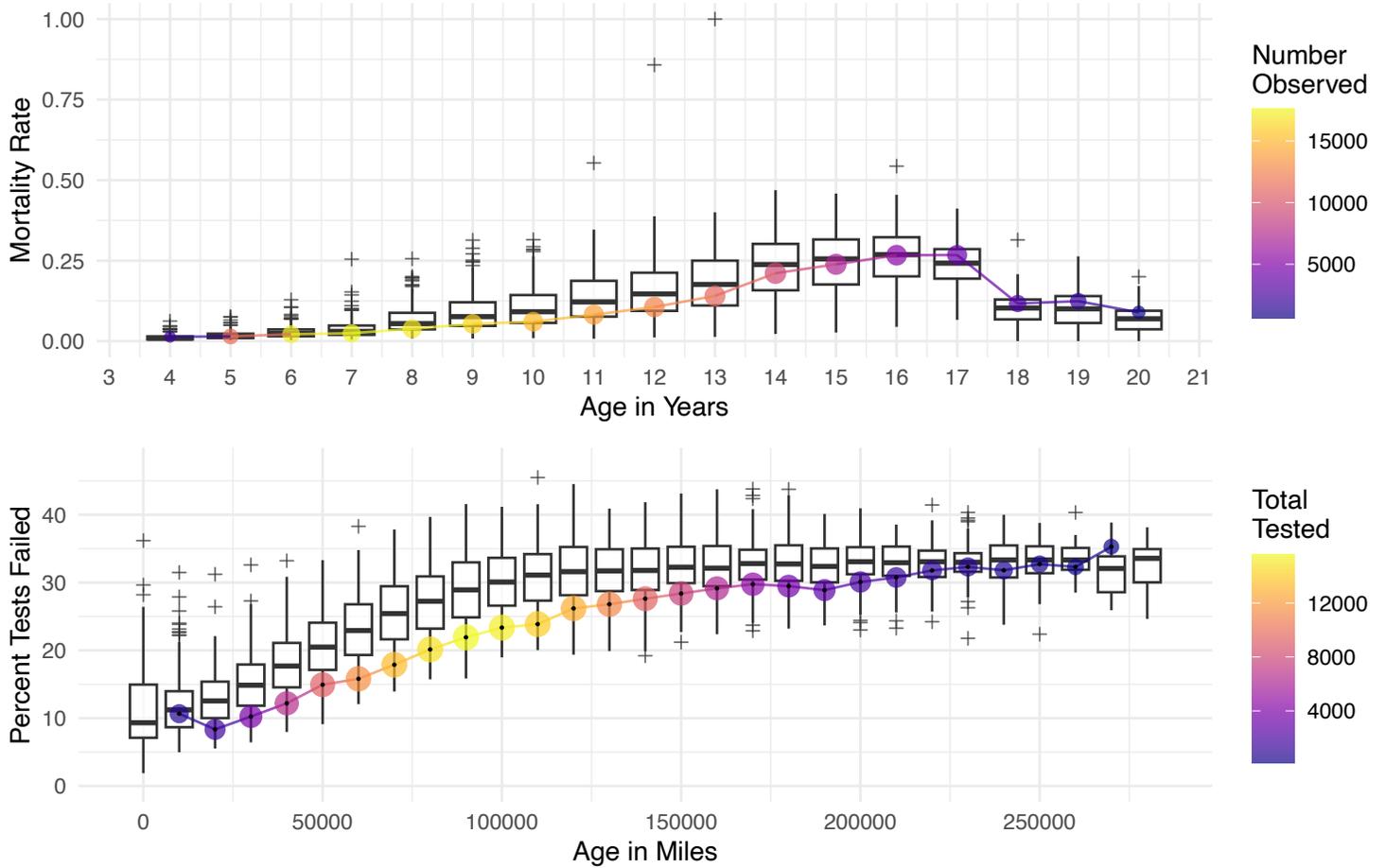

| Mortality rates | | | |
|---|---|---|---|
| Age in Years | Observed | Died | Mortality Rate |
| 4 | 1980 | 27 | 0.0136 |
| 5 | 11368 | 166 | 0.0146 |
| 6 | 17132 | 368 | 0.0215 |
| 7 | 17566 | 433 | 0.0246 |
| 8 | 16965 | 650 | 0.0383 |
| 9 | 16085 | 861 | 0.0535 |
| 10 | 14986 | 903 | 0.0603 |
| 11 | 13867 | 1140 | 0.0822 |
| 12 | 12522 | 1327 | 0.1060 |
| 13 | 10999 | 1544 | 0.1400 |
| 14 | 9211 | 1943 | 0.2110 |
| 15 | 7070 | 1687 | 0.2390 |
| 16 | 5294 | 1413 | 0.2670 |
| 17 | 3842 | 1028 | 0.2680 |
| 18 | 2636 | 308 | 0.1170 |
| 19 | 1749 | 217 | 0.1240 |
| 20 | 638 | 57 | 0.0893 |

| Mechanical Reliability Rates | | |
|---|---|---|
| Mileage at test | N tested | Pct failed |
| 10000 | 713 | 10.70 |
| 20000 | 1630 | 8.34 |
| 30000 | 3749 | 10.20 |
| 40000 | 6636 | 12.20 |
| 50000 | 9233 | 14.90 |
| 60000 | 11512 | 15.80 |
| 70000 | 13363 | 17.90 |
| 80000 | 14786 | 20.10 |
| 100000 | 15208 | 23.40 |
| 110000 | 14090 | 23.90 |
| 120000 | 12885 | 26.20 |
| 130000 | 11146 | 26.80 |
| 140000 | 9356 | 27.60 |
| 150000 | 7596 | 28.40 |
| 160000 | 6210 | 29.10 |
| 170000 | 4712 | 29.80 |
| 180000 | 3535 | 29.50 |

## Audi A4 2002

At 5 years of age, the mortality rate of a Audi A4 2002 (manufactured as a Car or Light Van) ranked number 106 out of 202 vehicles of the same age and type (any Car or Light Van constructed in 2002). One is the lowest (or best) and 202 the highest mortality rate. For vehicles reaching 120000 miles, its unreliability score (rate of failing an inspection) ranked 184 out of 193 vehicles of the same age, type, and mileage. One is the highest (or worst) and 193 the lowest rate of failing an inspection.

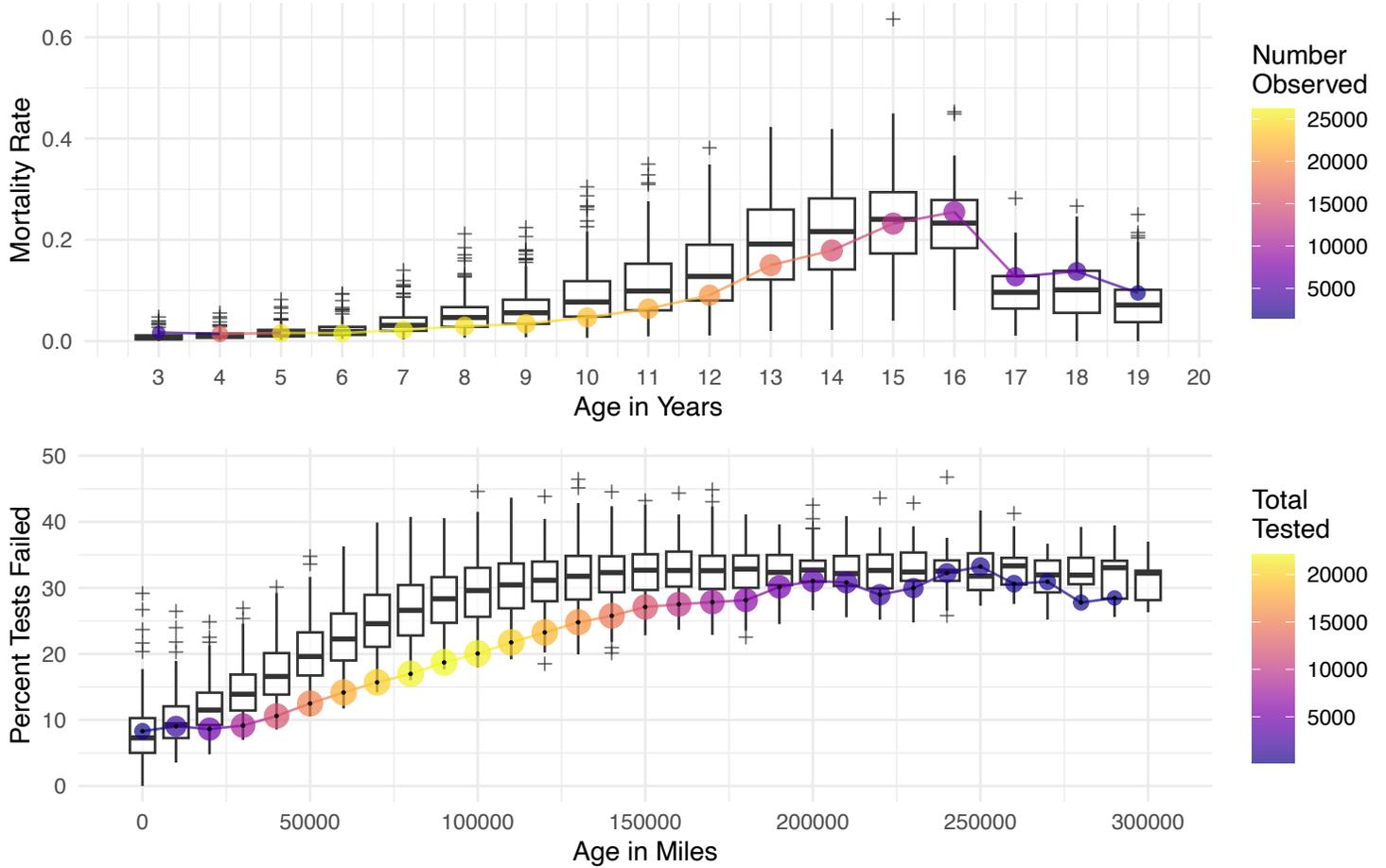

| Mortality rates | | | |
|---|---|---|---|
| Age in Years | Observed | Died | Mortality Rate |
| 3 | 3141 | 54 | 0.0172 |
| 4 | 16407 | 234 | 0.0143 |
| 5 | 25012 | 407 | 0.0163 |
| 6 | 26126 | 415 | 0.0159 |
| 7 | 25616 | 573 | 0.0224 |
| 8 | 24805 | 732 | 0.0295 |
| 9 | 23814 | 816 | 0.0343 |
| 10 | 22733 | 1064 | 0.0468 |
| 11 | 21429 | 1371 | 0.0640 |
| 12 | 19790 | 1793 | 0.0906 |
| 13 | 17547 | 2631 | 0.1500 |
| 14 | 14562 | 2608 | 0.1790 |
| 15 | 11751 | 2727 | 0.2320 |
| 16 | 8936 | 2279 | 0.2550 |
| 17 | 6316 | 803 | 0.1270 |
| 18 | 4299 | 592 | 0.1380 |
| 19 | 1514 | 144 | 0.0951 |

| Mechanical Reliability Rates | | |
|---|---|---|
| Mileage at test | N tested | Pct failed |
| 0 | 229 | 8.30 |
| 10000 | 1660 | 9.04 |
| 20000 | 4341 | 8.62 |
| 30000 | 8362 | 9.15 |
| 40000 | 12030 | 10.60 |
| 50000 | 15425 | 12.50 |
| 60000 | 18156 | 14.20 |
| 100000 | 21694 | 20.10 |
| 110000 | 20306 | 21.70 |
| 120000 | 18559 | 23.30 |
| 130000 | 16625 | 24.80 |
| 140000 | 14551 | 25.80 |
| 150000 | 12133 | 27.10 |
| 160000 | 9902 | 27.50 |
| 170000 | 7945 | 27.80 |
| 200000 | 3549 | 31.10 |
| 220000 | 1847 | 29.00 |

## Audi A4 2003

At 5 years of age, the mortality rate of a Audi A4 2003 (manufactured as a Car or Light Van) ranked number 81 out of 213 vehicles of the same age and type (any Car or Light Van constructed in 2003). One is the lowest (or best) and 213 the highest mortality rate. For vehicles reaching 120000 miles, its unreliability score (rate of failing an inspection) ranked 193 out of 205 vehicles of the same age, type, and mileage. One is the highest (or worst) and 205 the lowest rate of failing an inspection.

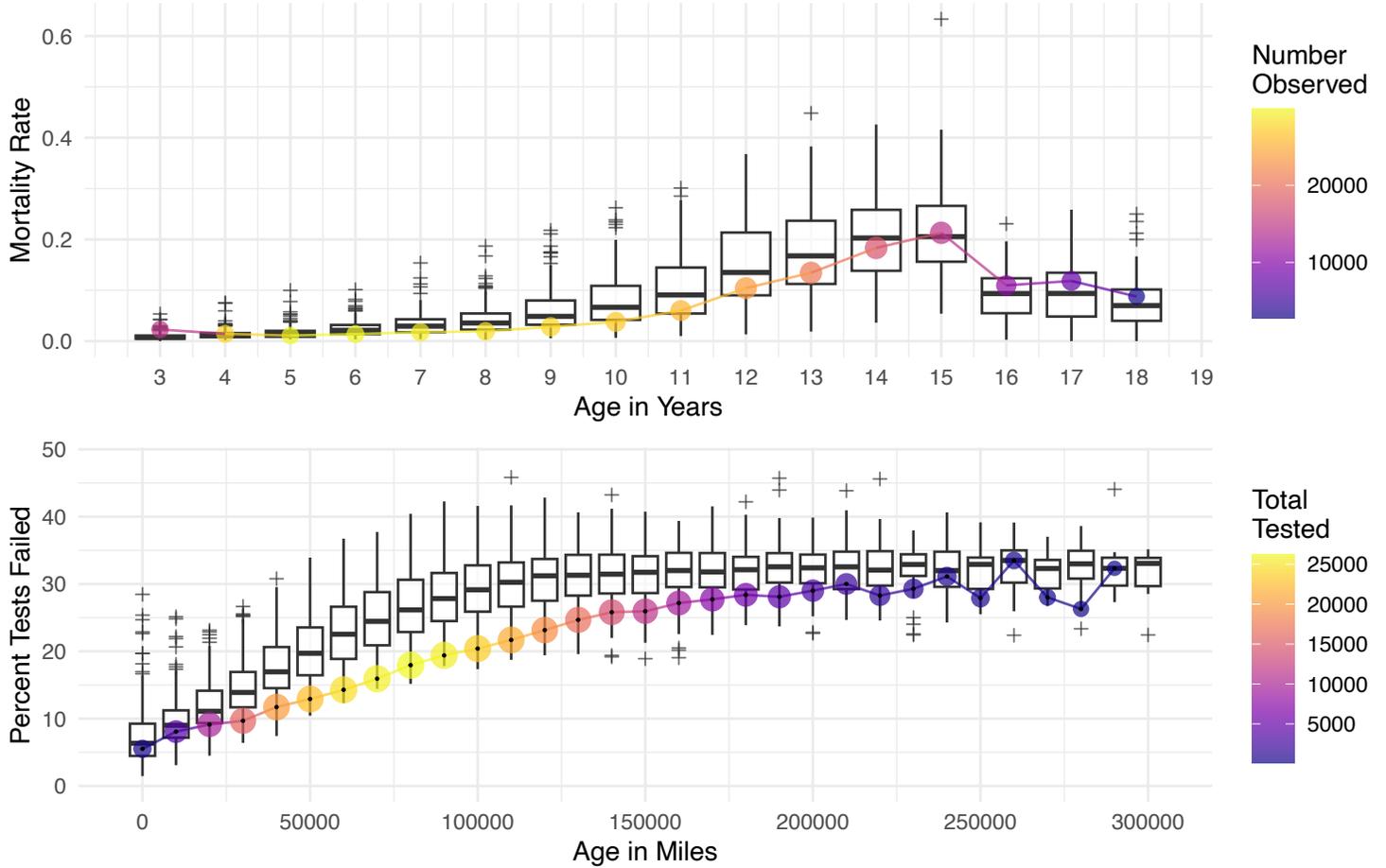

### Mortality rates

| Age in Years | Observed | Died | Mortality Rate |
|---|---|---|---|
| 3 | 14686 | 337 | 0.0229 |
| 4 | 27693 | 397 | 0.0143 |
| 5 | 29859 | 329 | 0.0110 |
| 6 | 29596 | 410 | 0.0139 |
| 7 | 29082 | 512 | 0.0176 |
| 8 | 28425 | 548 | 0.0193 |
| 9 | 27722 | 780 | 0.0281 |
| 10 | 26768 | 1016 | 0.0380 |
| 11 | 25541 | 1535 | 0.0601 |
| 12 | 23612 | 2461 | 0.1040 |
| 13 | 20772 | 2793 | 0.1340 |
| 14 | 17691 | 3244 | 0.1830 |
| 15 | 14309 | 3048 | 0.2130 |
| 16 | 10816 | 1186 | 0.1100 |
| 17 | 7571 | 896 | 0.1180 |
| 18 | 2791 | 245 | 0.0878 |

### Mechanical Reliability Rates

| Mileage at test | N tested | Pct failed |
|---|---|---|
| 0 | 471 | 5.52 |
| 10000 | 3531 | 8.07 |
| 20000 | 9763 | 9.13 |
| 30000 | 15824 | 9.68 |
| 40000 | 20176 | 11.70 |
| 50000 | 22781 | 12.90 |
| 60000 | 24696 | 14.30 |
| 70000 | 25760 | 15.90 |
| 80000 | 26253 | 17.90 |
| 90000 | 25607 | 19.40 |
| 100000 | 23603 | 20.40 |
| 110000 | 21377 | 21.70 |
| 120000 | 19232 | 23.10 |
| 130000 | 16677 | 24.70 |
| 140000 | 14026 | 25.80 |
| 150000 | 11619 | 25.90 |
| 160000 | 9356 | 27.20 |

## Audi A4 2004

At 5 years of age, the mortality rate of a Audi A4 2004 (manufactured as a Car or Light Van) ranked number 66 out of 229 vehicles of the same age and type (any Car or Light Van constructed in 2004). One is the lowest (or best) and 229 the highest mortality rate. For vehicles reaching 20000 miles, its unreliability score (rate of failing an inspection) ranked 172 out of 225 vehicles of the same age, type, and mileage. One is the highest (or worst) and 225 the lowest rate of failing an inspection.

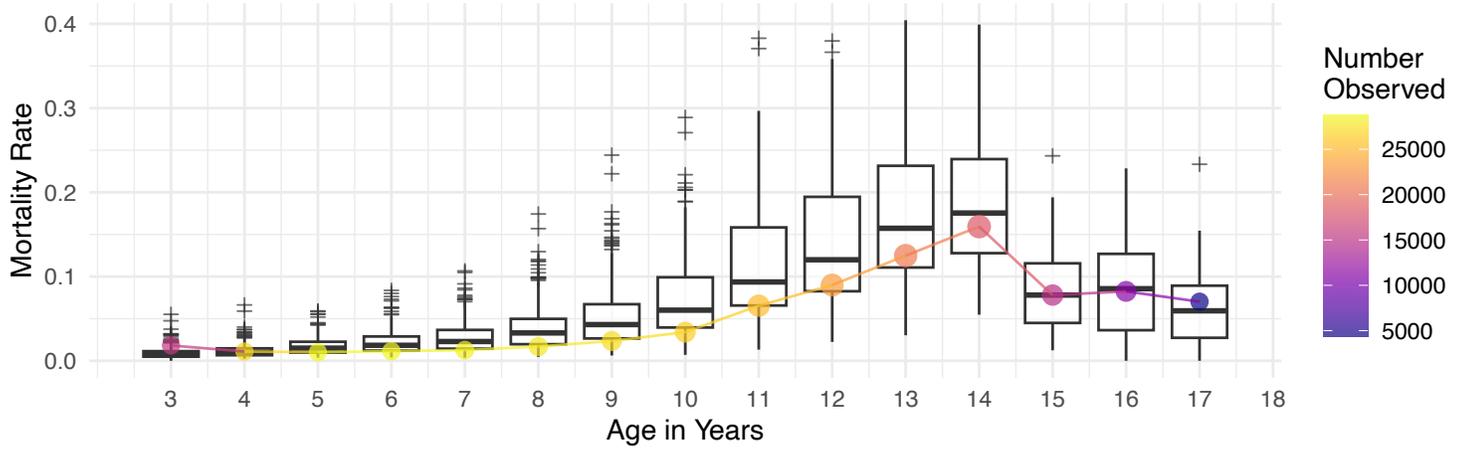

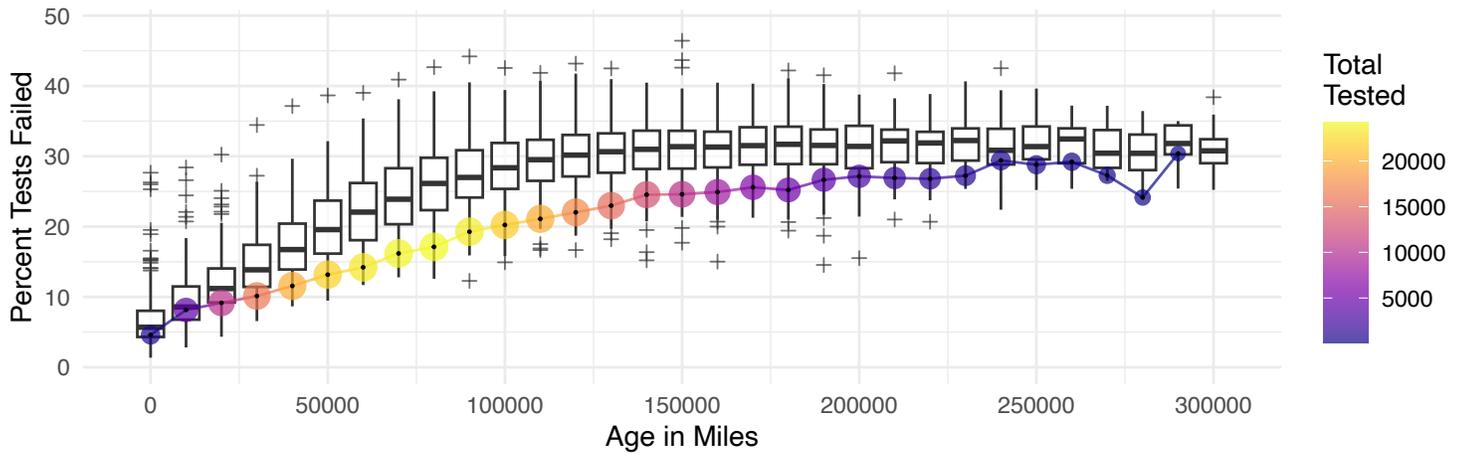

| Mortality rates | | | |
|---|---|---|---|
| Age in Years | Observed | Died | Mortality Rate |
| 3 | 15499 | 282 | 0.0182 |
| 4 | 26789 | 295 | 0.0110 |
| 5 | 28709 | 294 | 0.0102 |
| 6 | 28468 | 325 | 0.0114 |
| 7 | 28097 | 356 | 0.0127 |
| 8 | 27655 | 468 | 0.0169 |
| 9 | 27104 | 638 | 0.0235 |
| 10 | 26332 | 896 | 0.0340 |
| 11 | 25171 | 1647 | 0.0654 |
| 12 | 23199 | 2086 | 0.0899 |
| 13 | 20876 | 2607 | 0.1250 |
| 14 | 18118 | 2888 | 0.1590 |
| 15 | 14711 | 1148 | 0.0780 |
| 16 | 10969 | 904 | 0.0824 |
| 17 | 4376 | 307 | 0.0702 |

| Mechanical Reliability Rates | | |
|---|---|---|
| Mileage at test | N tested | Pct failed |
| 0 | 500 | 4.60 |
| 10000 | 3999 | 8.15 |
| 20000 | 10514 | 9.15 |
| 30000 | 16143 | 10.10 |
| 40000 | 19963 | 11.60 |
| 50000 | 21996 | 13.20 |
| 60000 | 23285 | 14.20 |
| 70000 | 24066 | 16.20 |
| 80000 | 24258 | 17.10 |
| 90000 | 23561 | 19.30 |
| 100000 | 21559 | 20.20 |
| 110000 | 19786 | 21.10 |
| 120000 | 17433 | 22.00 |
| 130000 | 15193 | 23.00 |
| 140000 | 12906 | 24.50 |
| 150000 | 10372 | 24.60 |
| 160000 | 8236 | 24.90 |

## Audi A4 2005

At 5 years of age, the mortality rate of a Audi A4 2005 (manufactured as a Car or Light Van) ranked number 93 out of 240 vehicles of the same age and type (any Car or Light Van constructed in 2005). One is the lowest (or best) and 240 the highest mortality rate. For vehicles reaching 40000 miles, its unreliability score (rate of failing an inspection) ranked 218 out of 236 vehicles of the same age, type, and mileage. One is the highest (or worst) and 236 the lowest rate of failing an inspection.

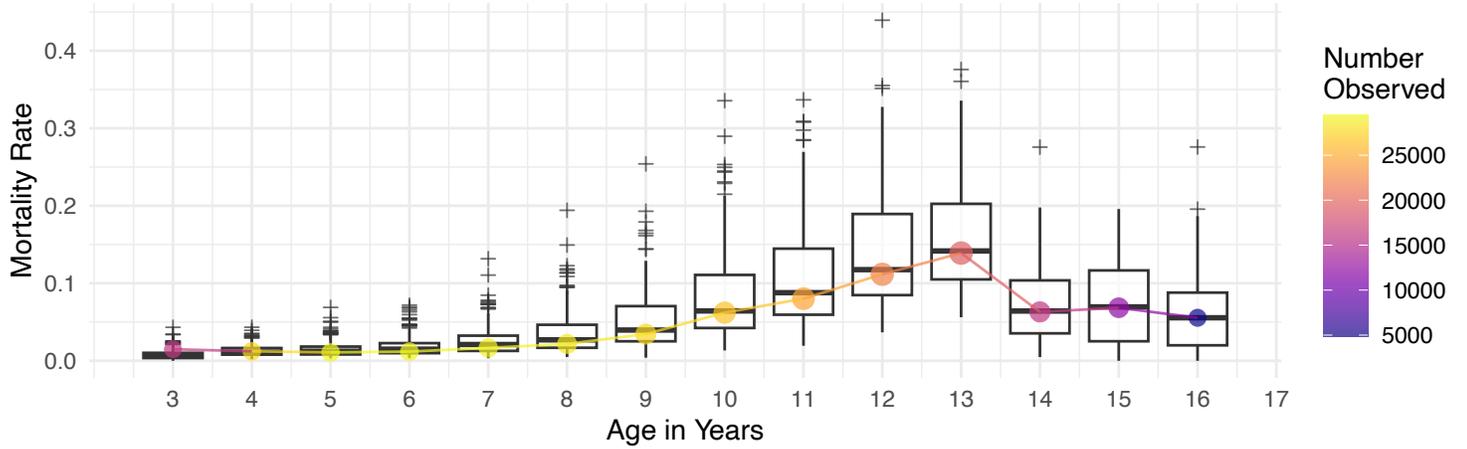

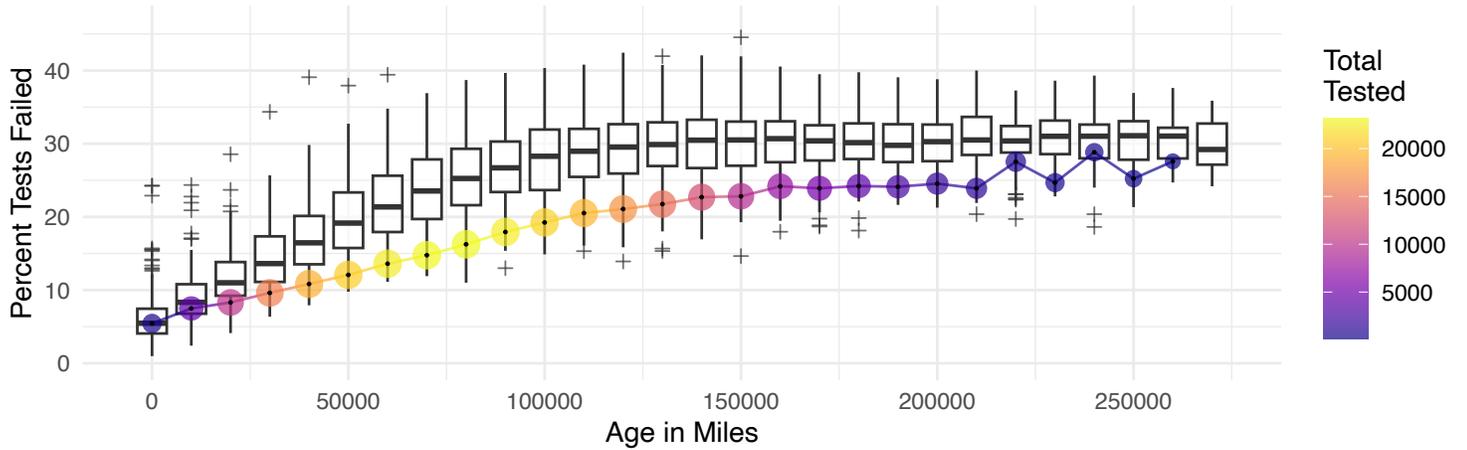

| Mortality rates | | | |
| --- | --- | --- | --- |
| Age in Years | Observed | Died | Mortality Rate |
| 3 | 16149 | 241 | 0.0149 |
| 4 | 27644 | 339 | 0.0123 |
| 5 | 29395 | 319 | 0.0109 |
| 6 | 29312 | 348 | 0.0119 |
| 7 | 28946 | 483 | 0.0167 |
| 8 | 28385 | 618 | 0.0218 |
| 9 | 27606 | 952 | 0.0345 |
| 10 | 26309 | 1635 | 0.0621 |
| 11 | 24274 | 1945 | 0.0801 |
| 12 | 21997 | 2455 | 0.1120 |
| 13 | 19316 | 2684 | 0.1390 |
| 14 | 16038 | 1012 | 0.0631 |
| 15 | 12127 | 830 | 0.0684 |
| 16 | 4865 | 270 | 0.0555 |

| Mechanical Reliability Rates | | |
| --- | --- | --- |
| Mileage at test | N tested | Pct failed |
| 0 | 497 | 5.43 |
| 10000 | 3953 | 7.49 |
| 20000 | 10293 | 8.31 |
| 30000 | 15997 | 9.61 |
| 40000 | 19106 | 10.80 |
| 50000 | 20916 | 12.10 |
| 60000 | 22494 | 13.60 |
| 70000 | 23119 | 14.80 |
| 80000 | 23192 | 16.30 |
| 90000 | 22564 | 17.90 |
| 100000 | 21101 | 19.20 |
| 110000 | 19570 | 20.50 |
| 120000 | 17330 | 21.10 |
| 130000 | 14741 | 21.70 |
| 140000 | 12266 | 22.70 |
| 150000 | 9547 | 22.80 |
| 160000 | 7352 | 24.20 |

## Audi A4 2006

At 5 years of age, the mortality rate of a Audi A4 2006 (manufactured as a Car or Light Van) ranked number 78 out of 225 vehicles of the same age and type (any Car or Light Van constructed in 2006). One is the lowest (or best) and 225 the highest mortality rate. For vehicles reaching 80000 miles, its unreliability score (rate of failing an inspection) ranked 202 out of 220 vehicles of the same age, type, and mileage. One is the highest (or worst) and 220 the lowest rate of failing an inspection.

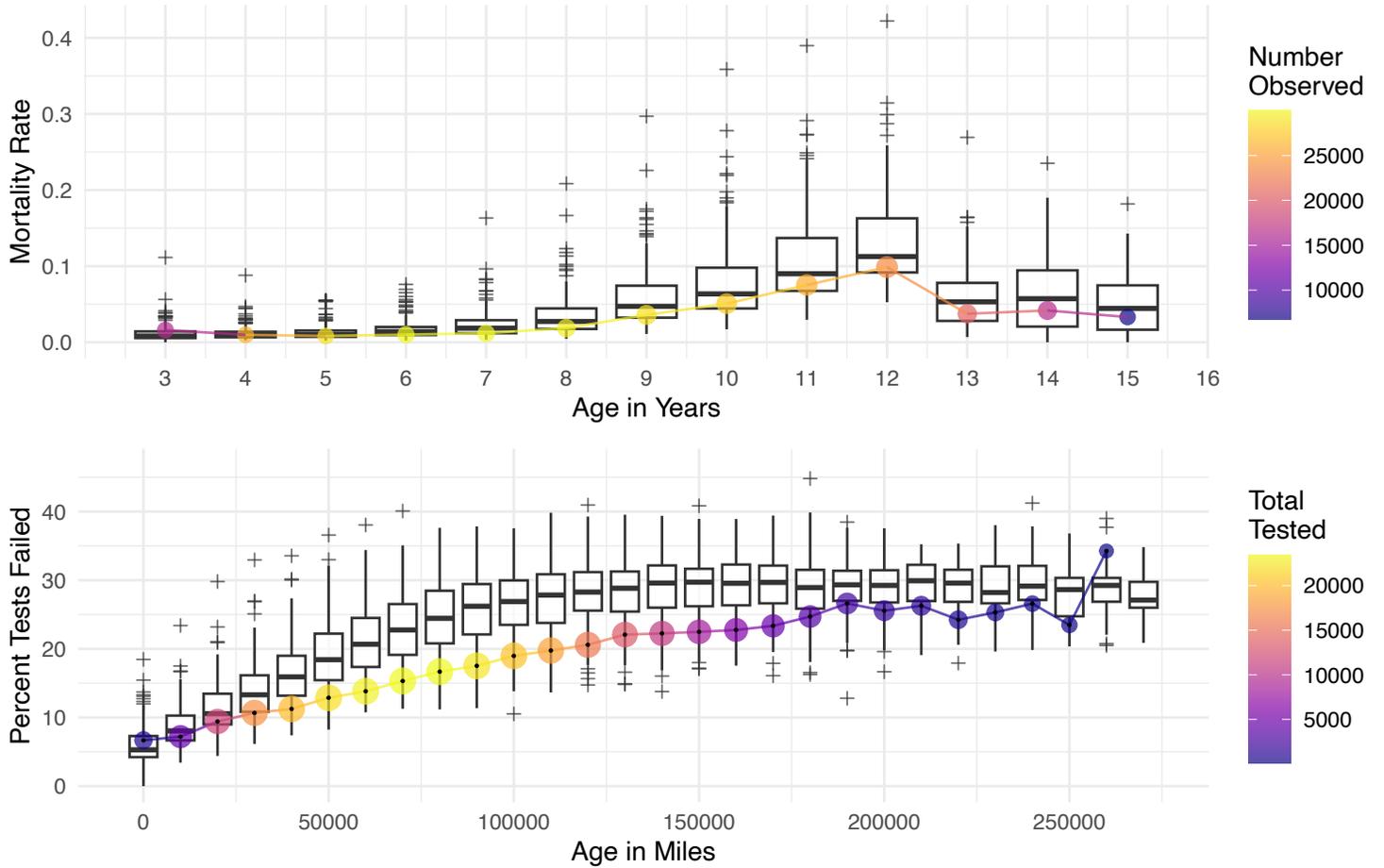

<table>
<tr><td colspan="4" align="center">Mortality rates</td></tr>
<tr><th>Age in Years</th><th>Observed</th><th>Died</th><th>Mortality Rate</th></tr>
<tr><td>3</td><td>16244</td><td>261</td><td>0.01610</td></tr>
<tr><td>4</td><td>26032</td><td>256</td><td>0.00983</td></tr>
<tr><td>5</td><td>28791</td><td>240</td><td>0.00834</td></tr>
<tr><td>6</td><td>29965</td><td>303</td><td>0.01010</td></tr>
<tr><td>7</td><td>29855</td><td>356</td><td>0.01190</td></tr>
<tr><td>8</td><td>29446</td><td>548</td><td>0.01860</td></tr>
<tr><td>9</td><td>28718</td><td>1033</td><td>0.03600</td></tr>
<tr><td>10</td><td>27454</td><td>1393</td><td>0.05070</td></tr>
<tr><td>11</td><td>25788</td><td>1942</td><td>0.07530</td></tr>
<tr><td>12</td><td>23650</td><td>2337</td><td>0.09880</td></tr>
<tr><td>13</td><td>20654</td><td>774</td><td>0.03750</td></tr>
<tr><td>14</td><td>16340</td><td>683</td><td>0.04180</td></tr>
<tr><td>15</td><td>6758</td><td>224</td><td>0.03310</td></tr>
</table>

<table>
<tr><td colspan="3" align="center">Mechanical Reliability Rates</td></tr>
<tr><th>Mileage at test</th><th>N tested</th><th>Pct failed</th></tr>
<tr><td>0</td><td>494</td><td>6.68</td></tr>
<tr><td>10000</td><td>4299</td><td>7.21</td></tr>
<tr><td>20000</td><td>11771</td><td>9.42</td></tr>
<tr><td>30000</td><td>17436</td><td>10.70</td></tr>
<tr><td>40000</td><td>20069</td><td>11.30</td></tr>
<tr><td>50000</td><td>22064</td><td>12.90</td></tr>
<tr><td>60000</td><td>23020</td><td>13.80</td></tr>
<tr><td>70000</td><td>23446</td><td>15.30</td></tr>
<tr><td>80000</td><td>23436</td><td>16.70</td></tr>
<tr><td>90000</td><td>22451</td><td>17.50</td></tr>
<tr><td>100000</td><td>20506</td><td>19.00</td></tr>
<tr><td>110000</td><td>18419</td><td>19.80</td></tr>
<tr><td>120000</td><td>15787</td><td>20.60</td></tr>
<tr><td>130000</td><td>13302</td><td>22.10</td></tr>
<tr><td>140000</td><td>10843</td><td>22.20</td></tr>
<tr><td>150000</td><td>8409</td><td>22.50</td></tr>
<tr><td>170000</td><td>4544</td><td>23.30</td></tr>
</table>

## Audi A4 2007

At 5 years of age, the mortality rate of a Audi A4 2007 (manufactured as a Car or Light Van) ranked number 60 out of 219 vehicles of the same age and type (any Car or Light Van constructed in 2007). One is the lowest (or best) and 219 the highest mortality rate. For vehicles reaching 20000 miles, its unreliability score (rate of failing an inspection) ranked 142 out of 214 vehicles of the same age, type, and mileage. One is the highest (or worst) and 214 the lowest rate of failing an inspection.

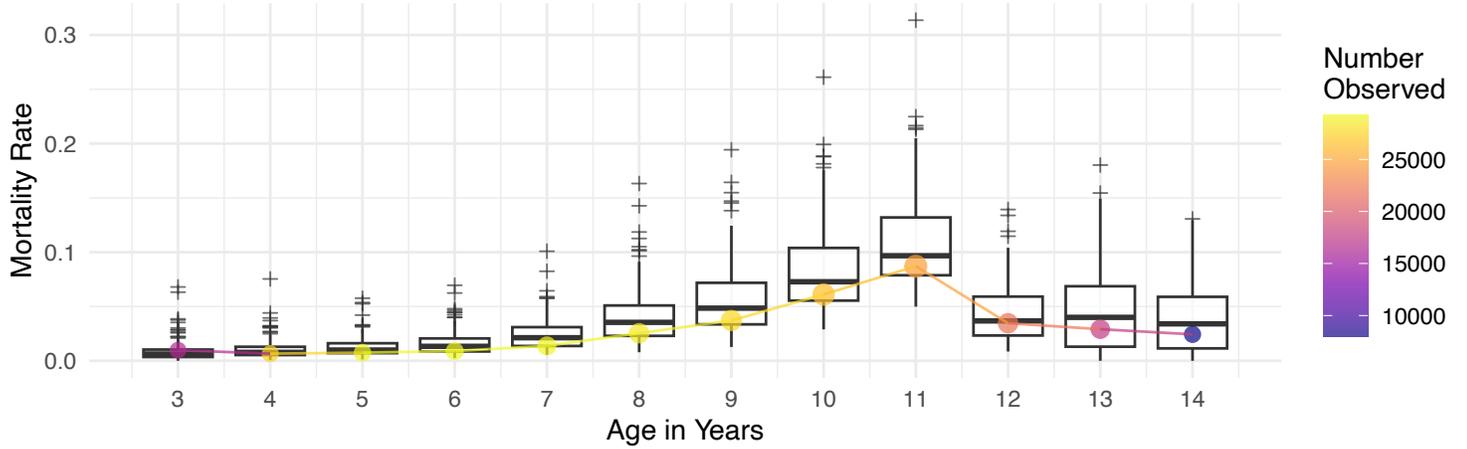

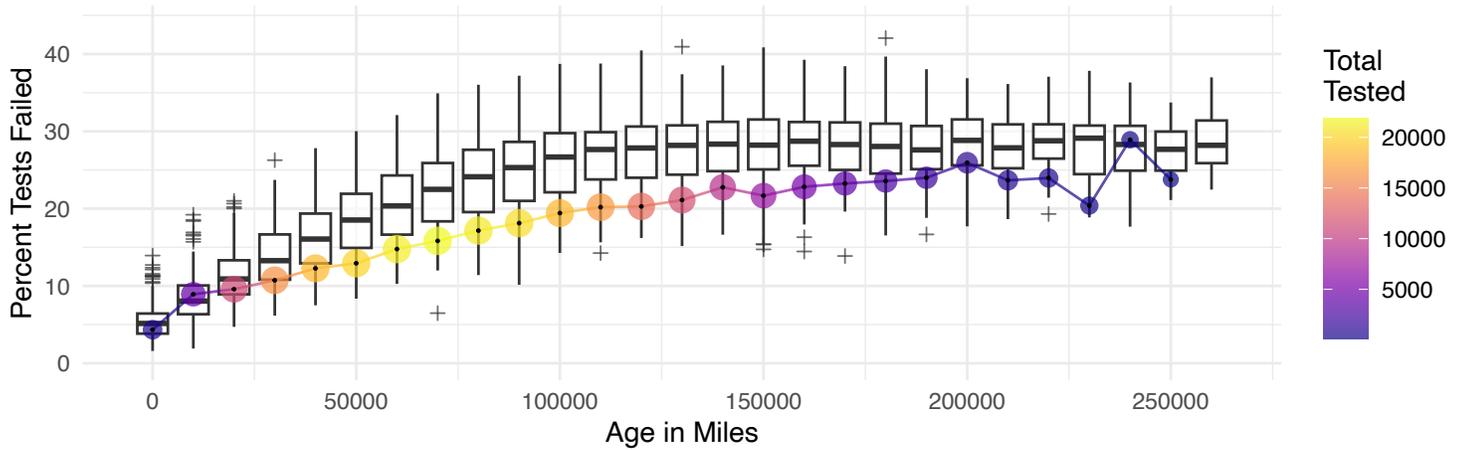

| Mortality rates | | | |
|---|---|---|---|
| Age in Years | Observed | Died | Mortality Rate |
| 3 | 16446 | 161 | 0.00979 |
| 4 | 27270 | 180 | 0.00660 |
| 5 | 29082 | 218 | 0.00750 |
| 6 | 29202 | 262 | 0.00897 |
| 7 | 28995 | 399 | 0.01380 |
| 8 | 28512 | 724 | 0.02540 |
| 9 | 27607 | 1025 | 0.03710 |
| 10 | 26385 | 1612 | 0.06110 |
| 11 | 24574 | 2140 | 0.08710 |
| 12 | 21851 | 756 | 0.03460 |
| 13 | 17797 | 518 | 0.02910 |
| 14 | 8006 | 195 | 0.02440 |

| Mechanical Reliability Rates | | |
|---|---|---|
| Mileage at test | N tested | Pct failed |
| 0 | 530 | 4.34 |
| 10000 | 4178 | 8.93 |
| 20000 | 11230 | 9.58 |
| 30000 | 16160 | 10.70 |
| 40000 | 18612 | 12.30 |
| 50000 | 20123 | 12.90 |
| 60000 | 21299 | 14.80 |
| 70000 | 21897 | 15.80 |
| 80000 | 21199 | 17.20 |
| 90000 | 20513 | 18.10 |
| 100000 | 18488 | 19.40 |
| 110000 | 16739 | 20.20 |
| 120000 | 13966 | 20.30 |
| 130000 | 11770 | 21.10 |
| 140000 | 9298 | 22.80 |
| 150000 | 7021 | 21.70 |
| 160000 | 5025 | 22.80 |

## Audi A4 2008

At 5 years of age, the mortality rate of a Audi A4 2008 (manufactured as a Car or Light Van) ranked number 137 out of 218 vehicles of the same age and type (any Car or Light Van constructed in 2008). One is the lowest (or best) and 218 the highest mortality rate. For vehicles reaching 20000 miles, its unreliability score (rate of failing an inspection) ranked 159 out of 212 vehicles of the same age, type, and mileage. One is the highest (or worst) and 212 the lowest rate of failing an inspection.

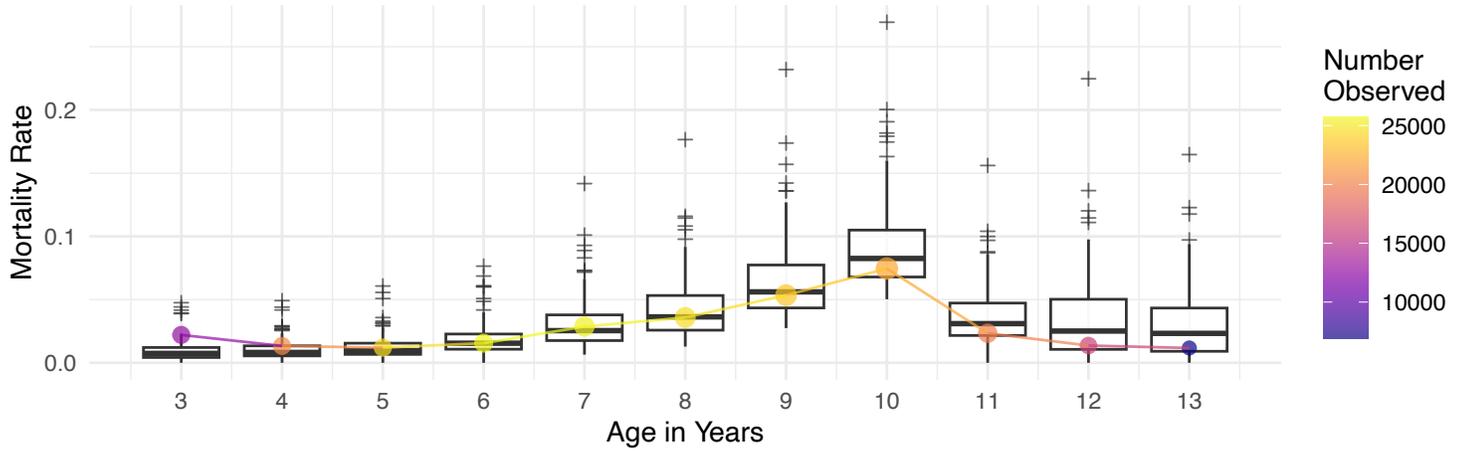

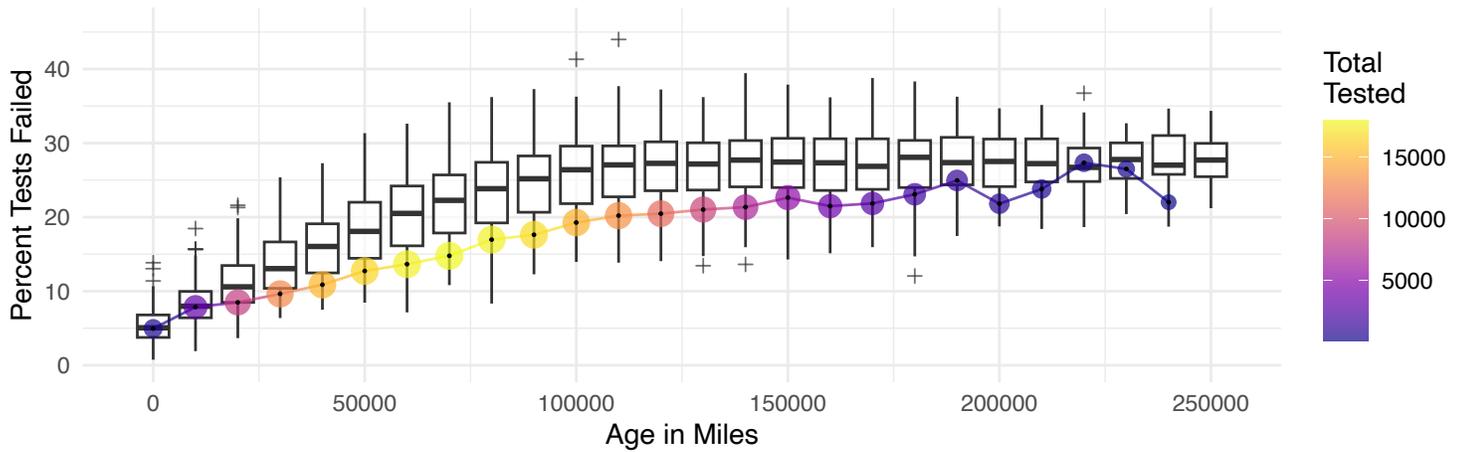

| Mortality rates | | | |
|---|---|---|---|
| Age in Years | Observed | Died | Mortality Rate |
| 3 | 12648 | 279 | 0.0221 |
| 4 | 20482 | 273 | 0.0133 |
| 5 | 24414 | 292 | 0.0120 |
| 6 | 25676 | 400 | 0.0156 |
| 7 | 25382 | 727 | 0.0286 |
| 8 | 24568 | 880 | 0.0358 |
| 9 | 23516 | 1258 | 0.0535 |
| 10 | 22050 | 1644 | 0.0746 |
| 11 | 19767 | 462 | 0.0234 |
| 12 | 16115 | 220 | 0.0137 |
| 13 | 6921 | 81 | 0.0117 |

| Mechanical Reliability Rates | | |
|---|---|---|
| Mileage at test | N tested | Pct failed |
| 0 | 424 | 4.95 |
| 10000 | 3293 | 7.87 |
| 20000 | 8504 | 8.49 |
| 30000 | 12784 | 9.64 |
| 40000 | 15143 | 10.90 |
| 50000 | 16503 | 12.70 |
| 60000 | 17653 | 13.70 |
| 70000 | 17978 | 14.80 |
| 80000 | 17619 | 17.00 |
| 90000 | 16800 | 17.60 |
| 100000 | 14945 | 19.30 |
| 110000 | 13311 | 20.20 |
| 120000 | 11269 | 20.50 |
| 130000 | 8907 | 21.00 |
| 140000 | 6915 | 21.40 |
| 150000 | 5336 | 22.60 |
| 160000 | 3736 | 21.50 |

## Audi A4 2009

At 5 years of age, the mortality rate of a Audi A4 2009 (manufactured as a Car or Light Van) ranked number 131 out of 205 vehicles of the same age and type (any Car or Light Van constructed in 2009). One is the lowest (or best) and 205 the highest mortality rate. For vehicles reaching 20000 miles, its unreliability score (rate of failing an inspection) ranked 172 out of 200 vehicles of the same age, type, and mileage. One is the highest (or worst) and 200 the lowest rate of failing an inspection.

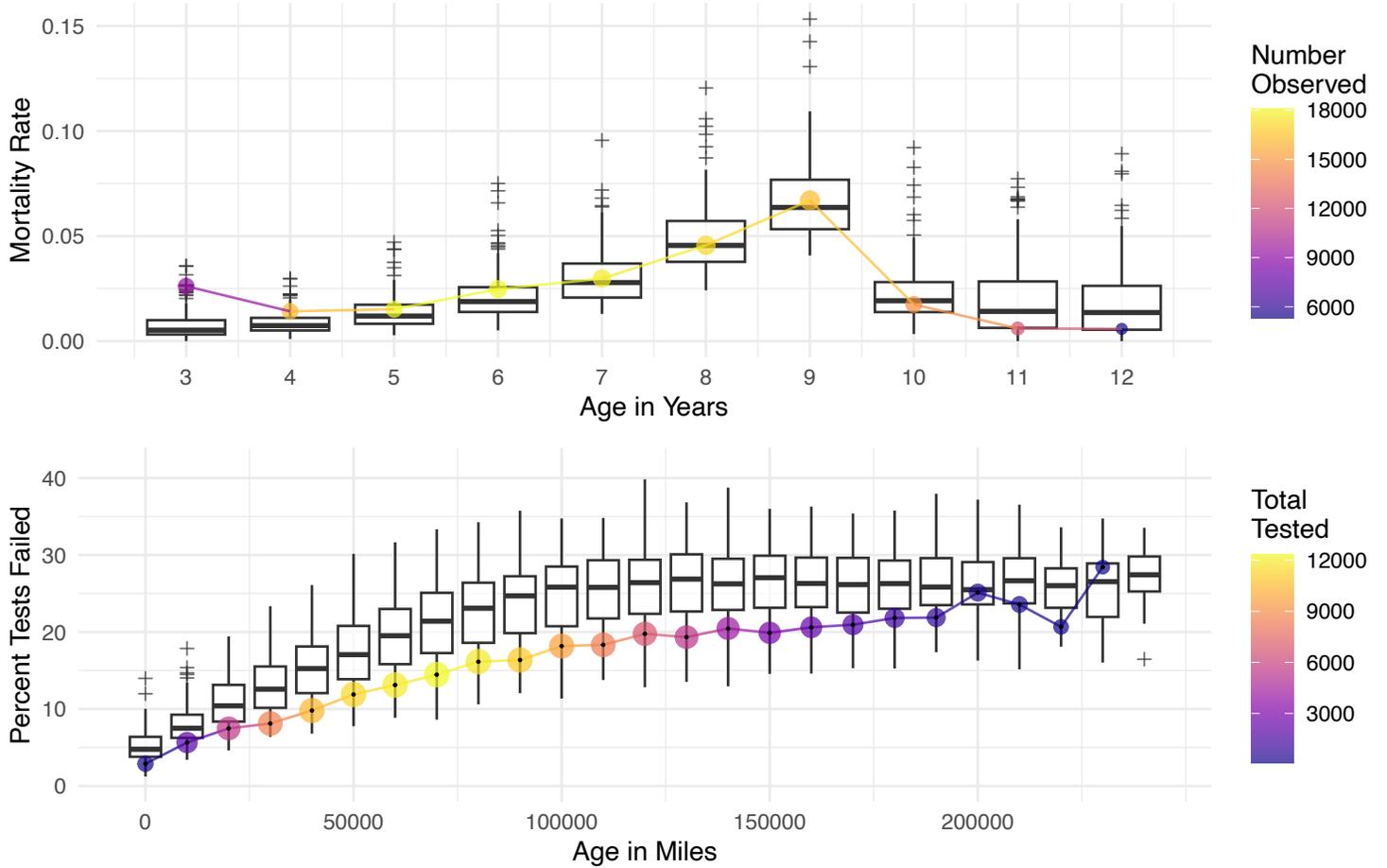

### Mortality rates

| Age in Years | Observed | Died | Mortality Rate |
|---|---|---|---|
| 3 | 9286 | 243 | 0.02620 |
| 4 | 16080 | 228 | 0.01420 |
| 5 | 18046 | 274 | 0.01520 |
| 6 | 18028 | 446 | 0.02470 |
| 7 | 17635 | 525 | 0.02980 |
| 8 | 17048 | 778 | 0.04560 |
| 9 | 16186 | 1084 | 0.06700 |
| 10 | 14672 | 256 | 0.01740 |
| 11 | 12132 | 73 | 0.00602 |
| 12 | 5317 | 31 | 0.00583 |

### Mechanical Reliability Rates

| Mileage at test | N tested | Pct failed |
|---|---|---|
| 0 | 208 | 2.88 |
| 10000 | 1897 | 5.64 |
| 20000 | 5536 | 7.48 |
| 30000 | 8545 | 8.11 |
| 40000 | 10425 | 9.80 |
| 50000 | 11383 | 11.90 |
| 60000 | 11995 | 13.10 |
| 70000 | 12352 | 14.50 |
| 80000 | 11910 | 16.10 |
| 90000 | 10993 | 16.40 |
| 100000 | 9671 | 18.20 |
| 110000 | 8373 | 18.30 |
| 120000 | 6837 | 19.80 |
| 130000 | 5425 | 19.30 |
| 140000 | 4150 | 20.50 |
| 150000 | 2951 | 19.90 |
| 160000 | 2232 | 20.60 |

## Audi A4 2010

At 5 years of age, the mortality rate of a Audi A4 2010 (manufactured as a Car or Light Van) ranked number 167 out of 206 vehicles of the same age and type (any Car or Light Van constructed in 2010). One is the lowest (or best) and 206 the highest mortality rate. For vehicles reaching 20000 miles, its unreliability score (rate of failing an inspection) ranked 183 out of 201 vehicles of the same age, type, and mileage. One is the highest (or worst) and 201 the lowest rate of failing an inspection.

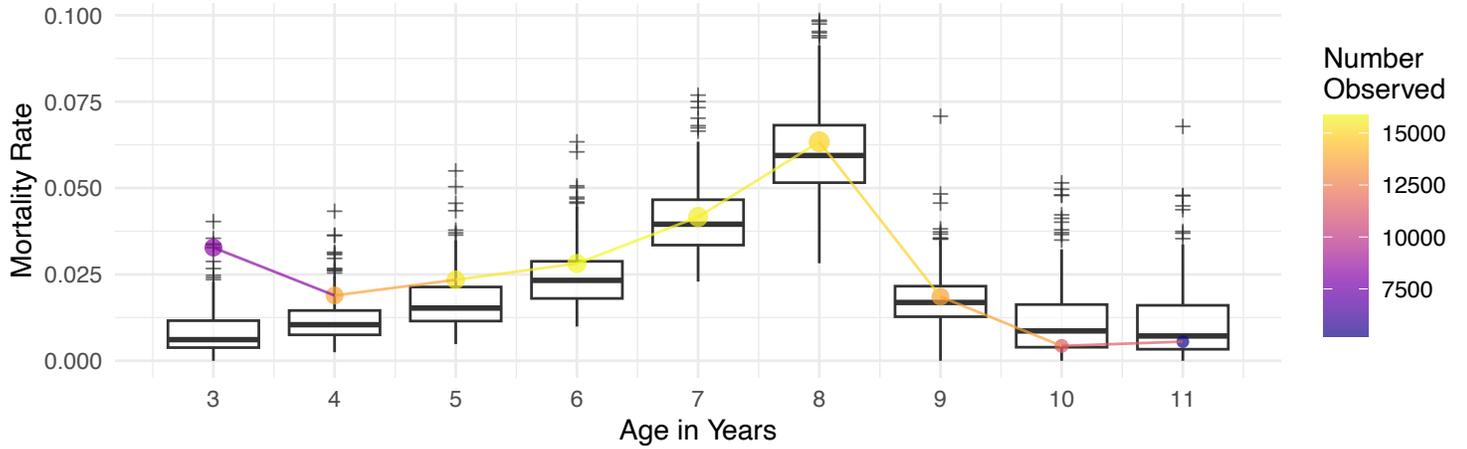

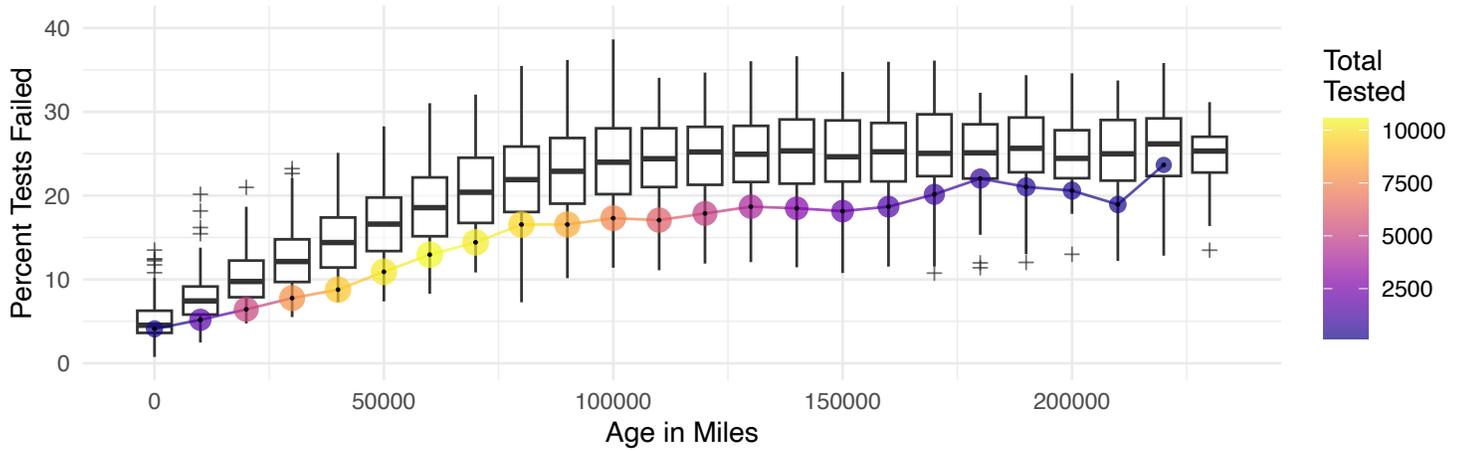

Mortality rates

| Age in Years | Observed | Died | Mortality Rate |
|---|---|---|---|
| 3 | 8241 | 270 | 0.03280 |
| 4 | 13594 | 257 | 0.01890 |
| 5 | 15441 | 362 | 0.02340 |
| 6 | 15838 | 446 | 0.02820 |
| 7 | 15542 | 647 | 0.04160 |
| 8 | 14909 | 945 | 0.06340 |
| 9 | 13578 | 252 | 0.01860 |
| 10 | 11381 | 49 | 0.00431 |
| 11 | 5222 | 29 | 0.00555 |

Mechanical Reliability Rates

| Mileage at test | N tested | Pct failed |
|---|---|---|
| 0 | 195 | 4.10 |
| 10000 | 1660 | 5.18 |
| 20000 | 5148 | 6.43 |
| 30000 | 7784 | 7.77 |
| 40000 | 9307 | 8.78 |
| 50000 | 10138 | 10.90 |
| 60000 | 10565 | 13.00 |
| 70000 | 10268 | 14.40 |
| 80000 | 9752 | 16.60 |
| 90000 | 8557 | 16.60 |
| 100000 | 7470 | 17.30 |
| 110000 | 6131 | 17.10 |
| 120000 | 4900 | 17.90 |
| 130000 | 4015 | 18.70 |
| 140000 | 2823 | 18.50 |
| 150000 | 2000 | 18.20 |
| 160000 | 1445 | 18.70 |

**Audi A4 2011**

At 5 years of age, the mortality rate of a Audi A4 2011 (manufactured as a Car or Light Van) ranked number 173 out of 211 vehicles of the same age and type (any Car or Light Van constructed in 2011). One is the lowest (or best) and 211 the highest mortality rate. For vehicles reaching 20000 miles, its unreliability score (rate of failing an inspection) ranked 175 out of 205 vehicles of the same age, type, and mileage. One is the highest (or worst) and 205 the lowest rate of failing an inspection.

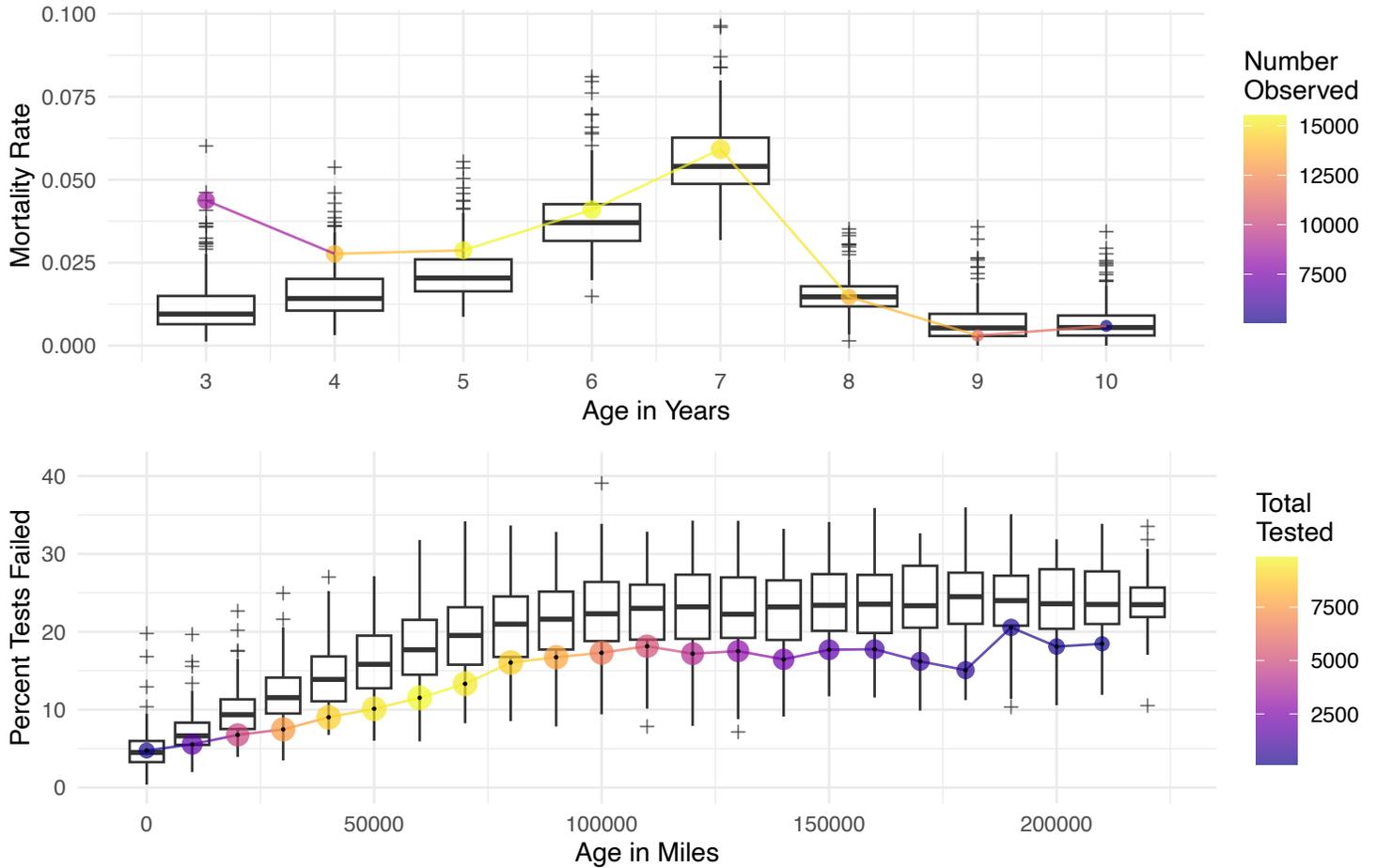

Mechanical Reliability Rates

| Mileage at test | N tested | Pct failed |
|---|---|---|
| 0 | 189 | 4.76 |
| 10000 | 1638 | 5.56 |
| 20000 | 4792 | 6.76 |
| 30000 | 7538 | 7.47 |
| 40000 | 8876 | 9.02 |
| 50000 | 9430 | 10.10 |
| 60000 | 9834 | 11.50 |
| 70000 | 9538 | 13.30 |
| 80000 | 8861 | 16.00 |
| 90000 | 7812 | 16.70 |
| 100000 | 6553 | 17.30 |
| 110000 | 5208 | 18.10 |
| 120000 | 3997 | 17.20 |
| 130000 | 2957 | 17.50 |
| 140000 | 2187 | 16.50 |
| 150000 | 1605 | 17.70 |
| 160000 | 1081 | 17.80 |

Mortality rates

| Age in Years | Observed | Died | Mortality Rate |
|---|---|---|---|
| 3 | 8662 | 379 | 0.04380 |
| 4 | 13952 | 386 | 0.02770 |
| 5 | 15463 | 444 | 0.02870 |
| 6 | 15528 | 636 | 0.04100 |
| 7 | 15035 | 890 | 0.05920 |
| 8 | 13827 | 203 | 0.01470 |
| 9 | 11552 | 35 | 0.00303 |
| 10 | 5053 | 30 | 0.00594 |

**Audi A4 2012**

At 5 years of age, the mortality rate of a Audi A4 2012 (manufactured as a Car or Light Van) ranked number 182 out of 212 vehicles of the same age and type (any Car or Light Van constructed in 2012). One is the lowest (or best) and 212 the highest mortality rate. For vehicles reaching 20000 miles, its unreliability score (rate of failing an inspection) ranked 180 out of 206 vehicles of the same age, type, and mileage. One is the highest (or worst) and 206 the lowest rate of failing an inspection.

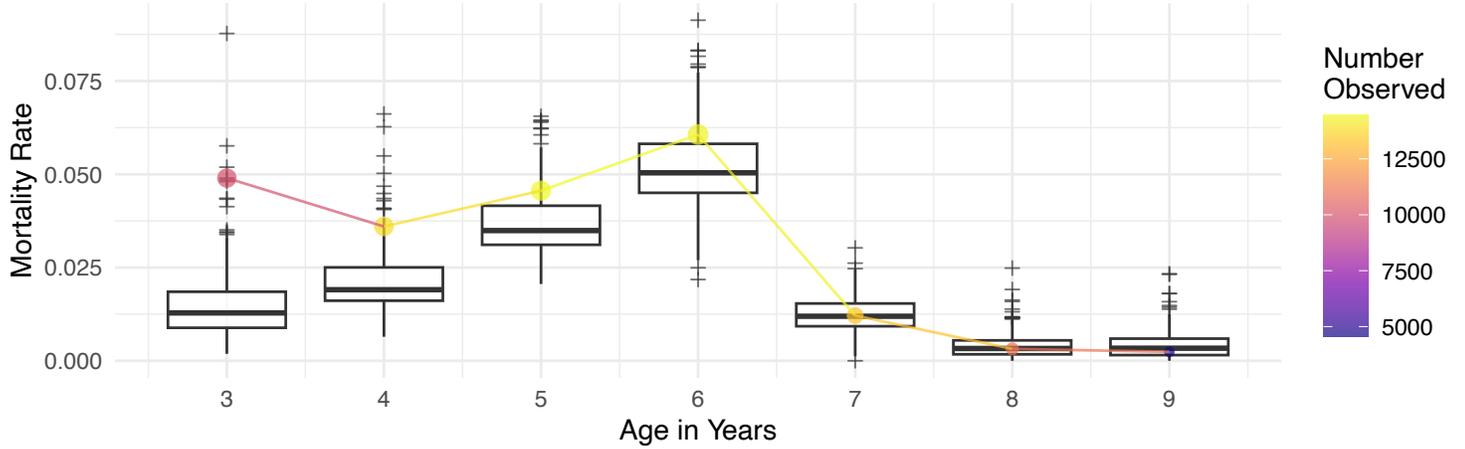

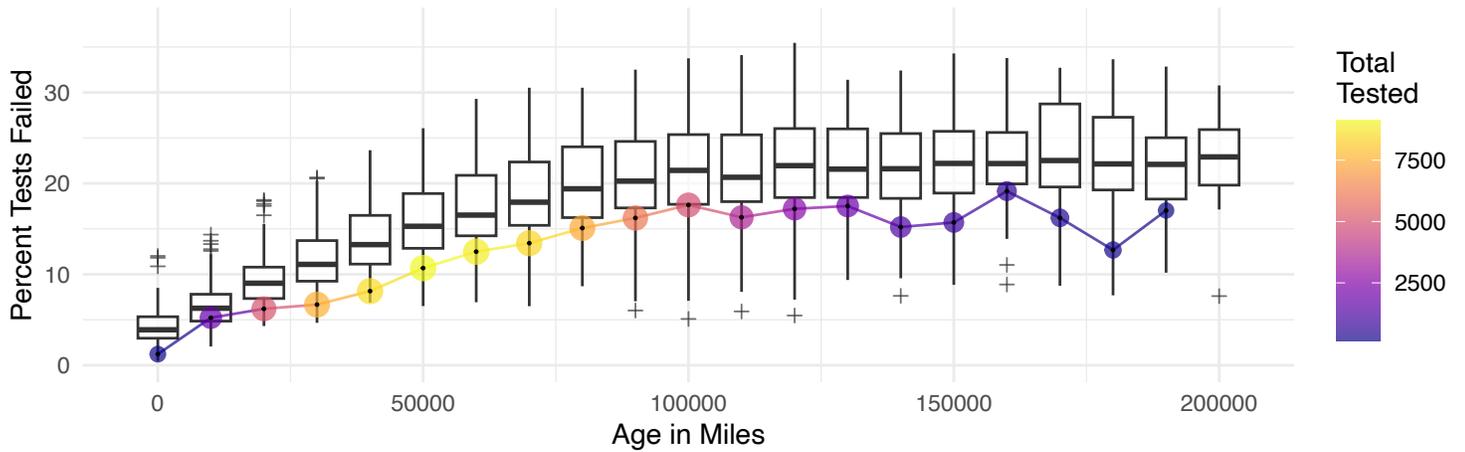

Mortality rates

| Age in Years | Observed | Died | Mortality Rate |
|---|---|---|---|
| 3 | 9920 | 486 | 0.04900 |
| 4 | 13779 | 496 | 0.03600 |
| 5 | 14431 | 658 | 0.04560 |
| 6 | 14263 | 865 | 0.06060 |
| 7 | 13186 | 160 | 0.01210 |
| 8 | 10953 | 35 | 0.00320 |
| 9 | 4566 | 11 | 0.00241 |

Mechanical Reliability Rates

| Mileage at test | N tested | Pct failed |
|---|---|---|
| 0 | 164 | 1.22 |
| 10000 | 1762 | 5.22 |
| 20000 | 5092 | 6.21 |
| 30000 | 7515 | 6.67 |
| 40000 | 8561 | 8.15 |
| 50000 | 9134 | 10.70 |
| 60000 | 8888 | 12.50 |
| 70000 | 8425 | 13.40 |
| 80000 | 7378 | 15.10 |
| 90000 | 6106 | 16.20 |
| 100000 | 4907 | 17.60 |
| 110000 | 3571 | 16.30 |
| 120000 | 2633 | 17.20 |
| 130000 | 1913 | 17.50 |
| 140000 | 1250 | 15.20 |
| 150000 | 866 | 15.70 |
| 160000 | 611 | 19.10 |

## Audi A4 2013

At 5 years of age, the mortality rate of a Audi A4 2013 (manufactured as a Car or Light Van) ranked number 182 out of 221 vehicles of the same age and type (any Car or Light Van constructed in 2013). One is the lowest (or best) and 221 the highest mortality rate. For vehicles reaching 20000 miles, its unreliability score (rate of failing an inspection) ranked 189 out of 215 vehicles of the same age, type, and mileage. One is the highest (or worst) and 215 the lowest rate of failing an inspection.

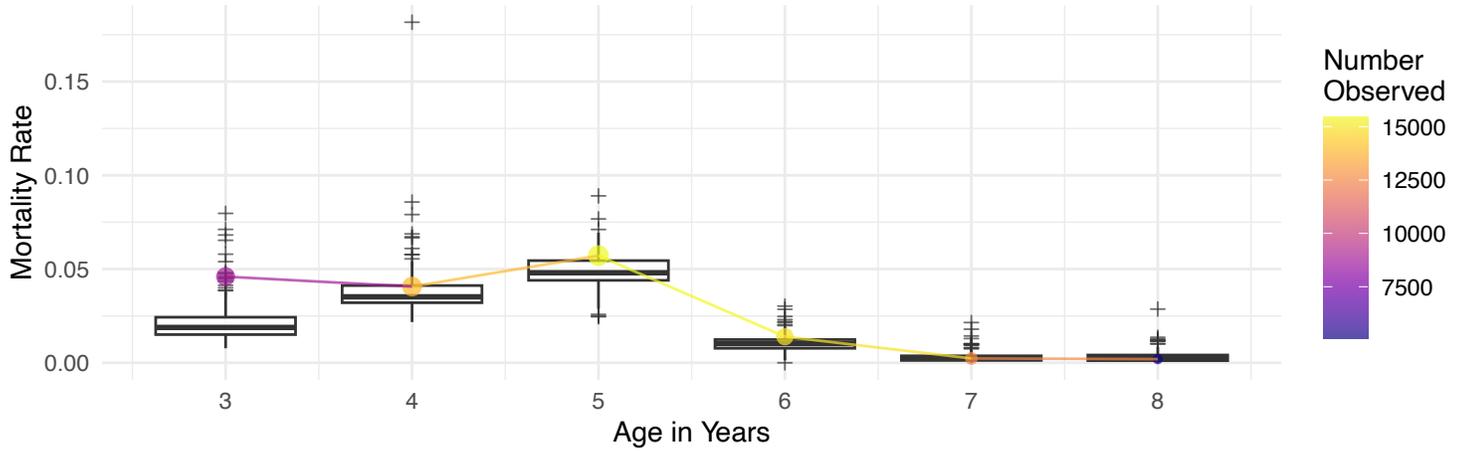

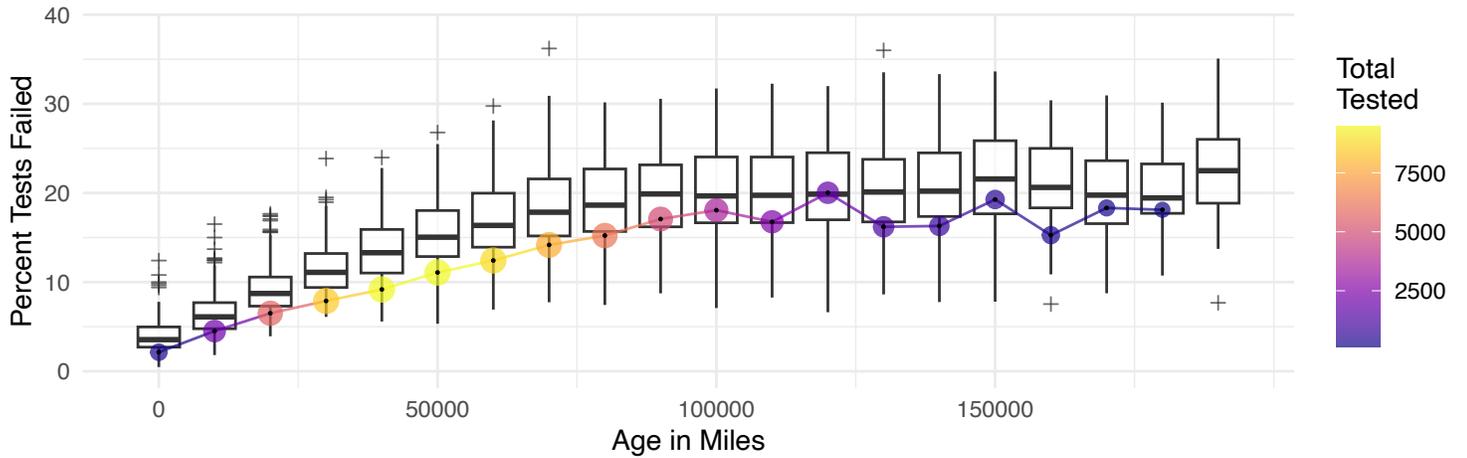

Mortality rates

| Age in Years | Observed | Died | Mortality Rate |
|---|---|---|---|
| 3 | 8790 | 404 | 0.04600 |
| 4 | 13873 | 564 | 0.04070 |
| 5 | 15422 | 881 | 0.05710 |
| 6 | 14846 | 207 | 0.01390 |
| 7 | 12451 | 29 | 0.00233 |
| 8 | 5084 | 10 | 0.00197 |

Mechanical Reliability Rates

| Mileage at test | N tested | Pct failed |
|---|---|---|
| 0 | 234 | 2.14 |
| 10000 | 2040 | 4.51 |
| 20000 | 5702 | 6.51 |
| 30000 | 8472 | 7.88 |
| 40000 | 9481 | 9.18 |
| 50000 | 9460 | 11.10 |
| 60000 | 8821 | 12.40 |
| 70000 | 7747 | 14.20 |
| 80000 | 6269 | 15.20 |
| 90000 | 4919 | 17.10 |
| 100000 | 3531 | 18.10 |
| 110000 | 2517 | 16.80 |
| 120000 | 1803 | 20.00 |
| 130000 | 1185 | 16.20 |
| 140000 | 773 | 16.30 |
| 150000 | 519 | 19.30 |
| 160000 | 321 | 15.30 |

# Audi A4 2014

At 5 years of age, the mortality rate of a Audi A4 2014 (manufactured as a Car or Light Van) ranked number 202 out of 236 vehicles of the same age and type (any Car or Light Van constructed in 2014). One is the lowest (or best) and 236 the highest mortality rate. For vehicles reaching 40000 miles, its unreliability score (rate of failing an inspection) ranked 206 out of 230 vehicles of the same age, type, and mileage. One is the highest (or worst) and 230 the lowest rate of failing an inspection.

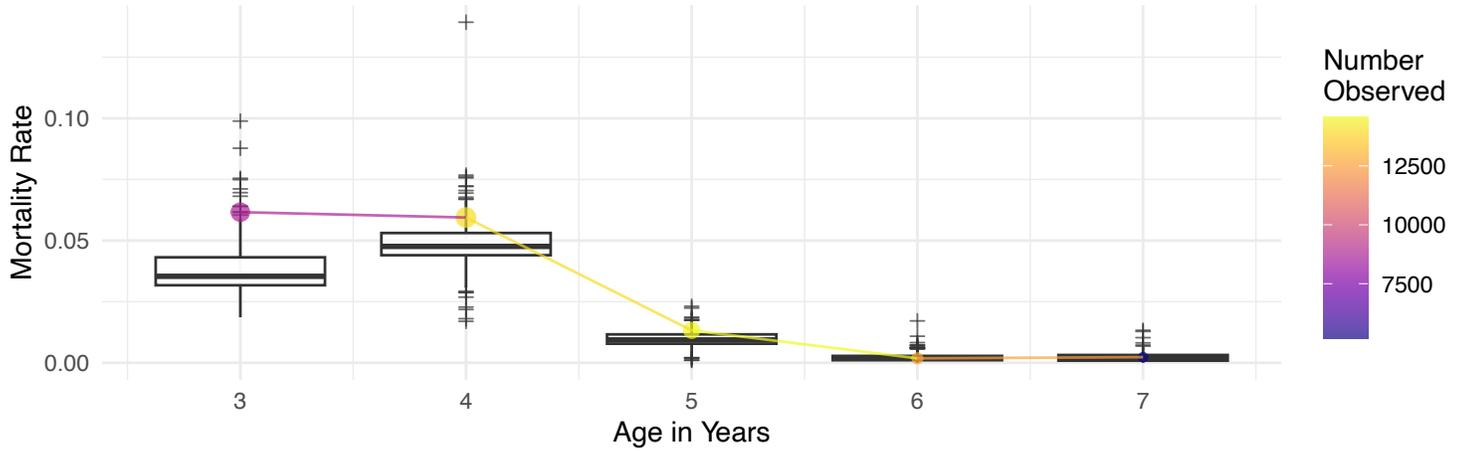

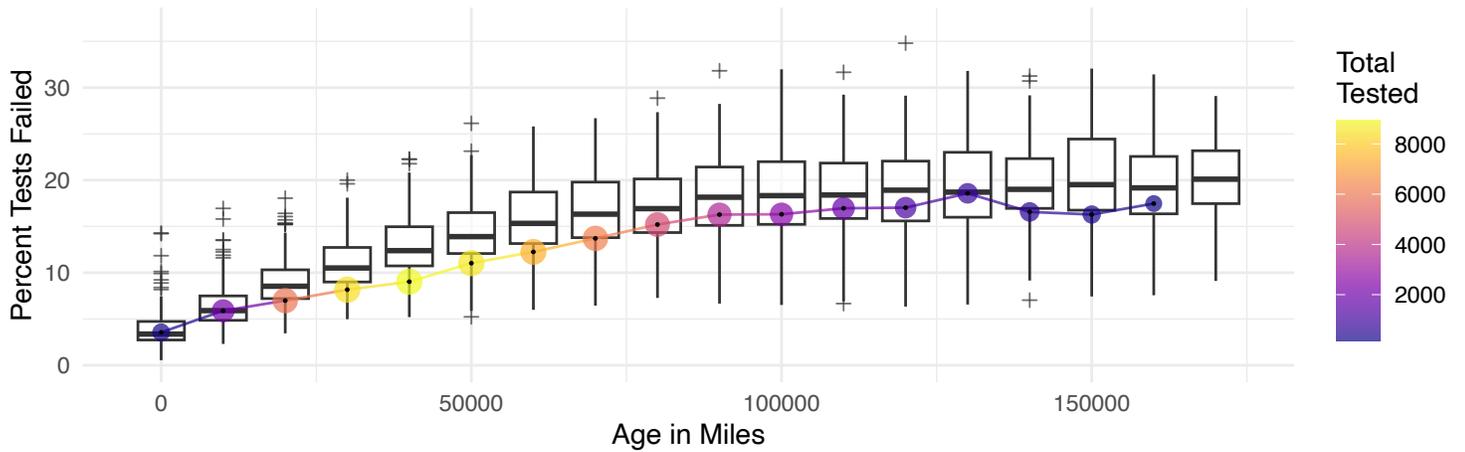

### Mortality rates

| Age in Years | Observed | Died | Mortality Rate |
|---|---|---|---|
| 3 | 8715 | 537 | 0.06160 |
| 4 | 13933 | 828 | 0.05940 |
| 5 | 14546 | 193 | 0.01330 |
| 6 | 12595 | 23 | 0.00183 |
| 7 | 5183 | 12 | 0.00232 |

### Mechanical Reliability Rates

| Mileage at test | N tested | Pct failed |
|---|---|---|
| 0 | 253 | 3.56 |
| 10000 | 2105 | 5.89 |
| 20000 | 5954 | 6.99 |
| 30000 | 8437 | 8.17 |
| 40000 | 8949 | 9.03 |
| 50000 | 8639 | 11.00 |
| 60000 | 7421 | 12.20 |
| 70000 | 6193 | 13.70 |
| 80000 | 4787 | 15.20 |
| 90000 | 3581 | 16.30 |
| 100000 | 2568 | 16.30 |
| 110000 | 1804 | 17.00 |
| 120000 | 1192 | 17.00 |
| 130000 | 802 | 18.60 |
| 140000 | 525 | 16.60 |
| 150000 | 307 | 16.30 |
| 160000 | 166 | 17.50 |

**Audi A4 2015**

At 5 years of age, the mortality rate of a Audi A4 2015 (manufactured as a Car or Light Van) ranked number 97 out of 247 vehicles of the same age and type (any Car or Light Van constructed in 2015). One is the lowest (or best) and 247 the highest mortality rate. For vehicles reaching 20000 miles, its unreliability score (rate of failing an inspection) ranked 186 out of 241 vehicles of the same age, type, and mileage. One is the highest (or worst) and 241 the lowest rate of failing an inspection.

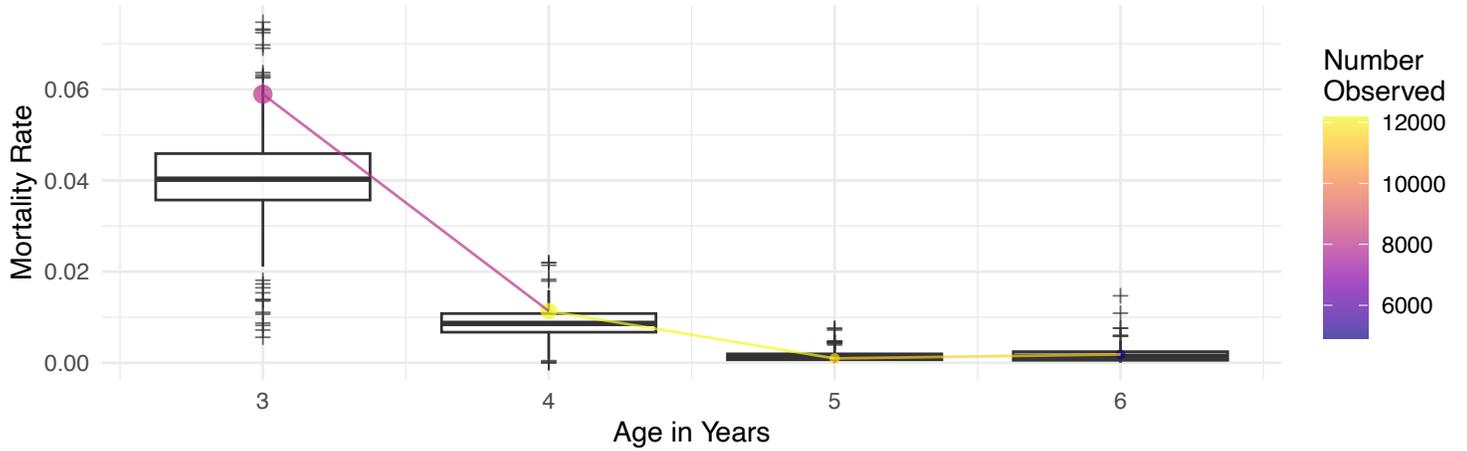

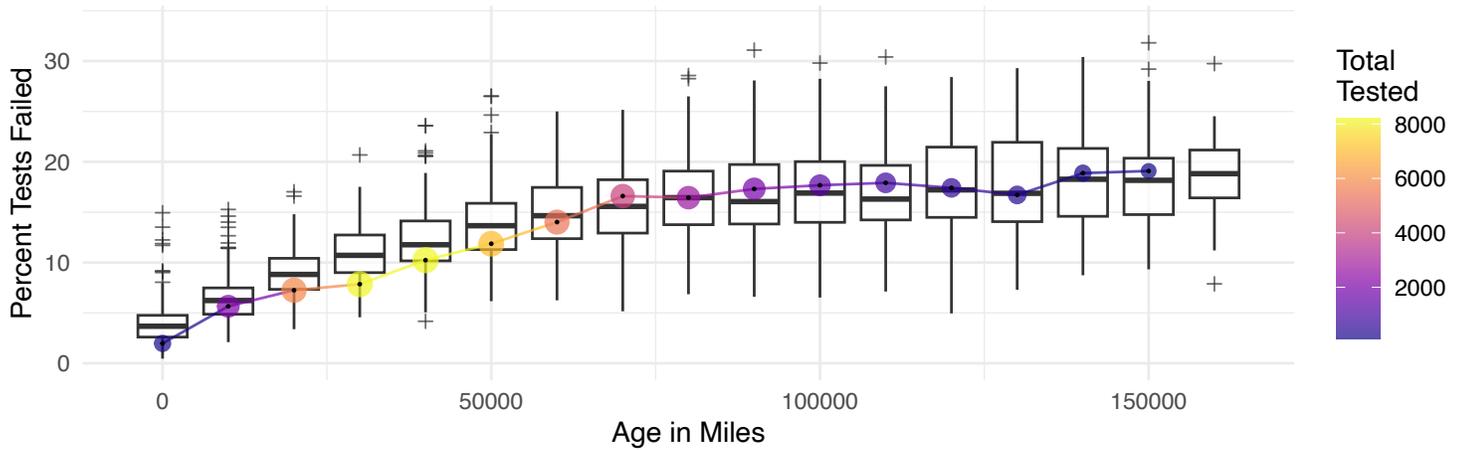

Mortality rates

| Age in Years | Observed | Died | Mortality Rate |
|---|---|---|---|
| 3 | 7992 | 471 | 0.058900 |
| 4 | 12156 | 138 | 0.011400 |
| 5 | 11433 | 11 | 0.000962 |
| 6 | 4918 | 9 | 0.001830 |

Mechanical Reliability Rates

| Mileage at test | N tested | Pct failed |
|---|---|---|
| 0 | 203 | 1.97 |
| 10000 | 2068 | 5.66 |
| 20000 | 5848 | 7.25 |
| 30000 | 8164 | 7.85 |
| 40000 | 8216 | 10.20 |
| 50000 | 7061 | 11.90 |
| 60000 | 5394 | 14.00 |
| 70000 | 3988 | 16.60 |
| 80000 | 2755 | 16.40 |
| 90000 | 1970 | 17.30 |
| 100000 | 1341 | 17.70 |
| 110000 | 843 | 17.90 |
| 120000 | 534 | 17.40 |
| 130000 | 365 | 16.70 |
| 140000 | 196 | 18.90 |
| 150000 | 110 | 19.10 |

**Audi A4 2016**

At 5 years of age, the mortality rate of a Audi A4 2016 (manufactured as a Car or Light Van) ranked number 124 out of 252 vehicles of the same age and type (any Car or Light Van constructed in 2016). One is the lowest (or best) and 252 the highest mortality rate. For vehicles reaching 20000 miles, its unreliability score (rate of failing an inspection) ranked 184 out of 246 vehicles of the same age, type, and mileage. One is the highest (or worst) and 246 the lowest rate of failing an inspection.

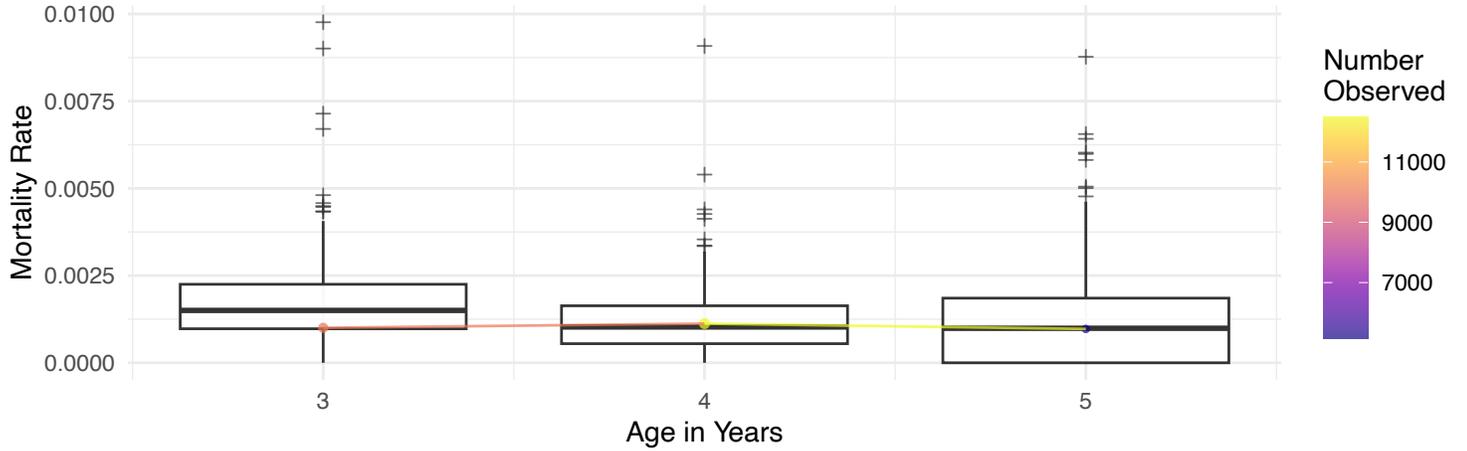

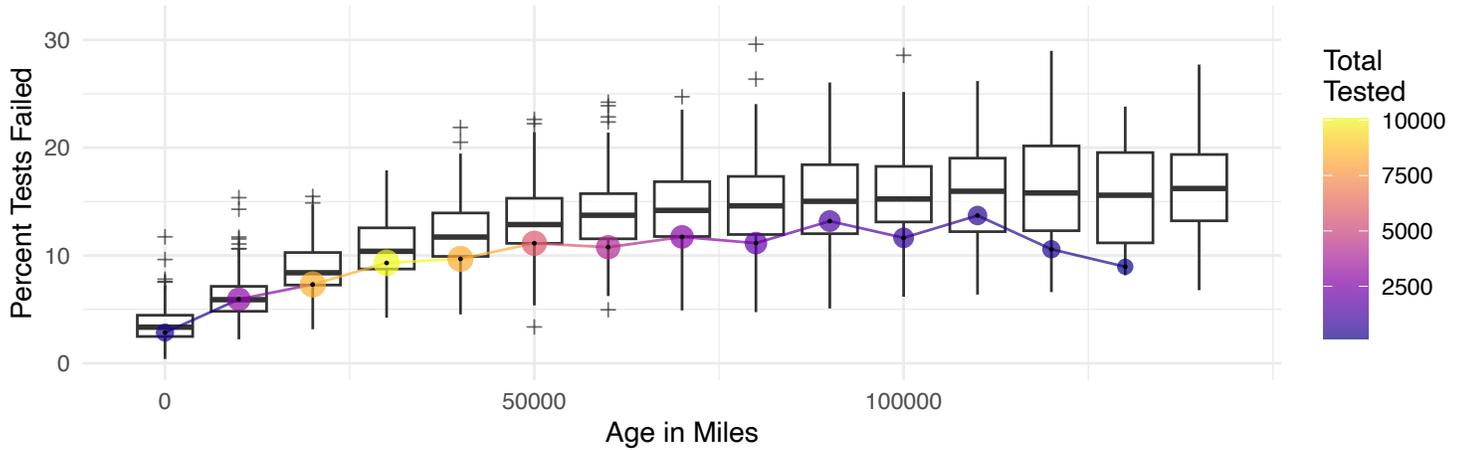

Mortality rates

| Age in Years | Observed | Died | Mortality Rate |
|---|---|---|---|
| 3 | 9969 | 10 | 0.00100 |
| 4 | 12470 | 14 | 0.00112 |
| 5 | 5152 | 5 | 0.00097 |

Mechanical Reliability Rates

| Mileage at test | N tested | Pct failed |
|---|---|---|
| 0 | 281 | 2.85 |
| 10000 | 2947 | 5.94 |
| 20000 | 8360 | 7.31 |
| 30000 | 10091 | 9.31 |
| 40000 | 8211 | 9.68 |
| 50000 | 5740 | 11.10 |
| 60000 | 4023 | 10.80 |
| 70000 | 2858 | 11.70 |
| 80000 | 1875 | 11.10 |
| 90000 | 1342 | 13.20 |
| 100000 | 825 | 11.60 |
| 110000 | 518 | 13.70 |
| 120000 | 312 | 10.60 |
| 130000 | 134 | 8.96 |

**Audi A4 2017**

At 3 years of age, the mortality rate of a Audi A4 2017 (manufactured as a Car or Light Van) ranked number 113 out of 247 vehicles of the same age and type (any Car or Light Van constructed in 2017). One is the lowest (or best) and 247 the highest mortality rate. For vehicles reaching 20000 miles, its unreliability score (rate of failing an inspection) ranked 176 out of 240 vehicles of the same age, type, and mileage. One is the highest (or worst) and 240 the lowest rate of failing an inspection.

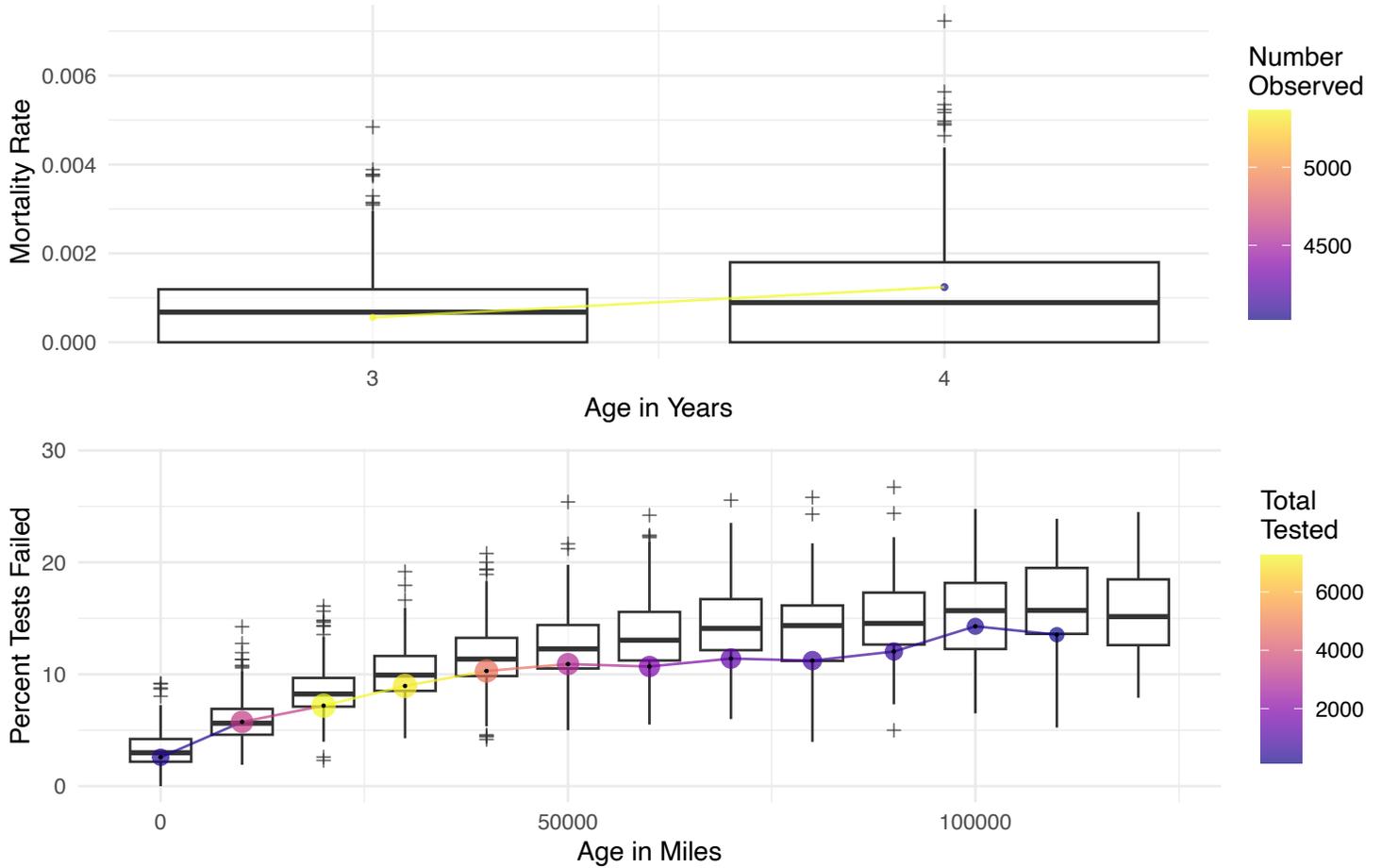

Mortality rates

| Age in Years | Observed | Died | Mortality Rate |
|---|---|---|---|
| 3 | 5360 | 3 | 0.00056 |
| 4 | 4027 | 5 | 0.00124 |

Mechanical Reliability Rates

| Mileage at test | N tested | Pct failed |
|---|---|---|
| 0 | 349 | 2.58 |
| 10000 | 3342 | 5.75 |
| 20000 | 7273 | 7.19 |
| 30000 | 7046 | 8.96 |
| 40000 | 4663 | 10.30 |
| 50000 | 2933 | 10.90 |
| 60000 | 1861 | 10.70 |
| 70000 | 1272 | 11.40 |
| 80000 | 811 | 11.20 |
| 90000 | 507 | 12.00 |
| 100000 | 294 | 14.30 |
| 110000 | 133 | 13.50 |

**Audi A4 2018**

At 3 years of age, the mortality rate of a Audi A4 2018 (manufactured as a Car or Light Van) ranked number 3 out of 222 vehicles of the same age and type (any Car or Light Van constructed in 2018). One is the lowest (or best) and 222 the highest mortality rate. For vehicles reaching 20000 miles, its unreliability score (rate of failing an inspection) ranked 93 out of 215 vehicles of the same age, type, and mileage. One is the highest (or worst) and 215 the lowest rate of failing an inspection.

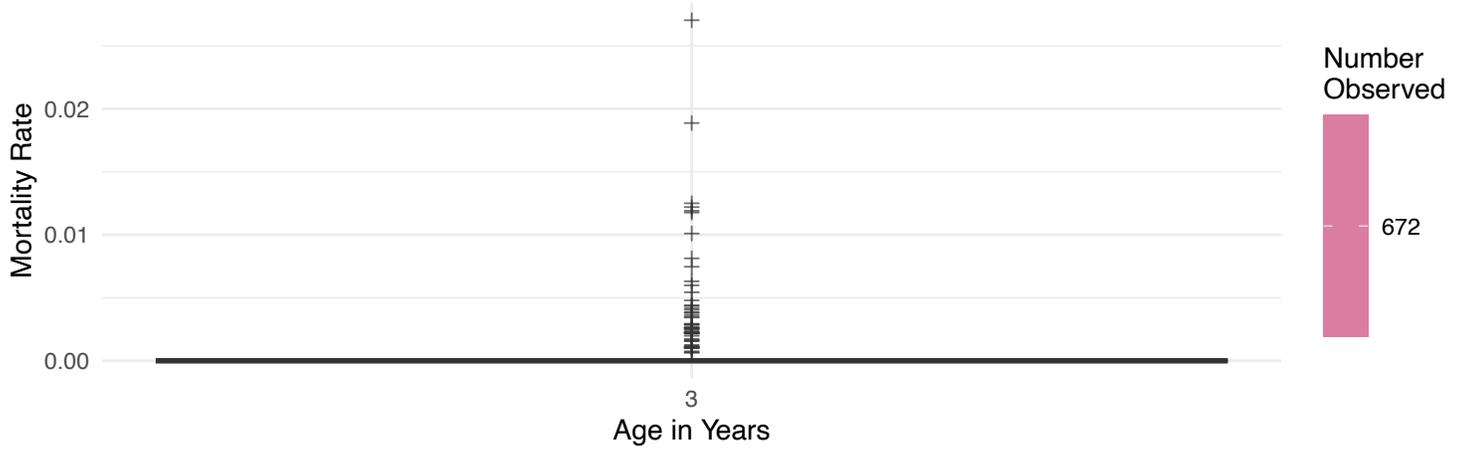

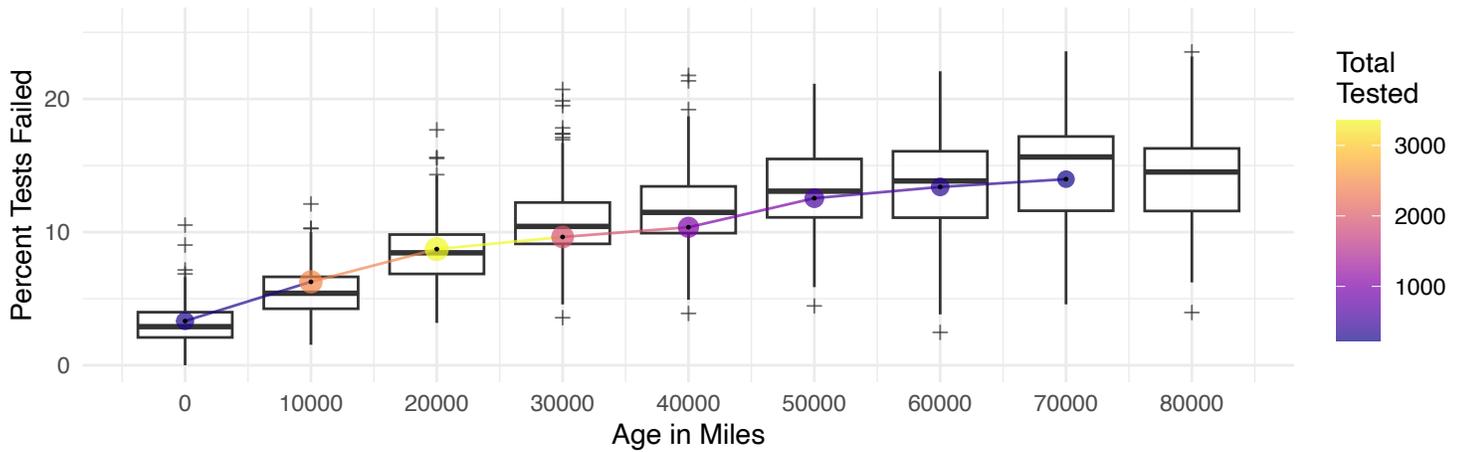

Mortality rates

| Age in Years | Observed | Died | Mortality Rate |
|---|---|---|---|
| 3 | 672 | 0 | 0 |

Mechanical Reliability Rates

| Mileage at test | N tested | Pct failed |
|---|---|---|
| 0 | 301 | 3.32 |
| 10000 | 2442 | 6.27 |
| 20000 | 3359 | 8.72 |
| 30000 | 1941 | 9.63 |
| 40000 | 1023 | 10.40 |
| 50000 | 566 | 12.50 |
| 60000 | 351 | 13.40 |
| 70000 | 229 | 14.00 |

# Audi A5 2007

At 5 years of age, the mortality rate of a Audi A5 2007 (manufactured as a Car or Light Van) ranked number 16 out of 219 vehicles of the same age and type (any Car or Light Van constructed in 2007). One is the lowest (or best) and 219 the highest mortality rate. For vehicles reaching 20000 miles, its unreliability score (rate of failing an inspection) ranked 135 out of 214 vehicles of the same age, type, and mileage. One is the highest (or worst) and 214 the lowest rate of failing an inspection.

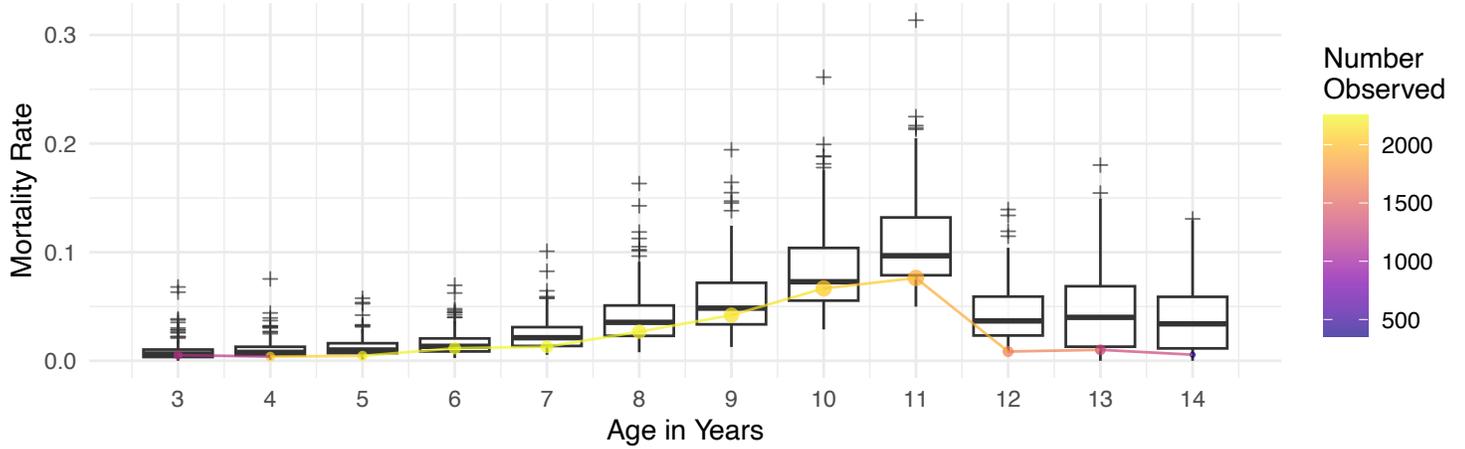

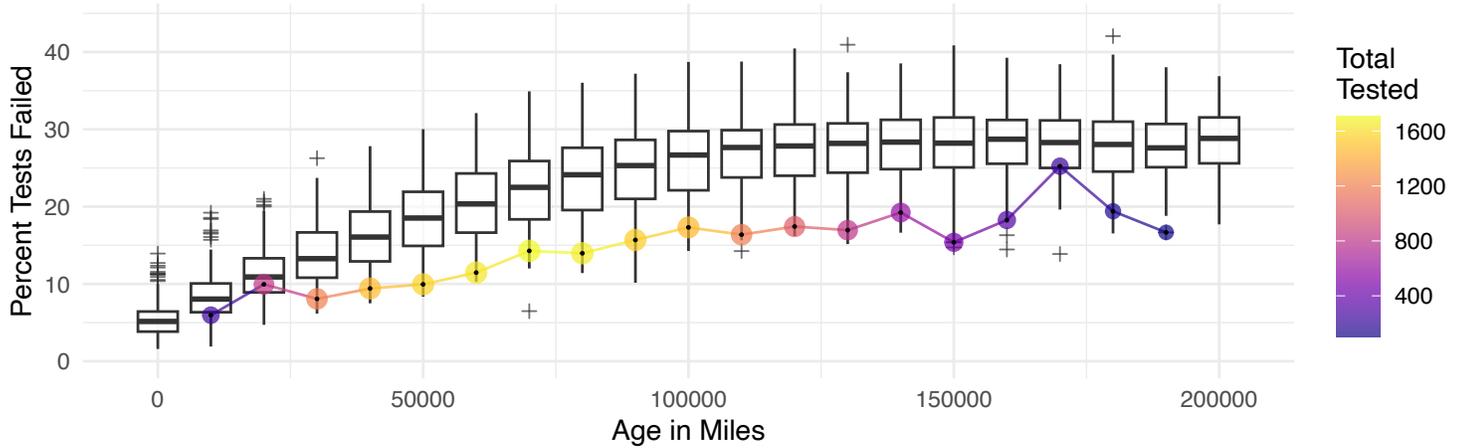

<table>
<tr><td colspan="4" align="center">Mortality rates</td></tr>
</table>

| Age in Years | Observed | Died | Mortality Rate |
|---|---|---|---|
| 3 | 1155 | 6 | 0.00519 |
| 4 | 2052 | 8 | 0.00390 |
| 5 | 2210 | 10 | 0.00452 |
| 6 | 2249 | 26 | 0.01160 |
| 7 | 2241 | 29 | 0.01290 |
| 8 | 2203 | 59 | 0.02680 |
| 9 | 2142 | 90 | 0.04200 |
| 10 | 2040 | 136 | 0.06670 |
| 11 | 1889 | 144 | 0.07620 |
| 12 | 1644 | 14 | 0.00852 |
| 13 | 1286 | 13 | 0.01010 |
| 14 | 356 | 2 | 0.00562 |

Mechanical Reliability Rates

| Mileage at test | N tested | Pct failed |
|---|---|---|
| 10000 | 235 | 5.96 |
| 20000 | 824 | 9.95 |
| 30000 | 1202 | 8.07 |
| 40000 | 1477 | 9.41 |
| 50000 | 1568 | 9.95 |
| 60000 | 1640 | 11.50 |
| 70000 | 1709 | 14.30 |
| 80000 | 1668 | 14.00 |
| 90000 | 1549 | 15.70 |
| 100000 | 1467 | 17.30 |
| 110000 | 1197 | 16.40 |
| 120000 | 1033 | 17.40 |
| 130000 | 802 | 17.00 |
| 140000 | 603 | 19.20 |
| 150000 | 455 | 15.40 |
| 160000 | 312 | 18.30 |
| 170000 | 218 | 25.20 |

## Audi A5 2008

At 5 years of age, the mortality rate of a Audi A5 2008 (manufactured as a Car or Light Van) ranked number 93 out of 218 vehicles of the same age and type (any Car or Light Van constructed in 2008). One is the lowest (or best) and 218 the highest mortality rate. For vehicles reaching 20000 miles, its unreliability score (rate of failing an inspection) ranked 171 out of 212 vehicles of the same age, type, and mileage. One is the highest (or worst) and 212 the lowest rate of failing an inspection.

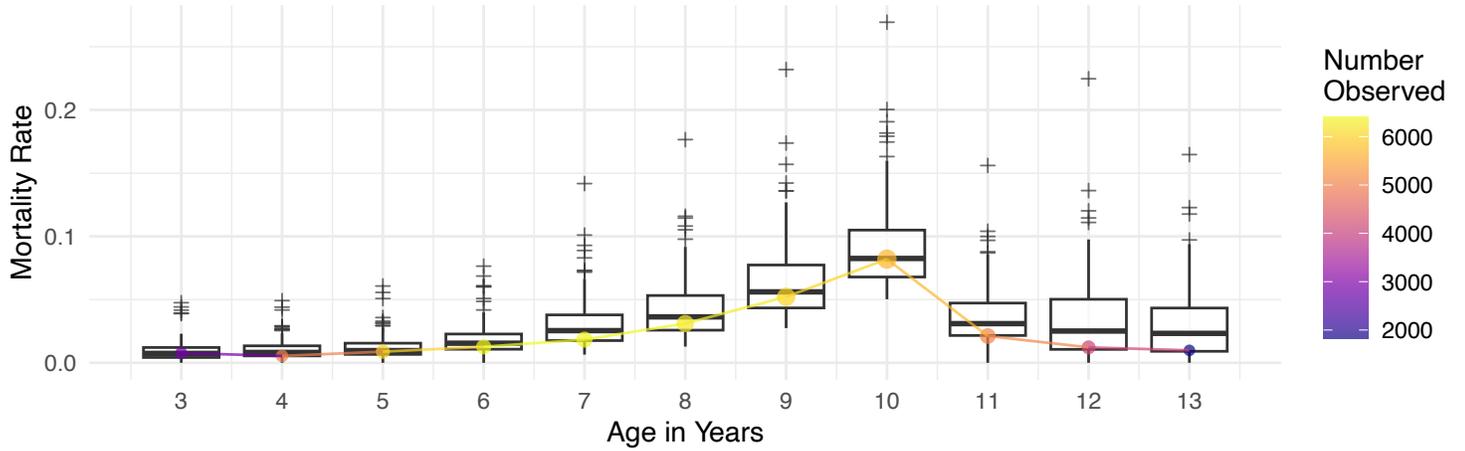

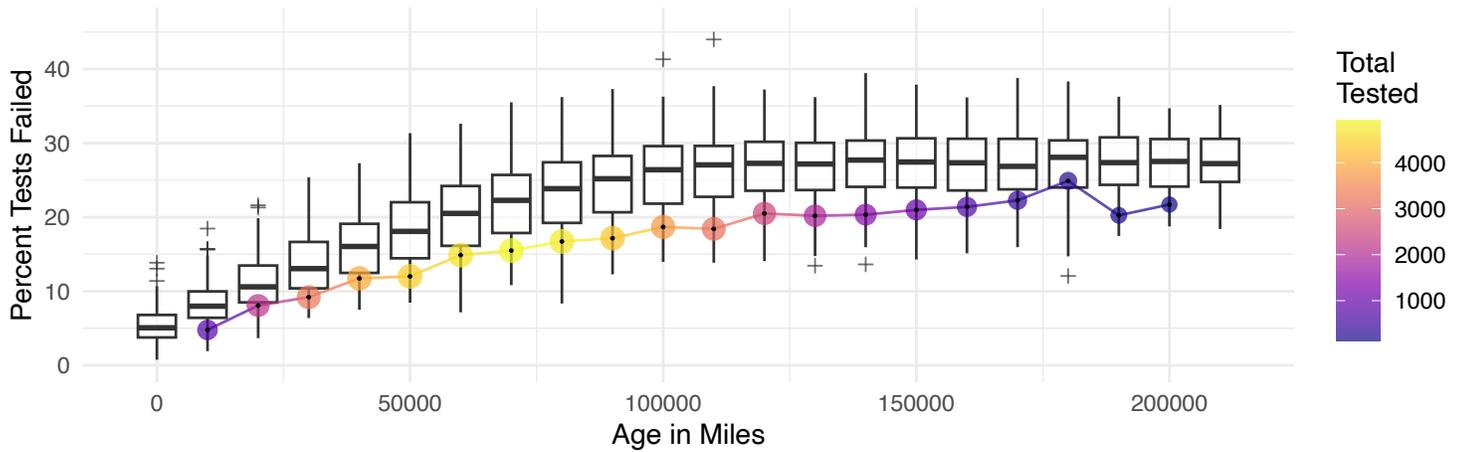

### Mortality rates

| Age in Years | Observed | Died | Mortality Rate |
|---|---|---|---|
| 3 | 2836 | 22 | 0.00776 |
| 4 | 4979 | 27 | 0.00542 |
| 5 | 6005 | 53 | 0.00883 |
| 6 | 6394 | 81 | 0.01270 |
| 7 | 6368 | 117 | 0.01840 |
| 8 | 6243 | 194 | 0.03110 |
| 9 | 6016 | 314 | 0.05220 |
| 10 | 5659 | 464 | 0.08200 |
| 11 | 5048 | 107 | 0.02120 |
| 12 | 4097 | 50 | 0.01220 |
| 13 | 1828 | 18 | 0.00985 |

### Mechanical Reliability Rates

| Mileage at test | N tested | Pct failed |
|---|---|---|
| 10000 | 817 | 4.77 |
| 20000 | 2182 | 8.07 |
| 30000 | 3295 | 9.20 |
| 40000 | 3989 | 11.70 |
| 50000 | 4439 | 12.00 |
| 60000 | 4684 | 14.90 |
| 70000 | 4926 | 15.50 |
| 80000 | 4754 | 16.70 |
| 90000 | 4336 | 17.20 |
| 100000 | 3753 | 18.70 |
| 110000 | 3176 | 18.40 |
| 120000 | 2483 | 20.50 |
| 130000 | 1902 | 20.20 |
| 140000 | 1421 | 20.30 |
| 150000 | 1015 | 21.00 |
| 160000 | 692 | 21.40 |
| 190000 | 163 | 20.20 |

## Audi A5 2009

At 5 years of age, the mortality rate of a Audi A5 2009 (manufactured as a Car or Light Van) ranked number 117 out of 205 vehicles of the same age and type (any Car or Light Van constructed in 2009). One is the lowest (or best) and 205 the highest mortality rate. For vehicles reaching 20000 miles, its unreliability score (rate of failing an inspection) ranked 165 out of 200 vehicles of the same age, type, and mileage. One is the highest (or worst) and 200 the lowest rate of failing an inspection.

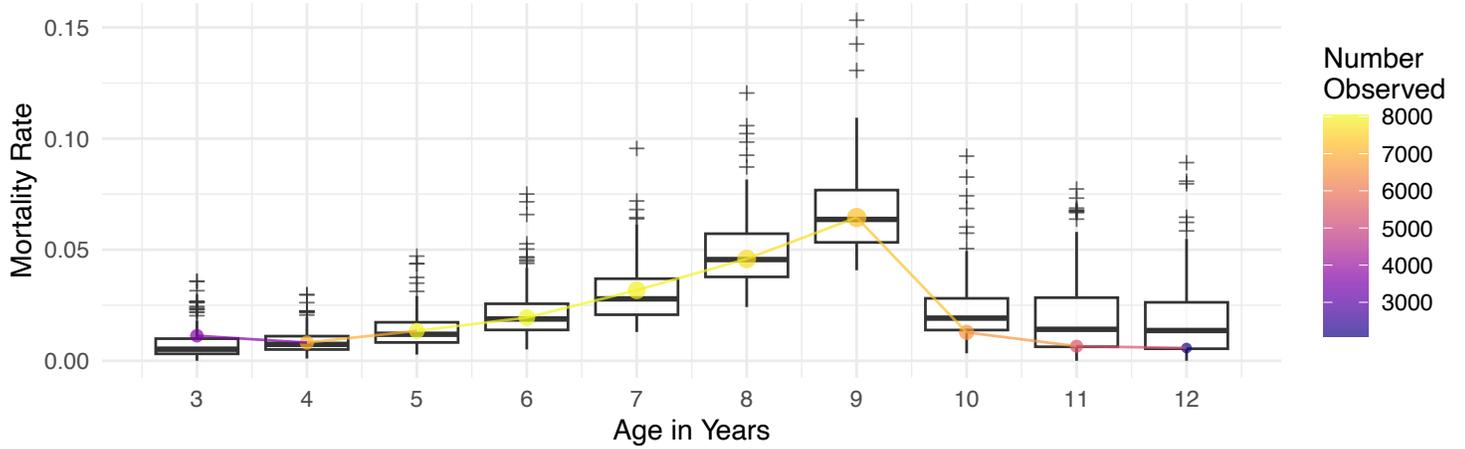

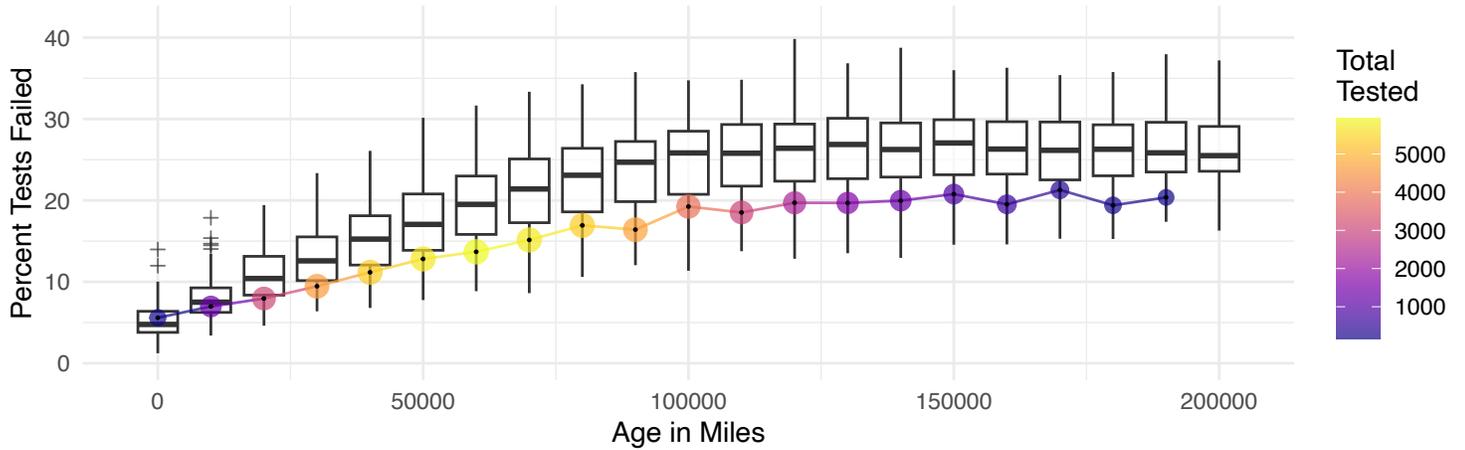

### Mortality rates

| Age in Years | Observed | Died | Mortality Rate |
|---|---|---|---|
| 3 | 3637 | 41 | 0.01130 |
| 4 | 6973 | 57 | 0.00817 |
| 5 | 7938 | 107 | 0.01350 |
| 6 | 8023 | 156 | 0.01940 |
| 7 | 7904 | 251 | 0.03180 |
| 8 | 7632 | 350 | 0.04590 |
| 9 | 7243 | 467 | 0.06450 |
| 10 | 6565 | 84 | 0.01280 |
| 11 | 5344 | 35 | 0.00655 |
| 12 | 2085 | 12 | 0.00576 |

### Mechanical Reliability Rates

| Mileage at test | N tested | Pct failed |
|---|---|---|
| 0 | 197 | 5.58 |
| 10000 | 1291 | 6.97 |
| 20000 | 3081 | 7.95 |
| 30000 | 4611 | 9.46 |
| 40000 | 5345 | 11.20 |
| 50000 | 5705 | 12.80 |
| 60000 | 5940 | 13.70 |
| 70000 | 5744 | 15.10 |
| 80000 | 5339 | 16.90 |
| 90000 | 4693 | 16.40 |
| 100000 | 3881 | 19.20 |
| 110000 | 3019 | 18.50 |
| 120000 | 2334 | 19.70 |
| 130000 | 1671 | 19.70 |
| 140000 | 1262 | 20.00 |
| 150000 | 876 | 20.80 |
| 160000 | 584 | 19.50 |

**Audi A5 2010**

At 5 years of age, the mortality rate of a Audi A5 2010 (manufactured as a Car or Light Van) ranked number 125 out of 206 vehicles of the same age and type (any Car or Light Van constructed in 2010). One is the lowest (or best) and 206 the highest mortality rate. For vehicles reaching 20000 miles, its unreliability score (rate of failing an inspection) ranked 159 out of 201 vehicles of the same age, type, and mileage. One is the highest (or worst) and 201 the lowest rate of failing an inspection.

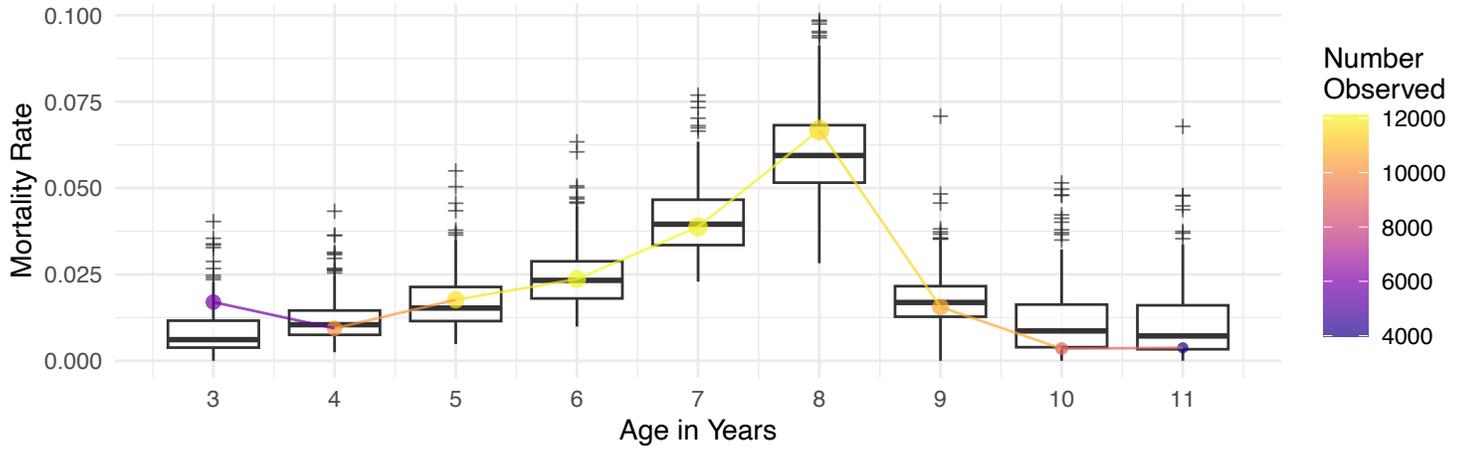

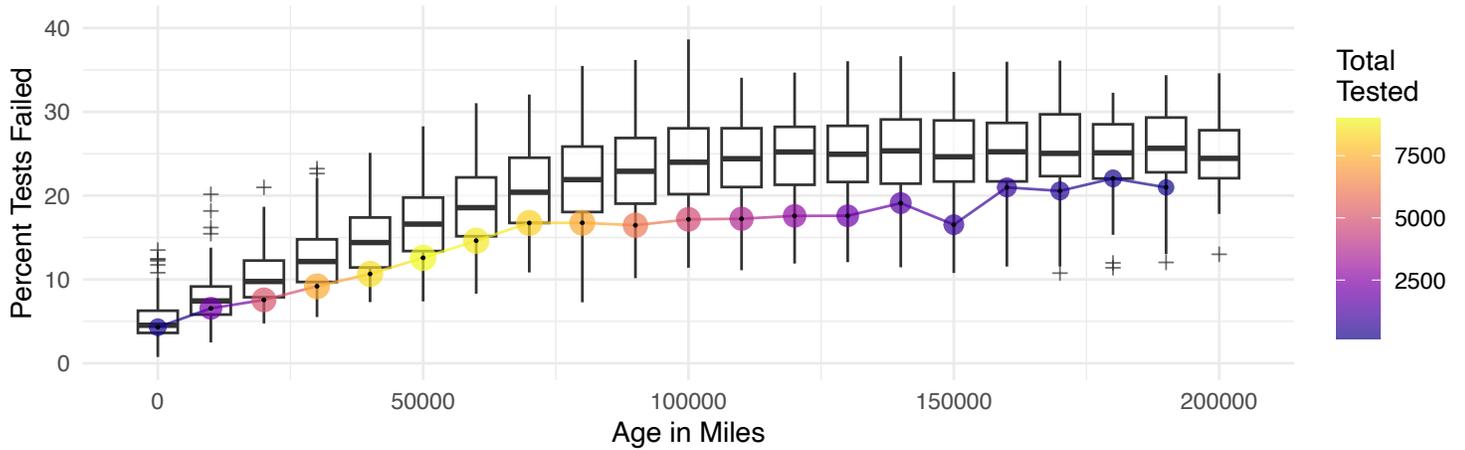

Mortality rates

| Age in Years | Observed | Died | Mortality Rate |
|---|---|---|---|
| 3 | 5766 | 98 | 0.01700 |
| 4 | 10114 | 95 | 0.00939 |
| 5 | 11716 | 207 | 0.01770 |
| 6 | 12098 | 286 | 0.02360 |
| 7 | 11940 | 463 | 0.03880 |
| 8 | 11497 | 768 | 0.06680 |
| 9 | 10522 | 164 | 0.01560 |
| 10 | 8792 | 31 | 0.00353 |
| 11 | 3977 | 15 | 0.00377 |

Mechanical Reliability Rates

| Mileage at test | N tested | Pct failed |
|---|---|---|
| 0 | 256 | 4.30 |
| 10000 | 2075 | 6.55 |
| 20000 | 5056 | 7.56 |
| 30000 | 7328 | 9.18 |
| 40000 | 8446 | 10.70 |
| 50000 | 9008 | 12.60 |
| 60000 | 8687 | 14.60 |
| 70000 | 8167 | 16.70 |
| 80000 | 7293 | 16.80 |
| 90000 | 6036 | 16.50 |
| 100000 | 4735 | 17.20 |
| 110000 | 3620 | 17.20 |
| 120000 | 2821 | 17.60 |
| 130000 | 1955 | 17.60 |
| 140000 | 1392 | 19.10 |
| 150000 | 877 | 16.50 |
| 160000 | 610 | 21.00 |

**Audi A5 2011**

At 5 years of age, the mortality rate of a Audi A5 2011 (manufactured as a Car or Light Van) ranked number 136 out of 211 vehicles of the same age and type (any Car or Light Van constructed in 2011). One is the lowest (or best) and 211 the highest mortality rate. For vehicles reaching 100000 miles, its unreliability score (rate of failing an inspection) ranked 154 out of 195 vehicles of the same age, type, and mileage. One is the highest (or worst) and 195 the lowest rate of failing an inspection.

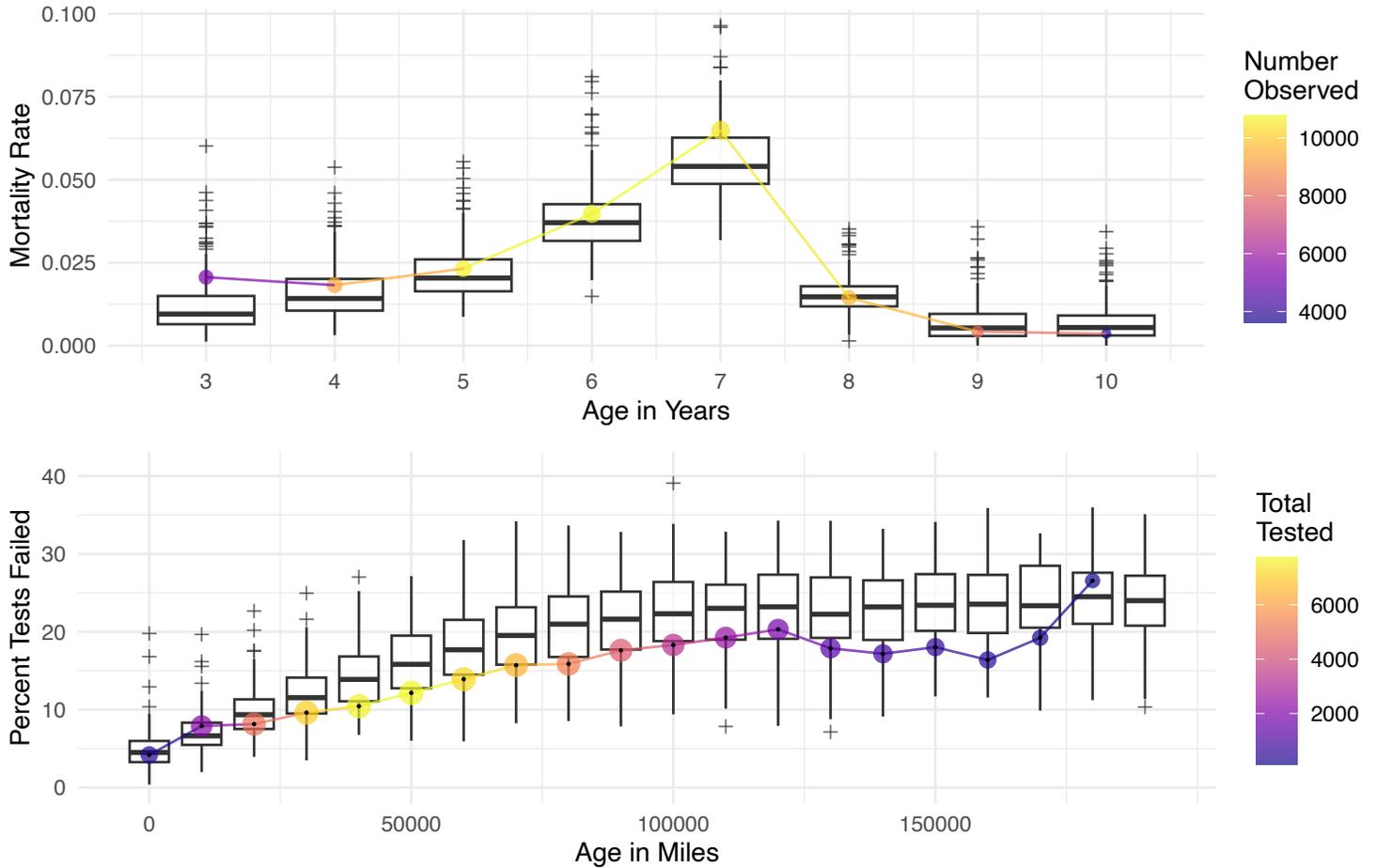

Mortality rates

| Age in Years | Observed | Died | Mortality Rate |
|---|---|---|---|
| 3 | 5335 | 110 | 0.02060 |
| 4 | 9382 | 171 | 0.01820 |
| 5 | 10697 | 248 | 0.02320 |
| 6 | 10779 | 428 | 0.03970 |
| 7 | 10453 | 678 | 0.06490 |
| 8 | 9627 | 139 | 0.01440 |
| 9 | 8069 | 34 | 0.00421 |
| 10 | 3616 | 13 | 0.00360 |

Mechanical Reliability Rates

| Mileage at test | N tested | Pct failed |
|---|---|---|
| 0 | 358 | 4.19 |
| 10000 | 2175 | 7.91 |
| 20000 | 5029 | 8.15 |
| 30000 | 6979 | 9.63 |
| 40000 | 7630 | 10.40 |
| 50000 | 7743 | 12.20 |
| 60000 | 7182 | 13.90 |
| 70000 | 6446 | 15.70 |
| 80000 | 5297 | 15.90 |
| 90000 | 4415 | 17.60 |
| 100000 | 3288 | 18.30 |
| 110000 | 2504 | 19.20 |
| 120000 | 1792 | 20.30 |
| 130000 | 1236 | 17.90 |
| 140000 | 815 | 17.20 |
| 160000 | 354 | 16.40 |
| 170000 | 208 | 19.20 |

**Audi A5 2012**

At 5 years of age, the mortality rate of a Audi A5 2012 (manufactured as a Car or Light Van) ranked number 195 out of 212 vehicles of the same age and type (any Car or Light Van constructed in 2012). One is the lowest (or best) and 212 the highest mortality rate. For vehicles reaching 100000 miles, its unreliability score (rate of failing an inspection) ranked 111 out of 183 vehicles of the same age, type, and mileage. One is the highest (or worst) and 183 the lowest rate of failing an inspection.

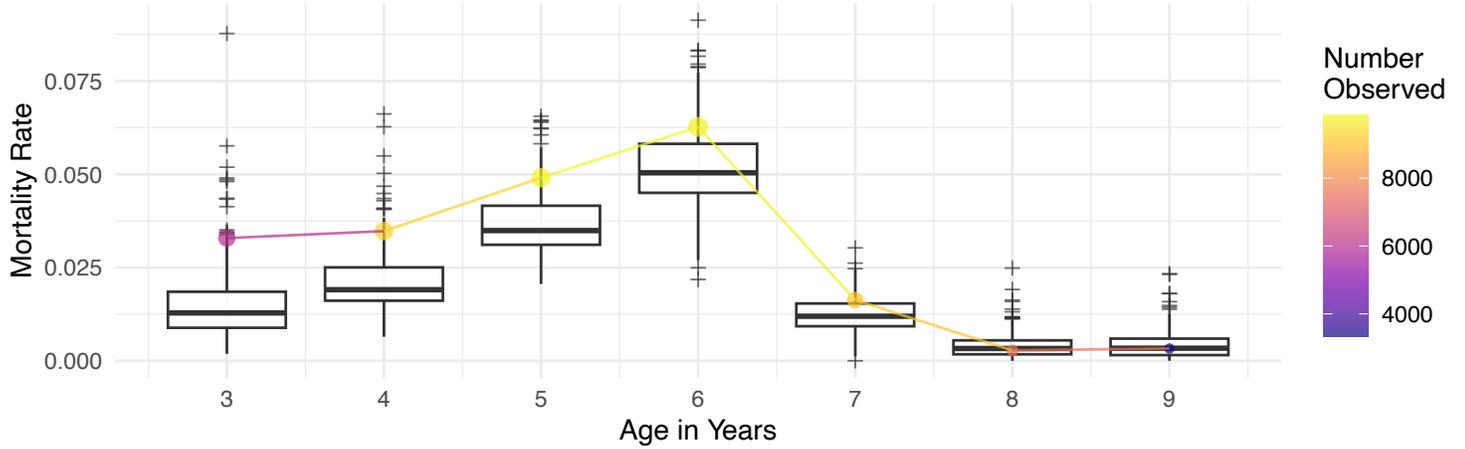

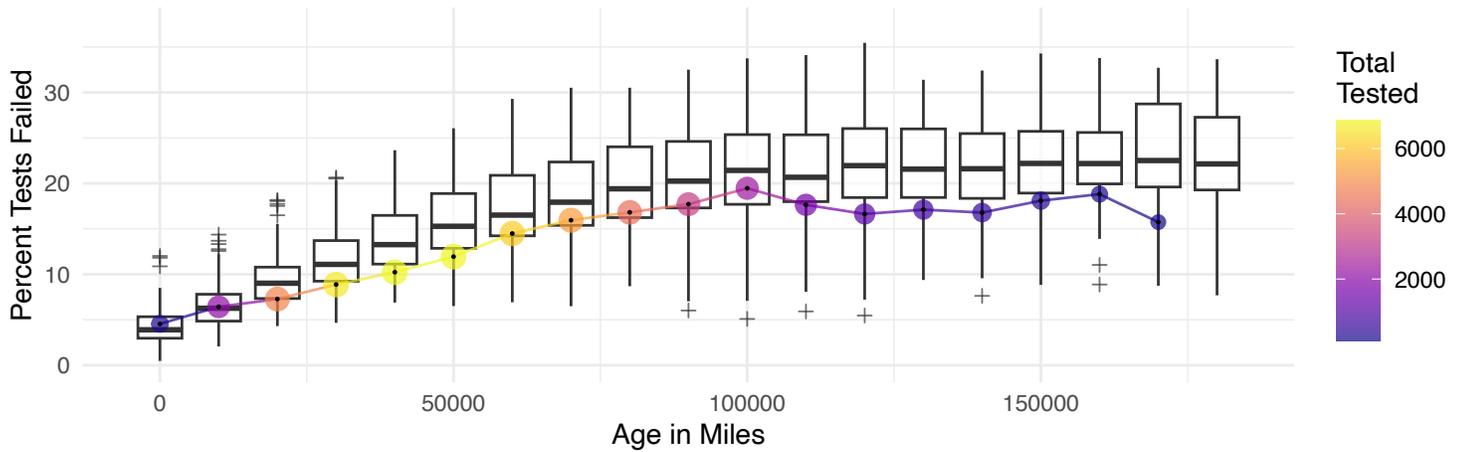

Mortality rates

| Age in Years | Observed | Died | Mortality Rate |
|---|---|---|---|
| 3 | 6020 | 198 | 0.03290 |
| 4 | 9262 | 322 | 0.03480 |
| 5 | 9845 | 483 | 0.04910 |
| 6 | 9743 | 611 | 0.06270 |
| 7 | 9031 | 147 | 0.01630 |
| 8 | 7539 | 21 | 0.00279 |
| 9 | 3338 | 11 | 0.00330 |

Mechanical Reliability Rates

| Mileage at test | N tested | Pct failed |
|---|---|---|
| 0 | 242 | 4.55 |
| 10000 | 2100 | 6.43 |
| 20000 | 4848 | 7.28 |
| 30000 | 6626 | 8.87 |
| 40000 | 6871 | 10.20 |
| 50000 | 6808 | 11.90 |
| 60000 | 6172 | 14.50 |
| 70000 | 5368 | 15.90 |
| 80000 | 4395 | 16.80 |
| 90000 | 3235 | 17.70 |
| 100000 | 2374 | 19.50 |
| 110000 | 1627 | 17.60 |
| 120000 | 1100 | 16.60 |
| 130000 | 713 | 17.10 |
| 140000 | 536 | 16.80 |
| 150000 | 326 | 18.10 |
| 170000 | 108 | 15.70 |

**Audi A5 2013**

At 5 years of age, the mortality rate of a Audi A5 2013 (manufactured as a Car or Light Van) ranked number 162 out of 221 vehicles of the same age and type (any Car or Light Van constructed in 2013). One is the lowest (or best) and 221 the highest mortality rate. For vehicles reaching 60000 miles, its unreliability score (rate of failing an inspection) ranked 137 out of 214 vehicles of the same age, type, and mileage. One is the highest (or worst) and 214 the lowest rate of failing an inspection.

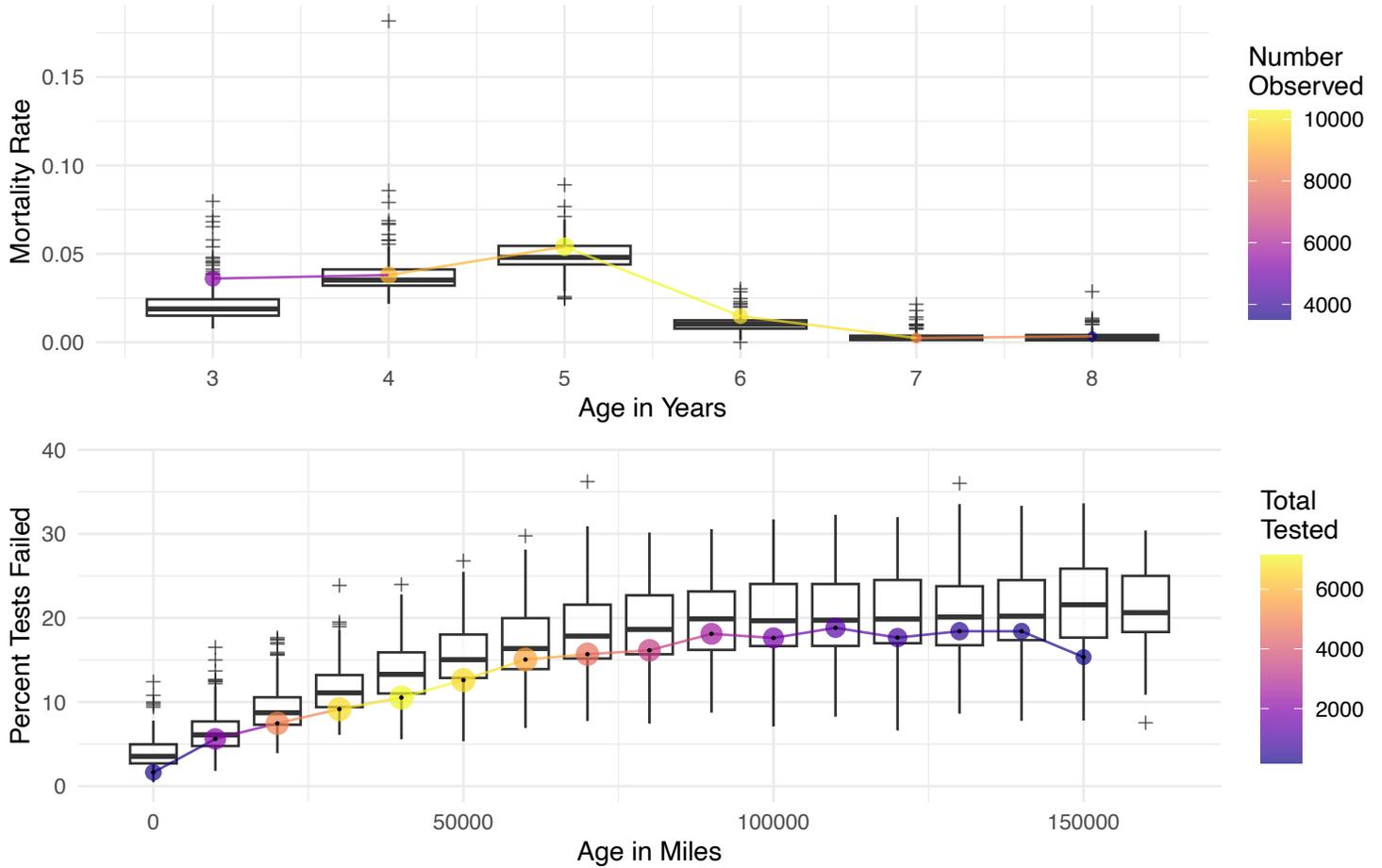

Mechanical Reliability Rates

| Mileage at test | N tested | Pct failed |
|---|---|---|
| 0 | 244 | 1.64 |
| 10000 | 2047 | 5.62 |
| 20000 | 5020 | 7.47 |
| 30000 | 6691 | 9.16 |
| 40000 | 7147 | 10.50 |
| 50000 | 6653 | 12.60 |
| 60000 | 5707 | 15.10 |
| 70000 | 4609 | 15.70 |
| 80000 | 3406 | 16.10 |
| 90000 | 2573 | 18.10 |
| 100000 | 1748 | 17.60 |
| 110000 | 1174 | 18.80 |
| 120000 | 804 | 17.70 |
| 130000 | 532 | 18.40 |
| 140000 | 277 | 18.40 |
| 150000 | 176 | 15.30 |

## Audi A5 2014

At 5 years of age, the mortality rate of a Audi A5 2014 (manufactured as a Car or Light Van) ranked number 211 out of 236 vehicles of the same age and type (any Car or Light Van constructed in 2014). One is the lowest (or best) and 236 the highest mortality rate. For vehicles reaching 20000 miles, its unreliability score (rate of failing an inspection) ranked 141 out of 230 vehicles of the same age, type, and mileage. One is the highest (or worst) and 230 the lowest rate of failing an inspection.

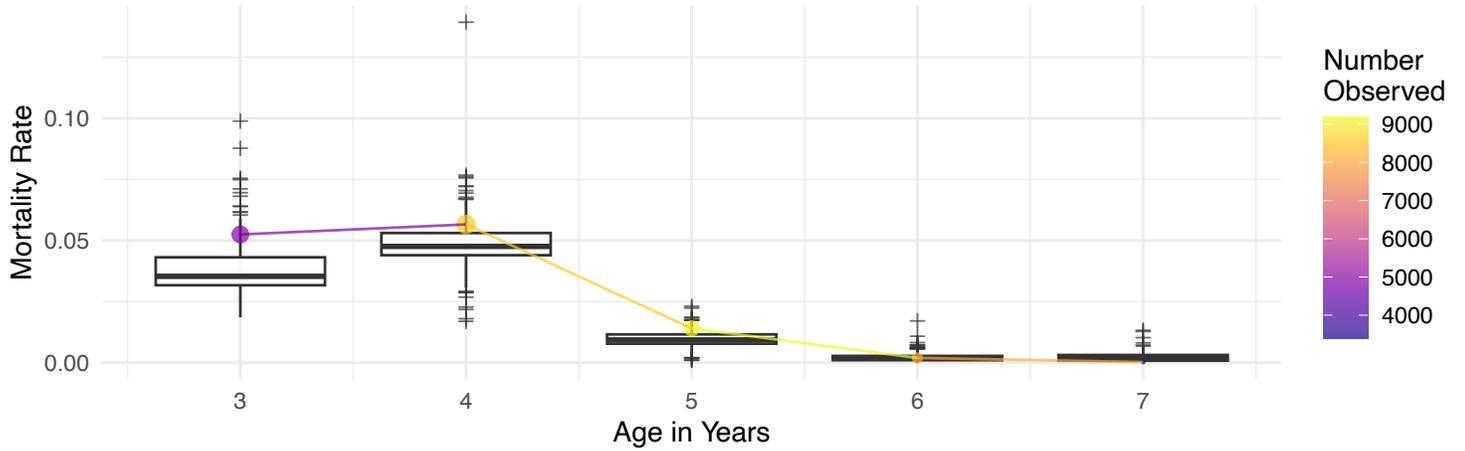

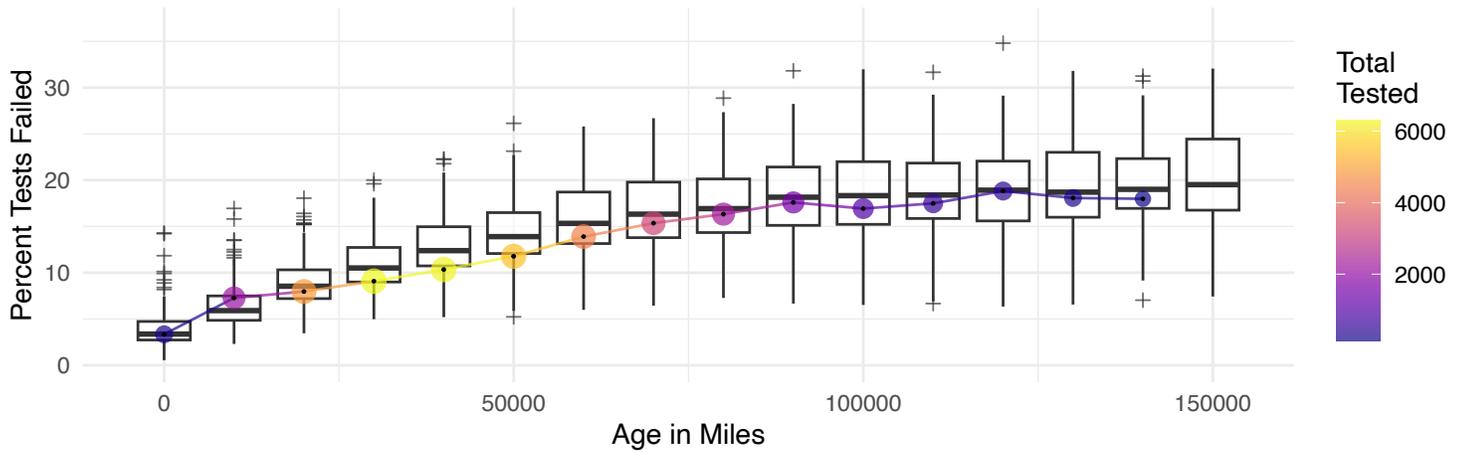

Mortality rates

| Age in Years | Observed | Died | Mortality Rate |
|---|---|---|---|
| 3 | 4915 | 258 | 0.052500 |
| 4 | 8534 | 483 | 0.056600 |
| 5 | 9169 | 129 | 0.014100 |
| 6 | 8031 | 16 | 0.001990 |
| 7 | 3396 | 1 | 0.000294 |

Mechanical Reliability Rates

| Mileage at test | N tested | Pct failed |
|---|---|---|
| 0 | 269 | 3.35 |
| 10000 | 2131 | 7.27 |
| 20000 | 4883 | 7.97 |
| 30000 | 6306 | 9.09 |
| 40000 | 6245 | 10.30 |
| 50000 | 5471 | 11.80 |
| 60000 | 4417 | 13.90 |
| 70000 | 3224 | 15.40 |
| 80000 | 2284 | 16.30 |
| 90000 | 1523 | 17.60 |
| 100000 | 945 | 16.90 |
| 110000 | 583 | 17.50 |
| 120000 | 409 | 18.80 |
| 130000 | 249 | 18.10 |
| 140000 | 139 | 18.00 |

**Audi A5 2015**

At 5 years of age, the mortality rate of a Audi A5 2015 (manufactured as a Car or Light Van) ranked number 123 out of 247 vehicles of the same age and type (any Car or Light Van constructed in 2015). One is the lowest (or best) and 247 the highest mortality rate. For vehicles reaching 40000 miles, its unreliability score (rate of failing an inspection) ranked 158 out of 240 vehicles of the same age, type, and mileage. One is the highest (or worst) and 240 the lowest rate of failing an inspection.

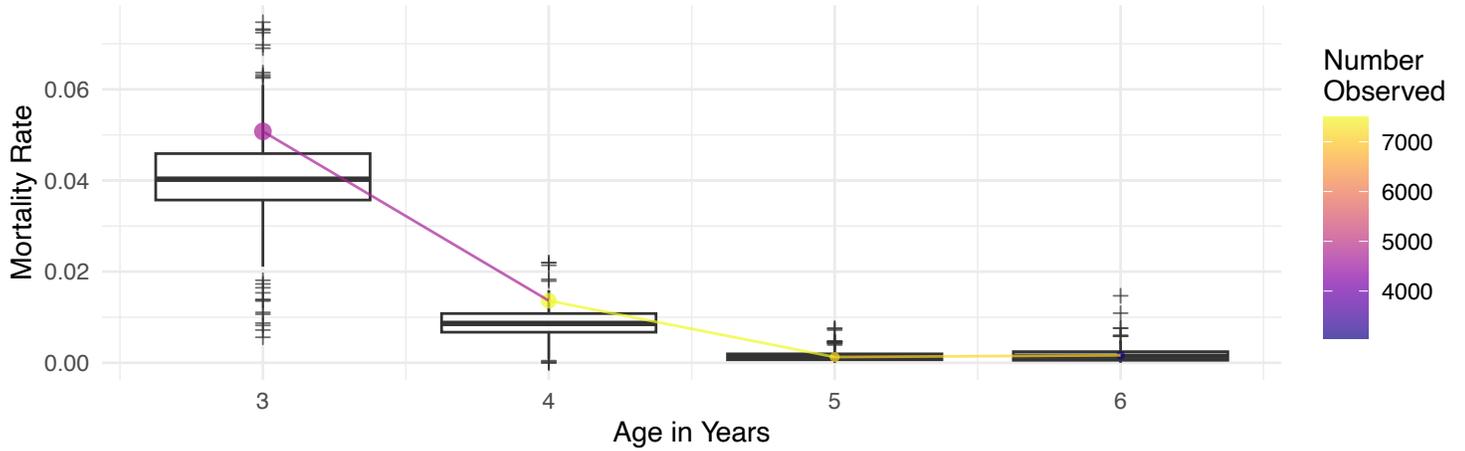

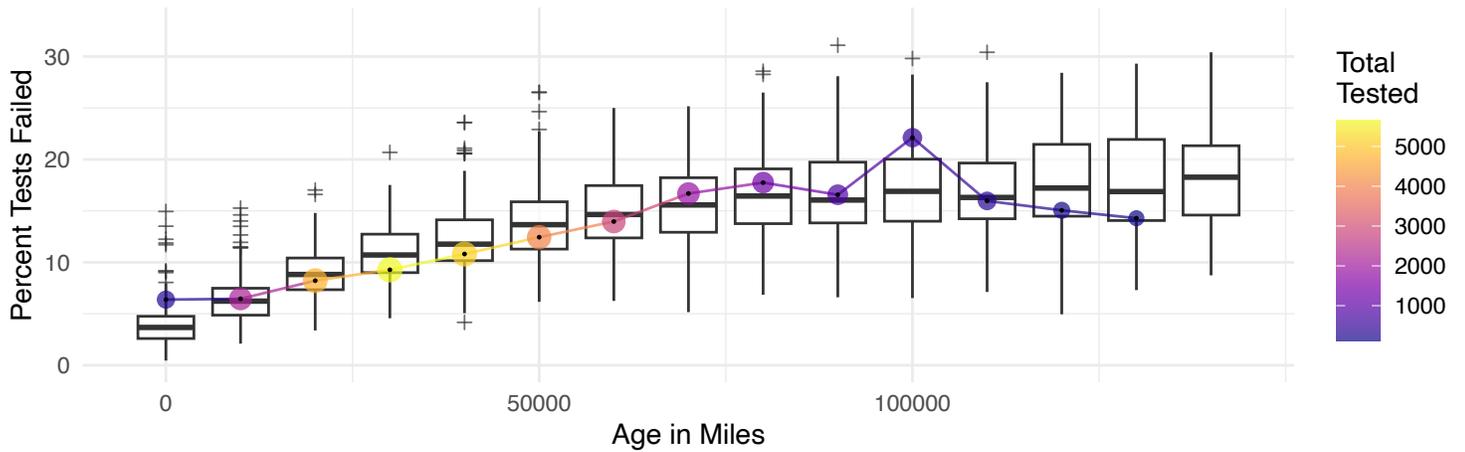

Mortality rates

| Age in Years | Observed | Died | Mortality Rate |
|---|---|---|---|
| 3 | 4688 | 238 | 0.05080 |
| 4 | 7486 | 102 | 0.01360 |
| 5 | 7124 | 9 | 0.00126 |
| 6 | 3050 | 5 | 0.00164 |

Mechanical Reliability Rates

| Mileage at test | N tested | Pct failed |
|---|---|---|
| 0 | 298 | 6.38 |
| 10000 | 2199 | 6.46 |
| 20000 | 4698 | 8.22 |
| 30000 | 5662 | 9.27 |
| 40000 | 5172 | 10.80 |
| 50000 | 4019 | 12.40 |
| 60000 | 2857 | 14.00 |
| 70000 | 1870 | 16.70 |
| 80000 | 1268 | 17.70 |
| 90000 | 827 | 16.60 |
| 100000 | 511 | 22.10 |
| 110000 | 357 | 16.00 |
| 120000 | 186 | 15.10 |
| 130000 | 119 | 14.30 |

**Audi A5 2016**

At 5 years of age, the mortality rate of a Audi A5 2016 (manufactured as a Car or Light Van) ranked number 115 out of 252 vehicles of the same age and type (any Car or Light Van constructed in 2016). One is the lowest (or best) and 252 the highest mortality rate. For vehicles reaching 20000 miles, its unreliability score (rate of failing an inspection) ranked 157 out of 246 vehicles of the same age, type, and mileage. One is the highest (or worst) and 246 the lowest rate of failing an inspection.

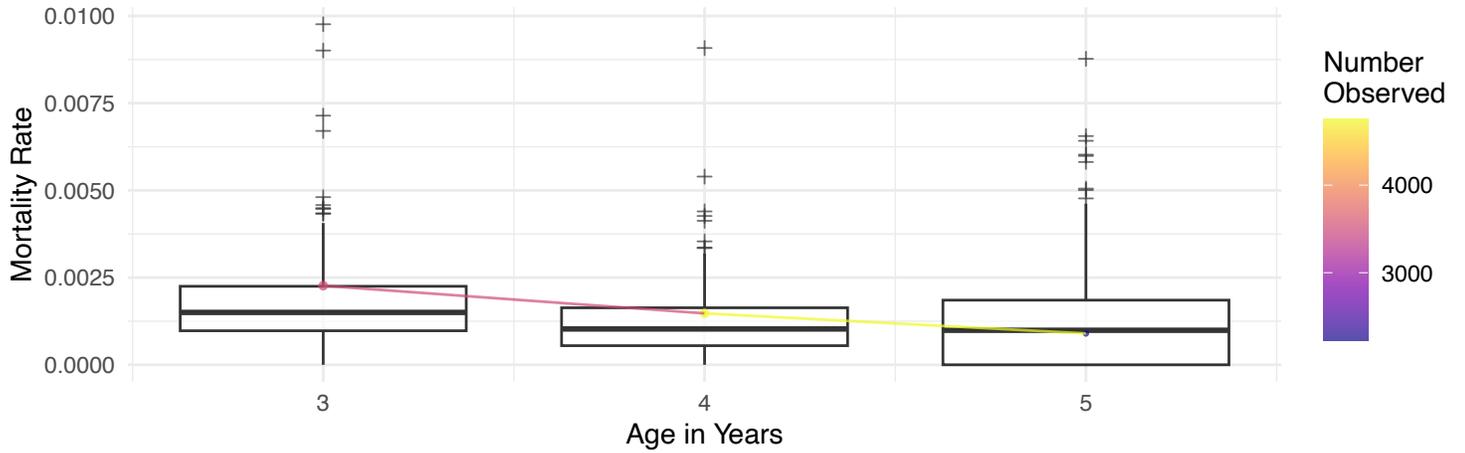

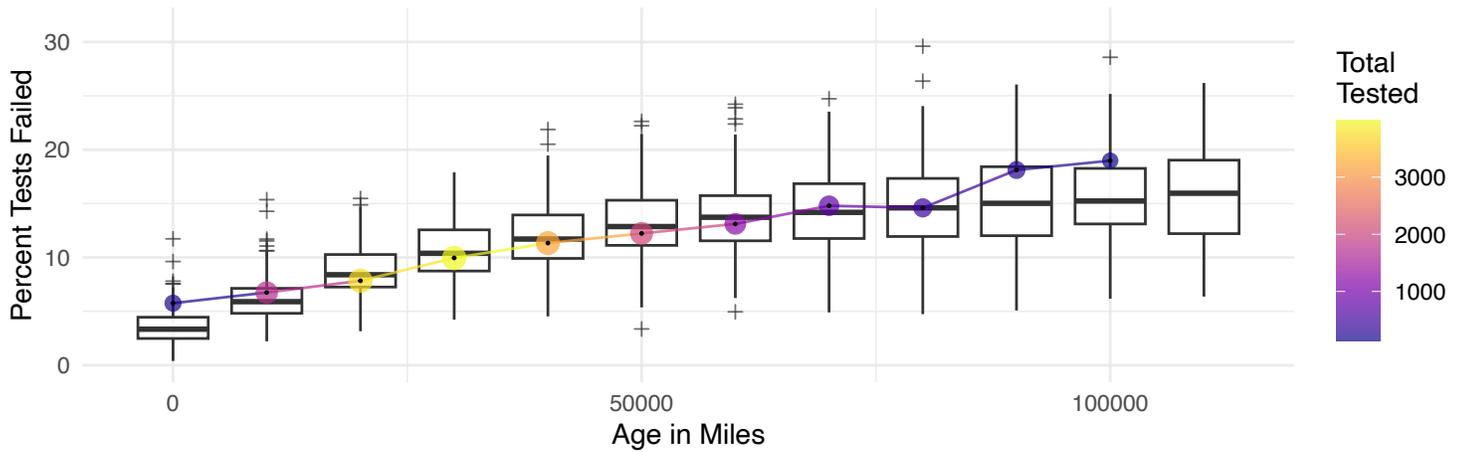

Mortality rates

| Age in Years | Observed | Died | Mortality Rate |
|---|---|---|---|
| 3 | 3529 | 8 | 0.002270 |
| 4 | 4748 | 7 | 0.001470 |
| 5 | 2228 | 2 | 0.000898 |

Mechanical Reliability Rates

| Mileage at test | N tested | Pct failed |
|---|---|---|
| 0 | 191 | 5.76 |
| 10000 | 1749 | 6.75 |
| 20000 | 3777 | 7.84 |
| 30000 | 3998 | 9.95 |
| 40000 | 3211 | 11.30 |
| 50000 | 2077 | 12.20 |
| 60000 | 1243 | 13.10 |
| 70000 | 784 | 14.80 |
| 80000 | 452 | 14.60 |
| 90000 | 265 | 18.10 |
| 100000 | 137 | 19.00 |

**Audi A5 2017**

At 3 years of age, the mortality rate of a Audi A5 2017 (manufactured as a Car or Light Van) ranked number 207 out of 247 vehicles of the same age and type (any Car or Light Van constructed in 2017). One is the lowest (or best) and 247 the highest mortality rate. For vehicles reaching 20000 miles, its unreliability score (rate of failing an inspection) ranked 153 out of 240 vehicles of the same age, type, and mileage. One is the highest (or worst) and 240 the lowest rate of failing an inspection.

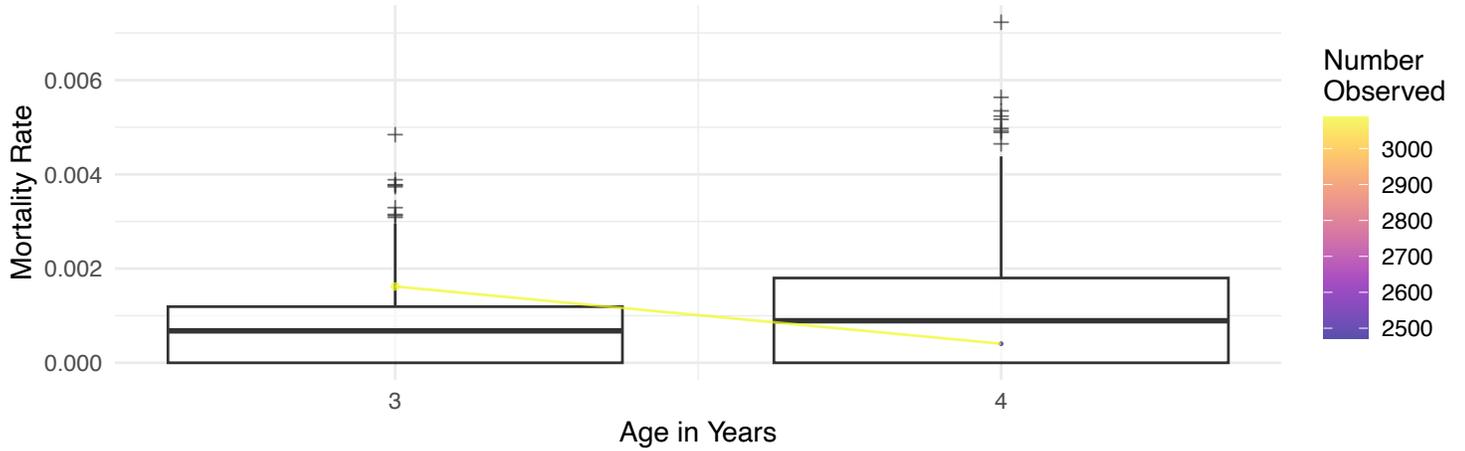

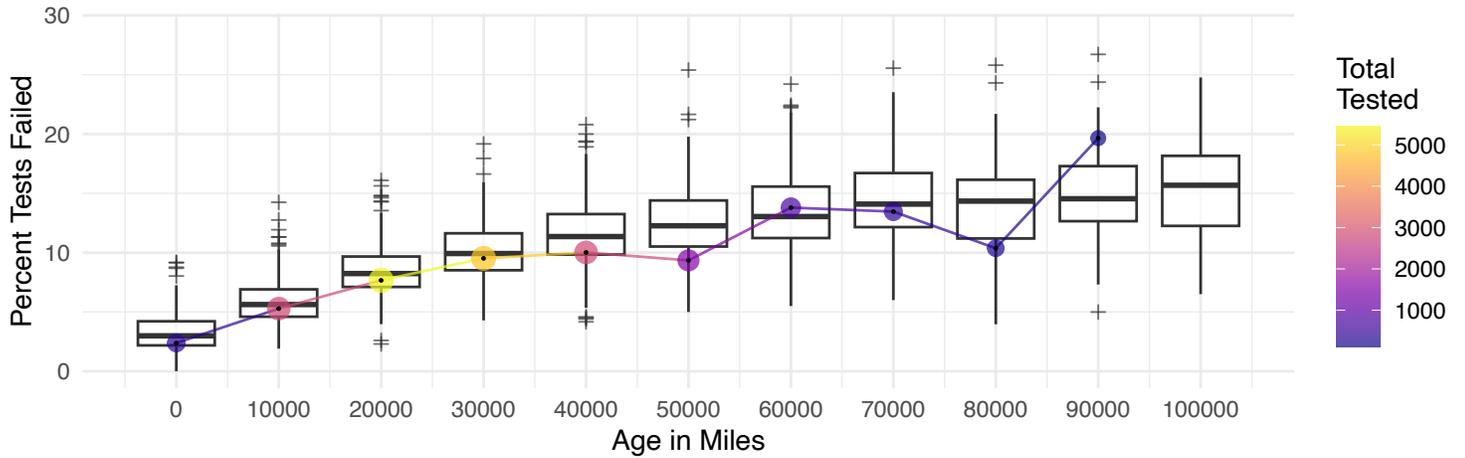

<table>
<tr><td colspan="4" align="center">Mortality rates</td></tr>
</table>

| Age in Years | Observed | Died | Mortality Rate |
| --- | --- | --- | --- |
| 3 | 3086 | 5 | 0.001620 |
| 4 | 2472 | 1 | 0.000405 |

Mechanical Reliability Rates

| Mileage at test | N tested | Pct failed |
| --- | --- | --- |
| 0 | 378 | 2.38 |
| 10000 | 2840 | 5.28 |
| 20000 | 5453 | 7.67 |
| 30000 | 4822 | 9.52 |
| 40000 | 2865 | 10.00 |
| 50000 | 1519 | 9.35 |
| 60000 | 761 | 13.80 |
| 70000 | 416 | 13.50 |
| 80000 | 241 | 10.40 |
| 90000 | 117 | 19.70 |

**Audi A5 2018**

At 3 years of age, the mortality rate of a Audi A5 2018 (manufactured as a Car or Light Van) ranked number 4 out of 222 vehicles of the same age and type (any Car or Light Van constructed in 2018). One is the lowest (or best) and 222 the highest mortality rate. For vehicles reaching 20000 miles, its unreliability score (rate of failing an inspection) ranked 98 out of 215 vehicles of the same age, type, and mileage. One is the highest (or worst) and 215 the lowest rate of failing an inspection.

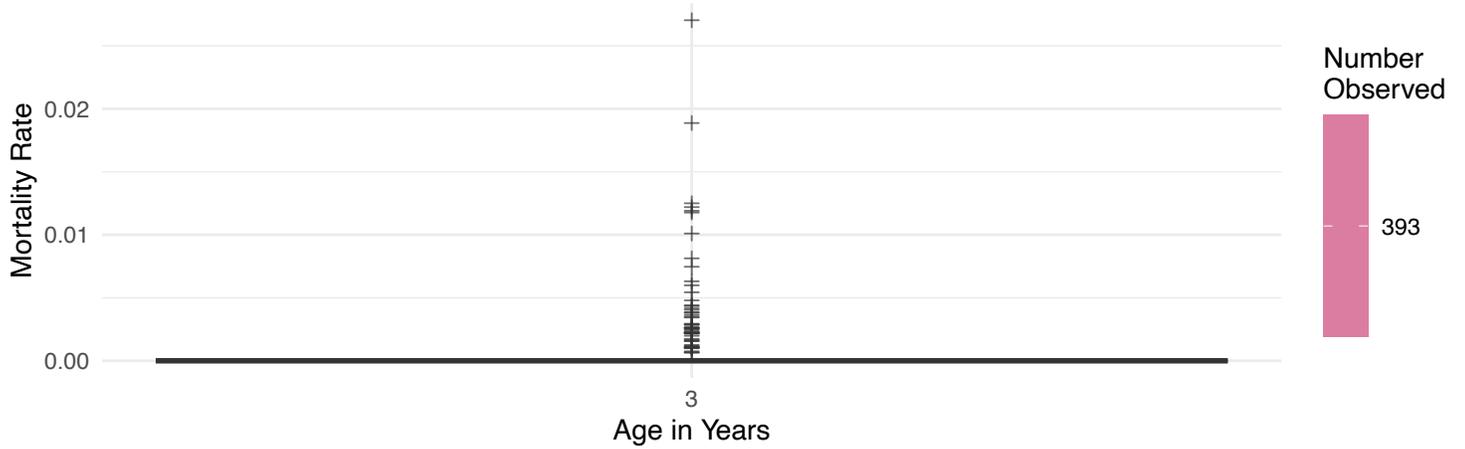

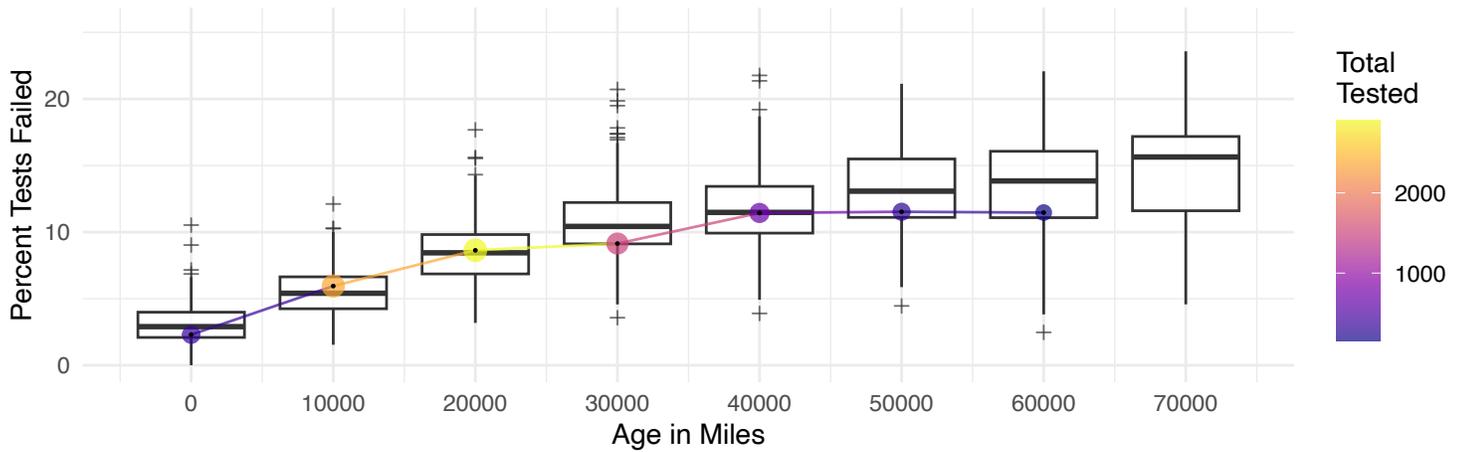

Mortality rates

| Age in Years | Observed | Died | Mortality Rate |
|---|---|---|---|
| 3 | 393 | 0 | 0 |

Mechanical Reliability Rates

| Mileage at test | N tested | Pct failed |
|---|---|---|
| 0 | 348 | 2.30 |
| 10000 | 2371 | 5.95 |
| 20000 | 2907 | 8.63 |
| 30000 | 1510 | 9.14 |
| 40000 | 682 | 11.40 |
| 50000 | 295 | 11.50 |
| 60000 | 157 | 11.50 |

## Audi A6 1994

At 15 years of age, the mortality rate of a Audi A6 1994 (manufactured as a Car or Light Van) ranked number 75 out of 120 vehicles of the same age and type (any Car or Light Van constructed in 1994). One is the lowest (or best) and 120 the highest mortality rate. For vehicles reaching 120000 miles, its unreliability score (rate of failing an inspection) ranked 82 out of 112 vehicles of the same age, type, and mileage. One is the highest (or worst) and 112 the lowest rate of failing an inspection.

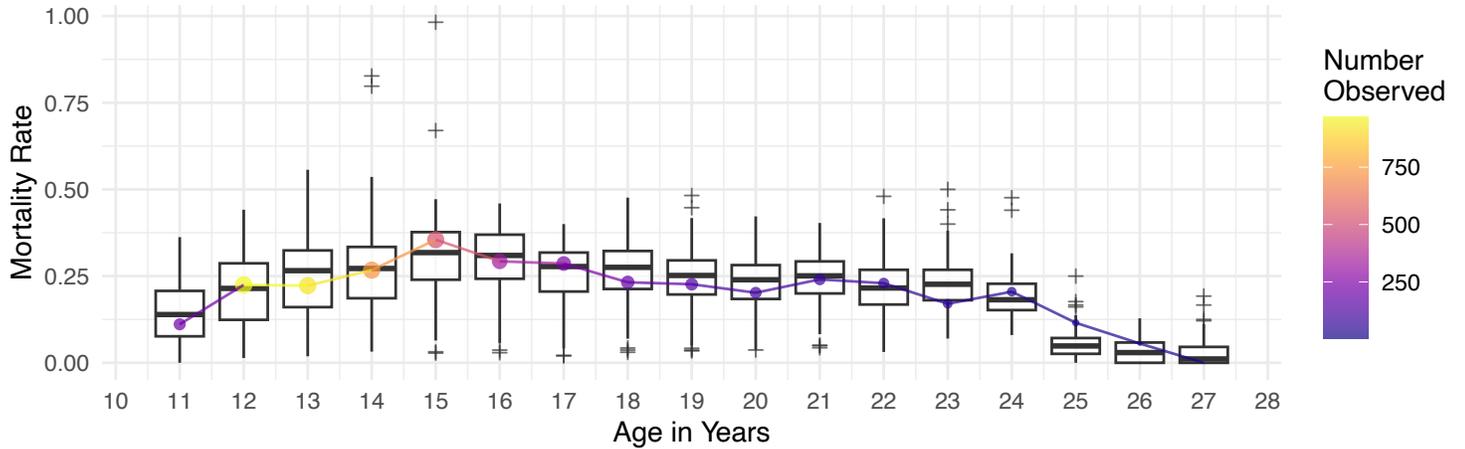

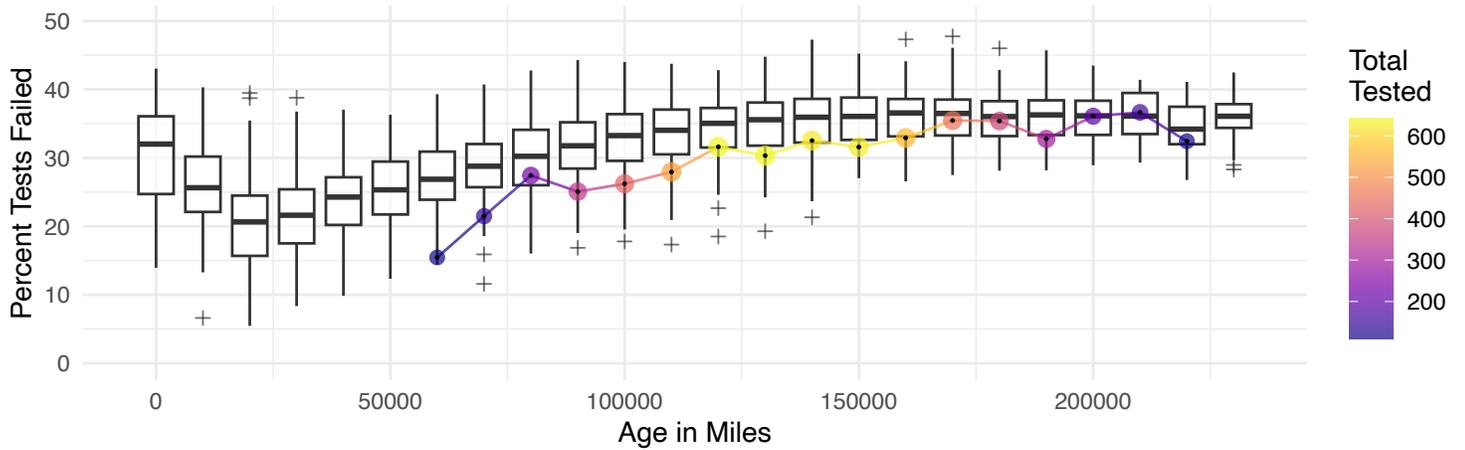

| Mortality rates | | | |
|---|---|---|---|
| Age in Years | Observed | Died | Mortality Rate |
| 11 | 207 | 23 | 0.1110 |
| 12 | 966 | 217 | 0.2250 |
| 13 | 921 | 205 | 0.2230 |
| 14 | 726 | 194 | 0.2670 |
| 15 | 524 | 186 | 0.3550 |
| 16 | 335 | 98 | 0.2930 |
| 17 | 234 | 67 | 0.2860 |
| 18 | 168 | 39 | 0.2320 |
| 19 | 128 | 29 | 0.2270 |
| 20 | 99 | 20 | 0.2020 |
| 21 | 79 | 19 | 0.2410 |
| 22 | 61 | 14 | 0.2300 |
| 23 | 47 | 8 | 0.1700 |
| 24 | 39 | 8 | 0.2050 |
| 25 | 26 | 3 | 0.1150 |
| 26 | 18 | 1 | 0.0556 |

| Mechanical Reliability Rates | | |
|---|---|---|
| Mileage at test | N tested | Pct failed |
| 60000 | 110 | 15.5 |
| 70000 | 135 | 21.5 |
| 80000 | 226 | 27.4 |
| 90000 | 343 | 25.1 |
| 100000 | 431 | 26.2 |
| 110000 | 526 | 27.9 |
| 120000 | 642 | 31.6 |
| 130000 | 643 | 30.3 |
| 140000 | 621 | 32.5 |
| 150000 | 627 | 31.6 |
| 160000 | 562 | 32.9 |
| 170000 | 454 | 35.5 |
| 180000 | 373 | 35.4 |
| 190000 | 296 | 32.8 |
| 200000 | 194 | 36.1 |
| 210000 | 161 | 36.6 |
| 220000 | 111 | 32.4 |

# Audi A6 1995

At 10 years of age, the mortality rate of a Audi A6 1995 (manufactured as a Car or Light Van) ranked number 62 out of 148 vehicles of the same age and type (any Car or Light Van constructed in 1995). One is the lowest (or best) and 148 the highest mortality rate. For vehicles reaching 120000 miles, its unreliability score (rate of failing an inspection) ranked 105 out of 135 vehicles of the same age, type, and mileage. One is the highest (or worst) and 135 the lowest rate of failing an inspection.

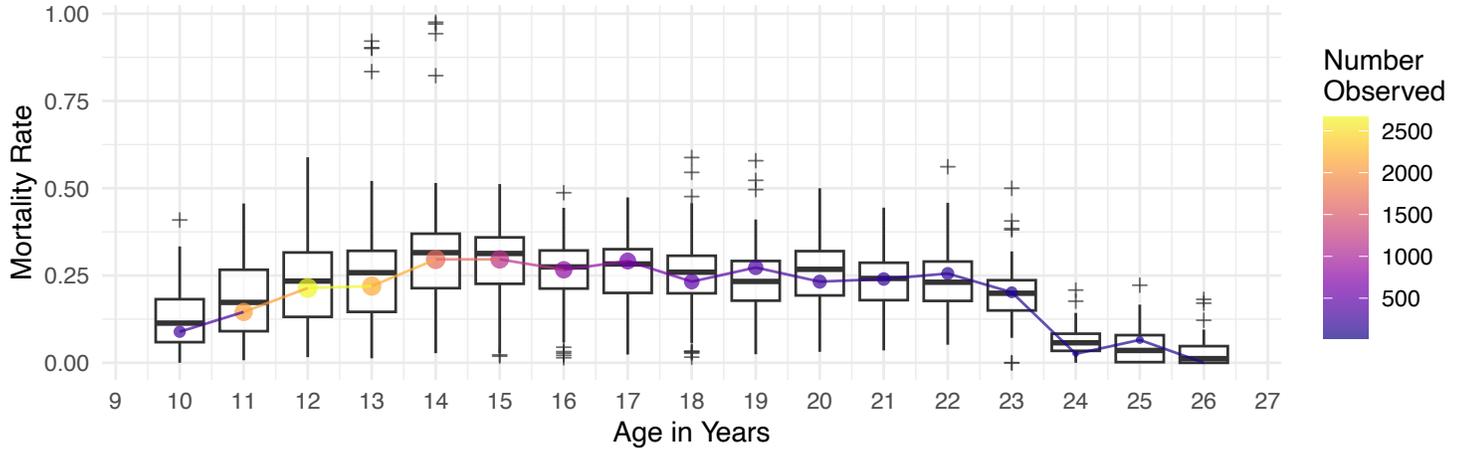

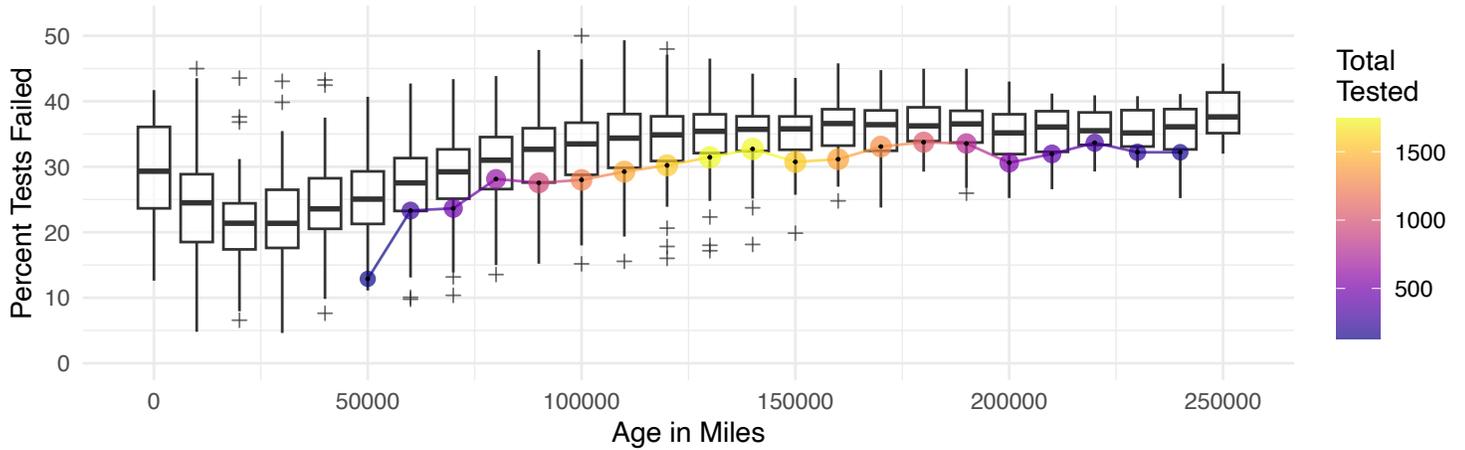

| Mortality rates | | | |
|---|---|---|---|
| Age in Years | Observed | Died | Mortality Rate |
| 10 | 270 | 24 | 0.0889 |
| 11 | 2073 | 302 | 0.1460 |
| 12 | 2657 | 569 | 0.2140 |
| 13 | 2158 | 474 | 0.2200 |
| 14 | 1661 | 492 | 0.2960 |
| 15 | 1154 | 342 | 0.2960 |
| 16 | 807 | 215 | 0.2660 |
| 17 | 590 | 172 | 0.2920 |
| 18 | 413 | 96 | 0.2320 |
| 19 | 315 | 86 | 0.2730 |
| 20 | 228 | 53 | 0.2320 |
| 21 | 175 | 42 | 0.2400 |
| 22 | 133 | 34 | 0.2560 |
| 23 | 99 | 20 | 0.2020 |
| 24 | 78 | 2 | 0.0256 |
| 25 | 61 | 4 | 0.0656 |
| 26 | 21 | 0 | 0.0000 |

| Mechanical Reliability Rates | | |
|---|---|---|
| Mileage at test | N tested | Pct failed |
| 50000 | 132 | 12.9 |
| 60000 | 266 | 23.3 |
| 100000 | 1254 | 28.0 |
| 110000 | 1429 | 29.3 |
| 120000 | 1569 | 30.2 |
| 130000 | 1746 | 31.4 |
| 140000 | 1739 | 32.7 |
| 150000 | 1571 | 30.7 |
| 160000 | 1406 | 31.2 |
| 170000 | 1306 | 33.1 |
| 180000 | 1046 | 33.7 |
| 190000 | 775 | 33.5 |
| 200000 | 539 | 30.6 |
| 210000 | 388 | 32.0 |
| 220000 | 303 | 33.7 |
| 230000 | 208 | 32.2 |
| 240000 | 149 | 32.2 |

## Audi A6 1996

At 10 years of age, the mortality rate of a Audi A6 1996 (manufactured as a Car or Light Van) ranked number 70 out of 162 vehicles of the same age and type (any Car or Light Van constructed in 1996). One is the lowest (or best) and 162 the highest mortality rate. For vehicles reaching 120000 miles, its unreliability score (rate of failing an inspection) ranked 136 out of 147 vehicles of the same age, type, and mileage. One is the highest (or worst) and 147 the lowest rate of failing an inspection.

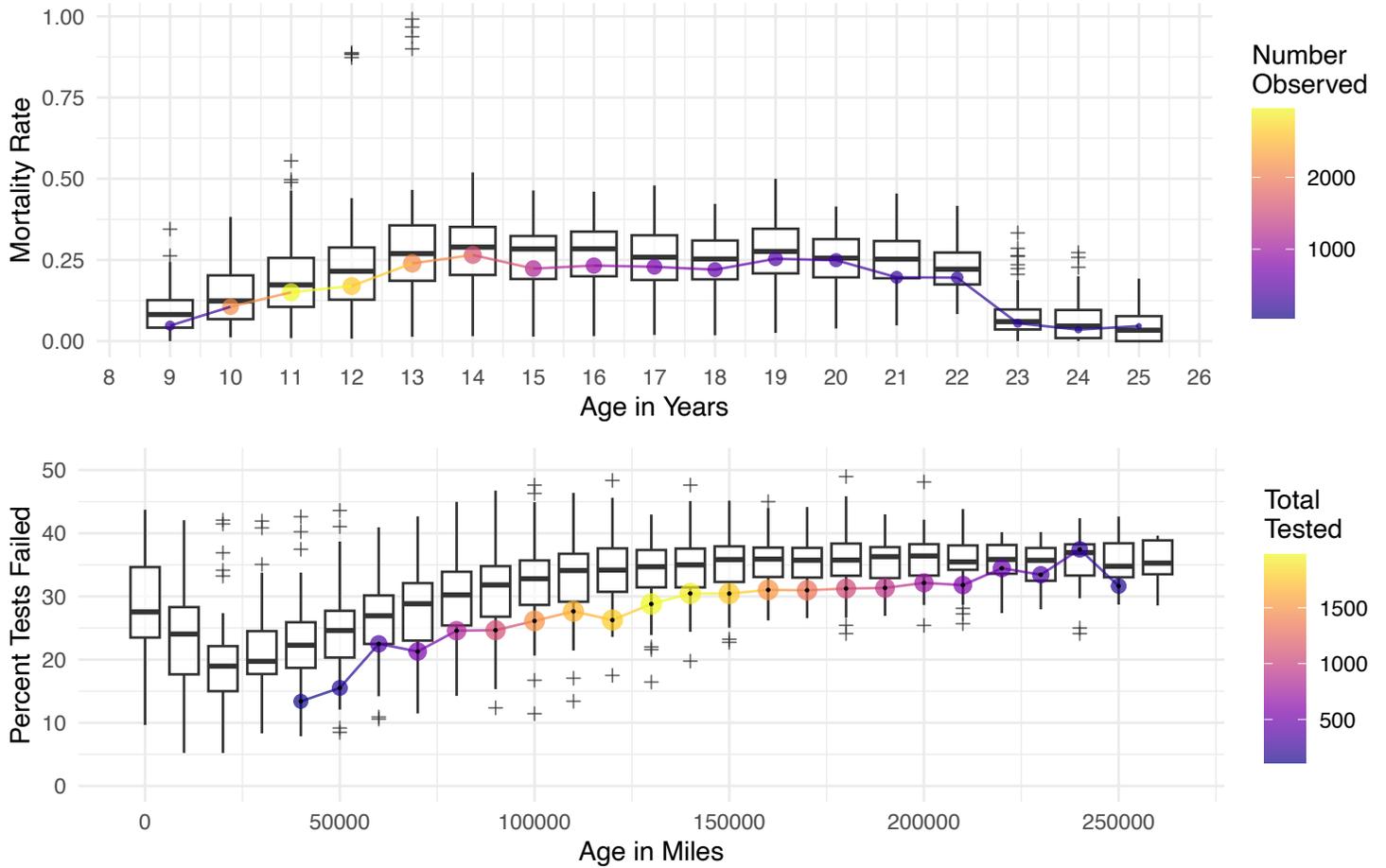

| Mortality rates | | | |
|---|---|---|---|
| Age in Years | Observed | Died | Mortality Rate |
| 9 | 273 | 13 | 0.0476 |
| 10 | 2146 | 227 | 0.1060 |
| 11 | 2945 | 443 | 0.1500 |
| 12 | 2610 | 443 | 0.1700 |
| 13 | 2139 | 512 | 0.2390 |
| 14 | 1598 | 424 | 0.2650 |
| 15 | 1155 | 258 | 0.2230 |
| 16 | 894 | 208 | 0.2330 |
| 17 | 686 | 157 | 0.2290 |
| 18 | 527 | 116 | 0.2200 |
| 19 | 410 | 104 | 0.2540 |
| 20 | 305 | 76 | 0.2490 |
| 21 | 229 | 45 | 0.1970 |
| 22 | 184 | 36 | 0.1960 |
| 23 | 145 | 8 | 0.0552 |
| 24 | 113 | 4 | 0.0354 |
| 25 | 43 | 2 | 0.0465 |

| Mechanical Reliability Rates | | |
|---|---|---|
| Mileage at test | N tested | Pct failed |
| 40000 | 112 | 13.4 |
| 100000 | 1493 | 26.1 |
| 110000 | 1668 | 27.6 |
| 120000 | 1786 | 26.3 |
| 130000 | 1979 | 28.8 |
| 140000 | 1940 | 30.5 |
| 150000 | 1777 | 30.4 |
| 160000 | 1508 | 31.0 |
| 170000 | 1314 | 31.0 |
| 180000 | 1056 | 31.2 |
| 190000 | 922 | 31.3 |
| 200000 | 731 | 32.1 |
| 210000 | 563 | 31.8 |
| 220000 | 470 | 34.5 |
| 230000 | 326 | 33.4 |
| 240000 | 227 | 37.4 |
| 250000 | 158 | 31.6 |

## Audi A6 1997

At 10 years of age, the mortality rate of a Audi A6 1997 (manufactured as a Car or Light Van) ranked number 80 out of 187 vehicles of the same age and type (any Car or Light Van constructed in 1997). One is the lowest (or best) and 187 the highest mortality rate. For vehicles reaching 120000 miles, its unreliability score (rate of failing an inspection) ranked 137 out of 167 vehicles of the same age, type, and mileage. One is the highest (or worst) and 167 the lowest rate of failing an inspection.

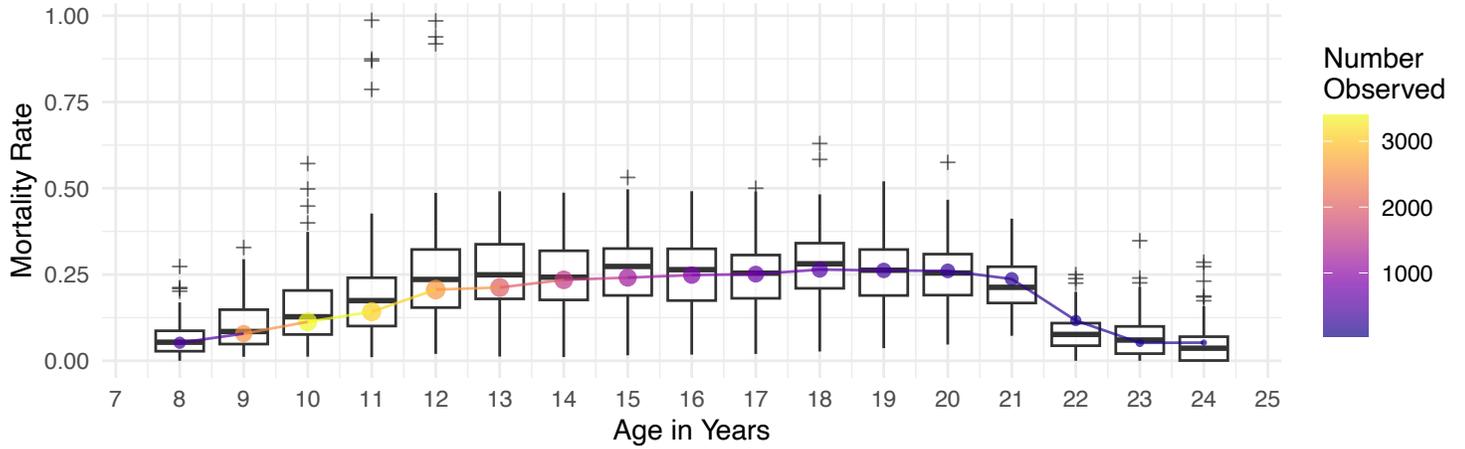

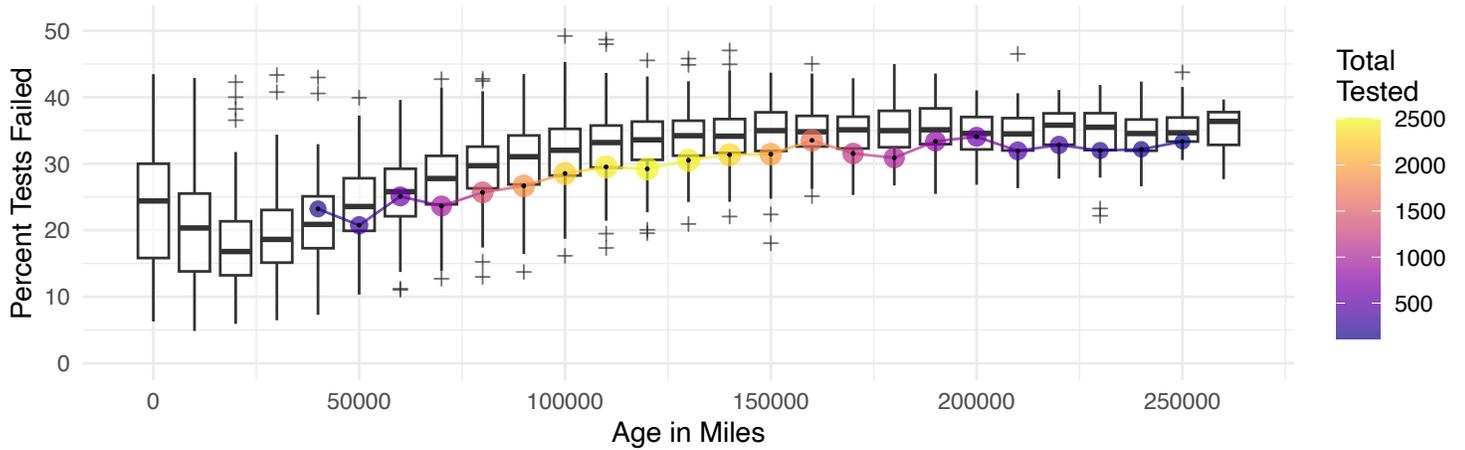

| Mortality rates | | | |
|---|---|---|---|
| Age in Years | Observed | Died | Mortality Rate |
| 8 | 458 | 24 | 0.0524 |
| 9 | 2528 | 199 | 0.0787 |
| 10 | 3389 | 383 | 0.1130 |
| 11 | 3074 | 437 | 0.1420 |
| 12 | 2583 | 533 | 0.2060 |
| 13 | 1996 | 424 | 0.2120 |
| 14 | 1543 | 362 | 0.2350 |
| 15 | 1166 | 281 | 0.2410 |
| 16 | 874 | 217 | 0.2480 |
| 17 | 653 | 164 | 0.2510 |
| 18 | 488 | 129 | 0.2640 |
| 19 | 359 | 94 | 0.2620 |
| 20 | 261 | 68 | 0.2610 |
| 21 | 194 | 46 | 0.2370 |
| 22 | 137 | 16 | 0.1170 |
| 23 | 96 | 5 | 0.0521 |
| 24 | 38 | 2 | 0.0526 |

| Mechanical Reliability Rates | | |
|---|---|---|
| Mileage at test | N tested | Pct failed |
| 40000 | 181 | 23.2 |
| 100000 | 2271 | 28.5 |
| 110000 | 2429 | 29.5 |
| 120000 | 2508 | 29.2 |
| 130000 | 2458 | 30.5 |
| 140000 | 2309 | 31.4 |
| 150000 | 1979 | 31.4 |
| 160000 | 1659 | 33.5 |
| 170000 | 1186 | 31.5 |
| 180000 | 1010 | 30.9 |
| 190000 | 809 | 33.4 |
| 200000 | 605 | 34.0 |
| 210000 | 404 | 31.9 |
| 220000 | 326 | 32.8 |
| 230000 | 197 | 32.0 |
| 240000 | 143 | 32.2 |
| 250000 | 114 | 33.3 |

**Audi A6 1998**

At 10 years of age, the mortality rate of a Audi A6 1998 (manufactured as a Car or Light Van) ranked number 82 out of 196 vehicles of the same age and type (any Car or Light Van constructed in 1998). One is the lowest (or best) and 196 the highest mortality rate. For vehicles reaching 120000 miles, its unreliability score (rate of failing an inspection) ranked 110 out of 172 vehicles of the same age, type, and mileage. One is the highest (or worst) and 172 the lowest rate of failing an inspection.

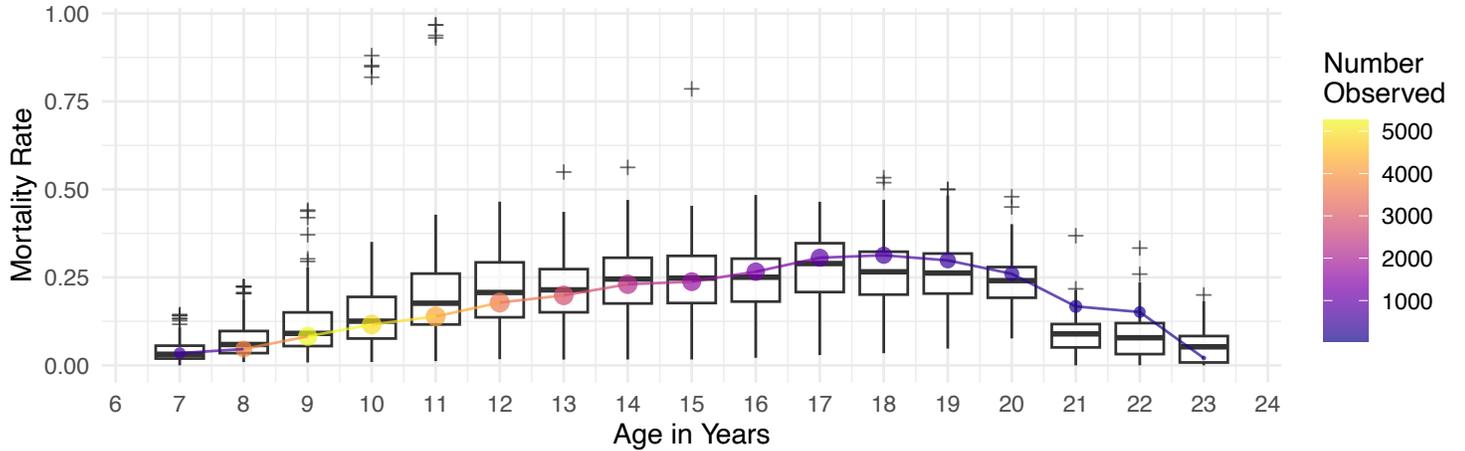

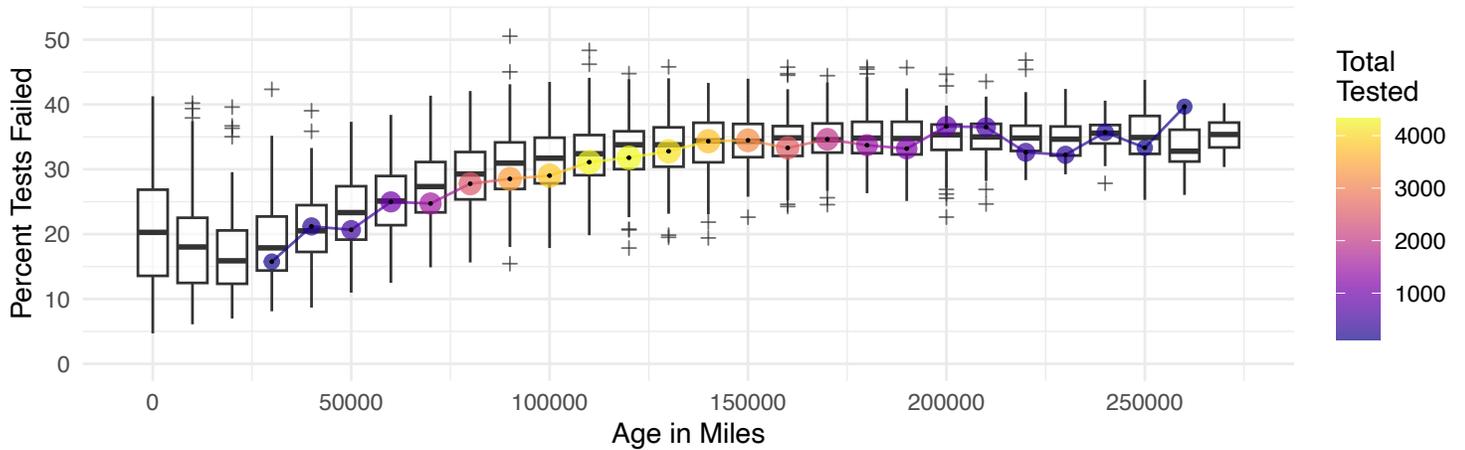

| Mortality rates | | | |
|---|---|---|---|
| Age in Years | Observed | Died | Mortality Rate |
| 7 | 576 | 20 | 0.0347 |
| 8 | 3801 | 176 | 0.0463 |
| 9 | 5259 | 434 | 0.0825 |
| 10 | 4909 | 570 | 0.1160 |
| 11 | 4210 | 585 | 0.1390 |
| 12 | 3508 | 626 | 0.1780 |
| 13 | 2815 | 560 | 0.1990 |
| 14 | 2200 | 507 | 0.2300 |
| 15 | 1665 | 396 | 0.2380 |
| 16 | 1247 | 332 | 0.2660 |
| 17 | 899 | 275 | 0.3060 |
| 18 | 617 | 193 | 0.3130 |
| 19 | 419 | 125 | 0.2980 |
| 20 | 292 | 76 | 0.2600 |
| 21 | 203 | 34 | 0.1670 |
| 22 | 132 | 20 | 0.1520 |
| 23 | 48 | 1 | 0.0208 |

| Mechanical Reliability Rates | | |
|---|---|---|
| Mileage at test | N tested | Pct failed |
| 30000 | 146 | 15.8 |
| 40000 | 302 | 21.2 |
| 100000 | 3841 | 29.0 |
| 110000 | 4282 | 31.1 |
| 120000 | 4323 | 31.8 |
| 130000 | 4102 | 32.8 |
| 140000 | 3757 | 34.3 |
| 150000 | 3165 | 34.5 |
| 160000 | 2561 | 33.3 |
| 170000 | 1978 | 34.6 |
| 180000 | 1474 | 33.7 |
| 190000 | 1160 | 33.2 |
| 200000 | 830 | 36.6 |
| 210000 | 578 | 36.5 |
| 220000 | 420 | 32.6 |
| 230000 | 304 | 32.2 |
| 240000 | 224 | 35.7 |

**Audi A6 1999**

At 10 years of age, the mortality rate of a Audi A6 1999 (manufactured as a Car or Light Van) ranked number 79 out of 201 vehicles of the same age and type (any Car or Light Van constructed in 1999). One is the lowest (or best) and 201 the highest mortality rate. For vehicles reaching 120000 miles, its unreliability score (rate of failing an inspection) ranked 125 out of 181 vehicles of the same age, type, and mileage. One is the highest (or worst) and 181 the lowest rate of failing an inspection.

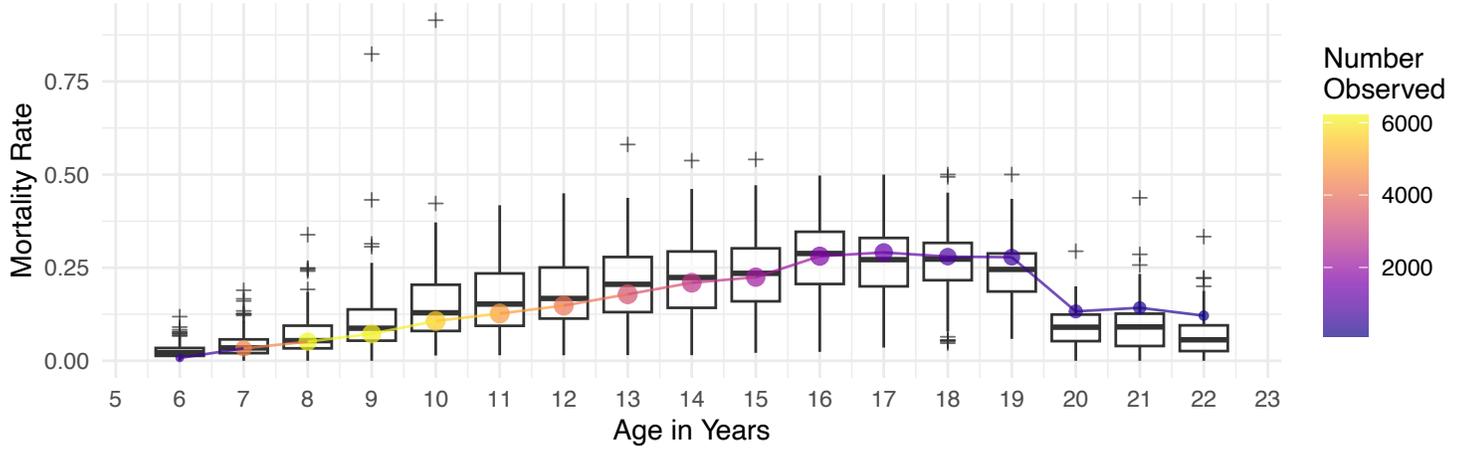

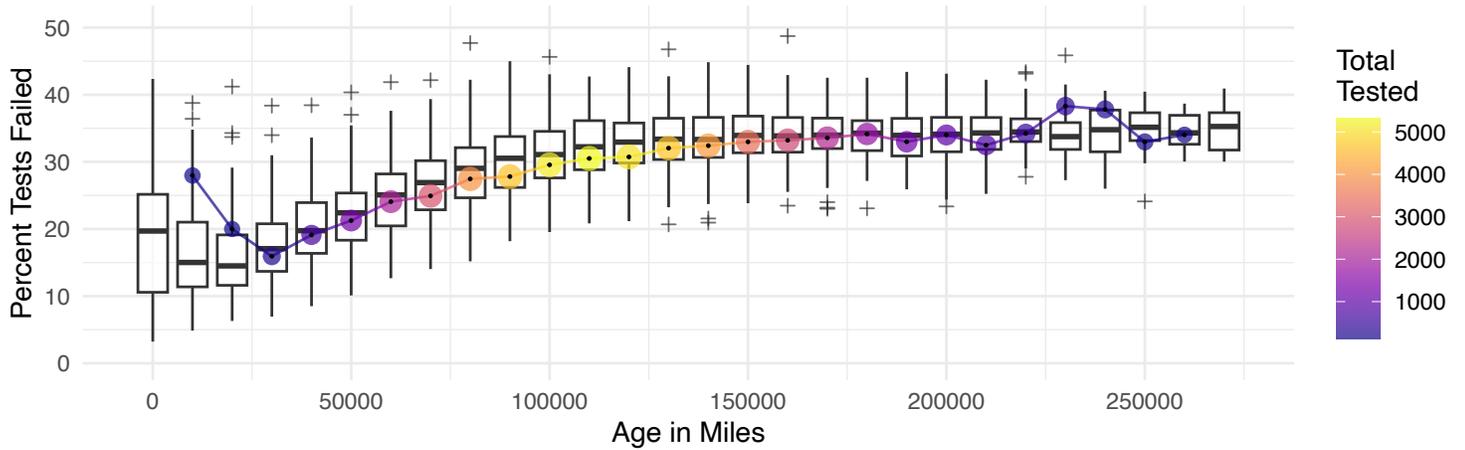

| Mortality rates | | | |
|---|---|---|---|
| Age in Years | Observed | Died | Mortality Rate |
| 6 | 664 | 5 | 0.00753 |
| 7 | 4373 | 146 | 0.03340 |
| 8 | 6191 | 318 | 0.05140 |
| 9 | 6055 | 439 | 0.07250 |
| 10 | 5519 | 590 | 0.10700 |
| 11 | 4793 | 608 | 0.12700 |
| 12 | 4083 | 605 | 0.14800 |
| 13 | 3409 | 609 | 0.17900 |
| 14 | 2749 | 576 | 0.21000 |
| 15 | 2147 | 482 | 0.22400 |
| 16 | 1631 | 458 | 0.28100 |
| 17 | 1155 | 336 | 0.29100 |
| 18 | 812 | 227 | 0.28000 |
| 19 | 586 | 163 | 0.27800 |
| 20 | 392 | 52 | 0.13300 |
| 21 | 260 | 37 | 0.14200 |
| 22 | 99 | 12 | 0.12100 |

| Mechanical Reliability Rates | | |
|---|---|---|
| Mileage at test | N tested | Pct failed |
| 10000 | 143 | 28.0 |
| 20000 | 125 | 20.0 |
| 30000 | 276 | 15.9 |
| 100000 | 5173 | 29.6 |
| 110000 | 5326 | 30.5 |
| 120000 | 5062 | 30.7 |
| 130000 | 4699 | 32.0 |
| 140000 | 4131 | 32.4 |
| 150000 | 3432 | 33.0 |
| 160000 | 2773 | 33.2 |
| 170000 | 2195 | 33.6 |
| 180000 | 1745 | 34.2 |
| 190000 | 1197 | 33.0 |
| 200000 | 941 | 34.0 |
| 210000 | 717 | 32.5 |
| 220000 | 514 | 34.2 |
| 240000 | 275 | 37.8 |

## Audi A6 2000

At 5 years of age, the mortality rate of a Audi A6 2000 (manufactured as a Car or Light Van) ranked number 94 out of 198 vehicles of the same age and type (any Car or Light Van constructed in 2000). One is the lowest (or best) and 198 the highest mortality rate. For vehicles reaching 120000 miles, its unreliability score (rate of failing an inspection) ranked 139 out of 184 vehicles of the same age, type, and mileage. One is the highest (or worst) and 184 the lowest rate of failing an inspection.

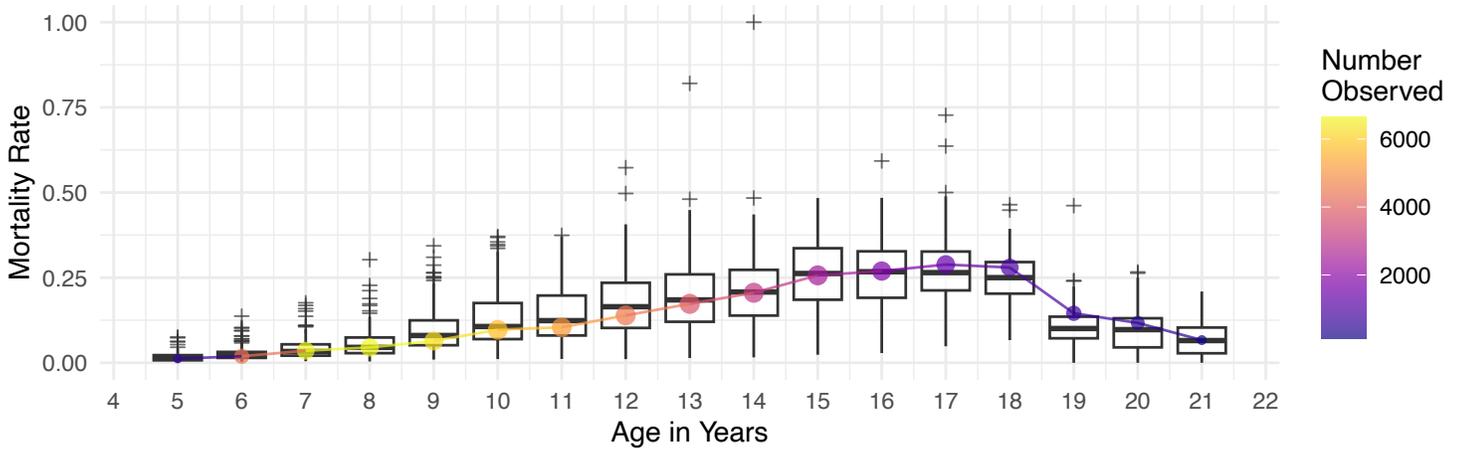

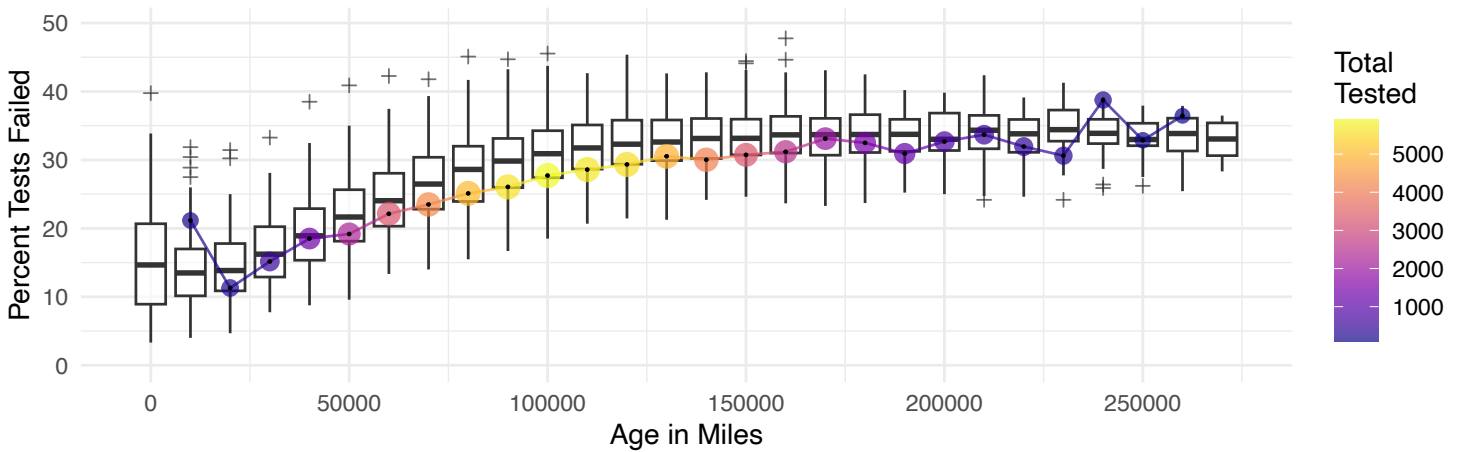

<table>
<tr><td colspan="4" align="center">Mortality rates</td></tr>
<tr><td>Age in Years</td><td>Observed</td><td>Died</td><td>Mortality Rate</td></tr>
</table>

| Age in Years | Observed | Died | Mortality Rate |
|---|---|---|---|
| 5 | 581 | 7 | 0.0120 |
| 6 | 4311 | 83 | 0.0193 |
| 7 | 6600 | 234 | 0.0355 |
| 8 | 6630 | 299 | 0.0451 |
| 9 | 6238 | 400 | 0.0641 |
| 10 | 5718 | 555 | 0.0971 |
| 11 | 5072 | 529 | 0.1040 |
| 12 | 4454 | 620 | 0.1390 |
| 13 | 3779 | 654 | 0.1730 |
| 14 | 3079 | 633 | 0.2060 |
| 15 | 2399 | 616 | 0.2570 |
| 16 | 1741 | 469 | 0.2690 |
| 17 | 1258 | 363 | 0.2890 |
| 18 | 890 | 249 | 0.2800 |
| 19 | 612 | 89 | 0.1450 |
| 20 | 410 | 48 | 0.1170 |
| 21 | 135 | 9 | 0.0667 |

Mechanical Reliability Rates

| Mileage at test | N tested | Pct failed |
|---|---|---|
| 10000 | 156 | 21.2 |
| 20000 | 248 | 11.3 |
| 30000 | 574 | 15.2 |
| 100000 | 5891 | 27.7 |
| 110000 | 5649 | 28.6 |
| 120000 | 5501 | 29.3 |
| 130000 | 4851 | 30.5 |
| 140000 | 4080 | 30.0 |
| 150000 | 3305 | 30.7 |
| 160000 | 2587 | 31.2 |
| 170000 | 1963 | 33.1 |
| 180000 | 1471 | 32.5 |
| 190000 | 1083 | 30.9 |
| 200000 | 780 | 32.7 |
| 210000 | 567 | 33.7 |
| 220000 | 410 | 32.0 |
| 230000 | 307 | 30.6 |

# Audi A6 2001

At 5 years of age, the mortality rate of a Audi A6 2001 (manufactured as a Car or Light Van) ranked number 125 out of 205 vehicles of the same age and type (any Car or Light Van constructed in 2001). One is the lowest (or best) and 205 the highest mortality rate. For vehicles reaching 120000 miles, its unreliability score (rate of failing an inspection) ranked 162 out of 194 vehicles of the same age, type, and mileage. One is the highest (or worst) and 194 the lowest rate of failing an inspection.

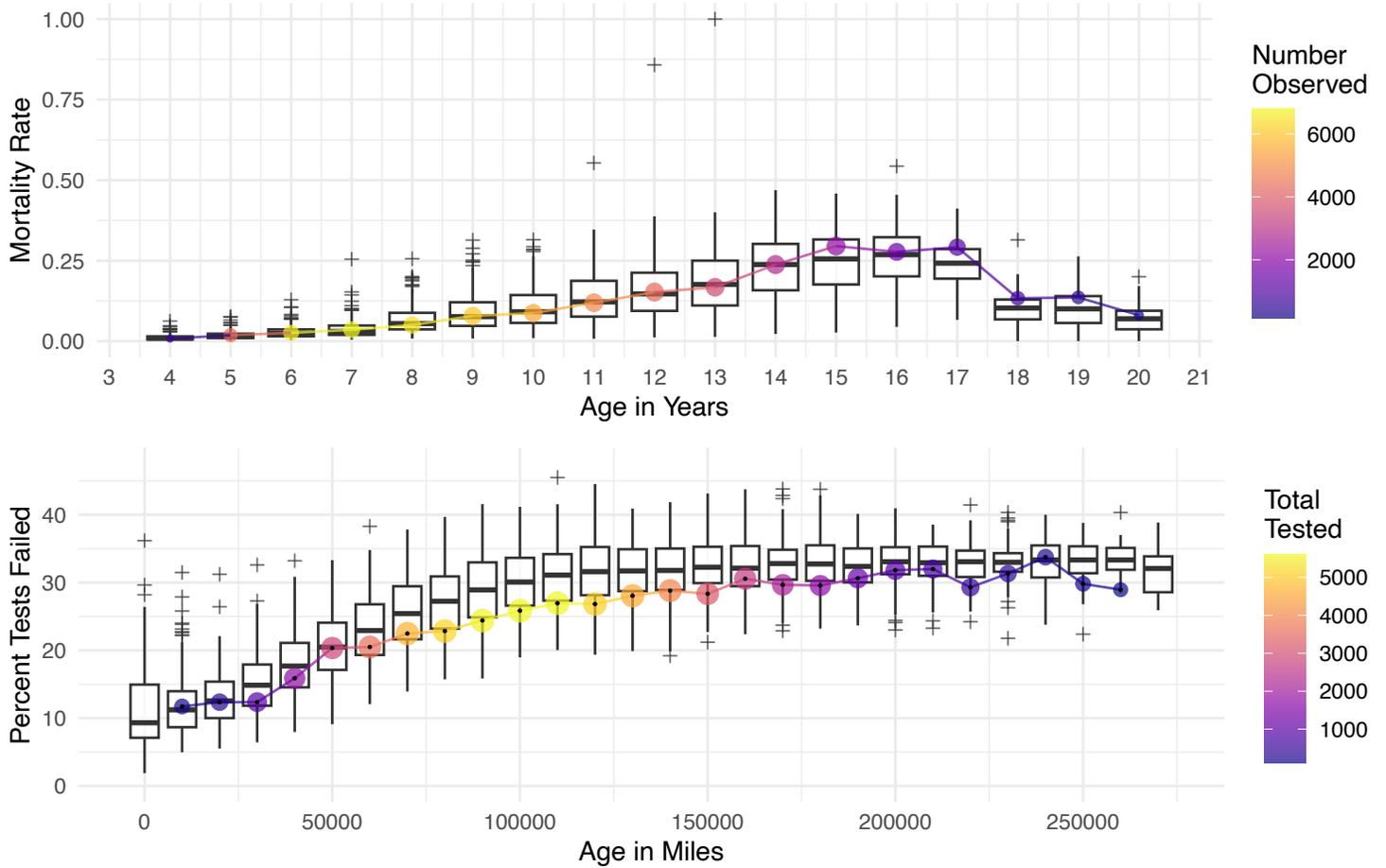

| Mortality rates | | | |
|---|---|---|---|
| Age in Years | Observed | Died | Mortality Rate |
| 4 | 650 | 5 | 0.00769 |
| 5 | 4258 | 79 | 0.01860 |
| 6 | 6658 | 172 | 0.02580 |
| 7 | 6781 | 239 | 0.03520 |
| 8 | 6441 | 323 | 0.05010 |
| 9 | 6004 | 471 | 0.07840 |
| 10 | 5424 | 472 | 0.08700 |
| 11 | 4848 | 578 | 0.11900 |
| 12 | 4202 | 640 | 0.15200 |
| 13 | 3501 | 586 | 0.16700 |
| 14 | 2832 | 673 | 0.23800 |
| 15 | 2110 | 623 | 0.29500 |
| 16 | 1461 | 405 | 0.27700 |
| 17 | 1042 | 304 | 0.29200 |
| 18 | 698 | 93 | 0.13300 |
| 19 | 464 | 63 | 0.13600 |
| 20 | 149 | 12 | 0.08050 |

| Mechanical Reliability Rates | | |
|---|---|---|
| Mileage at test | N tested | Pct failed |
| 10000 | 162 | 11.7 |
| 20000 | 356 | 12.4 |
| 30000 | 930 | 12.4 |
| 40000 | 1756 | 15.9 |
| 50000 | 2846 | 20.3 |
| 60000 | 3621 | 20.5 |
| 70000 | 4662 | 22.5 |
| 100000 | 5610 | 25.8 |
| 110000 | 5588 | 27.0 |
| 120000 | 5080 | 26.8 |
| 130000 | 4549 | 28.1 |
| 140000 | 3995 | 28.8 |
| 150000 | 3243 | 28.3 |
| 160000 | 2647 | 30.6 |
| 170000 | 2147 | 29.7 |
| 180000 | 1698 | 29.6 |
| 200000 | 908 | 31.8 |

**Audi A6 2002**

At 5 years of age, the mortality rate of a Audi A6 2002 (manufactured as a Car or Light Van) ranked number 127 out of 202 vehicles of the same age and type (any Car or Light Van constructed in 2002). One is the lowest (or best) and 202 the highest mortality rate. For vehicles reaching 120000 miles, its unreliability score (rate of failing an inspection) ranked 171 out of 193 vehicles of the same age, type, and mileage. One is the highest (or worst) and 193 the lowest rate of failing an inspection.

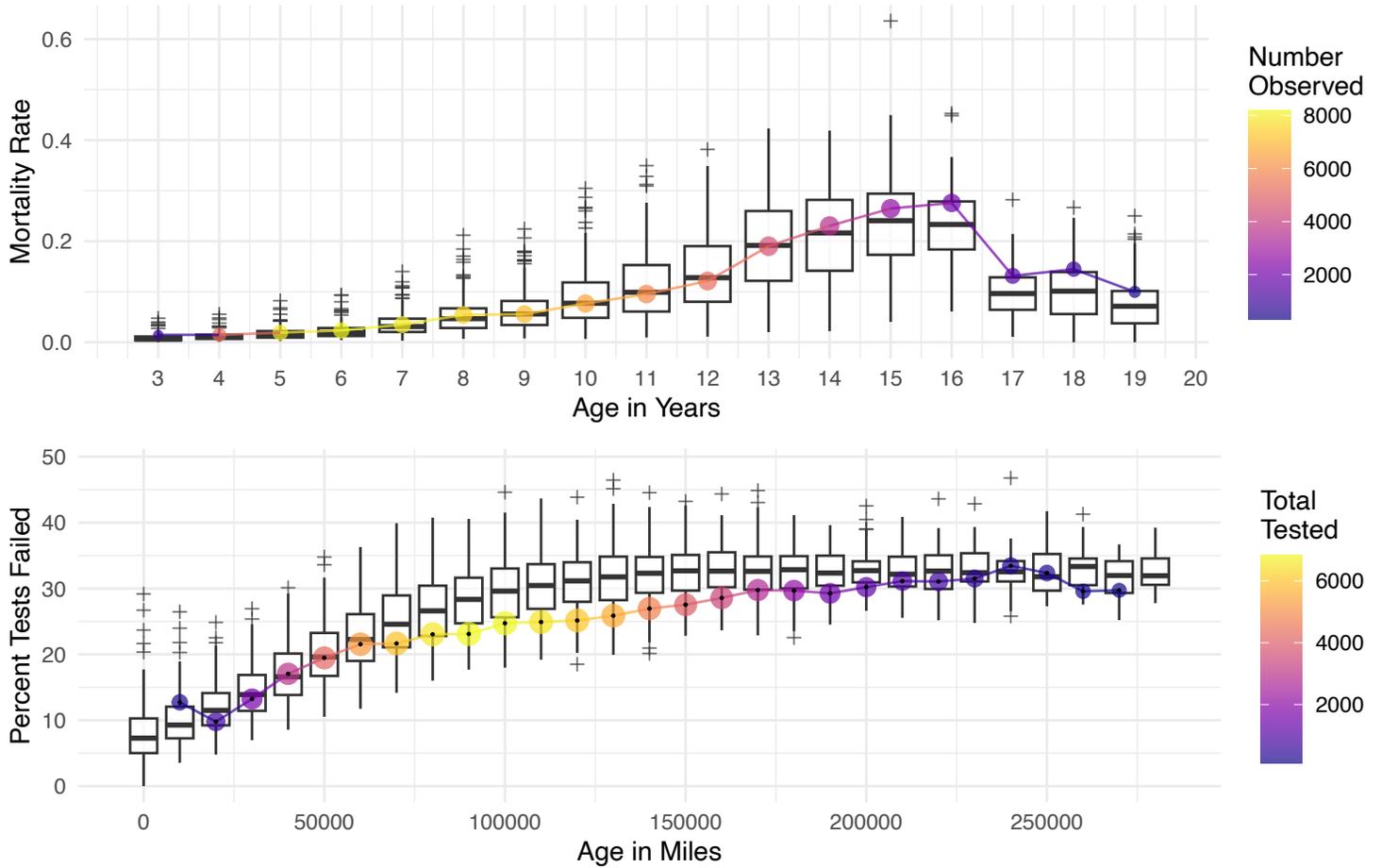

| Mortality rates | | | |
|---|---|---|---|
| Age in Years | Observed | Died | Mortality Rate |
| 3 | 875 | 13 | 0.0149 |
| 4 | 5185 | 76 | 0.0147 |
| 5 | 7936 | 152 | 0.0192 |
| 6 | 8167 | 191 | 0.0234 |
| 7 | 7900 | 276 | 0.0349 |
| 8 | 7512 | 407 | 0.0542 |
| 9 | 6999 | 389 | 0.0556 |
| 10 | 6504 | 499 | 0.0767 |
| 11 | 5917 | 564 | 0.0953 |
| 12 | 5257 | 637 | 0.1210 |
| 13 | 4498 | 855 | 0.1900 |
| 14 | 3557 | 819 | 0.2300 |
| 15 | 2690 | 712 | 0.2650 |
| 16 | 1951 | 538 | 0.2760 |
| 17 | 1347 | 177 | 0.1310 |
| 18 | 919 | 133 | 0.1450 |
| 19 | 321 | 32 | 0.0997 |

| Mechanical Reliability Rates | | |
|---|---|---|
| Mileage at test | N tested | Pct failed |
| 10000 | 220 | 12.70 |
| 20000 | 705 | 9.79 |
| 30000 | 1662 | 13.20 |
| 40000 | 3002 | 17.10 |
| 50000 | 4196 | 19.50 |
| 60000 | 5231 | 21.50 |
| 70000 | 5962 | 21.70 |
| 80000 | 6582 | 23.00 |
| 90000 | 6841 | 23.10 |
| 100000 | 6807 | 24.70 |
| 110000 | 6527 | 24.90 |
| 120000 | 6046 | 25.10 |
| 130000 | 5562 | 25.90 |
| 140000 | 4726 | 27.00 |
| 150000 | 3998 | 27.50 |
| 160000 | 3423 | 28.60 |
| 170000 | 2775 | 29.80 |

## Audi A6 2003

At 5 years of age, the mortality rate of a Audi A6 2003 (manufactured as a Car or Light Van) ranked number 141 out of 213 vehicles of the same age and type (any Car or Light Van constructed in 2003). One is the lowest (or best) and 213 the highest mortality rate. For vehicles reaching 100000 miles, its unreliability score (rate of failing an inspection) ranked 187 out of 208 vehicles of the same age, type, and mileage. One is the highest (or worst) and 208 the lowest rate of failing an inspection.

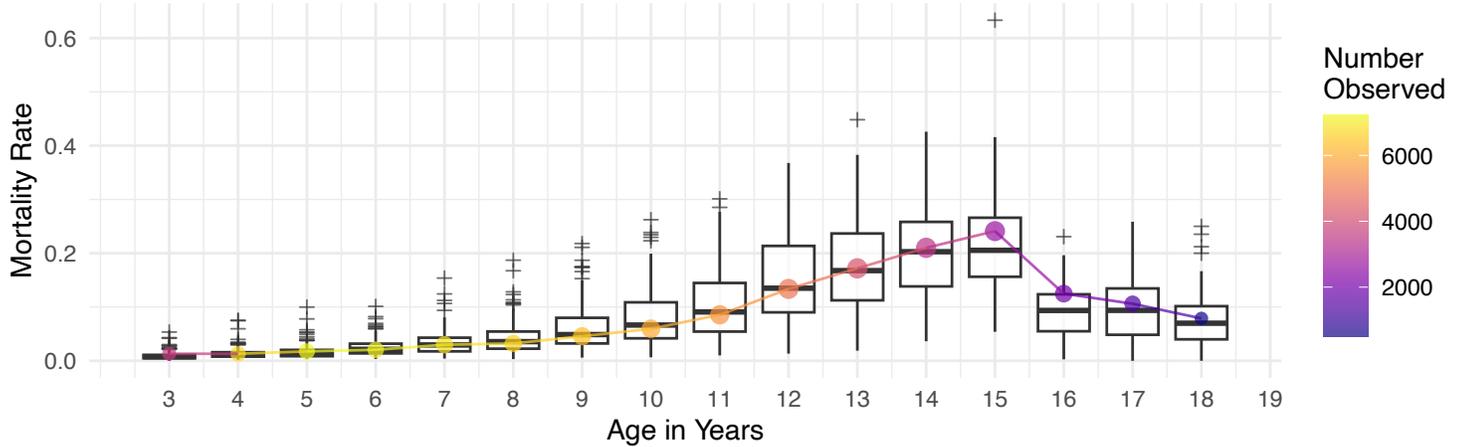

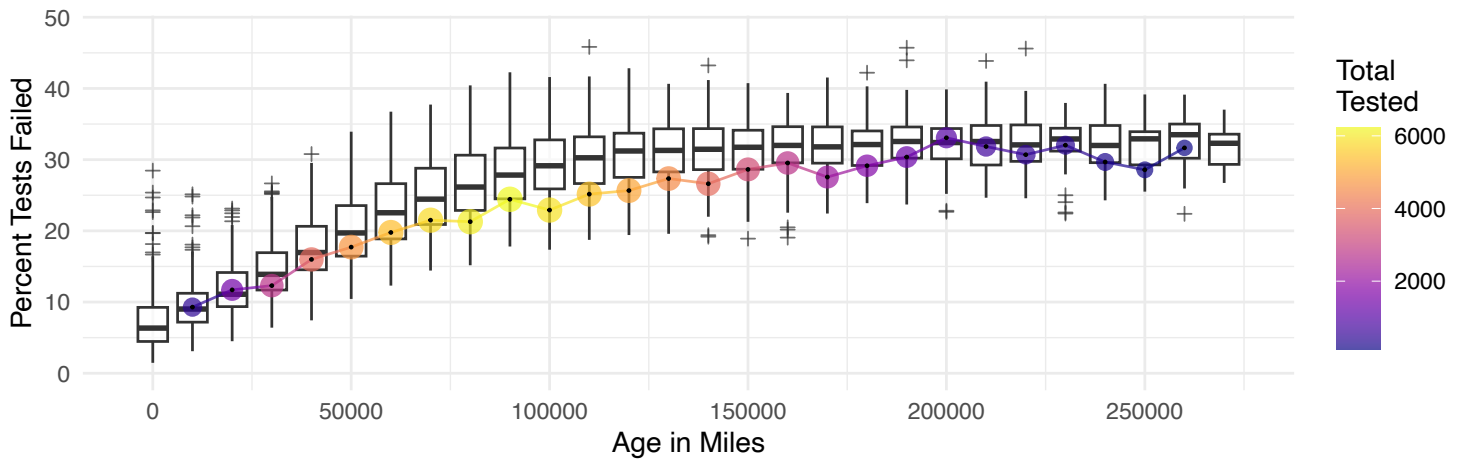

| Mortality rates | | | |
|---|---|---|---|
| Age in Years | Observed | Died | Mortality Rate |
| 3 | 3657 | 48 | 0.0131 |
| 4 | 6741 | 88 | 0.0131 |
| 5 | 7207 | 126 | 0.0175 |
| 6 | 7080 | 145 | 0.0205 |
| 7 | 6870 | 205 | 0.0298 |
| 8 | 6610 | 213 | 0.0322 |
| 9 | 6326 | 289 | 0.0457 |
| 10 | 5964 | 355 | 0.0595 |
| 11 | 5520 | 472 | 0.0855 |
| 12 | 4923 | 658 | 0.1340 |
| 13 | 4189 | 720 | 0.1720 |
| 14 | 3418 | 718 | 0.2100 |
| 15 | 2667 | 643 | 0.2410 |
| 16 | 1938 | 242 | 0.1250 |
| 17 | 1329 | 141 | 0.1060 |
| 18 | 508 | 40 | 0.0787 |

| Mechanical Reliability Rates | | |
|---|---|---|
| Mileage at test | N tested | Pct failed |
| 10000 | 463 | 9.29 |
| 20000 | 1452 | 11.70 |
| 30000 | 2755 | 12.30 |
| 40000 | 3903 | 16.00 |
| 50000 | 4702 | 17.70 |
| 60000 | 5361 | 19.80 |
| 70000 | 5801 | 21.50 |
| 80000 | 6065 | 21.30 |
| 90000 | 6213 | 24.40 |
| 100000 | 5904 | 22.90 |
| 110000 | 5442 | 25.20 |
| 120000 | 4981 | 25.70 |
| 130000 | 4513 | 27.30 |
| 140000 | 3915 | 26.60 |
| 150000 | 3287 | 28.60 |
| 160000 | 2763 | 29.50 |
| 170000 | 2174 | 27.60 |

## Audi A6 2004

At 5 years of age, the mortality rate of a Audi A6 2004 (manufactured as a Car or Light Van) ranked number 136 out of 229 vehicles of the same age and type (any Car or Light Van constructed in 2004). One is the lowest (or best) and 229 the highest mortality rate. For vehicles reaching 20000 miles, its unreliability score (rate of failing an inspection) ranked 156 out of 225 vehicles of the same age, type, and mileage. One is the highest (or worst) and 225 the lowest rate of failing an inspection.

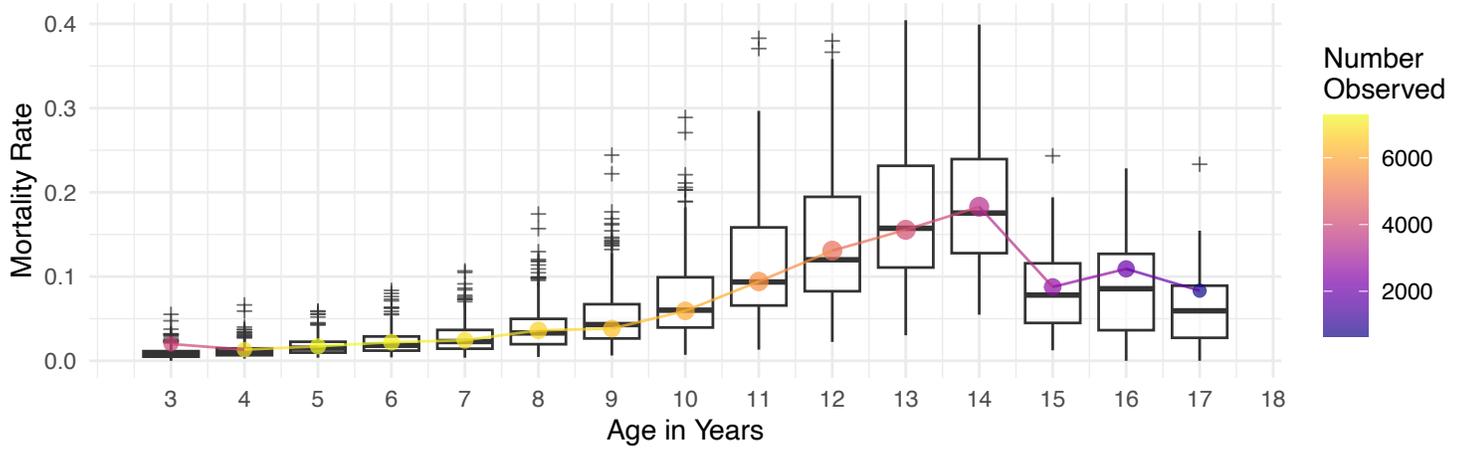

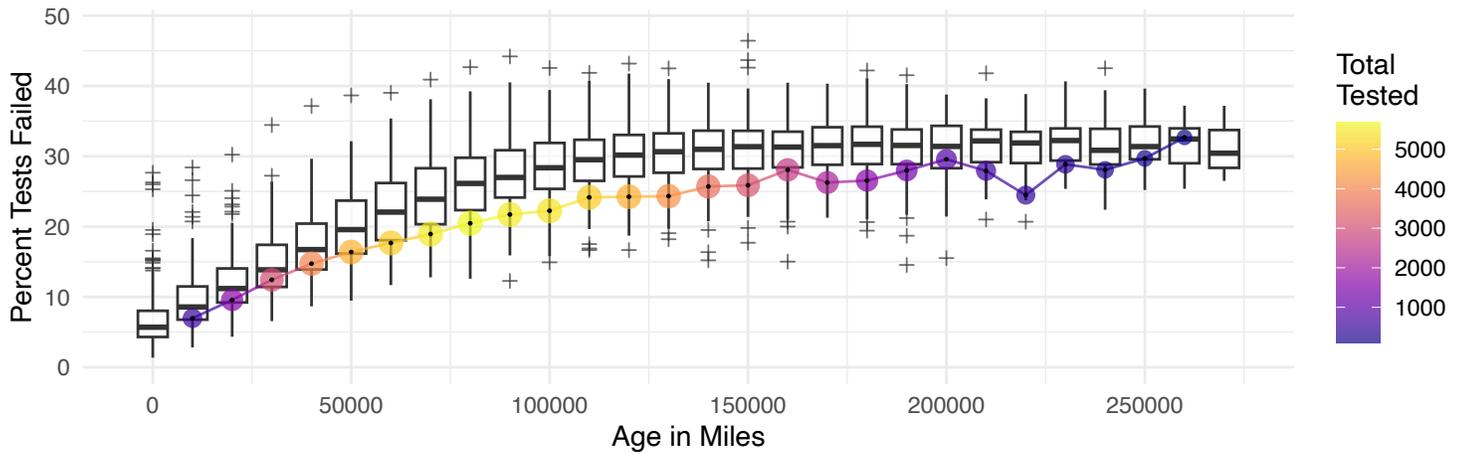

| Mortality rates | | | |
|---|---|---|---|
| Age in Years | Observed | Died | Mortality Rate |
| 3 | 3956 | 79 | 0.0200 |
| 4 | 6866 | 91 | 0.0133 |
| 5 | 7266 | 125 | 0.0172 |
| 6 | 7098 | 157 | 0.0221 |
| 7 | 6895 | 168 | 0.0244 |
| 8 | 6666 | 241 | 0.0362 |
| 9 | 6350 | 243 | 0.0383 |
| 10 | 6037 | 358 | 0.0593 |
| 11 | 5534 | 521 | 0.0941 |
| 12 | 4871 | 637 | 0.1310 |
| 13 | 4115 | 640 | 0.1560 |
| 14 | 3407 | 623 | 0.1830 |
| 15 | 2646 | 232 | 0.0877 |
| 16 | 1878 | 205 | 0.1090 |
| 17 | 649 | 54 | 0.0832 |

| Mechanical Reliability Rates | | |
|---|---|---|
| Mileage at test | N tested | Pct failed |
| 10000 | 531 | 6.97 |
| 20000 | 1630 | 9.57 |
| 30000 | 3032 | 12.40 |
| 40000 | 4046 | 14.70 |
| 50000 | 4718 | 16.40 |
| 60000 | 5108 | 17.70 |
| 70000 | 5497 | 18.90 |
| 80000 | 5685 | 20.50 |
| 90000 | 5516 | 21.70 |
| 100000 | 5424 | 22.30 |
| 110000 | 5121 | 24.20 |
| 120000 | 4693 | 24.20 |
| 130000 | 4217 | 24.30 |
| 140000 | 3637 | 25.70 |
| 150000 | 3169 | 25.90 |
| 160000 | 2636 | 28.10 |
| 170000 | 2178 | 26.30 |

## Audi A6 2005

At 5 years of age, the mortality rate of a Audi A6 2005 (manufactured as a Car or Light Van) ranked number 160 out of 240 vehicles of the same age and type (any Car or Light Van constructed in 2005). One is the lowest (or best) and 240 the highest mortality rate. For vehicles reaching 20000 miles, its unreliability score (rate of failing an inspection) ranked 146 out of 235 vehicles of the same age, type, and mileage. One is the highest (or worst) and 235 the lowest rate of failing an inspection.

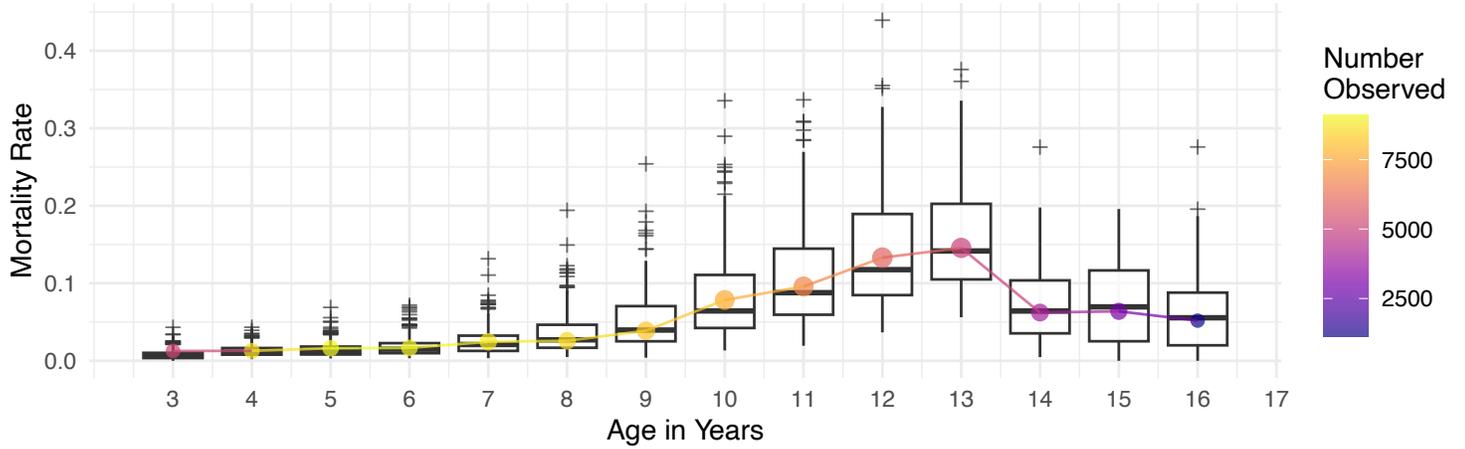

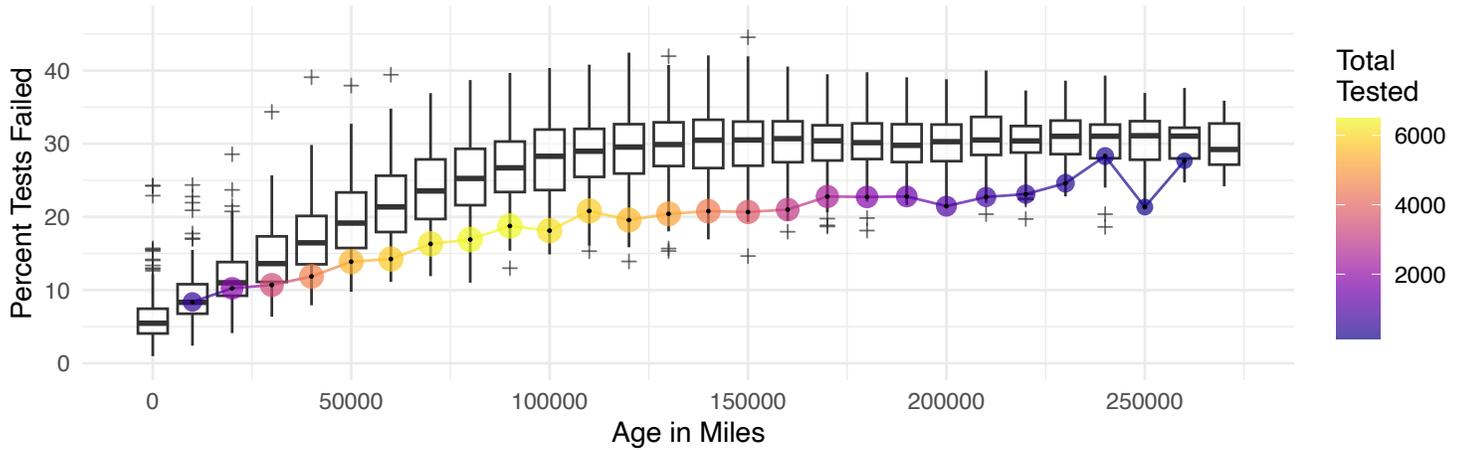

| Mortality rates | | | |
|---|---|---|---|
| Age in Years | Observed | Died | Mortality Rate |
| 3 | 5114 | 65 | 0.0127 |
| 4 | 8652 | 111 | 0.0128 |
| 5 | 9084 | 147 | 0.0162 |
| 6 | 8965 | 149 | 0.0166 |
| 7 | 8773 | 215 | 0.0245 |
| 8 | 8484 | 219 | 0.0258 |
| 9 | 8137 | 316 | 0.0388 |
| 10 | 7609 | 597 | 0.0785 |
| 11 | 6772 | 649 | 0.0958 |
| 12 | 5898 | 785 | 0.1330 |
| 13 | 4977 | 725 | 0.1460 |
| 14 | 4050 | 252 | 0.0622 |
| 15 | 2977 | 190 | 0.0638 |
| 16 | 1139 | 59 | 0.0518 |

| Mechanical Reliability Rates | | |
|---|---|---|
| Mileage at test | N tested | Pct failed |
| 10000 | 525 | 8.38 |
| 20000 | 1865 | 10.20 |
| 30000 | 3405 | 10.70 |
| 40000 | 4544 | 11.90 |
| 50000 | 5463 | 13.90 |
| 60000 | 5814 | 14.20 |
| 70000 | 6267 | 16.30 |
| 80000 | 6478 | 16.90 |
| 90000 | 6494 | 18.80 |
| 100000 | 6159 | 18.10 |
| 110000 | 5872 | 20.80 |
| 120000 | 5514 | 19.60 |
| 130000 | 5172 | 20.40 |
| 140000 | 4446 | 20.80 |
| 150000 | 3764 | 20.70 |
| 160000 | 2981 | 21.00 |
| 180000 | 1831 | 22.70 |

# Audi A6 2006

At 5 years of age, the mortality rate of a Audi A6 2006 (manufactured as a Car or Light Van) ranked number 113 out of 225 vehicles of the same age and type (any Car or Light Van constructed in 2006). One is the lowest (or best) and 225 the highest mortality rate. For vehicles reaching 20000 miles, its unreliability score (rate of failing an inspection) ranked 185 out of 220 vehicles of the same age, type, and mileage. One is the highest (or worst) and 220 the lowest rate of failing an inspection.

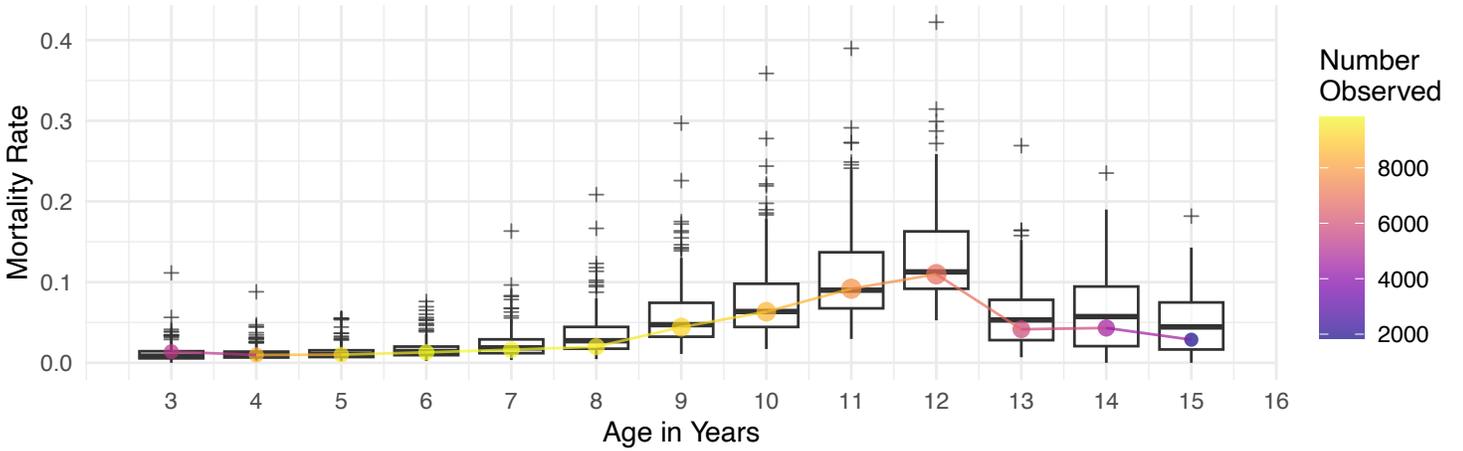

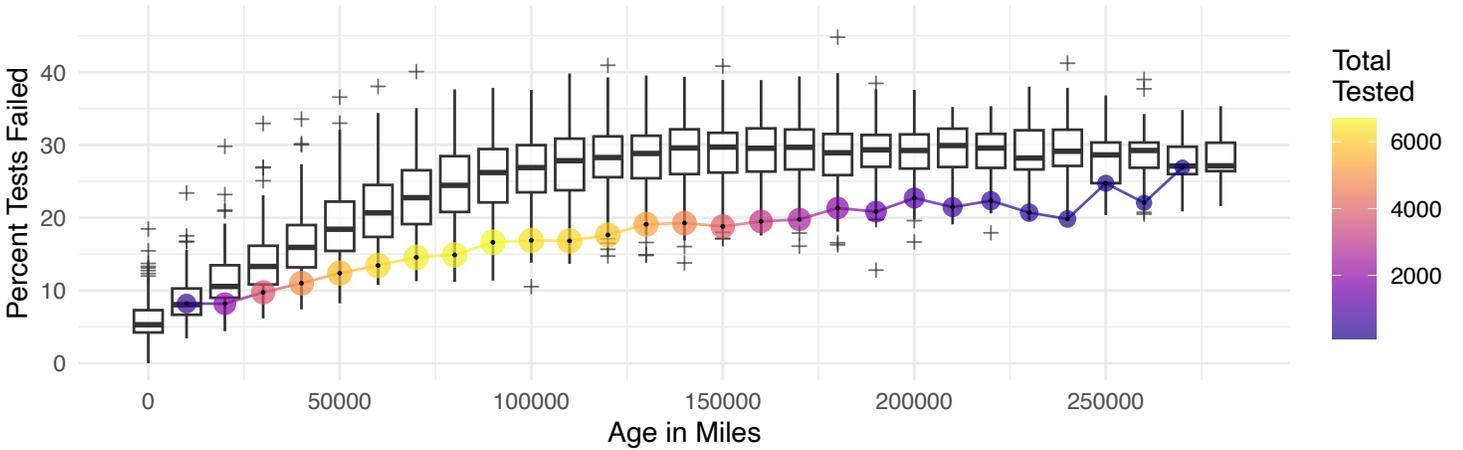

Mortality rates

| Age in Years | Observed | Died | Mortality Rate |
|---|---|---|---|
| 3 | 5337 | 74 | 0.01390 |
| 4 | 8544 | 83 | 0.00971 |
| 5 | 9466 | 96 | 0.01010 |
| 6 | 9809 | 127 | 0.01290 |
| 7 | 9709 | 157 | 0.01620 |
| 8 | 9474 | 189 | 0.01990 |
| 9 | 9107 | 403 | 0.04430 |
| 10 | 8513 | 539 | 0.06330 |
| 11 | 7732 | 710 | 0.09180 |
| 12 | 6866 | 754 | 0.11000 |
| 13 | 5825 | 242 | 0.04150 |
| 14 | 4581 | 198 | 0.04320 |
| 15 | 1850 | 53 | 0.02860 |

Mechanical Reliability Rates

| Mileage at test | N tested | Pct failed |
|---|---|---|
| 10000 | 573 | 8.20 |
| 20000 | 2029 | 8.18 |
| 30000 | 3775 | 9.75 |
| 40000 | 4901 | 11.00 |
| 50000 | 5701 | 12.40 |
| 60000 | 6183 | 13.40 |
| 70000 | 6534 | 14.60 |
| 80000 | 6613 | 14.90 |
| 90000 | 6707 | 16.60 |
| 100000 | 6340 | 16.90 |
| 110000 | 6125 | 16.80 |
| 120000 | 5880 | 17.60 |
| 130000 | 5307 | 19.10 |
| 140000 | 4735 | 19.30 |
| 150000 | 3924 | 18.80 |
| 160000 | 3044 | 19.50 |
| 170000 | 2474 | 19.80 |

## Audi A6 2007

At 5 years of age, the mortality rate of a Audi A6 2007 (manufactured as a Car or Light Van) ranked number 66 out of 219 vehicles of the same age and type (any Car or Light Van constructed in 2007). One is the lowest (or best) and 219 the highest mortality rate. For vehicles reaching 20000 miles, its unreliability score (rate of failing an inspection) ranked 187 out of 214 vehicles of the same age, type, and mileage. One is the highest (or worst) and 214 the lowest rate of failing an inspection.

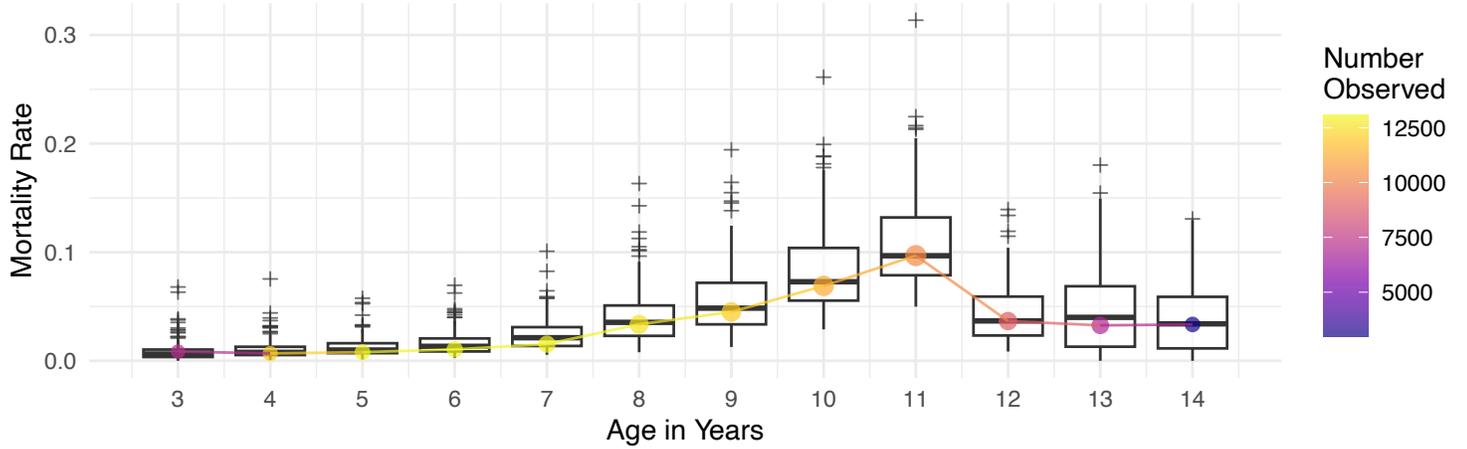

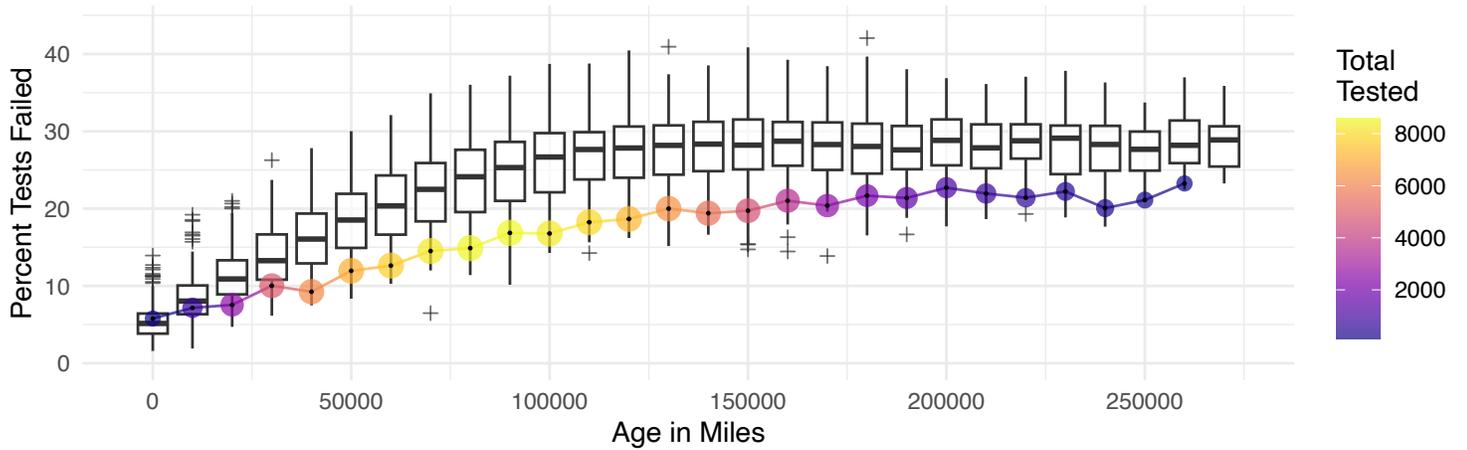

<table>
<tr><td colspan="4" align="center">Mortality rates</td></tr>
<tr><th>Age in Years</th><th>Observed</th><th>Died</th><th>Mortality Rate</th></tr>
<tr><td>3</td><td>7155</td><td>61</td><td>0.00853</td></tr>
<tr><td>4</td><td>12041</td><td>82</td><td>0.00681</td></tr>
<tr><td>5</td><td>12972</td><td>103</td><td>0.00794</td></tr>
<tr><td>6</td><td>13047</td><td>137</td><td>0.01050</td></tr>
<tr><td>7</td><td>12896</td><td>198</td><td>0.01540</td></tr>
<tr><td>8</td><td>12557</td><td>422</td><td>0.03360</td></tr>
<tr><td>9</td><td>11933</td><td>537</td><td>0.04500</td></tr>
<tr><td>10</td><td>11160</td><td>770</td><td>0.06900</td></tr>
<tr><td>11</td><td>10201</td><td>988</td><td>0.09690</td></tr>
<tr><td>12</td><td>8861</td><td>324</td><td>0.03660</td></tr>
<tr><td>13</td><td>6987</td><td>228</td><td>0.03260</td></tr>
<tr><td>14</td><td>2990</td><td>100</td><td>0.03340</td></tr>
</table>

Mechanical Reliability Rates

| Mileage at test | N tested | Pct failed |
|---|---|---|
| 0 | 121 | 5.79 |
| 10000 | 726 | 7.16 |
| 20000 | 2622 | 7.55 |
| 30000 | 4868 | 10.00 |
| 40000 | 6193 | 9.25 |
| 50000 | 7093 | 12.00 |
| 60000 | 7714 | 12.60 |
| 70000 | 8295 | 14.50 |
| 80000 | 8543 | 14.90 |
| 90000 | 8600 | 16.90 |
| 100000 | 8340 | 16.80 |
| 110000 | 7922 | 18.20 |
| 120000 | 7281 | 18.70 |
| 130000 | 6400 | 20.00 |
| 140000 | 5586 | 19.40 |
| 150000 | 4612 | 19.70 |
| 160000 | 3596 | 21.00 |

## Audi A6 2008

At 5 years of age, the mortality rate of a Audi A6 2008 (manufactured as a Car or Light Van) ranked number 102 out of 218 vehicles of the same age and type (any Car or Light Van constructed in 2008). One is the lowest (or best) and 218 the highest mortality rate. For vehicles reaching 20000 miles, its unreliability score (rate of failing an inspection) ranked 154 out of 212 vehicles of the same age, type, and mileage. One is the highest (or worst) and 212 the lowest rate of failing an inspection.

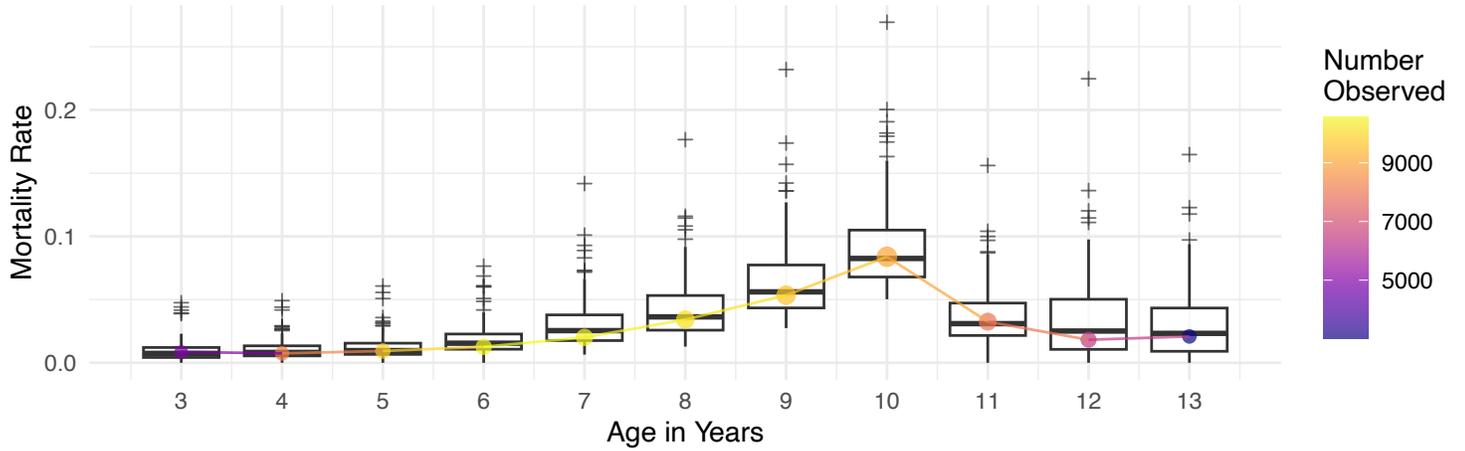

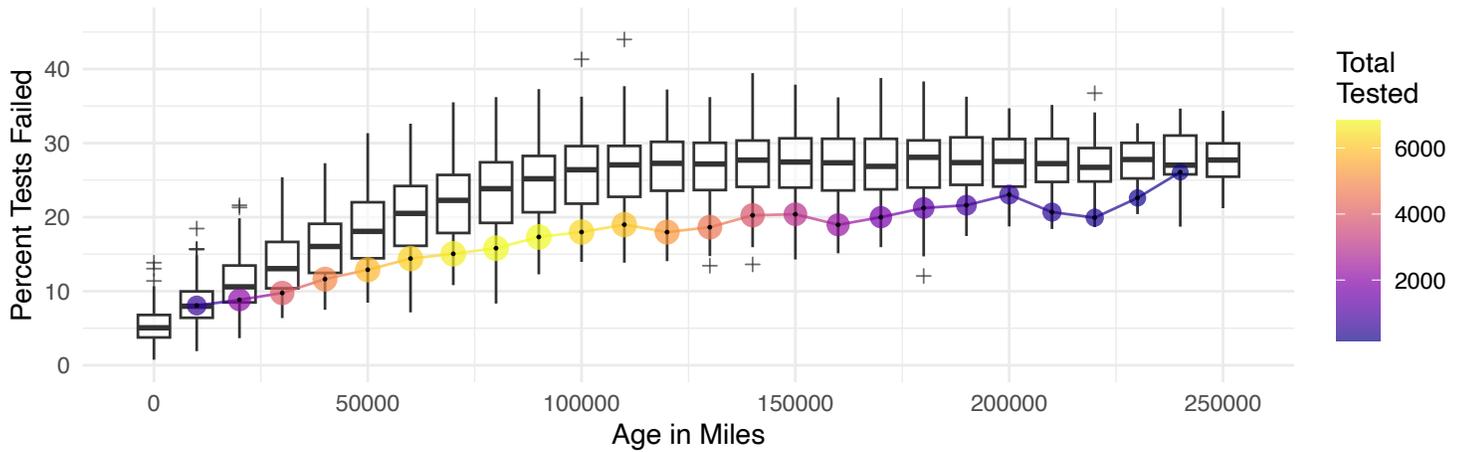

Mortality rates

| Age in Years | Observed | Died | Mortality Rate |
|---|---|---|---|
| 3 | 4909 | 42 | 0.00856 |
| 4 | 8297 | 62 | 0.00747 |
| 5 | 9862 | 92 | 0.00933 |
| 6 | 10536 | 133 | 0.01260 |
| 7 | 10386 | 209 | 0.02010 |
| 8 | 10121 | 346 | 0.03420 |
| 9 | 9627 | 515 | 0.05350 |
| 10 | 8990 | 754 | 0.08390 |
| 11 | 7961 | 259 | 0.03250 |
| 12 | 6482 | 118 | 0.01820 |
| 13 | 3020 | 63 | 0.02090 |

Mechanical Reliability Rates

| Mileage at test | N tested | Pct failed |
|---|---|---|
| 10000 | 607 | 8.07 |
| 20000 | 2094 | 8.83 |
| 30000 | 4006 | 9.76 |
| 40000 | 4892 | 11.60 |
| 50000 | 5823 | 12.90 |
| 60000 | 6232 | 14.40 |
| 70000 | 6658 | 15.00 |
| 80000 | 6846 | 15.80 |
| 90000 | 6834 | 17.30 |
| 100000 | 6364 | 18.00 |
| 110000 | 5982 | 19.00 |
| 120000 | 5260 | 18.00 |
| 130000 | 4564 | 18.60 |
| 140000 | 3841 | 20.30 |
| 150000 | 2991 | 20.40 |
| 160000 | 2293 | 19.00 |
| 180000 | 1243 | 21.20 |

# Audi A6 2009

At 5 years of age, the mortality rate of a Audi A6 2009 (manufactured as a Car or Light Van) ranked number 125 out of 205 vehicles of the same age and type (any Car or Light Van constructed in 2009). One is the lowest (or best) and 205 the highest mortality rate. For vehicles reaching 40000 miles, its unreliability score (rate of failing an inspection) ranked 173 out of 200 vehicles of the same age, type, and mileage. One is the highest (or worst) and 200 the lowest rate of failing an inspection.

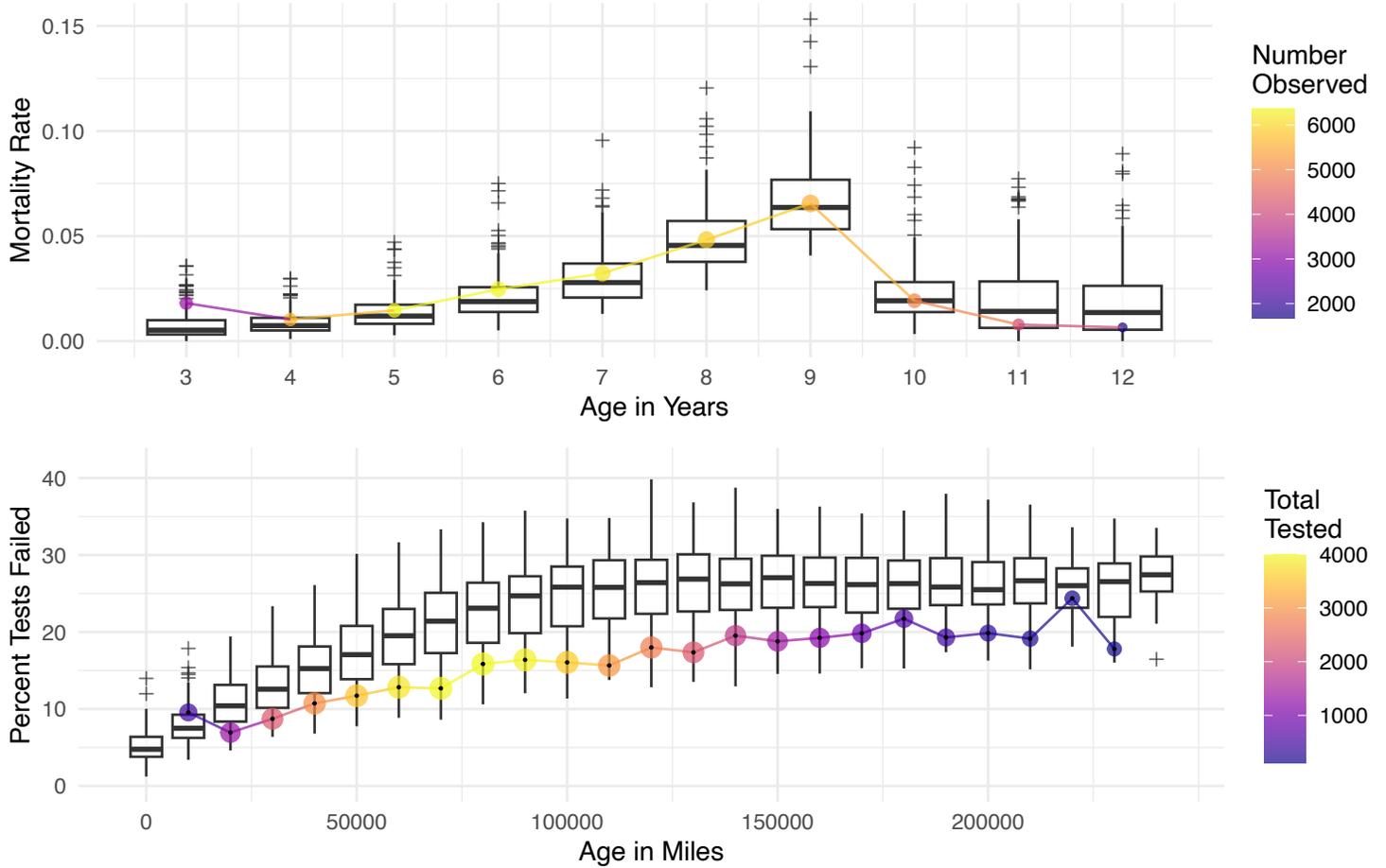

| Mortality rates | | | |
|---|---|---|---|
| Age in Years | Observed | Died | Mortality Rate |
| 3 | 3153 | 57 | 0.01810 |
| 4 | 5634 | 58 | 0.01030 |
| 5 | 6350 | 93 | 0.01460 |
| 6 | 6321 | 156 | 0.02470 |
| 7 | 6141 | 198 | 0.03220 |
| 8 | 5868 | 283 | 0.04820 |
| 9 | 5509 | 361 | 0.06550 |
| 10 | 4947 | 95 | 0.01920 |
| 11 | 4062 | 32 | 0.00788 |
| 12 | 1678 | 11 | 0.00656 |

| Mechanical Reliability Rates | | |
|---|---|---|
| Mileage at test | N tested | Pct failed |
| 10000 | 450 | 9.56 |
| 20000 | 1341 | 6.94 |
| 30000 | 2280 | 8.73 |
| 40000 | 2955 | 10.70 |
| 50000 | 3505 | 11.70 |
| 60000 | 3835 | 12.80 |
| 70000 | 3989 | 12.70 |
| 80000 | 4001 | 15.80 |
| 90000 | 3943 | 16.40 |
| 100000 | 3570 | 16.10 |
| 110000 | 3042 | 15.60 |
| 120000 | 2746 | 18.00 |
| 130000 | 2259 | 17.40 |
| 140000 | 1823 | 19.50 |
| 150000 | 1421 | 18.80 |
| 160000 | 1159 | 19.20 |
| 190000 | 451 | 19.30 |

## Audi A6 2010

At 5 years of age, the mortality rate of a Audi A6 2010 (manufactured as a Car or Light Van) ranked number 144 out of 206 vehicles of the same age and type (any Car or Light Van constructed in 2010). One is the lowest (or best) and 206 the highest mortality rate. For vehicles reaching 20000 miles, its unreliability score (rate of failing an inspection) ranked 118 out of 201 vehicles of the same age, type, and mileage. One is the highest (or worst) and 201 the lowest rate of failing an inspection.

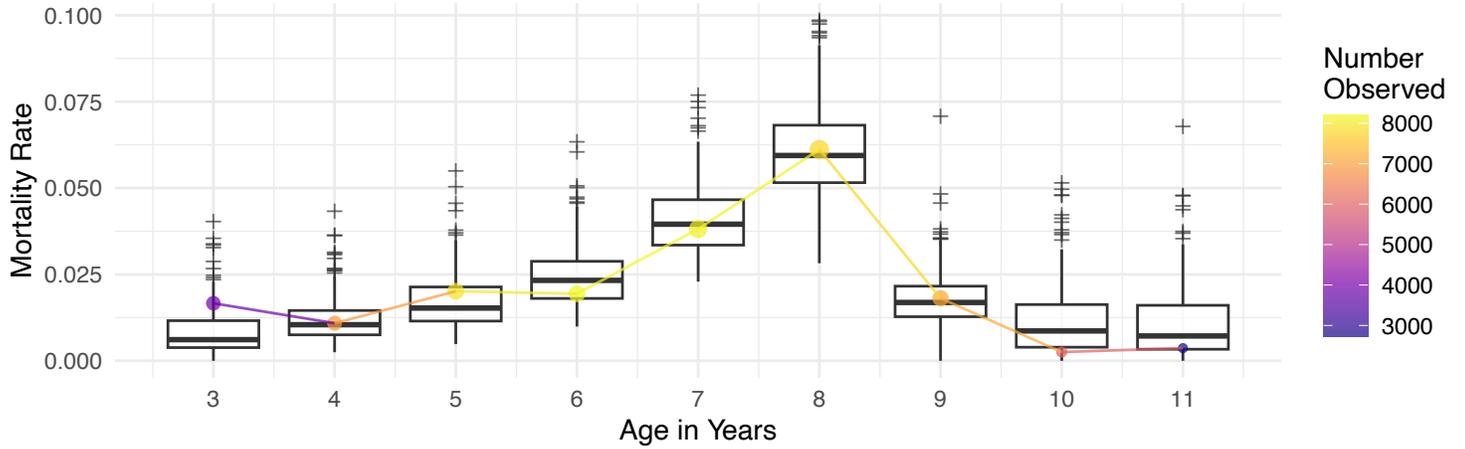

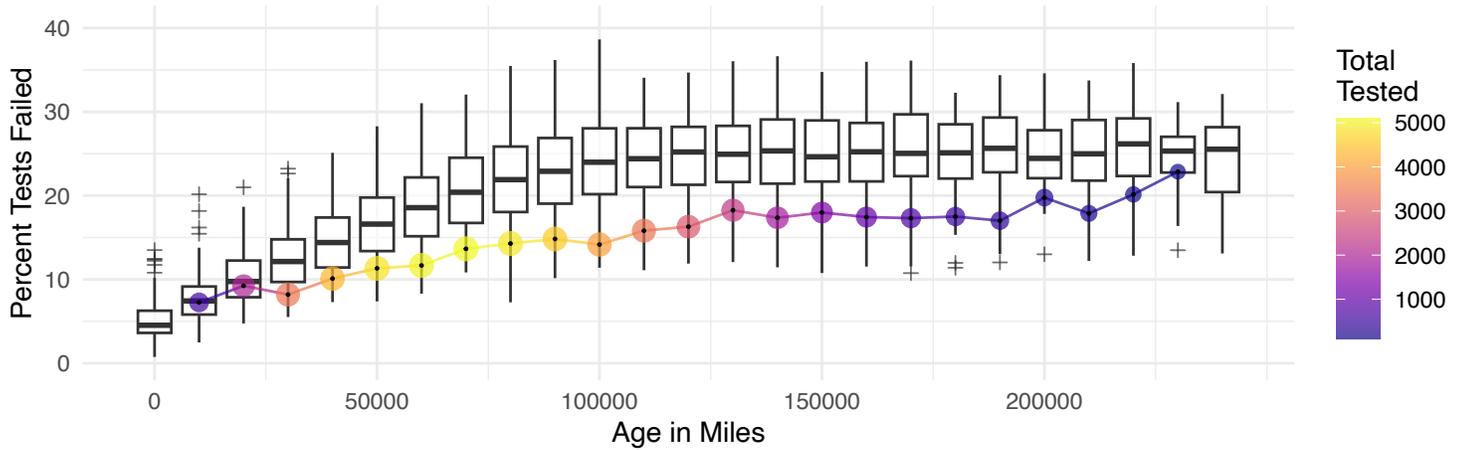

### Mechanical Reliability Rates

| Mileage at test | N tested | Pct failed |
|-----------------|----------|------------|
| 10000 | 565 | 7.26 |
| 20000 | 2002 | 9.24 |
| 30000 | 3393 | 8.19 |
| 40000 | 4305 | 10.10 |
| 50000 | 4847 | 11.30 |
| 60000 | 5040 | 11.70 |
| 70000 | 5105 | 13.70 |
| 80000 | 4928 | 14.30 |
| 90000 | 4552 | 14.80 |
| 110000 | 3377 | 15.80 |
| 120000 | 2915 | 16.30 |
| 130000 | 2377 | 18.30 |
| 140000 | 1849 | 17.40 |
| 150000 | 1379 | 18.00 |
| 160000 | 969 | 17.40 |
| 170000 | 699 | 17.30 |
| 180000 | 497 | 17.50 |

### Mortality rates

| Age in Years | Observed | Died | Mortality Rate |
|--------------|----------|------|----------------|
| 3 | 3843 | 64 | 0.01670 |
| 4 | 6890 | 75 | 0.01090 |
| 5 | 7951 | 160 | 0.02010 |
| 6 | 8188 | 159 | 0.01940 |
| 7 | 8099 | 309 | 0.03820 |
| 8 | 7766 | 475 | 0.06120 |
| 9 | 7128 | 129 | 0.01810 |
| 10 | 5927 | 15 | 0.00253 |
| 11 | 2732 | 10 | 0.00366 |

**Audi A6 2011**

At 5 years of age, the mortality rate of a Audi A6 2011 (manufactured as a Car or Light Van) ranked number 139 out of 211 vehicles of the same age and type (any Car or Light Van constructed in 2011). One is the lowest (or best) and 211 the highest mortality rate. For vehicles reaching 20000 miles, its unreliability score (rate of failing an inspection) ranked 162 out of 205 vehicles of the same age, type, and mileage. One is the highest (or worst) and 205 the lowest rate of failing an inspection.

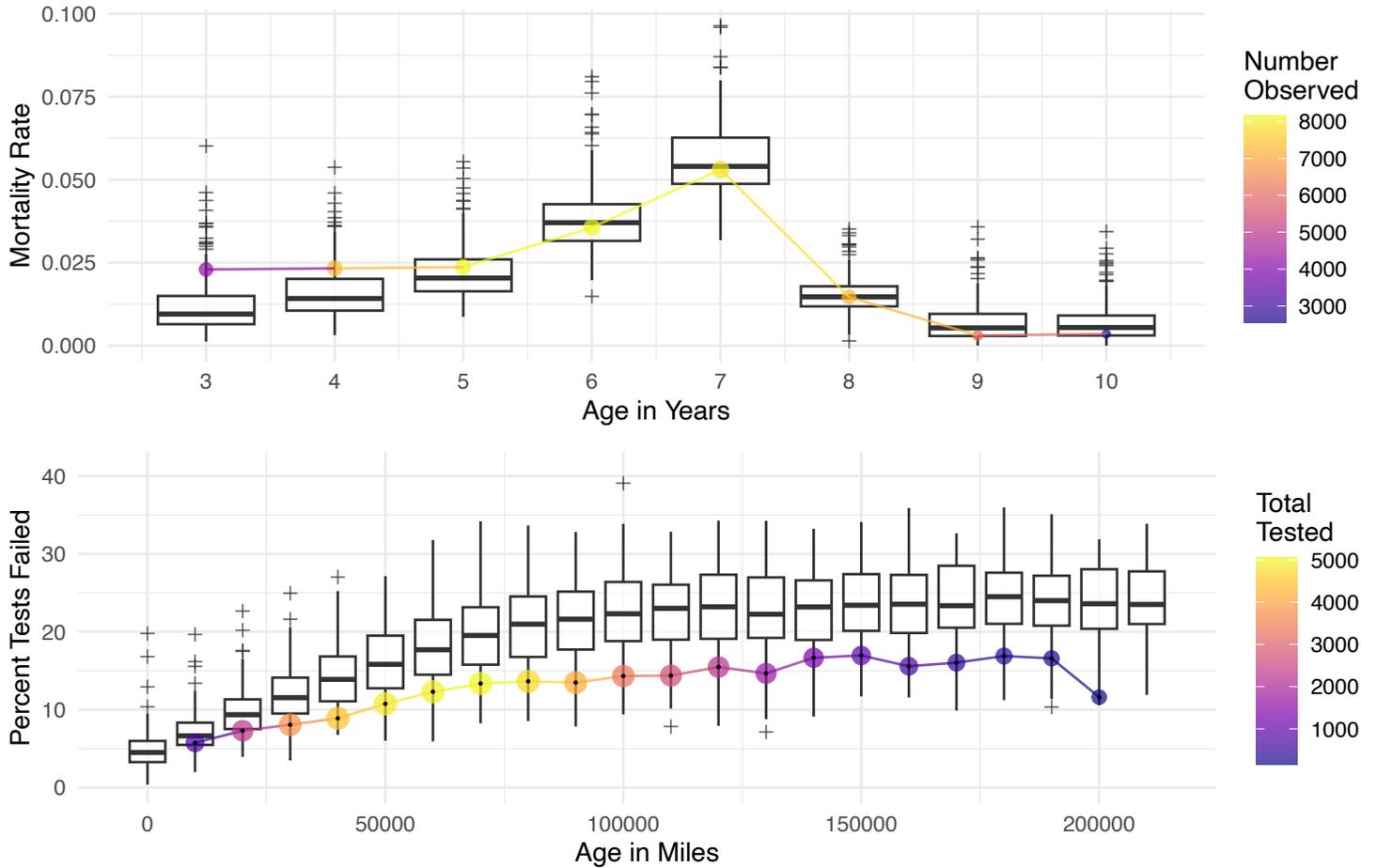

Mortality rates

| Age in Years | Observed | Died | Mortality Rate |
| --- | --- | --- | --- |
| 3 | 4227 | 97 | 0.02290 |
| 4 | 7302 | 170 | 0.02330 |
| 5 | 8158 | 193 | 0.02370 |
| 6 | 8168 | 291 | 0.03560 |
| 7 | 7882 | 418 | 0.05300 |
| 8 | 7243 | 106 | 0.01460 |
| 9 | 5939 | 18 | 0.00303 |
| 10 | 2549 | 9 | 0.00353 |

Mechanical Reliability Rates

| Mileage at test | N tested | Pct failed |
| --- | --- | --- |
| 10000 | 679 | 5.74 |
| 20000 | 2248 | 7.30 |
| 30000 | 3757 | 8.06 |
| 40000 | 4441 | 8.89 |
| 50000 | 4911 | 10.80 |
| 60000 | 4968 | 12.30 |
| 70000 | 5061 | 13.40 |
| 80000 | 4662 | 13.60 |
| 90000 | 4147 | 13.50 |
| 100000 | 3332 | 14.30 |
| 110000 | 2699 | 14.40 |
| 120000 | 2210 | 15.50 |
| 130000 | 1726 | 14.70 |
| 140000 | 1272 | 16.70 |
| 150000 | 861 | 17.00 |
| 170000 | 424 | 16.00 |
| 180000 | 302 | 16.90 |

**Audi A6 2012**

At 5 years of age, the mortality rate of a Audi A6 2012 (manufactured as a Car or Light Van) ranked number 188 out of 212 vehicles of the same age and type (any Car or Light Van constructed in 2012). One is the lowest (or best) and 212 the highest mortality rate. For vehicles reaching 20000 miles, its unreliability score (rate of failing an inspection) ranked 164 out of 206 vehicles of the same age, type, and mileage. One is the highest (or worst) and 206 the lowest rate of failing an inspection.

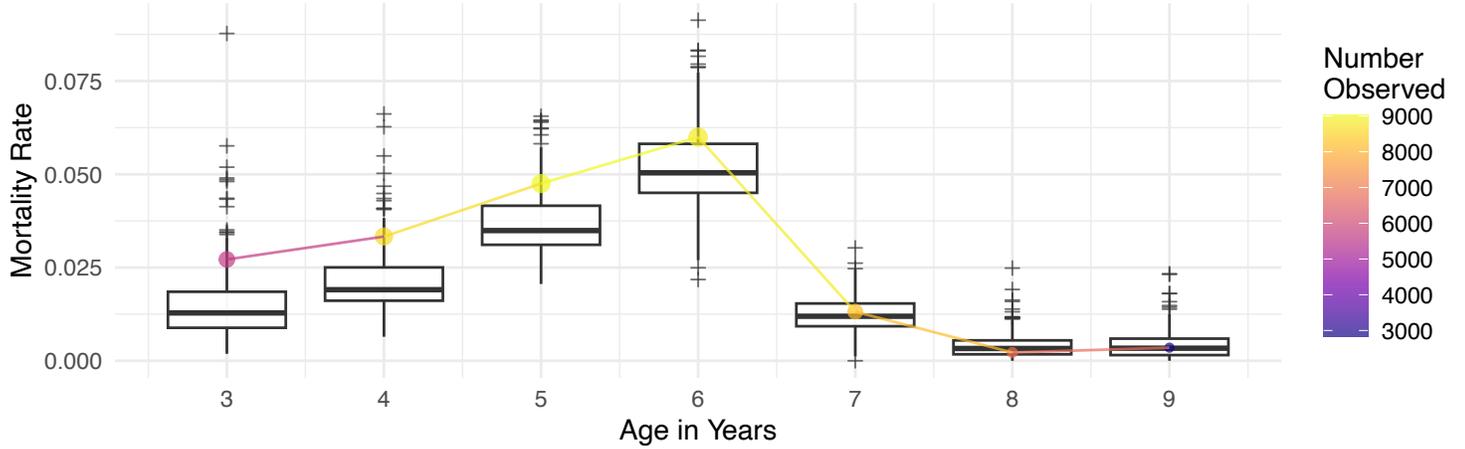

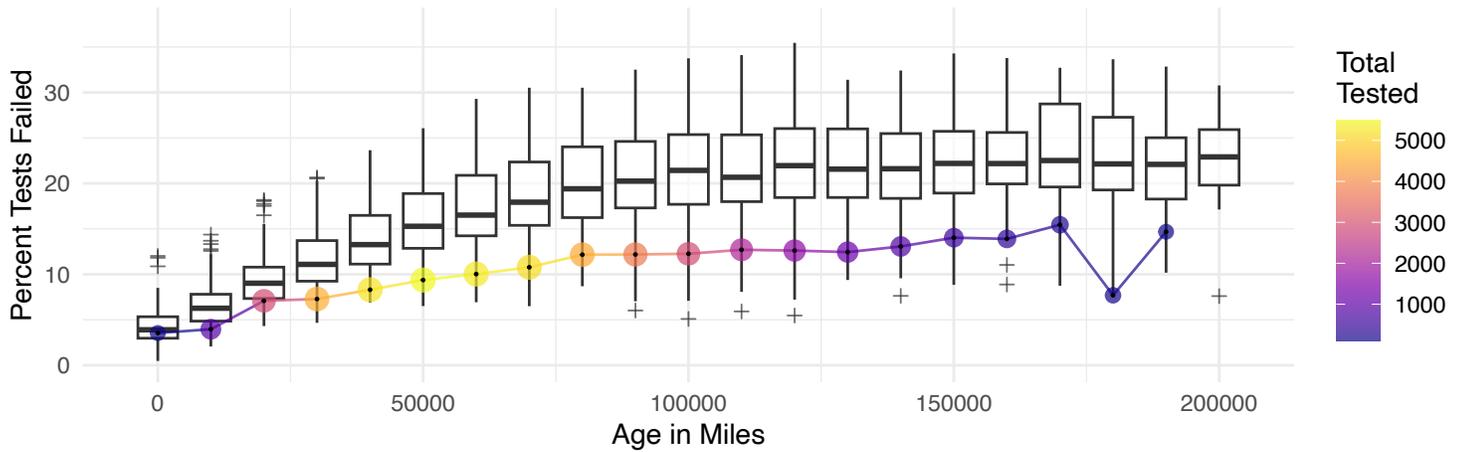

Mortality rates

| Age in Years | Observed | Died | Mortality Rate |
|---|---|---|---|
| 3 | 5668 | 154 | 0.02720 |
| 4 | 8586 | 286 | 0.03330 |
| 5 | 9009 | 428 | 0.04750 |
| 6 | 8822 | 529 | 0.06000 |
| 7 | 8132 | 107 | 0.01320 |
| 8 | 6754 | 15 | 0.00222 |
| 9 | 2838 | 10 | 0.00352 |

Mechanical Reliability Rates

| Mileage at test | N tested | Pct failed |
|---|---|---|
| 0 | 113 | 3.54 |
| 10000 | 883 | 3.96 |
| 20000 | 2893 | 7.09 |
| 30000 | 4529 | 7.29 |
| 40000 | 5227 | 8.30 |
| 50000 | 5495 | 9.37 |
| 60000 | 5315 | 10.00 |
| 70000 | 5142 | 10.80 |
| 80000 | 4457 | 12.20 |
| 90000 | 3727 | 12.20 |
| 100000 | 2896 | 12.30 |
| 110000 | 2181 | 12.70 |
| 120000 | 1689 | 12.60 |
| 130000 | 1133 | 12.40 |
| 140000 | 819 | 13.10 |
| 150000 | 556 | 14.00 |
| 160000 | 367 | 13.90 |

## Audi A6 2013

At 5 years of age, the mortality rate of a Audi A6 2013 (manufactured as a Car or Light Van) ranked number 199 out of 221 vehicles of the same age and type (any Car or Light Van constructed in 2013). One is the lowest (or best) and 221 the highest mortality rate. For vehicles reaching 20000 miles, its unreliability score (rate of failing an inspection) ranked 140 out of 215 vehicles of the same age, type, and mileage. One is the highest (or worst) and 215 the lowest rate of failing an inspection.

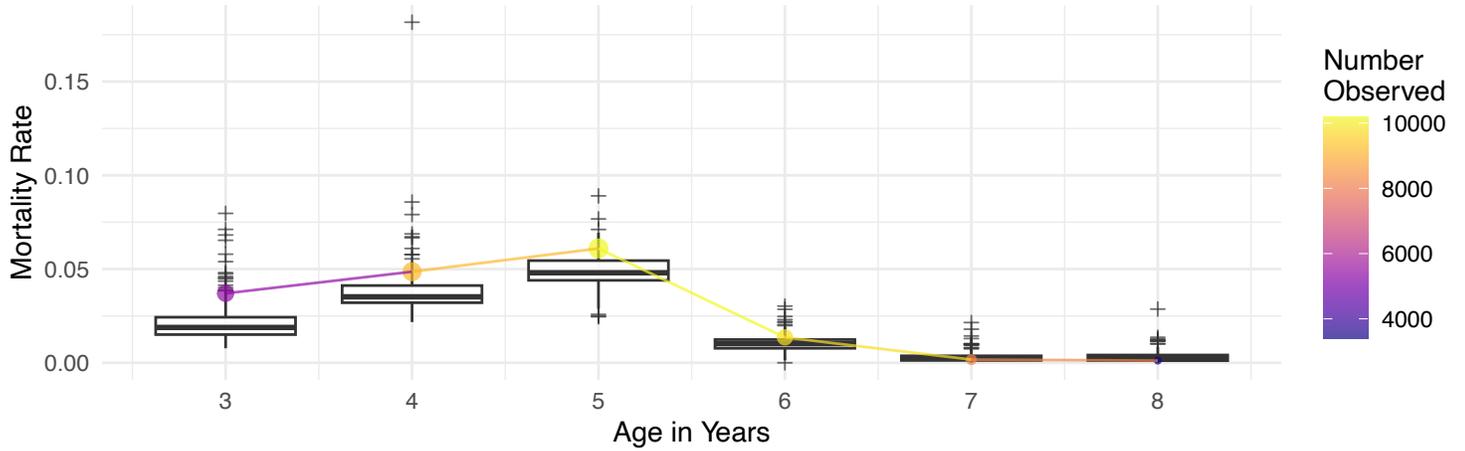

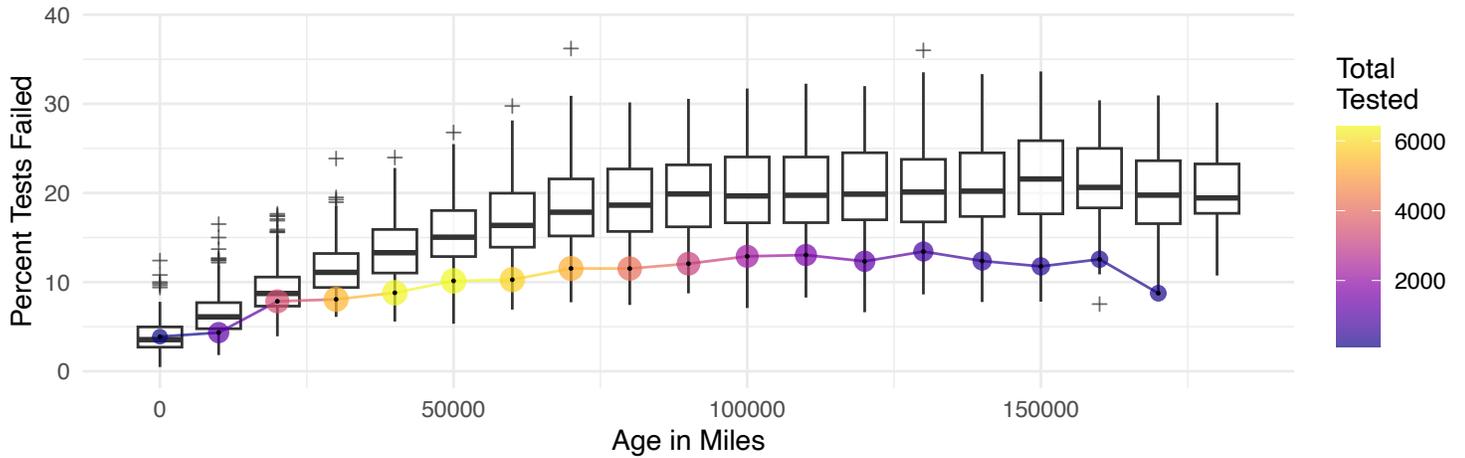

Mortality rates

| Age in Years | Observed | Died | Mortality Rate |
|---|---|---|---|
| 3 | 5428 | 201 | 0.03700 |
| 4 | 9138 | 444 | 0.04860 |
| 5 | 10167 | 620 | 0.06100 |
| 6 | 9731 | 132 | 0.01360 |
| 7 | 8175 | 14 | 0.00171 |
| 8 | 3403 | 4 | 0.00118 |

Mechanical Reliability Rates

| Mileage at test | N tested | Pct failed |
|---|---|---|
| 0 | 103 | 3.88 |
| 10000 | 1085 | 4.33 |
| 20000 | 3467 | 7.85 |
| 30000 | 5406 | 8.07 |
| 40000 | 6374 | 8.80 |
| 50000 | 6422 | 10.10 |
| 60000 | 5881 | 10.30 |
| 70000 | 5264 | 11.50 |
| 80000 | 4149 | 11.50 |
| 90000 | 3133 | 12.10 |
| 100000 | 2234 | 12.90 |
| 110000 | 1603 | 13.00 |
| 120000 | 1127 | 12.30 |
| 130000 | 685 | 13.40 |
| 140000 | 477 | 12.40 |
| 150000 | 323 | 11.80 |
| 170000 | 160 | 8.75 |

## Audi A6 2014

At 5 years of age, the mortality rate of a Audi A6 2014 (manufactured as a Car or Light Van) ranked number 184 out of 236 vehicles of the same age and type (any Car or Light Van constructed in 2014). One is the lowest (or best) and 236 the highest mortality rate. For vehicles reaching 20000 miles, its unreliability score (rate of failing an inspection) ranked 177 out of 230 vehicles of the same age, type, and mileage. One is the highest (or worst) and 230 the lowest rate of failing an inspection.

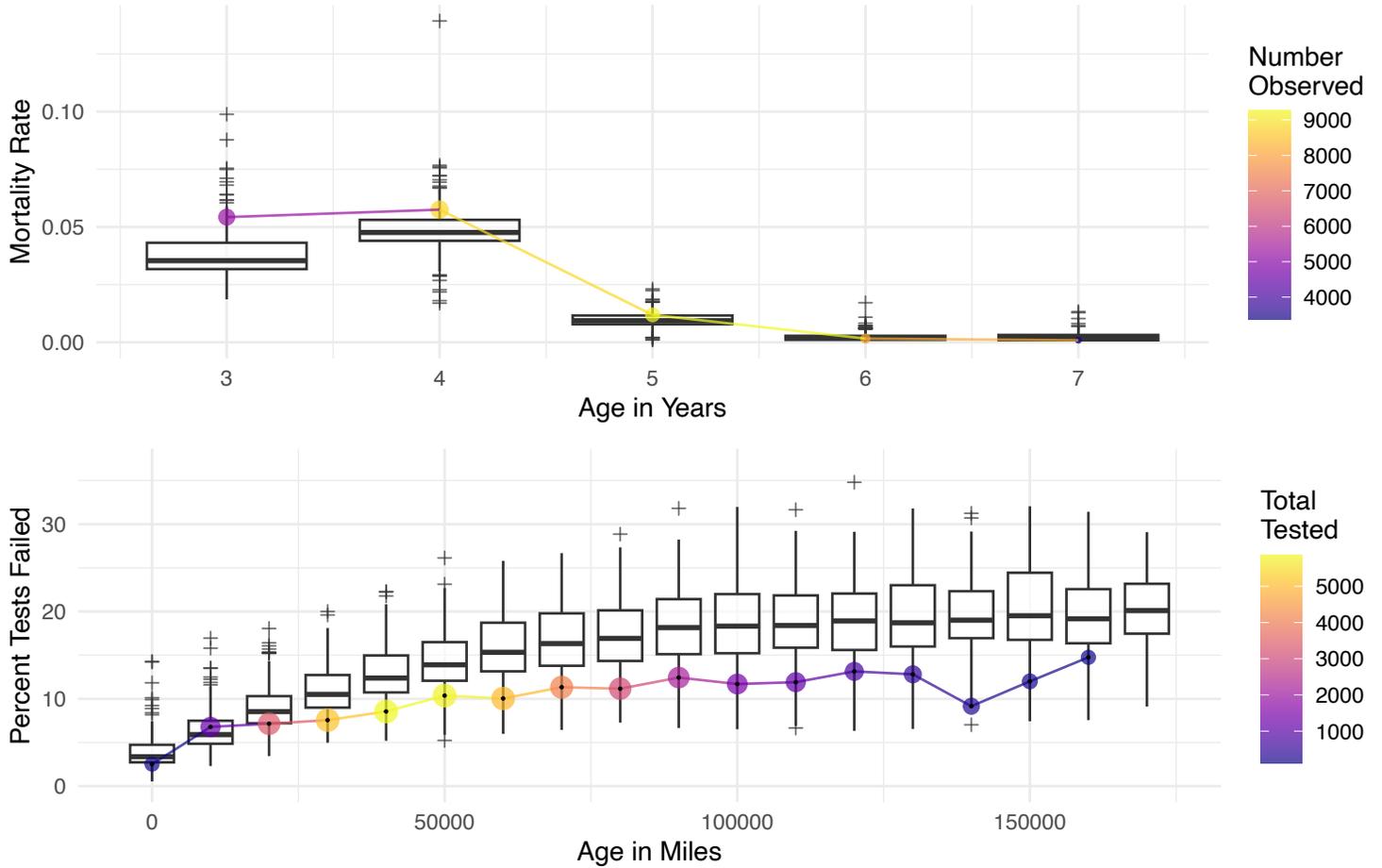

Mortality rates

| Age in Years | Observed | Died | Mortality Rate |
|---|---|---|---|
| 3 | 5233 | 284 | 0.054300 |
| 4 | 8780 | 505 | 0.057500 |
| 5 | 9254 | 110 | 0.011900 |
| 6 | 8107 | 13 | 0.001600 |
| 7 | 3377 | 3 | 0.000888 |

Mechanical Reliability Rates

| Mileage at test | N tested | Pct failed |
|---|---|---|
| 0 | 119 | 2.52 |
| 10000 | 1179 | 6.79 |
| 20000 | 3399 | 7.15 |
| 30000 | 5119 | 7.56 |
| 40000 | 5868 | 8.55 |
| 50000 | 5792 | 10.40 |
| 60000 | 4999 | 10.00 |
| 70000 | 4077 | 11.30 |
| 80000 | 3064 | 11.20 |
| 90000 | 2211 | 12.40 |
| 100000 | 1479 | 11.70 |
| 110000 | 1100 | 11.90 |
| 120000 | 700 | 13.10 |
| 130000 | 445 | 12.80 |
| 140000 | 284 | 9.15 |
| 150000 | 175 | 12.00 |
| 160000 | 122 | 14.80 |

**Audi A6 2015**

At 5 years of age, the mortality rate of a Audi A6 2015 (manufactured as a Car or Light Van) ranked number 42 out of 247 vehicles of the same age and type (any Car or Light Van constructed in 2015). One is the lowest (or best) and 247 the highest mortality rate. For vehicles reaching 20000 miles, its unreliability score (rate of failing an inspection) ranked 183 out of 241 vehicles of the same age, type, and mileage. One is the highest (or worst) and 241 the lowest rate of failing an inspection.

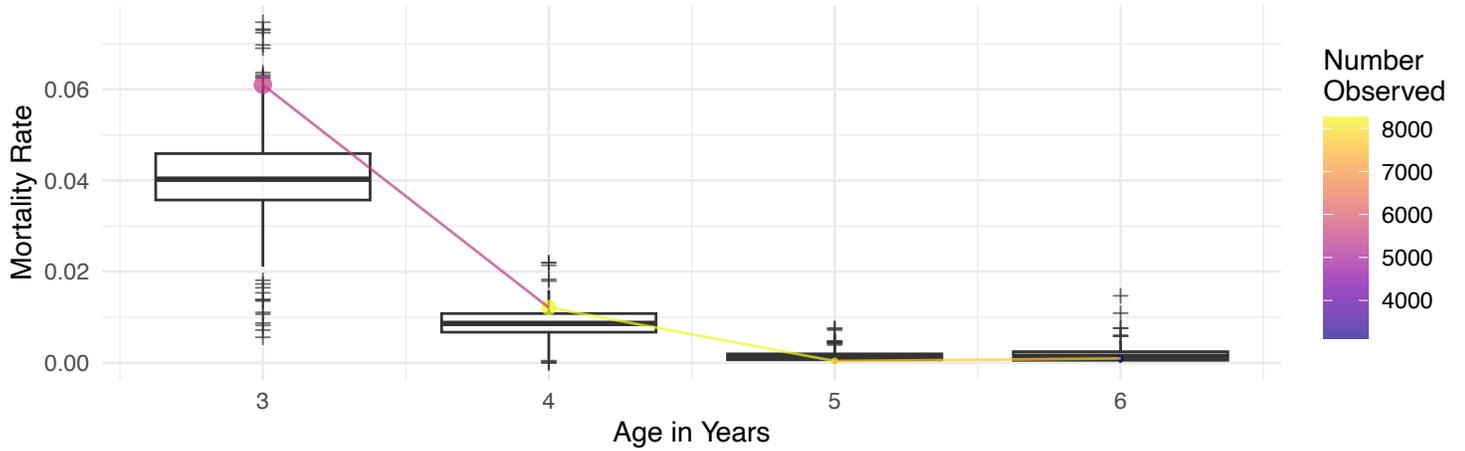

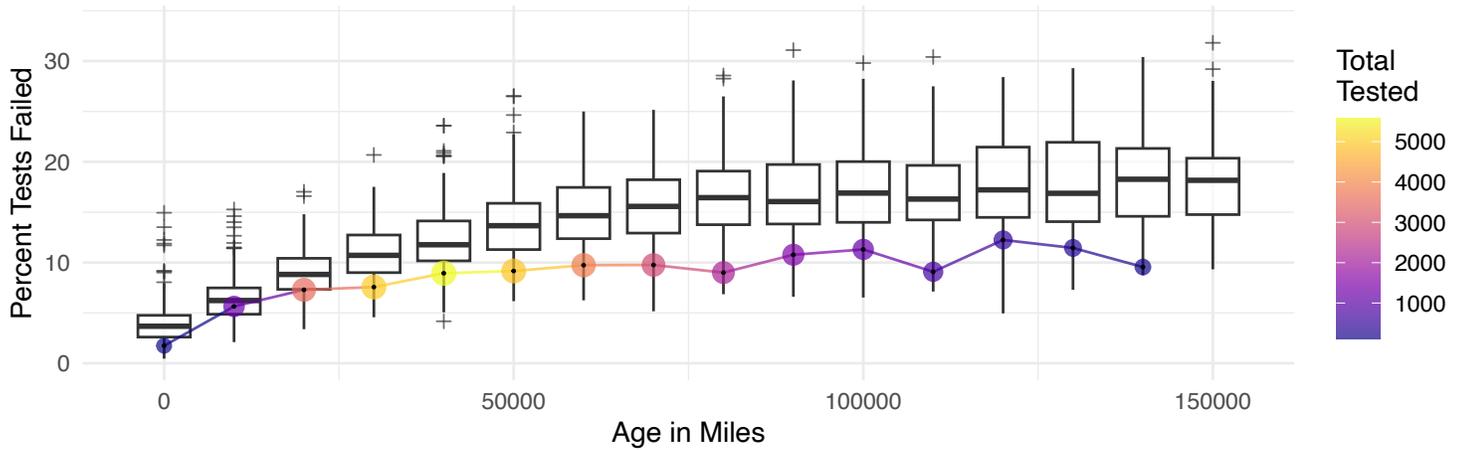

Mortality rates

| Age in Years | Observed | Died | Mortality Rate |
|---|---|---|---|
| 3 | 5426 | 331 | 0.061000 |
| 4 | 8253 | 100 | 0.012100 |
| 5 | 7729 | 3 | 0.000388 |
| 6 | 3118 | 3 | 0.000962 |

Mechanical Reliability Rates

| Mileage at test | N tested | Pct failed |
|---|---|---|
| 0 | 115 | 1.74 |
| 10000 | 1170 | 5.64 |
| 20000 | 3539 | 7.29 |
| 30000 | 5011 | 7.56 |
| 40000 | 5584 | 8.94 |
| 50000 | 4913 | 9.16 |
| 60000 | 3851 | 9.74 |
| 70000 | 2921 | 9.76 |
| 80000 | 2133 | 9.00 |
| 90000 | 1504 | 10.80 |
| 100000 | 1106 | 11.30 |
| 110000 | 694 | 9.08 |
| 120000 | 441 | 12.20 |
| 130000 | 262 | 11.50 |
| 140000 | 157 | 9.55 |

**Audi A6 2016**

At 5 years of age, the mortality rate of a Audi A6 2016 (manufactured as a Car or Light Van) ranked number 156 out of 252 vehicles of the same age and type (any Car or Light Van constructed in 2016). One is the lowest (or best) and 252 the highest mortality rate. For vehicles reaching 20000 miles, its unreliability score (rate of failing an inspection) ranked 140 out of 246 vehicles of the same age, type, and mileage. One is the highest (or worst) and 246 the lowest rate of failing an inspection.

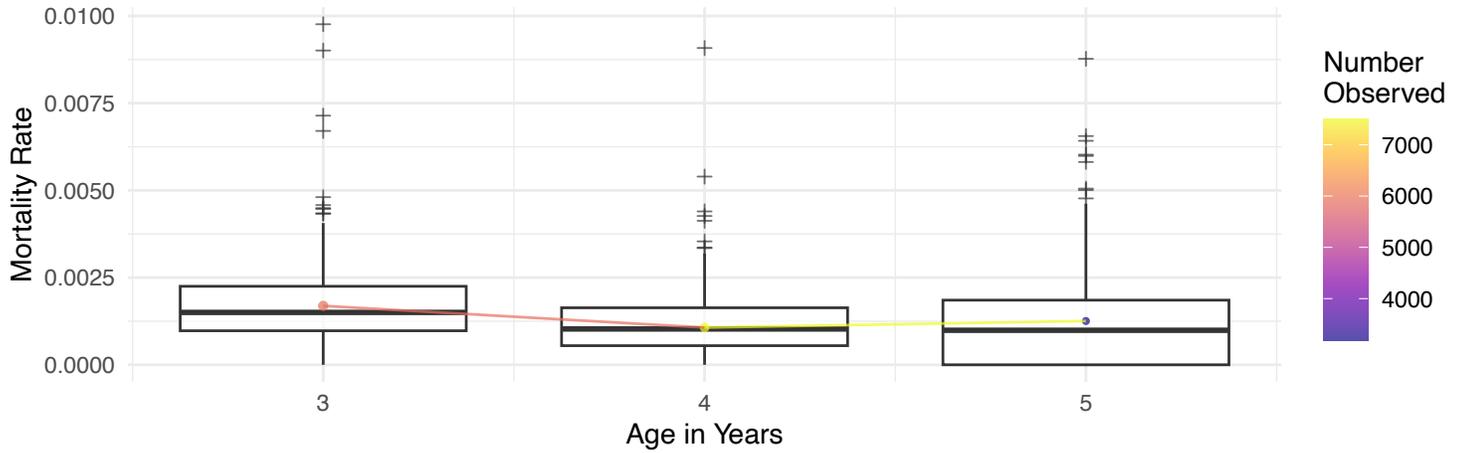

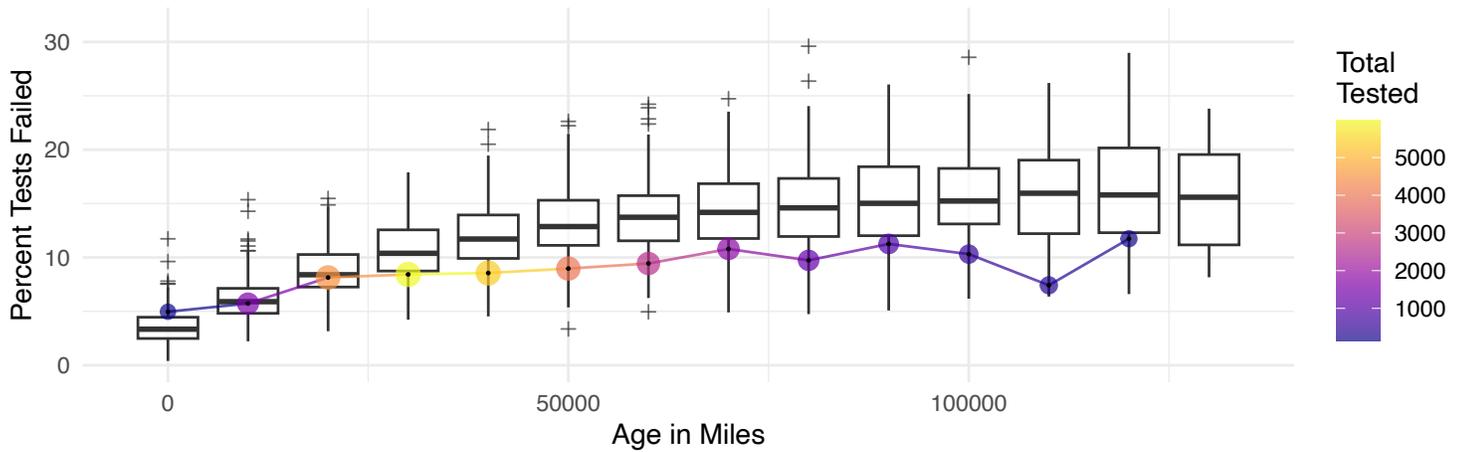

Mortality rates

| Age in Years | Observed | Died | Mortality Rate |
|---|---|---|---|
| 3 | 5918 | 10 | 0.00169 |
| 4 | 7489 | 8 | 0.00107 |
| 5 | 3187 | 4 | 0.00126 |

Mechanical Reliability Rates

| Mileage at test | N tested | Pct failed |
|---|---|---|
| 0 | 141 | 4.96 |
| 10000 | 1554 | 5.73 |
| 20000 | 4511 | 8.14 |
| 30000 | 5993 | 8.43 |
| 40000 | 5387 | 8.56 |
| 50000 | 3991 | 8.97 |
| 60000 | 2648 | 9.44 |
| 70000 | 1789 | 10.80 |
| 80000 | 1202 | 9.73 |
| 90000 | 844 | 11.30 |
| 100000 | 514 | 10.30 |
| 110000 | 310 | 7.42 |
| 120000 | 196 | 11.70 |

**Audi A6 2017**

At 3 years of age, the mortality rate of a Audi A6 2017 (manufactured as a Car or Light Van) ranked number 146 out of 247 vehicles of the same age and type (any Car or Light Van constructed in 2017). One is the lowest (or best) and 247 the highest mortality rate. For vehicles reaching 20000 miles, its unreliability score (rate of failing an inspection) ranked 171 out of 240 vehicles of the same age, type, and mileage. One is the highest (or worst) and 240 the lowest rate of failing an inspection.

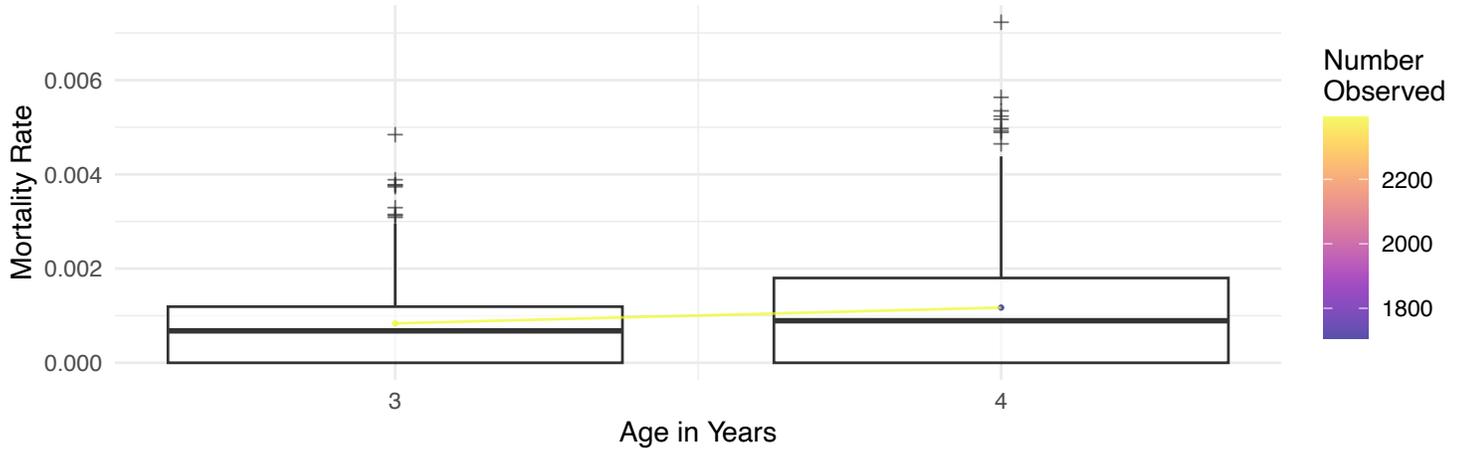

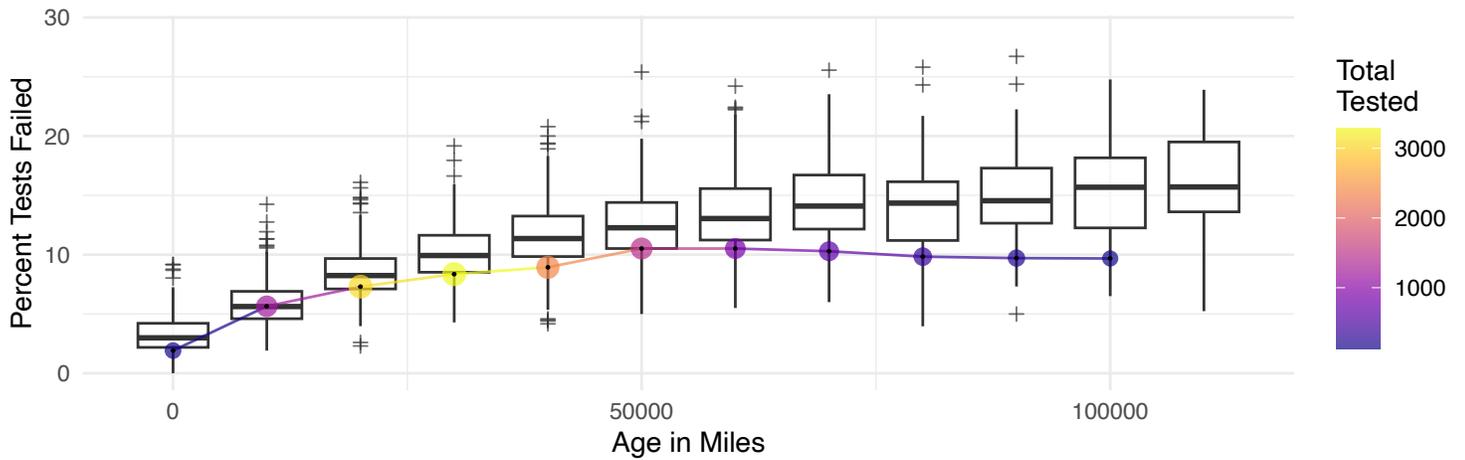

Mortality rates

| Age in Years | Observed | Died | Mortality Rate |
|---|---|---|---|
| 3 | 2391 | 2 | 0.000836 |
| 4 | 1707 | 2 | 0.001170 |

Mechanical Reliability Rates

| Mileage at test | N tested | Pct failed |
|---|---|---|
| 0 | 157 | 1.91 |
| 10000 | 1222 | 5.65 |
| 20000 | 3082 | 7.30 |
| 30000 | 3290 | 8.36 |
| 40000 | 2373 | 8.93 |
| 50000 | 1455 | 10.50 |
| 60000 | 875 | 10.50 |
| 70000 | 622 | 10.30 |
| 80000 | 356 | 9.83 |
| 90000 | 206 | 9.71 |
| 100000 | 124 | 9.68 |

**Audi A6 2018**

At 3 years of age, the mortality rate of a Audi A6 2018 (manufactured as a Car or Light Van) ranked number 5 out of 222 vehicles of the same age and type (any Car or Light Van constructed in 2018). One is the lowest (or best) and 222 the highest mortality rate. For vehicles reaching 20000 miles, its unreliability score (rate of failing an inspection) ranked 105 out of 215 vehicles of the same age, type, and mileage. One is the highest (or worst) and 215 the lowest rate of failing an inspection.

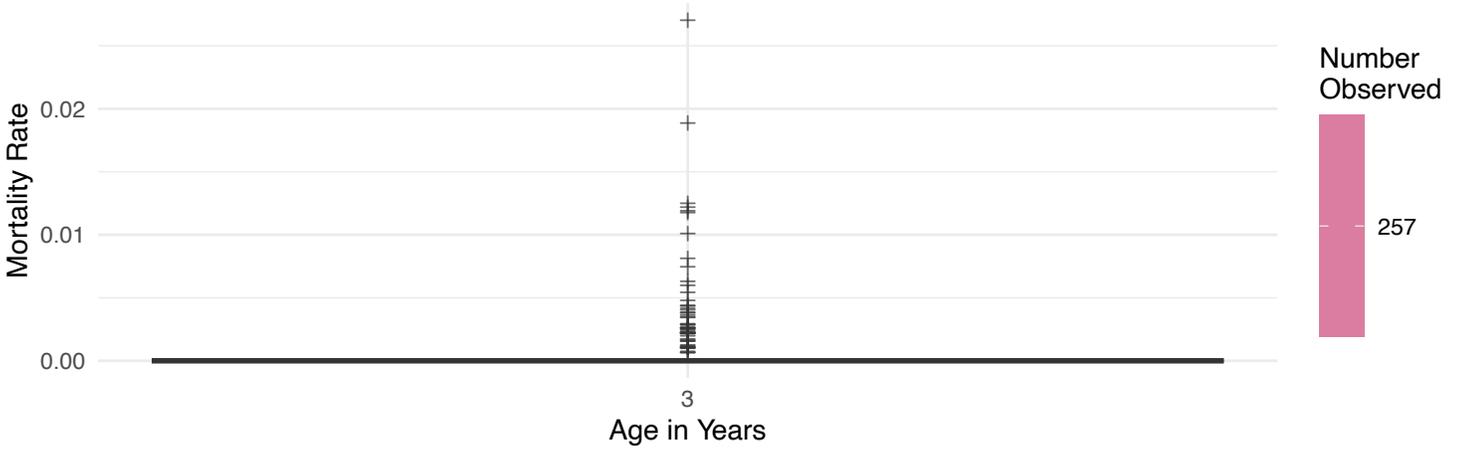

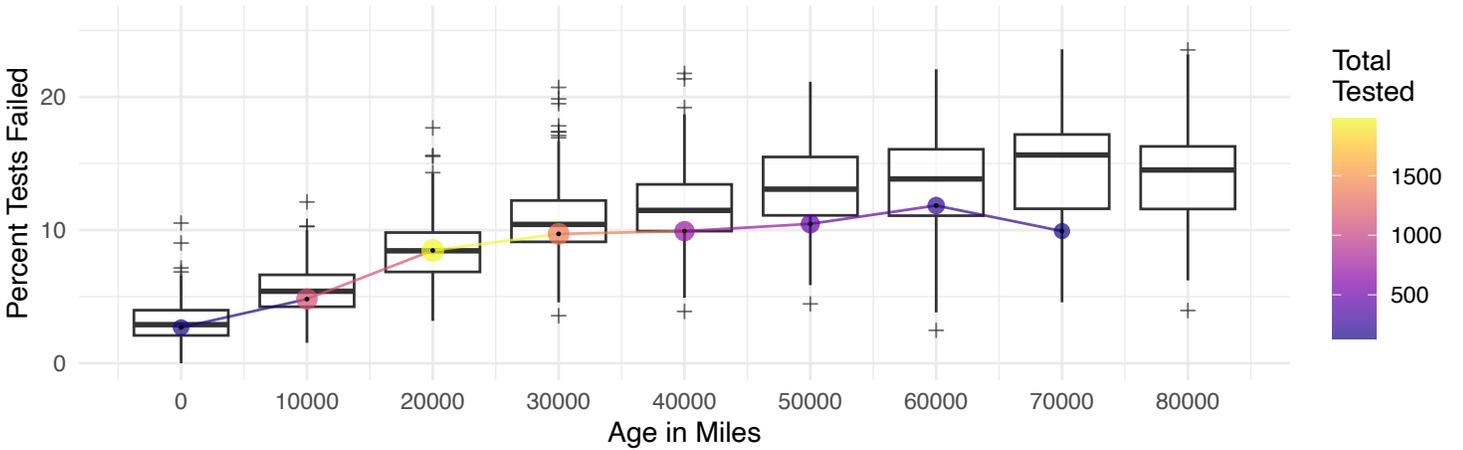

Mortality rates

| Age in Years | Observed | Died | Mortality Rate |
|---|---|---|---|
| 3 | 257 | 0 | 0 |

Mechanical Reliability Rates

| Mileage at test | N tested | Pct failed |
|---|---|---|
| 0 | 149 | 2.68 |
| 10000 | 1121 | 4.82 |
| 20000 | 1983 | 8.47 |
| 30000 | 1369 | 9.72 |
| 40000 | 736 | 9.92 |
| 50000 | 449 | 10.50 |
| 60000 | 228 | 11.80 |
| 70000 | 131 | 9.92 |

# Audi A7 2011

At 5 years of age, the mortality rate of a Audi A7 2011 (manufactured as a Car or Light Van) ranked number 171 out of 211 vehicles of the same age and type (any Car or Light Van constructed in 2011). One is the lowest (or best) and 211 the highest mortality rate. For vehicles reaching 100000 miles, its unreliability score (rate of failing an inspection) ranked 186 out of 195 vehicles of the same age, type, and mileage. One is the highest (or worst) and 195 the lowest rate of failing an inspection.

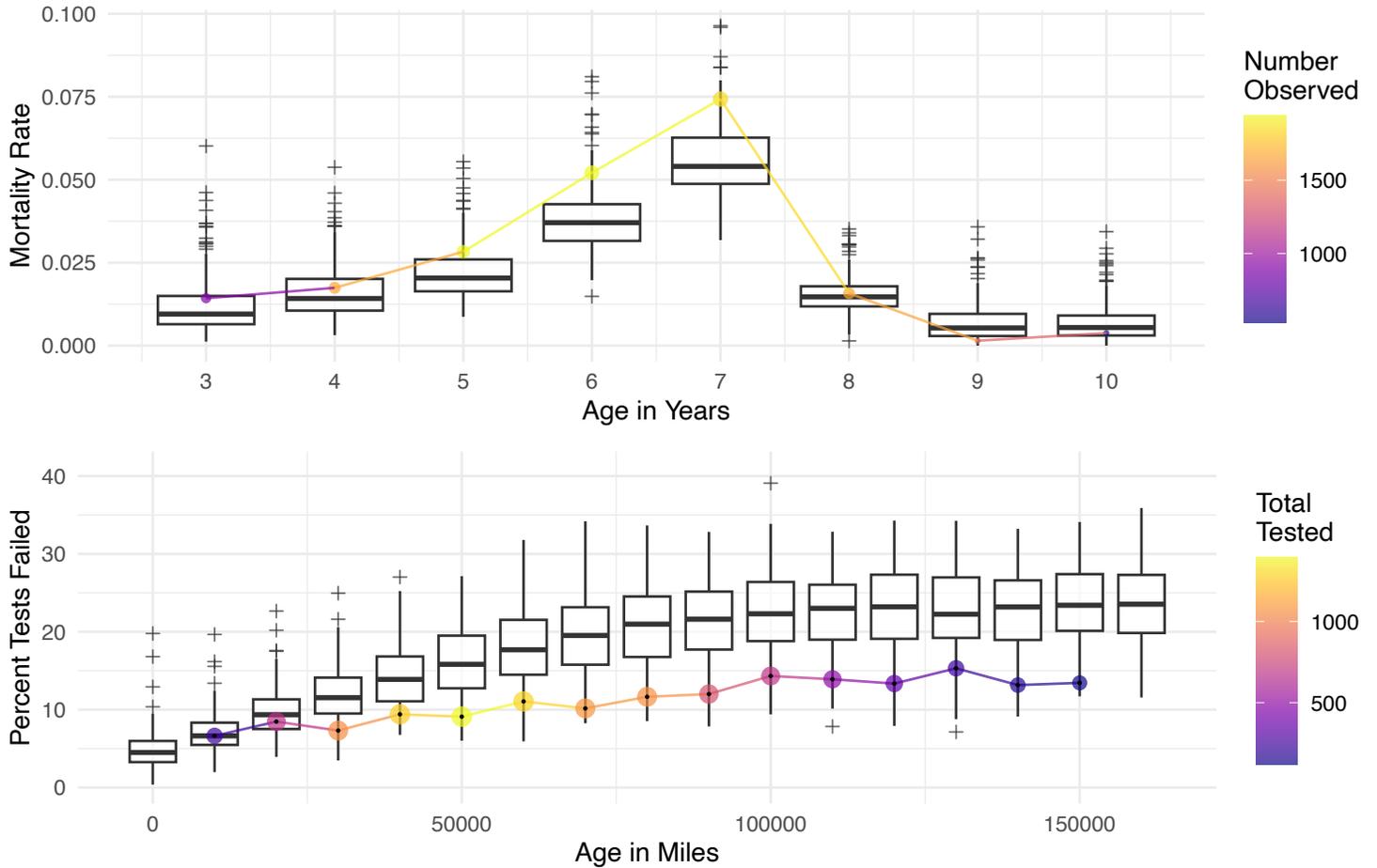

Mortality rates

| Age in Years | Observed | Died | Mortality Rate |
|---|---|---|---|
| 3 | 911 | 13 | 0.01430 |
| 4 | 1665 | 29 | 0.01740 |
| 5 | 1943 | 55 | 0.02830 |
| 6 | 1939 | 101 | 0.05210 |
| 7 | 1844 | 137 | 0.07430 |
| 8 | 1650 | 26 | 0.01580 |
| 9 | 1358 | 2 | 0.00147 |
| 10 | 528 | 2 | 0.00379 |

Mechanical Reliability Rates

| Mileage at test | N tested | Pct failed |
|---|---|---|
| 10000 | 226 | 6.64 |
| 20000 | 673 | 8.47 |
| 30000 | 1052 | 7.32 |
| 40000 | 1265 | 9.41 |
| 50000 | 1395 | 9.10 |
| 60000 | 1284 | 11.10 |
| 70000 | 1082 | 10.20 |
| 80000 | 1064 | 11.70 |
| 90000 | 816 | 12.00 |
| 100000 | 677 | 14.30 |
| 110000 | 489 | 13.90 |
| 120000 | 337 | 13.40 |
| 130000 | 222 | 15.30 |
| 140000 | 152 | 13.20 |
| 150000 | 119 | 13.40 |

**Audi A7 2012**

At 5 years of age, the mortality rate of a Audi A7 2012 (manufactured as a Car or Light Van) ranked number 145 out of 212 vehicles of the same age and type (any Car or Light Van constructed in 2012). One is the lowest (or best) and 212 the highest mortality rate. For vehicles reaching 20000 miles, its unreliability score (rate of failing an inspection) ranked 170 out of 206 vehicles of the same age, type, and mileage. One is the highest (or worst) and 206 the lowest rate of failing an inspection.

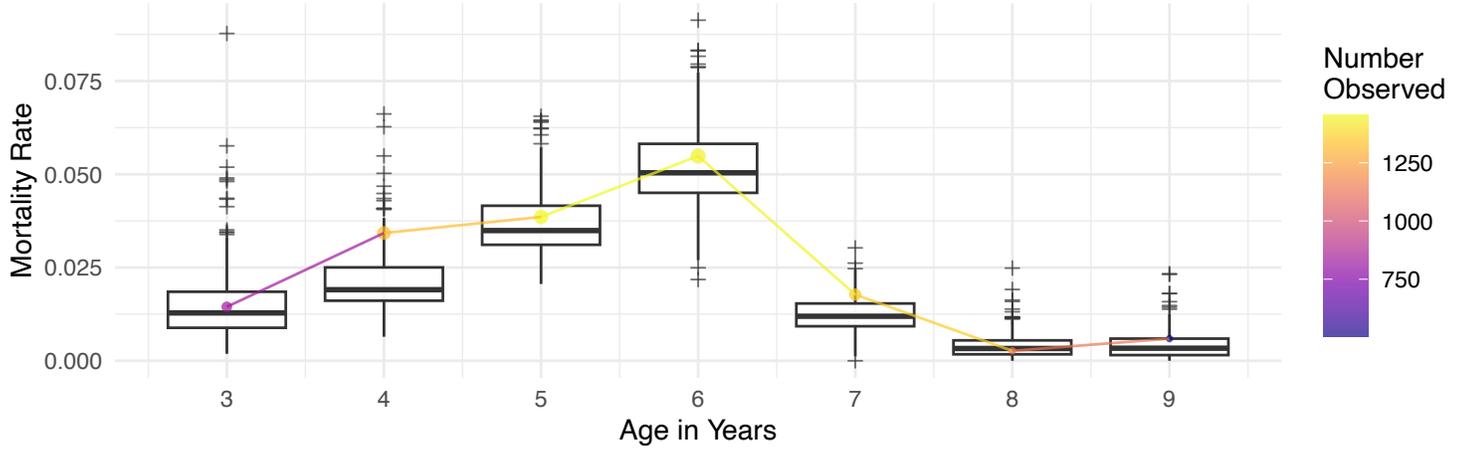

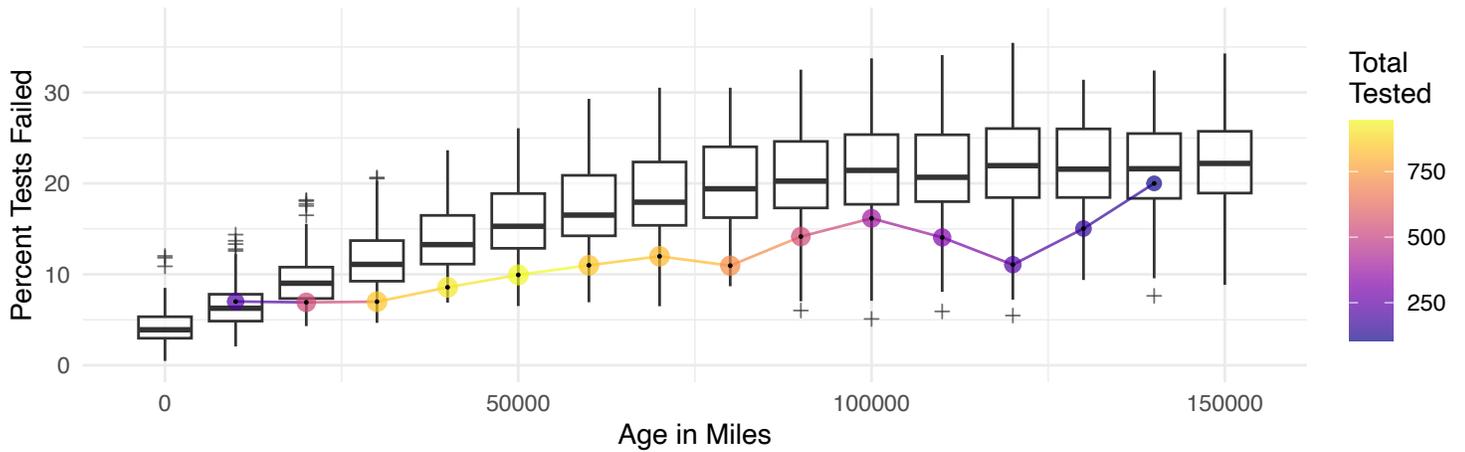

Mortality rates

| Age in Years | Observed | Died | Mortality Rate |
|---|---|---|---|
| 3 | 829 | 12 | 0.01450 |
| 4 | 1314 | 45 | 0.03420 |
| 5 | 1452 | 56 | 0.03860 |
| 6 | 1439 | 79 | 0.05490 |
| 7 | 1352 | 24 | 0.01780 |
| 8 | 1125 | 3 | 0.00267 |
| 9 | 505 | 3 | 0.00594 |

Mechanical Reliability Rates

| Mileage at test | N tested | Pct failed |
|---|---|---|
| 10000 | 214 | 7.01 |
| 20000 | 535 | 6.92 |
| 30000 | 844 | 6.99 |
| 40000 | 910 | 8.57 |
| 50000 | 946 | 9.94 |
| 60000 | 846 | 11.00 |
| 70000 | 818 | 12.00 |
| 80000 | 712 | 11.00 |
| 90000 | 537 | 14.20 |
| 100000 | 390 | 16.20 |
| 110000 | 313 | 14.10 |
| 120000 | 217 | 11.10 |
| 130000 | 153 | 15.00 |
| 140000 | 105 | 20.00 |

## Audi A7 2013

At 5 years of age, the mortality rate of a Audi A7 2013 (manufactured as a Car or Light Van) ranked number 131 out of 221 vehicles of the same age and type (any Car or Light Van constructed in 2013). One is the lowest (or best) and 221 the highest mortality rate. For vehicles reaching 20000 miles, its unreliability score (rate of failing an inspection) ranked 116 out of 215 vehicles of the same age, type, and mileage. One is the highest (or worst) and 215 the lowest rate of failing an inspection.

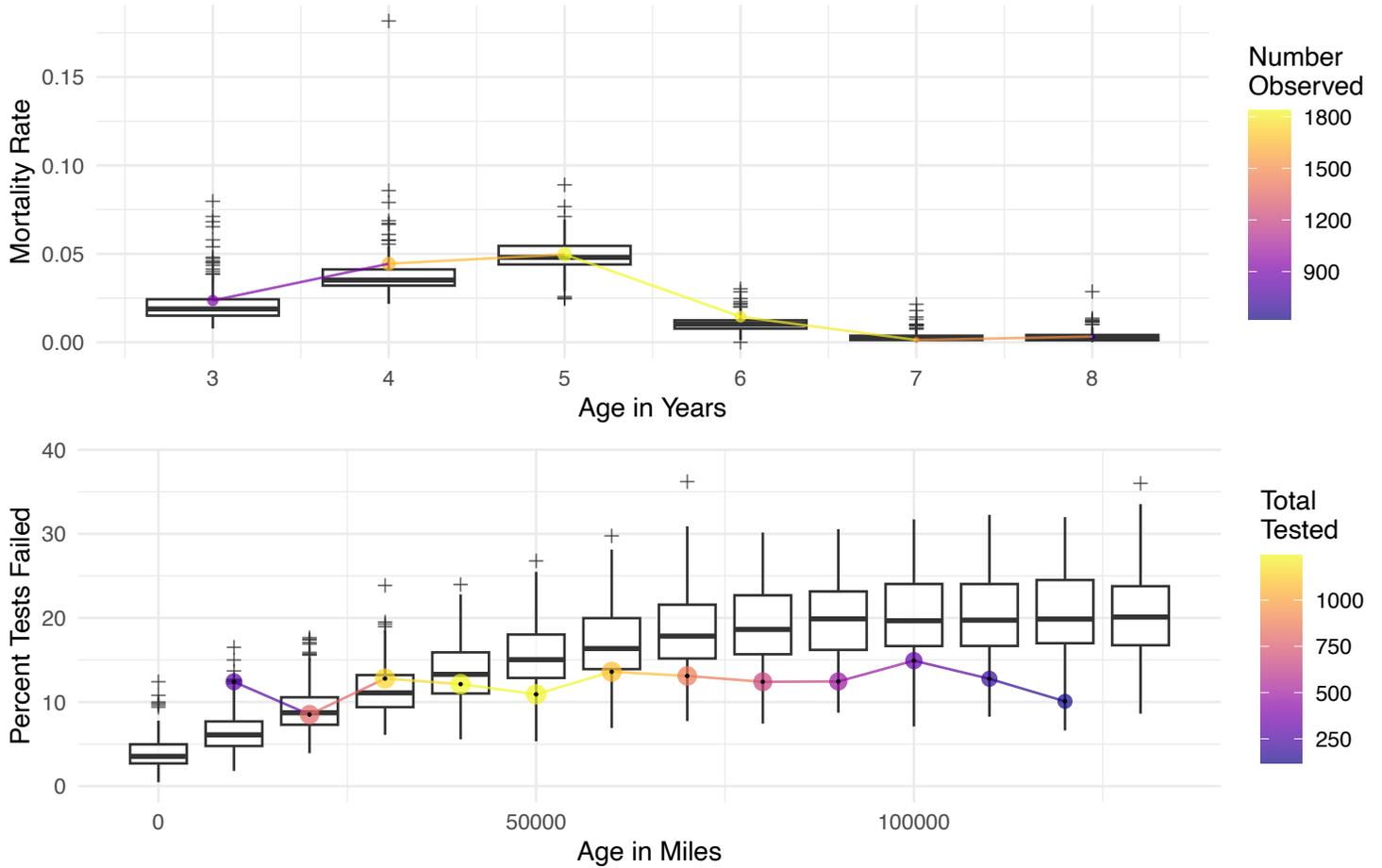

### Mortality rates

| Age in Years | Observed | Died | Mortality Rate |
|---|---|---|---|
| 3 | 931 | 22 | 0.02360 |
| 4 | 1644 | 73 | 0.04440 |
| 5 | 1833 | 91 | 0.04960 |
| 6 | 1797 | 26 | 0.01450 |
| 7 | 1520 | 2 | 0.00132 |
| 8 | 623 | 2 | 0.00321 |

### Mechanical Reliability Rates

| Mileage at test | N tested | Pct failed |
|---|---|---|
| 10000 | 282 | 12.40 |
| 20000 | 799 | 8.51 |
| 30000 | 1157 | 12.80 |
| 40000 | 1245 | 12.10 |
| 50000 | 1236 | 10.90 |
| 60000 | 1088 | 13.60 |
| 70000 | 870 | 13.10 |
| 80000 | 661 | 12.40 |
| 90000 | 482 | 12.40 |
| 100000 | 315 | 14.90 |
| 110000 | 188 | 12.80 |
| 120000 | 119 | 10.10 |

## Audi A7 2014

At 5 years of age, the mortality rate of a Audi A7 2014 (manufactured as a Car or Light Van) ranked number 20 out of 236 vehicles of the same age and type (any Car or Light Van constructed in 2014). One is the lowest (or best) and 236 the highest mortality rate. For vehicles reaching 20000 miles, its unreliability score (rate of failing an inspection) ranked 75 out of 230 vehicles of the same age, type, and mileage. One is the highest (or worst) and 230 the lowest rate of failing an inspection.

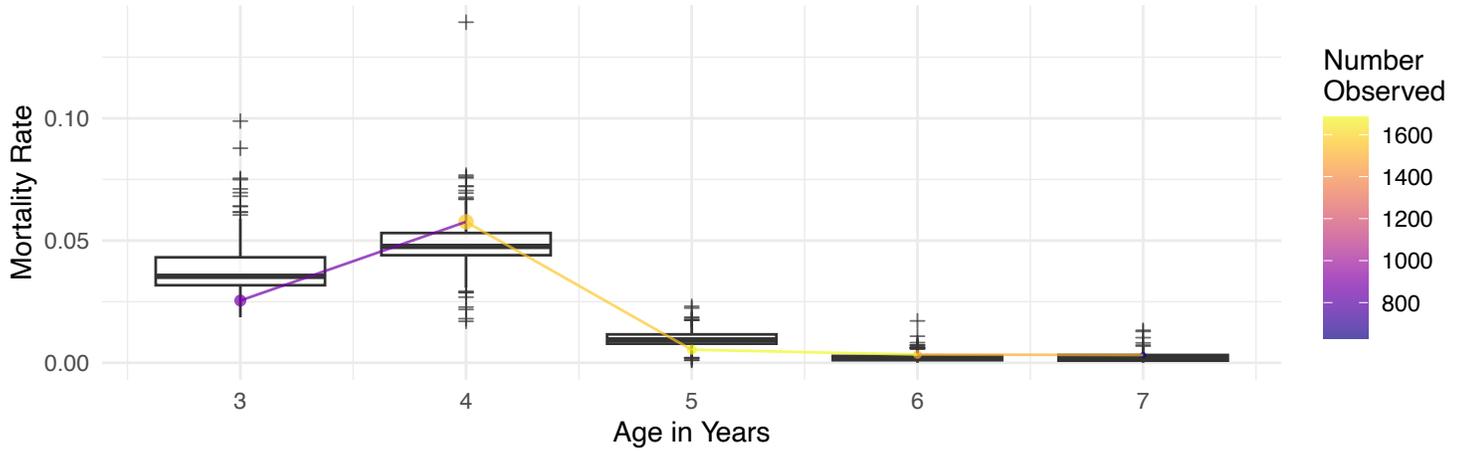

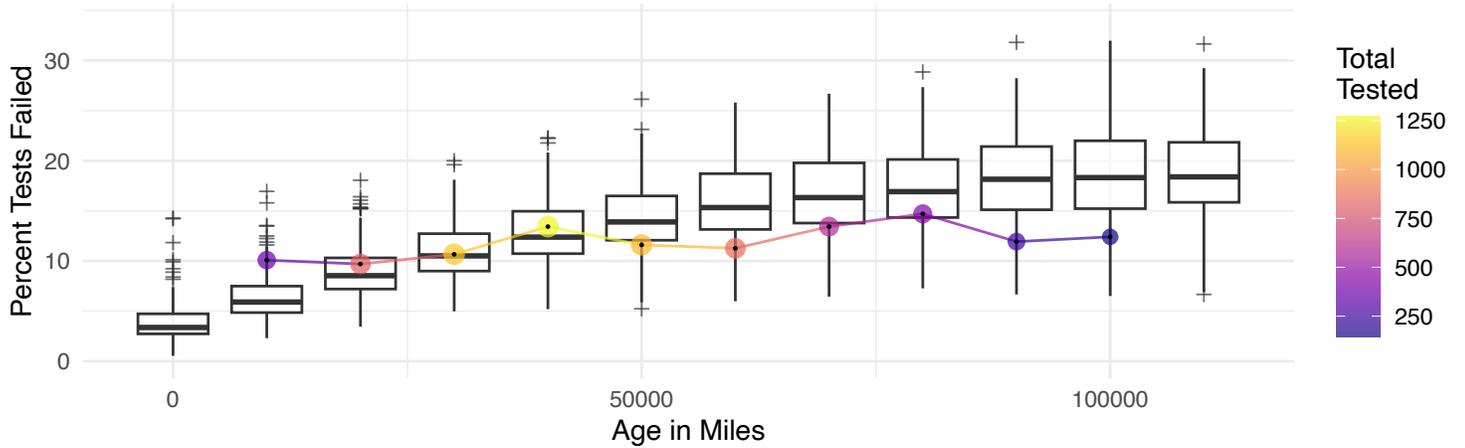

Mortality rates

| Age in Years | Observed | Died | Mortality Rate |
|---|---|---|---|
| 3 | 864 | 22 | 0.02550 |
| 4 | 1561 | 90 | 0.05770 |
| 5 | 1681 | 9 | 0.00535 |
| 6 | 1481 | 5 | 0.00338 |
| 7 | 630 | 2 | 0.00317 |

Mechanical Reliability Rates

| Mileage at test | N tested | Pct failed |
|---|---|---|
| 10000 | 337 | 10.1 |
| 20000 | 804 | 9.7 |
| 30000 | 1143 | 10.7 |
| 40000 | 1273 | 13.4 |
| 50000 | 1094 | 11.6 |
| 60000 | 843 | 11.3 |
| 70000 | 588 | 13.4 |
| 80000 | 442 | 14.7 |
| 90000 | 243 | 11.9 |
| 100000 | 145 | 12.4 |

# Audi A7 2015

At 5 years of age, the mortality rate of a Audi A7 2015 (manufactured as a Car or Light Van) ranked number 143 out of 247 vehicles of the same age and type (any Car or Light Van constructed in 2015). One is the lowest (or best) and 247 the highest mortality rate. For vehicles reaching 20000 miles, its unreliability score (rate of failing an inspection) ranked 24 out of 241 vehicles of the same age, type, and mileage. One is the highest (or worst) and 241 the lowest rate of failing an inspection.

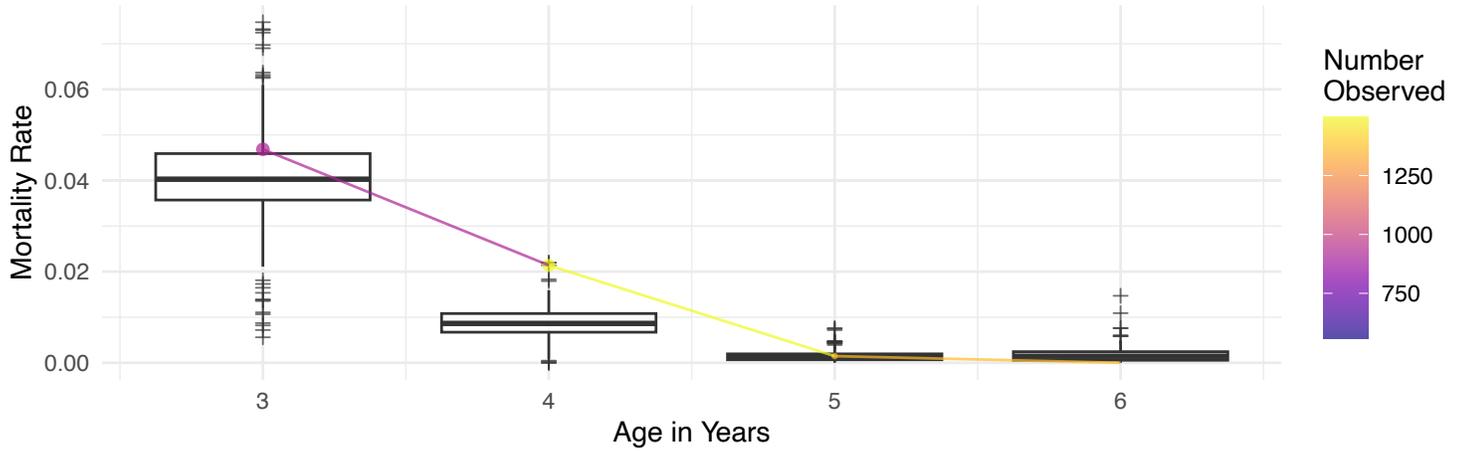

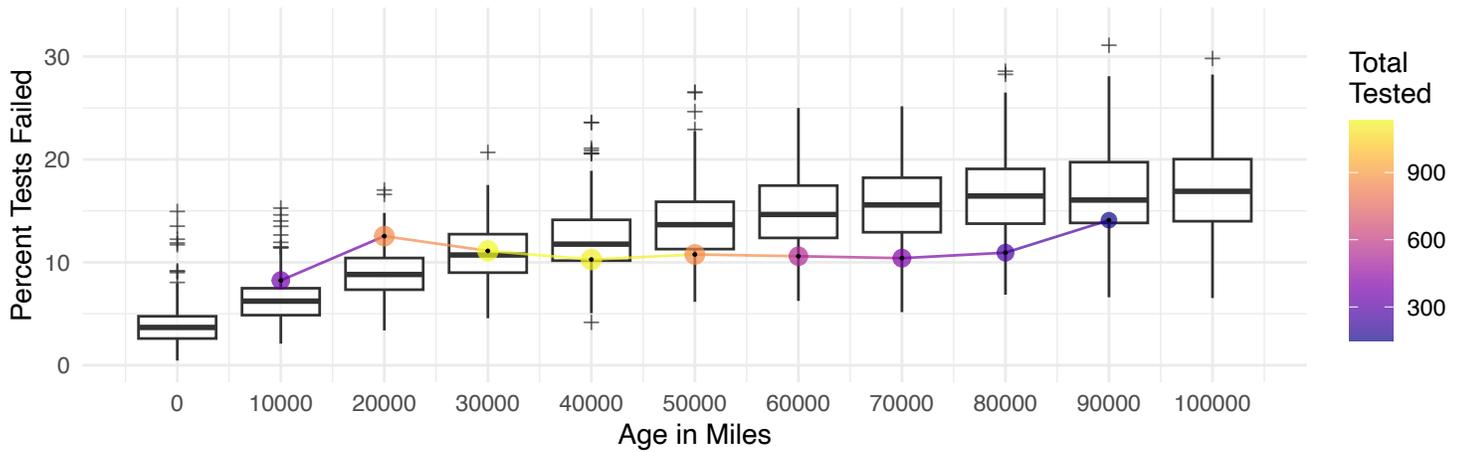

Mortality rates

| Age in Years | Observed | Died | Mortality Rate |
|---|---|---|---|
| 3 | 918 | 43 | 0.04680 |
| 4 | 1496 | 32 | 0.02140 |
| 5 | 1362 | 2 | 0.00147 |
| 6 | 559 | 0 | 0.00000 |

Mechanical Reliability Rates

| Mileage at test | N tested | Pct failed |
|---|---|---|
| 10000 | 352 | 8.24 |
| 20000 | 853 | 12.50 |
| 30000 | 1134 | 11.10 |
| 40000 | 1110 | 10.30 |
| 50000 | 864 | 10.80 |
| 60000 | 538 | 10.60 |
| 70000 | 394 | 10.40 |
| 80000 | 256 | 10.90 |
| 90000 | 149 | 14.10 |

## Audi A7 2016

At 5 years of age, the mortality rate of a Audi A7 2016 (manufactured as a Car or Light Van) ranked number 185 out of 252 vehicles of the same age and type (any Car or Light Van constructed in 2016). One is the lowest (or best) and 252 the highest mortality rate. For vehicles reaching 20000 miles, its unreliability score (rate of failing an inspection) ranked 100 out of 246 vehicles of the same age, type, and mileage. One is the highest (or worst) and 246 the lowest rate of failing an inspection.

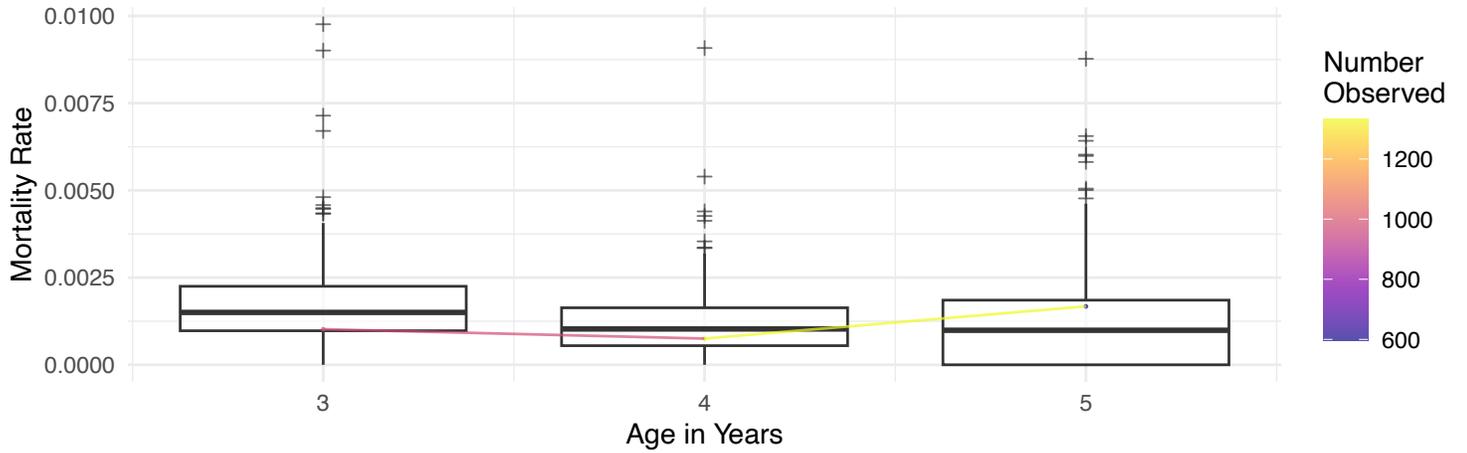

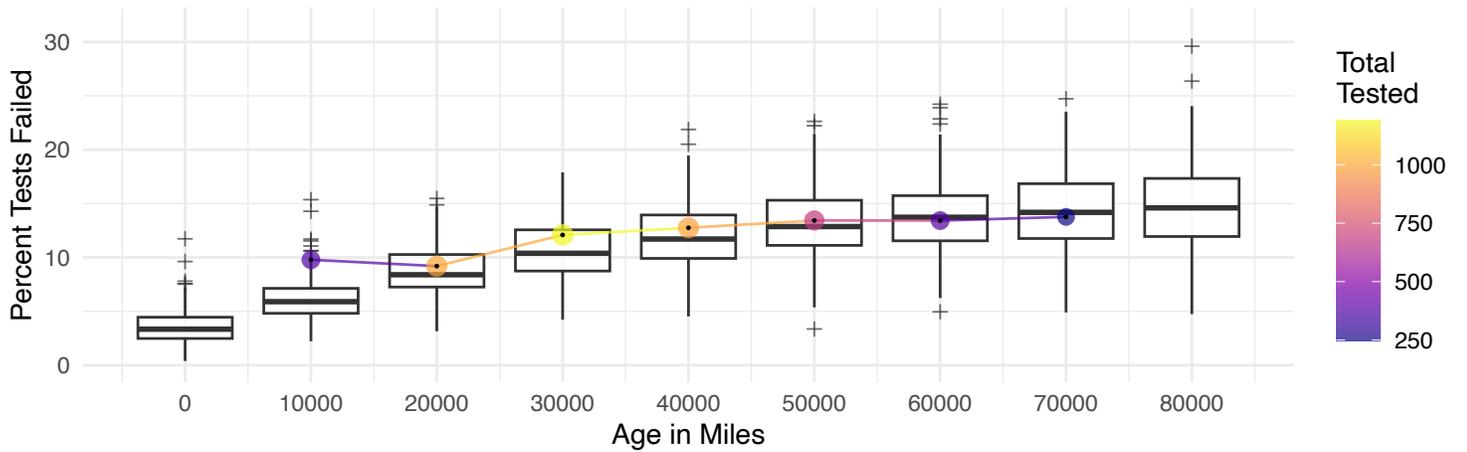

Mortality rates

| Age in Years | Observed | Died | Mortality Rate |
|---|---|---|---|
| 3 | 981 | 1 | 0.001020 |
| 4 | 1331 | 1 | 0.000751 |
| 5 | 597 | 1 | 0.001680 |

Mechanical Reliability Rates

| Mileage at test | N tested | Pct failed |
|---|---|---|
| 10000 | 378 | 9.79 |
| 20000 | 1000 | 9.20 |
| 30000 | 1191 | 12.10 |
| 40000 | 988 | 12.80 |
| 50000 | 685 | 13.40 |
| 60000 | 380 | 13.40 |
| 70000 | 247 | 13.80 |

**Audi A7 2017**

At 3 years of age, the mortality rate of a Audi A7 2017 (manufactured as a Car or Light Van) ranked number 3 out of 247 vehicles of the same age and type (any Car or Light Van constructed in 2017). One is the lowest (or best) and 247 the highest mortality rate. For vehicles reaching 20000 miles, its unreliability score (rate of failing an inspection) ranked 66 out of 240 vehicles of the same age, type, and mileage. One is the highest (or worst) and 240 the lowest rate of failing an inspection.

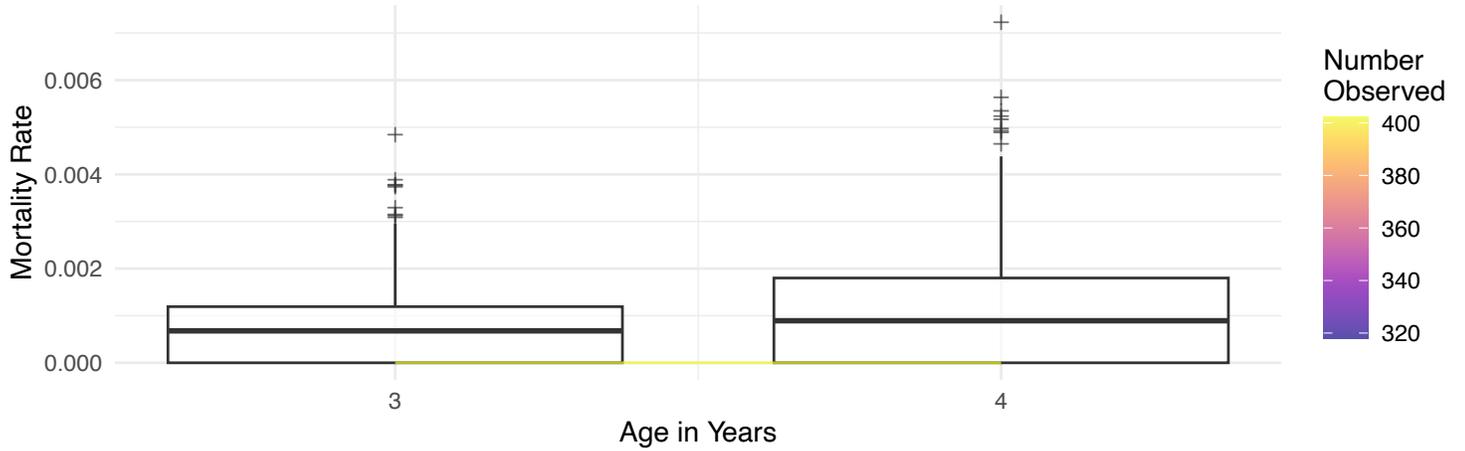

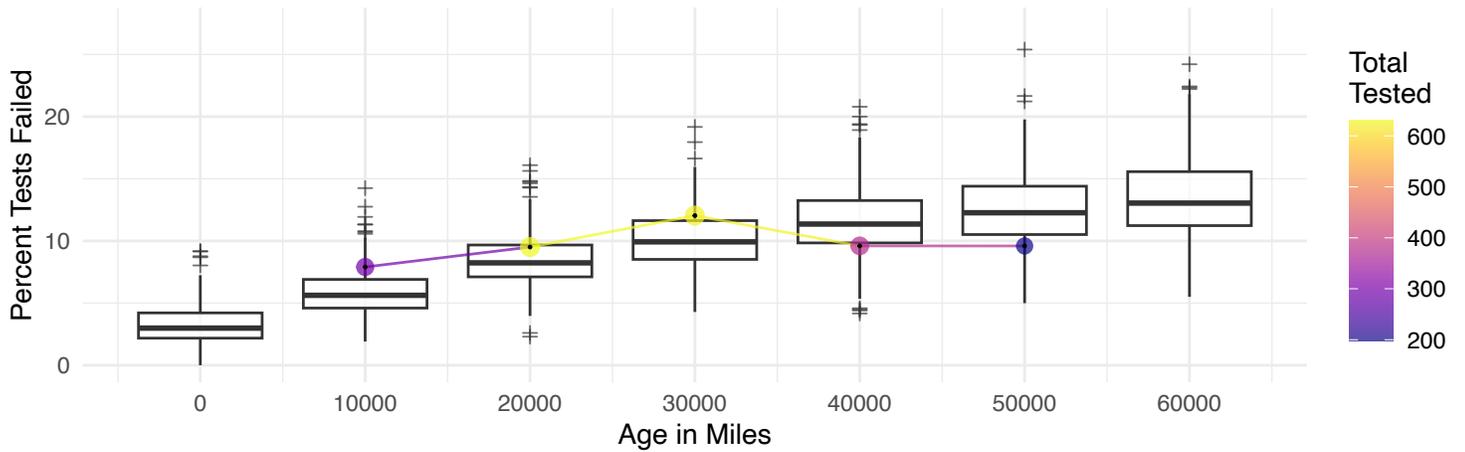

<table>
<tr><th colspan="4">Mortality rates</th></tr>
<tr><th>Age in Years</th><th>Observed</th><th>Died</th><th>Mortality Rate</th></tr>
<tr><td>3</td><td>402</td><td>0</td><td>0</td></tr>
<tr><td>4</td><td>318</td><td>0</td><td>0</td></tr>
</table>

| Mechanical Reliability Rates | | |
|---|---|---|
| Mileage at test | N tested | Pct failed |
| 10000 | 291 | 7.90 |
| 20000 | 631 | 9.51 |
| 30000 | 623 | 12.00 |
| 40000 | 375 | 9.60 |
| 50000 | 198 | 9.60 |

**Audi A8 2004**

At 5 years of age, the mortality rate of a Audi A8 2004 (manufactured as a Car or Light Van) ranked number 188 out of 229 vehicles of the same age and type (any Car or Light Van constructed in 2004). One is the lowest (or best) and 229 the highest mortality rate. For vehicles reaching 20000 miles, its unreliability score (rate of failing an inspection) ranked 211 out of 225 vehicles of the same age, type, and mileage. One is the highest (or worst) and 225 the lowest rate of failing an inspection.

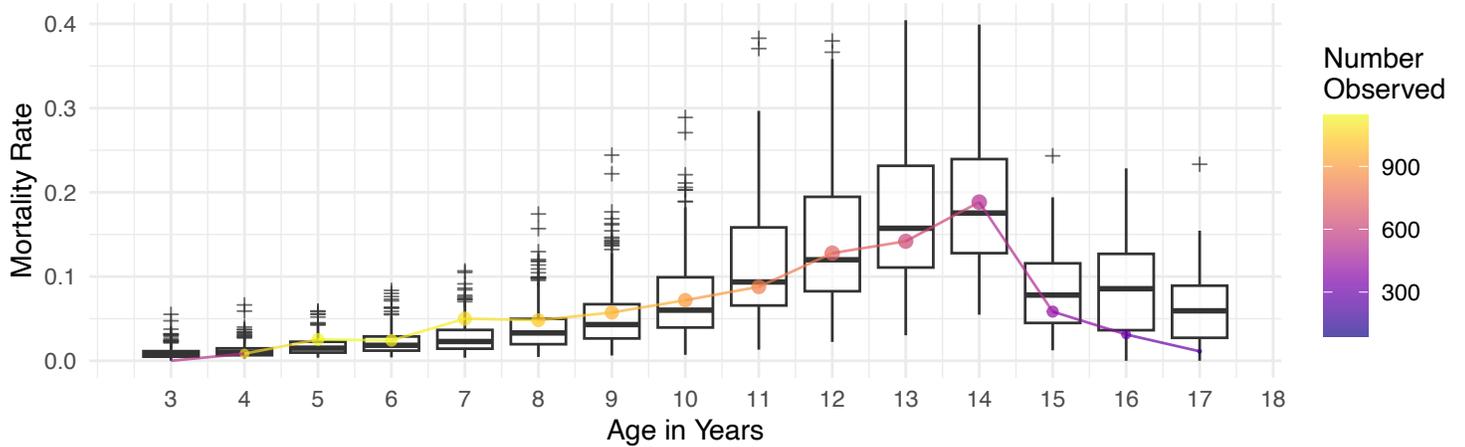

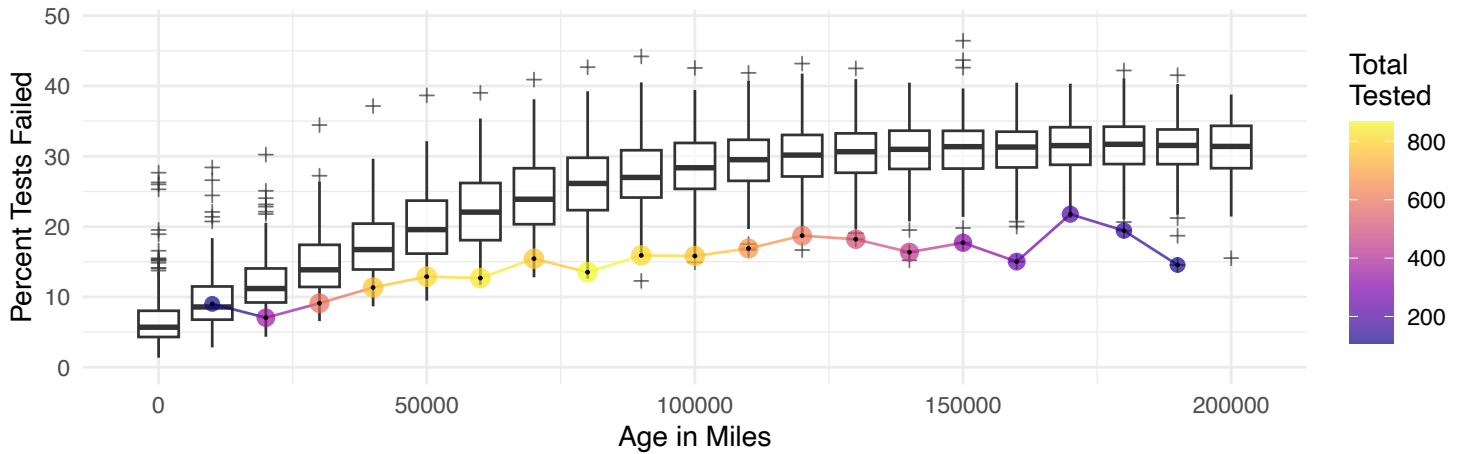

<table>
<tr><td colspan="4" align="center">Mortality rates</td></tr>
<tr><th>Age in Years</th><th>Observed</th><th>Died</th><th>Mortality Rate</th></tr>
<tr><td>3</td><td>531</td><td>0</td><td>0.00000</td></tr>
<tr><td>4</td><td>1063</td><td>9</td><td>0.00847</td></tr>
<tr><td>5</td><td>1142</td><td>29</td><td>0.02540</td></tr>
<tr><td>6</td><td>1114</td><td>27</td><td>0.02420</td></tr>
<tr><td>7</td><td>1080</td><td>54</td><td>0.05000</td></tr>
<tr><td>8</td><td>1017</td><td>49</td><td>0.04820</td></tr>
<tr><td>9</td><td>959</td><td>55</td><td>0.05740</td></tr>
<tr><td>10</td><td>890</td><td>64</td><td>0.07190</td></tr>
<tr><td>11</td><td>808</td><td>71</td><td>0.08790</td></tr>
<tr><td>12</td><td>712</td><td>91</td><td>0.12800</td></tr>
<tr><td>13</td><td>613</td><td>87</td><td>0.14200</td></tr>
<tr><td>14</td><td>515</td><td>97</td><td>0.18800</td></tr>
<tr><td>15</td><td>394</td><td>23</td><td>0.05840</td></tr>
<tr><td>16</td><td>288</td><td>9</td><td>0.03120</td></tr>
<tr><td>17</td><td>89</td><td>1</td><td>0.01120</td></tr>
</table>

<table>
<tr><td colspan="3" align="center">Mechanical Reliability Rates</td></tr>
<tr><th>Mileage at test</th><th>N tested</th><th>Pct failed</th></tr>
<tr><td>10000</td><td>111</td><td>9.01</td></tr>
<tr><td>20000</td><td>355</td><td>7.04</td></tr>
<tr><td>30000</td><td>593</td><td>9.11</td></tr>
<tr><td>40000</td><td>767</td><td>11.30</td></tr>
<tr><td>50000</td><td>807</td><td>12.90</td></tr>
<tr><td>60000</td><td>836</td><td>12.70</td></tr>
<tr><td>70000</td><td>771</td><td>15.40</td></tr>
<tr><td>80000</td><td>866</td><td>13.50</td></tr>
<tr><td>90000</td><td>805</td><td>15.90</td></tr>
<tr><td>100000</td><td>765</td><td>15.80</td></tr>
<tr><td>110000</td><td>669</td><td>16.90</td></tr>
<tr><td>120000</td><td>604</td><td>18.70</td></tr>
<tr><td>130000</td><td>500</td><td>18.20</td></tr>
<tr><td>140000</td><td>440</td><td>16.40</td></tr>
<tr><td>150000</td><td>322</td><td>17.70</td></tr>
<tr><td>160000</td><td>253</td><td>15.00</td></tr>
<tr><td>190000</td><td>110</td><td>14.50</td></tr>
</table>

## Audi A8 2005

At 5 years of age, the mortality rate of a Audi A8 2005 (manufactured as a Car or Light Van) ranked number 157 out of 240 vehicles of the same age and type (any Car or Light Van constructed in 2005). One is the lowest (or best) and 240 the highest mortality rate. For vehicles reaching 20000 miles, its unreliability score (rate of failing an inspection) ranked 220 out of 235 vehicles of the same age, type, and mileage. One is the highest (or worst) and 235 the lowest rate of failing an inspection.

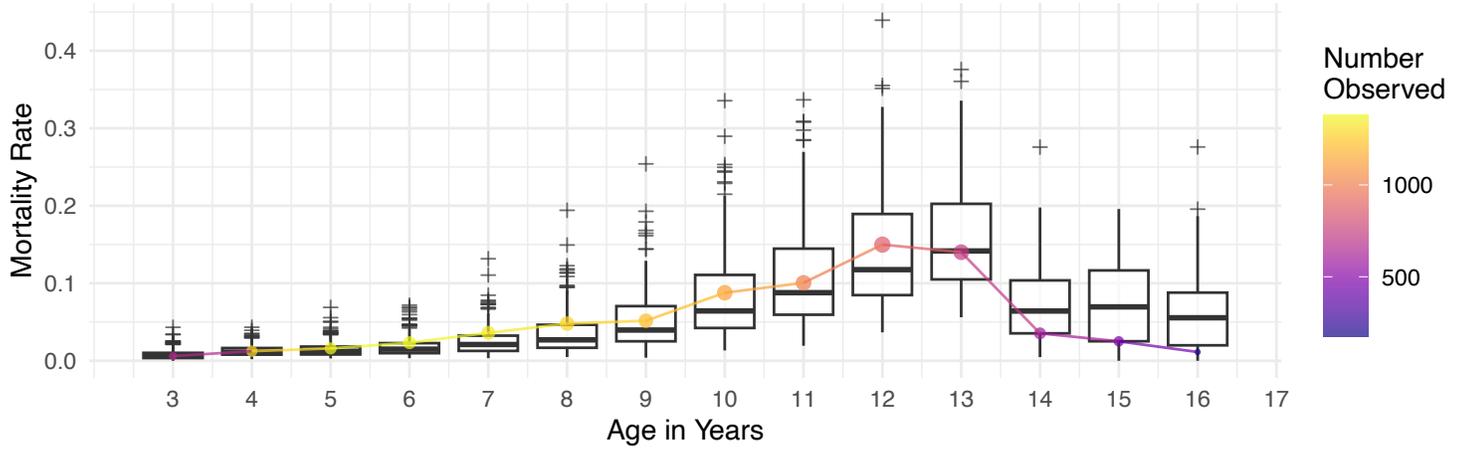

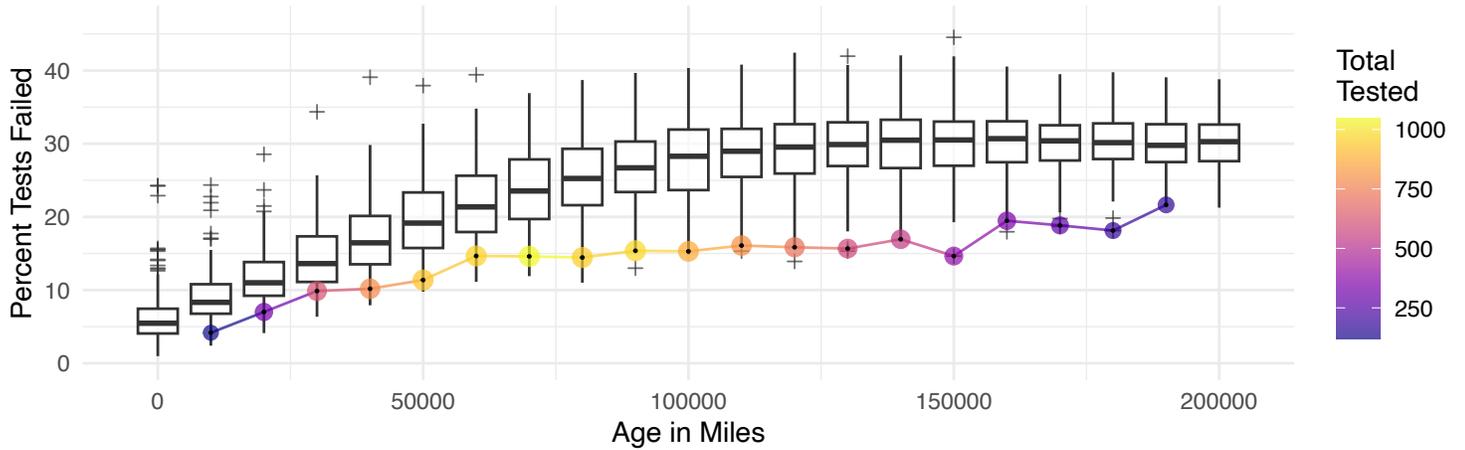

| Mortality rates | | | |
|---|---|---|---|
| Age in Years | Observed | Died | Mortality Rate |
| 3 | 661 | 4 | 0.00605 |
| 4 | 1285 | 16 | 0.01250 |
| 5 | 1377 | 22 | 0.01600 |
| 6 | 1371 | 32 | 0.02330 |
| 7 | 1330 | 48 | 0.03610 |
| 8 | 1272 | 61 | 0.04800 |
| 9 | 1200 | 62 | 0.05170 |
| 10 | 1116 | 98 | 0.08780 |
| 11 | 994 | 100 | 0.10100 |
| 12 | 874 | 131 | 0.15000 |
| 13 | 728 | 102 | 0.14000 |
| 14 | 593 | 21 | 0.03540 |
| 15 | 438 | 11 | 0.02510 |
| 16 | 176 | 2 | 0.01140 |

| Mechanical Reliability Rates | | |
|---|---|---|
| Mileage at test | N tested | Pct failed |
| 10000 | 120 | 4.17 |
| 20000 | 314 | 7.01 |
| 30000 | 618 | 9.87 |
| 40000 | 786 | 10.20 |
| 50000 | 932 | 11.40 |
| 60000 | 962 | 14.70 |
| 70000 | 1048 | 14.60 |
| 80000 | 955 | 14.50 |
| 90000 | 976 | 15.40 |
| 100000 | 870 | 15.30 |
| 110000 | 802 | 16.10 |
| 120000 | 713 | 15.80 |
| 130000 | 587 | 15.70 |
| 140000 | 543 | 16.90 |
| 150000 | 396 | 14.60 |
| 160000 | 339 | 19.50 |
| 170000 | 276 | 18.80 |

## Audi A8 2006

At 5 years of age, the mortality rate of a Audi A8 2006 (manufactured as a Car or Light Van) ranked number 194 out of 225 vehicles of the same age and type (any Car or Light Van constructed in 2006). One is the lowest (or best) and 225 the highest mortality rate. For vehicles reaching 40000 miles, its unreliability score (rate of failing an inspection) ranked 188 out of 220 vehicles of the same age, type, and mileage. One is the highest (or worst) and 220 the lowest rate of failing an inspection.

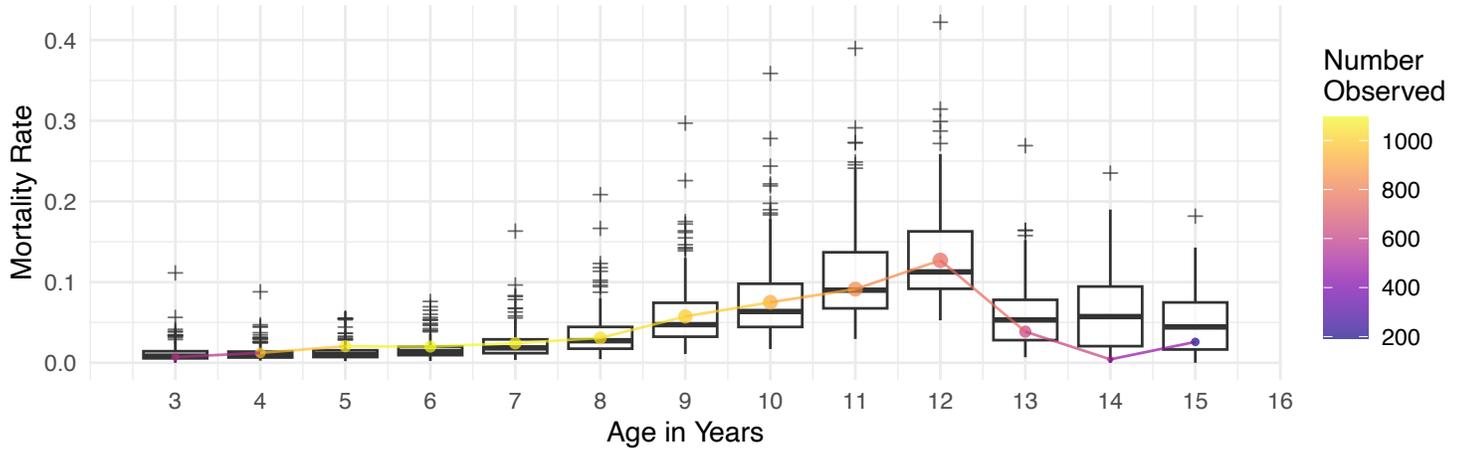

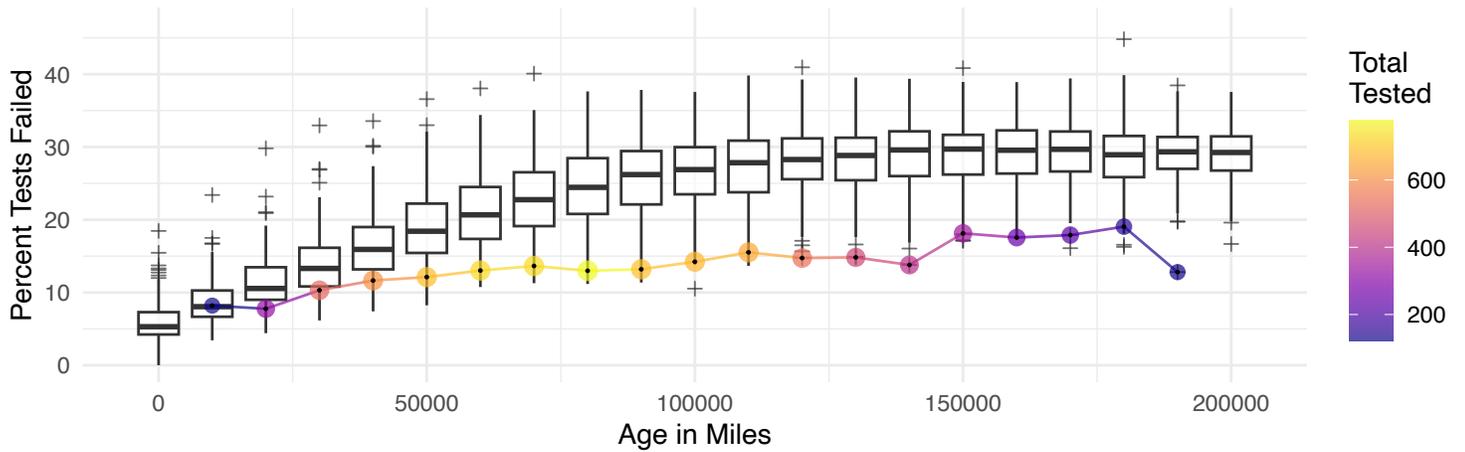

Mortality rates

| Age in Years | Observed | Died | Mortality Rate |
|---|---|---|---|
| 3 | 567 | 4 | 0.00705 |
| 4 | 972 | 12 | 0.01230 |
| 5 | 1073 | 22 | 0.02050 |
| 6 | 1093 | 22 | 0.02010 |
| 7 | 1072 | 26 | 0.02430 |
| 8 | 1037 | 32 | 0.03090 |
| 9 | 992 | 57 | 0.05750 |
| 10 | 920 | 69 | 0.07500 |
| 11 | 832 | 76 | 0.09130 |
| 12 | 748 | 95 | 0.12700 |
| 13 | 621 | 24 | 0.03860 |
| 14 | 467 | 2 | 0.00428 |
| 15 | 193 | 5 | 0.02590 |

Mechanical Reliability Rates

| Mileage at test | N tested | Pct failed |
|---|---|---|
| 10000 | 122 | 8.20 |
| 20000 | 322 | 7.76 |
| 30000 | 524 | 10.30 |
| 40000 | 610 | 11.60 |
| 50000 | 685 | 12.10 |
| 60000 | 730 | 13.00 |
| 70000 | 733 | 13.60 |
| 80000 | 778 | 13.00 |
| 90000 | 690 | 13.20 |
| 100000 | 675 | 14.20 |
| 110000 | 651 | 15.50 |
| 120000 | 543 | 14.70 |
| 130000 | 452 | 14.80 |
| 140000 | 399 | 13.80 |
| 150000 | 342 | 18.10 |
| 170000 | 218 | 17.90 |
| 180000 | 147 | 19.00 |

# Audi A8 2007

At 5 years of age, the mortality rate of a Audi A8 2007 (manufactured as a Car or Light Van) ranked number 170 out of 219 vehicles of the same age and type (any Car or Light Van constructed in 2007). One is the lowest (or best) and 219 the highest mortality rate. For vehicles reaching 20000 miles, its unreliability score (rate of failing an inspection) ranked 189 out of 214 vehicles of the same age, type, and mileage. One is the highest (or worst) and 214 the lowest rate of failing an inspection.

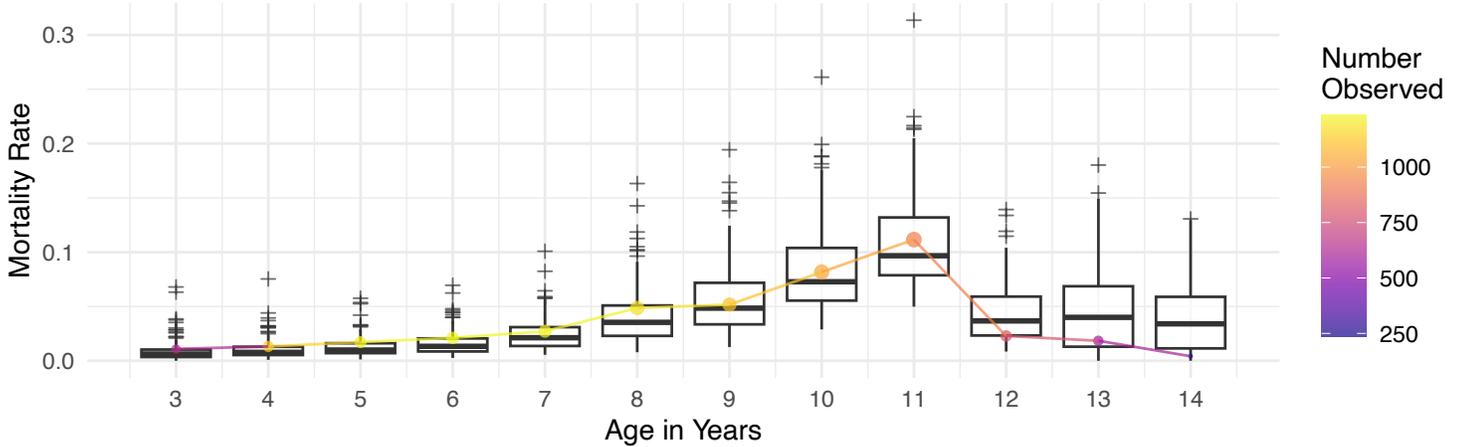

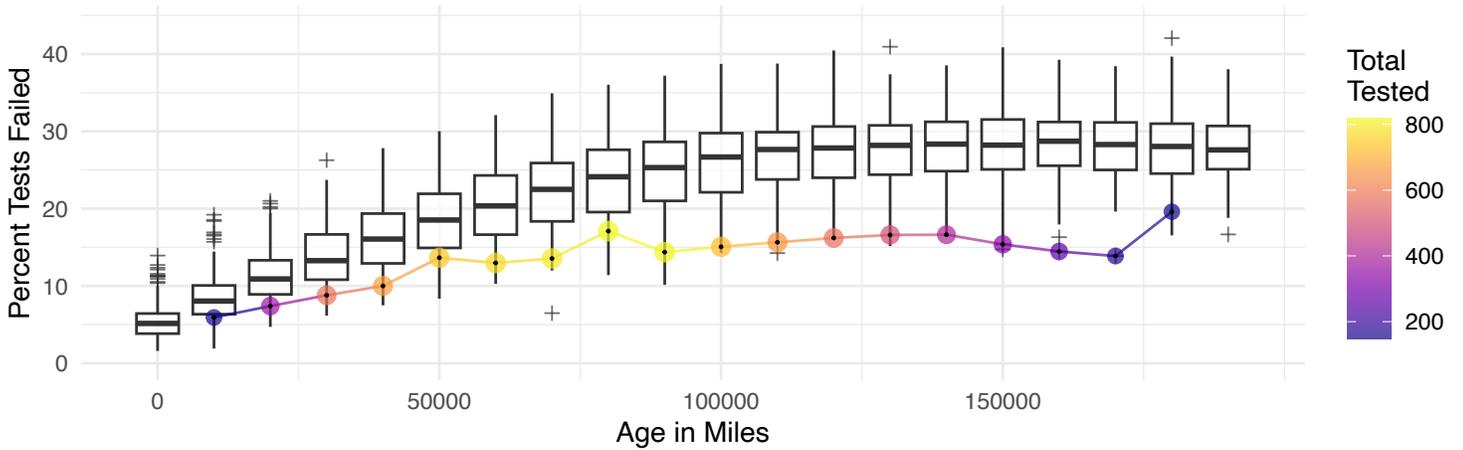

### Mortality rates

| Age in Years | Observed | Died | Mortality Rate |
|---|---|---|---|
| 3 | 639 | 7 | 0.01100 |
| 4 | 1137 | 15 | 0.01320 |
| 5 | 1222 | 21 | 0.01720 |
| 6 | 1232 | 26 | 0.02110 |
| 7 | 1213 | 33 | 0.02720 |
| 8 | 1173 | 57 | 0.04860 |
| 9 | 1102 | 57 | 0.05170 |
| 10 | 1028 | 84 | 0.08170 |
| 11 | 923 | 103 | 0.11200 |
| 12 | 788 | 18 | 0.02280 |
| 13 | 597 | 11 | 0.01840 |
| 14 | 237 | 1 | 0.00422 |

### Mechanical Reliability Rates

| Mileage at test | N tested | Pct failed |
|---|---|---|
| 10000 | 152 | 5.92 |
| 20000 | 365 | 7.40 |
| 30000 | 580 | 8.79 |
| 40000 | 670 | 10.00 |
| 50000 | 740 | 13.60 |
| 60000 | 771 | 13.00 |
| 70000 | 797 | 13.60 |
| 80000 | 819 | 17.10 |
| 90000 | 814 | 14.40 |
| 100000 | 717 | 15.10 |
| 110000 | 652 | 15.60 |
| 120000 | 574 | 16.20 |
| 130000 | 500 | 16.60 |
| 140000 | 409 | 16.60 |
| 150000 | 338 | 15.40 |
| 170000 | 173 | 13.90 |
| 180000 | 148 | 19.60 |

**Audi A8 2008**

At 5 years of age, the mortality rate of a Audi A8 2008 (manufactured as a Car or Light Van) ranked number 179 out of 218 vehicles of the same age and type (any Car or Light Van constructed in 2008). One is the lowest (or best) and 218 the highest mortality rate. For vehicles reaching 20000 miles, its unreliability score (rate of failing an inspection) ranked 147 out of 212 vehicles of the same age, type, and mileage. One is the highest (or worst) and 212 the lowest rate of failing an inspection.

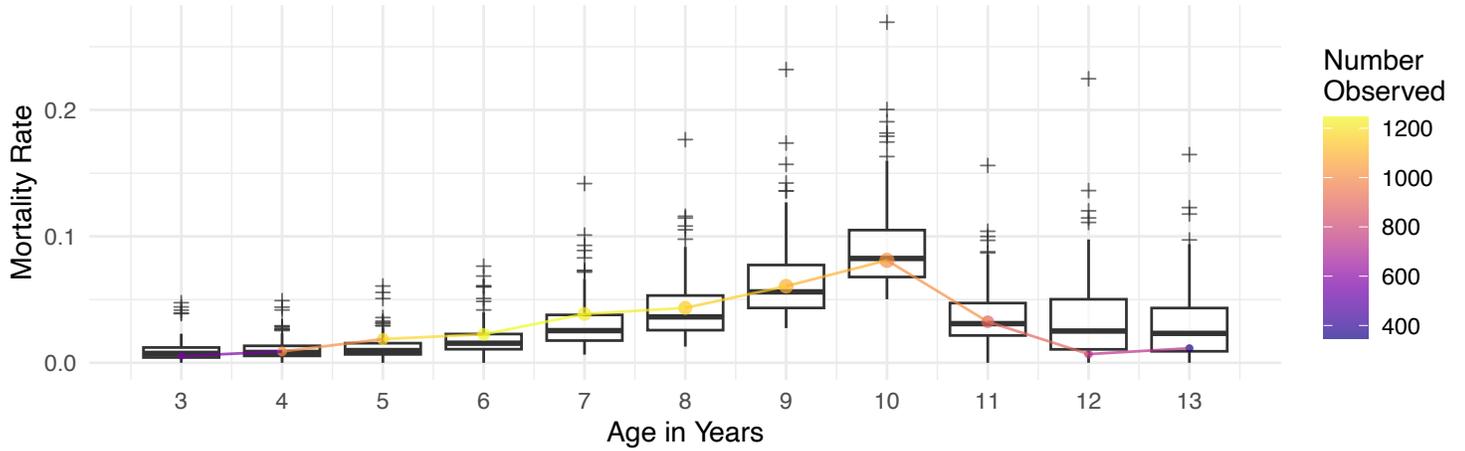

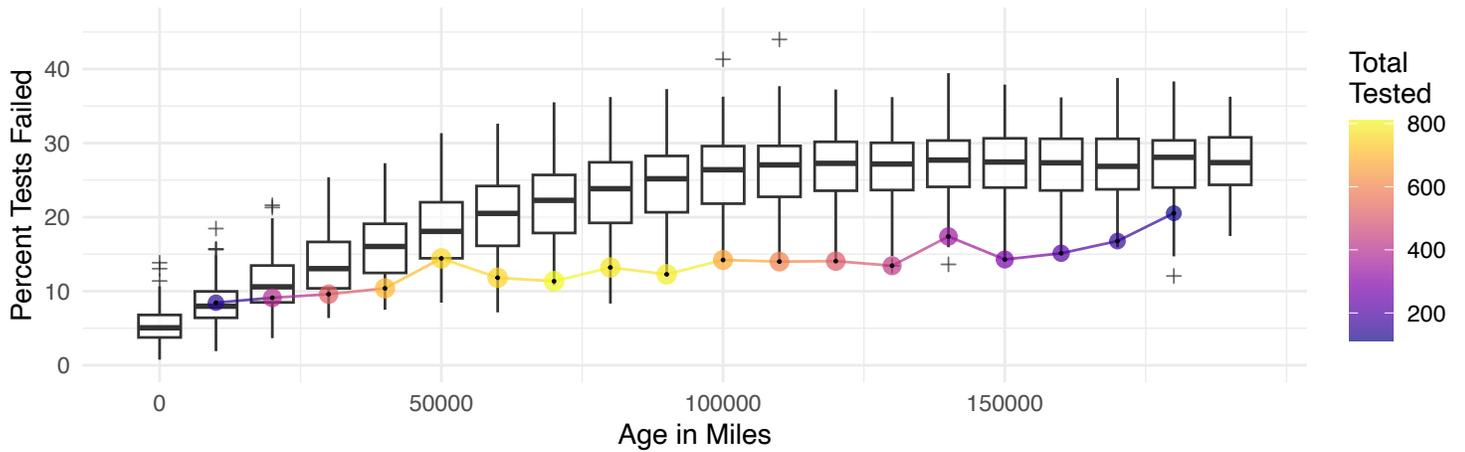

Mortality rates

| Age in Years | Observed | Died | Mortality Rate |
|---|---|---|---|
| 3 | 562 | 3 | 0.00534 |
| 4 | 1003 | 9 | 0.00897 |
| 5 | 1173 | 22 | 0.01880 |
| 6 | 1243 | 28 | 0.02250 |
| 7 | 1217 | 47 | 0.03860 |
| 8 | 1151 | 50 | 0.04340 |
| 9 | 1089 | 66 | 0.06060 |
| 10 | 1012 | 82 | 0.08100 |
| 11 | 893 | 29 | 0.03250 |
| 12 | 727 | 5 | 0.00688 |
| 13 | 348 | 4 | 0.01150 |

Mechanical Reliability Rates

| Mileage at test | N tested | Pct failed |
|---|---|---|
| 10000 | 142 | 8.45 |
| 20000 | 395 | 9.11 |
| 30000 | 521 | 9.60 |
| 40000 | 703 | 10.40 |
| 50000 | 755 | 14.40 |
| 60000 | 753 | 11.80 |
| 70000 | 811 | 11.30 |
| 80000 | 780 | 13.20 |
| 90000 | 790 | 12.30 |
| 100000 | 675 | 14.20 |
| 110000 | 600 | 14.00 |
| 120000 | 512 | 14.10 |
| 130000 | 424 | 13.40 |
| 140000 | 345 | 17.40 |
| 150000 | 280 | 14.30 |
| 160000 | 205 | 15.10 |
| 170000 | 155 | 16.80 |

## Audi A8 2011

At 5 years of age, the mortality rate of a Audi A8 2011 (manufactured as a Car or Light Van) ranked number 170 out of 211 vehicles of the same age and type (any Car or Light Van constructed in 2011). One is the lowest (or best) and 211 the highest mortality rate. For vehicles reaching 20000 miles, its unreliability score (rate of failing an inspection) ranked 205 out of 205 vehicles of the same age, type, and mileage. One is the highest (or worst) and 205 the lowest rate of failing an inspection.

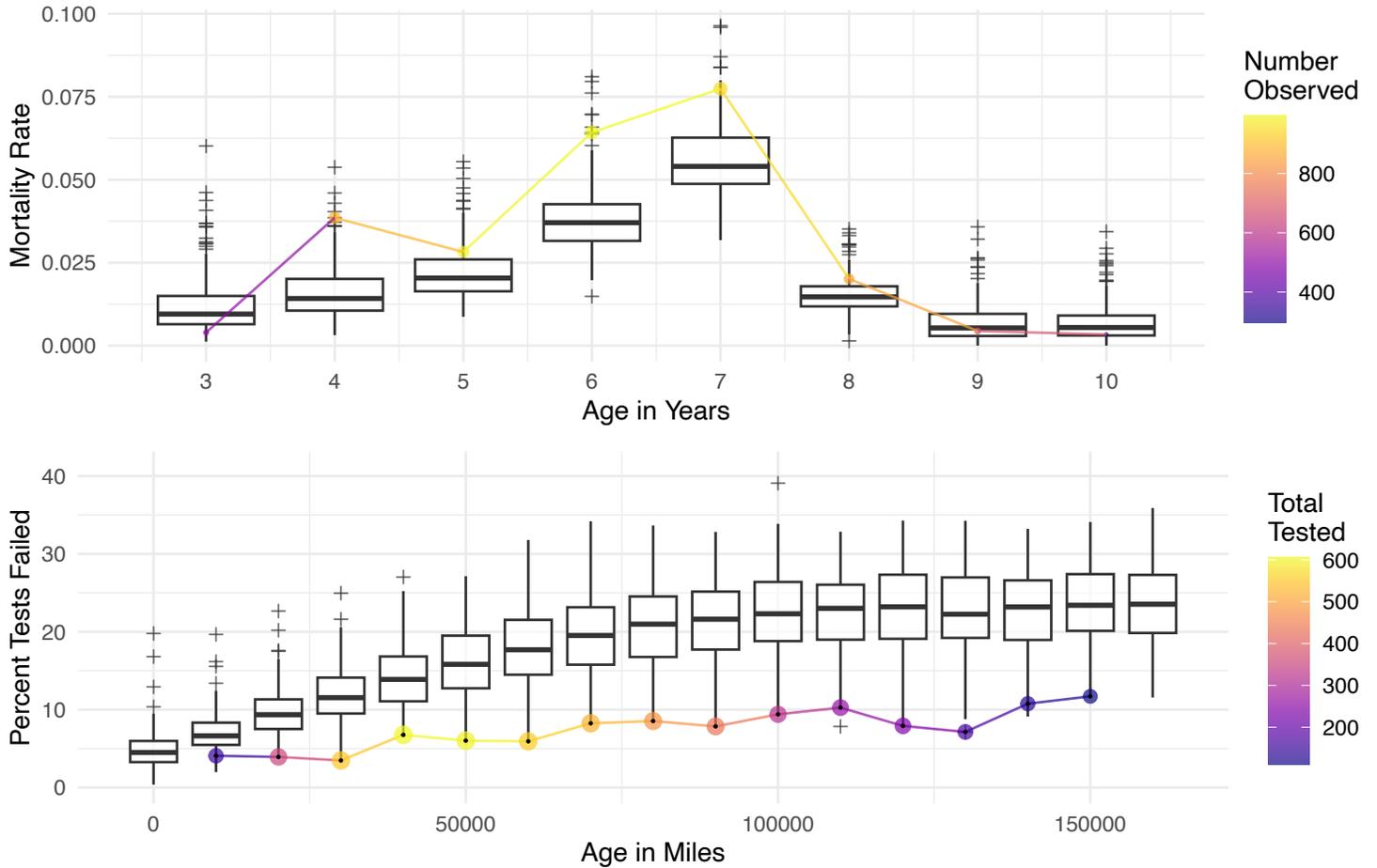

<table>
<tr><td colspan="4" align="center">Mortality rates</td></tr>
</table>

| Age in Years | Observed | Died | Mortality Rate |
|:---:|:---:|:---:|:---:|
| 3 | 505 | 2 | 0.00396 |
| 4 | 882 | 34 | 0.03850 |
| 5 | 992 | 28 | 0.02820 |
| 6 | 996 | 64 | 0.06430 |
| 7 | 943 | 73 | 0.07740 |
| 8 | 845 | 17 | 0.02010 |
| 9 | 681 | 3 | 0.00441 |
| 10 | 297 | 1 | 0.00337 |

<table>
<tr><td colspan="3" align="center">Mechanical Reliability Rates</td></tr>
</table>

| Mileage at test | N tested | Pct failed |
|:---:|:---:|:---:|
| 10000 | 147 | 4.08 |
| 20000 | 356 | 3.93 |
| 30000 | 546 | 3.48 |
| 40000 | 606 | 6.77 |
| 50000 | 598 | 6.02 |
| 60000 | 556 | 5.94 |
| 70000 | 521 | 8.25 |
| 80000 | 480 | 8.54 |
| 90000 | 433 | 7.85 |
| 100000 | 319 | 9.40 |
| 110000 | 263 | 10.30 |
| 120000 | 240 | 7.92 |
| 130000 | 168 | 7.14 |
| 140000 | 130 | 10.80 |
| 150000 | 111 | 11.70 |

## Audi A8 2012

At 5 years of age, the mortality rate of a Audi A8 2012 (manufactured as a Car or Light Van) ranked number 187 out of 212 vehicles of the same age and type (any Car or Light Van constructed in 2012). One is the lowest (or best) and 212 the highest mortality rate. For vehicles reaching 20000 miles, its unreliability score (rate of failing an inspection) ranked 206 out of 206 vehicles of the same age, type, and mileage. One is the highest (or worst) and 206 the lowest rate of failing an inspection.

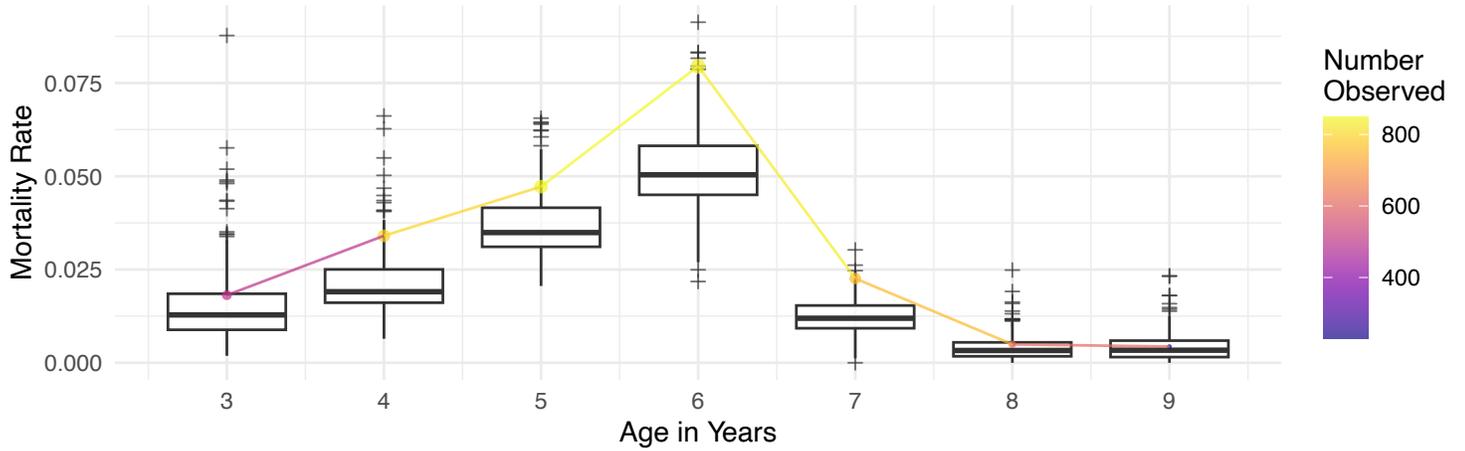

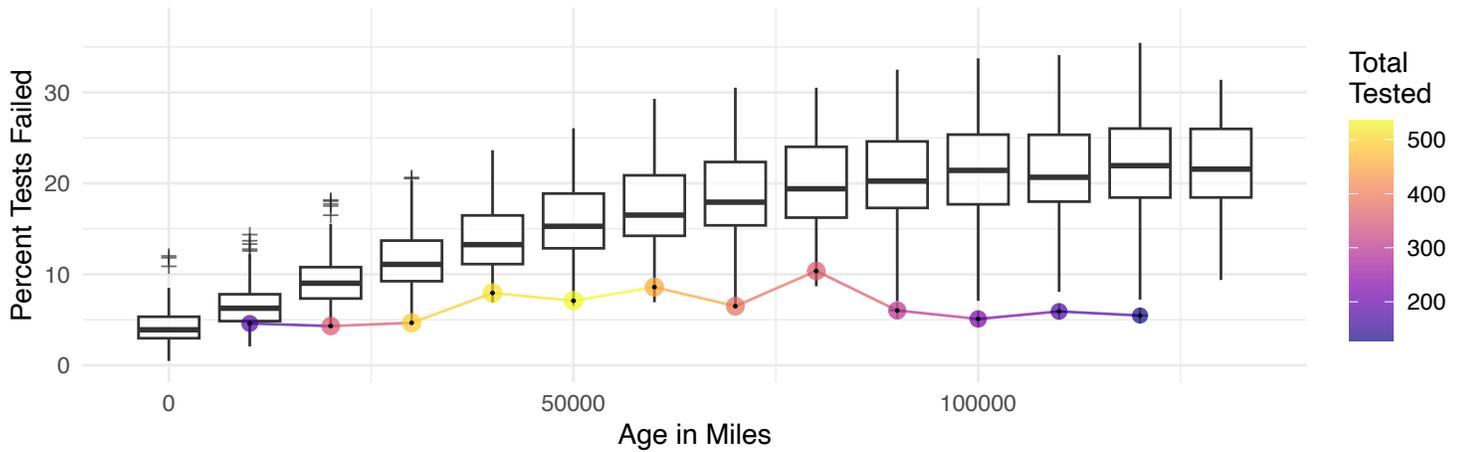

### Mortality rates

| Age in Years | Observed | Died | Mortality Rate |
|---|---|---|---|
| 3 | 496 | 9 | 0.01810 |
| 4 | 793 | 27 | 0.03400 |
| 5 | 847 | 40 | 0.04720 |
| 6 | 830 | 66 | 0.07950 |
| 7 | 752 | 17 | 0.02260 |
| 8 | 607 | 3 | 0.00494 |
| 9 | 230 | 1 | 0.00435 |

### Mechanical Reliability Rates

| Mileage at test | N tested | Pct failed |
|---|---|---|
| 10000 | 174 | 4.60 |
| 20000 | 348 | 4.31 |
| 30000 | 493 | 4.67 |
| 40000 | 516 | 7.95 |
| 50000 | 536 | 7.09 |
| 60000 | 454 | 8.59 |
| 70000 | 385 | 6.49 |
| 80000 | 357 | 10.40 |
| 90000 | 299 | 6.02 |
| 100000 | 216 | 5.09 |
| 110000 | 169 | 5.92 |
| 120000 | 128 | 5.47 |

# Audi A8 2014

At 5 years of age, the mortality rate of a Audi A8 2014 (manufactured as a Car or Light Van) ranked number 108 out of 236 vehicles of the same age and type (any Car or Light Van constructed in 2014). One is the lowest (or best) and 236 the highest mortality rate. For vehicles reaching 20000 miles, its unreliability score (rate of failing an inspection) ranked 206 out of 230 vehicles of the same age, type, and mileage. One is the highest (or worst) and 230 the lowest rate of failing an inspection.

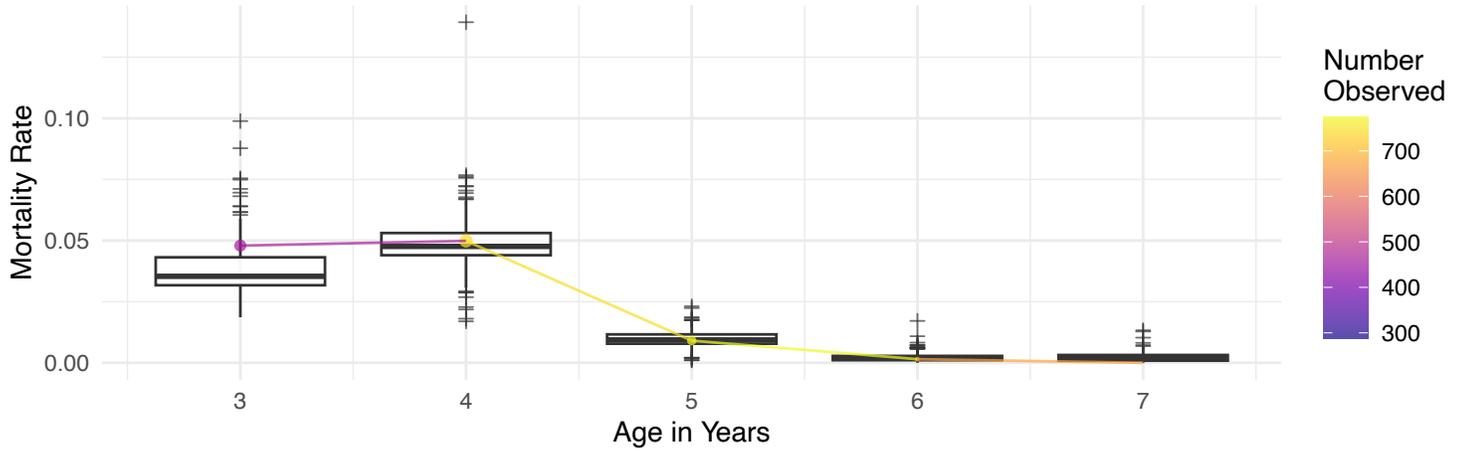

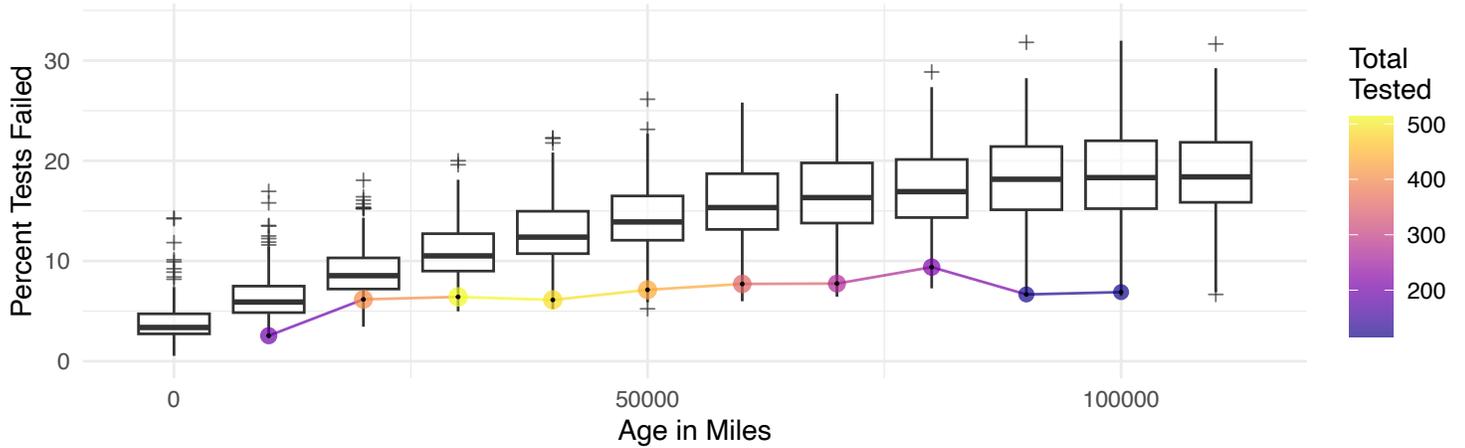

| Mortality rates | | | |
|---|---|---|---|
| Age in Years | Observed | Died | Mortality Rate |
| 3 | 459 | 22 | 0.04790 |
| 4 | 742 | 37 | 0.04990 |
| 5 | 773 | 7 | 0.00906 |
| 6 | 673 | 1 | 0.00149 |
| 7 | 288 | 0 | 0.00000 |

| Mechanical Reliability Rates | | |
|---|---|---|
| Mileage at test | N tested | Pct failed |
| 10000 | 196 | 2.55 |
| 20000 | 405 | 6.17 |
| 30000 | 514 | 6.42 |
| 40000 | 490 | 6.12 |
| 50000 | 435 | 7.13 |
| 60000 | 350 | 7.71 |
| 70000 | 271 | 7.75 |
| 80000 | 213 | 9.39 |
| 90000 | 120 | 6.67 |
| 100000 | 116 | 6.90 |

## Audi A8 2015

At 5 years of age, the mortality rate of a Audi A8 2015 (manufactured as a Car or Light Van) ranked number 1 out of 247 vehicles of the same age and type (any Car or Light Van constructed in 2015). One is the lowest (or best) and 247 the highest mortality rate. For vehicles reaching 20000 miles, its unreliability score (rate of failing an inspection) ranked 139 out of 241 vehicles of the same age, type, and mileage. One is the highest (or worst) and 241 the lowest rate of failing an inspection.

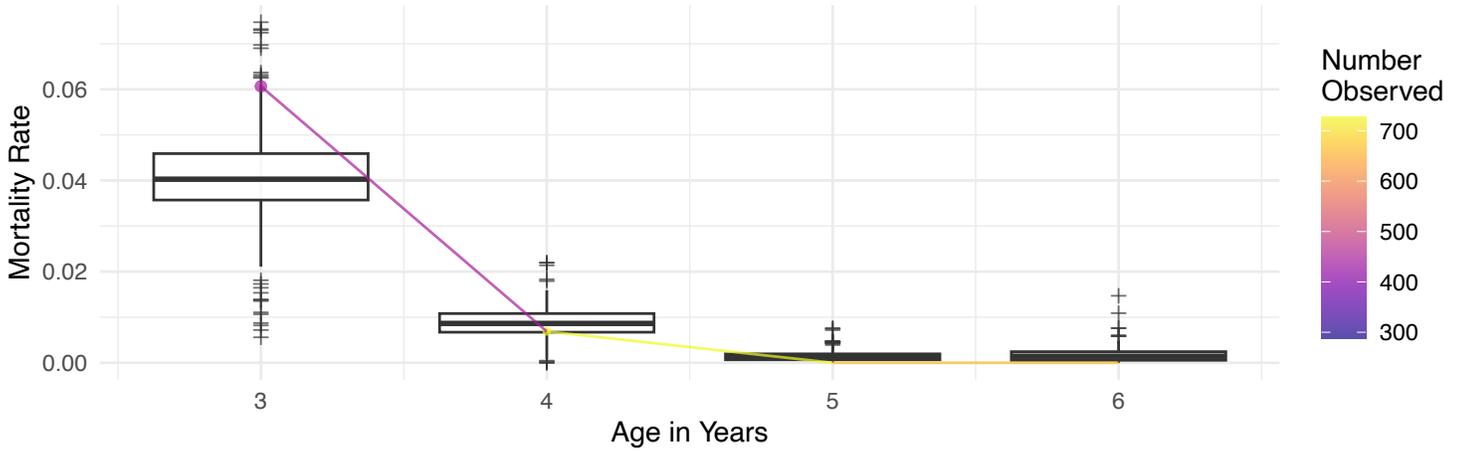

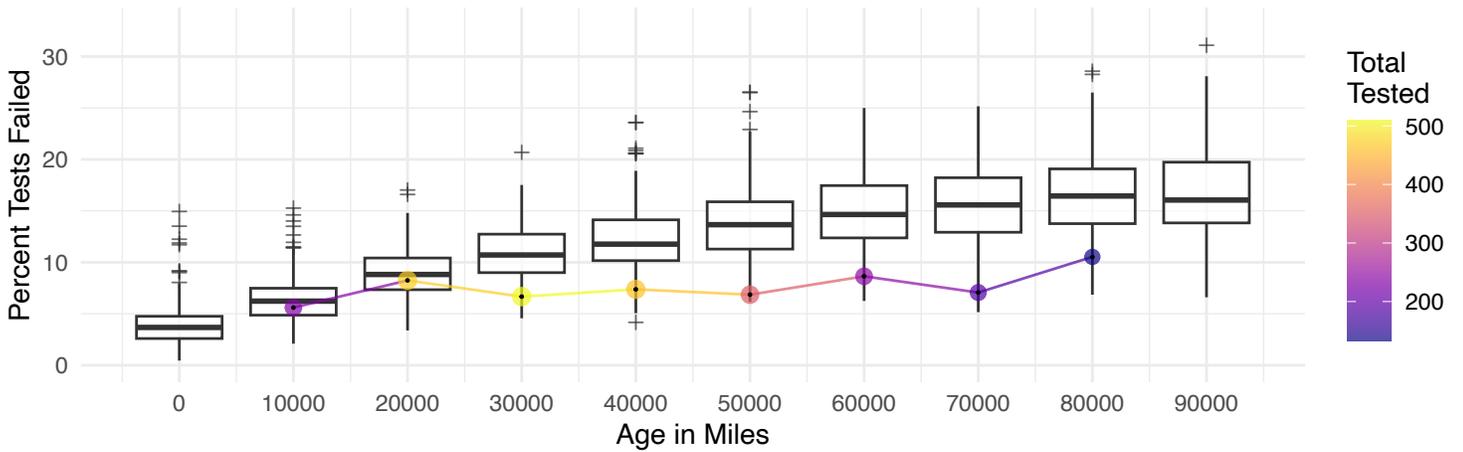

Mortality rates

| Age in Years | Observed | Died | Mortality Rate |
|---|---|---|---|
| 3 | 445 | 27 | 0.06070 |
| 4 | 726 | 5 | 0.00689 |
| 5 | 665 | 0 | 0.00000 |
| 6 | 288 | 0 | 0.00000 |

Mechanical Reliability Rates

| Mileage at test | N tested | Pct failed |
|---|---|---|
| 10000 | 232 | 5.60 |
| 20000 | 473 | 8.25 |
| 30000 | 510 | 6.67 |
| 40000 | 461 | 7.38 |
| 50000 | 350 | 6.86 |
| 60000 | 243 | 8.64 |
| 70000 | 184 | 7.07 |
| 80000 | 133 | 10.50 |

## Audi Coupe 1994

At 15 years of age, the mortality rate of a Audi Coupe 1994 (manufactured as a Car or Light Van) ranked number 28 out of 120 vehicles of the same age and type (any Car or Light Van constructed in 1994). One is the lowest (or best) and 120 the highest mortality rate. For vehicles reaching 120000 miles, its unreliability score (rate of failing an inspection) ranked 52 out of 112 vehicles of the same age, type, and mileage. One is the highest (or worst) and 112 the lowest rate of failing an inspection.

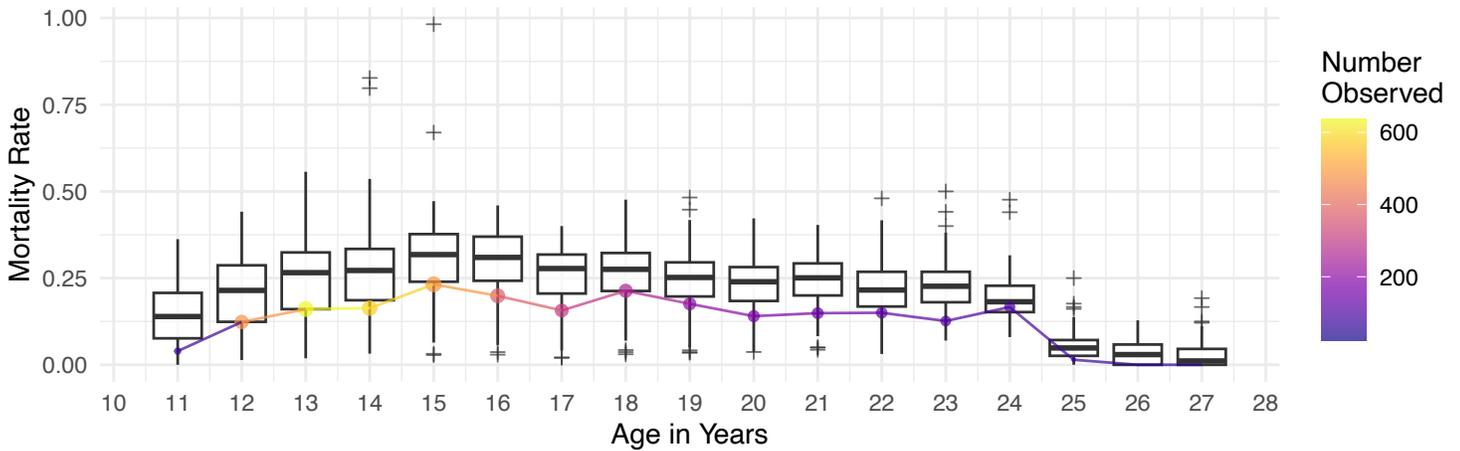

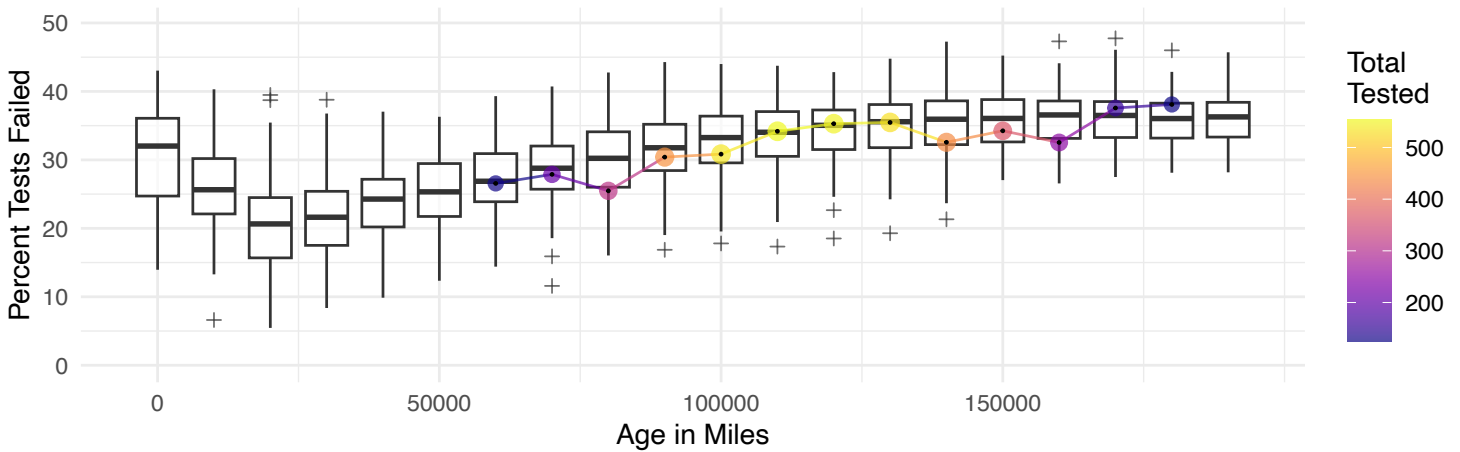

<table>
<tr><td colspan="4" align="center">Mortality rates</td></tr>
<tr><th>Age in Years</th><th>Observed</th><th>Died</th><th>Mortality Rate</th></tr>
<tr><td>11</td><td>76</td><td>3</td><td>0.0395</td></tr>
<tr><td>12</td><td>478</td><td>59</td><td>0.1230</td></tr>
<tr><td>13</td><td>635</td><td>102</td><td>0.1610</td></tr>
<tr><td>14</td><td>575</td><td>94</td><td>0.1630</td></tr>
<tr><td>15</td><td>487</td><td>113</td><td>0.2320</td></tr>
<tr><td>16</td><td>377</td><td>75</td><td>0.1990</td></tr>
<tr><td>17</td><td>300</td><td>47</td><td>0.1570</td></tr>
<tr><td>18</td><td>253</td><td>54</td><td>0.2130</td></tr>
<tr><td>19</td><td>199</td><td>35</td><td>0.1760</td></tr>
<tr><td>20</td><td>164</td><td>23</td><td>0.1400</td></tr>
<tr><td>21</td><td>141</td><td>21</td><td>0.1490</td></tr>
<tr><td>22</td><td>120</td><td>18</td><td>0.1500</td></tr>
<tr><td>23</td><td>103</td><td>13</td><td>0.1260</td></tr>
<tr><td>24</td><td>90</td><td>15</td><td>0.1670</td></tr>
<tr><td>25</td><td>69</td><td>1</td><td>0.0145</td></tr>
<tr><td>26</td><td>57</td><td>0</td><td>0.0000</td></tr>
<tr><td>27</td><td>24</td><td>0</td><td>0.0000</td></tr>
</table>

| | Mechanical Reliability Rates | |
| --- | --- | --- |
| Mileage at test | N tested | Pct failed |
| 60000 | 128 | 26.6 |
| 70000 | 208 | 27.9 |
| 80000 | 314 | 25.5 |
| 90000 | 444 | 30.4 |
| 100000 | 535 | 30.8 |
| 110000 | 553 | 34.2 |
| 120000 | 547 | 35.3 |
| 130000 | 530 | 35.5 |
| 140000 | 436 | 32.6 |
| 150000 | 365 | 34.2 |
| 160000 | 252 | 32.5 |
| 170000 | 157 | 37.6 |
| 180000 | 126 | 38.1 |

## Audi Coupe 1995

At 10 years of age, the mortality rate of a Audi Coupe 1995 (manufactured as a Car or Light Van) ranked number 55 out of 148 vehicles of the same age and type (any Car or Light Van constructed in 1995). One is the lowest (or best) and 148 the highest mortality rate. For vehicles reaching 120000 miles, its unreliability score (rate of failing an inspection) ranked 87 out of 135 vehicles of the same age, type, and mileage. One is the highest (or worst) and 135 the lowest rate of failing an inspection.

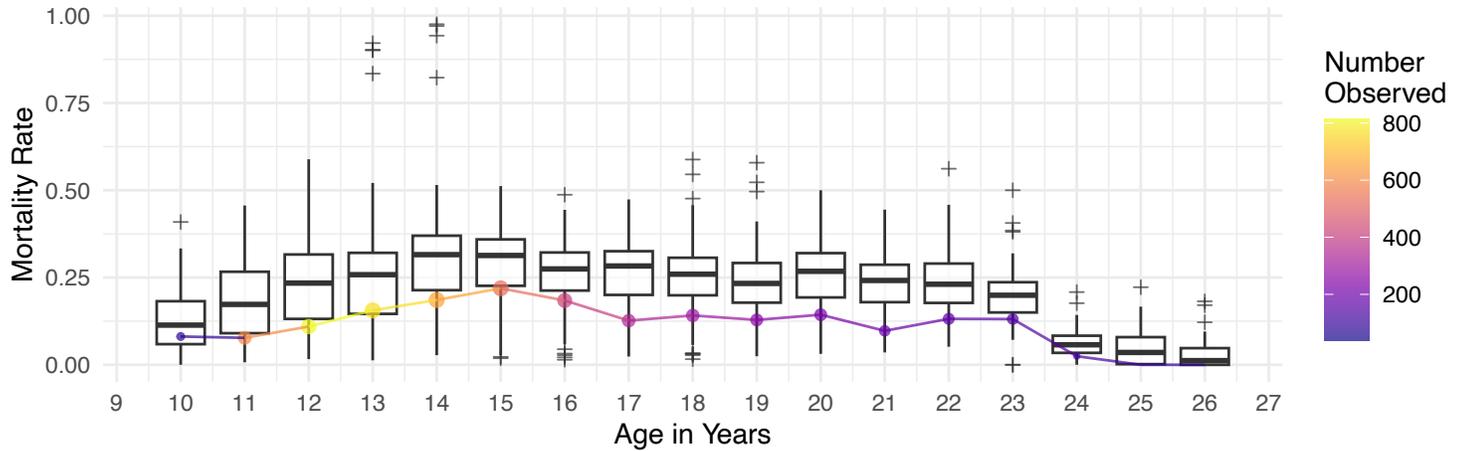

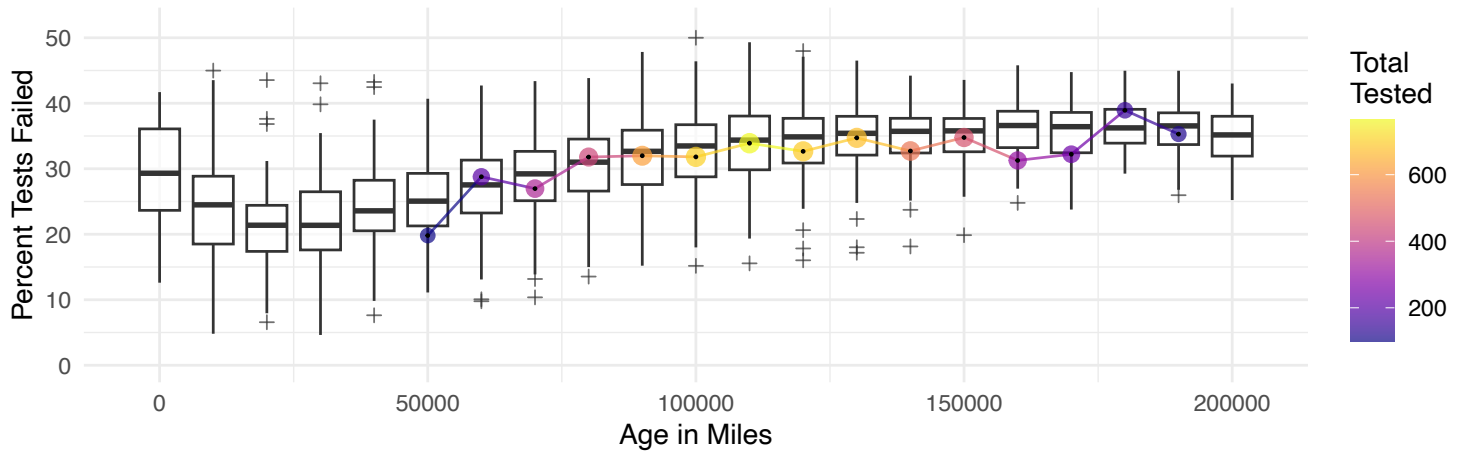

Mortality rates

| Age in Years | Observed | Died | Mortality Rate |
|---|---|---|---|
| 10 | 74 | 6 | 0.0811 |
| 11 | 596 | 46 | 0.0772 |
| 12 | 812 | 89 | 0.1100 |
| 13 | 764 | 119 | 0.1560 |
| 14 | 646 | 120 | 0.1860 |
| 15 | 525 | 115 | 0.2190 |
| 16 | 407 | 75 | 0.1840 |
| 17 | 332 | 42 | 0.1270 |
| 18 | 290 | 41 | 0.1410 |
| 19 | 249 | 32 | 0.1290 |
| 20 | 216 | 31 | 0.1440 |
| 21 | 185 | 18 | 0.0973 |
| 22 | 167 | 22 | 0.1320 |
| 23 | 145 | 19 | 0.1310 |
| 24 | 122 | 3 | 0.0246 |
| 25 | 94 | 0 | 0.0000 |
| 26 | 39 | 0 | 0.0000 |

Mechanical Reliability Rates

| Mileage at test | N tested | Pct failed |
|---|---|---|
| 50000 | 101 | 19.8 |
| 60000 | 205 | 28.8 |
| 70000 | 356 | 27.0 |
| 80000 | 431 | 31.8 |
| 90000 | 619 | 32.0 |
| 100000 | 701 | 31.8 |
| 110000 | 764 | 33.9 |
| 120000 | 692 | 32.7 |
| 130000 | 674 | 34.7 |
| 140000 | 538 | 32.7 |
| 150000 | 463 | 34.8 |
| 160000 | 307 | 31.3 |
| 170000 | 239 | 32.2 |
| 180000 | 149 | 38.9 |
| 190000 | 119 | 35.3 |

**Audi Q2 2017**

At 3 years of age, the mortality rate of a Audi Q2 2017 (manufactured as a Car or Light Van) ranked number 122 out of 247 vehicles of the same age and type (any Car or Light Van constructed in 2017). One is the lowest (or best) and 247 the highest mortality rate. For vehicles reaching 20000 miles, its unreliability score (rate of failing an inspection) ranked 229 out of 240 vehicles of the same age, type, and mileage. One is the highest (or worst) and 240 the lowest rate of failing an inspection.

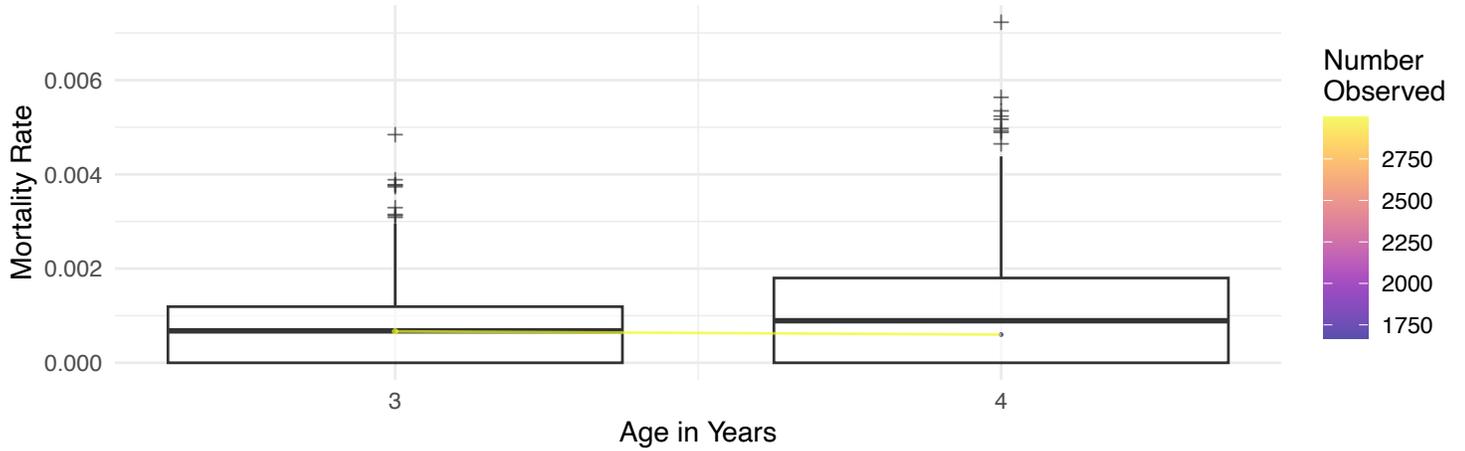

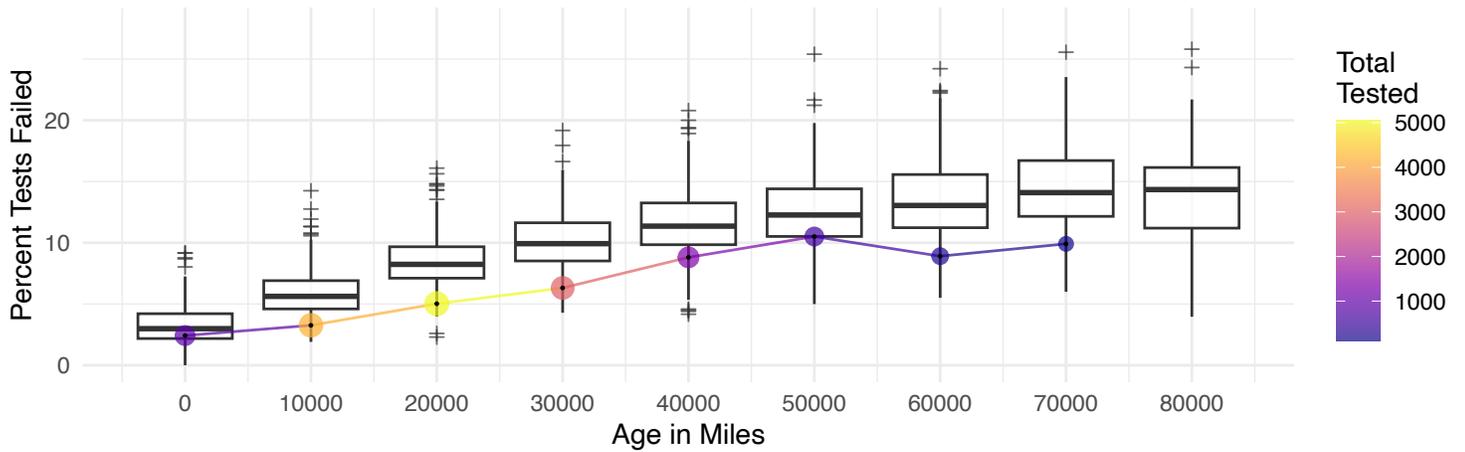

Mortality rates

| Age in Years | Observed | Died | Mortality Rate |
|---|---|---|---|
| 3 | 2999 | 2 | 0.000667 |
| 4 | 1669 | 1 | 0.000599 |

Mechanical Reliability Rates

| Mileage at test | N tested | Pct failed |
|---|---|---|
| 0 | 909 | 2.42 |
| 10000 | 4173 | 3.26 |
| 20000 | 5058 | 5.02 |
| 30000 | 3010 | 6.31 |
| 40000 | 1351 | 8.81 |
| 50000 | 590 | 10.50 |
| 60000 | 247 | 8.91 |
| 70000 | 111 | 9.91 |

**Audi Q2 2018**

At 3 years of age, the mortality rate of a Audi Q2 2018 (manufactured as a Car or Light Van) ranked number 6 out of 222 vehicles of the same age and type (any Car or Light Van constructed in 2018). One is the lowest (or best) and 222 the highest mortality rate. For vehicles reaching 20000 miles, its unreliability score (rate of failing an inspection) ranked 173 out of 215 vehicles of the same age, type, and mileage. One is the highest (or worst) and 215 the lowest rate of failing an inspection.

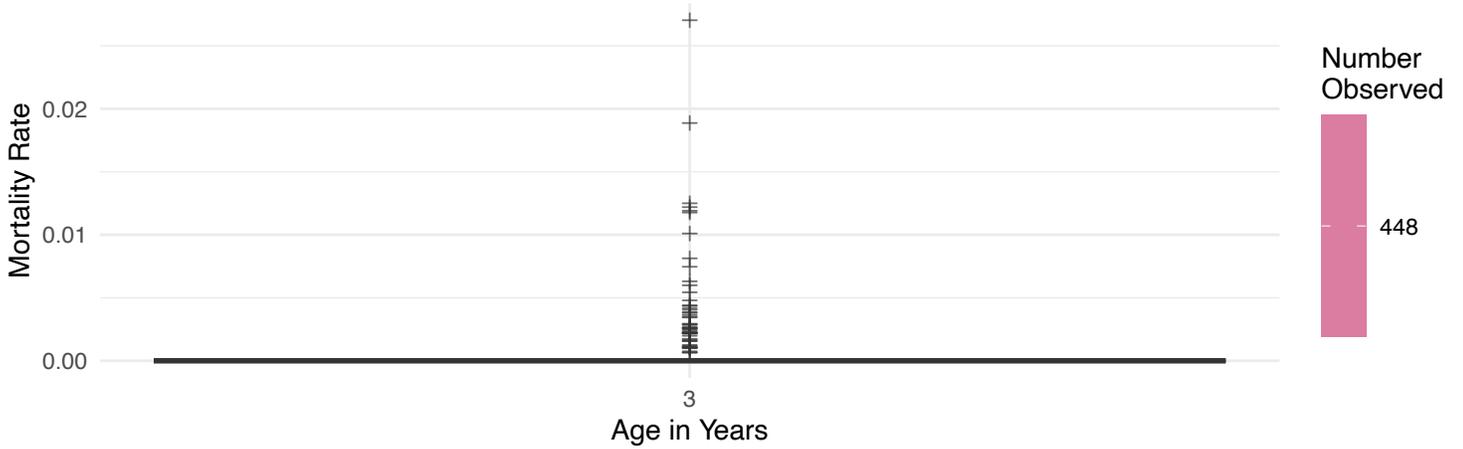

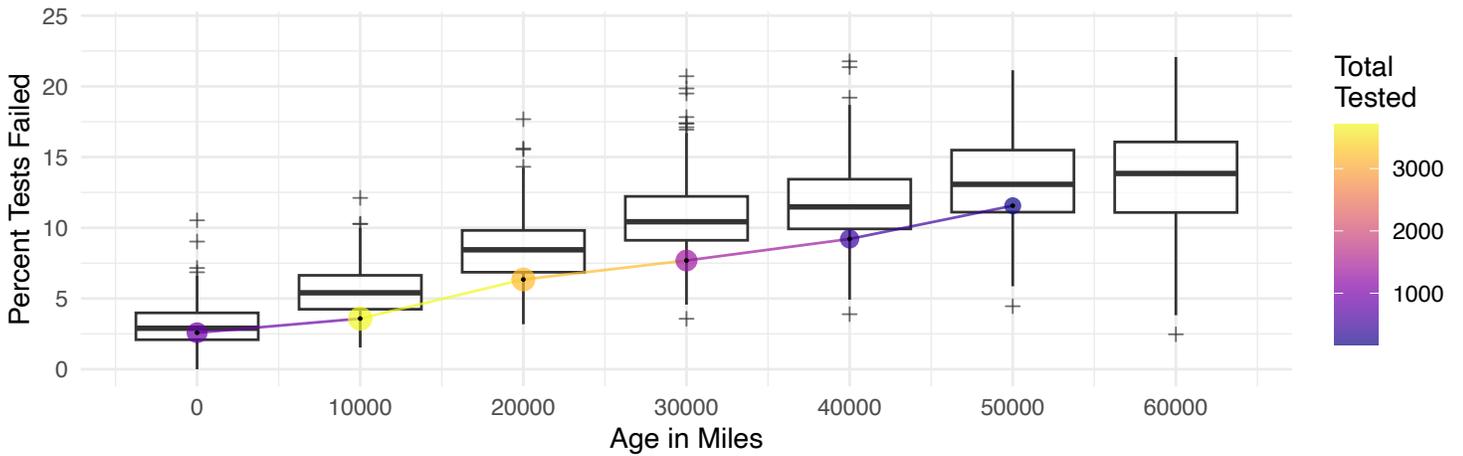

Mortality rates

| Age in Years | Observed | Died | Mortality Rate |
|---|---|---|---|
| 3 | 448 | 0 | 0 |

Mechanical Reliability Rates

| Mileage at test | N tested | Pct failed |
|---|---|---|
| 0 | 891 | 2.58 |
| 10000 | 3714 | 3.58 |
| 20000 | 3198 | 6.35 |
| 30000 | 1379 | 7.69 |
| 40000 | 456 | 9.21 |
| 50000 | 173 | 11.60 |

**Audi Q3 2012**

At 5 years of age, the mortality rate of a Audi Q3 2012 (manufactured as a Car or Light Van) ranked number 74 out of 212 vehicles of the same age and type (any Car or Light Van constructed in 2012). One is the lowest (or best) and 212 the highest mortality rate. For vehicles reaching 20000 miles, its unreliability score (rate of failing an inspection) ranked 199 out of 206 vehicles of the same age, type, and mileage. One is the highest (or worst) and 206 the lowest rate of failing an inspection.

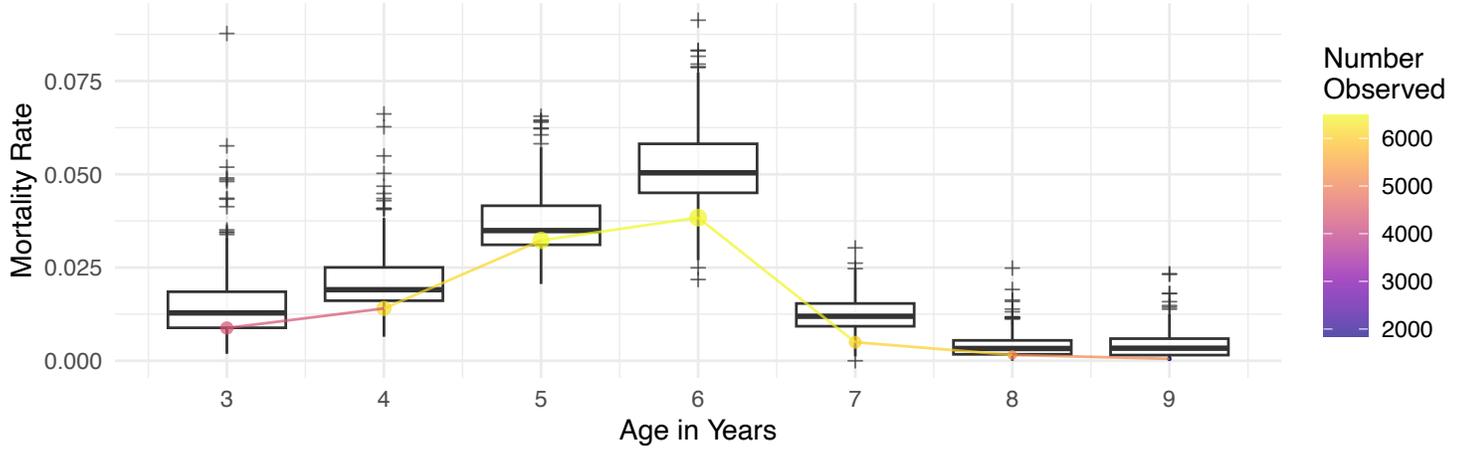

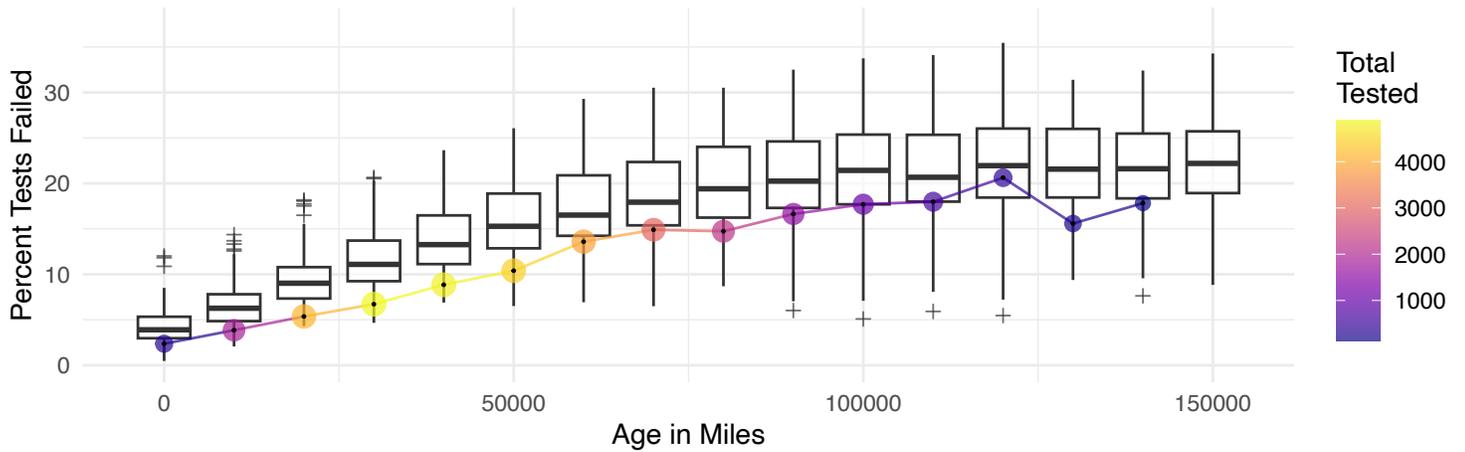

<table>
<tr><td colspan="4" align="center">Mortality rates</td></tr>
</table>

| Age in Years | Observed | Died | Mortality Rate |
|:---:|:---:|:---:|:---:|
| 3 | 4324 | 38 | 0.008790 |
| 4 | 6129 | 86 | 0.014000 |
| 5 | 6475 | 209 | 0.032300 |
| 6 | 6437 | 247 | 0.038400 |
| 7 | 6023 | 30 | 0.004980 |
| 8 | 5071 | 8 | 0.001580 |
| 9 | 1848 | 1 | 0.000541 |

Mechanical Reliability Rates

| Mileage at test | N tested | Pct failed |
|:---:|:---:|:---:|
| 0 | 296 | 2.36 |
| 10000 | 1894 | 3.85 |
| 20000 | 4139 | 5.36 |
| 30000 | 4899 | 6.72 |
| 40000 | 4803 | 8.85 |
| 50000 | 4464 | 10.40 |
| 60000 | 3872 | 13.60 |
| 70000 | 3027 | 14.90 |
| 80000 | 2219 | 14.70 |
| 90000 | 1575 | 16.60 |
| 100000 | 1028 | 17.70 |
| 110000 | 634 | 18.00 |
| 120000 | 383 | 20.60 |
| 130000 | 231 | 15.60 |
| 140000 | 129 | 17.80 |

# Audi Q3 2013

At 5 years of age, the mortality rate of a Audi Q3 2013 (manufactured as a Car or Light Van) ranked number 77 out of 221 vehicles of the same age and type (any Car or Light Van constructed in 2013). One is the lowest (or best) and 221 the highest mortality rate. For vehicles reaching 20000 miles, its unreliability score (rate of failing an inspection) ranked 197 out of 215 vehicles of the same age, type, and mileage. One is the highest (or worst) and 215 the lowest rate of failing an inspection.

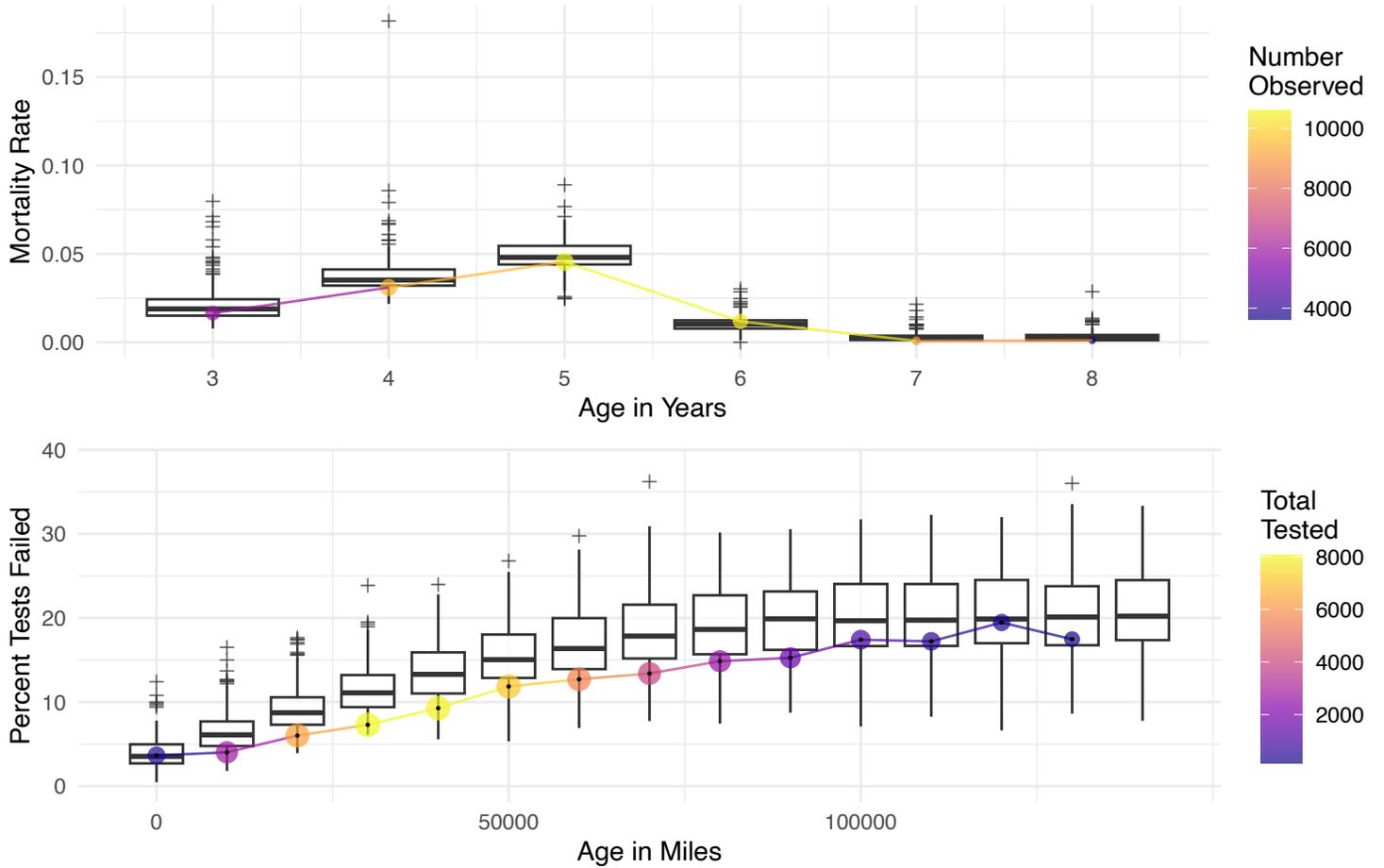

Mortality rates

| Age in Years | Observed | Died | Mortality Rate |
|---|---|---|---|
| 3 | 5835 | 96 | 0.016500 |
| 4 | 9461 | 293 | 0.031000 |
| 5 | 10582 | 482 | 0.045500 |
| 6 | 10331 | 122 | 0.011800 |
| 7 | 8810 | 7 | 0.000795 |
| 8 | 3633 | 4 | 0.001100 |

Mechanical Reliability Rates

| Mileage at test | N tested | Pct failed |
|---|---|---|
| 0 | 410 | 3.66 |
| 10000 | 2835 | 4.02 |
| 20000 | 6386 | 6.01 |
| 30000 | 8063 | 7.29 |
| 40000 | 7927 | 9.27 |
| 50000 | 7155 | 11.90 |
| 60000 | 5611 | 12.70 |
| 70000 | 3956 | 13.40 |
| 80000 | 2605 | 14.90 |
| 90000 | 1632 | 15.30 |
| 100000 | 965 | 17.40 |
| 110000 | 575 | 17.20 |
| 120000 | 329 | 19.50 |
| 130000 | 166 | 17.50 |

**Audi Q3 2014**

At 5 years of age, the mortality rate of a Audi Q3 2014 (manufactured as a Car or Light Van) ranked number 116 out of 236 vehicles of the same age and type (any Car or Light Van constructed in 2014). One is the lowest (or best) and 236 the highest mortality rate. For vehicles reaching 20000 miles, its unreliability score (rate of failing an inspection) ranked 217 out of 230 vehicles of the same age, type, and mileage. One is the highest (or worst) and 230 the lowest rate of failing an inspection.

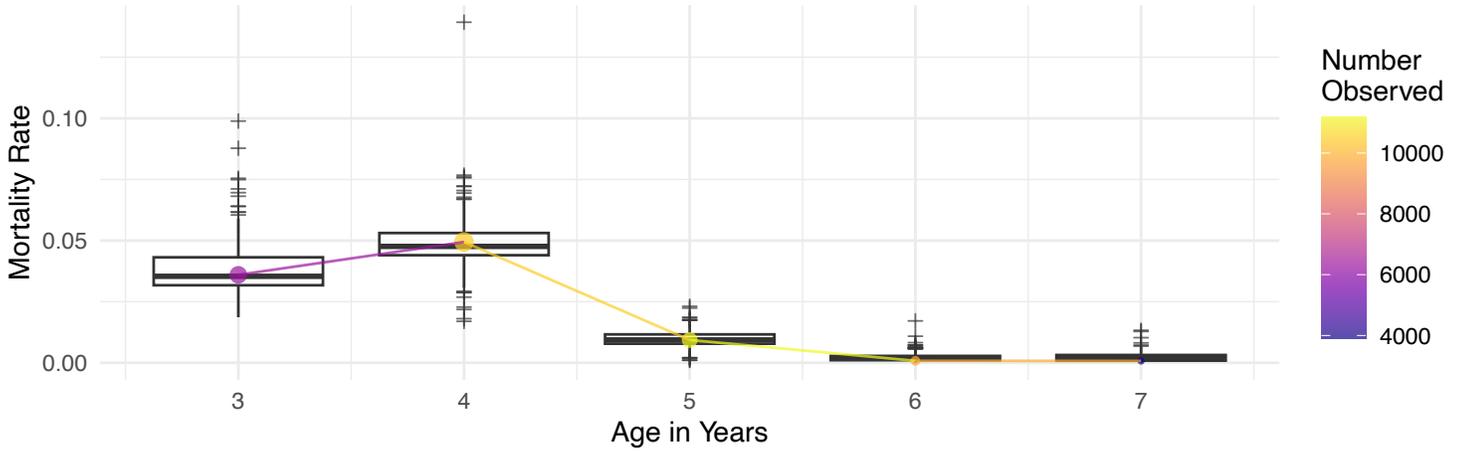

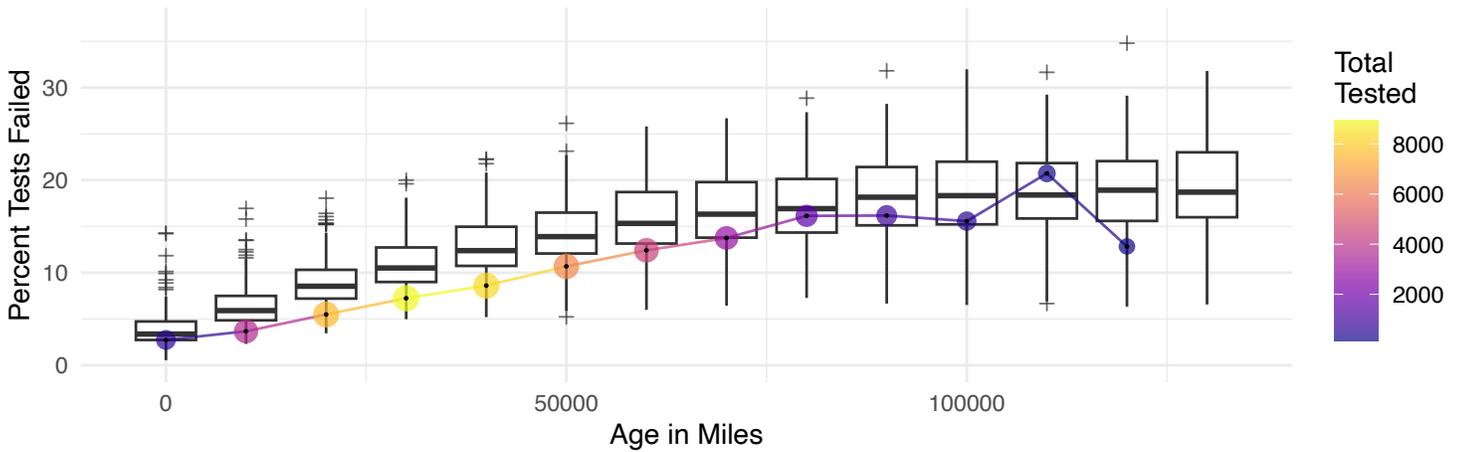

Mortality rates

| Age in Years | Observed | Died | Mortality Rate |
|---|---|---|---|
| 3 | 6309 | 227 | 0.036000 |
| 4 | 10402 | 514 | 0.049400 |
| 5 | 11163 | 104 | 0.009320 |
| 6 | 9725 | 8 | 0.000823 |
| 7 | 3907 | 3 | 0.000768 |

Mechanical Reliability Rates

| Mileage at test | N tested | Pct failed |
|---|---|---|
| 0 | 586 | 2.73 |
| 10000 | 3680 | 3.67 |
| 20000 | 7600 | 5.49 |
| 30000 | 8951 | 7.24 |
| 40000 | 8296 | 8.61 |
| 50000 | 6411 | 10.70 |
| 60000 | 4412 | 12.40 |
| 70000 | 2908 | 13.80 |
| 80000 | 1561 | 16.10 |
| 90000 | 921 | 16.20 |
| 100000 | 488 | 15.60 |
| 110000 | 246 | 20.70 |
| 120000 | 148 | 12.80 |

**Audi Q3 2015**

At 5 years of age, the mortality rate of a Audi Q3 2015 (manufactured as a Car or Light Van) ranked number 87 out of 247 vehicles of the same age and type (any Car or Light Van constructed in 2015). One is the lowest (or best) and 247 the highest mortality rate. For vehicles reaching 20000 miles, its unreliability score (rate of failing an inspection) ranked 213 out of 241 vehicles of the same age, type, and mileage. One is the highest (or worst) and 241 the lowest rate of failing an inspection.

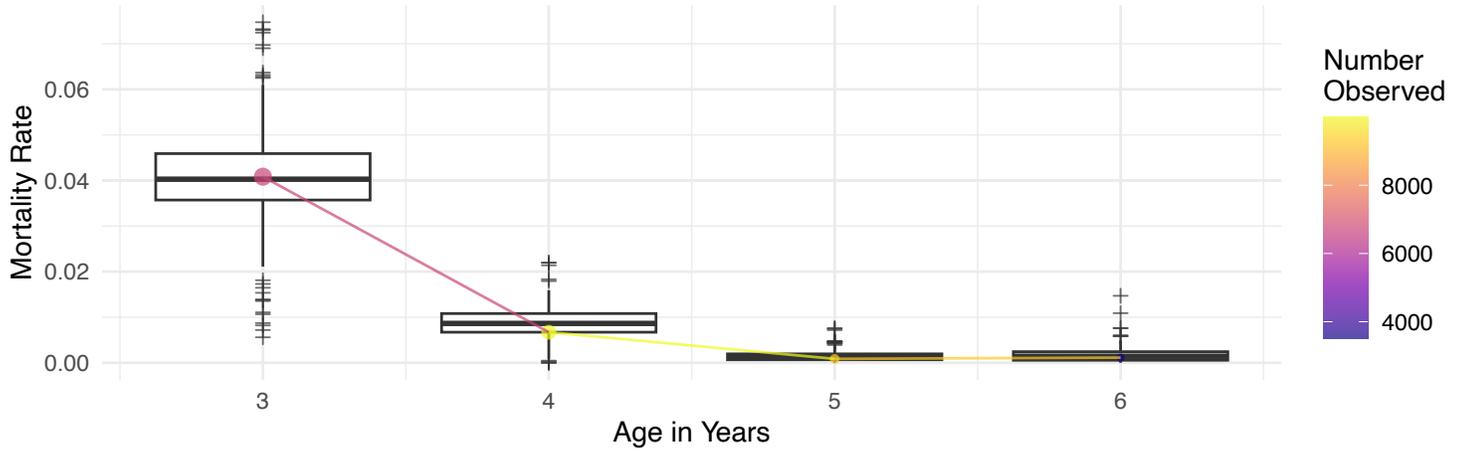

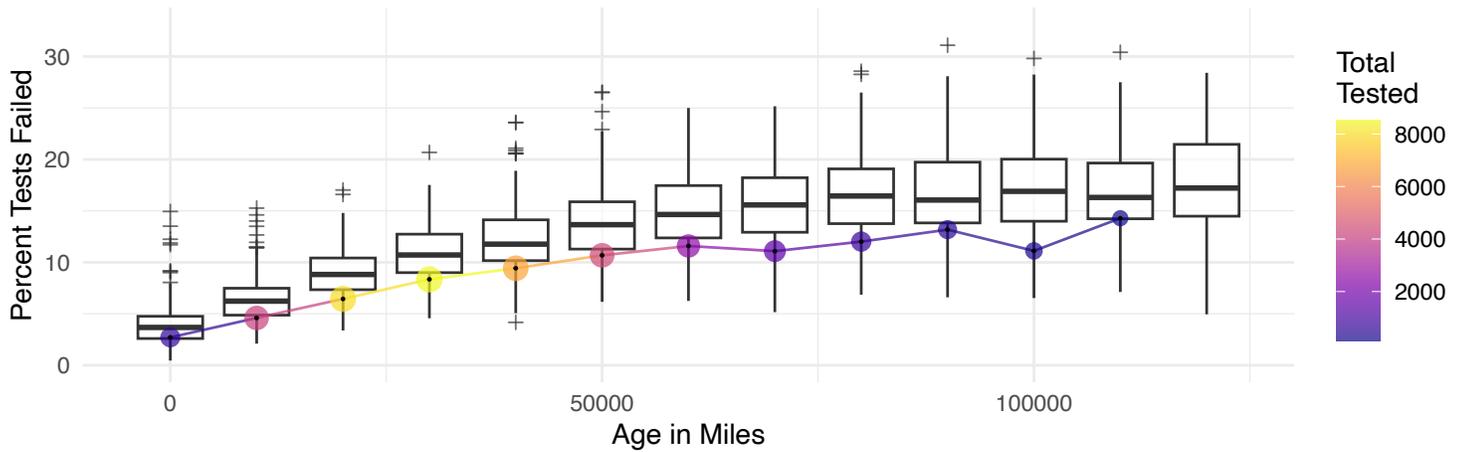

Mortality rates

| Age in Years | Observed | Died | Mortality Rate |
|---|---|---|---|
| 3 | 6710 | 274 | 0.040800 |
| 4 | 9982 | 67 | 0.006710 |
| 5 | 9245 | 8 | 0.000865 |
| 6 | 3514 | 4 | 0.001140 |

Mechanical Reliability Rates

| Mileage at test | N tested | Pct failed |
|---|---|---|
| 0 | 628 | 2.71 |
| 10000 | 4156 | 4.60 |
| 20000 | 8021 | 6.45 |
| 30000 | 8530 | 8.35 |
| 40000 | 6813 | 9.42 |
| 50000 | 4420 | 10.70 |
| 60000 | 2546 | 11.60 |
| 70000 | 1398 | 11.10 |
| 80000 | 716 | 12.00 |
| 90000 | 425 | 13.20 |
| 100000 | 225 | 11.10 |
| 110000 | 126 | 14.30 |

## Audi Q3 2016

At 5 years of age, the mortality rate of a Audi Q3 2016 (manufactured as a Car or Light Van) ranked number 109 out of 252 vehicles of the same age and type (any Car or Light Van constructed in 2016). One is the lowest (or best) and 252 the highest mortality rate. For vehicles reaching 20000 miles, its unreliability score (rate of failing an inspection) ranked 199 out of 246 vehicles of the same age, type, and mileage. One is the highest (or worst) and 246 the lowest rate of failing an inspection.

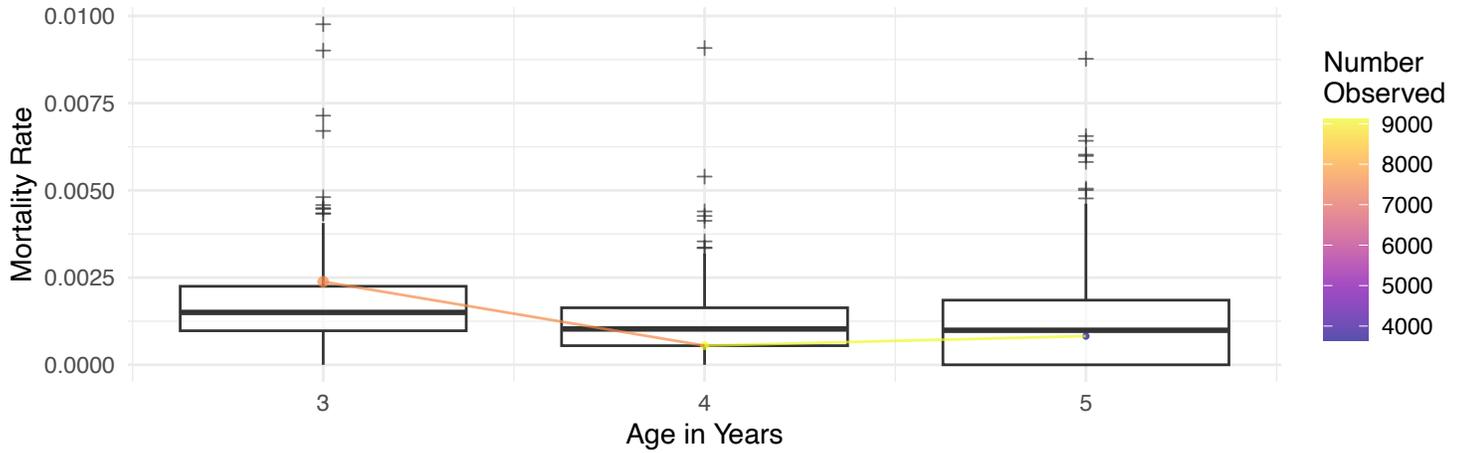

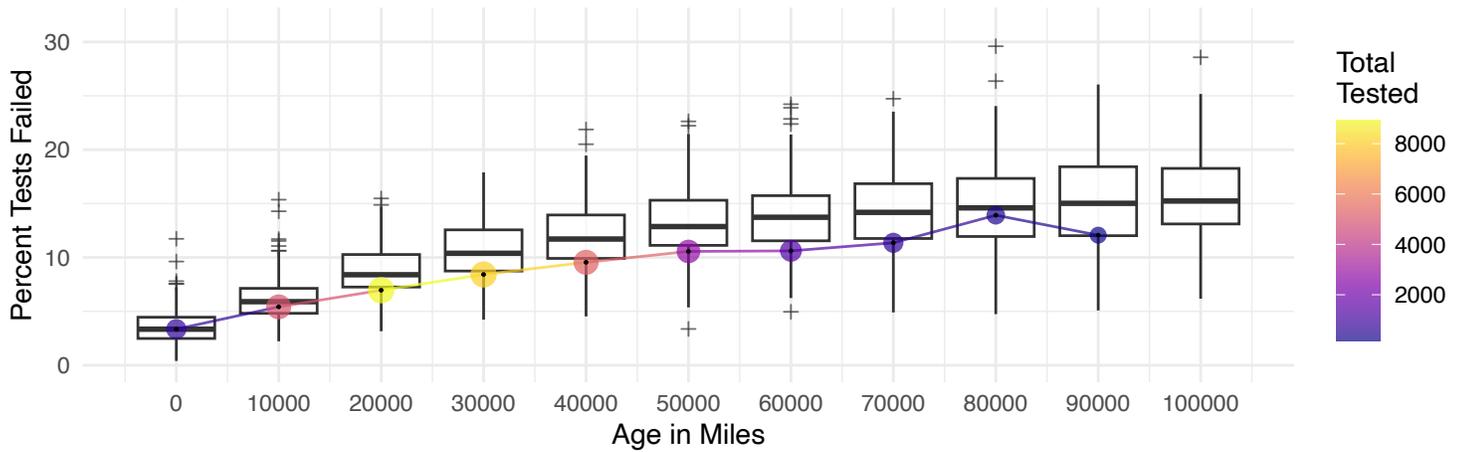

Mortality rates

| Age in Years | Observed | Died | Mortality Rate |
|---|---|---|---|
| 3 | 7557 | 18 | 0.002380 |
| 4 | 9101 | 5 | 0.000549 |
| 5 | 3648 | 3 | 0.000822 |

Mechanical Reliability Rates

| Mileage at test | N tested | Pct failed |
|---|---|---|
| 0 | 717 | 3.35 |
| 10000 | 4946 | 5.42 |
| 20000 | 8935 | 6.96 |
| 30000 | 8110 | 8.42 |
| 40000 | 5340 | 9.55 |
| 50000 | 2794 | 10.60 |
| 60000 | 1385 | 10.60 |
| 70000 | 669 | 11.40 |
| 80000 | 352 | 13.90 |
| 90000 | 174 | 12.10 |

**Audi Q3 2017**

At 3 years of age, the mortality rate of a Audi Q3 2017 (manufactured as a Car or Light Van) ranked number 104 out of 247 vehicles of the same age and type (any Car or Light Van constructed in 2017). One is the lowest (or best) and 247 the highest mortality rate. For vehicles reaching 20000 miles, its unreliability score (rate of failing an inspection) ranked 208 out of 240 vehicles of the same age, type, and mileage. One is the highest (or worst) and 240 the lowest rate of failing an inspection.

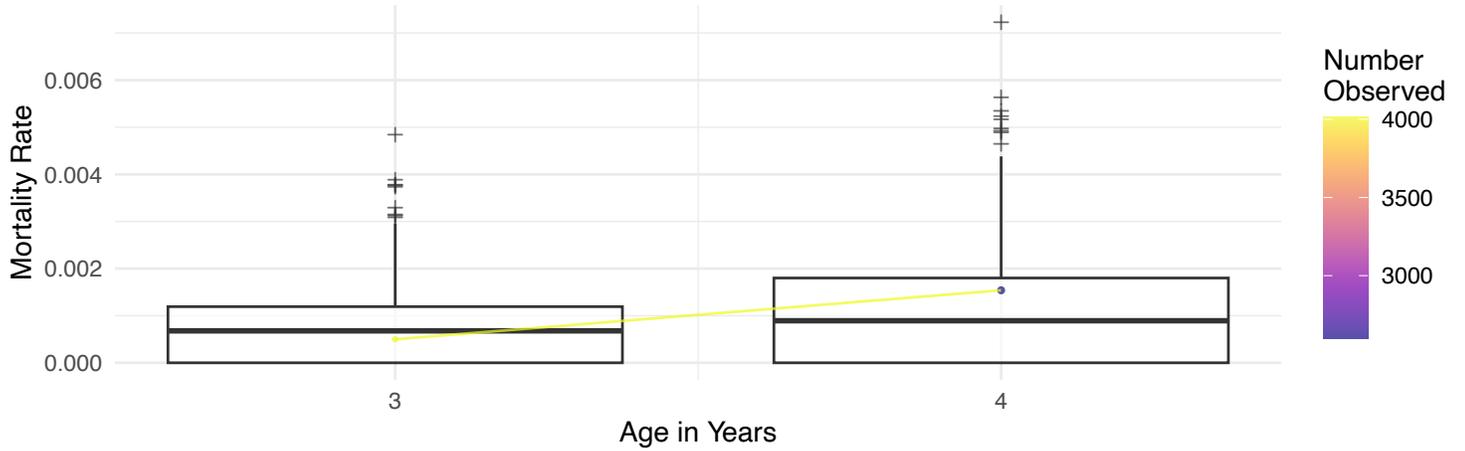

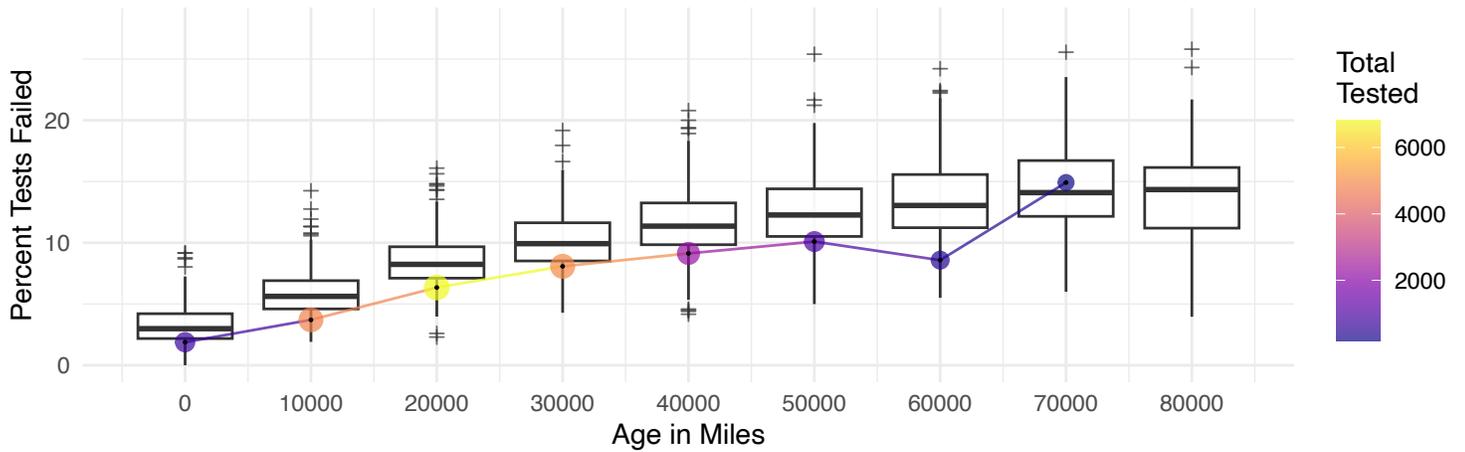

<table>
<tr><td colspan="4" align="center">Mortality rates</td></tr>
<tr><th>Age in Years</th><th>Observed</th><th>Died</th><th>Mortality Rate</th></tr>
<tr><td>3</td><td>4011</td><td>2</td><td>0.000499</td></tr>
<tr><td>4</td><td>2599</td><td>4</td><td>0.001540</td></tr>
</table>

| Mechanical Reliability Rates | | |
|---|---|---|
| Mileage at test | N tested | Pct failed |
| 0 | 744 | 1.88 |
| 10000 | 4829 | 3.71 |
| 20000 | 6824 | 6.35 |
| 30000 | 4966 | 8.07 |
| 40000 | 2289 | 9.13 |
| 50000 | 931 | 10.10 |
| 60000 | 408 | 8.58 |
| 70000 | 181 | 14.90 |

**Audi Q3 2018**

At 3 years of age, the mortality rate of a Audi Q3 2018 (manufactured as a Car or Light Van) ranked number 189 out of 222 vehicles of the same age and type (any Car or Light Van constructed in 2018). One is the lowest (or best) and 222 the highest mortality rate. For vehicles reaching 40000 miles, its unreliability score (rate of failing an inspection) ranked 92 out of 175 vehicles of the same age, type, and mileage. One is the highest (or worst) and 175 the lowest rate of failing an inspection.

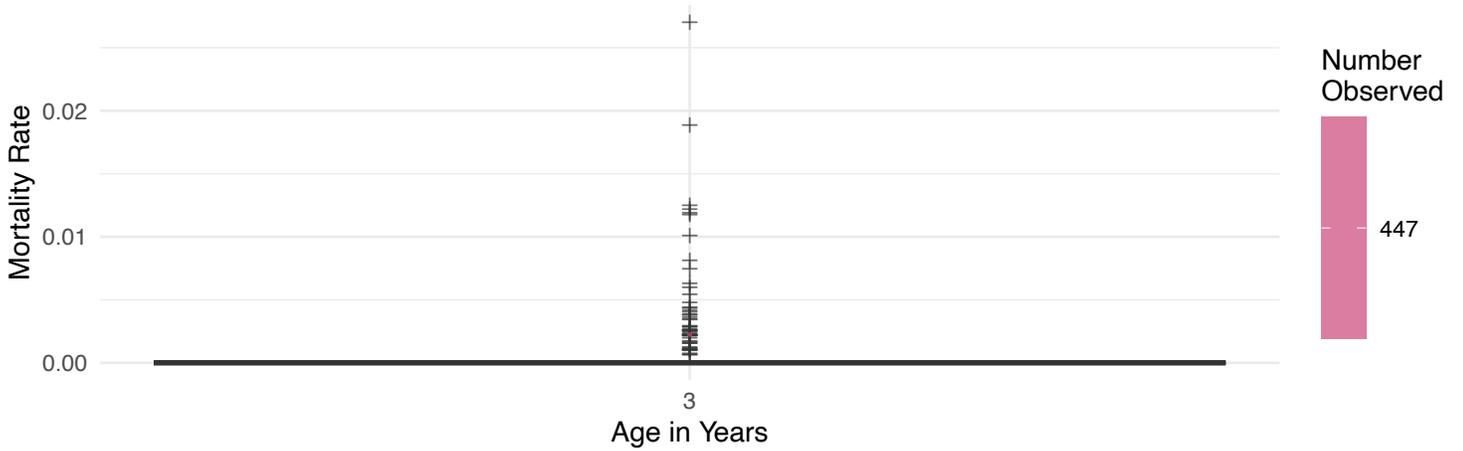

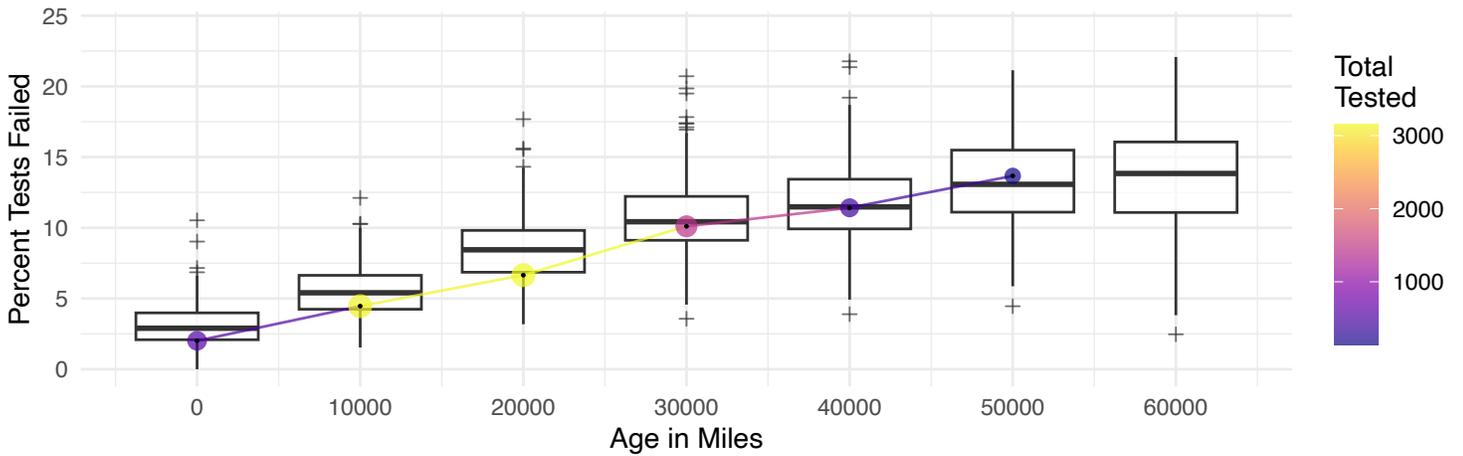

Mortality rates

| Age in Years | Observed | Died | Mortality Rate |
|---|---|---|---|
| 3 | 447 | 1 | 0.00224 |

Mechanical Reliability Rates

| Mileage at test | N tested | Pct failed |
|---|---|---|
| 0 | 547 | 2.01 |
| 10000 | 3115 | 4.46 |
| 20000 | 3157 | 6.65 |
| 30000 | 1425 | 10.10 |
| 40000 | 403 | 11.40 |
| 50000 | 139 | 13.70 |

**Audi Q5 2009**

At 5 years of age, the mortality rate of a Audi Q5 2009 (manufactured as a Car or Light Van) ranked number 39 out of 205 vehicles of the same age and type (any Car or Light Van constructed in 2009). One is the lowest (or best) and 205 the highest mortality rate. For vehicles reaching 120000 miles, its unreliability score (rate of failing an inspection) ranked 151 out of 185 vehicles of the same age, type, and mileage. One is the highest (or worst) and 185 the lowest rate of failing an inspection.

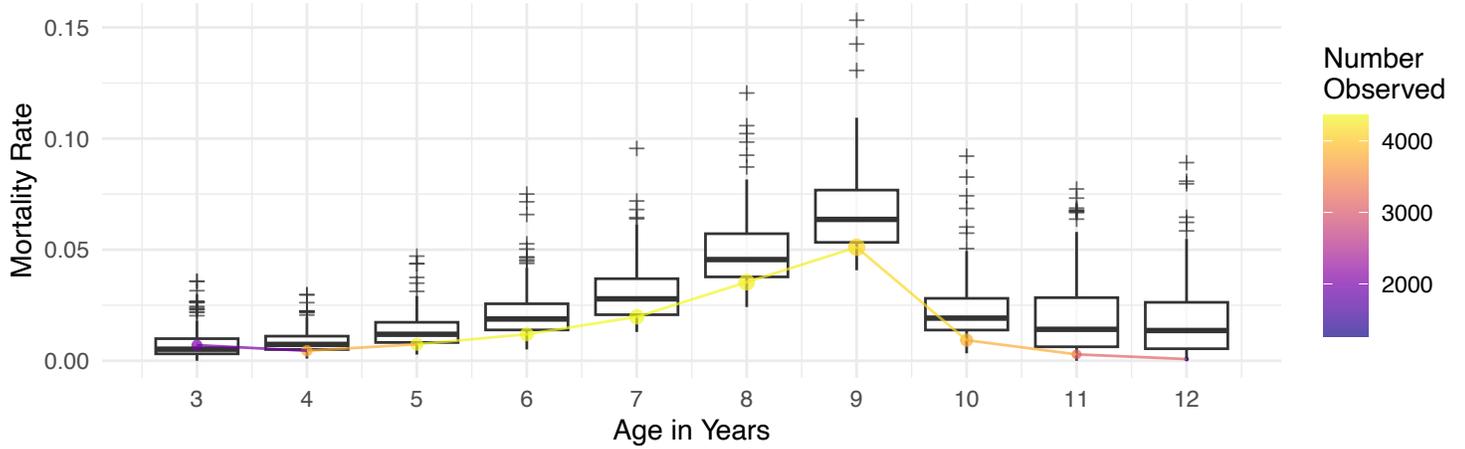

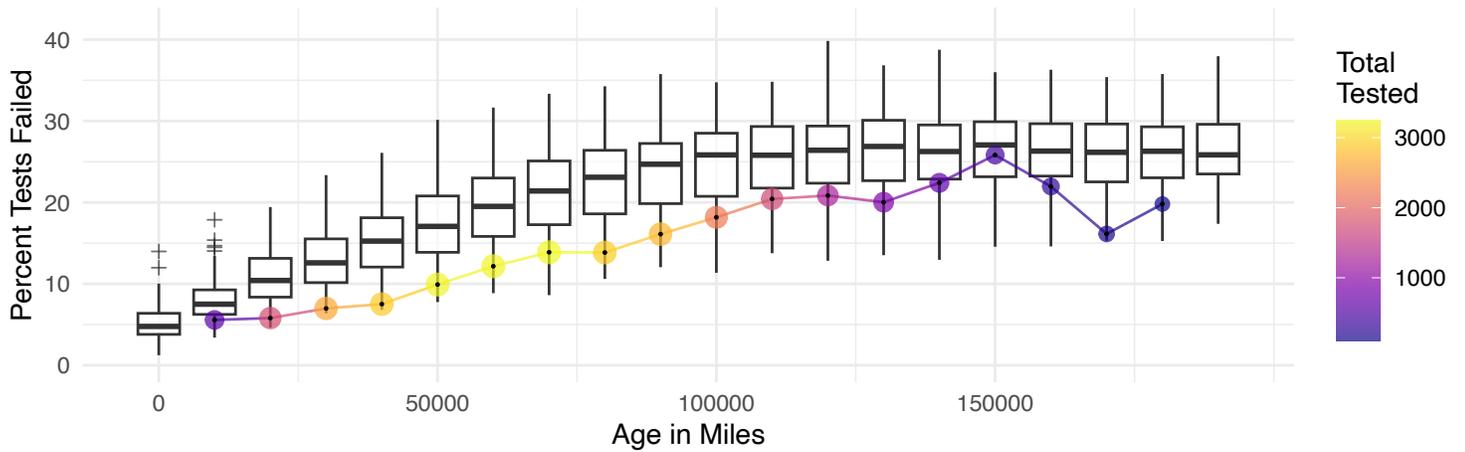

Mortality rates

| Age in Years | Observed | Died | Mortality Rate |
|---|---|---|---|
| 3 | 1987 | 14 | 0.007050 |
| 4 | 3764 | 17 | 0.004520 |
| 5 | 4301 | 32 | 0.007440 |
| 6 | 4353 | 52 | 0.011900 |
| 7 | 4349 | 86 | 0.019800 |
| 8 | 4268 | 151 | 0.035400 |
| 9 | 4110 | 210 | 0.051100 |
| 10 | 3774 | 35 | 0.009270 |
| 11 | 3103 | 9 | 0.002900 |
| 12 | 1267 | 1 | 0.000789 |

Mechanical Reliability Rates

| Mileage at test | N tested | Pct failed |
|---|---|---|
| 10000 | 557 | 5.57 |
| 20000 | 1743 | 5.79 |
| 30000 | 2623 | 6.98 |
| 40000 | 2902 | 7.51 |
| 50000 | 3219 | 9.91 |
| 60000 | 3224 | 12.20 |
| 70000 | 3250 | 13.90 |
| 80000 | 2926 | 13.80 |
| 90000 | 2757 | 16.10 |
| 100000 | 2196 | 18.20 |
| 110000 | 1699 | 20.40 |
| 120000 | 1257 | 20.80 |
| 130000 | 879 | 20.00 |
| 140000 | 616 | 22.40 |
| 150000 | 395 | 25.80 |
| 160000 | 255 | 22.00 |
| 170000 | 155 | 16.10 |

# Audi Q5 2010

At 5 years of age, the mortality rate of a Audi Q5 2010 (manufactured as a Car or Light Van) ranked number 67 out of 206 vehicles of the same age and type (any Car or Light Van constructed in 2010). One is the lowest (or best) and 206 the highest mortality rate. For vehicles reaching 20000 miles, its unreliability score (rate of failing an inspection) ranked 194 out of 201 vehicles of the same age, type, and mileage. One is the highest (or worst) and 201 the lowest rate of failing an inspection.

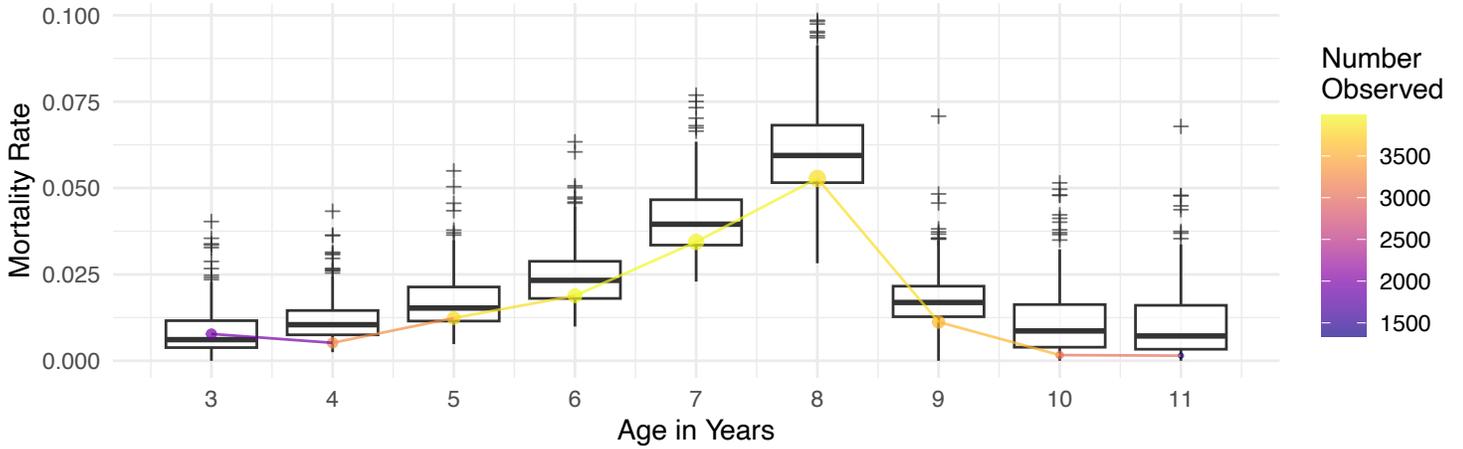

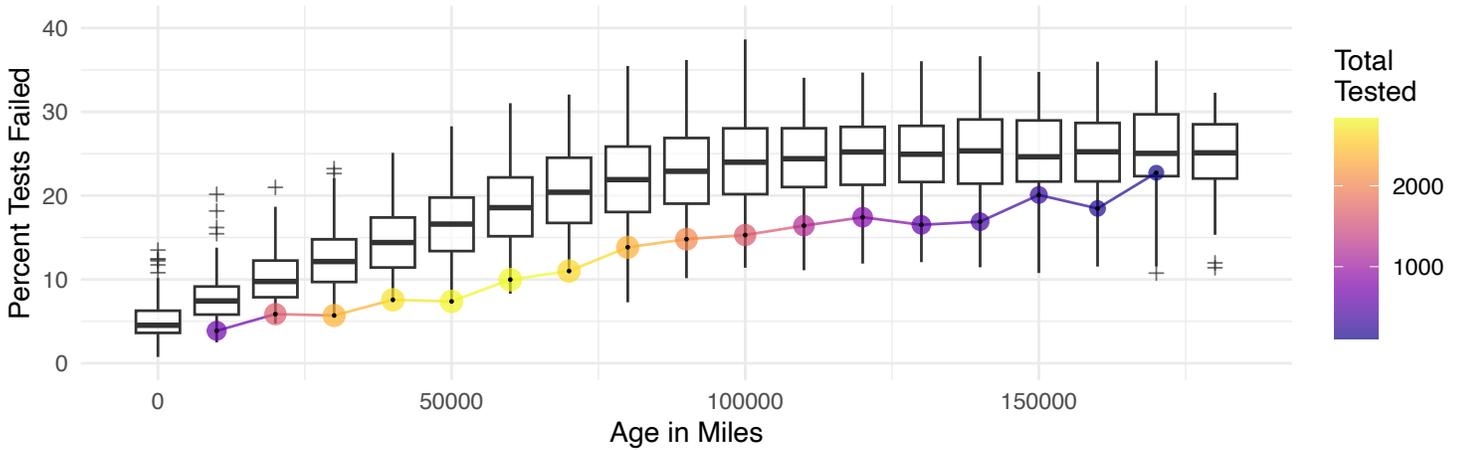

Mortality rates

| Age in Years | Observed | Died | Mortality Rate |
|---|---|---|---|
| 3 | 1930 | 15 | 0.00777 |
| 4 | 3270 | 17 | 0.00520 |
| 5 | 3798 | 47 | 0.01240 |
| 6 | 3983 | 75 | 0.01880 |
| 7 | 3951 | 136 | 0.03440 |
| 8 | 3827 | 202 | 0.05280 |
| 9 | 3558 | 40 | 0.01120 |
| 10 | 3017 | 5 | 0.00166 |
| 11 | 1339 | 2 | 0.00149 |

Mechanical Reliability Rates

| Mileage at test | N tested | Pct failed |
|---|---|---|
| 10000 | 646 | 3.87 |
| 20000 | 1672 | 5.86 |
| 30000 | 2369 | 5.70 |
| 40000 | 2695 | 7.57 |
| 50000 | 2806 | 7.38 |
| 60000 | 2836 | 9.98 |
| 70000 | 2636 | 11.00 |
| 80000 | 2356 | 13.80 |
| 90000 | 2040 | 14.80 |
| 100000 | 1634 | 15.30 |
| 110000 | 1194 | 16.40 |
| 120000 | 838 | 17.40 |
| 130000 | 545 | 16.50 |
| 140000 | 367 | 16.90 |
| 150000 | 254 | 20.10 |
| 160000 | 157 | 18.50 |
| 170000 | 110 | 22.70 |

**Audi Q5 2011**

At 5 years of age, the mortality rate of a Audi Q5 2011 (manufactured as a Car or Light Van) ranked number 85 out of 211 vehicles of the same age and type (any Car or Light Van constructed in 2011). One is the lowest (or best) and 211 the highest mortality rate. For vehicles reaching 20000 miles, its unreliability score (rate of failing an inspection) ranked 200 out of 205 vehicles of the same age, type, and mileage. One is the highest (or worst) and 205 the lowest rate of failing an inspection.

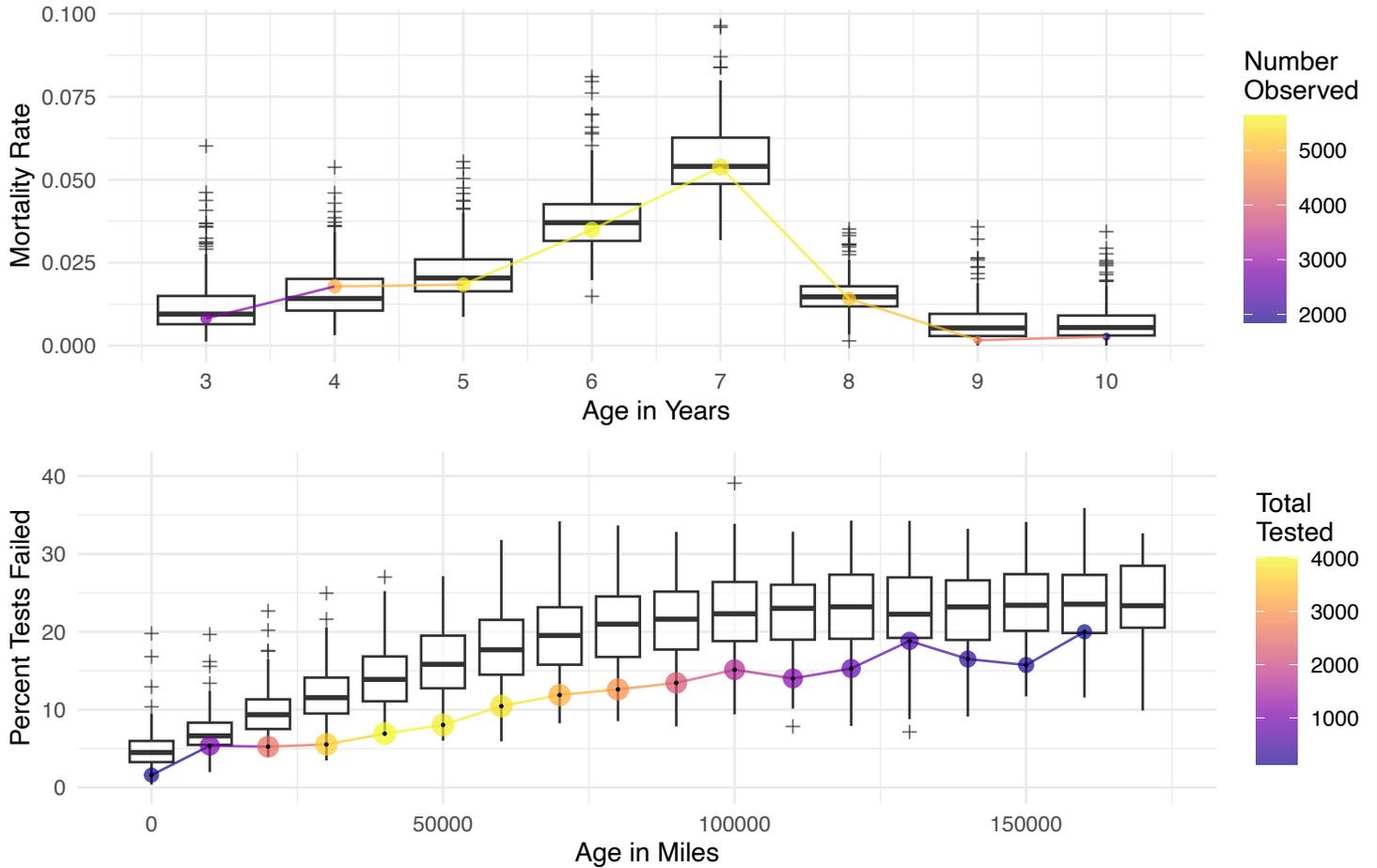

Mortality rates

| Age in Years | Observed | Died | Mortality Rate |
|---|---|---|---|
| 3 | 2821 | 23 | 0.00815 |
| 4 | 4930 | 88 | 0.01780 |
| 5 | 5554 | 102 | 0.01840 |
| 6 | 5639 | 197 | 0.03490 |
| 7 | 5523 | 297 | 0.05380 |
| 8 | 5102 | 72 | 0.01410 |
| 9 | 4231 | 7 | 0.00165 |
| 10 | 1850 | 5 | 0.00270 |

Mechanical Reliability Rates

| Mileage at test | N tested | Pct failed |
|---|---|---|
| 0 | 125 | 1.60 |
| 10000 | 1024 | 5.37 |
| 20000 | 2533 | 5.25 |
| 30000 | 3626 | 5.54 |
| 40000 | 4017 | 6.92 |
| 50000 | 3916 | 8.04 |
| 60000 | 3818 | 10.50 |
| 70000 | 3315 | 11.90 |
| 80000 | 2975 | 12.60 |
| 90000 | 2269 | 13.40 |
| 100000 | 1639 | 15.10 |
| 110000 | 1149 | 14.00 |
| 120000 | 785 | 15.30 |
| 130000 | 510 | 18.80 |
| 140000 | 303 | 16.50 |
| 150000 | 197 | 15.70 |
| 160000 | 125 | 20.00 |

## Audi Q5 2012

At 5 years of age, the mortality rate of a Audi Q5 2012 (manufactured as a Car or Light Van) ranked number 164 out of 212 vehicles of the same age and type (any Car or Light Van constructed in 2012). One is the lowest (or best) and 212 the highest mortality rate. For vehicles reaching 20000 miles, its unreliability score (rate of failing an inspection) ranked 205 out of 206 vehicles of the same age, type, and mileage. One is the highest (or worst) and 206 the lowest rate of failing an inspection.

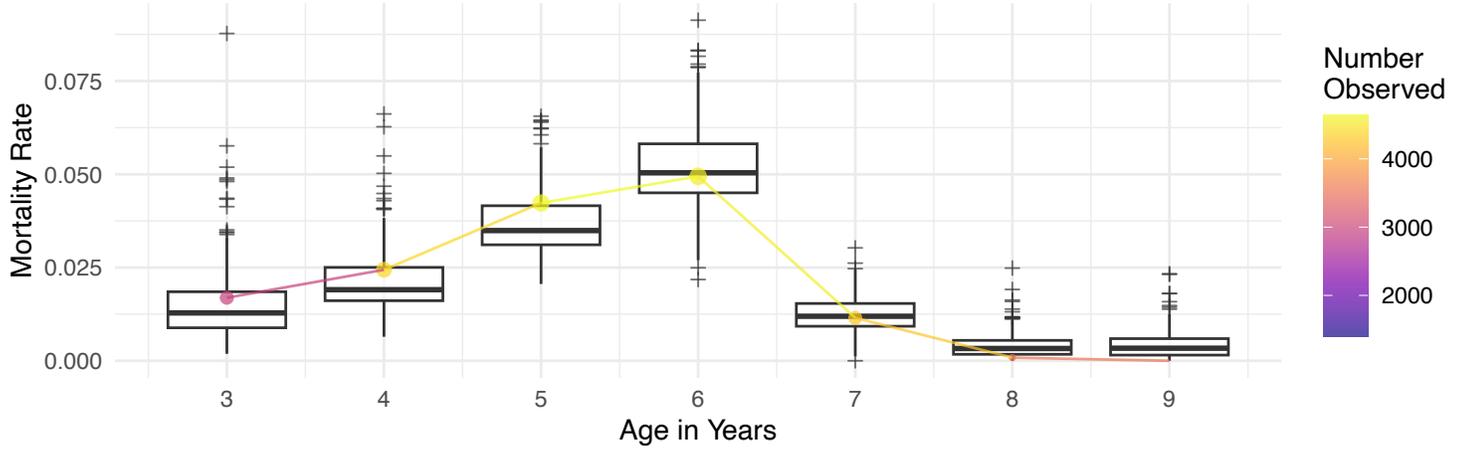

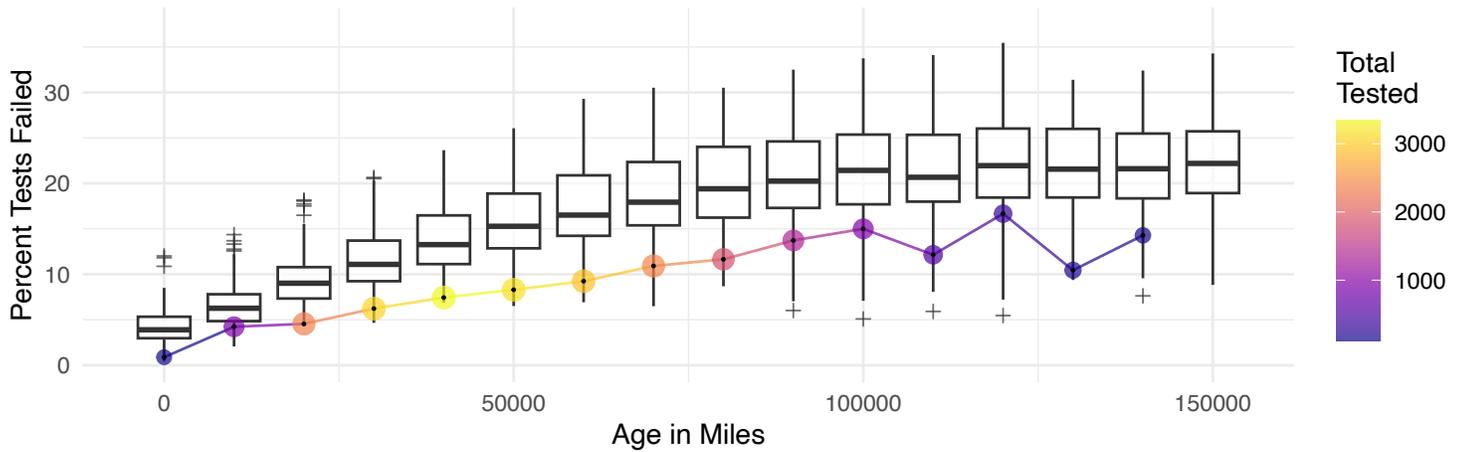

Mortality rates

| Age in Years | Observed | Died | Mortality Rate |
|---|---|---|---|
| 3 | 2958 | 50 | 0.016900 |
| 4 | 4385 | 107 | 0.024400 |
| 5 | 4633 | 196 | 0.042300 |
| 6 | 4551 | 225 | 0.049400 |
| 7 | 4254 | 49 | 0.011500 |
| 8 | 3504 | 3 | 0.000856 |
| 9 | 1401 | 0 | 0.000000 |

Mechanical Reliability Rates

| Mileage at test | N tested | Pct failed |
|---|---|---|
| 0 | 113 | 0.885 |
| 10000 | 922 | 4.230 |
| 20000 | 2351 | 4.550 |
| 30000 | 3082 | 6.230 |
| 40000 | 3346 | 7.440 |
| 50000 | 3181 | 8.300 |
| 60000 | 2931 | 9.250 |
| 70000 | 2400 | 10.900 |
| 80000 | 1854 | 11.700 |
| 90000 | 1355 | 13.700 |
| 100000 | 920 | 15.000 |
| 110000 | 535 | 12.100 |
| 120000 | 354 | 16.700 |
| 130000 | 201 | 10.400 |
| 140000 | 147 | 14.300 |

**Audi Q5 2013**

At 5 years of age, the mortality rate of a Audi Q5 2013 (manufactured as a Car or Light Van) ranked number 161 out of 221 vehicles of the same age and type (any Car or Light Van constructed in 2013). One is the lowest (or best) and 221 the highest mortality rate. For vehicles reaching 20000 miles, its unreliability score (rate of failing an inspection) ranked 203 out of 215 vehicles of the same age, type, and mileage. One is the highest (or worst) and 215 the lowest rate of failing an inspection.

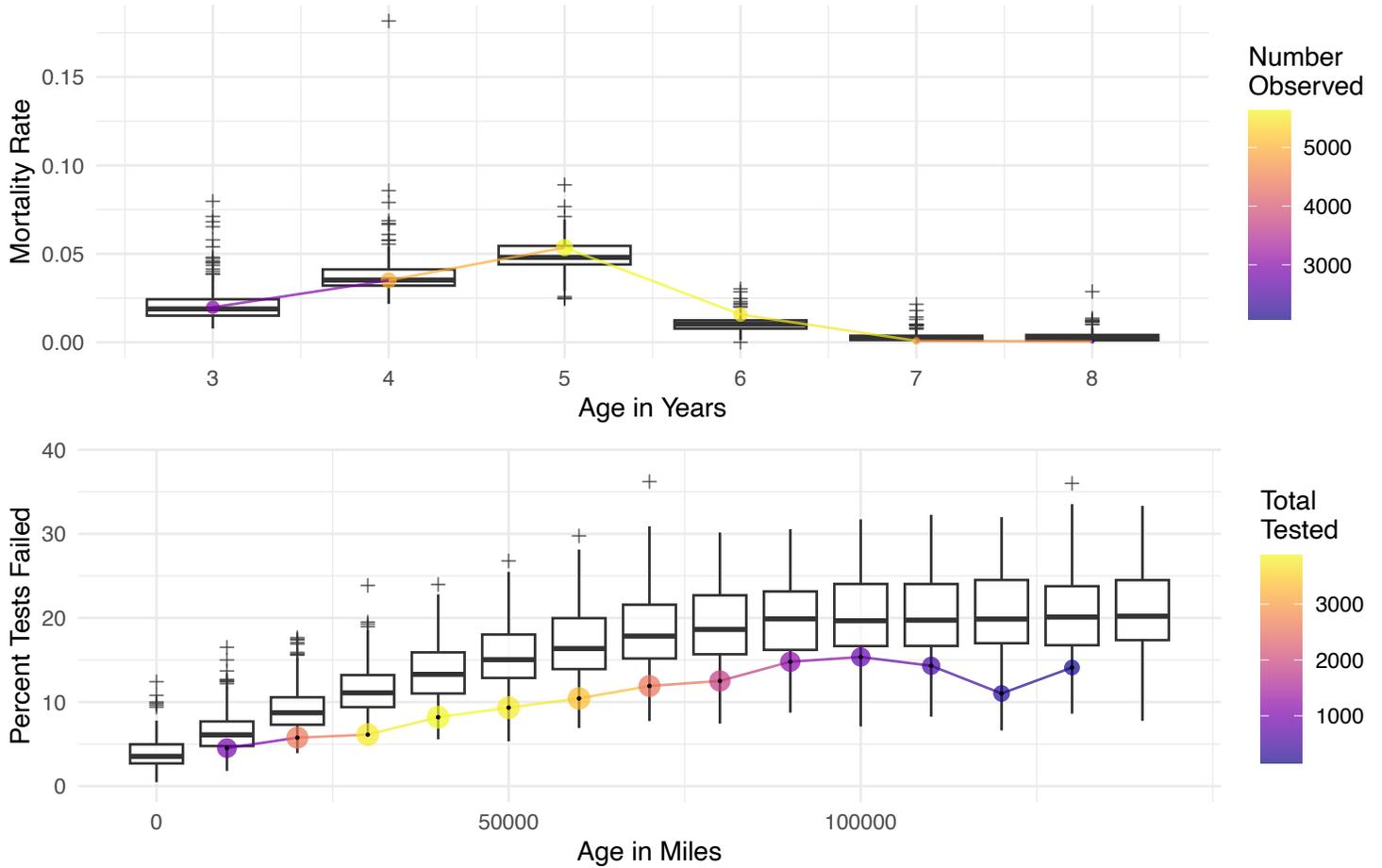

Mortality rates

| Age in Years | Observed | Died | Mortality Rate |
|---|---|---|---|
| 3 | 2980 | 59 | 0.019800 |
| 4 | 4966 | 174 | 0.035000 |
| 5 | 5622 | 301 | 0.053500 |
| 6 | 5473 | 86 | 0.015700 |
| 7 | 4668 | 4 | 0.000857 |
| 8 | 2073 | 1 | 0.000482 |

Mechanical Reliability Rates

| Mileage at test | N tested | Pct failed |
|---|---|---|
| 10000 | 950 | 4.53 |
| 20000 | 2601 | 5.77 |
| 30000 | 3756 | 6.12 |
| 40000 | 3880 | 8.20 |
| 50000 | 3770 | 9.34 |
| 60000 | 3355 | 10.40 |
| 70000 | 2561 | 11.90 |
| 80000 | 1871 | 12.50 |
| 90000 | 1210 | 14.80 |
| 100000 | 801 | 15.40 |
| 110000 | 503 | 14.30 |
| 120000 | 245 | 11.00 |
| 130000 | 156 | 14.10 |

**Audi Q5 2014**

At 5 years of age, the mortality rate of a Audi Q5 2014 (manufactured as a Car or Light Van) ranked number 125 out of 236 vehicles of the same age and type (any Car or Light Van constructed in 2014). One is the lowest (or best) and 236 the highest mortality rate. For vehicles reaching 120000 miles, its unreliability score (rate of failing an inspection) ranked 76 out of 113 vehicles of the same age, type, and mileage. One is the highest (or worst) and 113 the lowest rate of failing an inspection.

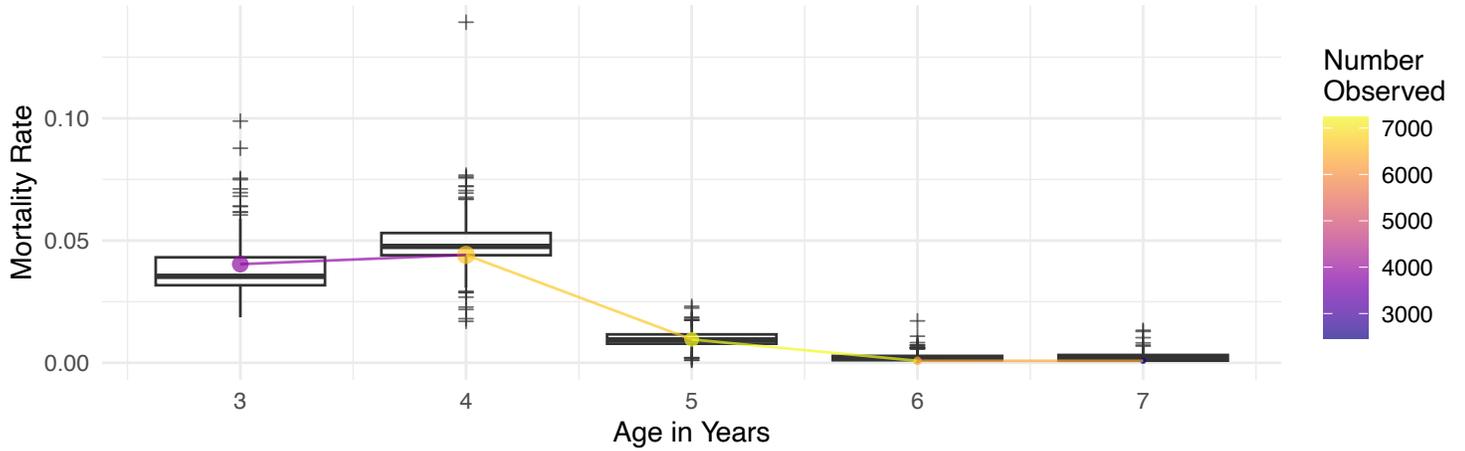

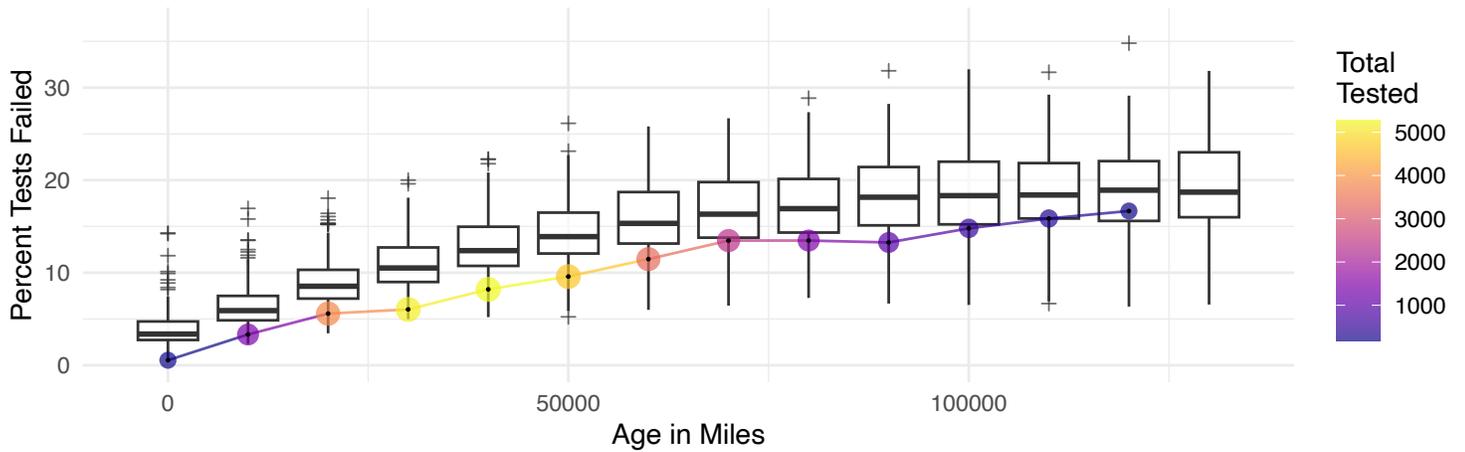

Mortality rates

| Age in Years | Observed | Died | Mortality Rate |
|---|---|---|---|
| 3 | 3890 | 157 | 0.040400 |
| 4 | 6687 | 295 | 0.044100 |
| 5 | 7225 | 69 | 0.009550 |
| 6 | 6269 | 5 | 0.000798 |
| 7 | 2466 | 2 | 0.000811 |

Mechanical Reliability Rates

| Mileage at test | N tested | Pct failed |
|---|---|---|
| 0 | 184 | 0.543 |
| 10000 | 1413 | 3.330 |
| 20000 | 3817 | 5.580 |
| 30000 | 5153 | 6.040 |
| 40000 | 5281 | 8.200 |
| 50000 | 4685 | 9.580 |
| 60000 | 3383 | 11.500 |
| 70000 | 2383 | 13.500 |
| 80000 | 1477 | 13.500 |
| 90000 | 927 | 13.300 |
| 100000 | 507 | 14.800 |
| 110000 | 309 | 15.900 |
| 120000 | 180 | 16.700 |

```
## Supplementary code for manuscript.

## Please note — much of the second half of the analysis will still
function
## if you have only the life table data. However, much of the
analysis will
## not be reproducible without the sensitive personal data  —
especially the
## construction of the mortality models. I have made these available
## within the limits that are permissible for sensitive government
data.

####################################################

## The Whip of Theseus project....

####################################################

## loads packages and data

require(doMC)
require(beepr)
require(ggplot2)
require(plyr)
require(rpart)
require(randomForest)
require(data.table)
require(fmsb)
require(readr)

require(rvest)
require(htmltools)

require(xml2)

setDTthreads(threads = detectCores()-2)

## loads the inferred models years for VINs. Check your paths,
obviously.
MY_inferred<-readRDS("/Users/sauley/Documents/R_workspaces/Analyses/
Automobile_survival_UK/data/MY_inferred1.rds")

###############################################################
#####

## Lists all the dirs. Obviously you have to designate your paths.
MOT_dirs<-list.dirs("/Volumes/PHOTO_DRIVE/Cars/unzipped_MOTs")
MOT_dirs<-MOT_dirs[-c(1)]

use_ncores<-detectCores()-1

ticktock<-Sys.time()
```

```r
for(i in 1:length(MOT_dirs)){

  ## list directories
  subvec2<-list.files(MOT_dirs[[i]])

  ## assembles a single-year tibble of locations, dates, fates
  YearTibble<-NULL
  counter_n<-NULL
  for(j in 1:length(subvec2)){

    ## reads in the DVSA data
    subframe<-read_delim(paste0(MOT_dirs[[i]],"/", subvec2[[j]]),
                         trim_ws = TRUE, name_repair = "minimal",
                         na = c("^\\?\\?$","^\\?$", "NA", "^\\*$"),
skip_empty_rows = TRUE,
                         num_threads = use_ncores)

    ## removes data we don't need right now
    subframe<-subframe[,c(colnames(subframe) %in%
c("unique_vehicle",
                                                        ## cause of
death
                                                        "is_exported",
"is_scrapped", "is_cod", # "is_prs",

"outer_postcode",

"odometer_reading_units","odometer_reading",
                                                        ## test date and
birthdate

"test_date","first_use_date", "age_at_test"))]

    ## corrects odometer readings to (shudder) miles
    corrected_odometer<-subframe$odometer_reading
    corrected_odometer[subframe$odometer_reading_units=="km"]<-
round(corrected_odometer[subframe$odometer_reading_units=="km"]*0.62
1371192)
    ## assumes missing data are in miles (>99% of measures are in
miles)
    corrected_odometer[subframe$odometer_reading_units=="??"]<-
round(corrected_odometer[subframe$odometer_reading_units=="??"]*0.62
1371192)

    mean(corrected_odometer<=0)

    ## makes zero/negative numbers into NAs
    corrected_odometer[which(corrected_odometer<=0)]<-NA

    ## Now removes/replaces unnecessary columns
    subframe$corrected_odometer<-corrected_odometer
    subframe<-subframe[,!(colnames(subframe) %in%
c("odometer_reading","odometer_reading_units"))]
```

```r
    ## condenses scrapped/exported/is_COD (certificate of
destruction) to one vector
    fate<-subframe$is_scrapped
    fate[which(fate==0)]<-"survived"
    fate[which(fate==1)]<-"scrapped"
    fate[which(subframe$is_exported==1)]<-"exported"
    fate[which(subframe$is_cod==1)]<-"cert_destroyed"

    ## Now removes/replaces unnecessary columns
    subframe$fate<-fate
    subframe<-subframe[,!(colnames(subframe) %in% c("is_exported",
"is_scrapped", "is_cod"))]

    ## masks fucked up dates/ages
    subframe$age_at_test[which(subframe$first_use_date==10000101)]<-
NA

subframe$first_use_date[which(subframe$first_use_date==10000101)]<-
NA

    YearTibble[[j]]<-subframe

    ## if successful, counts
    if(length(subframe)>0){counter_n<-c(counter_n, j)}

    print(j)
    print(max(subframe$test_date, na.rm = T))
    rm(subframe)

  }

  YearTibble<-ldply(YearTibble, as.data.frame)

  if(length(counter_n) != length(subvec2)){break()}

  write.csv(YearTibble, file = paste0("/Volumes/PHOTO_DRIVE/Cars/
LifeLocs/",

tail(unlist(strsplit(MOT_dirs[[i]], split = "/")), 1),
                                 "_frame.csv"))

  saveRDS(YearTibble, file =  paste0("/Volumes/PHOTO_DRIVE/Cars/
LifeLocs/",

tail(unlist(strsplit(MOT_dirs[[i]], split = "/")), 1),
                                 "_frame.rds"))

  ## saves a local copy: takes ~10Gb space
  saveRDS(YearTibble, file =  paste0("/Users/sauley/Documents/
R_workspaces/Analyses/Automobile_survival_UK/data/LifeLocs/",

tail(unlist(strsplit(MOT_dirs[[i]], split = "/")), 1),
```

```
                              "_frame.rds"))

  beep(2)
  print(Sys.time()-ticktock)
}

beep(8)
rm(YearTibble, fate,corrected_odometer)

##############################################

## now pulls up these data and keeps only the first and last
observation of each ID

##############################################

## Loads year tibble
MOT_dirs2<-list.files("/Users/sauley/Documents/R_workspaces/
Analyses/Automobile_survival_UK/data/LifeLocs/")

## keeps a count by type etc
N_total<-NULL
N_tables_by_type<-NULL

LifeStart<-NULL
ticktock<-Sys.time()

for(i in 1:length(MOT_dirs2)){

  Ytib<-readRDS(paste0("/Users/sauley/Documents/R_workspaces/
Analyses/Automobile_survival_UK/data/LifeLocs/",MOT_dirs2[[i]]))

  N_tables_by_type[[i]]<-table(Ytib$test_class)
  N_total<-c(N_total,nrow(Ytib))

  ## adds to the existing data table
  #Ytib<-rbind(Ytib, simpleLife)
  Ydate<-as.numeric(as.Date(Ytib$test_date, format = "%Y-%m-%d",
origin = "1800-01-01"))

  ## orders by unique ID, then by date of test, age at test, then
odometer reading,
  Ytib<-Ytib[order(Ytib$unique_vehicle, Ydate, Ytib$fate,
                   Ytib$age_at_test, Ytib$corrected_odometer,
decreasing = T),]

  ## and keeps only the first instance observed in this year
  First_Y<-Ytib[which(!duplicated(Ytib$unique_vehicle, fromLast =
T)),]

  ## cuts unnecessary guff
  First_Y<-First_Y[,c("unique_vehicle","first_use_date",
"test_date","corrected_odometer","outer_postcode")]
```

```
  rm(Ytib, Ydate)
  gc()
  ## attaches to other tables, eliminates any junk.

  ## keeps first instance of each ID only.
  if(i==1){
    LifeStart<-First_Y
  }

  if(i>1){
    ## adds any vehicles not already observed
    LifeStart<-rbind(LifeStart,
                     First_Y[which(!(First_Y$unique_vehicle %in%
LifeStart$unique_vehicle)),])
  }

  rm(First_Y)

  print(i)
  # print(dim(LifeStop))
  print(dim(LifeStart))
  beep(2)
  print(Sys.time()-ticktock)

}

N_tables_by_type<-ldply(N_tables_by_type, as.data.frame)

saveRDS(LifeStart, file = "/Users/sauley/Documents/R_workspaces/
Analyses/Automobile_survival_UK/data/LifeStart.rds")
rm(LifeStart)

## runs consecutively to avoid killing the nodes
LifeStop<-NULL
ticktock<-Sys.time()

for(i in 1:length(MOT_dirs2)){

  Ytib<-readRDS(paste0("/Users/sauley/Documents/R_workspaces/
Analyses/Automobile_survival_UK/data/LifeLocs/",MOT_dirs2[[i]]))

  ## adds to the existing data table
  #Ytib<-rbind(Ytib, simpleLife)
  Ydate<-as.numeric(as.Date(Ytib$test_date, format = "%Y-%m-%d",
origin = "1800-01-01"))

  ## orders by unique ID, then by date of test, age at test, fate,
and then odometer reading
  Ytib<-Ytib[order(Ytib$unique_vehicle, Ydate, Ytib$fate,
                   Ytib$age_at_test, Ytib$corrected_odometer,
decreasing = T),]

  ## ordering
  Last_Y<-Ytib[which(!duplicated(Ytib$unique_vehicle, fromLast =
```

```
F)),]
   Last_Y<-
Last_Y[,c("unique_vehicle","test_date","age_at_test","corrected_odom
eter", "fate")]

  rm(Ytib, Ydate)
  gc()
  ## attaches to other tables, eliminates any junk.

  ## keeps last instance of each ID
  if(i==1){
    LifeStop<-Last_Y
  }

  if(i>1){
    ## replaces any IDs observed again
    LifeStop<-rbind(LifeStop, Last_Y)
    ## the fromLast=T is all-important here
    LifeStop<-LifeStop[which(!duplicated(LifeStop$unique_vehicle,
fromLast = T)),]
  }

  rm(Last_Y)

  print(i)
  print(dim(LifeStop))
  #  print(dim(LifeStart))
  beep(2)
  print(Sys.time()-ticktock)

}

LifeStop<-as.data.table(LifeStop)

colnames(LifeStop)[2:4]<-paste0("Death_", c(colnames(LifeStop)
[2:4]))

saveRDS(LifeStop, file = "/Users/sauley/Documents/R_workspaces/
Analyses/Automobile_survival_UK/data/LifeStop.rds")
#rm(LifeStop)

##############################################################

## compiles vehicle stats, and a VIN table, for all unique IDs

##############################################################

## run once...

require(readr)
require(doMC)
require(beepr)
require(ggplot2)
require(plyr)
```

```r
## Lists all the dirs. Obviously you have to designate your paths.
MOT_dirs<-list.dirs("/Volumes/PHOTO_DRIVE/Cars/unzipped_MOTs")
MOT_dirs<-MOT_dirs[-c(1)]

use_ncores<-detectCores()-1

Ticker<-Sys.time()

for(i in 1:length(MOT_dirs)){

  ## list directories
  subvec2<-list.files(MOT_dirs[[i]])

  ## assembles into a single-year tibble
  YearTibble<-NULL
  counter_n<-NULL
  for(j in 1:length(subvec2)){

    ## reads in the DVSA data with make+model+engine+fuel+VIN
    subframe<-read_delim(paste0(MOT_dirs[[i]],"/", subvec2[[j]]),
                         trim_ws = TRUE, name_repair = "minimal",
                         na = c("^??$","^?$", "NA", "^\\*$"),
skip_empty_rows = TRUE,
                         num_threads = use_ncores)

    ## removes data we don't need right now
    subframe<-subframe[,c(colnames(subframe) %in%
c("unique_vehicle", "vin11",
                                              ## keeps types
of breakdowns
"result","is_prs", "test_class",
                                                "make", "model",
"engine_cc", "fuel_name"))]
    ## compresses test result into one vector
    subframe$test_result<-subframe$result
    subframe$test_result[subframe$is_prs=="PRS FAIL"]<-"PRS_FAIL"

    subframe<-subframe[,which(!(colnames(subframe) %in% c("result",
"is_prs")))]

    YearTibble[[j]]<-subframe

    ## if successful, counts
    if(length(subframe)>0){counter_n<-c(counter_n, j)}

    print(j)
    print(max(subframe$test_date, na.rm = T))
    rm(subframe)

  }
  YearTibble<-ldply(YearTibble, as.data.frame)
```

```
   ## saves a local copy: takes ~8Gb
   saveRDS(YearTibble, file =  paste0("/Users/sauley/Documents/
R_workspaces/Analyses/Automobile_survival_UK/data/Survival_data/",

tail(unlist(strsplit(MOT_dirs[[i]], split = "/")), 1),
                                     "_frame.rds"))

   beep(2)
   print(Sys.time()-Ticker)
}

beep(8)

################################################################
##

##  Now builds table of unique IDs, with some stats.
MOT_dirs3<-list.files("/Users/sauley/Documents/R_workspaces/
Analyses/Automobile_survival_UK/data/Survival_data/")
Vehicle_properties<-NULL
ticktock<-Sys.time()

for(i in 1:length(MOT_dirs3)){

   ## points to temp local folder
   Ytib<-readRDS(paste0("/Users/sauley/Documents/R_workspaces/
Analyses/Automobile_survival_UK/data/
Survival_data/",MOT_dirs3[[i]]))

   ## keeps non-duplicated instances
   Ytib<-Ytib[!duplicated(Ytib$unique_vehicle),]

   Vehicle_properties<-rbind(Vehicle_properties,
                             Ytib)

   Vehicle_properties<-Vehicle_properties[!
duplicated(Vehicle_properties$unique_vehicle),]

   print(i)
   beep(2)
   print(Sys.time()-ticktock)

}
dim(Vehicle_properties)

## Adds year first used
saveRDS(Vehicle_properties, file = "/Users/qtnzsne/Documents/
R_workspaces/Analyses/Automobile_survival_UK/data/Survival_data/
Vehicle_properties.rds")

rm(Vehicle_properties)
```

```
################################################################
##

## loads up the start/stop survival data, and builds lifetables by
VIN

################################################################
##

#LifeVec<-readRDS("/Users/sauley/Documents/R_workspaces/Analyses/
Automobile_survival_UK/data/LifeStart.rds")
LifeVec<-readRDS("/Users/qtnzsne/Documents/R_workspaces/Analyses/
Automobile_survival_UK/data/LifeStart.rds")

##
LifeStop<-readRDS("/Users/qtnzsne/Documents/R_workspaces/Analyses/
Automobile_survival_UK/data/LifeStop.rds")
#LifeStop<-readRDS("/Users/sauley/Documents/R_workspaces/Analyses/
Automobile_survival_UK/data/LifeStop.rds")

## aligns (faster than merge)
LifeVec<-LifeVec[order(LifeVec$unique_vehicle),]
LifeStop<-LifeStop[order(LifeStop$unique_vehicle),]

sum(LifeVec$unique_vehicle==LifeStop$unique_vehicle)
sum(LifeVec$unique_vehicle!=LifeStop$unique_vehicle)

if(sum(LifeVec$unique_vehicle!=LifeStop$unique_vehicle)>0)
{print("Screwup in alignment")}

LifeVec<-data.table(data.frame(LifeVec, LifeStop))
#LifeVec<-data.table(merge(LifeVec, LifeStop, by.x =
"unique_vehicle", by.y = "unique_vehicle", all = T))

set.seed(20983)
sampVec<-sample(seq(1,nrow(LifeVec)), size = 100000, replace = F)

rm(LifeStop)
gc()
## Splits into 'life tables by time' and 'life tables by physical
age (mileage)'
LifeVec_time<-
LifeVec[,c("unique_vehicle","first_use_date","test_date","Death_test
_date",
                    "Death_age_at_test", "fate")]

###############################

## passes some QC:

###############################
```

```r
## Calculates the -age- at which the vehicle was first observed
## to exclude immortal time bias

## relative to the age recorded by the DVSA

firstObs_days<-as.Date(as.character(LifeVec_time$test_date),
                       format = "%Y-%m-%d", origin = "1800-01-01")

firstObs_days<-as.numeric(firstObs_days-
as.Date(as.character(LifeVec_time$first_use_date),
                                         format = "%Y%m%d",
origin = "1800-01-01"))

## Logs some more strict QC measures
FUD<-as.numeric(as.Date(as.character(LifeVec$first_use_date),
                        format = "%Y%m%d", origin = "1800-01-01"))

DOD<-as.numeric(as.Date(LifeVec$Death_test_date,  origin =
"1800-01-01"))

## diff for FUD-DOD
Lifespan_diff<-c(DOD-FUD)/365.25
Lifespan_diff<-Lifespan_diff-LifeVec$Death_age_at_test

## How many are off by 3 months (includes cars re-registered later
on)
sum(abs(Lifespan_diff)>0.25, na.rm = T)
mean(abs(Lifespan_diff)>0.25, na.rm = T)

## flags QC fails for any vehicles that have a difference of 3
months (0.25 year) or more
## between DVLA lifespan and (first use date -  death test date)
LifeVec$HighQ<-rep(1, time = nrow(LifeVec))

LifeVec$HighQ[which(abs(Lifespan_diff)>0.25)]<-0
LifeVec$HighQ[which(is.na(Lifespan_diff))]<-0

rm(FUD, DOD,Lifespan_diff)

## adds a flag for any vehicles where they go for MOT before age
zero
LifeVec$HighQ[which(firstObs_days<0)]<-0

## Adds QC vector back to 'life tables by time'
LifeVec_time<-
LifeVec[,c("unique_vehicle","first_use_date","test_date","Death_test
_date",
                        "Death_age_at_test", "fate","HighQ")]

## Outputs the QC  vector for the emissions project.
```

```r
HighQ_ids<-data.frame(unique_vehicle = LifeVec_time$unique_vehicle,
                      HighQ = LifeVec_time$HighQ)

saveRDS(HighQ_ids, file = "data/Emissions_QC_list.rds")

## loads and aligns vehicle properties
Vehicle_properties<-readRDS(file = "/Users/sauley/Documents/
R_workspaces/Analyses/Automobile_survival_UK/data/Survival_data/
Vehicle_properties.rds")

sum(Vehicle_properties$unique_vehicle!=HighQ_ids$unique_vehicle)

Vehicle_properties<-
Vehicle_properties[order(Vehicle_properties$unique_vehicle),]

## makes a Make-Model index to remove Taxis for QC
MM_index<-paste0(toupper(Vehicle_properties$make),"__",
toupper(Vehicle_properties$model))

## removes Taxis from this vector, for the emissions mapping
pipeline
HighQ_ids_taxi_free<-HighQ_ids
## loads vehicle properties
HighQ_ids_taxi_free$HighQ[grep("TAXI",MM_index)]<-0

saveRDS(HighQ_ids_taxi_free, file = "data/
Emissions_QC_list_noTaxis.rds")

rm(HighQ_ids_taxi_free, Vehicle_properties,MM_index)
gc()

###########################

## Splits into 'life tables by time' and 'life tables by physical
age (mileage)'
LifeVec_miles<-LifeVec[,c("unique_vehicle","corrected_odometer",
                          "Death_corrected_odometer","fate",
"HighQ")]

rm(LifeVec)

## creates a 'missing presumed dead' value for anything not
## MOT-ed since 1/6/2019 (last eighteen months of records)
MPD<-LifeVec_time$Death_test_date<=as.numeric(as.Date("2019-06-01",
format = "%Y-%m-%d",  origin = "1800-01-01"))
## updates the 'fate' vector to include MPD
LifeVec_time$fate[which(MPD & LifeVec_time$fate=="survived")]<-"MPD"
LifeVec_miles$fate[which(MPD &
LifeVec_miles$fate=="survived")]<-"MPD"

## Keeps the 97.3% of IDs passing tight QC
HighQs<-LifeVec_time$unique_vehicle[which(LifeVec_time$HighQ==1)]

saveRDS(LifeVec_time, "data/LifeVec_time.rds")
```

```
saveRDS(LifeVec_miles, "data/LifeVec_miles.rds")

rm(LifeVec, LifeVec_miles)
gc()

rm(MPD)

table(LifeVec_time$fate)

Vehicle_properties<-readRDS("/Users/sauley/Documents/R_workspaces/
Analyses/Automobile_survival_UK/data/Survival_data/
Vehicle_properties.rds")

## lists top 1000 most common and models, by VIN
table(Vehicle_properties$vin11)

#Lead_vin11s<-head(table(Vehicle_properties$vin11)
[order(table(Vehicle_properties$vin11), decreasing = T)], 1000)

## keeps all with >1000 vehicles in the cohort
Lead_vin11s<-table(Vehicle_properties$vin11)
[table(Vehicle_properties$vin11)>=1000]

## cuts the ?? VIN
Lead_vin11s<-Lead_vin11s[-c(grep("\\?", names(Lead_vin11s)))]

## removes some junk VINs
Lead_vin11s<-Lead_vin11s[-c(grep("DVLASWA3971",
names(Lead_vin11s)))]
Lead_vin11s<-Lead_vin11s[-c(grep("GV73BXXXXXX",
names(Lead_vin11s)))]
Lead_vin11s<-Lead_vin11s[-c(grep("XXXXXXXXXXX",
names(Lead_vin11s)))]

Vin11_test_classes<-
table(Vehicle_properties$vin11[which(Vehicle_properties$vin11 %in%
names(Lead_vin11s))],
                              by =
Vehicle_properties$test_class[which(Vehicle_properties$vin11 %in%
names(Lead_vin11s))])

## and if we ignore the vehicle assembly plant?
vin10<-substr(Vehicle_properties$vin11, 1, 10)

head(table(vin10)[order(table(vin10), decreasing = T)], 1000)
Lead_vin10s<-head(table(vin10)[order(table(vin10), decreasing = T)],
5000)
Lead_vin10s<-table(vin10)[order(table(vin10), decreasing = T)]

## keeps all with >1000 vehicles in the cohort
Lead_vin10s<-table(vin10)[table(vin10)>=1000]

## removes crap
```

```
Lead_vin10s<-Lead_vin10s[which(names(Lead_vin10s)!="??")]
Lead_vin10s<-Lead_vin10s[-c(grep("DVLASWA3971",
names(Lead_vin10s)))]
Lead_vin10s<-Lead_vin10s[-c(grep("GV73BXXXXXX",
names(Lead_vin10s)))]
Lead_vin10s<-Lead_vin10s[-c(grep("XXXXXXXXXXX",
names(Lead_vin10s)))]

## orders by decreasing frequency
Lead_vin10s<-Lead_vin10s[order(Lead_vin10s, decreasing  =T)]
#rm(vin10)

##################################################################
##################

## Measures the failures per test at each age, for each MMY or VIN11

##################################################################
##################

## measures 'failures per test', or fpt, by age. For MMYs.
#rm(Vehicle_properties)

## retains only common MMYs (otherwise table is too massive + full
of sampling errors)
MOT_dirs3<-list.files("/Users/sauley/Documents/R_workspaces/
Analyses/Automobile_survival_UK/data/Survival_data/")
MOT_dirs3<-MOT_dirs3[-c(grep("Fail",MOT_dirs3)]]
MOT_dirs3<-MOT_dirs3[-c(grep("Vehicle",MOT_dirs3)]]

FPT_MMY<-NULL
FPT_MMY_miles<-NULL
FPT_MMY_age<-NULL

## keeps only data on failures for the first test of the year
## (retesting depends on not scrapping the thing after failure)
FPT_MMY_miles_firstTest<-NULL
FPT_MMY_age_firstTest<-NULL

ticktock<-Sys.time()
MMY_set<-MMY_index_1k

for(i in 1:length(MOT_dirs3)){

  ## loads one table...
  Ytib<-readRDS(paste0("/Users/sauley/Documents/R_workspaces/
Analyses/Automobile_survival_UK/data/LifeLocs/",MOT_dirs2[[i]]))

  ## keeps what we want..
  Ytib<-Ytib[,c(1,4,5,6)]
```

```
## loads the next...
Ytib2<-readRDS(paste0("/Users/sauley/Documents/R_workspaces/
Analyses/Automobile_survival_UK/data/
Survival_data/",MOT_dirs3[[i]]))

## checks...
if(sum(Ytib2$unique_vehicle!=Ytib$unique_vehicle)>0){
  break()
  print(paste0("stopped at ",i))
}

Ytib2<-cbind(Ytib,Ytib2[,-c(1)])
rm(Ytib)

## Cuts anything that fails QC
Ytib2<-Ytib2[which(Ytib2$unique_vehicle %in% HighQs),]

## gets the rounded-down age at test of each vehicle
## (floor() is how annual mortality risk is binned)
Ytib2$age_at_test_year<-floor(Ytib2$age_at_test)

## Also rounds down mileage to nearest 10k miles
Ytib2$mileage_10k<-c(floor(Ytib2$corrected_odometer/10000)*10000)

## makes MMY vector for matching

## converts ages in years to ages in days
AAT_days_Ysub<-ceiling(Ytib2$age_at_test*365.25)

## subtracts this from the test date
Birthday_est_Ysub<-c(Ytib2$test_date-AAT_days_Ysub)
## and keeps the year
Birthday_est_Ysub<-year(Birthday_est_Ysub)

## gets the make-model- or make-model-year index
Ytib2$MMY<-paste(Ytib2$make, Ytib2$model, Birthday_est_Ysub, sep =
"__")
rm(Birthday_est_Ysub, AAT_days_Ysub)

## cuts dates and non-indexed mmy data
Ytib2<-Ytib2[,-c(5:10)]

## slims down MMYs
Ytib<-Ytib2[which(Ytib2$MMY %in% MMY_set),]
Ytib2<-Ytib2[which(Ytib2$MMY %in% MMY_set),]

## cuts anyone over a million miles
Ytib2<-Ytib2[which(Ytib2$mileage_10k<=1e+06),]

## and over 110 years
Ytib<-Ytib[which(Ytib$age_at_test_year<=110),]
## or less than -2
Ytib<-Ytib[which(Ytib$age_at_test_year>=(-2)),]
```

```
  VIN_risk_age_all<-table(Ytib$age_at_test_year[!
is.na(Ytib$age_at_test_year)],
                          by = Ytib$test_result[!
is.na(Ytib$age_at_test_year)])

  VIN_risk_miles_all<-table(Ytib2$mileage_10k[!
is.na(Ytib2$mileage_10k)],
                            by = Ytib2$test_result[!
is.na(Ytib2$mileage_10k)])

  VIN_risk_age_all<-reshape(data.frame(VIN_risk_age_all), direction
= "wide",
                            idvar = "Var1", timevar = "by")

  ## risk by MMY + age

  ## Now does this by MMy
  VIN_risk_age<-table(paste(Ytib$MMY[!is.na(Ytib$age_at_test_year)],
                            Ytib$age_at_test_year[!
is.na(Ytib$age_at_test_year)], sep = "__"),
                      by = Ytib$test_result[!
is.na(Ytib$age_at_test_year)])

  VIN_risk_age<-reshape(data.frame(VIN_risk_age), direction =
"wide",
                        idvar = "Var1", timevar = "by")

  ## Keeps
  FPT_MMY_age[[i]]<-VIN_risk_age

  ## Now does this by MMY + mileage (binned)
  VIN_risk_miles<-table(paste(Ytib2$MMY[!is.na(Ytib2$mileage_10k)],
                              Ytib2$mileage_10k[!
is.na(Ytib2$mileage_10k)], sep = "__"),
                        by = Ytib2$test_result[!
is.na(Ytib2$mileage_10k)])

  VIN_risk_miles<-reshape(data.frame(VIN_risk_miles), direction =
"wide",
                          idvar = "Var1", timevar = "by")

  ## Keeps
  FPT_MMY_miles[[i]]<-VIN_risk_miles

  ## re-aggregates, using only the first test of the year
  Ytib<-Ytib[order(Ytib$age_at_test,Ytib$corrected_odometer),]
  Ytib<-Ytib[which(!duplicated(Ytib$unique_vehicle)),]

  Ytib2<-Ytib2[order(Ytib2$corrected_odometer, Ytib2$age_at_test),]
```

```
  Ytib2<-Ytib2[which(!duplicated(Ytib2$unique_vehicle)),]

  VIN_risk_miles<-NULL
  VIN_risk_age<-NULL

  ## Now does this by VIN
  VIN_risk_age<-table(paste(Ytib$MMY[!is.na(Ytib$age_at_test_year)],
                            Ytib$age_at_test_year[!
is.na(Ytib$age_at_test_year)], sep = "__"),
                      by = Ytib$test_result[!
is.na(Ytib$age_at_test_year)])

  VIN_risk_age<-reshape(data.frame(VIN_risk_age), direction =
"wide",
                        idvar = "Var1", timevar = "by")

  ## Keeps
  FPT_MMY_age_firstTest[[i]]<-VIN_risk_age

  ## Now does this by VIN + mileage (binned)
  VIN_risk_miles<-table(paste(Ytib2$MMY[!is.na(Ytib2$mileage_10k)],
                              Ytib2$mileage_10k[!
is.na(Ytib2$mileage_10k)], sep = "__"),
                        by = Ytib2$test_result[!
is.na(Ytib2$mileage_10k)])

  VIN_risk_miles<-reshape(data.frame(VIN_risk_miles), direction =
"wide",
                          idvar = "Var1", timevar = "by")

  ## Keeps
  FPT_MMY_miles_firstTest[[i]]<-VIN_risk_miles

  ## Cleans workspace
  VIN_risk_miles<-NULL
  VIN_risk_age<-NULL
  rm(Ytib, Ytib2)

  beep()
  print(i)
}

FPT_MMY_miles<-ldply(FPT_MMY_miles,as.data.frame)
FPT_MMY_age<-ldply(FPT_MMY_age,as.data.frame)

FPT_MMY_miles_firstTest<-
ldply(FPT_MMY_miles_firstTest,as.data.frame)
FPT_MMY_age_firstTest<-ldply(FPT_MMY_age_firstTest,as.data.frame)
```

```
saveRDS(FPT_MMY_miles_firstTest, "data/FPT_MMY_miles_firstTest.rds")
saveRDS(FPT_MMY_age_firstTest, "data/FPT_MMY_age_firstTest.rds")

rm(FPT_MMY_miles_firstTest,FPT_MMY_age_firstTest)

## Aggregates repeated estimates for the same VIN + age or VIN +
mileage
FPT_MMY_age<-data.frame(N_passed =
aggregate(FPT_MMY_age$Freq.PASSED,
                                                       by =
list(FPT_MMY_age$Var1), sum, na.rm = T),
                        N_failed =
c(aggregate(FPT_MMY_age$Freq.FAILED,
                                                       by =
list(FPT_MMY_age$Var1), sum, na.rm = T)[,2]),
                        N_prs_failed =
c(aggregate(FPT_MMY_age$Freq.PRS_FAIL,
                                                          by =
list(FPT_MMY_age$Var1), sum, na.rm = T)[,2]))

colnames(FPT_MMY_age)[1:2]<-c("Var1", "N_passed")

## resplits mileage and MMY and makes columns
Agemarker<-lapply(as.character(FPT_MMY_age$Var1), strsplit, split =
"__")
Agemarker<-lapply(Agemarker, unlist)

FPT_MMY_age$MMY<-lapply(seq(1, length(Agemarker)), function(i)
paste(Agemarker[i][[1]][[1]],

Agemarker[i][[1]][[2]],

Agemarker[i][[1]][[3]], sep = "__"))

#FPT_MMY_age$MMY<-unlist(lapply(Agemarker, head, 1))
FPT_MMY_age$age_at_test_year<-unlist(lapply(Agemarker, tail, 1))
FPT_MMY_age$age_at_test_year<-
as.numeric(FPT_MMY_age$age_at_test_year)

FPT_MMY_miles<-data.frame(N_passed =
aggregate(FPT_MMY_miles$Freq.PASSED,
                                                         by =
list(FPT_MMY_miles$Var1), sum, na.rm = T),
                          N_failed =
c(aggregate(FPT_MMY_miles$Freq.FAILED,
                                                           by =
list(FPT_MMY_miles$Var1), sum, na.rm = T)[,2]),
                          N_prs_failed =
c(aggregate(FPT_MMY_miles$Freq.PRS_FAIL,
                                                            by =
list(FPT_MMY_miles$Var1), sum, na.rm = T)[,2]))
colnames(FPT_MMY_miles)[1:2]<-c("Var1", "N_passed")
```

```
## resplits mileage and MMY and makes columns
Milemarker<-lapply(as.character(FPT_MMY_miles$Var1), strsplit, split
= "__")
Milemarker<-lapply(Milemarker, unlist)

FPT_MMY_miles$MMY<-lapply(seq(1, length(Milemarker)), function(i)
paste(Milemarker[i][[1]][[1]],

Milemarker[i][[1]][[2]],

Milemarker[i][[1]][[3]], sep = "__"))

#FPT_MMY_miles$MMY<-unlist(lapply(Milemarker, head, 1))
FPT_MMY_miles$miles<-unlist(lapply(Milemarker, tail, 1))
FPT_MMY_miles$miles<-as.numeric(FPT_MMY_miles$miles)

rm(Milemarker)

## adds total testing numbers
FPT_MMY_age$N_tests<-
c(FPT_MMY_age$N_passed+FPT_MMY_age$N_failed+FPT_MMY_age$N_prs_failed
)
FPT_MMY_miles$N_tests<-
c(FPT_MMY_miles$N_passed+FPT_MMY_miles$N_failed+FPT_MMY_miles$N_prs_
failed)

saveRDS(FPT_MMY_miles, "data/FPT_MMY_miles_clean.rds")
saveRDS(FPT_MMY_age, "data/FPT_MMY_age_clean.rds")

################################################################
###############
################################################################
###############

################################################################
###############

## measures 'failures per test', or fpt, by age, for VIN11s
rm(Vehicle_properties)

## retains only common VINs (otherwise table is too massive + full
of sampling errors)
MOT_dirs3<-list.files("/Users/sauley/Documents/R_workspaces/
Analyses/Automobile_survival_UK/data/Survival_data/")
MOT_dirs3<-MOT_dirs3[-c(grep("Fail",MOT_dirs3))]
MOT_dirs3<-MOT_dirs3[-c(grep("Vehicle",MOT_dirs3))]

FPT_vin11<-NULL
FPT_vin11_miles<-NULL
```

```
FPT_vin11_age<-NULL

## keeps only data on failures for the first test of the year
## (retesting depends on not scrapping the thing after failure)
FPT_vin11_miles_firstTest<-NULL
FPT_vin11_age_firstTest<-NULL

ticktock<-Sys.time()
vin_set<-names(Lead_vin11s)

for(i in 1:length(MOT_dirs3)){

  ## loads one table...
  Ytib<-readRDS(paste0("/Users/sauley/Documents/R_workspaces/
Analyses/Automobile_survival_UK/data/LifeLocs/",MOT_dirs2[[i]]))

  ## keeps what we want..
  Ytib<-Ytib[,c(1,5,6)]

  ## loads the next...
  Ytib2<-readRDS(paste0("/Users/sauley/Documents/R_workspaces/
Analyses/Automobile_survival_UK/data/
Survival_data/",MOT_dirs3[[i]]))

  ## checks...
  if(sum(Ytib2$unique_vehicle!=Ytib$unique_vehicle)>0){
    break()
    print(paste0("stopped at ",i))
  }

  Ytib2<-cbind(Ytib,Ytib2[,-c(1)])
  rm(Ytib)

  ## Cuts anything that fails QC
  Ytib2<-Ytib2[which(Ytib2$unique_vehicle %in% HighQs),]

  ## gets the rounded-down age at test of each vehicle
  ## (floor() is how annual mortality risk is binned)
  Ytib2$age_at_test_year<-floor(Ytib2$age_at_test)

  ## Also rounds down mileage to nearest 10k miles
  Ytib2$mileage_10k<-c(floor(Ytib2$corrected_odometer/10000)*10000)

  ## slims down vin11s
  Ytib<-Ytib2[which(Ytib2$vin11 %in% vin_set),]
  Ytib2<-Ytib2[which(Ytib2$vin11 %in% vin_set),]

  ## cuts anyone over a million miles
  Ytib2<-Ytib2[which(Ytib2$mileage_10k<=1e+06),]

  ## and over 110 years
  Ytib<-Ytib[which(Ytib$age_at_test_year<=110),]
  ## or less than -2
  Ytib<-Ytib[which(Ytib$age_at_test_year>=(-2)),]
```

```
  VIN_risk_age_all<-table(Ytib$age_at_test_year[!
is.na(Ytib$age_at_test_year)],
                          by = Ytib$test_result[!
is.na(Ytib$age_at_test_year)])

  VIN_risk_miles_all<-table(Ytib2$mileage_10k[!
is.na(Ytib2$mileage_10k)],
                            by = Ytib2$test_result[!
is.na(Ytib2$mileage_10k)])

  VIN_risk_age_all<-reshape(data.frame(VIN_risk_age_all), direction
= "wide",
                            idvar = "Var1", timevar = "by")

  ## risk by VIN + age

  ## Now does this by VIN
  VIN_risk_age<-table(paste(Ytib$vin11[!
is.na(Ytib$age_at_test_year)],
                            Ytib$age_at_test_year[!
is.na(Ytib$age_at_test_year)], sep = "__"),
                      by = Ytib$test_result[!
is.na(Ytib$age_at_test_year)])

  VIN_risk_age<-reshape(data.frame(VIN_risk_age), direction =
"wide",
                        idvar = "Var1", timevar = "by")

  ## Keeps
  FPT_vin11_age[[i]]<-VIN_risk_age

  ## Now does this by VIN + mileage (binned)
  VIN_risk_miles<-table(paste(Ytib2$vin11[!
is.na(Ytib2$mileage_10k)],
                              Ytib2$mileage_10k[!
is.na(Ytib2$mileage_10k)], sep = "__"),
                        by = Ytib2$test_result[!
is.na(Ytib2$mileage_10k)])

  VIN_risk_miles<-reshape(data.frame(VIN_risk_miles), direction =
"wide",
                          idvar = "Var1", timevar = "by")

  ## Keeps
  FPT_vin11_miles[[i]]<-VIN_risk_miles

  ## re-aggregates, using only the first test of the year
  Ytib<-Ytib[order(Ytib$age_at_test,Ytib$corrected_odometer),]
  Ytib<-Ytib[which(!duplicated(Ytib$unique_vehicle)),]
```

```
    Ytib2<-Ytib2[order(Ytib2$corrected_odometer, Ytib2$age_at_test),]
    Ytib2<-Ytib2[which(!duplicated(Ytib2$unique_vehicle)),]

    VIN_risk_miles<-NULL
    VIN_risk_age<-NULL

    ## Now does this by VIN
    VIN_risk_age<-table(paste(Ytib$vin11[!
is.na(Ytib$age_at_test_year)],
                              Ytib$age_at_test_year[!
is.na(Ytib$age_at_test_year)], sep = "__"),
                        by = Ytib$test_result[!
is.na(Ytib$age_at_test_year)])

    VIN_risk_age<-reshape(data.frame(VIN_risk_age), direction =
"wide",
                          idvar = "Var1", timevar = "by")

    ## Keeps
    FPT_vin11_age_firstTest[[i]]<-VIN_risk_age

    ## Now does this by VIN + mileage (binned)
    VIN_risk_miles<-table(paste(Ytib2$vin11[!
is.na(Ytib2$mileage_10k)],
                                Ytib2$mileage_10k[!
is.na(Ytib2$mileage_10k)], sep = "__"),
                          by = Ytib2$test_result[!
is.na(Ytib2$mileage_10k)])

    VIN_risk_miles<-reshape(data.frame(VIN_risk_miles), direction =
"wide",
                            idvar = "Var1", timevar = "by")

    ## Keeps
    FPT_vin11_miles_firstTest[[i]]<-VIN_risk_miles

    ## Cleans workspace
    VIN_risk_miles<-NULL
    VIN_risk_age<-NULL
    rm(Ytib, Ytib2)

    beep()
    print(i)
}

FPT_vin11_miles<-ldply(FPT_vin11_miles,as.data.frame)
FPT_vin11_age<-ldply(FPT_vin11_age,as.data.frame)
```

```
FPT_vin11_miles_firstTest<-
ldply(FPT_vin11_miles_firstTest,as.data.frame)
FPT_vin11_age_firstTest<-
ldply(FPT_vin11_age_firstTest,as.data.frame)

saveRDS(FPT_vin11_miles_firstTest, "data/
FPT_vin11_miles_firstTest.rds")
saveRDS(FPT_vin11_age_firstTest, "data/FPT_vin11_age_firstTest.rds")

rm(FPT_vin11_miles_firstTest,FPT_vin11_age_firstTest)

## Aggregates repeated estimates for the same VIN + age or VIN +
mileage
FPT_vin11_age<-data.frame(N_passed =
aggregate(FPT_vin11_age$Freq.PASSED,
                                              by =
list(FPT_vin11_age$Var1), sum, na.rm = T),
                          N_failed =
c(aggregate(FPT_vin11_age$Freq.FAILED,
                                              by =
list(FPT_vin11_age$Var1), sum, na.rm = T)[,2]),
                          N_prs_failed =
c(aggregate(FPT_vin11_age$Freq.PRS_FAIL,
                                              by =
list(FPT_vin11_age$Var1), sum, na.rm = T)[,2]))

colnames(FPT_vin11_age)[1:2]<-c("Var1", "N_passed")

## resplits mileage and vin and makes columns
Agemarker<-lapply(as.character(FPT_vin11_age$Var1), strsplit, split
= "__")
Agemarker<-lapply(Agemarker, unlist)

FPT_vin11_age$vin11<-unlist(lapply(Agemarker, head, 1))
FPT_vin11_age$age_at_test_year<-unlist(lapply(Agemarker, tail, 1))
FPT_vin11_age$age_at_test_year<-
as.numeric(FPT_vin11_age$age_at_test_year)

FPT_vin11_miles<-data.frame(N_passed =
aggregate(FPT_vin11_miles$Freq.PASSED,
                                              by =
list(FPT_vin11_miles$Var1), sum, na.rm = T),
                          N_failed =
c(aggregate(FPT_vin11_miles$Freq.FAILED,
                                              by =
list(FPT_vin11_miles$Var1), sum, na.rm = T)[,2]),
                          N_prs_failed =
c(aggregate(FPT_vin11_miles$Freq.PRS_FAIL,
                                              by =
list(FPT_vin11_miles$Var1), sum, na.rm = T)[,2]))
colnames(FPT_vin11_miles)[1:2]<-c("Var1", "N_passed")
```

```
## resplits mileage and vin and makes columns
Milemarker<-lapply(as.character(FPT_vin11_miles$Var1), strsplit,
split = "__")
Milemarker<-lapply(Milemarker, unlist)

FPT_vin11_miles$vin11<-unlist(lapply(Milemarker, head, 1))
FPT_vin11_miles$miles<-unlist(lapply(Milemarker, tail, 1))
FPT_vin11_miles$miles<-as.numeric(FPT_vin11_miles$miles)

rm(Milemarker)

## adds total testing numbers
FPT_vin11_age$N_tests<-
c(FPT_vin11_age$N_passed+FPT_vin11_age$N_failed+FPT_vin11_age$N_prs_
failed)
FPT_vin11_miles$N_tests<-
c(FPT_vin11_miles$N_passed+FPT_vin11_miles$N_failed+FPT_vin11_miles$
N_prs_failed)

saveRDS(FPT_vin11_miles, "data/FPT_vin11_miles_clean.rds")
saveRDS(FPT_vin11_age, "data/FPT_vin11_age_clean.rds")

################################################################
####
################################################################
####

## Makes a life table of everything, subdivided by test class

################################################################
####
################################################################
####

## gets the unique IDs we want
sub_index<-
Vehicle_properties$unique_vehicle[which(Vehicle_properties$test_clas
s==4)]

## throws out low-quality
sub_index<-sub_index[which(sub_index %in% HighQs)]

## builds a data frame from these
sub_frame<-LifeVec_time[which(LifeVec_time$unique_vehicle %in%
sub_index),]
## builds a life table for chronological time

## builds a vector of vehicles at risk + deaths
mx_vec<-NULL
age_vec<-c(0:60)
dx_vec<-NULL
```

```r
expo_vec<-NULL
censored_x<-NULL
scraps<-NULL
exports<-NULL

for(i in age_vec){

  ## ignores any vehicles first observed after the start of
observation period.
  sub_frame2<-sub_frame[which(sub_frame$firstObs_years <= i),]

  ## how many of these went to scrappage/export/destruction *during*
this period?
  lil_fates<-sub_frame2$fate[which(sub_frame2$Death_age_at_test>=i &
                                   sub_frame2$Death_age_at_test
<c(i+1))]
  exported_sum<-0
  exported_sum<-sum(lil_fates=="exported")

  scraps<-c(scraps,sum(lil_fates %in%
c("scrapped","cert_destroyed")))
  exports<-c(exports,exported_sum)

 Non_exported<-c(length(sub_frame2$fate)-
sum(lil_fates=="exported"))

  ## If all were exported (rare), fills with NA
  if(Non_exported==0){
    censored_x<-c(censored_x, NA)
    mx_vec<-c(mx_vec,NA)
    expo_vec<-c(expo_vec, NA)
    dx_vec<-c(dx_vec, NA)
  }

  if(Non_exported > 0){
    ## removes out-migrants during period from calculations of
exposure to/observation of mortality risk
    #   sub_frame2<-sub_frame2[-
c(which(sub_frame2$Death_age_at_test>=i &
    #                              sub_frame2$Death_age_at_test
<c(i+1) &
    #                              sub_frame2$fate=="exported")),]

    ## counts all those dying after the observation period starts
(the pop exposed to risk)
    # at_risk<-sum(Age_at_death>=floor(i*365.25), na.rm = T)
    #  at_risk<-sum(c(sub_frame$firstObs_years)>=i, na.rm = T)
    #  at_risk<-sum(sub_frame2$Death_age_at_test>=i, na.rm = T)

    at_risk<-c(sum(sub_frame2$Death_age_at_test>=i, na.rm = T)-
exported_sum)

    ## how many deaths during the observation period?
    died<-c(length(which(sub_frame2$fate!="survived" &
```

```
                          sub_frame2$Death_age_at_test>=i
                        & sub_frame2$Death_age_at_test <c(i+1)))-
exported_sum)

    #    died<-sum(c(Age_at_death[which(sub_frame2$fate!
="survived")] %in% c(floor(i*365.25):c(floor((i+1)*365.25)-1))),
na.rm = T)

    censored_x<-c(censored_x, c(nrow(sub_frame)-nrow(sub_frame2))-
exported_sum)
    mx_vec<-c(mx_vec,c(died/at_risk))
    expo_vec<-c(expo_vec, at_risk)
    dx_vec<-c(dx_vec, died)
  }

}

qx_vec<-mxtoqx(mx_vec,ax = 0.5, n = 1)

###############################################################
##################

## Builds lifetables by VIN

###############################################################
##################

Vehicle_properties<-readRDS("/Users/sauley/Documents/R_workspaces/
Analyses/Automobile_survival_UK/data/Survival_data/
Vehicle_properties.rds")
LifeVec_time<-readRDS("data/LifeVec_time.rds")
## Filters out low-QC vehicles
#Vehicle_properties<-
Vehicle_properties[which(Vehicle_properties$unique_vehicle %in%
HighQs),]

Lifetab_v11s<-NULL
Lifetab_v11s_stats<-NULL
ticktock<-Sys.time()

for(j in 1:length(Lead_vin11s)){

  ## gets the unique IDs we want
  sub_index<-
Vehicle_properties$unique_vehicle[which(Vehicle_properties$vin11==na
mes(Lead_vin11s)[[j]])]

  ## builds a data frame from these
  sub_frame<-LifeVec_time[which(LifeVec_time$unique_vehicle %in%
sub_index),]

  ## adds test classes
  # sub_frame$test_class<-
```

```
Vehicle_properties$test_class[which(Vehicle_properties$vin11==names(
Lead_vin11s)[[j]])]

  ## removes the QC fails
  sub_frame<-sub_frame[which(sub_frame$HighQ==1),]

  ## fixes the date formats to numeric, origin 1/1/1800
  sub_frame$first_use_date<-
as.Date(as.character(sub_frame$first_use_date),
                                  format = "%Y%m%d", origin =
"1800-01-01")
  sub_frame$test_date<-as.Date(as.character(sub_frame$test_date),
                            format = "%Y-%m-%d", origin =
"1800-01-01")
  sub_frame$Death_test_date<-
as.numeric(as.Date(sub_frame$Death_test_date,
                                          origin =
"1800-01-01"))

  ## Counts number of vehicles with >6 month difference between
stated age at death
  ## and the age at death estimated from testing data alone
  N_death_gaps<-sum(abs(c(sub_frame$Death_test_date-
                          as.numeric(sub_frame$first_use_date))/
365.25-
                        sub_frame$Death_age_at_test)>0.5, na.rm
=T)

  ## how many days between the first and last test?
  ObsTime<-c(sub_frame$Death_test_date-
              as.numeric(sub_frame$test_date))

  ## subtracts this from age at death to get age at first
observation in years
  sub_frame$firstObs_years<-c(sub_frame$Death_age_at_test-c(ObsTime/
365.25))

  ## keeps some numbers
  Lifetab_v11s_stats[[j]]<-c(table(sub_frame$fate), N_death_gaps)

  ## Gets age at death/ltf
  sub_frame$Age_at_death<-sub_frame$Death_test_date-
as.numeric(sub_frame$first_use_date)

  ## screens out any vehicles whose first use date is > 4 years
after the median
  ## first use date of the make+model (excludes mostly imports,
errors)
  ## NB this doesn't make sense for long production runs with the
  ## same VIN for non-11 character VINS. e.g. LADA has a 40 year
production run of the same make+model
  #  N_late_regs<-
```

```
sum(sub_frame$first_use_date>median(sub_frame$first_use_date, na.rm
= T)+(365*4), na.rm = T)
  #  Pct_late_regs<-
mean(sub_frame$first_use_date>median(sub_frame$first_use_date, na.rm
= T)+(365*4), na.rm = T)*100

  ## builds a life table for chronological time

  ## builds a vector of vehicles at risk + deaths
  mx_vec<-NULL
  age_vec<-c(0:60)
  dx_vec<-NULL
  expo_vec<-NULL
  censored_x<-NULL
  scraps<-NULL
  exports<-NULL

  for(i in age_vec){

    ## ignores any vehicles first observed after the start of
observation period.
    sub_frame2<-sub_frame[which(sub_frame$firstObs_years <= i),]

    ## how many of these went to scrappage/export/destruction
*during* this period?
    lil_fates<-sub_frame2$fate[which(sub_frame2$Death_age_at_test>=i
&
                                     sub_frame2$Death_age_at_test
<c(i+1))]
    exported_sum<-0
    exported_sum<-sum(lil_fates=="exported")

    scraps<-c(scraps,sum(lil_fates %in%
c("scrapped","cert_destroyed")))
    exports<-c(exports,exported_sum)

    Non_exported<-c(length(sub_frame2$fate)-
sum(lil_fates=="exported"))

    ## If all were exported (rare), fills with NA
    if(Non_exported==0){
      censored_x<-c(censored_x, NA)
      mx_vec<-c(mx_vec,NA)
      expo_vec<-c(expo_vec, NA)
      dx_vec<-c(dx_vec, NA)
    }

    if(Non_exported > 0){
      ## removes out-migrants during period from calculations of
exposure to/observation of mortality risk
      #    sub_frame2<-sub_frame2[-
c(which(sub_frame2$Death_age_at_test>=i &
      #                                  sub_frame2$Death_age_at_test
```

```
<c(i+1) &
      #
sub_frame2$fate=="exported")),]

    ## counts all those dying after the observation period starts
(the pop exposed to risk)
    # at_risk<-sum(Age_at_death>=floor(i*365.25), na.rm = T)
    #  at_risk<-sum(c(sub_frame$firstObs_years)>=i, na.rm = T)
    # at_risk<-sum(sub_frame2$Death_age_at_test>=i, na.rm = T)

    at_risk<-c(sum(sub_frame2$Death_age_at_test>=i, na.rm = T)-
exported_sum)

    ## how many deaths during the observation period?
    died<-c(length(which(sub_frame2$fate!="survived" &
                         sub_frame2$Death_age_at_test>=i
                     & sub_frame2$Death_age_at_test <c(i+1)))-
exported_sum)

    #   died<-sum(c(Age_at_death[which(sub_frame2$fate!
="survived")] %in% c(floor(i*365.25):c(floor((i+1)*365.25)-1))),
na.rm = T)

    censored_x<-c(censored_x, c(nrow(sub_frame)-nrow(sub_frame2))-
exported_sum)
    mx_vec<-c(mx_vec,c(died/at_risk))
    expo_vec<-c(expo_vec, at_risk)
    dx_vec<-c(dx_vec, died)
  }

}

qx_vec<-mxtoqx(mx_vec,ax = 0.5, n = 1)

Lifetab_v11s[[j]]<-data.frame(index = rep(j, times =
length(mx_vec)),
                             age = age_vec,
                             vin11 = rep(names(Lead_vin11s)[[j]],
times = length(mx_vec)),
                             scraps,
                             exports,
                             censored_x,
                             mx_vec,
                             expo_vec,
                             dx_vec,
                             qx_vec)

if(j %in% seq(1, 10000, by = 100)){
  print(j/9626)
  beep()
  print(Sys.time()-ticktock)
```

```
  }

}
Lifetab_v11s<-ldply(Lifetab_v11s, as.data.frame)
saveRDS(Lifetab_v11s, "Lifetab_v11s.rds")

## gets the consensus test classes, and an out-of-class rate
Test_class_freq<-
Vin11_test_classes[match(Lifetab_v11s$vin11,rownames(Vin11_test_clas
ses)),]

Tclass_vin11<-as.numeric(colnames(Test_class_freq)
[max.col(Test_class_freq,ties.method="first")])

## subtracts row sums to get out-of-class error (highest for small
van/car mixes)
Tclass_vin11_OOB<-c(1-(apply(Test_class_freq,1,max)/
rowSums(Test_class_freq)))*100

## adds test class, test error rate, other metadata
Lifetab_v11s_meta<-data.frame(Lifetab_v11s,
                              Tclass = Tclass_vin11,
                              Tclass_error = Tclass_vin11_OOB)

saveRDS(Lifetab_v11s_meta, "Lifetab_v11s_meta.rds")

################################################################
##################

## Builds lifetables by Make-Model-Year

################################################################
##################

Vehicle_properties<-readRDS("/Users/sauley/Documents/R_workspaces/
Analyses/Automobile_survival_UK/data/Survival_data/
Vehicle_properties.rds")
LifeVec_time<-readRDS("data/LifeVec_time.rds")

Vehicle_properties<-
Vehicle_properties[order(Vehicle_properties$unique_vehicle),]

sum(Vehicle_properties$unique_vehicle==LifeVec_time$unique_vehicle)
sum(Vehicle_properties$unique_vehicle!=LifeVec_time$unique_vehicle)

## converts ages in years to ages in days
AAT_days<-ceiling(LifeVec_time$Death_age_at_test*365.25)

## subtracts this from the test date
Birthday_est<-c(LifeVec_time$Death_test_date-AAT_days)
## and keeps the year
BirthYear_est<-year(Birthday_est)

## gets the make-model- or make-model-year index
```

```r
MMY_index<-paste(Vehicle_properties$make, Vehicle_properties$model,
BirthYear_est, sep = "__")

## fixes the irregular spacing of "Land rover" vs "Landrover"
MMY_index<-gsub("LAND ROVER", "LANDROVER", MMY_index)

## same for MMY
MMY_index_1k<-names(table(MMY_index))[which(table(MMY_index)>=1000)]

## removes NA-tail MMYs: "FORD____NA" N=2827    "PEUGEOT____NA" N
=1114 "VAUXHALL____NA" N = 1734
MMY_index_1k<-MMY_index_1k[-c(grep("NA$",MMY_index_1k))]

## and 51 years and make combos with no model names (e.g.
"HONDA____1971" N = 1010)
MMY_index_1k<-MMY_index_1k[-c(grep("____",MMY_index_1k))]

## what was the most common?
tail(table(MMY_index)[order(table(MMY_index))])

gc()

Lifetab_MMY<-NULL
Lifetab_MMY_stats<-NULL
Old_liars<-NULL

ticktock<-Sys.time()

gc()

for(j in 1:length(MMY_index_1k)){

  ## gets the unique IDs we want
  #  sub_index<-
Vehicle_properties_sub[which(MMY_index_short==c(MMY_index_1k[[j]]))]
  #  sub_index<-
Vehicle_properties_sub[which(MMY_index==c(MMY_index_1k[[j]]))]

  ## builds a data frame from these
  #  sub_frame<-LifeVec_time[which(LifeVec_time$unique_vehicle %in%
sub_index),]

  ## gets the unique IDs we want
  sub_frame<-LifeVec_time[which(MMY_index==c(MMY_index_1k[[j]])),]

  ## counts and removes low-quality vehicles
  N_crap<-sum(sub_frame$HighQ==0)

  sub_frame<-sub_frame[which(sub_frame$HighQ==1),]

  ## fixes the date formats to numeric, origin 1/1/1800
```

```r
  sub_frame$first_use_date<-
as.Date(as.character(sub_frame$first_use_date),
                                  format = "%Y%m%d", origin =
"1800-01-01")
  sub_frame$test_date<-as.Date(as.character(sub_frame$test_date),
                            format = "%Y-%m-%d", origin =
"1800-01-01")
  sub_frame$Death_test_date<-
as.numeric(as.Date(sub_frame$Death_test_date,
                                        origin =
"1800-01-01"))

  ## shortens search vectors
  #  Vehicle_properties_sub<-
Vehicle_properties_sub[which(MMY_index_short!=c(MMY_index_1k[[j]]))]

  ## Counts number of vehicles with >6 month difference between
stated age at death
  ## and the age at death estimated from testing data alone
  N_death_gaps<-sum(abs(c(sub_frame$Death_test_date-
                          as.numeric(sub_frame$first_use_date))/
365.25-
                          sub_frame$Death_age_at_test)>0.5, na.rm
=T)

  ## how many days between the first and last test?
  ObsTime<-c(sub_frame$Death_test_date-
              as.numeric(sub_frame$test_date))

  # ObsTime<-c(sub_frame$Death_test_date-
  #              as.numeric(sub_frame$first_use_date))

  ## subtracts this from age at death to get age at first
observation in years
  sub_frame$firstObs_years<-c(sub_frame$Death_age_at_test-c(ObsTime/
365.25))

  ## Gets age at death/ltf
  sub_frame$Age_at_death<-sub_frame$Death_test_date-
as.numeric(sub_frame$first_use_date)

  ## resplits the MMY back down
  Lifetab_MMY_Year<-as.numeric(tail(strsplit(MMY_index_1k[[j]],
split = "__")[[1]],1))

  ## counts cases where model year and age are impossible (typos)
  Old_liars[[j]]<-
sum(c(Lifetab_MMY_Year+sub_frame$Death_age_at_test)>2023)

  ## if any, removes them.
  sub_frame<-
sub_frame[which(c(Lifetab_MMY_Year+sub_frame$Death_age_at_test)<=202
```

```
3),]

  ## keeps some numbers
  Lifetab_MMY_stats[[j]]<-c(table(sub_frame$fate), N_death_gaps,
N_crap)

  ## builds a life table for chronological time

  ## builds a vector of vehicles at risk + deaths
  mx_vec<-NULL
  age_vec<-c(0:c(2022-Lifetab_MMY_Year))
  dx_vec<-NULL
  expo_vec<-NULL
  censored_x<-NULL
  scraps<-NULL
  exports<-NULL

  for(i in age_vec){

    ## ignores any vehicles first observed after the start of
observation period.
    sub_frame2<-sub_frame[which(sub_frame$firstObs_years <= i),]

    ## how many of these went to scrappage/export/destruction
*during* this period?
    lil_fates<-sub_frame2$fate[which(sub_frame2$Death_age_at_test>=i
&
                                     sub_frame2$Death_age_at_test
<c(i+1))]
    exported_sum<-0
    exported_sum<-sum(lil_fates=="exported")

    scraps<-c(scraps,sum(lil_fates %in%
c("scrapped","cert_destroyed")))
    exports<-c(exports,exported_sum)

    Non_exported<-c(length(sub_frame2$fate)-
sum(lil_fates=="exported"))

    ## If all were exported (rare), fills with NA
    if(Non_exported==0){
      censored_x<-c(censored_x, NA)
      mx_vec<-c(mx_vec,NA)
      expo_vec<-c(expo_vec, NA)
      dx_vec<-c(dx_vec, NA)
    }

    if(Non_exported > 0){
      ## removes out-migrants during period from calculations of
exposure to/observation of mortality risk
      #   sub_frame2<-sub_frame2[-
c(which(sub_frame2$Death_age_at_test>=i &
      #                                sub_frame2$Death_age_at_test
<c(i+1) &
```

```
        #
sub_frame2$fate=="exported")),]

        ## counts all those dying after the observation period starts
(the pop exposed to risk)
        # at_risk<-sum(Age_at_death>=floor(i*365.25), na.rm = T)
        #  at_risk<-sum(c(sub_frame$firstObs_years)>=i, na.rm = T)
        # at_risk<-sum(sub_frame2$Death_age_at_test>=i, na.rm = T)

        at_risk<-c(sum(sub_frame2$Death_age_at_test>=i, na.rm = T)-
exported_sum)

        ## how many deaths during the observation period?
        died<-c(length(which(sub_frame2$fate!="survived" &
                            sub_frame2$Death_age_at_test>=i
                    & sub_frame2$Death_age_at_test <c(i+1)))-
exported_sum)

        #   died<-sum(c(Age_at_death[which(sub_frame2$fate!
="survived")] %in% c(floor(i*365.25):c(floor((i+1)*365.25)-1))),
na.rm = T)

        censored_x<-c(censored_x, c(nrow(sub_frame)-nrow(sub_frame2))-
exported_sum)
        mx_vec<-c(mx_vec,c(died/at_risk))
        expo_vec<-c(expo_vec, at_risk)
        dx_vec<-c(dx_vec, died)
      }

    }

  qx_vec<-mxtoqx(mx_vec,ax = 0.5, n = 1)

  Lifetab_MMY[[j]]<-data.frame(index = rep(j, times =
length(mx_vec)),
                              age = age_vec,
                              MMY = rep(MMY_index_1k[[j]], times =
length(mx_vec)),
                              YOB = Lifetab_MMY_Year,
                              scraps,
                              exports,
                              censored_x,
                              mx_vec,
                              expo_vec,
                              dx_vec,
                              qx_vec)

  if(j %in% seq(1, 10000, by = 100)){
    print(j/6281)
    beep()
    print(Sys.time()-ticktock)
  }

}
```

```r
saveRDS(Lifetab_MMY, "Lifetab_MMY.rds")

Lifetab_MMY<-ldply(Lifetab_MMY, as.data.frame)

Lifetab_MMY<-Lifetab_MMY[which(Lifetab_MMY$expo_vec!=0),]

## how many life-years of data are we capturing?
sum(Lifetab_MMY$expo_vec)

## 339,099,004 life-years of exposure to risk

## aggregates the fraction of MMY combinations in each test class
MMY_test_class<-table(MMY_index, by = Vehicle_properties$test_class)
MMY_test_class<-as.data.frame.matrix(MMY_test_class)

## counts a consensus call (majority rule)
Max_class<-colnames(MMY_test_class)[apply(MMY_test_class,
1,which.max)]
Max_class<-as.numeric(as.character(Max_class))
xobj1<-rowSums(MMY_test_class)
xobj2<-apply(MMY_test_class, 1,which.max)
xobj2<-lapply(seq(1, length(xobj2)), function(i)
MMY_test_class[i,xobj2[i]])
Err_vec<-unlist(xobj2)/xobj1

## attaches a consensus and an OOB rate to the data frame
Lifetab_MMY$test_class<-Max_class[match(Lifetab_MMY$MMY,
rownames(MMY_test_class))]
Lifetab_MMY$test_otherclass<-c(100*(1-Err_vec[match(Lifetab_MMY$MMY,
names(Err_vec))]))

saveRDS(Lifetab_MMY, "Lifetab_MMY_meta.rds")

################

## loads vehicle props table again
Vehicle_properties<-readRDS("/Users/sauley/Documents/R_workspaces/
Analyses/Automobile_survival_UK/data/Survival_data/
Vehicle_properties.rds")

## reorders to match MMY index
Vehicle_properties<-
Vehicle_properties[order(Vehicle_properties$unique_vehicle),]

##############################################################
##########

## generates probabilities of survival by mileage

##############################################################
##########
```

```r
## MMY index first...

## a life table by mileage, by MMY
LifeVec_miles<-readRDS("data/LifeVec_miles.rds")
LifeVec_miles<-LifeVec_miles[order(LifeVec_miles$unique_vehicle),]

## checks alignment
sum(LifeVec_miles$unique_vehicle!=Vehicle_properties$unique_vehicle)
sum(LifeVec_miles$unique_vehicle==Vehicle_properties$unique_vehicle)

Lifetab_MMY_miles<-NULL
Lifetab_MMY_stats_miles<-NULL
ticktock<-Sys.time()

for(j in 1:length(MMY_index_1k)){

  ## gets the unique IDs we want
  sub_frame<-LifeVec_miles[which(MMY_index==c(MMY_index_1k[[j]])),]

  # sub_index<-
Vehicle_properties$unique_vehicle[which(MMY_index==c(MMY_index_10k[[
j]]))]

  ## builds a data frame from these
  #  sub_frame<-LifeVec_miles[which(LifeVec_miles$unique_vehicle
%in% sub_index),]

  ## Screens out QC fails
  sub_frame<-sub_frame[which(sub_frame$HighQ==1),]

  Lifetab_MMY_stats_miles[[j]]<-table(sub_frame$fate)

  ## builds a life table for 'physical age'

  ## builds a vector of vehicles at risk + deaths
  mx_vec<-NULL
  age_vec<-seq(0, 1000000, by = 1000)
  dx_vec<-NULL
  expo_vec<-NULL
  censored_x<-NULL
  scraps<-NULL
  exports<-NULL

  for(i in age_vec){

    ## counts all those dying after the observation period starts
(the pop exposed to risk)
    at_risk<-sum(sub_frame$Death_corrected_odometer>=i, na.rm = T)
    ## counts those not yet exposed to risk (immortal time bias)
    left_censored<-sum(sub_frame$corrected_odometer<i, na.rm = T)

    ## how many deaths during the observation period?
```

```
    died<-length(which(sub_frame$Death_corrected_odometer %in% c(i:
(i+999))))

    ## how many of these went to scrappage/export/destruction during
follow-up?
    lil_fates<-
sub_frame$fate[which(sub_frame$Death_corrected_odometer>=i &

sub_frame$Death_corrected_odometer<(i+999))]

    exported_sum<-0
    exported_sum<-sum(lil_fates=="exported")

    Non_exported<-c(at_risk-sum(lil_fates=="exported"))

    scraps<-c(scraps,sum(lil_fates %in%
c("scrapped","cert_destroyed")))
    exports<-c(exports,exported_sum)

    ## subtracts the number of vehicles exported (after reaching
this mileage)
    ## from the population exposed to risk
    at_risk<-c(at_risk-exported_sum)
    died<-c(died-exported_sum)

    if(Non_exported==0){
      censored_x<-c(censored_x, NA)
      mx_vec<-c(mx_vec,NA)
      expo_vec<-c(expo_vec, NA)
      dx_vec<-c(dx_vec, NA)
    }

    if(Non_exported>0){
      mx_vec<-c(mx_vec,c(died/at_risk))
      expo_vec<-c(expo_vec, at_risk)
      dx_vec<-c(dx_vec, died)
      censored_x<-c(censored_x, left_censored)
    }

  }

  qx_vec<-mxtoqx(mx_vec,ax = 0.5, n = 1)

  Lifetab_MMY_miles[[j]]<-data.frame(index = rep(j, times =
length(mx_vec)),

                                    age = age_vec,
                                    MMY = rep(MMY_index_1k[[j]],
times = length(mx_vec)),

                                    scraps,
                                    exports,
                                    censored_x,
                                    mx_vec,
                                    expo_vec,
                                    dx_vec,
```

```
                                qx_vec)

  if(j %in% seq(1, 10000, by = 100)){
    print(j/6281)
    print(j)
    beep()
    print(Sys.time()-ticktock)
  }

}

Lifetab_MMY_miles<-ldply(Lifetab_MMY_miles, as.data.frame)

saveRDS(Lifetab_MMY_miles, "Lifetab_MMY_miles.rds")

MMY_test_classes<-table(MMY_index[which(MMY_index %in%
MMY_index_1k)],
                        by =
Vehicle_properties$test_class[which(MMY_index %in% MMY_index_1k)])

## gets the consensus test classes, and an out-of-class rate
Test_class_freq<-
MMY_test_classes[match(Lifetab_MMY_miles$MMY,rownames(MMY_test_class
es)),]

Tclass_MMY<-as.numeric(colnames(Test_class_freq)
[max.col(Test_class_freq,ties.method="first")])

## subtracts row sums to get out-of-class error (highest for small
van/car mixes)
Tclass_MMY_OOB<-c(1-(apply(Test_class_freq,1,max)/
rowSums(Test_class_freq)))*100

## adds test class, test error rate, other metadata
Lifetab_MMY_miles_meta<-data.frame(Lifetab_MMY_miles,
                                   Tclass = Tclass_MMY,
                                   Tclass_error = Tclass_MMY_OOB)

## adds the year to MMY_miles_meta
xobj<-lapply(Lifetab_MMY_miles_meta$MMY, strsplit, split = "__")
xobj<-lapply(xobj, unlist)
xobj<-lapply(xobj,tail, 1)

Lifetab_MMY_miles_meta$YOB<-as.numeric(unlist(xobj))
rm(xobj)
gc()

saveRDS(Lifetab_MMY_miles_meta, "Lifetab_MMY_miles_meta.rds")

## attaches vehicle properties to each life table
```

```
###############################################

## a life table by mileage, by VIN11
LifeVec_miles<-readRDS("data/LifeVec_miles.rds")

## just does VIN >25k obs
Lead_vin11s_25k<-Lead_vin11s[which(Lead_vin11s>=25000)]
Lead_vin11s_10k<-Lead_vin11s[which(Lead_vin11s>=10000)]

Lifetab_v11s_miles<-NULL
Lifetab_v11s_stats_miles<-NULL
ticktock<-Sys.time()

for(j in 1:length(Lead_vin11s_10k)){

  ## gets the unique IDs we want
  sub_index<-
Vehicle_properties$unique_vehicle[which(Vehicle_properties$vin11==na
mes(Lead_vin11s_10k)[[j]])]

  ## builds a data frame from these
  sub_frame<-LifeVec_miles[which(LifeVec_miles$unique_vehicle %in%
sub_index),]

  ## Screens out QC fails
  sub_frame<-sub_frame[which(sub_frame$HighQ==1),]

  Lifetab_v11s_stats_miles[[j]]<-table(sub_frame$fate)

  ## builds a life table for 'physical age'

  ## builds a vector of vehicles at risk + deaths
  mx_vec<-NULL
  age_vec<-seq(0, 1000000, by = 1000)
  dx_vec<-NULL
  expo_vec<-NULL
  censored_x<-NULL
  scraps<-NULL
  exports<-NULL

  for(i in age_vec){

    ## counts all those dying after the observation period starts
(the pop exposed to risk)
    at_risk<-sum(sub_frame$Death_corrected_odometer>=i, na.rm = T)
    ## counts those not yet exposed to risk (immortal time bias)
    left_censored<-sum(sub_frame$corrected_odometer<i, na.rm = T)

    ## how many deaths during the observation period?
    died<-length(which(sub_frame$Death_corrected_odometer %in% c(i:
(i+999))))

    ## how many of these went to scrappage/export/destruction during
```

```
follow-up?
    lil_fates<-
sub_frame$fate[which(sub_frame$Death_corrected_odometer>=i &

sub_frame$Death_corrected_odometer<(i+999))]

    exported_sum<-0
    exported_sum<-sum(lil_fates=="exported")

    Non_exported<-c(at_risk-sum(lil_fates=="exported"))

    scraps<-c(scraps,sum(lil_fates %in%
c("scrapped","cert_destroyed")))
    exports<-c(exports,exported_sum)

    ## subtracts the number of vehicles exported (after reaching
this mileage)
    ## from the population exposed to risk
    at_risk<-c(at_risk-exported_sum)
    died<-c(died-exported_sum)

    if(Non_exported==0){
      censored_x<-c(censored_x, NA)
      mx_vec<-c(mx_vec,NA)
      expo_vec<-c(expo_vec, NA)
      dx_vec<-c(dx_vec, NA)
    }

    if(Non_exported>0){
      mx_vec<-c(mx_vec,c(died/at_risk))
      expo_vec<-c(expo_vec, at_risk)
      dx_vec<-c(dx_vec, died)
      censored_x<-c(censored_x, left_censored)
    }

  }

  qx_vec<-mxtoqx(mx_vec,ax = 0.5, n = 1)

  Lifetab_v11s_miles[[j]]<-data.frame(index = rep(j, times =
length(mx_vec)),
                                     age = age_vec,
                                     vin11 =
rep(names(Lead_vin11s_10k)[[j]], times = length(mx_vec)),
                                     scraps,
                                     exports,
                                     censored_x,
                                     mx_vec,
                                     expo_vec,
                                     dx_vec,
                                     qx_vec)

  if(j %in% seq(1, 10000, by = 100)){
    print(j/996)
```

```
    beep()
    print(Sys.time()-ticktock)
  }

}

Lifetab_v11s_miles<-ldply(Lifetab_v11s_miles, as.data.frame)

saveRDS(Lifetab_v11s_miles, "Lifetab_v11s_miles.rds")

## gets the consensus test classes, and an out-of-class rate
Test_class_freq<-
Vin11_test_classes[match(Lifetab_v11s_miles$vin11,rownames(Vin11_tes
t_classes)),]

Tclass_vin11<-as.numeric(colnames(Test_class_freq)
[max.col(Test_class_freq,ties.method="first")])

## subtracts row sums to get out-of-class error (highest for small
van/car mixes)
Tclass_vin11_OOB<-c(1-(apply(Test_class_freq,1,max)/
rowSums(Test_class_freq)))*100

## adds test class, test error rate, other metadata
Lifetab_v11s_miles_meta<-data.frame(Lifetab_v11s_miles,
                                    Tclass = Tclass_vin11,
                                    Tclass_error = Tclass_vin11_OOB)

saveRDS(Lifetab_v11s_miles_meta, "Lifetab_v11s_miles_meta.rds")

############################################

## Merges on the failures per test for VIN11s
FPT_vin11_miles<-readRDS("data/FPT_vin11_miles_clean.rds")

FPT_lifetab_miles<-
FPT_vin11_miles[match(paste(Lifetab_v11s_miles_meta$vin11,

Lifetab_v11s_miles_meta$age, sep = "__"),
                                     FPT_vin11_miles$Var1),]

cor.test(FPT_lifetab_miles$N_failed/c(FPT_lifetab_miles$N_tests),
         Lifetab_v11s_miles$qx_vec)

cor.test(c(FPT_lifetab_miles$N_failed/c(FPT_lifetab_miles$N_tests))
[which(Lifetab_v11s_miles$qx_vec<1)],

c(Lifetab_v11s_miles$qx_vec[which(Lifetab_v11s_miles$qx_vec<1)]))

cor.test(xobj$N_failed/
c(xobj$N_passed+xobj$N_failed+xobj$N_prs_failed),
         Lifetab_v11s_miles$mx_vec)
```

```r
## What is the correlation between breakdown rates at a given
mileage,
## and survival rates over the next 10,000 miles?

cor_vec1<-NULL
cor_vec2<-NULL

for(i in seq(0,1000000, by = 10000)){

  vec1<-c(c(FPT_lifetab_miles$N_failed/c(FPT_lifetab_miles$N_tests))
[which(Lifetab_v11s_miles_meta$Tclass==4 &

Lifetab_v11s_miles_meta$expo_vec>10 &

Lifetab_v11s_miles_meta$age==i)])
  vec2<-log(c(Lifetab_v11s_miles$qx_vec)
[which(Lifetab_v11s_miles_meta$Tclass==4
                                        &
Lifetab_v11s_miles_meta$expo_vec>10
                                        &
Lifetab_v11s_miles_meta$age==i)])

  if(sum(!is.na(vec1+vec2))<100){
    cor_vec1<-c(cor_vec1,NA)
    cor_vec2<-rbind(cor_vec2,c(sum(!is.na(vec1+vec2)),NA,NA))
  }

  if(sum(!is.na(vec1+vec2))>=100){
    cor_vec1<-c(cor_vec1,cor.test(vec1,vec2)$estimate)
    cor_vec2<-rbind(cor_vec2, c(sum(!
is.na(vec1+vec2)),cor.test(vec1,vec2)$conf.int[1:2]))
  }
}

plot(cor_vec1~seq(0,1000000, by = 10000),
     type = "l", ylim = c(-1,1), xlab = "Miles accumulated", ylab =
"Rsq breakdown ~ mortality rate")

abline(h = seq(-1,1, by = 0.2), col = "grey")
abline(h = 0, lty = 3, col= "red")

points(c(cor_vec2[,2],cor_vec2[,3])~c(seq(0,1000000, by =
10000),seq(0,1000000, by = 10000)),
       type = "l", col = "lightblue")
points(c(cor_vec2[,2],cor_vec2[,3])~c(seq(0,1000000, by =
10000),seq(0,1000000, by = 10000)),
       pch = 20, cex = 0.3)

points(cor_vec1~seq(0,1000000, by = 10000),  pch = 20, col =
"darkblue")

#points(cor_vec2[,2]~seq(0,1000000, by = 10000),  pch = "-")
```

```
#points(cor_vec2[,3]~seq(0,1000000, by = 10000),  pch = "-")

#text(y= cor_vec1+0.3, x = seq(0,1000000, by = 10000),labels
=cor_vec2[,1])

## After the initial 10k miles, during which 'lemons' are killed
off, it is weakly negative.

boxplot(log(Lifetab_v11s_miles_meta$mx_vec[Lifetab_v11s_miles_meta$e
xpo_vec>100])~Lifetab_v11s_miles_meta$age[Lifetab_v11s_miles_meta$ex
po_vec>100],
        xlab = "miles on odometer",
        ylab = "log probability of death",
        pch = "-", col = "orange")

plot(log10(Lifetab_v11s_miles_meta$qx_vec[Lifetab_v11s_miles_meta$in
dex==1])~

Lifetab_v11s_miles_meta$age[Lifetab_v11s_miles_meta$index==1],
     type= "l", col = rgb(0,0,0.5,0.2),
     xlim = c(0,1000000),ylim = c(-6,0),
     xlab = "miles driven", ylab = "probability of death")

for(i in 1:max(Lifetab_v11s_miles_meta$index)){

points(log10(Lifetab_v11s_miles_meta$qx_vec[Lifetab_v11s_miles_meta$
index==i])~

Lifetab_v11s_miles_meta$age[Lifetab_v11s_miles_meta$index==i],
        type= "l", col = rgb(0,0,0.5,0.2))
}

## keeps expo >100 vehicles

plot(log10(Lifetab_v10s_miles$qx_vec[which(Lifetab_v10s_miles$index=
=1 &

Lifetab_v10s_miles$expo_vec>=1000)])~
      Lifetab_v10s_miles$age[which(Lifetab_v10s_miles$index==1 &

Lifetab_v10s_miles$expo_vec>=1000)],
     type= "l", col = rgb(0,0,0,0.2),
     xlim = c(0,1000000),ylim = c(-6,0),
     xlab = "miles driven", ylab = "probability of death")

for(i in 1:max(Lifetab_v10s_miles$index)){

points(log10(Lifetab_v10s_miles$qx_vec[which(Lifetab_v10s_miles$inde
x==i &

Lifetab_v10s_miles$expo_vec>=1000)])~
```

```
            Lifetab_v10s_miles$age[which(Lifetab_v10s_miles$index==i
&

Lifetab_v10s_miles$expo_vec>=1000)],
        type= "l", col = rgb(0,0,0,0.2))
}

###############################################################
##################

## Builds lifetables by Make-Model-Year

###############################################################
##################

## converts ages in years to ages in days
AAT_days<-ceiling(LifeVec_time$Death_age_at_test*365.25)

## subtracts this from the test date
Birthday_est<-c(LifeVec_time$Death_test_date-AAT_days)
## and keeps the year
BirthYear_est<-year(Birthday_est)

## gets the make-model- or make-model-year index
MMY_index<-paste0(Vehicle_properties$make, Vehicle_properties$model,
BirthYear_est)

MM_vec<-paste0(Vehicle_properties$make, Vehicle_properties$model)
MM_index<-table(MM_vec)

## looks at only those with >10k vehicles
MM_index_10k<-MM_index[which(MM_index>=10000)]

## same for MMY
table(MMY_index)[which(table(MMY_index>=10000))]

Lifetab_MM<-NULL
Lifetab_MM_stats<-NULL
ticktock<-Sys.time()

ticktock<-Sys.time()

for(j in 1:length(Lead_vin10s)){

  ## gets the unique IDs we want
  sub_index<-
Vehicle_properties$unique_vehicle[which(vin10==names(Lead_vin10s)
[[j]])]

  ## builds a data frame from these
```

```r
    sub_frame<-LifeVec_time[which(LifeVec_time$unique_vehicle %in%
sub_index),]

  ## fixes the date formats to numeric, origin 1/1/1800
    sub_frame$first_use_date<-
as.Date(as.character(sub_frame$first_use_date),
                                    format = "%Y%m%d", origin =
"1800-01-01")
    sub_frame$test_date<-as.Date(as.character(sub_frame$test_date),
                            format = "%Y-%m-%d", origin =
"1800-01-01")
    sub_frame$Death_test_date<-
as.numeric(as.Date(sub_frame$Death_test_date,
                                                origin =
"1800-01-01"))

  Lifetab_v10s_stats[[j]]<-table(sub_frame$fate)

  ## calculates age at first MOT (age first observed)
    sub_frame$firstObs<-as.numeric(sub_frame$test_date-
sub_frame$first_use_date)

  ## Gets age at death/ltf
    sub_frame$Age_at_death<-sub_frame$Death_test_date-
as.numeric(sub_frame$first_use_date)

  ## screens out any vehicles whose first use date is > 4 years
after the median
  ## first use date of the make+model (excludes mostly imports,
errors)
    N_late_regs<-
sum(sub_frame$first_use_date>median(sub_frame$first_use_date, na.rm
= T)+(365*4), na.rm = T)
    Pct_late_regs<-
mean(sub_frame$first_use_date>median(sub_frame$first_use_date, na.rm
= T)+(365*4), na.rm = T)*100

    sub_frame<-sub_frame[-
c(which(sub_frame$first_use_date>median(sub_frame$first_use_date,
na.rm = T)+(365*4))),]

  ## builds a life table for chronological time

  ## builds a vector of vehicles at risk + deaths
  mx_vec<-NULL
  age_vec<-c(0:60)
  dx_vec<-NULL
  expo_vec<-NULL
  censored_x<-NULL

  for(i in age_vec){
```

```r
    ## removes any vehicles falling entirely after the observation
period.
    sub_frame2<-sub_frame[which(sub_frame$firstObs <=
floor((i+1)*365.25)),]

    Age_at_death<-sub_frame2$Death_test_date-
as.numeric(sub_frame2$first_use_date)

    ## counts all those dying after the observation period starts
(the pop exposed to risk)
    at_risk<-sum(Age_at_death>=floor(i*365.25), na.rm = T)

    ## how many deaths during the observation period?
    died<-sum(c(Age_at_death[which(sub_frame2$fate!="survived")]
%in% c(floor(i*365.25):c(floor((i+1)*365.25)-1))), na.rm = T)

    censored_x<-c(censored_x, nrow(sub_frame)-nrow(sub_frame2))
    mx_vec<-c(mx_vec,c(died/at_risk))
    expo_vec<-c(expo_vec, at_risk)
    dx_vec<-c(dx_vec, died)

  }

  qx_vec<-mxtoqx(mx_vec,ax = 0.5, n = 1)

  Lifetab_MM[[j]]<-data.frame(index = rep(j, times =
length(mx_vec)),
                             age = age_vec,
                             vin10 = rep(names(Lead_vin10s)[[j]],
times = length(mx_vec)),
                             censored_x,
                             mx_vec,
                             expo_vec,
                             dx_vec,
                             qx_vec)

  if(j %in% seq(1, 10000, by = 100)){
    print(j/8704)
    beep()
    print(Sys.time()-ticktock)
  }

}

saveRDS(Lifetab_MM, "data/Lifetab_MM.rds")

Lifetab_MM<-ldply(Lifetab_MM, as.data.frame)

####################################################################
###################

## Assembles make-model prices and depreciation rates

####################################################################
```

```
##################

## Gets past data on new-car prices.
Prices_hist<-read.csv("data/metadata/Price_data/
Price_indices_Jan_2004_and_2014.csv")

Prices_hist$Make<-toupper(trimws(Prices_hist$Make))
Prices_hist$Model<-toupper(trimws(Prices_hist$Model))

## tries to get engine size and fuel type
Prices_hist$diesel_prob<-Prices_hist$Diesel
Prices_hist$diesel_prob[Prices_hist$diesel_prob==""]<-NA
Prices_hist$diesel_prob[grep("iesel",tolower(Prices_hist$Description
))]<-"Diesel"
Prices_hist$diesel_prob[grep("tdi",tolower(Prices_hist$Description))
]<-"Diesel"
Prices_hist$diesel_prob[grep("
tdsl",tolower(Prices_hist$Description))]<-"Diesel"
Prices_hist$diesel_prob[grep("hdi",tolower(Prices_hist$Description))
]<-"Diesel"
Prices_hist$diesel_prob[grep("tdsl",tolower(Prices_hist$Description)
)]<-"Diesel"
Prices_hist$diesel_prob[grep("tdci",tolower(Prices_hist$Description)
)]<-"Diesel"
Prices_hist$diesel_prob[grep("d-4d",tolower(Prices_hist$Description)
)]<-"Diesel"
Prices_hist$diesel_prob[grep("\\.
[0-9]tid",tolower(Prices_hist$Description))]<-"Diesel"
Prices_hist$diesel_prob[grep("\\.
[0-9]dtr",tolower(Prices_hist$Description))]<-"Diesel"
Prices_hist$diesel_prob[grep("\\.[0-9]
dtr",tolower(Prices_hist$Description))]<-"Diesel"
Prices_hist$diesel_prob[grep("\\.[0-9]
dci",tolower(Prices_hist$Description))]<-"Diesel"
Prices_hist$diesel_prob[grep("\\.[0-9]
jtd",tolower(Prices_hist$Description))]<-"Diesel"
Prices_hist$diesel_prob[grep("\\.[0-9]
dsl",tolower(Prices_hist$Description))]<-"Diesel"
Prices_hist$diesel_prob[grep("
td4",tolower(Prices_hist$Description))]<-"Diesel"
Prices_hist$diesel_prob[grep("
td5",tolower(Prices_hist$Description))]<-"Diesel"
Prices_hist$diesel_prob[grep("
td6",tolower(Prices_hist$Description))]<-"Diesel"

Prices_hist$diesel_prob[grep(" [0-9]\\.[0-9]d
",tolower(Prices_hist$Description))]<-"Diesel"

Prices_hist$diesel_prob[grep(" [0-9]\\.
[0-9]td",tolower(Prices_hist$Description))]<-"Diesel"
Prices_hist$diesel_prob[grep(" [0-9]\\.[0-9]
td",tolower(Prices_hist$Description))]<-"Diesel"
```

```r
Prices_hist$diesel_prob[grep("etrol",tolower(Prices_hist$Description
))]<-"Petrol"

Prices_hist$diesel_prob[grep("electric",tolower(Prices_hist$Descript
ion))]<-"Electric"

Prices_hist$diesel_prob[grep("hybrid",tolower(Prices_hist$Descriptio
n))]<-"Hybrid Electric (Clean)"
Prices_hist$diesel_prob[grep("diesel electric
hybrid",tolower(Prices_hist$Description))]<-"Electric Diesel"
Prices_hist$diesel_prob[grep("hdi
electric",tolower(Prices_hist$Description))]<-"Electric Diesel"
Prices_hist$diesel_prob[grep("electric
pack",tolower(Prices_hist$Description))]<-"Diesel"

## the volvo bi-fuels
Prices_hist$diesel_prob[grep("bi-
fuel",tolower(Prices_hist$Description))]<-"Gas Bi-Fuel"

## everything else is likely petrol
Prices_hist$diesel_prob[which(is.na(Prices_hist$diesel_prob))]<-"Pet
rol"

## gets putative engine size
Prices_hist$eng_size_prob<-Prices_hist$CC
Prices_hist$eng_size_prob[which(Prices_hist$eng_size_prob=="")]<-NA
Prices_hist$eng_size_prob<-
as.numeric(trimws(as.character(Prices_hist$eng_size_prob)))

## finds engine size designations, substrings
require(stringr)
Eng_substr<-lapply(str_extract_all(Prices_hist$Description, " [0-9]\
\.[0-9]"), head, 1)

## gets any leading engine size designations
Eng_substr2<-lapply(str_extract_all(Prices_hist$Description,"^[0-9]\
\.[0-9]"), head, 1)

## infills any empty strings with other estimate, of possible
Eng_substr<-lapply(seq(1, length(Eng_substr)),
                   function(i) ifelse(length(Eng_substr[[i]]==0),
Eng_substr[[i]], Eng_substr2[[i]]))

## backfills with NAs
Eng_substr<-lapply(seq(1, length(Eng_substr)), function(i)
ifelse(length(Eng_substr[[i]]==0), Eng_substr[[i]], NA))
## makes into ccs
Eng_substr<-as.numeric(trimws(unlist(Eng_substr)))*1000
Prices_hist$eng_size_prob[which(is.na(Prices_hist$eng_size_prob))]<-
Eng_substr[which(is.na(Prices_hist$eng_size_prob))]

Prices_hist$eng_size_prob[grep("689cc",tolower(Prices_hist$Descripti
```

```r
on))]<-689

sum(!is.na(Prices_hist$eng_size_prob))
mean(!is.na(Prices_hist$eng_size_prob))

sum(!is.na(Prices_hist$diesel_prob))
mean(!is.na(Prices_hist$diesel_prob))

## checks how many unique new-vehicle make+model prices we have
length(unique(paste0(Prices_hist$Year_priced,
Prices_hist$Make,Prices_hist$Model)))

## and with engine size variants?
length(unique(paste0(Prices_hist$Year_priced,
Prices_hist$Make,Prices_hist$Model,
                     floor(Prices_hist$eng_size_prob/100))))

length(unique(paste0(Prices_hist$Year_priced,
Prices_hist$Make,Prices_hist$Model,
                     Prices_hist$diesel_prob,
                     floor(Prices_hist$eng_size_prob/100))))

## sees how many directly match vehicles in the vehicle properties
table
sum(unique(paste(Prices_hist$Make,Prices_hist$Model,
Prices_hist$Year_priced, sep = "__")) %in%
      MMY_index)

## sees how many directly match vehicles in the vehicle properties
table
sum(unique(paste(Prices_hist$Make,Prices_hist$Model,
Prices_hist$Year_priced, sep = "__")) %in%
      MMY_index_1k)

## Great.

## Changes "MERCEDES BENZ" to "MERCEDES-BENZ" to allow matches
Prices_hist$Make<-gsub("MERCEDES BENZ","MERCEDES-
BENZ",Prices_hist$Make)
Prices_hist$Make[Prices_hist$Make=="LAND ROVER"]<-"LANDROVER"
Prices_hist$Make[Prices_hist$Make=="ALFA"]<-"ALFA ROMEO"

####################
####################

## Loads up April 3rd data

Prices_now<-list.files("data/metadata/Price_data/
Prices_Apr_3_2023/")
price_frame_now<-NULL

## pulls in each file, gets price data and make+model+year etc
for(i in 1:length(Prices_now)){
  tpage<-NULL
```

```r
   ## must wrap in unlist(as.character())
   tpage<-read_html(unlist(as.character(paste0("data/metadata/
Price_data/Prices_Apr_3_2023/",
                                              Prices_now[[i]]]))))

   ## Grabs price
   price_sub<-html_text(html_nodes(tpage,'.product-card-pricing'))
   price_sub<-gsub("\n","",price_sub)
   price_sub<-gsub(" ","",price_sub)
   price_sub<-gsub(",","",price_sub)
   price_sub<-gsub("£","",price_sub)

   deets<-html_text(html_nodes(tpage,'.product-card-details'))
   #strsplit(deets[[1]], split = "\n")

   Makemod<-html_text(html_nodes(tpage,'.product-card-
details__title'))
   Makemod<-gsub("\n","",Makemod)
   Makemod<-trimws(Makemod)

   #  selInf<-html_text(html_nodes(tpage,"product-card-seller-info"))

   Keyspecs<-lapply(html_text(html_nodes(tpage,'.listing-key-
specs')), strsplit, split = "\n")
   if(length(Keyspecs)>0){
     Keyspecs<-lapply(Keyspecs, unlist)
     Keyspecs<-lapply(Keyspecs, trimws)
     ## pads to length 25
     Keyspecs<-lapply(seq(1,length(Keyspecs)), function(j)
c(Keyspecs[[j]],rep(NA, 25 - length(Keyspecs[[j]]))))
     Keyspecs<-lapply(Keyspecs, unlist)

     Keyspecs<-matrix(unlist(Keyspecs), ncol =25, byrow = T)

     #Keyspecs<-lapply(html_text(html_nodes(tpage,'.listing-key-
specs')), strsplit, split = "\n")

     ## if keyspecs is too short (when there is an ad at the end)
trims off the tail end

     ## removes makemodel and price for any ads under finance
     lease_ads<-NULL
     lease_ads<-grep("permonth", tolower(price_sub))

     if(length(lease_ads)>0){
       Makemod<-Makemod[-c(lease_ads)]
       price_sub<-price_sub[-c(lease_ads)]
     }

     price_frame_now[[i]]<-data.frame(model=Makemod,
                                      Keyspecs,
                                      #
location=html_text(html_nodes(tpage,'.seller-location')),
                                      price=price_sub)
```

```r
  }
}
beep(5)

price_frame_now<-ldply(price_frame_now, as.data.frame)

## splits out the year
price_frame_now$MYear<-as.numeric(substr(start = 1,stop = 4,

as.character(price_frame_now$X1)))
## the make model
price_frame_now$Make<-
unlist(lapply(lapply(price_frame_now$model,strsplit, split =
" "),unlist), head, 1))

price_frame_now$Make<-toupper(price_frame_now$Make)

## corrects the MAZDA[somenumber] mistakes
price_frame_now$Make<-gsub("MAZDA[0-9]", "MAZDA",
price_frame_now$Make)

## couple of others
price_frame_now$Make[which(price_frame_now$Make=="LAND")]<-"LANDROVE
R"
price_frame_now$Make[which(price_frame_now$Make=="ALFA")]<-"ALFA
ROMEO"

saveRDS(price_frame_now, "price_frame_now.rds")

## works through the MMY index, matching all vehicles possible to
their
## historic and current price
# MMY_index_1k
Make_1k<-unlist(lapply(strsplit(MMY_index_1k, split = "__"),
head,1))
Mod_1k<-unlist(lapply(lapply(strsplit(MMY_index_1k, split = "__"),
head,2), tail, 1))
Year_1k<-as.numeric(unlist(lapply(strsplit(MMY_index_1k, split =
"__"), tail,1)))

price_all<-NULL

for(i in 1:length(MMY_index_1k)){

  MY_match<-NULL

  ## is there a make+year match?
  MY_match<-which(price_frame_now$Make %in% Make_1k[i] &
price_frame_now$MYear %in% Year_1k[i])
```

```r
  if(length(MY_match)==0){
    price_all[[i]]<-c(NA,NA,NA,NA)
  }

  if(length(MY_match)>0){
    ## Gets the prospective matches
    sub_prices<-price_frame_now[MY_match,]

    ## only keeps those where the model description contains the
model name
    sub_prices<-sub_prices[grep(Mod_1k[i], sub_prices$model),]

    ## keeps the median, range, sample size
    price_all[[i]]<-c(median(as.numeric(sub_prices$price), na.rm =
T),
                      range(as.numeric(sub_prices$price), na.rm =
T),
                      sum(!is.na(sub_prices$price)))
  }

  ## finds if there are current prices, also using grep (fuzzy
matching) on models
  MY_match<-NULL
  MY_match<-which(Prices_hist$Make %in% Make_1k[i] &
Prices_hist$Year_priced %in% Year_1k[i])

  if(length(MY_match)==0){
    price_all[[i]]<-c(unlist(price_all[[i]]),NA,NA,NA,NA)
  }

  if(length(MY_match)>0){
    ## Gets the prospective matches
    sub_prices<-Prices_hist[MY_match,]

    ## only keeps those where the model description contains the
model name
    sub_prices<-sub_prices[grep(Mod_1k[i], sub_prices$Model),]

    ## keeps the median, range, sample size
    price_all[[i]]<-c(unlist(price_all[[i]]),
                      median(as.numeric(sub_prices$Price), na.rm =
T),
                      range(as.numeric(sub_prices$Price), na.rm =
T),
                      sum(!is.na(sub_prices$Price)))
  }

}

price_all<-matrix(unlist(price_all),ncol = 8, byrow = T)
```

```r
beep()

price_all[which(price_all==-Inf)]<-NA
price_all[which(price_all==Inf)]<-NA
price_all<-as.data.frame(price_all)

colnames(price_all)<-c("Median_now", "Upper_now", "Lower_now",
"N_price_now",
                       "Median_new", "Upper_new", "Lower_new",
"N_price_new")

## calculates fractional depreciation rates where possible
price_all$Depreciation<-c(price_all[,1]/price_all[,5])

## and total depreciation rates per year
price_all$Depreciation_PPY<-c(price_all[,1]-price_all[,5])/c(2021-
Year_1k)

price_all$MMY<-MMY_index_1k

## attaches sale prices and current prices to lifetables
Price_frame1<-price_all[match(Lifetab_MMY_meta$MMY,price_all$MMY),]
Price_frame2<-
price_all[match(Lifetab_MMY_miles_meta$MMY,price_all$MMY),]

## new price and mortality rates?
cor.test(Lifetab_MMY_meta$mx_vec[Lifetab_MMY_meta$age==10],
         Price_frame1$Median_new[Lifetab_MMY_meta$age==10])$estimate

cor.test(Lifetab_MMY_miles_meta$mx_vec[Lifetab_MMY_miles_meta$age==1
00000],

Price_frame2$Median_new[Lifetab_MMY_miles_meta$age==100000])
$estimate

cor.test(Lifetab_MMY_meta$mx_vec[Lifetab_MMY_meta$age==10],
         Price_frame1$Depreciation_PPY[Lifetab_MMY_meta$age==10])
$estimate

for(i in 1:nrow(Vehicle_properties)){
  ## gets the ith price
  sub_frame1<-Prices_hist[i,]

  if(sub_frame1$Year_priced %in% c(2004, 2014)){

    price_frame_now_sub<-
price_frame_now[which(toupper(price_frame_now$Make)==sub_frame1$Make
                                       &
price_frame_now$MYear==sub_frame1$Year_priced),]
    ## gets the first model match
    price_frame_now_sub
```

```r
      }

}

################################################################
####

## Attaches prices to the MMY data frames

## loads MMY life tables
Lifetab_MMY<-readRDS("Lifetab_MMY_meta.rds")
#Lifetab_MMY<-ldply(Lifetab_MMY, as.data.frame)
Lifetab_MMY<-Lifetab_MMY[which(Lifetab_MMY$expo_vec!=0),]
Lifetab_MMY_Year<-unlist(as.numeric(lapply(strsplit(Lifetab_MMY$MMY,
split = "__"), tail, 1)))
Old_liars<-which(c(Lifetab_MMY_Year+Lifetab_MMY$age)>2023)

## finds all matches for MMY, regardless of engine size

head(unique(Lifetab_MMY$MMY))

## how many historic prices match a life table?
mean(paste(toupper(Prices_hist$Make),
           toupper(Prices_hist$Model),
           Prices_hist$Year_priced, sep = "__") %in%
unique(Lifetab_MMY$MMY))

sum(paste(toupper(Prices_hist$Make),
          toupper(Prices_hist$Model),
          Prices_hist$Year_priced, sep = "__") %in%
unique(Lifetab_MMY$MMY))

sum(unique(Lifetab_MMY$MMY) %in% paste(toupper(Prices_hist$Make),
                                       toupper(Prices_hist$Model),
                                       Prices_hist$Year_priced, sep
= "__") )

## 45%; N = 2537 prices, 243 make-model-year combinations

## attaches historic prices to the data frame
Prices_hist_1<-Prices_hist[match(Lifetab_MMY$MMY,
                                 paste(toupper(Prices_hist$Make),
                                       toupper(Prices_hist$Model),
                                       Prices_hist$Year_priced, sep
= "__")),]

Age10<-Lifetab_MMY[which(Lifetab_MMY$age==10),]

q10_vec<-
as.numeric(Age10$qx_vec[match(paste(toupper(Prices_hist$Make),

toupper(Prices_hist$Model),
```

```
Prices_hist$Year_priced, sep = "__"), Age10$MMY)])

## filter by year
qplot(as.numeric(Prices_hist$Price[which(Prices_hist$Year_priced==20
04)]),
      q10_vec[which(Prices_hist$Year_priced==2004)])+
  geom_smooth(method = "lm")

cor.test(as.numeric(Prices_hist$Price[which(Prices_hist$Year_priced=
=2004)]),
         q10_vec[which(Prices_hist$Year_priced==2004)])

qplot(as.numeric(Prices_hist$Price[which(Prices_hist$Year_priced==20
14)]),
      q10_vec[which(Prices_hist$Year_priced==2014)])+
  geom_smooth(method = "lm")

cor.test(as.numeric(Prices_hist$Price[which(Prices_hist$Year_priced=
=2014)]),
         q10_vec[which(Prices_hist$Year_priced==2014)])

qplot(q10_vec[!is.na(q10_vec)],
      as.numeric(Prices_hist$Price[!is.na(q10_vec)]))+
  geom_smooth(method = "lm")

cor.test(q10_vec, log10(as.numeric(Prices_hist$Price)))

## attaches probability of death at age 10 to price data
Prices_hist$q10<-list(NA, length = nrow(Prices_hist))

paste(toupper(Prices_hist$Make), toupper(Prices_hist$Model),
Prices_hist$Year_priced, sep = "__")

## Attaches the historic and current prices to the two respective
cohorts
## as columns in the vehicle properties data frame.

qplot(as.numeric(substr(start = 1,stop =
4,as.character(price_frame_now$X1))),
      as.numeric(price_frame_now$price), alpha = I(0.2))+
  geom_smooth()+
  theme_minimal()

qplot(jitter(as.numeric(substr(start = 1,stop =
4,as.character(price_frame_now$X1)))),
      log10(as.numeric(price_frame_now$price)), alpha = I(0.2), xlab
= "year", ylab = "price")+
  geom_smooth()+
  theme_minimal()
```

```
boxplot(log10(as.numeric(price_frame_now$price))~c(as.numeric(substr
(start = 1,stop = 4,as.character(price_frame_now$X1)))),
          xlab = "year", ylab = "price", pch  ="-")

#######################################################
#######################################################

## very simply, aggregates the major/minor failure rates per MOT
## per week for the entire dataset

#######################################################
#######################################################

Fails_by_age_crossectional<-
aggregate(c(Vehicle_properties$test_result!="PASSED"),
                                  by =
list(floor(LifeVec_time$Death_age_at_test)),
                                        mean, na.rm = T)

Fails_by_age_crossectional<-
Fails_by_age_crossectional[which(Fails_by_age_crossectional>0),]
Fails_by_age_crossectional<-
Fails_by_age_crossectional[which(Fails_by_age_crossectional<120),]

## instead builds "major fails by age" for annual cohorts

## the failure rate by age, all data. Cross-sectional.
boxplot(c(Vehicle_properties$test_result!="PASSED")
[which(LifeVec_time$Death_age_at_test>=1 &

LifeVec_time$Death_age_at_test<110)]~

floor(LifeVec_time$Death_age_at_test[which(LifeVec_time$Death_age_at
_test>=1 &

LifeVec_time$Death_age_at_test<110)]))

## builds a vector of vehicles at risk + deaths
mx_vec<-NULL
age_vec<-c(0:60) ## in years...
dx_vec<-NULL
expo_vec<-NULL
censored_x<-NULL

for(i in c(1:length(age_vec))){

  ## How many vehicles are first observed after the observation
period (are not at risk?)
  nonrisk_pop<-length(which(sub_frame$firstObs>floor((age_vec[i]
+1)*365.25)))
```

```
   ## how many were alive at start of each age?
   at_risk<-sum(sub_frame$Age_at_death>=floor(age_vec[i]*365.25),
na.rm = T)

   ## how many were observed to die during this period?
   died_all<-
length(which(sub_frame$Age_at_death>=floor(age_vec[i]*365.25) &

sub_frame$Age_at_death<floor((age_vec[i+1])*365.25)))

   dx_vec[[i]]<-died_all
   expo_vec[[i]]<-at_risk
   mx_vec[[i]]<-c(died_all/at_risk)

   #  mx_vec[[i]]<-c(died_all/c(at_risk-nonrisk_pop))

}

## subtracts all those who were unobservable because their
## first MOT fell *entirely* after the the observation period
unobservables<-sum(sub_frame$firstObs>=c((i+1)*365.25))
at_risk<-at_risk-unobservables

## how many deaths during the observation period?
died<-sum(c(Age_at_death[sub_frame$fate!="Survived"] %in%
c(floor(i*365.25):c(floor((i+1)*365.25)-1))), na.rm = T)

censored_x<-c(censored_x, unobservables)
mx_vec<-c(mx_vec,c(died/at_risk))
expo_vec<-c(expo_vec, at_risk)
dx_vec<-c(dx_vec, died)

## picks out just the Nissan Micra 1992 MY (an incredible 37k sold
in UK)
## coming out of the Tyne and Wear Nissan factory

## has a peek to check if VIN is good on naming etc.
head(Vehicle_properties[which(Vehicle_properties$vin11=="SJNEAAK11U0
"),], 100)

## Yep all 1L Nissan Micras in petrol. Rad.
## what's the failure rate per test, independent of age?
mean(Vehicle_properties$test_result[which(Vehicle_properties$vin11==
"SJNEAAK11U0")]=="PASSED")
```

```r
## minor repairs?
mean(Vehicle_properties$test_result[which(Vehicle_properties$vin11==
"SJNEAAK11U0")]=="PRS_FAIL")

## great.

## gets the unique IDs we want
sub_index<-
Vehicle_properties$unique_vehicle[which(Vehicle_properties$vin11="S
JNEAAK11U0")]
sub_frame<-LifeVec_time[which(LifeVec_time$unique_vehicle %in%
sub_index),]

table(sub_frame$fate)

## Gets age at death/ltf
Age_at_death<-sub_frame$Death_test_date-
as.numeric(sub_frame$first_use_date)

## screens out any vehicles whose first use date is > 4 years after
the median
## first use date of each make+model (excludes most imports)
N_late_regs<-
sum(sub_frame$first_use_date>median(sub_frame$first_use_date, na.rm
= T)+(365*4), na.rm = T)
Pct_late_regs<-
mean(sub_frame$first_use_date>median(sub_frame$first_use_date, na.rm
= T)+(365*4), na.rm = T)*100

sub_frame<-sub_frame[-
c(which(sub_frame$first_use_date>median(sub_frame$first_use_date,
na.rm = T)+(365*4))),]

## builds a vector of vehicles at risk + deaths
mx_vec<-NULL
age_vec<-c(0:100)
dx_vec<-NULL
expo_vec<-NULL
for(i in age_vec){

  at_risk<-sum(Age_at_death>=c(i*365.25), na.rm = T)
  died<-sum(c(Age_at_death[sub_frame$fate!="Survived"] %in%
c(floor(i*365.25):c(floor((i+1)*365.25)-1))), na.rm = T)

  mx_vec<-c(mx_vec,c(died/at_risk))
  expo_vec<-c(expo_vec, at_risk)
  dx_vec<-c(dx_vec, died)

}
mxtoqx(mx_vec)
Ltab<-data.frame(mx_vec,expo_vec,dx_vec)
```

```
##########################################################

## very simply, aggregates the scrappage rates per MOT
## per week for the entire dataset

## a rough measure of reliability

require(lubridate)

scrap_rates<-NULL
export_rates<-NULL
for(i in 13:17){
  ## gets the year we want
  YearTibble<-readRDS(paste0("/Users/sauley/Documents/R_workspaces/
Analyses/Automobile_survival_UK/data/Emission_moves_data/",
                             MOT_dirs2[[c(i)]]]))

  ## keeps only cars
  YearTibble<-YearTibble[which(YearTibble$test_class==4),]

  mean_scrappage_rate<-aggregate(YearTibble$is_scrapped==1,
                                  by =
list(floor_date(YearTibble$test_date, unit = "week")),
                                  mean, na.rm = T)

  scrap_rates<-rbind(scrap_rates,mean_scrappage_rate)

  ## gets export rates

  mean_scrappage_rate<-aggregate(YearTibble$is_exported==1,
                                  by =
list(floor_date(YearTibble$test_date, unit = "week")),
                                  mean, na.rm = T)

  export_rates<-rbind(export_rates,mean_scrappage_rate)

  print(i)
}

## looks at 5 and 10 year old vehicles only

## a rough measure of reliability

require(lubridate)
```

```
scrap_rates_5Y<-NULL
scrap_rates_10Y<-NULL
export_rates<-NULL
for(i in 13:17){
  ## gets the year we want
  YearTibble<-readRDS(paste0("/Users/sauley/Documents/R_workspaces/
Analyses/Automobile_survival_UK/data/Emission_moves_data/",
                            MOT_dirs2[[c(i)]]]))

  ## keeps 5 Y/O and 10 Y/O vehicles
  #  YearTibble<-paste0("/Volumes/PHOTO_DRIVE/Cars/LifeLocs/",
  #                     tail(unlist(strsplit(MOT_dirs[[i]], split =
"/")), 1),
  #                     "_frame.rds")

  YearTibble<-paste0("/Users/sauley/Documents/R_workspaces/Analyses/
Automobile_survival_UK/data/Survival_data/",
                     tail(unlist(strsplit(MOT_dirs[[i]], split =
"/")), 1),
                     "_frame.rds")
  ## keeps only cars
  # YearTibble<-YearTibble[which(YearTibble$test_class==4),]

  mean_scrappage_rate<-aggregate(YearTibble$is_scrapped==1,
                                 by =
list(floor_date(YearTibble$test_date, unit = "week")),
                                 mean, na.rm = T)

  scrap_rates<-rbind(scrap_rates,mean_scrappage_rate)

  ## gets export rates

  mean_scrappage_rate<-aggregate(YearTibble$is_exported==1,
                                 by =
list(floor_date(YearTibble$test_date, unit = "week")),
                                 mean, na.rm = T)

  export_rates<-rbind(export_rates,mean_scrappage_rate)

  print(i)
}

####################################################################
##

## Builds counts of MOT failure rates over time / by age
```

```
################################################################
#

Fail_rates<-NULL
ticktock<-Sys.time()
MOT_dirs2<-list.files("/Users/sauley/Documents/R_workspaces/
Analyses/Automobile_survival_UK/data/Survival_data/")
MOT_dirs2<-MOT_dirs2[-c(grep("Vehicle_properties", MOT_dirs2))]

for(i in 1:length(MOT_dirs2)){

   Ytib<-readRDS(file =  paste0("/Users/sauley/Documents/
R_workspaces/Analyses/Automobile_survival_UK/data/
Survival_data/",MOT_dirs2[[i]]))

   Fail_rates<-rbind(Fail_rates,
Ytib[,c("unique_vehicle","test_result")])

   beep()
   print(i)
   print(Sys.time()-ticktock)

}

Fail_rates<-table(Fail_rates$unique_vehicle, by =
Fail_rates$test_result)

saveRDS(Fail_rates, file = "data/Survival_data/Fail_rates.rds")

rm(Fail_rates)

##################################################

## This ends the part of the analysis strictly reading the database
provided
## by the Department for Transport. Below here, findings should be
more accessible
## as they depend on processed data that has been publicly shared.

##################################################

## loads regional data

require(survival)
require(survminer)
require(randomForest)
require(fastDummies)

require(randomForestSRC)
```

```
Pcodes<-read.csv("data/metadata/postcodes.csv")

## removes variables we don't need.
Pcodes<-Pcodes[,c("MSOA.Code","Postcode.area", "Postcode.district",
                  "Constituency",
                  "Latitude", "Longitude",
                  "National.Park","Population","Households",

"Built.up.area","Built.up.sub.division","Rural.urban",
                  "Region","Altitude","London.zone",
                  "Index.of.Multiple.Deprivation","Quality",

"Distance.to.station","Average.Income","Travel.To.Work.Area","UPRNs"
,"Distance.to.sea")]

## note: UPRNs are unique property IDs

## adds income data
## small-area incomes etc 2011 and 2018
Small_incomes_2011<-read.csv("data/metadata/Regional_data/
Incomes_smallArea_2011.csv")
Small_incomes_2011<-
Small_incomes_2011[match(Pcodes$MSOA.Code,Small_incomes_2011$MSOA.co
de),]

## keeps what we want
Pcodes$Weekly_income<-Small_incomes_2011$Total.weekly.income....
Pcodes$Weekly_income_CI<-Small_incomes_2011$Confidence.interval....

## net income...
Pcodes$NIBH<-Small_incomes_2011$Net.weekly.income....

## net income before housing...
Pcodes$NIBH<-Small_incomes_2011$Net.income.before.housing.costs....

## Net income after housing
Pcodes$NIAH<-Small_incomes_2011$Net.income.after.housing.costs....

rm(Small_incomes_2011)

## Adds poverty

Small_poverty_2011<-read.csv("data/metadata/Regional_data/
Household_poverty_2011.csv")
Small_poverty_2011<-
Small_poverty_2011[match(Pcodes$MSOA.Code,Small_poverty_2011$MSOA.co
de),]

## keeps what we want: pct households in poverty, before and after
housing costs
Pcodes$Pct_in_poverty_AH<-
Small_poverty_2011$Percentage.of.Households.Below.60..of.the.Median.
Income...after.housing.costs.
Pcodes$Pct_in_poverty_BH<-
```

```
Small_poverty_2011$Percentage.of.Households.Below.60..of.the.Median.
Income...before.housing.costs.

## gets data from the lookup tables for electoral regions, and
region type classifications
Pcode_lookup<-read.csv("data/metadata/Regional_data/
National_Statistics_Postcode_Lookup_UK.csv")

Pcode_lookup<-
Pcode_lookup[,c("Lower.Super.Output.Area.Code","Parliamentary.Consti
tuency.Code",
                        "Output.Area.Classification.Code",
"Output.Area.Classification.Name")]

## gets the vote share in the 2010 general election
VoteShare<-read.csv("data/metadata/Regional_data/GE2010-results-
flatfile-website.csv")

## reduces to major parties, bins "other" voters
VS2<-VoteShare[,-c(1:6)]

## 'major parties' are anyone with >0.5% share of national vote
## Lab = Labor Lab, Con = conservative Con, SNP  = scot national
party,
## LD = Lib Dems , UKIP = United Kingdom Independence Party morons,
## Grn  =Greens, BNP = British National Party, PC        = Plaid
Cymru
## SF = Sinn Fein

Other_Votes<-rowSums(VS2[,which(!(colnames(VS2) %in% c("Lab",
"Con","LD",
                                                        "SNP",
"UKIP", "DUP", "BNP", "PC", "Grn", "SF")))], na.rm  =T)

VS2<-VS2[,which(colnames(VS2) %in% c("Lab", "Con","LD",
                                        "SNP", "UKIP", "DUP", "BNP",
"PC", "Grn", "SF"))]

VS2<-VS2/VoteShare$Votes

VS2$Other<-Other_Votes/VoteShare$Votes

VoteShare<-data.frame(VoteShare[,c(1:6)], VS2)

## corrected constituency names
VoteShare$CCN<-VoteShare$Constituency.Name
Pcodes$CCN<-Pcodes$Constituency

VoteShare$CCN<-gsub(" \\& ", " and ",VoteShare$CCN)
VoteShare$CCN<-gsub(", ", " ",VoteShare$CCN)
VoteShare$CCN<-gsub("-", " ",VoteShare$CCN)

Pcodes$CCN<-gsub(" \\& ", " and ",Pcodes$CCN)
Pcodes$CCN<-gsub(", ", " ",Pcodes$CCN)
```

```
Pcodes$CCN<-gsub("-", " ",Pcodes$CCN)

VoteShare$CCN<-gsub("Wrekin The", "The Wrekin",VoteShare$CCN,
ignore.case = T)
VoteShare$CCN<-gsub("Cotswolds The", "The Cotswolds",VoteShare$CCN,
ignore.case = T)
VoteShare$CCN<-gsub("Chester City Of", "City Of
Chester",VoteShare$CCN, ignore.case = T)
VoteShare$CCN<-gsub("Durham City Of", "City Of
Durham",VoteShare$CCN, ignore.case = T)

## NB order of replacement matters
replace_vec<-c("South West","South East","North West","North East",
               "North", "South", "East", "West", "Central", "Mid ")
for(i in 1:length(replace_vec)){

  Pcodes$CCN[grep(paste0("^",replace_vec[[i]]), Pcodes$CCN,
ignore.case  =T)]<-
paste(Pcodes$CCN[grep(paste0("^",replace_vec[[i]]),

Pcodes$CCN, ignore.case  =T)],

paste0(" ",replace_vec[[i]]))
  Pcodes$CCN<-gsub(paste0("^",replace_vec[[i]]), "",Pcodes$CCN)

  VoteShare$CCN[grep(paste0("^",replace_vec[[i]]), VoteShare$CCN,
ignore.case  =T)]<-
paste(VoteShare$CCN[grep(paste0("^",replace_vec[[i]]),

VoteShare$CCN, ignore.case  =T)],

paste0(" ",replace_vec[[i]]))
  VoteShare$CCN<-gsub(paste0("^",replace_vec[[i]]),
"",VoteShare$CCN)

}

VoteShare$CCN<-trimws(VoteShare$CCN)
Pcodes$CCN<-trimws(Pcodes$CCN)

for(i in 1:3){
  VoteShare$CCN<-gsub("  ", " ",VoteShare$CCN)
  Pcodes$CCN<-gsub("  ", " ",Pcodes$CCN)
}

## fixes remainder manually
VoteShare$CCN[toupper(VoteShare$CCN)=="ABERDEENSHIRE WEST AND
KINCARDINE"]<-"ABERDEENSHIRE AND KINCARDINE WEST"
VoteShare$CCN[toupper(VoteShare$CCN)=="AMPTONSHIRE SOUTH
NORTH"]<-"NORTHAMPTONSHIRE SOUTH"
VoteShare$CCN[toupper(VoteShare$CCN)=="AYRSHIRE NORTH AND
ARRAN"]<-"AYRSHIRE AND ARRAN NORTH"
VoteShare$CCN[toupper(VoteShare$CCN)=="BASILDON SOUTH AND THURROCK
```

```r
EAST"]<-"BASILDON AND EAST THURROCK SOUTH"
VoteShare$CCN[toupper(VoteShare$CCN)=="BRIDGWATER AND SOMERSET
WEST"]<-"BRIDGWATER AND WEST SOMERSET"
VoteShare$CCN[toupper(VoteShare$CCN)=="CARMARTHEN WEST AND
PEMBROKESHIRE SOUTH"]<-"CARMARTHEN WEST AND SOUTH PEMBROKESHIRE"
VoteShare$CCN[toupper(VoteShare$CCN)=="DEVON WEST AND
TORRIDGE"]<-"TORRIDGE AND WEST DEVON"
VoteShare$CCN[toupper(VoteShare$CCN)=="DORSET MID AND POOLE
NORTH"]<-"DORSET AND NORTH POOLE MID"
VoteShare$CCN[toupper(VoteShare$CCN)=="DUNFERMLINE AND FIFE
WEST"]<-"DUNFERMLINE AND WEST FIFE"
VoteShare$CCN[toupper(VoteShare$CCN)=="FAVERSHAM AND KENT
MID"]<-"FAVERSHAM AND MID KENT"
VoteShare$CCN[toupper(VoteShare$CCN)=="HEREFORD AND HEREFORDSHIRE
SOUTH"]<-"HEREFORD AND SOUTH HEREFORDSHIRE"
VoteShare$CCN[toupper(VoteShare$CCN)=="HULL EAST"]<-"KINGSTON UPON
HULL EAST"
VoteShare$CCN[toupper(VoteShare$CCN)=="HULL NORTH"]<-"KINGSTON UPON
HULL NORTH"
VoteShare$CCN[toupper(VoteShare$CCN)=="HULL WEST AND
HESSLE"]<-"KINGSTON UPON HULL WEST AND HESSLE"
VoteShare$CCN[toupper(VoteShare$CCN)=="LINLITHGOW AND FALKIRK
EAST"]<-"LINLITHGOW AND EAST FALKIRK"
VoteShare$CCN[toupper(VoteShare$CCN)=="MIDDLESBROUGH SOUTH AND
CLEVELAND EAST"]<-"MIDDLESBROUGH SOUTH AND EAST CLEVELAND"
VoteShare$CCN[toupper(VoteShare$CCN)=="NA H EILEANAN AN IAR (WESTERN
ISLES)"]<-"NA H EILEANAN AN IAR"
VoteShare$CCN[toupper(VoteShare$CCN)=="SUFFOLK CENTRAL AND IPSWICH
NORTH"]<-"SUFFOLK AND NORTH IPSWICH CENTRAL"
VoteShare$CCN[toupper(VoteShare$CCN)== "UXBRIDGE AND RUISLIP
SOUTH"]<-"UXBRIDGE AND SOUTH RUISLIP"
VoteShare$CCN[toupper(VoteShare$CCN)=="WORTHING EAST AND
SHOREHAM"]<-"WORTHING AND SHOREHAM EAST"
VoteShare$CCN[toupper(VoteShare$CCN)=="YNYS MON"]<-"YNYS MÔN"

set1<-unique(toupper(Pcodes$CCN)[which(!(toupper(Pcodes$CCN) %in%
toupper(VoteShare$CCN)))])
set2<-unique(toupper(VoteShare$CCN[which(!(toupper(VoteShare$CCN)
%in% toupper(Pcodes$CCN)))]))

cbind(set1[order(set1)], set2[order(set2)])

## maps vote share onto the Pcodes data frame
Votes<-
VoteShare[c(match(toupper(Pcodes$CCN),toupper(VoteShare$CCN))),c(6:1
7)]

## coerces non-candidate regions to zero votes
Votes[is.na(Votes)]<-0

Pcodes<-data.frame(Pcodes, Votes)
```

```
rm(Votes, VoteShare, VS2)
gc()

saveRDS(Pcodes, "Pcodes.rds")

## aggregates these metrics up to the postcode region level, for all
numeric vectors

Pcodes_brief<-aggregate(Pcodes$Population, by =
list(Pcodes$Postcode.district), sum, na.rm =T)
Pcodes_brief$Households<-aggregate(Pcodes$Households, by =
list(Pcodes$Postcode.district), sum, na.rm =T)[,2]

Pcodes_brief$Altitude<-aggregate(Pcodes$Altitude, by =
list(Pcodes$Postcode.district), median, na.rm =T)[,2]
Pcodes_brief$IMD<-aggregate(Pcodes$Index.of.Multiple.Deprivation, by
= list(Pcodes$Postcode.district), median, na.rm =T)[,2]
Pcodes_brief$Distance_Station<-aggregate(Pcodes$Distance.to.station,
by = list(Pcodes$Postcode.district), median, na.rm =T)[,2]
Pcodes_brief$Distance_Sea<-aggregate(Pcodes$Distance.to.sea, by =
list(Pcodes$Postcode.district), median, na.rm =T)[,2]

Pcodes_brief$Average_income<-aggregate(Pcodes$Average.Income, by =
list(Pcodes$Postcode.district), mean, na.rm =T)[,2]
Pcodes_brief$Average_wk_income<-aggregate(Pcodes$Weekly_income, by =
list(Pcodes$Postcode.district), mean, na.rm =T)[,2]
Pcodes_brief$NIBH<-aggregate(Pcodes$NIBH, by =
list(Pcodes$Postcode.district), mean, na.rm =T)[,2]
Pcodes_brief$NIAH<-aggregate(Pcodes$NIAH, by =
list(Pcodes$Postcode.district), mean, na.rm =T)[,2]

Pcodes_brief$Pct_poverty_AH<-aggregate(Pcodes$Pct_in_poverty_AH, by
= list(Pcodes$Postcode.district), median, na.rm =T)[,2]
Pcodes_brief$Pct_poverty_BH<-aggregate(Pcodes$Pct_in_poverty_BH, by
= list(Pcodes$Postcode.district), median, na.rm =T)[,2]

Pcodes_brief$minQuality<-aggregate(Pcodes$Quality, by =
list(Pcodes$Postcode.district), min, na.rm =T)[,2]
Pcodes_brief$maxQuality<-aggregate(Pcodes$Quality, by =
list(Pcodes$Postcode.district), max, na.rm =T)[,2]

Pcodes_brief$maxQuality[Pcodes_brief$minQuality==-Inf]<-NA
Pcodes_brief$minQuality[Pcodes_brief$minQuality==Inf]<-NA

Pcodes_brief$vote_Lab<-aggregate(Pcodes$Lab, by =
list(Pcodes$Postcode.district), median, na.rm =T)[,2]
Pcodes_brief$vote_LibDem<-aggregate(Pcodes$LD, by =
list(Pcodes$Postcode.district), median, na.rm =T)[,2]
Pcodes_brief$vote_Con<-aggregate(Pcodes$Con, by =
list(Pcodes$Postcode.district), median, na.rm =T)[,2]
Pcodes_brief$vote_UKIP<-aggregate(Pcodes$UKIP, by =
list(Pcodes$Postcode.district), median, na.rm =T)[,2]
Pcodes_brief$vote_Grn<-aggregate(Pcodes$Grn, by =
list(Pcodes$Postcode.district), median, na.rm =T)[,2]
```

```
Pcodes_brief$vote_PC<-aggregate(Pcodes$PC, by =
list(Pcodes$Postcode.district), median, na.rm =T)[,2]
Pcodes_brief$vote_DUP<-aggregate(Pcodes$DUP, by =
list(Pcodes$Postcode.district), median, na.rm =T)[,2]
Pcodes_brief$vote_SNP<-aggregate(Pcodes$SNP, by =
list(Pcodes$Postcode.district), median, na.rm =T)[,2]
Pcodes_brief$vote_SF<-aggregate(Pcodes$SF, by =
list(Pcodes$Postcode.district), median, na.rm =T)[,2]
Pcodes_brief$vote_BNP<-aggregate(Pcodes$BNP, by =
list(Pcodes$Postcode.district), median, na.rm =T)[,2]
Pcodes_brief$vote_Other<-aggregate(Pcodes$Other, by =
list(Pcodes$Postcode.district), median, na.rm =T)[,2]

## encodes the regional classifications as presence/absence
Rural_frame<-NULL
for(i in 1:length(unique(Pcodes$Rural.urban))){
  Rural_code<-unique(Pcodes$Rural.urban)[[i]]

  Rural_sub<-aggregate(as.numeric(Pcodes$Rural.urban==Rural_code),
by = list(Pcodes$Postcode.district), sum, na.rm =T)[,2]
  Rural_frame[[i]]<-as.numeric(Rural_sub>0)

}

Rural_frame<-matrix(unlist(Rural_frame), ncol = length(Rural_frame),
byrow = F)
xobj<-unique(Pcodes$Rural.urban)
xobj[xobj==""]<-"Unclassified"
xobj<-gsub(" ", "_",xobj)
xobj<-paste0("Rural_urban_",xobj)

colnames(Rural_frame)<-xobj
Rural_frame<-as.data.frame(Rural_frame)
Rural_frame$N_Classifications<-rowSums(Rural_frame)

Pcodes_brief<-data.frame(Pcodes_brief,Rural_frame)

## the same with London zones
## encodes the London zones as presence/absence
xobj<-Pcodes$London.zone
xobj[is.na(xobj)]<-10
## changes from NAs to arbitrary values
London_frame<-NULL
for(i in 1:length(unique(xobj))){
  London_code<-unique(xobj)[[i]]

  London_sub<-aggregate(as.numeric(xobj==London_code),
                        by = list(Pcodes$Postcode.district), sum,
na.rm =T)[,2]
  London_frame[[i]]<-as.numeric(London_sub>0)

}

London_frame<-matrix(unlist(London_frame), ncol =
```

```
length(London_frame), byrow = F)
London_frame<-London_frame[,!is.na(unique(Pcodes$London.zone))]

xobj<-unique(Pcodes$London.zone)
xobj<-xobj[!is.na(xobj)]
xobj<-gsub(" ", "_",xobj)
xobj<-paste0("London_zone_",xobj)
colnames(London_frame)<-xobj

London_frame<-as.data.frame(London_frame)
#London_frame$N_Zones<-rowSums(London_frame)
Pcodes_brief<-data.frame(Pcodes_brief,London_frame)

## national parks present?
Pcodes_brief$NatParks<-
aggregate(as.numeric(Pcodes$National.Park==""), by =
list(Pcodes$Postcode.district), sum, na.rm =T)[,2]
Pcodes_brief$NatParks<-as.numeric(Pcodes_brief$NatParks>0)

# nb: only 17 regions are not national park adjacent

## and adds approx. lat/longs
Pcodes_brief$approx_Lat<-aggregate(Pcodes$Latitude, by =
list(Pcodes$Postcode.district), mean, na.rm =T)[,2]
Pcodes_brief$approx_Long<-aggregate(Pcodes$Longitude, by =
list(Pcodes$Postcode.district), mean, na.rm =T)[,2]

colnames(Pcodes_brief)[1:2]<-c("Postcode", "Population")

saveRDS(Pcodes_brief, file = "Pcodes_brief.rds")
saveRDS(Pcodes, file = "Pcodes.rds")
rm(Pcodes)
gc()

##################################################################
####
##################################################################
####

## gets all the vehicles tested in 2011 (the census year)

## Loads year tibble
#Ytib<-readRDS("/Users/sauley/Documents/R_workspaces/Analyses/
Automobile_survival_UK/data/LifeLocs/e1_2011_frame.rds")
Ytib<-readRDS("/Users/qtnzsne/Documents/R_workspaces/Analyses/
Automobile_survival_UK/data/LifeLocs/e1_2011_frame.rds")

head(Ytib)

## loads vehicle props table again
Vehicle_properties<-readRDS("/Users/qtnzsne/Documents/R_workspaces/
Analyses/Automobile_survival_UK/data/Survival_data/
Vehicle_properties.rds")
```

```r
## Matches to vehicle ID, which also removes the low-quality calls
Year_2011_IDs<-
Vehicle_properties[which(Vehicle_properties$unique_vehicle %in%
Ytib$unique_vehicle),]

Year_2011_IDs$MMY<-MMY_index[which(Vehicle_properties$unique_vehicle
%in% Ytib$unique_vehicle)]

## how many are <10 frequency?
length(which(table(Year_2011_IDs$MMY)<10))

## 176063 vehicles

## keeps vehicle types appearing at least 10 times (removes lots of
typos)
Year_2011_IDs<-Year_2011_IDs[which(Year_2011_IDs$MMY %in%
rownames(table(Year_2011_IDs$MMY))
[which(table(Year_2011_IDs$MMY)>=10)]),]
table(Year_2011_IDs$MMY)[which(table(Year_2011_IDs$MMY)>=10)]

## cuts down to common make-model-years only
#Year_2011_IDs<-Year_2011_IDs[which(Year_2011_IDs$MMY %in%
MMY_index_1k),]

length(unique(Year_2011_IDs$MMY[which(Year_2011_IDs$MMY %in%
MMY_index_1k)]))

## what are the most common makes and models?
tail(table(Year_2011_IDs$MMY)[order(table(Year_2011_IDs$MMY))],100)

## how many of them are there?
sum(tail(table(Year_2011_IDs$MMY)
[order(table(Year_2011_IDs$MMY))],100))

## For all these vehicles, counts how many previous MOTs were
conducted, and failed or passed.

## Loads year tibble
#MOT_dirs3<-list.files("/Users/sauley/Documents/R_workspaces/
Analyses/Automobile_survival_UK/data/Survival_data/")
MOT_dirs3<-list.files("/Users/qtnzsne/Documents/R_workspaces/
Analyses/Automobile_survival_UK/data/Survival_data/")

## keeps a count by type etc
N_tests_2011<-NULL
N_fails_2011<-NULL
N_prs_fails_2011<-NULL
ticktock<-Sys.time()

for(i in 1:6){

  #Ytib<-readRDS(paste0("/Users/sauley/Documents/R_workspaces/
Analyses/Automobile_survival_UK/data/
```

```r
Survival_data/",MOT_dirs3[[i]]]))
  Ytib<-readRDS(paste0("/Users/qtnzsne/Documents/R_workspaces/
Analyses/Automobile_survival_UK/data/
Survival_data/",MOT_dirs3[[i]]]))

  Ytib<-Ytib[which(Ytib$unique_vehicle %in%
Year_2011_IDs$unique_vehicle),]

  ## aggregates N tests for each ID
  N_tests_sub<-table(Ytib$unique_vehicle)
  N_fails_sub<-
table(Ytib$unique_vehicle[Ytib$test_result=="FAILED"])
  N_prs_sub<-
table(Ytib$unique_vehicle[Ytib$test_result=="PRS_FAIL"])

  ## merges these into a uniform data frame
  N_tests_sub<-as.data.frame(N_tests_sub)
  N_fails_sub<-as.data.frame(N_fails_sub)
  N_prs_sub<-as.data.frame(N_prs_sub)

  N_tests_2011[[i]]<-c(N_tests =
N_tests_sub[match(Year_2011_IDs$unique_vehicle, N_tests_sub$Var1),],
                       N_fails =
N_fails_sub[match(Year_2011_IDs$unique_vehicle, N_fails_sub$Var1),],
                       N_prs =
N_prs_sub[match(Year_2011_IDs$unique_vehicle, N_prs_sub$Var1),])

  rm(N_tests_sub,N_fails_sub,N_prs_sub)
  print(i)
  print(ticktock-Sys.time())
}

## kinda dumb to loop this but...
Test_count_2011<-cbind(N_tests_2011[[1]]$N_tests.Freq,
N_tests_2011[[1]]$N_fails.Freq, N_tests_2011[[1]]$N_prs.Freq)
Test_count_2011[is.na(Test_count_2011)]<-0
for(i in 2:6){
  xobj<-cbind(N_tests_2011[[i]]$N_tests.Freq, N_tests_2011[[i]]
$N_fails.Freq, N_tests_2011[[i]]$N_prs.Freq)
  xobj[is.na(xobj)]<-0
  Test_count_2011<-Test_count_2011+xobj
}
Test_count_2011<-as.data.frame(Test_count_2011)
colnames(Test_count_2011)<-c("N_tests", "N_failed", "N_prs_fail")
Test_count_2011$fraction_passed<-c(Test_count_2011[,1]/
rowSums(Test_count_2011))

Test_count_2011$fraction_passed[is.nan(Test_count_2011$fraction_pass
ed)]<-NA

rm(N_tests_2011)
rm(Ytib)
gc()
```

```
## Merges on
Year_2011_IDs<-data.frame(Year_2011_IDs, Test_count_2011)

rm(Test_count_2011)
gc()

## re-loads ytib data
#Ytib<-readRDS("/Users/sauley/Documents/R_workspaces/Analyses/
Automobile_survival_UK/data/LifeLocs/e1_2011_frame.rds")
Ytib<-readRDS("/Users/qtnzsne/Documents/R_workspaces/Analyses/
Automobile_survival_UK/data/LifeLocs/e1_2011_frame.rds")

## adds age at test data, odometer at test, location of test, etc.
Ytib<-Ytib[match(Year_2011_IDs$unique_vehicle,
Ytib$unique_vehicle),]

Year_2011_IDs$postcode<-Ytib$outer_postcode
Year_2011_IDs$age_at_test<-Ytib$age_at_test
Year_2011_IDs$test_date<-Ytib$test_date
Year_2011_IDs$first_use_date<-Ytib$first_use_date
Year_2011_IDs$corrected_odometer<-Ytib$corrected_odometer
Year_2011_IDs$fate<-Ytib$fate

Year_2011_IDs$test_date<-as.Date(Year_2011_IDs$test_date,
                                 format = "%Y%m%d", origin =
"1800-01-01")

## Makes the zero engine_cc values into NA, unless they're electric
Year_2011_IDs$engine_cc[which(Year_2011_IDs$engine_cc==0 &
Year_2011_IDs$fuel_name!="Electric")]<-NA

## Keeps only the last test of the year
Year_2011_IDs<-Year_2011_IDs[order(Year_2011_IDs$test_date),]

## counts and removes exports
sum(Year_2011_IDs$fate=="exported") ## N = 434,571
#Year_2011_IDs<-Year_2011_IDs[which(Year_2011_IDs$fate!
="exported"),]

## Loads vehicles tested in 2011
#LifeStart<-readRDS("/Users/sauley/Documents/R_workspaces/Analyses/
Automobile_survival_UK/data/LifeStart.rds")
LifeStart<-readRDS("/Users/qtnzsne/Documents/R_workspaces/Analyses/
Automobile_survival_UK/data/LifeStart.rds")

LifeStart<-LifeStart[which(LifeStart$unique_vehicle %in%
Year_2011_IDs$unique_vehicle),]

## keeps the first postcode, first odometer reading
Year_2011_IDs$first_postcode<-
LifeStart$outer_postcode[match(Year_2011_IDs$unique_vehicle,
LifeStart$unique_vehicle)]
Year_2011_IDs$first_odo<-
LifeStart$corrected_odometer[match(Year_2011_IDs$unique_vehicle,
```

```
LifeStart$unique_vehicle)]

rm(LifeStart)

## gets death
#LifeStop<-readRDS("/Users/sauley/Documents/R_workspaces/Analyses/
Automobile_survival_UK/data/LifeStop.rds")
LifeStop<-readRDS("/Users/qtnzsne/Documents/R_workspaces/Analyses/
Automobile_survival_UK/data/LifeStop.rds")

LifeStop<-LifeStop[which(LifeStop$unique_vehicle %in%
Year_2011_IDs$unique_vehicle),]

## keeps death odometer, age, and cause
Year_2011_IDs$death_age<-
LifeStop$Death_age_at_test[match(Year_2011_IDs$unique_vehicle,
LifeStop$unique_vehicle)]
Year_2011_IDs$death_odometer<-
LifeStop$Death_corrected_odometer[match(Year_2011_IDs$unique_vehicle
, LifeStop$unique_vehicle)]
Year_2011_IDs$death_fate<-
LifeStop$fate[match(Year_2011_IDs$unique_vehicle,
LifeStop$unique_vehicle)]

rm(LifeStop)

## cleans two oddities
sum(is.na(Year_2011_IDs$age_at_test))
Year_2011_IDs<-Year_2011_IDs[which(!
is.na(Year_2011_IDs$age_at_test)),]

## Survival time, in years
Year_2011_IDs$SurvTime<-c(Year_2011_IDs$death_age-
Year_2011_IDs$age_at_test)

## removes any impossible survival times (accounting for rounding
errors)
sum(Year_2011_IDs$SurvTime<(-0.1))  ## 4731 errors
mean(Year_2011_IDs$SurvTime<(-0.1)) ##  fraction errors =
0.0002623886

sum(Year_2011_IDs$SurvTime>12)  ## 143 errors observed post-2021
mean(Year_2011_IDs$SurvTime>12) ##  fraction errors =  7.931003e-06

## removes the 1.4% of vehicles where the odometer goes down –
## sometimes happens when odometers exceed maximum values of 1
million or 100k
## but more often is likely just typos

Year_2011_IDs$SurvTime_odometer<-c(Year_2011_IDs$death_odometer-
Year_2011_IDs$corrected_odometer)
sum(Year_2011_IDs$SurvTime_odometer<0, na.rm = T)  ## 249,576 errors
observed
mean(Year_2011_IDs$SurvTime_odometer<0, na.rm = T) ##  fraction
```

```
errors = 0.01399643

Year_2011_IDs<-Year_2011_IDs[which(!
(Year_2011_IDs$SurvTime<(-0.1))),]
Year_2011_IDs<-Year_2011_IDs[which(!(Year_2011_IDs$SurvTime>12)),]
Year_2011_IDs<-Year_2011_IDs[which(!
(Year_2011_IDs$SurvTime_odometer<0)),]

## for stratification by testing date, adds a "Julian Days at
Testing" variable
## indicating the day of the year testing occurred, as this will
obviously be
## confounded with survival.

#c(as.Date("01-01-2011",
#        format = "%d-%m-%Y", origin = "01-01-1800")-
Year_2011_IDs$test_date)

## Julian day at Testing, (seasonal purchase+expenditure effects)
Year_2011_IDs$JDAT<-as.numeric(Year_2011_IDs$test_date-
as.Date("01-01-2011",

format = "%d-%m-%Y", origin = "01-01-1800"))+1

## loads vehicles passing QC
HighQ_ids<-readRDS(file = "data/Emissions_QC_list.rds")

HighQ_ids<-HighQ_ids$unique_vehicle[HighQ_ids$HighQ==1]

## removes and counts QC failures
sum(Year_2011_IDs$unique_vehicle %in% HighQ_ids)
sum(!(Year_2011_IDs$unique_vehicle %in% HighQ_ids))

## 17,138,392 high-quality vehicles left, 439,370 low-quality
vehicles removed
Year_2011_IDs<-Year_2011_IDs[which(Year_2011_IDs$unique_vehicle %in%
HighQ_ids),]

rm(HighQ_ids)
gc()

## and reconstructs the 'missing presumed dead' value for anything
not
## MOT-ed since 1/6/2019 (last eighteen months of records)
Year_2011_IDs$fate<-
LifeVec_time$fate[match(Year_2011_IDs$unique_vehicle,
LifeVec_time$unique_vehicle)]

table(Year_2011_IDs$fate)

##cert_destroyed      exported        MPD       scrapped
survived
```

```
##       3260433        392759        8047842        56829
5380529

Year_2011_IDs$died<-c(Year_2011_IDs$fate!="survived")

###################

## Adds an interesting extra: Baseline frequency of the vehicle-
## rare vehicles have higher perceived social value but lower
reliability/survival:
## they become rare because of
## worse mortality rates / lower reliability.

Base_freq<-table(Year_2011_IDs$MMY)

Year_2011_IDs$Base_freq<-
Base_freq[match(Year_2011_IDs$MMY,rownames(Base_freq))]

## now we have everything we need.
saveRDS(Year_2011_IDs, "Year_2011_IDs.rds")

###############################################################
#########
###############################################################
#########

## constructs mortality models for single vehicle types.

###############################################################
#########
###############################################################
#########

Freq_2011<-table(Year_2011_IDs$MMY)[order(table(Year_2011_IDs$MMY))]
tail(Freq_2011, 50)

## gets the 100 most common cars, 20 most common vans and motorbikes
Freq_Cars_2011<-
table(Year_2011_IDs$MMY[which(Year_2011_IDs$test_class==4)])
[order(table(Year_2011_IDs$MMY[which(Year_2011_IDs$test_class==4)]))
]
Freq_Moto_2011<-
table(Year_2011_IDs$MMY[which(Year_2011_IDs$test_class<=2)])
[order(table(Year_2011_IDs$MMY[which(Year_2011_IDs$test_class<=2)]))
]
Freq_Vans_2011<-
table(Year_2011_IDs$MMY[which(Year_2011_IDs$test_class==7)])
[order(table(Year_2011_IDs$MMY[which(Year_2011_IDs$test_class==7)]))
]

#target_vehicles<-names(tail(Freq_Cars_2011, 100))
target_vehicles2<-names(tail(Freq_Moto_2011, 20))
target_vehicles3<-names(tail(Freq_Vans_2011, 20))
```

```r
## gets every vehicle with over N=10,000 instances on the road
target_vehicles<-names(which(Freq_Cars_2011>=10000))

## Builds one huge model, stratified by vehicle type (fixed effects)
sub_frame<-Year_2011_IDs[which(Year_2011_IDs$MMY %in%
target_vehicles),]
## removes LTFs
sub_frame<-sub_frame[which(sub_frame$death_fate!="exported"),]

## Merges on relevant vehicle properties for each 2011 vehicle
#Vehicle_properties<-readRDS("/Users/sauley/Documents/R_workspaces/
Analyses/Automobile_survival_UK/data/Survival_data/
Vehicle_properties.rds")
Vehicle_properties<-readRDS("/Users/qtnzsne/Documents/R_workspaces/
Analyses/Automobile_survival_UK/data/Survival_data/
Vehicle_properties.rds")

## Just keeps make and fuel name
Vehicle_properties<-Vehicle_properties[,c("unique_vehicle","make",
"fuel_name")]

## cut the ~58 million we don't need, keep fuel name
sub_frame$fuel_name<-
Vehicle_properties$fuel_name[match(sub_frame$unique_vehicle,Vehicle_
properties$unique_vehicle)]

rm(Vehicle_properties)
gc()

## fuel types?
table(sub_frame$fuel_name)

## Bins minor fuel types (N<200) into "other"
sub_frame$fuel_name[sub_frame$fuel_name %in% c("CNG", "LNG","Gas Bi-
Fuel",
                                                "Steam", "Gas", "Fuel
Cells")]<-"Other"

table(sub_frame$fuel_name)

## merges on social data
sub_frame<-data.frame(sub_frame,
                      Pcodes_brief[match(sub_frame$postcode,
Pcodes_brief$Postcode),])

## makes age at test into a strata
sub_frame$AAT<-floor(sub_frame$age_at_test)
## removes vehicles that died on the date of testing
sub_frame<-
sub_frame[which(sub_frame$death_age>sub_frame$age_at_test),]

## Converts mileage and engine size into easier-to-comprehend units
```

```
- per Liter and per 10k miles
sub_frame$engine_cc<-c(sub_frame$engine_cc/1000)
sub_frame$corrected_odometer<-c(sub_frame$corrected_odometer/10000)

## does the same with income metrics
## turns net income  / income per week into 1 unit = 100 pounds
sub_frame$NIBH<-sub_frame$NIBH/100
sub_frame$NIAH<-sub_frame$NIAH/100
sub_frame$Average_wk_income<-sub_frame$Average_wk_income/100

## turns overall income into 10k pound increments
sub_frame$Average_income<-sub_frame$Average_income/10000

## rescales IMD to max = 10
sub_frame$IMD<-sub_frame$IMD/3251.2

## what fraction is missing? 0.73% or 4,015,863 NA values
mean(is.na(sub_frame))

saveRDS(sub_frame,"throwaway.rds")

require("randomForestSRC")
set.seed(293)
system.time(sub_frame<-impute(data = sub_frame, blocks = 50,
                              splitrule = "random",
                              nodesize = 500, fast = TRUE, seed =
-19338))

## random sample imputation (less accurate but unbiased)
#set.seed(9373)
#for(i in c(1:28,30:ncol(sub_frame))){

#  if(sum(is.na(sub_frame[,i]))>0){
#
#    rand_vec<-sample(sub_frame[!is.na(sub_frame[,i]),i],
#                     replace =T)
#                     size = sum(is.na(sub_frame[,i]))),
#    sub_frame[is.na(sub_frame[,i]),i]<-rand_vec
#    rm(rand_vec)
#  }
#}

## Gets cars only
sub_frame_cars<-sub_frame[which(sub_frame$test_class==4),]

dim(sub_frame_cars)

set.seed(101)

## Doesn't stratify by make-model, but instead predicts make effects
## independent of odometer readings, geographic location, and urban/
```

nonurban divide
```
system.time(CPH_makes<-coxph(Surv(age_at_test,death_age, died) ~

JDAT+corrected_odometer+approx_Lat*approx_Long+

Rural_urban_Urban_major_conurbation+make, data = sub_frame_cars))

## builds model to estimate baseline concordance
system.time(CPH_basecase<-coxph(Surv(age_at_test,death_age, died) ~
                                JDAT+
                                strata(MMY), data =
sub_frame_cars))

## NB fuel type makes a difference (electric is better)
system.time(CPH_physical<-coxph(Surv(age_at_test,death_age, died) ~
                          JDAT+corrected_odometer+engine_cc+

N_tests*N_failed+fuel_name+approx_Lat*approx_Long+
                                strata(MMY), data = sub_frame_cars))

## Minimal social predictors: income and lat/long only
system.time(CPH_basic_social<-coxph(Surv(age_at_test,death_age,
died) ~

JDAT+IMD+approx_Lat*approx_Long+
                                        strata(MMY), data =
sub_frame_cars))

## Social+spatiotemporal predictors only
system.time(CPH_full_social<-coxph(Surv(age_at_test,death_age, died)
~

JDAT+IMD+NIBH+NIAH+Population+Pct_poverty_BH+Distance_Station+
                                vote_Lab+vote_Con+vote_LibDem+

Rural_urban_Urban_major_conurbation+approx_Lat*approx_Long+
                                strata(MMY), data = sub_frame_cars))

system.time(CPH_full<-coxph(Surv(age_at_test,death_age, died) ~

JDAT+corrected_odometer+engine_cc+N_tests*N_failed+

IMD+NIBH+NIAH+Population+Pct_poverty_BH+Distance_Station+
                                vote_Lab+vote_Con+vote_LibDem+

Rural_urban_Urban_major_conurbation+approx_Lat*approx_Long+
                                strata(MMY), data = sub_frame_cars))

summary(CPH_physical)
summary(CPH_full_social)

cox.zph(CPH_full)
cox.zph(CPH_physical)
```

```r
cox.zph(CPH_full_social)

#plot(cox.zph(CPH_full))

###################################################################
##########

###################################################################
##########

## constructs the same model, but for progressively older vehicles
car_age_coefs<-NULL
omnibus_t<-NULL

for(i in 1:14){
  sub_frame_cars_t<-sub_frame_cars[grep(paste0(i+1994, "$"),
sub_frame_cars$MMY),]

  system.time(CPH_full_t<-coxph(Surv(age_at_test,death_age, died) ~

JDAT+corrected_odometer+engine_cc+N_tests*N_failed+

NIBH+IMD+NIAH+Population+Pct_poverty_BH+Distance_Station+
                                vote_Lab+vote_Con+vote_LibDem+

Rural_urban_Urban_major_conurbation+approx_Lat*approx_Long+
                                strata(MMY), data =
sub_frame_cars_t))
  xobj2<-summary(CPH_full_t)

  ## income indicators:
  CPH_social_income<-coxph(Surv(age_at_test,death_age, died) ~

NIBH+IMD+NIAH+Pct_poverty_BH+approx_Lat*approx_Long+
                                strata(MMY), data = sub_frame_cars_t)

  ## now physical indicators
  CPH_miles<-coxph(Surv(age_at_test,death_age, died) ~
corrected_odometer+engine_cc+fuel_name+N_tests*N_failed+
                        approx_Lat*approx_Long+
                        strata(MMY), data = sub_frame_cars_t)

  cph_t_z<-cox.zph(CPH_full_t)

  omnibus_t[[i]]<-cph_t_z$table[nrow(cph_t_z$table),3]

  car_age_coefs[[i]]<-data.frame(model = rep(i, times = 19),
                                Year = rep(i+1994, times = 19),
                                sample_N = rep(xobj2$n, times =
19),
                                Concordance =
rep(xobj2$concordance[1], times = 19),
                                Concordance_soc =
```

```
                         rep(summary(CPH_social_income)$concordance[1], times = 19),
                                            Concordance_mech =
rep(summary(CPH_miles)$concordance[1], times = 19),
                                            Var = rownames(xobj2$coefficients),
                                            xobj2$coefficients)
    print(i+1994)
}
car_age_coefs<-ldply(car_age_coefs, as.data.frame)

plot(car_age_coefs$Year[car_age_coefs$Var=="NIAH"],
car_age_coefs$z[car_age_coefs$Var=="NIAH"])
plot(car_age_coefs$Year[car_age_coefs$Var=="corrected_odometer"],
      car_age_coefs$z[car_age_coefs$Var=="corrected_odometer"])

## constructs the same model, but for progressively higher mileage
vehicles
q10<-quantile(sub_frame_cars$corrected_odometer, seq(0,1,by = 0.1))

car_mile_coefs<-NULL
for(i in 1:10){
    sub_frame_cars_t<-
sub_frame_cars[which(sub_frame_cars$corrected_odometer>=q10[i] &

sub_frame_cars$corrected_odometer<q10[(i+1)]),]
    system.time(CPH_full_t<-coxph(Surv(age_at_test,death_age, died) ~

JDAT+corrected_odometer+engine_cc+N_tests*N_failed+

NIBH+IMD+NIAH+Population+Pct_poverty_BH+Distance_Station+
                                    vote_Lab+vote_Con+vote_LibDem+

Rural_urban_Urban_major_conurbation+approx_Lat*approx_Long+
                                    strata(MMY), data =
sub_frame_cars_t))
    xobj2<-summary(CPH_full_t)

    ## income indicators:
    CPH_social_income<-coxph(Surv(age_at_test,death_age, died) ~

NIBH+IMD+NIAH+Pct_poverty_BH+approx_Lat*approx_Long+
                            strata(MMY), data = sub_frame_cars_t)

    ## now physical indicators
    CPH_miles<-coxph(Surv(age_at_test,death_age, died) ~
corrected_odometer+engine_cc+fuel_name+N_tests*N_failed+
                    approx_Lat*approx_Long+
                    strata(MMY), data = sub_frame_cars_t)

    car_mile_coefs[[i]]<-data.frame(model = rep(i, times = 19),
                                    q10 = rep(i, times = 19),
                                    sample_N = rep(xobj2$n, times =
19),
                                    Concordance =
```

```
                rep(xobj2$concordance[1], times = 19),
                                    Concordance_soc =
        rep(summary(CPH_social_income)$concordance[1], times = 19),
                                    Concordance_mech =
        rep(summary(CPH_miles)$concordance[1], times = 19),
                                    Var = rownames(xobj2$coefficients),
                                    xobj2$coefficients)
    print(i)
}
car_mile_coefs<-ldply(car_mile_coefs, as.data.frame)

plot(car_mile_coefs$q10[car_age_coefs$Var=="corrected_odometer"],
        car_mile_coefs$z[car_age_coefs$Var=="corrected_odometer"])

plot(car_mile_coefs$q10[car_age_coefs$Var=="corrected_odometer"],
        car_mile_coefs$exp.coef.
[car_age_coefs$Var=="corrected_odometer"], ylim = c(0,2))
points(car_mile_coefs$q10[car_age_coefs$Var=="corrected_odometer"],

exp(car_mile_coefs$coef[car_age_coefs$Var=="corrected_odometer"]+
            car_mile_coefs$se.coef.
[car_age_coefs$Var=="corrected_odometer"]), pch = "-")

    points(car_mile_coefs$q10[car_age_coefs$Var=="corrected_odometer"],

exp(car_mile_coefs$coef[car_age_coefs$Var=="corrected_odometer"]-
                car_mile_coefs$se.coef.
[car_age_coefs$Var=="corrected_odometer"]), pch = "-")
abline(h = 1, col = "red")

###################################

## builds standalone models for each common MMY to compare the
stability of coefficients
set.seed(2038)
Cars_model_out<-NULL
## keeps the omnibus test
omni_MMY_cox<-NULL
for(i in 1:length(target_vehicles)){

    ## constructs a traditional survival model, nested, to evaluate
the
    ## relative predictive capacity of wear-and-tear versus social
variables

    ## gets subset
    sub_frame2<-
sub_frame_cars[which(sub_frame_cars$MMY==target_vehicles[[i]]),]

    ## builds full model
    system.time(CPH_full_sub<-coxph(Surv(age_at_test,death_age, died)
~

JDAT+corrected_odometer+engine_cc+N_tests*N_failed+
```

```r
NIBH+IMD+NIAH+Population+Pct_poverty_BH+Distance_Station+
                                vote_Lab+vote_Con+vote_LibDem+

Rural_urban_Urban_major_conurbation+approx_Lat*approx_Long,
                              data = sub_frame2))
  xobj2<-summary(CPH_full_sub)

  # Okay let's do it.
  ## basic social: just net income (before housing), and nothing
else
 # CPH_social<-coxph(Surv(age_at_test,death_age, died) ~ NIBH, data
= sub_frame2)

  ## income indicators:
  CPH_social_income<-coxph(Surv(age_at_test,death_age, died) ~

NIBH+IMD+NIAH+Pct_poverty_BH+approx_Lat*approx_Long, data =
sub_frame2)

  ## now physical indicators
  CPH_miles<-coxph(Surv(age_at_test,death_age, died) ~
corrected_odometer+engine_cc+fuel_name+N_tests*N_failed+
                    approx_Lat*approx_Long, data = sub_frame2)

  ## the zph test
  cph_obj<-matrix(rep(NA, times = 19*2), ncol = 2)
  try(cph_obj<-cox.zph(CPH_full_sub)[[1]][1:19,c(1,3)])

  #cph_t_z<-cox.zph(CPH_full_t)
  cph_omni_sub<-NA
  try(cph_omni_sub<-cox.zph(CPH_full_sub)[[1]][20,3])
  omni_MMY_cox[[i]]<-cph_omni_sub

  Cars_model_out[[i]]<-data.frame(model = rep(i, times = 19),
                                MMY = rep(target_vehicles[[i]],
times = 19),
                                sample_N = rep(xobj2$n, times =
19),
                                Concordance =
rep(xobj2$concordance[1], times = 19),
                                Concordance_soc =
rep(summary(CPH_social_income)$concordance[1], times = 19),
                                Concordance_mech =
rep(summary(CPH_miles)$concordance[1], times = 19),
                                Var =
rownames(xobj2$coefficients),
                                xobj2$coefficients,
                                cph_obj)

  rm(CPH_miles, CPH_social_income,CPH_full_sub,sub_frame2)
  xobj2<-NULL
```

```
  print(i)
}

Car_mods<-ldply(Cars_model_out, as.data.frame)

boxplot(abs(Car_mods$z)~Car_mods$Var, las = 2, horizontal = F, pch =
"-")
boxplot((Car_mods$z)~Car_mods$Var, las = 2, horizontal = F, pch =
"-")

## orders by effect size, predictor type
Car_mods2<-Car_mods

#Re_order<-with(Car_mods2, reorder(Var, z, median , na.rm=T,
decreasing = F))
#Car_mods2<-Car_mods2[order(Car_mods2$z, decreasing = T),]

Car_mods2$Variable<-factor(Car_mods2$Var, levels = c("JDAT",

"corrected_odometer","engine_cc",
                                                  "N_tests",
"N_failed","N_tests:N_failed",

"IMD","NIBH","NIAH","Population","Pct_poverty_BH",

"vote_Con","vote_Lab","vote_LibDem",

"Distance_Station", "Rural_urban_Urban_major_conurbation",

"approx_Lat","approx_Long","approx_Lat:approx_Long"))
par(mar = c(4.1,12.1,2.1,2.1))

boxplot((Car_mods2$z)~Car_mods2$Variable,
        col = c("grey",rep("orange", times = 5),
                rep("lightgreen", times = 8),rep("lightblue", times
= 5)),
        las = 2, horizontal = T, pch = 3, cex = 0.5, xlab = "z-
score", ylab = "", axes = F)

axis(1, at = c(-25,-10,-5,0,5,10,25,50,100))
axis(2, labels = c("Time of Test (Seasonal)",
                   "Odometer","Engine capacity",
                   "N tests", "N failed tests","N tests:N failed",
                   "Indices of Multiple Deprivation","Net Income
Before Housing","Net Income After Housing",
                   "Population Size","Pct in Poverty Before
Housing",
                   "Cons. Vote Share","Labor Vote Share","LibDem
Vote Share",
                   "Train Station Distance", "Major Urban Area",
                   "Latitude","Longitude","Latitude x Longitude"),
at =1:19, las =2)

abline(v = c(-25,-10,-5,0,5,10,25,50,100), lty = 3, col =
```

```r
"darkgrey")
abline(v =0, col = "red")
abline(h = seq(1.5, 19.5, by = 2), lty = 3, col = "darkgrey")

par(mar = c(5.1,4.1,4.1,2.1))

saveRDS(sub_frame, "sub_frame_temp.rds")

###############################################################
#####

## repeats the entire process, but for the 20 most common vans and
motorbikes
sum(Year_2011_IDs$MMY %in% target_vehicles2)
sum(Year_2011_IDs$MMY %in% target_vehicles3)

## Builds one huge model, stratified by vehicle type (fixed effects)
sub_frame<-Year_2011_IDs[which(Year_2011_IDs$MMY %in%
c(target_vehicles2, target_vehicles3)),]
## removes LTFs
sub_frame<-sub_frame[which(sub_frame$death_fate!="exported"),]

## Merges on relevant vehicle properties for each 2011 vehicle
#Vehicle_properties<-readRDS("/Users/sauley/Documents/R_workspaces/
Analyses/Automobile_survival_UK/data/Survival_data/
Vehicle_properties.rds")
Vehicle_properties<-readRDS("/Users/qtnzsne/Documents/R_workspaces/
Analyses/Automobile_survival_UK/data/Survival_data/
Vehicle_properties.rds")

## Just keeps make and fuel name
Vehicle_properties<-Vehicle_properties[,c("unique_vehicle","make",
"fuel_name")]

## cut the ~58 million we don't need, keep fuel name
sub_frame$fuel_name<-
Vehicle_properties$fuel_name[match(sub_frame$unique_vehicle,Vehicle_
properties$unique_vehicle)]

rm(Vehicle_properties)
gc()

## fuel types?
table(sub_frame$fuel_name)

## Bins minor fuel types (N<200) into "other"
sub_frame$fuel_name[sub_frame$fuel_name %in% c("CNG", "LNG","Gas Bi-
Fuel",
                                               "Steam", "Gas", "Fuel
Cells")]<-"Other"

table(sub_frame$fuel_name)
```

```r
## merges on social data
sub_frame<-data.frame(sub_frame,
                      Pcodes_brief[match(sub_frame$postcode,
Pcodes_brief$Postcode),])

## makes age at test into a strata
sub_frame$AAT<-floor(sub_frame$age_at_test)
## removes vehicles that died on the date of testing
sub_frame<-
sub_frame[which(sub_frame$death_age>sub_frame$age_at_test),]

## Converts mileage and engine size into easier-to-comprehend units
- per Liter and per 10k miles
sub_frame$engine_cc<-c(sub_frame$engine_cc/1000)
sub_frame$corrected_odometer<-c(sub_frame$corrected_odometer/10000)

## does the same with income metrics
## turns net income  / income per week into 1 unit = 100 pounds
sub_frame$NIBH<-sub_frame$NIBH/100
sub_frame$NIAH<-sub_frame$NIAH/100
sub_frame$Average_wk_income<-sub_frame$Average_wk_income/100

## turns overall income into 10k pound increments
sub_frame$Average_income<-sub_frame$Average_income/10000

## rescales IMD to max = 10
sub_frame$IMD<-sub_frame$IMD/3251.2

## Imputes
set.seed(293)
system.time(sub_frame<-impute(data = sub_frame, blocks = 50,
                              splitrule = "random",
                              nodesize = 500, fast = TRUE, seed =
-19338))

## random sample imputation (less accurate but unbiased)
#set.seed(9373)
#for(i in c(1:28,30:ncol(sub_frame))){

#  if(sum(is.na(sub_frame[,i]))>0){
#
#    rand_vec<-sample(sub_frame[!is.na(sub_frame[,i]),i],
#                     size = sum(is.na(sub_frame[,i])),
#                     replace =T)
#    sub_frame[is.na(sub_frame[,i]),i]<-rand_vec
#    rm(rand_vec)
#  }
#}

sub_frame_vans<-sub_frame[which(sub_frame$test_class==7),]

## Knocks out a stratified single model; for vans, identical model
structure to cars
```

```r
system.time(CPH_full_vans<-coxph(Surv(age_at_test,death_age, died) ~

JDAT+corrected_odometer+engine_cc+N_tests*N_failed+

NIBH+IMD+NIAH+Population+Pct_poverty_BH+Distance_Station+
                                vote_Lab+vote_Con+vote_LibDem+

Rural_urban_Urban_major_conurbation+approx_Lat*approx_Long+
                                strata(MMY), data = sub_frame_vans))

Van_model_out<-NULL
cox_omni_van<-NULL
for(i in 1:length(target_vehicles3)){

  ## constructs a traditional survival model, nested, to evaluate
the
  ## relative predictive capacity of wear-and-tear versus social
variables

  ## gets subset
  sub_frame2<-
sub_frame_cars[which(sub_frame_cars$MMY==target_vehicles[[i]]),]

  ## builds full model
  system.time(CPH_full_sub<-coxph(Surv(age_at_test,death_age, died)
~

JDAT+corrected_odometer+engine_cc+N_tests*N_failed+

NIBH+IMD+NIAH+Population+Pct_poverty_BH+Distance_Station+
                                vote_Lab+vote_Con+vote_LibDem+

Rural_urban_Urban_major_conurbation+approx_Lat*approx_Long,
                                data = sub_frame2))
  xobj2<-summary(CPH_full_sub)

  # Okay let's do it.
  ## basic social: just net income (before housing), and nothing
else
  # CPH_social<-coxph(Surv(age_at_test,death_age, died) ~ NIBH, data
= sub_frame2)

  ## income indicators:
  CPH_social_income<-coxph(Surv(age_at_test,death_age, died) ~

NIBH+IMD+NIAH+Pct_poverty_BH+approx_Lat*approx_Long, data =
sub_frame2)

  ## now physical indicators
  CPH_miles<-coxph(Surv(age_at_test,death_age, died) ~
corrected_odometer+engine_cc+fuel_name+N_tests*N_failed+
                    approx_Lat*approx_Long, data = sub_frame2)
```

```
## the zph test
cph_obj<-matrix(rep(NA, times = 19*2), ncol = 2)
try(cph_obj<-cox.zph(CPH_full_sub)[[1]][1:19,c(1,3)])

#cph_t_z<-cox.zph(CPH_full_t)
cph_omni_sub<-NA
try(cph_omni_sub<-cox.zph(CPH_full_sub)[[1]][20,3])
cox_omni_van[[i]]<-cph_omni_sub

Van_model_out[[i]]<-data.frame(model = rep(i, times = 19),
                               MMY = rep(target_vehicles[[i]],
times = 19),
                               sample_N = rep(xobj2$n, times =
19),
                               Concordance =
rep(xobj2$concordance[1], times = 19),
                               Concordance_soc =
rep(summary(CPH_social_income)$concordance[1], times = 19),
                               Concordance_mech =
rep(summary(CPH_miles)$concordance[1], times = 19),
                               Var = rownames(xobj2$coefficients),
                               xobj2$coefficients,
                               cph_obj)

  rm(CPH_miles, CPH_social_income,CPH_full_sub,sub_frame2)
  xobj2<-NULL
  print(i)
}

Van_mods<-ldply(Van_model_out, as.data.frame)

boxplot(log(Van_mods$z)~Van_mods$Var, las = 2)

sub_frame_bikes<-sub_frame[which(sub_frame$test_class<=2),]

## removes the three (obvious error) diesel bikes
sub_frame_bikes<-sub_frame_bikes[-
c(which(sub_frame_bikes$fuel_name=="Diesel")),]

## Knocks out a stratified single model; for bikes, identical model
structure to cars
system.time(CPH_full_bikes<-coxph(Surv(age_at_test,death_age, died)
~

JDAT+corrected_odometer+engine_cc+N_tests*N_failed+

NIBH+IMD+NIAH+Population+Pct_poverty_BH+Distance_Station+
                                  vote_Lab+vote_Con+vote_LibDem+

Rural_urban_Urban_major_conurbation+approx_Lat*approx_Long+
                                  strata(MMY), data =
```

```
    sub_frame_bikes))

Bike_model_out<-NULL
cox_omni_bikes<-NULL
for(i in 1:length(target_vehicles2)){

  ## constructs a traditional survival model, nested, to evaluate
the
  ## relative predictive capacity of wear-and-tear versus social
variables

  ## gets subset
  sub_frame2<-
sub_frame_cars[which(sub_frame_cars$MMY==target_vehicles[[i]]),]

  ## builds full model
  system.time(CPH_full_sub<-coxph(Surv(age_at_test,death_age, died)
~
JDAT+corrected_odometer+engine_cc+N_tests*N_failed+

NIBH+IMD+NIAH+Population+Pct_poverty_BH+Distance_Station+
                            vote_Lab+vote_Con+vote_LibDem+

Rural_urban_Urban_major_conurbation+approx_Lat*approx_Long,
                            data = sub_frame2))
  xobj2<-summary(CPH_full_sub)

  # Okay let's do it.
  ## basic social: just net income (before housing), and nothing
else
  # CPH_social<-coxph(Surv(age_at_test,death_age, died) ~ NIBH, data
= sub_frame2)

  ## income indicators:
  CPH_social_income<-coxph(Surv(age_at_test,death_age, died) ~

NIBH+IMD+NIAH+Pct_poverty_BH+approx_Lat*approx_Long, data =
sub_frame2)

  ## now physical indicators
  CPH_miles<-coxph(Surv(age_at_test,death_age, died) ~
corrected_odometer+engine_cc+fuel_name+N_tests*N_failed+
                    approx_Lat*approx_Long, data = sub_frame2)

  ## the zph test
  cph_obj<-matrix(rep(NA, times = 19*2), ncol = 2)
  try(cph_obj<-cox.zph(CPH_full_sub)[[1]][1:19,c(1,3)])

  #cph_t_z<-cox.zph(CPH_full_t)
  cph_omni_sub<-NA
  try(cph_omni_sub<-cox.zph(CPH_full_sub)[[1]][20,3])
  cox_omni_bikes[[i]]<-cph_omni_sub
```

```
   Bike_model_out[[i]]<-data.frame(model = rep(i, times = 19),
                                    MMY = rep(target_vehicles[[i]],
times = 19),
                                    sample_N = rep(xobj2$n, times =
19),
                                    Concordance =
rep(xobj2$concordance[1], times = 19),
                                    Concordance_soc =
rep(summary(CPH_social_income)$concordance[1], times = 19),
                                    Concordance_mech =
rep(summary(CPH_miles)$concordance[1], times = 19),
                                    Var =
rownames(xobj2$coefficients),
                                    xobj2$coefficients,
                                    cph_obj)

  rm(CPH_miles, CPH_social_income,CPH_full_sub,sub_frame2)
  xobj2<-NULL
  print(i)
}

Bike_mods<-ldply(Bike_model_out, as.data.frame)

boxplot(log(Bike_mods$z)~Bike_mods$Var)

## how many in all / each class fail the cph.z (omnibus) test
sum(unlist(cox_omni_van)<=0.05, na.rm = T)
sum(unlist(cox_omni_bikes)<=0.05, na.rm = T)
sum(unlist(omni_MMY_cox)<=0.05, na.rm = T)
sum(unlist(omni_MMY_cox)>0.05, na.rm = T)

mean(c(unlist(cox_omni_van),unlist(cox_omni_bikes),unlist(omni_MMY_c
ox))<=0.05, na.rm = T)

mean(c(unlist(cox_omni_van),unlist(cox_omni_bikes),unlist(omni_MMY_c
ox))>0.05, na.rm = T)

## and what fraction of coefficients fail?
mean(Car_mods$p<=0.05, na.rm = T)
mean(Van_mods$p<=0.05)
mean(Bike_mods$p<=0.05)

## save some stuff
saveRDS(Car_mods, "Car_mods.rds")
saveRDS(Van_mods, "Van_mods.rds")
saveRDS(Bike_mods, "Bike_mods.rds")

saveRDS(sub_frame_cars, "Car_frame.rds")
saveRDS(sub_frame_vans, "Van_frame.rds")
saveRDS(sub_frame_bikes, "Bike_frame.rds")

rm(sub_frame_cars, sub_frame_vans,sub_frame_bikes)
```

```
#####################

## Makes a figure showing the model coefficients, and notable
## differences between vans / bikes / cars

par(mfrow = c(1,2), mar = c(5.1,6.1,4.1,2.1))

plot(summary(CPH_full)$coef[,2],(c(1:19)),
     pch = 20, col = "orange", axes = F,
     ylab = "", xlab = "HR", xlim = c(0.7,1.7), cex = 0.5)

abline(v = seq(0.7, 1.7, by = 0.2), lty = 3, col ="grey")
abline(h = seq(1,19, by = 2), col ="grey")

for(i in 1:19){
  points(c(summary(CPH_full)$coef[i,2]+c(summary(CPH_full)
$coef[i,3]),
           summary(CPH_full)$coef[i,2]-c(summary(CPH_full)
$coef[i,3])),
         c(i,i),
         type = "l")
}

abline(v = 1, lty = 3)

axis(side= 1)
axis(side = 2, labels = rownames(summary(CPH_full)$coef), at =
c(1:19), las =2)

## adds p value, z score, and coef. labels
#text(rep(1.7, times = 19),
#      c(1:19), paste0("z = ",signif(summary(CPH_full)$coef[,4],
digits = 2)))
par(mar = c(5.1,3.1,4.1,2.1))

plot(summary(CPH_full)$coef[,4],(c(1:19)),
     pch = 20, col = "orange", axes = F,
     ylab = "", xlab = "z-score", cex = 0.5)

for(i in 1:19){
  points(c(0,
           summary(CPH_full)$coef[i,4]),
         c(i,i),
         type = "l")

  if(summary(CPH_full)$coef[i,5]>0.05){placetext<-"NS"}
  if(summary(CPH_full)$coef[i,5]<=0.05){placetext<-"*"}
  if(summary(CPH_full)$coef[i,5]<0.01){placetext<-"**"}
  if(summary(CPH_full)$coef[i,5]<0.001){placetext<-"***"}
  if(summary(CPH_full)$coef[i,5]<0.0001){placetext<-"****"}
  text(x = summary(CPH_full)$coef[i,4], y = i+0.5, labels =
```

```
placetext)

#   text(x = summary(CPH_full)$coef[i,4]*1.1, y = i+0.4, cex = 0.8,
#   labels = signif(summary(CPH_full)$coef[i,4], digits = 3))

}
abline(v = 1, lty = 3)

axis(side = 1)
axis(side = 2, labels = rep("", times = 19), at = c(1:19))

par(mfrow = c(1,1), mar = c(5.1,7.1,4.1,2.1))

#barplot(summary(CPH_full_bikes)
$coef~rownames(summary(CPH_full_bikes)$coef))
plot((c(1:19)-0.25),summary(CPH_full_bikes)$coef[,2],
     pch = 20, col = "orange", axes = F,
     xlab = "Variable", ylab = "HR")

points((c(1:19)+0.25),summary(CPH_full_vans)$coef[,2],
        pch = 20, col = "lightblue")

points(c(1:19),summary(CPH_full)$coef[,2],
        pch = 20, col = "hotpink")

points(c(1:19,1:19)-0.25,c(summary(CPH_full_bikes)
$coef[,2]+c(summary(CPH_full_bikes)$coef[,3]),
                    summary(CPH_full_bikes)$coef[,2]-
c(summary(CPH_full_bikes)$coef[,3])),
        pch = "-")

axis(side= 2)
abline(h = 1, lty = 3)

abline(h = 1, lty = 3)

############################################################
############################################################

## Subsamples and builds random forest models of survival

#require(randomForestSRC)
#require(ICcforest)
require(fastDummies)

#head(sub_frame)

#sub_frame<-readRDS("sub_frame_temp.rds")
#sub_frameRF<-readRDS("sub_frame_temp.rds")
```

```r
## Rebuilds sub_frame, but does not filter to keep common (>10k)
vehicles only

rm(sub_frame, sub_frame_cars)
gc()

## Builds one huge model with no stratification for vehicle type,
## but otherwise identical filters + transforms + featurisation to
the Cox PH model

Year_2011_IDs<-readRDS("Year_2011_IDs.rds")

## removes LTFs
sub_frameRF<-Year_2011_IDs[which(Year_2011_IDs$death_fate!
="exported"),]

## Merges on relevant vehicle properties for each 2011 vehicle
#Vehicle_properties<-readRDS("/Users/sauley/Documents/R_workspaces/
Analyses/Automobile_survival_UK/data/Survival_data/
Vehicle_properties.rds")
Vehicle_properties<-readRDS("/Users/qtnzsne/Documents/R_workspaces/
Analyses/Automobile_survival_UK/data/Survival_data/
Vehicle_properties.rds")

## Just keeps make and fuel name
Vehicle_properties<-Vehicle_properties[,c("unique_vehicle","make",
"fuel_name")]

## cut the millions of vehicles we don't need, keep fuel name
sub_frameRF$fuel_name<-
Vehicle_properties$fuel_name[match(sub_frameRF$unique_vehicle,Vehicl
e_properties$unique_vehicle)]

rm(Vehicle_properties)
gc()

## fuel types?
table(sub_frameRF$fuel_name)

## Bins minor fuel types (N<200) into "other"
sub_frameRF$fuel_name[sub_frameRF$fuel_name %in% c("CNG", "LNG","Gas
Bi-Fuel",
                                                   "Steam", "Gas", "Fuel
Cells",
                                                   "Electric
Diesel","Gas Diesel")]<-"Other"

table(sub_frameRF$fuel_name)

# Diesel            Electric Hybrid   Electric (Clean)
LPG
# 5045604                     9084                7760
16410
```

```
# Other                 Petrol
# 6467                  10178357

## merges on social data
sub_frameRF<-data.frame(sub_frameRF,
                        Pcodes_brief[match(sub_frameRF$postcode,
Pcodes_brief$Postcode),])

## makes age at test into a strata
sub_frameRF$AAT<-floor(sub_frameRF$age_at_test)
## removes vehicles that died on the date of testing
sub_frameRF<-
sub_frameRF[which(sub_frameRF$death_age>sub_frameRF$age_at_test),]

## Converts mileage and engine size into easier-to-comprehend units
- per Liter and per 10k miles
sub_frameRF$engine_cc<-c(sub_frameRF$engine_cc/1000)
sub_frameRF$corrected_odometer<-c(sub_frameRF$corrected_odometer/
10000)

## does the same with income metrics
## turns net income  / income per week into 1 unit = 100 pounds
sub_frameRF$NIBH<-sub_frameRF$NIBH/100
sub_frameRF$NIAH<-sub_frameRF$NIAH/100
sub_frameRF$Average_wk_income<-sub_frameRF$Average_wk_income/100

## turns overall income into 10k pound increments
sub_frameRF$Average_income<-sub_frameRF$Average_income/10000

## rescales IMD to max = 10
sub_frameRF$IMD<-sub_frameRF$IMD/3251.2

## what fraction is missing? 0.73% or 4,015,863 NA values
mean(is.na(sub_frameRF))

saveRDS(sub_frameRF,"throwaway2.rds")

## First, dummy-codes make, for all makes with >=1000 vehicles,
## and codes the rest as "RARE_MAKE"
common_makes_2011<-table(sub_frameRF$make)
[which(table(sub_frameRF$make)>=1000)]

sub_make1<-sub_frameRF$make

## a few name variants...
sub_make1[sub_make1=="MERCEDES"]<-"MERCEDES-BENZ"
sub_make1[sub_make1=="MERCEDES BENZ"]<-"MERCEDES-BENZ"
sub_make1[sub_make1=="HARLEY DAVIDSON"]<-"HARLEY-DAVIDSON"

## dummy codes all rare makes as the same thing
sub_make1[which(!(sub_make1 %in%
rownames(common_makes_2011)))]<-"RARE_MAKE"

xobj<-dummy_cols(sub_make1)
```

```
colnames(xobj)<-gsub("\\.data", "make",colnames(xobj))

for(i in 1:ncol(xobj)){xobj[,i]<-as.numeric(xobj[,i])}

sub_frameRF<-data.frame(sub_frameRF, xobj[,-c(1)])

## The fuel type
xobj<-dummy_cols(sub_frameRF$fuel_name)
colnames(xobj)<-gsub("\\.data", "fuel_name",colnames(xobj))

sub_frameRF<-data.frame(sub_frameRF, xobj[,-c(1)])

## The vehicle type
xobj<-dummy_cols(sub_frameRF$test_class)
colnames(xobj)<-gsub("\\.data", "test_class",colnames(xobj))

write.csv(colSums(xobj), "common_makes_2011.csv")

sub_frameRF<-data.frame(sub_frameRF, xobj[,-c(1)])

## manually dummy-codes because fastDummy can't hack millions of
rows of data
## only dummy-codes the 326 make-model-year combos with >10k
observations
xobj_vec<-rownames(table(sub_frameRF$MMY))
[table(sub_frameRF$MMY)>10000]
#xobj_vec<-unique(sub_frameRF$MMY)
xobj<-c(sub_frameRF$MMY==(xobj_vec[1]))

for(i in 2:length(xobj_vec)){
  xobj<-data.frame(xobj, as.numeric(sub_frameRF$MMY==(xobj_vec[i])))
  colnames(xobj)[i]<-xobj_vec[i]
  if(i %in% seq(1,1000, by = 10)){print(i)}
}
colnames(xobj)[1]<-xobj_vec[1]

sub_frameRF<-data.frame(sub_frameRF, xobj)

length(unique(sub_frameRF$MMY))

## and the thousands of postcodes? aggregates them up to postcode
regions, by necessity
xobj_vec<-unique(gsub("[0-9]", "", sub_frameRF$Postcode))
xobj2<-gsub("[0-9]", "", sub_frameRF$Postcode)
xobj<-c(xobj2==(xobj_vec[1]))

for(i in 2:length(xobj_vec)){
  xobj<-data.frame(xobj, as.numeric(xobj2==(xobj_vec[i])))
  colnames(xobj)[i]<-paste0("Pcode_",xobj_vec[i])
  if(i %in% seq(1,1000, by = 10)){print(i)}
}
colnames(xobj)[1]<-paste0("Pcode_",xobj_vec[1])

## removes the NA postcode crap
```

```
xobj<-xobj[,which(colnames(xobj)!="Pcode_NA")]

sub_frameRF<-data.frame(sub_frameRF, xobj)

#saveRDS(sub_frameRF, "sub_frameRF_temp.rds")

## Keeps the index
RF_uniq_index<-sub_frameRF$unique_vehicle
#saveRDS(RF_uniq_index, "temp_indx1.rds")
#saveRDS(sub_frameRF$MMY, "temp_indx2.rds")
#saveRDS(sub_frameRF$vin11, "temp_indx3.rds")

## removes shit we don't need: unique IDs, the current pass/fail,
JDAP, factorial stuff we just coded
sub_frameRF<-sub_frameRF[,-c(1:4,8,19,20,23:24,26,30)]

## what columns carry NAs?
which(colSums(is.na(sub_frameRF[1:10000,]))>0)

## What percentage of rows carry NAs?
mean(rowSums(is.na(sub_frameRF[1:10000,]))>0)

## for columns containing geographic classes, postcodes, makes, and
MMYs, infills NAs with zeroes (false)
for(i in c(45:63,489:ncol(sub_frameRF))){
  if(sum(is.na(sub_frameRF[,i]))>0){

    sub_frameRF[is.na(sub_frameRF[,i]),i]<-0
    print(i)
  }
}

## What percentage of rows carry NAs?
mean(rowSums(is.na(sub_frameRF[1:10000,]))>0)

## in what columns?
colMeans(is.na(sub_frameRF[1:10000,]))
[which(colMeans(is.na(sub_frameRF[1:10000,]))>0)]

## fraction passed, which defaults to NA if it's never been tested
before
mean(is.na(sub_frameRF$fraction_passed))

## Removes this indicator as it's contained in N tests and N fails
already
sub_frameRF<-sub_frameRF[,-
c(which(colnames(sub_frameRF)=="fraction_passed"))]

## keeps an index of MMY
sub_frameRF_MMY<-sub_frameRF$MMY
tail(table(sub_frameRF_MMY)[order(table(sub_frameRF_MMY))], 20)

## The '98 ford escort is the most common post-QC
```

```
## and removes the stuff we coded, plus the redundant test_date
## encoded by JDAT, and the redundant age at test
sub_frameRF<-sub_frameRF[,-c(which(colnames(sub_frameRF) %in%
                                   c("fuel_name", "test_class",
"test_date", "MMY",
                                         "postcode", "Postcode",
"AAT")))]

#saveRDS(sub_frameRF, "sub_frameRF_prep.rds")
rm(sub_make1)
rm(xobj, xobj2)
gc()

## Great. Now moves toward a model.

## removes vehicles above age 100 - highly enriched for coding
errors, w. a few real ones
## exhausts vector memory, so does this logically during sampling
below

sub_SurvTime<-sub_frameRF$SurvTime

sub_frameRF<-sub_frameRF[,-c(which(colnames(sub_frameRF) %in%
c("SurvTime", "death_age")))]

sub_frameRF$died<-as.factor(sub_frameRF$died)

## Takes a smaller random sample, 'only' 1m random cases,
## with exactly half surviving

## Gets cars only for the primary model: sample is 95% cars anyway
set.seed(203)
subset_vec<-c(sample(seq(1,nrow(sub_frameRF))
[which(sub_frameRF$died==TRUE &

sub_frameRF$test_class_4==1 & sub_frameRF$age_at_test<=100)], size =
500000, replace = F),
              sample(seq(1,nrow(sub_frameRF))
[which(sub_frameRF$died==FALSE &

sub_frameRF$test_class_4==1 & sub_frameRF$age_at_test<=100)], size =
500000, replace = F))

subset_vec2<-c(sample(seq(1,nrow(sub_frameRF))
[which(sub_frameRF$died==TRUE & sub_frameRF$test_class_4==1 &
sub_frameRF$age_at_test<=100)], size = 100000, replace = F),
               sample(seq(1,nrow(sub_frameRF))
[which(sub_frameRF$died==FALSE & sub_frameRF$test_class_4==1 &
sub_frameRF$age_at_test<=100)], size = 100000, replace = F))

subset_vec3<-c(sample(seq(1,nrow(sub_frameRF))
[which(sub_frameRF$died==TRUE & sub_frameRF$test_class_4==1 &
```

```
sub_frameRF$age_at_test<=100)], size = 25000, replace = F),
              sample(seq(1,nrow(sub_frameRF))
[which(sub_frameRF$died==FALSE & sub_frameRF$test_class_4==1 &
sub_frameRF$age_at_test<=100)], size = 25000, replace = F))

## for the largest sample, also restricts the test set to 1 million
cases
## oversamples to 2 million cases...
subset_vec5<-sample(seq(1,nrow(sub_frameRF))
[which(sub_frameRF$died==TRUE & sub_frameRF$test_class_4==1 &
sub_frameRF$age_at_test<=100)], size = 1000000, replace = F)
subset_vec6<-sample(seq(1,nrow(sub_frameRF))
[which(sub_frameRF$died==FALSE & sub_frameRF$test_class_4==1 &
sub_frameRF$age_at_test<=100)], size = 1000000, replace = F)

## keeps first 500k of each that are not in the training set
subset_vec5<-subset_vec5[which(!(subset_vec5 %in% subset_vec))]
subset_vec6<-subset_vec6[which(!(subset_vec6 %in% subset_vec))]

subset_vec4<-c(subset_vec5[1:500000],subset_vec6[1:500000])

## Builds a sample where
## higher mileage vehicles are deliberately over-sampled

## N>=200,000 miles?    96646 but only 10639 alive: balancing is an
issue for binary-predictive RF
sum(sub_frameRF$corrected_odometer>=20)
table(sub_frameRF$died[which(sub_frameRF$corrected_odometer>=20)])

## gets an UNBALANCED set by mortality, 50k of them, keeping oldest
vehicles
set.seed(30984)
subset_vec_highMiles<-c(sample(seq(1,nrow(sub_frameRF))
[which(sub_frameRF$corrected_odometer>=20 &
sub_frameRF$test_class_4==1)],
                        size = 50000, replace = F))

## and high years (balanced)
table(floor(sub_frameRF$age_at_test))
sum(sub_frameRF$age_at_test>=18)

## 18 and over vehicles n = 358,113
## with 300k cars (88k dead)

## gets a balanced 75k of each, keeping oldest vehicles
set.seed(274)
subset_vec_highYears<-c(sample(seq(1,nrow(sub_frameRF))
[which(sub_frameRF$died==TRUE &

sub_frameRF$age_at_test>=18 &

sub_frameRF$test_class_4==1)],
                        size = 75000, replace = F),
                   sample(seq(1,nrow(sub_frameRF))
```

```
                                [which(sub_frameRF$died==FALSE &

sub_frameRF$age_at_test>=18 &

sub_frameRF$test_class_4==1)],
                                        size = 75000, replace = F))

## Splits off the training data of each size, removes the massive
data frame from the workspace

sub_frameRF_train_1m<-sub_frameRF[subset_vec,]
sub_frameRF_train_200k<-sub_frameRF[subset_vec2,]
sub_frameRF_train_50k<-sub_frameRF[subset_vec3,]

sub_frameRF_train_highMiles<-sub_frameRF[subset_vec_highMiles,]
sub_frameRF_train_highYears<-sub_frameRF[subset_vec_highYears,]

## Takes 750k, equally weighted by the three main test classes
## e.g. 250k cars, 250k motorbikes each type, 250k Vans

set.seed(3984)
subset_vec_eqN<-c( sample(seq(1,nrow(sub_frameRF))
[which(sub_frameRF$test_class_2==1)], size = 250000, replace = F),
             sample(seq(1,nrow(sub_frameRF))
[which(sub_frameRF$test_class_4==1)], size = 250000, replace = F),
             sample(seq(1,nrow(sub_frameRF))
[which(sub_frameRF$test_class_7==1)], size = 250000, replace = F))

sub_frameRF_train_equalN<-sub_frameRF[subset_vec_eqN,]

rm(sub_frameRF)
gc()

## imputes missing data in each
## using large node sizes and blocks of 20k-25k observations for
training

set.seed(293)
system.time(sub_frameRF_train_1m<-impute(data =
sub_frameRF_train_1m, blocks = 50,
                                splitrule = "random",
                                nodesize = 500, fast = TRUE, seed =
-19338))

set.seed(293)
system.time(sub_frameRF_train_200k<-impute(data =
sub_frameRF_train_200k, blocks = 10,
                                    splitrule = "random",
                                    nodesize = 500, fast =
TRUE, seed = -19338))
```

```r
set.seed(293)
system.time(sub_frameRF_train_50k<-impute(data =
sub_frameRF_train_50k, blocks = 2,
                                    splitrule = "random",
                                    nodesize = 500, fast =
TRUE, seed = -19338))

set.seed(293)
system.time(sub_frameRF_train_highMiles<-impute(data =
sub_frameRF_train_highMiles,
                                    blocks = 2,
                                    splitrule = "random",
                                    nodesize = 500, fast =
TRUE, seed = -19338))

set.seed(293)
system.time(sub_frameRF_train_highYears<-impute(data =
sub_frameRF_train_highYears,
                                    blocks = 10,
                                    splitrule =
"random",
                                    nodesize = 500, fast
= TRUE, seed = -19338))
set.seed(293)
system.time(sub_frameRF_train_equalN<-impute(data =
sub_frameRF_train_equalN,
                                    blocks = 10,
                                    splitrule =
"random",
                                    nodesize = 500, fast
= TRUE, seed = -19338))

saveRDS(sub_frameRF_train_1m,"oneM_temp.rds")
saveRDS(sub_frameRF_train_200k,"twohund_temp.rds")
saveRDS(sub_frameRF_train_50k,"fiddy_temp.rds")
saveRDS(sub_frameRF_train_highMiles,"highMiles_temp.rds")
saveRDS(sub_frameRF_train_highYears,"highYears_temp.rds")
saveRDS(sub_frameRF_train_equalN,"sub_frameRF_train_equalN.rds")

## saves indexes
saveRDS(subset_vec,"subset_vec.rds")
saveRDS(subset_vec2,"subset_vec2.rds")
saveRDS(subset_vec3,"subset_vec3.rds")
saveRDS(subset_vec_highMiles,"subset_vec_highMiles.rds")
saveRDS(subset_vec_highYears,"subset_vec_highYears.rds")
saveRDS(subset_vec_eqN,"subset_vec_equalN.rds")

sub_frameRF_train_1m<-readRDS("oneM_temp.rds")
```

```
sub_frameRF_train_200k<-readRDS("twohund_temp.rds")
sub_frameRF_train_50k<-readRDS("fiddy_temp.rds")
sub_frameRF_train_highMiles<-readRDS("highMiles_temp.rds")
sub_frameRF_train_highYears<-readRDS("highYears_temp.rds")
sub_frameRF_train_eqN<-readRDS("subset_vec_equalN.rds")

#######################################################

sub_frameRF_train<-sub_frameRF_train_1m

sub_frameRF_train2<-sub_frameRF_train_50k
died_tf2<-sub_frameRF_train2$died
sub_frameRF_train2<-
sub_frameRF_train2[,which(colnames(sub_frameRF_train2)!="died")]

## Builds a survival forest
require(randomForestSRC)

require(doMC)
options(rf.cores=5, mc.cores=5)

set.seed(29383)

## Builds the all-in-one

sub_frameRF_train$SurvTime<-sub_SurvTime[subset_vec]
sub_frameRF_train$died<-c(as.numeric(sub_frameRF_train$died)-1)

require(parallel)

## trains a large model with tuned parameters from the subsample

set.seed(29383)
system.time(Survival_rf_all_lge<-rfsrc.fast(Surv(SurvTime, died)~.,
                                    data = sub_frameRF_train,
nodedepth = 10,
                                    nodesize = 160,
                                    sampsize = 5000,
                                    # restricts time points to a
smaller number: this is easy because mortality rates are 'lumpy' by
year
                                    ntime = 30,
                                    ntree = 100, mtry = 381,
save.memory = TRUE,
                                    importance = FALSE, forest =
TRUE))

## runs rapid variable selection, strict criteria
system.time(vsel_1<-var.select(Survival_rf_all_lge, conservative =
"high"))

saveRDS(Survival_rf_all_lge, "Survival_rf_all_lge.rds")

#rm(Survival_rf_all_lge)
```

```
gc()

sub_frameRF_train2<-sub_frameRF_train_50k
sub_frameRF_train2$SurvTime<-sub_SurvTime[subset_vec3]
sub_frameRF_train2$died<-c(as.numeric(sub_frameRF_train2$died)-1)

## and, given the high computational load, only
# runs variable importance on variables that are
## frequently used for splits
system.time(vimp_rf_lge<-vimp(Survival_rf_all_lge, seed = -19338,
                    ## runs on test data to increase train times
                    newdata = sub_frameRF_train2,
                    xvar.names = vsel_1$topvars,
                    importance="permute",# block.size = 25,
                    save.memory =T, do.trace = 60))

saveRDS(vimp_rf_lge, "vimp_rf_lge.rds")

## checks accuracy on random holdout cases
xpred_vals<-predict(sub_SurvTime[subset_vec3], newdata =
sub_frameRF_train2,
                    type="quantile", jitt=FALSE)

rm(vimp_rf_lge, Survival_rf_all_lge)
gc()

################################

## Builds RF using oversamples of old vehicles - old defined by
years and miles
sub_frameRF_train<-sub_frameRF_train_highMiles
#sub_frameRF_train<-sub_frameRF[subset_vec_highMiles,]

sub_frameRF_train$SurvTime<-sub_SurvTime[subset_vec_highMiles]

sub_frameRF_train$died<-c(as.numeric(sub_frameRF_train$died)-1)

## rapidly trains over shallow trees, large samples, large
(generalizable) terminal node sizes
set.seed(287484)
system.time(tune_tree_sub2<-tune.rfsrc(Surv(SurvTime, died)~.,
                                 data = sub_frameRF_train,
                                 sampsize = 5000,
                                 ntreeTry = 10, nodedepth = 4,
#strikeout = 3,
                                 nodesizeTry = seq(100, 200, by
= 20)))
## $optimal
## nodesize     mtry
##  200         613

## basically just a tree, so...
```

```
set.seed(287484)
system.time(tune_tree_sub2<-tune.rfsrc(Surv(SurvTime, died)~.,
                                       data = sub_frameRF_train,
                                       sampsize = 5000,
                                       ntreeTry = 10, nodedepth = 4,
#strikeout = 3,
                                       nodesizeTry = seq(200, 500,
by = 100)))
## $optimal
# nodesize      mtry
# 180        159

## these are really, really flat

set.seed(29383)
system.time(Survival_rf_highMiles<-rfsrc.fast(Surv(SurvTime,
died)~.,
                                       data =
sub_frameRF_train, nodedepth = 10,
                                       nodesize = 180, mtry =
159, do.trace =60,
                                       sampsize = 5000,
                                       # restricts time points
to a smaller number: this is easy because mortality rates are
'lumpy' by year
                                       ntime = 30,
                                       ntree = 100,
save.memory = TRUE,
                                       importance = FALSE,
forest = TRUE))

plot.variable(Survival_rf_highMiles,
              xvar.names = c("age_at_test","corrected_odometer"),
partial = TRUE, npts = 100)

saveRDS(Survival_rf_highMiles, "Survival_rf_highMiles.rds")

## runs rapid variable selection, strict criteria
system.time(vsel_highMiles<-var.select(Survival_rf_highMiles,
conservative = "high"))

#vimp_rf_highMiles<-vimp(Survival_rf_highMiles,
#                   seed = -19338,importance="permute",#
block.size = 25,
#     save.memory =T, do.trace = 60)

## uses only first N rows to test vimp
sub_frameRF_train2<-sub_frameRF_train[1:10000,]

## and, given the high computational load, only
# runs variable importance on variables that are
```

```
## frequently used for splits
system.time(vimp_rf_highMiles<-vimp(Survival_rf_highMiles, seed =
-19338,
                                ## runs on test data to reduce train
times
                                newdata = sub_frameRF_train2,
                                xvar.names = vsel_highMiles$topvars,
                                importance="permute",# block.size =
25,
                                save.memory =T, do.trace = 60))

##############

## runs older by year

#sub_frameRF_train<-sub_frameRF[subset_vec_highYears,]
sub_frameRF_train<-sub_frameRF_train_highYears
sub_frameRF_train$SurvTime<-sub_SurvTime[subset_vec_highYears]
sub_frameRF_train$died<-c(as.numeric(sub_frameRF_train$died)-1)

## rapidly trains over shallow trees, large samples, large
(generalizable) terminal node sizes
set.seed(287484)
system.time(tune_tree_sub2<-tune.rfsrc(Surv(SurvTime, died)~.,
                                    data = sub_frameRF_train,
                                    sampsize = 5000,
                                    ntreeTry = 10, nodedepth = 4,
#strikeout = 3,
                                    nodesizeTry = seq(100, 200,
by = 20)))

# $optimal
# nodesize    mtry
# 120        148
set.seed(29383)
system.time(Survival_rf_highYears<-rfsrc.fast(Surv(SurvTime,
died)~.,
                                            data =
sub_frameRF_train, nodedepth = 10,
                                            nodesize = 120, mtry =
148, do.trace = 60,
                                            sampsize = 5000,
                                            # restricts time
points to a smaller number: this is easy because mortality rates are
'lumpy' by year
                                            ntime = 30,
                                            ntree = 100,
save.memory = TRUE,
                                            importance = FALSE,
forest = TRUE))
```

```
plot.variable(Survival_rf_highYears,
              xvar.names = c("age_at_test","corrected_odometer"),
partial = TRUE, npts = 100)

## runs rapid variable selection, strict criteria
system.time(vsel_highYears<-var.select(Survival_rf_highYears,
conservative = "high"))

## uses only first N rows to test vimp
sub_frameRF_train2<-sub_frameRF_train[1:10000,]

## variable importance
system.time(vimp_rf_highYears<-vimp(Survival_rf_highYears, seed =
-19338,
                                    ## runs on test data to reduce
train times
                                    newdata = sub_frameRF_train2,
                                    xvar.names =
vsel_highYears$topvars,
                                    importance="permute",#
block.size = 25,
                                    save.memory =T, do.trace = 60))

saveRDS(Survival_rf_highYears, "Survival_rf_highYears.rds")
saveRDS(vimp_rf_highYears, "vimp_rf_highYears.rds")

############################################################

##############################

## Builds RF using equally weighted samples by motorbike/van/car
together

#sub_frameRF_train<-sub_frameRF_train_eqN

#sub_frameRF_train<-sub_frameRF[sub_frameRF_train_eqN,]

sub_frameRF_train<-readRDS("sub_frameRF_train_equalN.rds")

## removes dumb things
#sub_frameRF_train<-sub_frameRF_train[,-
c(which(colnames(sub_frameRF_train) %in% ("death_age")))]
sub_frameRF_train$SurvTime<-sub_SurvTime[subset_vec_eqN]

sub_frameRF_train$died<-c(as.numeric(sub_frameRF_train$died)-1)

## rapidly trains over shallow trees, large samples, large
(generalizable) terminal node sizes
set.seed(287484)
system.time(tune_tree_sub2<-tune.rfsrc(Surv(SurvTime, died)~.,
                                       data = sub_frameRF_train,
                                       sampsize = 5000,
                                       ntreeTry = 10, nodedepth = 4,
```

```
#strikeout = 3,
                                    nodesizeTry = seq(100, 200, by
= 20)))
## $optimal
## nodesize      mtry
##  120        247

set.seed(29383)
system.time(Survival_rf_eqN<-rfsrc.fast(Surv(SurvTime, died)~.,
                                    data =
sub_frameRF_train, nodedepth = 10,
                                    nodesize = 120, mtry =
247, do.trace =60,
                                    sampsize = 5000,
                                    # restricts time points
to a smaller number: this is easy because mortality rates are
'lumpy' by year
                                    ntime = 30,
                                    ntree = 100,
save.memory = TRUE,
                                    importance = FALSE,
forest = TRUE))

plot.variable(Survival_rf_eqN,
            xvar.names = c("age_at_test","corrected_odometer"),
partial = TRUE, npts = 100)

saveRDS(Survival_rf_eqN, "Survival_rf_eqN.rds")

## runs rapid variable selection, strict criteria
system.time(vsel_eqN<-var.select(Survival_rf_eqN, conservative =
"high"))

## uses only first N rows to test vimp
sub_frameRF_train2<-sub_frameRF_train[1:10000,]

## and, given the high computational load, only
# runs variable importance on variables that are
## frequently used for splits
system.time(vimp_rf_eqN<-vimp(Survival_rf_eqN, seed = -19338,
                                ## runs on test data to reduce train
times
                                newdata = sub_frameRF_train2,
                                xvar.names = vsel_eqN$topvars,
                                importance="permute",# block.size =
25,
                                save.memory =T, do.trace = 60))

barplot(tail(abs(vimp_rf_eqN$importance[order(abs(vimp_rf_eqN$import
ance))]), 30), las = 2)

saveRDS(vimp_rf_eqN, "vimp_rf_eqN.rds")
```

```
##############

rm(Survival_rf_highYears,Survival_rf_highMiles, Survival_rf_lge,
    vimp_rf_highMiles,vimp_rf_highYears,vimp_rf_lge,Survival_rf_eqN)

rm(sub_frameRF_train_1m,sub_frameRF_train_200k,
    sub_frameRF_train_50k,sub_frameRF_train_highMiles,
    sub_frameRF_train_highYears,
    sub_frameRF_train_eqN)

gc()

####################################################
####################################################

## From now on, just builds plots for the manuscript (and my own
curiosity)

####################################################
####################################################

## figures for the paper
require(ggplot2)
require(grid)
require(gridExtra)
require(viridis)

require(cowplot)
require(fitdistrplus)

## Makes a Weibull plot (effectively just a QQ plot of the CDF)
## showing huge divergences between Weibull and reality

#LifeVec_miles<-readRDS("data/LifeVec_miles.rds")
Lifetab_v11s<-readRDS("Lifetab_v11s.rds")
Lifetab_v11s_miles<-readRDS("Lifetab_v11s_miles.rds")

## loads a Weibull distribution, CDF by percentiles
Weibull_dist<-qweibull(seq(0,0.99, by = 0.01), shape=2, scale = 2)

## generates the hazard function of the Weibull dist.
require(reliaR)

## for each vector of qx:
## converts qx to lx values,
## converts these lx values to estimated quantiles of survival by
diviing by 100k (fraction surviving at each age)
## calculates the exact same quantiles of survival for a weibull
distribution with shape 5 scale 1,
## fits linear regression to test if the observed distribution looks
like Weibull CDF,
## and plots the entire thing.
```

```
require(fitdistrplus)
#require(pracma)
Lifetab_MMY_meta<-readRDS("Lifetab_MMY_meta.rds")
Weibull_fits<-NULL
Weibull_qq<-NULL

for(i in 1:max(Lifetab_MMY_meta$index)){

  qx_vector<-
Lifetab_MMY_meta$qx_vec[which(Lifetab_MMY_meta$index==i)]

  ## only fits a regression for >=10 fully observed years (ignores
tails for qx)
  if(c(length(qx_vector)-sum(qx_vector%in% c(0,1)))>=12){

    lx_vector<-qxtolx(qx_vector)

    Obs_quantiles<-lx_vector/100000
    ## removes zeroes, ones
    Obs_quantiles<-Obs_quantiles[which(Obs_quantiles!=0)]
    Obs_quantiles<-Obs_quantiles[which(Obs_quantiles!=1)]

    # Expected_quantiles<-cumsum(pweibull(Obs_quantiles,5,1))
    Expected_quantiles<-pweibull(Obs_quantiles,5,1)

    ## must regress Obs as y, expected (x) is exactly specified w.
zero error
    FitReg_x<-lm(Obs_quantiles~Expected_quantiles)
    ## orthogonal distance regression
    #  FitReg_x<-
odregress(as.numeric(Expected_quantiles),as.numeric(Obs_quantiles))

    ## finds a best-fit Weibull for the data
    fitWb<-fitdist(Obs_quantiles, "weibull", method = "mle")
    ## best-fit lognormal and normal dist
    fitLnorm<-fitdist(Obs_quantiles, "lnorm", method = "mle")
    fitnorm<-fitdist(Obs_quantiles, "norm", method = "mle")

    ## retains the A-D test of how well this fits,
    ## compared to simple normal and lognormal distributions

    AD_fits<-rep(NA, times = 3)
    AIC_fits<-rep(NA, times = 3)

    #   try(AD_fits<-gofstat(list(fitWb, fitLnorm, fitnorm))[8])
    #   try(AIC_fits<-gofstat(list(fitWb, fitLnorm, fitnorm))[12])

    try(AD_fits<-c(gofstat(fitWb)[8], gofstat(fitLnorm)
[8],gofstat(fitnorm)[8]))
    try(AIC_fits<-c(gofstat(fitWb)[12], gofstat(fitLnorm)
[12],gofstat(fitnorm)[12]))
```

```
      Expected_quantiles_best<-pweibull(Obs_quantiles,
                                        shape = fitWb$estimate[1],
                                        scale = fitWb$estimate[2])

      ## must regress Obs as y, expected (x) is exactly specified w.
zero error
      FitReg_best<-lm(Obs_quantiles~Expected_quantiles_best)

      Weibull_fits[[i]]<-data.frame(index_vehicle = i,
                                    totalDeaths =
max(Lifetab_MMY_meta$dx_vec[which(Lifetab_MMY_meta$index==i)]),
                                    peakExpo =
max(Lifetab_MMY_meta$expo_vec[which(Lifetab_MMY_meta$index==i)]),
                                    totalObsAges =
length(Obs_quantiles),
                                    Coef1 =
FitReg_x$coefficients[[1]],
                                    Coef2 =
FitReg_x$coefficients[[2]],
                                    pval = summary(FitReg_x)[4][[1]]
[2,c(4)],
                                    rsq = summary(FitReg_x)[8],
                                    rsq_best = summary(FitReg_best)
[[8]],
                                    rsq_adjusted = summary(FitReg_x)
[9],
                                    tval = summary(FitReg_x)[4][[1]]
[2,c(3)],
                                    fstat = summary(FitReg_x)[10][[1]]
[[1]],
                                    AD_wb = AD_fits[[1]][1],
                                    AD_lognorm = AD_fits[[1]][2],
                                    AD_norm = AD_fits[[1]][3],
                                    AC_wb =  AIC_fits[[1]][1],
                                    AC_lognorm =  AIC_fits[[1]][2],
                                    AC_norm =  AIC_fits[[1]][3])

      Weibull_qq[[i]]<-data.frame(index_vehicle = rep(i, times =
length(Obs_quantiles)),
                                  Ages =
                                    Obs_quantiles,
                                  Expected_quantiles)

   }

}

Weibull_fits<-ldply(Weibull_fits, as.data.frame)
Weibull_qq<-ldply(Weibull_qq, as.data.frame)

Weibull_fits<-Weibull_fits[!duplicated(Weibull_fits),]
```

```
Weibull_qq<-Weibull_fits[!duplicated(Weibull_qq),]

Weibull_fits$log10_expo<-log10(Weibull_fits$peakExpo)
Weibull_fits$log10_deaths<-log10(Weibull_fits$totalDeaths)

## makes obs ages as fatcor for boxplots
Weibull_fits$totalObsAges_fac<-as.factor(Weibull_fits$totalObsAges)

## plots sample sizes against R2 values  - increasing N does not
result in approx. of the
## Weibull function: R2 evens out around R2=0.87
DeathFit<-ggplot(aes(log10_deaths,adj.r.squared), data =
Weibull_fits)+
  geom_hex(bins = 50)+
  geom_smooth(col = "orange", bg = "orange")+
  scale_fill_viridis(option  = "D", name = "Vehicle Types\n
Observed")+
  geom_hline(yintercept = 0.9, col ="red")+
  scale_x_continuous(breaks = 1:4, labels = c("10","100", "1000",
"10k"))+
  scale_y_continuous(limits=c(0.4,1))+
  theme_minimal()+
  theme(legend.position = c(0.45,
0.25),legend.direction="horizontal")+
  labs(y= "Adjusted R squared", x = "Deaths Observed")

set.seed(2093)
YearsFit<-ggplot(aes(totalObsAges_fac,adj.r.squared,group =
totalObsAges_fac), data = Weibull_fits)+
  geom_jitter(width = 0.3, alpha = I(0.5), col ="lightblue", cex =
0.5)+
  geom_boxplot(varwidth = TRUE,outlier.shape = NA, notch = F, bg =
"orange")+
  geom_hline(yintercept = 0.9, col ="red")+
  ## direct labels sample sizes
  annotate("text", x = 1, y = 0.5, label = paste0("N =
",table(Weibull_fits$totalObsAges)[[1]]), angle = 90)+
  annotate("text", x = 2, y = 0.5, label = paste0("N =
",table(Weibull_fits$totalObsAges)[[2]]), angle = 90)+
  annotate("text", x = 3, y = 0.5, label = paste0("N =
",table(Weibull_fits$totalObsAges)[[3]]), angle = 90)+
  annotate("text", x = 4, y = 0.5, label = paste0("N =
",table(Weibull_fits$totalObsAges)[[4]]), angle = 90)+
  annotate("text", x = 5, y = 0.5, label = paste0("N =
",table(Weibull_fits$totalObsAges)[[5]]), angle = 90)+
  annotate("text", x = 6, y = 0.5, label = paste0("N =
",table(Weibull_fits$totalObsAges)[[6]]), angle = 90)+
  theme_minimal()+
  labs(y= "Adjusted R squared", x = "Years Observed")+
  scale_y_continuous(limits=c(0.4,1))

## direct labels sample sizes
# geom_text(x = c(1:6), y = rep(1.05, times = 6), label =
table(Weibull_fits$totalObsAges))
```

```
FigS_Weibull_Fit<-plot_grid(DeathFit,YearsFit, nrow = 1,  rel_widths
= c(3/5,2/5))

#ggsave(filename = "Publication_docs/FigS_Weibull_Fit.eps",
FigS_Weibull_Fit, device = "Cairo")

ggsave(plot = FigS_Weibull_Fit,filename = "Publication_docs/
FigS_Weibull_Fit.eps",
       device=cairo_ps, height = 5, width = 7.5)

ggsave(FigS_Weibull_Fit,filename = "Publication_docs/
FigS_Weibull_Fit.tiff",
       device=cairo_ps, height = 5, width = 7, dpi = 330)

## How many 'best-fit' models failed to exceed R2 = 0.9?
mean(Weibull_fits$rsq_best<=0.9)  # 19.4% ; N = 713
mean(Weibull_fits$r.squared<=0.9) # 62.1% ; N = 2283
mean(Weibull_fits$adj.r.squared<=0.9) ## 63.4% ; N = 2342

## plots observed ages against R2 values  - increasing observation
time
## does not result in closer approx. of the Weibull function
boxplot(Weibull_fits$adj.r.squared~Weibull_fits$totalObsAges,
       xlab = "Full Years Observed", ylab = "R squared, adjusted",
       pch ="-", col = "lightblue", ylim = c(0,1), varwidth =T)

abline(h = seq(0,1, by = 0.2), lty = 3)
abline(h = 0.9, lty = 3, col = "red")

## adds supplementary x-axis with sample sizes
axis(side = 3, labels = table(Weibull_fits$totalObsAges), at = 1:6)

## Generates a supplementary figure showing qq plot w. Weibull

par(mfrow = c(1,3), bty= "n")

## plots observed ages against R2 values  - increasing observation
time
## does not result in closer approx. of the Weibull function
boxplot(Weibull_fits$adj.r.squared~Weibull_fits$totalObsAges,
       xlab = "Full Years Observed", ylab = "R squared, adjusted",
       pch ="-", col = "lightblue", ylim = c(0,1))
abline(h = seq(0,1, by = 0.2), lty = 3)

par(mfrow = c(1,1), bty= "n")

## Figure 1  - Lifespan and age-specific mortality rates
par(bty = "n", mfrow = c(1,2))

## population age distributions (pop pyramids)
```

```r
## loads 2019 as example, keeps vehicle test class

## Lists all the dirs. Obviously you have to designate your paths.
## list directories
subvec2<-list.files("/Volumes/PHOTO_DRIVE/Cars/unzipped_MOTs/
e1_2019/")

## assembles into a single-year tibble
YearTibble<-NULL
counter_n<-NULL
for(j in 1:length(subvec2)){

  ## reads in the DVSA data with make+model+engine+fuel+VIN
  subframe<-read_delim(paste0("/Volumes/PHOTO_DRIVE/Cars/
unzipped_MOTs/e1_2019/",
                        subvec2[[j]]),
                    trim_ws = TRUE, name_repair = "minimal",
                 na = c("^??$","^?$", "NA", "^\\*$"),
skip_empty_rows = TRUE,
                    num_threads = use_ncores)

  ## removes data we don't need right now
  subframe<-subframe[,c(colnames(subframe) %in% c("unique_vehicle",
"vin11",
                                              ## keeps types of
breakdowns

"age_at_test","test_class","test_type",
                                              "make", "model",
"engine_cc", "fuel_name"))]
  ## compresses test result into one vector
  # subframe$test_result<-subframe$result
  #  subframe$test_result[subframe$is_prs=="PRS FAIL"]<-"PRS_FAIL"

  # subframe<-subframe[,which(!(colnames(subframe) %in% c("result",
"is_prs")))]

  YearTibble[[j]]<-subframe

  ## if successful, counts
  if(length(subframe)>0){counter_n<-c(counter_n, j)}

  print(j)
  print(max(subframe$test_date, na.rm = T))
  rm(subframe)

}
YearTibble<-ldply(YearTibble, as.data.frame)

#saveRDS(YearTibble, "YearTibble_temp.rds")

## removes retests
YearTibble<-YearTibble[which(YearTibble$test_type=="Normal Test"),]
```

```r
## randomly sorts, keeps one random test age for each unique ID
set.seed(133)
#YearTibble<-YearTibble[sample(c(1:nrow(YearTibble)), replace =F),]
#YearTibble<-YearTibble[which(!
duplicated(YearTibble$unique_vehicle)),]

freq_table_all<-
table(floor(YearTibble$age_at_test[which(YearTibble$age_at_test>=0 &

YearTibble$age_at_test<=120)]))

freq_table_moto<-
table(floor(YearTibble$age_at_test[which(YearTibble$age_at_test>=0 &

YearTibble$age_at_test<=120 &

YearTibble$test_class<=2)]))

freq_table_vans<-
table(floor(YearTibble$age_at_test[which(YearTibble$age_at_test>=0 &

YearTibble$age_at_test<=120 &

YearTibble$test_class==7)]))

freq_table_cars<-
table(floor(YearTibble$age_at_test[which(YearTibble$age_at_test>=0 &

YearTibble$age_at_test<=120 &

YearTibble$test_class==4)]))

## merges them all
freq_table_all<-data.frame(age = as.numeric(names(freq_table_all)),
                           All = as.numeric(freq_table_all))

freq_table_all<-merge(freq_table_all, data.frame(age =
as.numeric(names(freq_table_vans)),
                                                 Vans =
as.numeric(freq_table_vans)),
                      by.x = "age", by.y = "age", all = T)

freq_table_all<-merge(freq_table_all, data.frame(age =
as.numeric(names(freq_table_moto)),
                                                 motorbikes =
as.numeric(freq_table_moto)),
                      by.x = "age", by.y = "age", all = T)

freq_table_all<-merge(freq_table_all, data.frame(age =
as.numeric(names(freq_table_cars)),
                                                 cars =
```

```
                      as.numeric(freq_table_cars)),
                              by.x = "age", by.y = "age", all = T)

    freq_table_2019<-freq_table_all

    ## Now builds the same for 2005

    ## Lists all the dirs. Obviously you have to designate your paths.
    ## list directories
    subvec2<-list.files("/Volumes/PHOTO_DRIVE/Cars/unzipped_MOTs/
    e1_2005/")

    ## assembles into a single-year tibble
    YearTibble<-NULL
    counter_n<-NULL
    for(j in 1:length(subvec2)){

      ## reads in the DVSA data with make+model+engine+fuel+VIN
      subframe<-read_delim(paste0("/Volumes/PHOTO_DRIVE/Cars/
    unzipped_MOTs/e1_2005/",
                              subvec2[[j]]),
                         trim_ws = TRUE, name_repair = "minimal",
                         na = c("^??$","^?$", "NA", "^\\*$"),
    skip_empty_rows = TRUE,
                         num_threads = use_ncores)

      ## removes data we don't need right now
      subframe<-subframe[,c(colnames(subframe) %in% c("unique_vehicle",
    "vin11",
                                                      ## keeps types of
    breakdowns

    "age_at_test","test_class","test_type",
                                                      "make", "model",
    "engine_cc", "fuel_name"))]
      ## compresses test result into one vector
      # subframe$test_result<-subframe$result
      #  subframe$test_result[subframe$is_prs=="PRS FAIL"]<-"PRS_FAIL"

      # subframe<-subframe[,which(!(colnames(subframe) %in% c("result",
    "is_prs")))]

      YearTibble[[j]]<-subframe

      ## if successful, counts
      if(length(subframe)>0){counter_n<-c(counter_n, j)}

      print(j)
      print(max(subframe$test_date, na.rm = T))
      rm(subframe)

    }
    YearTibble<-ldply(YearTibble, as.data.frame)
```

```r
#saveRDS(YearTibble, "YearTibble_temp.rds")

## removes retests
YearTibble<-YearTibble[which(YearTibble$test_type=="Normal Test"),]

## randomly sorts, keeps one random test age for each unique ID
set.seed(133)
#YearTibble<-YearTibble[sample(c(1:nrow(YearTibble)), replace =F),]
#YearTibble<-YearTibble[which(!
duplicated(YearTibble$unique_vehicle)),]

freq_table_all<-
table(floor(YearTibble$age_at_test[which(YearTibble$age_at_test>=0 &

YearTibble$age_at_test<=120)]))

freq_table_moto<-
table(floor(YearTibble$age_at_test[which(YearTibble$age_at_test>=0 &

YearTibble$age_at_test<=120 &

YearTibble$test_class<=2)]))

freq_table_vans<-
table(floor(YearTibble$age_at_test[which(YearTibble$age_at_test>=0 &

YearTibble$age_at_test<=120 &

YearTibble$test_class==7)]))

freq_table_cars<-
table(floor(YearTibble$age_at_test[which(YearTibble$age_at_test>=0 &

YearTibble$age_at_test<=120 &

YearTibble$test_class==4)]))

## merges them all
freq_table_all<-data.frame(age = as.numeric(names(freq_table_all)),
                           All = as.numeric(freq_table_all))

freq_table_all<-merge(freq_table_all, data.frame(age =
as.numeric(names(freq_table_vans)),
                                                 Vans =
as.numeric(freq_table_vans)),
                      by.x = "age", by.y = "age", all = T)

freq_table_all<-merge(freq_table_all, data.frame(age =
as.numeric(names(freq_table_moto)),
                                                 motorbikes =
```

```
                            as.numeric(freq_table_moto)),
                                     by.x = "age", by.y = "age", all = T)

freq_table_all<-merge(freq_table_all, data.frame(age =
as.numeric(names(freq_table_cars)),
                                                 cars =
as.numeric(freq_table_cars)),
                            by.x = "age", by.y = "age", all = T)

freq_table_2005<-freq_table_all

freq_table_2005[is.na(freq_table_2005)]<-0
freq_table_2019[is.na(freq_table_2019)]<-0

saveRDS(freq_table_2005, "freq_table_2005.rds")
saveRDS(freq_table_2019, "freq_table_2019.rds")

freq_table_2005<-readRDS("freq_table_2005.rds")
freq_table_2019<-readRDS("freq_table_2019.rds")

freq_table_2005$Year<-rep(2005, times = nrow(freq_table_2005))
freq_table_2019$Year<-rep(2019, times = nrow(freq_table_2019))

## stacks to make back-to-back barplots
freq_table_all<-rbind(freq_table_2005,freq_table_2019)

freq_table_all$Year<-as.character(freq_table_all$Year)

## log-transforms

## log10 transforms
freq_table_all$log_cars<-log10(freq_table_all$cars)
freq_table_all$log_vans<-log10(freq_table_all$Vans)
freq_table_all$log_motorbikes<-log10(freq_table_all$motorbikes)

## makes 2005 values negative, to invert coords
freq_table_all$cars[which(freq_table_all$Year=="2005")]<-c(-
freq_table_all$cars[which(freq_table_all$Year=="2005")])
freq_table_all$Vans[which(freq_table_all$Year=="2005")]<-c(-
freq_table_all$Vans[which(freq_table_all$Year=="2005")])
freq_table_all$motorbikes[which(freq_table_all$Year=="2005")]<-c(-
freq_table_all$motorbikes[which(freq_table_all$Year=="2005")])

freq_table_all$log_cars[which(freq_table_all$Year=="2005")]<-c(-
freq_table_all$log_cars[which(freq_table_all$Year=="2005")])
freq_table_all$log_vans[which(freq_table_all$Year=="2005")]<-c(-
freq_table_all$log_vans[which(freq_table_all$Year=="2005")])
freq_table_all$log_motorbikes[which(freq_table_all$Year=="2005")]<-
c(-
freq_table_all$log_motorbikes[which(freq_table_all$Year=="2005")])
```

```
ggplot(freq_table_all, aes(x=age, y= log_cars, fill=Year)) +
  facet_wrap(~ Year, scales = "free_x") +
  geom_col() +
  coord_flip() +
  scale_y_continuous(expand = c(0, 0),
                     labels = function(x) signif(abs(x), 3)) +
  theme_minimal()+
  theme(panel.spacing.x = unit(0, "mm"))

## now makes two ggplots, and facet wraps with no gap
plot_2005<-
ggplot(freq_table_all[which(freq_table_all$Year=="2005"),],
                aes(x=age, y= log_cars)) +
  facet_wrap(~ Year, scales = "free_x") +
  geom_col(colour= "#17A589", fill = "#17A589") +
  coord_flip() +
  #  scale_y_reverse(name= "axis1",expand = expansion(mult=
c(c(6,0)))) +
  scale_y_continuous(expand = c(0, 0), limits = c(-6.5,0),
                     breaks = seq(-6,0, by = 1),
                     labels = c("1M","100k","10k","1000","100","10",
"0"))+
  #   labels = function(x) signif(abs(x), 3)) +
  scale_x_continuous(limits = c(3,120))+
  theme_minimal()+
  # theme(panel.spacing.x = unit(0, "mm"))+
  theme(plot.margin = unit(c(5.5, 0, 5.5, 5.5), "pt"))+
  ylab("log10 cars")

plot_2019<-
ggplot(freq_table_all[which(freq_table_all$Year=="2019"),],
                aes(x=age, y= log_cars)) +
  facet_wrap(~ Year, scales = "free_x") +
  geom_col(colour= "#3498DB", fill = "#3498DB")+
  coord_flip() +
  #  scale_y_reverse(name= "axis1",expand = expansion(mult=
c(c(6,0)))) +
  scale_y_continuous(expand = c(0, 0), limits = c(0, 6.5),
                     breaks = seq(0,6, by = 1),
                     labels = c("","10","100","1000","10k","100k",
"1M"))+
  # labels = function(x) signif(abs(x), 3)) +
  scale_x_continuous(limits = c(3,120))+
  theme_minimal()+
  # theme(panel.spacing.x = unit(0, "mm"))+
  theme(axis.title.y=element_blank(), axis.text.y=element_blank(),
        axis.line.y = element_blank(), axis.ticks.y=element_blank(),
        plot.margin = unit(c(5.5, 5.5, 5.5, -3.5), "pt"))+
  ylab("")

require(grid)
grid.newpage()
grid.draw(cbind(ggplotGrob(plot_2005), ggplotGrob(plot_2019), size =
```

```
        "last"))

## now does cars and motorbikes in classic pop pyramid colours
freq_table_comparison2<-data.frame(Age =
c(freq_table_2019$age,freq_table_2019$age),
                                     Freq =
c(log10(freq_table_2019$cars),
                                              c(-
log10(freq_table_2019$motorbikes))),
                                     Type = c(rep("Cars", times =
nrow(freq_table_2019)),
                                              rep("Motorbikes", times
= nrow(freq_table_2019))))

## adds colour change for recession years
## older (technical, two-quarter) recessions =
## 1919-1921
## 1930-31
## 1956
## 1961
## 1973-74-75
## 1980-81
## 1990-1991
## 2008-2009
freq_table_comparison2$recession<-c(freq_table_comparison2$Age %in%
c(2019-c(1919:1921, 1930,1931, 1956, 1961, 1973:1975, 1980,1981,
1990,1991,2008,2009)))

## now makes two ggplots, and facet wraps with no gap
plot_M<-
ggplot(freq_table_comparison2[which(freq_table_comparison2$Type=="Mo
torbikes"),],
                aes(x=Age, y= Freq)) +
  # geom_col(colour= "#3498db", fill = "#3498db") +
  coord_flip() +
  #  scale_y_reverse(name= "axis1",expand = expansion(mult=
c(c(6,0)))) +
  scale_y_continuous(expand = c(0, 0), limits = c(-6.5,0),
                      breaks = seq(-6,0, by = 1),
                      labels = c("1M","100k","10k","1k","100","10",
"0"))+
  #   labels = function(x) signif(abs(x), 3)) +
  scale_x_continuous(limits = c(3,120))+
  #  facet_wrap(~ Type, scales = "free_x") +
  geom_col(colour= c(c("#3498db", "#2874a6")
                    [c(as.numeric(freq_table_comparison2$recession[

which(freq_table_comparison2$Type=="Motorbikes")])+1)])[-c(1:3)],
          fill = c(c("#3498db", "#2874a6")
                    [c(as.numeric(freq_table_comparison2$recession[

which(freq_table_comparison2$Type=="Motorbikes")])+1)])[-c(1:3)])+
  theme_minimal(base_size = 14)+
```

```
    # theme(panel.spacing.x = unit(0, "mm"))+
    theme(plot.margin = unit(c(5.5, 0, 5.5, 5.5), "pt"))+
    ylab("Motorbikes")+
    xlab("Age in Years")+
    ## adds a horizontal line for when MOTs cease to be mandatory
    geom_vline(xintercept = 40, lty = 3)

plot_C<-
ggplot(freq_table_comparison2[which(freq_table_comparison2$Type=="Ca
rs"),],
                aes(x=Age, y= Freq)) +
  #  facet_wrap(~ Year, scales = "free_x") +
  #  geom_col(colour= "#e74c3c", fill = "#e74c3c")+
  coord_flip() +
  #  scale_y_reverse(name= "axis1",expand = expansion(mult=
c(c(6,0)))) +
  scale_y_continuous(expand = c(0, 0), limits = c(0, 6.5),
                     breaks = seq(0,6, by = 1),
                     labels = c("","10","100","1k","10k","100k",
"1M"))+
  # labels = function(x) signif(abs(x), 3)) +
  scale_x_continuous(limits = c(3,120))+
  ## colour vector with first 3 vals trimmed
  geom_col(colour= c(c("#e74c3c", "#b03a2e")
                     [c(as.numeric(freq_table_comparison2$recession[
                       which(freq_table_comparison2$Type=="Cars")])
+1)])[-c(1:3)],
           fill = c(c("#e74c3c", "#b03a2e")
                    [c(as.numeric(freq_table_comparison2$recession[
                      which(freq_table_comparison2$Type=="Cars")])
+1)])[-c(1:3)])+
  theme_minimal(base_size = 14)+
  # theme(panel.spacing.x = unit(0, "mm"))+
  theme(axis.title.y=element_blank(), axis.text.y=element_blank(),
        axis.line.y = element_blank(), axis.ticks.y=element_blank(),
        plot.margin = unit(c(5.5, 5.5, 5.5, -3.5), "pt"))+
  ylab("Cars and Light Vans")+
  ## WWI from 14-18, 101-105 years before
  annotate("text", label = "WWI", x = 103, y = 1.5) +
  ## WWII hole in car production, dominated by jeeps, from 74-80
  annotate("text", label = "WWII", x = 76, y = 2.5) +
  ## fuel crisis and stagflation era '73-81, 38 to 46 years before
2019
  annotate("text", label = "Fuel Crisis", x = 46, y = 5) +
  ## adds a horizontal line for when MOTs cease to be mandatory
  geom_vline(xintercept = 40, lty = 3) # +
# annotate("text", label = "MOTs mandatory V", x = 39, y = 5)

grid.newpage()
grid.draw(cbind(ggplotGrob(plot_M), ggplotGrob(plot_C), size =
"last"))
```

```
#ggsave(filename = "Publication_docs/Fig_1a_eps.eps",
#        device=cairo_ps, height = 6.5, width = 4.7)

cairo_ps("Publication_docs/Fig_1a_eps.eps", height = 6.5, width =
4.7)
grid.newpage()
grid.draw(cbind(ggplotGrob(plot_M), ggplotGrob(plot_C), size =
"last"))

dev.off()

## base_size in theme sets font baseline +theme_minimal(base_size =
18)

## Saves this great little figure
tiff("Publication_docs/Fig_1a.tiff", height = 6.5, width = 4.7,
units = "in", res = 300)
grid.draw(cbind(ggplotGrob(plot_M),
                ggplotGrob(plot_C), size = "last"))
dev.off()

#ggsave(file = "Fig_1a.tiff", device = "tiff", height = 7.5, width =
5,dpi = 300)

rm(YearTibble)
gc()

################################################################
###################

## Builds a compound figure of mortality by age for different
vehicle classes

## By MMY or 11-digit VIN is the question
Lifetab_v11s_meta<-readRDS("Lifetab_v11s_meta.rds")

Lifetab_v11s_sub<-
Lifetab_v11s_meta[which(Lifetab_v11s_meta$expo_vec>=100),]
Lifetab_v11s_sub$log_mx<-log10(Lifetab_v11s_sub$mx_vec)

Lifetab_v11s_sub<-Lifetab_v11s_sub[which(Lifetab_v11s_sub$log_mx !=-
Inf),]

## Builds figure 1b

#tiff(filename  = "Publication_docs/Fig_1b.tiff",
#     height = 7, width = 6.5, units = "in", res  =300)

cairo_ps("Publication_docs/Fig_1b.eps", height = 7, width = 7.5)

## runs MMY
```

```
Lifetab_MMY<-readRDS("Lifetab_MMY_meta.rds")

Lifetab_MMY_sub<-Lifetab_MMY[which(Lifetab_MMY$expo_vec>=100),]
Lifetab_MMY_sub$log_mx<-log10(Lifetab_MMY_sub$mx_vec)
Lifetab_MMY_sub<-Lifetab_MMY_sub[which(Lifetab_MMY_sub$log_mx !=-
Inf),]

set.seed(52)

par(bty = "n", mfrow = c(1,1), mar = c(4.6,5.1,4.1,0.1))

#layout(mat = matrix(c(1,2),
#                    nrow = 1,
#                    ncol = 2),
#       heights = c(5, 5),    # Heights of the two rows
#       widths = c(3, 1))

boxplot(Lifetab_MMY_sub$log_mx[which(Lifetab_MMY_sub$test_class==4 &
Lifetab_MMY_sub$age>=3 & Lifetab_MMY_sub$age<=47)]~
         Lifetab_MMY_sub$age[which(Lifetab_MMY_sub$test_class==4
&Lifetab_MMY_sub$age>=3 & Lifetab_MMY_sub$age<=47)],
        col = "#e74c3c", notch = F, varwidth = F,
        xlim = c(0,53),
        ylim = c(-4,0),
        pch = 20, outcex =  0.5, outcol = rgb(1,0,0),  ## Match
this to below
        xlab = "Age in Years", ylab = "Probability of Death \n",
cex.lab  =1.2,
        axes = F)
abline(h = seq(-4,0, by = 1), lty = 3, col = "darkgrey")

abline(v = seq(1,60, by = 2), lty = 3, col = "darkgrey")

table(Lifetab_MMY_sub$age)
## for ages where N<=20, ages 47 and over, plots actual data
subvec_jit<-
jitter(Lifetab_MMY_sub$age[which(Lifetab_MMY_sub$test_class==4 &
Lifetab_MMY_sub$age>47)]-2)

points(Lifetab_MMY_sub$log_mx[which(Lifetab_MMY_sub$test_class==4 &
Lifetab_MMY_sub$age>47)]~
        subvec_jit,
       pch = 20, cex = 0.5, col = rgb(1,0.5,0,0.7))

## connects the dots
# points(Lifetab_v11s_sub$log_mx[which(Lifetab_v11s_sub$Tclass==4 &
Lifetab_v11s_sub$age>47)]~
#           subvec_jit,
#        type = "l", col = rgb(0,0,0,0.4))

par(new = T)
boxplot(Lifetab_MMY_sub$log_mx[which(Lifetab_MMY_sub$test_class==4 &
```

```
Lifetab_MMY_sub$age>=3 & Lifetab_MMY_sub$age<=47)]~
        Lifetab_MMY_sub$age[which(Lifetab_MMY_sub$test_class==4
&Lifetab_MMY_sub$age>=3 & Lifetab_MMY_sub$age<=47)],
        col = "orange", #col = "#e74c3c",
        notch = F, varwidth = F,
        xlim = c(0,53),
        ylim = c(-4,0),
        pch = 20, outcex =  0.5, outcol = rgb(1,0.5,0,0.4),  ##
Match this to below
        xlab = "", ylab = "",
        axes = F)

axis(side = 1, at = c(seq(3,50, by = 5)), labels= c(seq(5,50, by =
5)),
      cex.axis = 1.2)
axis(side = 2, at = c(seq(-4,0, by = 1)),
      labels = c("1/10,000", #"one in ten",
                 "1/1,000",
                 "1/100",
                 "1/10",
                 "1/1"), las  =1,
      cex.axis = 1.2)

axis(side = 3, at = c(1:51), las = 2,
      cex.axis = 1.2,
      labels = table(!
is.na(Lifetab_MMY_sub$log_mx[which(Lifetab_MMY_sub$test_class==4 &
Lifetab_MMY_sub$age>=3)]), by =

Lifetab_MMY_sub$age[which(Lifetab_MMY_sub$test_class==4 &
Lifetab_MMY_sub$age>=3)]))

dev.off()

## Now to rule out reporting lag biases, scraps all observations
after 2011 and plots again –
## allowing for a 10+ year delay (to end of 2021) in reporting

cairo_ps("Publication_docs/Fig_S_reporting_lag_bias.eps", height =
7, width = 7.5)

Lifetab_MMY_sub_reportBias<-
Lifetab_MMY_sub[which(c(Lifetab_MMY$YOB+Lifetab_MMY$age)<=2011),]

set.seed(52)

table(Lifetab_MMY_sub_reportBias$age)
## cutoff for N<20 is now 33

par(bty = "n", mfrow = c(1,1), mar = c(4.6,5.1,4.1,0.1))

#layout(mat = matrix(c(1,2),
#                    nrow = 1,
```

```
#                    ncol = 2),
#        heights = c(5, 5),    # Heights of the two rows
#        widths = c(3, 1))

boxplot(Lifetab_MMY_sub_reportBias$log_mx[which(Lifetab_MMY_sub_repo
rtBias$test_class==4 & Lifetab_MMY_sub_reportBias$age>=3 &
Lifetab_MMY_sub_reportBias$age<=32)]~

Lifetab_MMY_sub_reportBias$age[which(Lifetab_MMY_sub_reportBias$test
_class==4 &Lifetab_MMY_sub_reportBias$age>=3 &
Lifetab_MMY_sub_reportBias$age<=32)],
        col = "#e74c3c", notch = F, varwidth = F,
        xlim = c(0,53),
        ylim = c(-4,0),
        pch = 20, outcex =  0.5, outcol = rgb(1,0,0,0),  ## Match
this to below
        xlab = "Age in Years", ylab = "Probability of Death \n",
cex.lab  =1.2,
        axes = F)
abline(h = seq(-4,0, by = 1), lty = 3, col = "darkgrey")

abline(v = seq(1,60, by = 2), lty = 3, col = "darkgrey")

table(Lifetab_MMY_sub_reportBias$age)
## for ages where N<=20, ages 47 and over, plots actual data
subvec_jit<-
jitter(Lifetab_MMY_sub_reportBias$age[which(Lifetab_MMY_sub_reportBi
as$test_class==4 & Lifetab_MMY_sub_reportBias$age>32)]-2)

points(Lifetab_MMY_sub_reportBias$log_mx[which(Lifetab_MMY_sub_repor
tBias$test_class==4 & Lifetab_MMY_sub_reportBias$age>32)]~
        subvec_jit,
        pch = 20, cex = 0.5, col = rgb(1,0.5,0,0.7))

## connects the dots
# points(Lifetab_v11s_sub$log_mx[which(Lifetab_v11s_sub$Tclass==4 &
Lifetab_v11s_sub$age>47)]~
#        subvec_jit,
#        type = "l", col = rgb(0,0,0,0.4))

par(new = T)
boxplot(Lifetab_MMY_sub_reportBias$log_mx[which(Lifetab_MMY_sub_repo
rtBias$test_class==4 & Lifetab_MMY_sub_reportBias$age>=3 &
Lifetab_MMY_sub_reportBias$age<=32)]~

Lifetab_MMY_sub_reportBias$age[which(Lifetab_MMY_sub_reportBias$test
_class==4 &Lifetab_MMY_sub_reportBias$age>=3 &
Lifetab_MMY_sub_reportBias$age<=32)],
        col = "orange", #col = "#e74c3c",
        notch = F, varwidth = F,
        xlim = c(0,53),
        ylim = c(-4,0),
        pch = 20, outcex =  0.5, outcol = rgb(1,0.5,0,0.4),  ##
```

Match this to below
```
        xlab = "", ylab = "",
        axes = F)

axis(side = 1, at = c(seq(3,50, by = 5)), labels= c(seq(5,50, by =
5)),
     cex.axis = 1.2)
axis(side = 2, at = c(seq(-4,0, by = 1)),
     labels = c("1/10,000", #"one in ten",
                "1/1,000",
                "1/100",
                "1/10",
                "1/1"), las  =1,
     cex.axis = 1.2)

axis(side = 3, at = c(1:50), las = 2,
     cex.axis = 1.2,
     labels = table(!
is.na(Lifetab_MMY_sub_reportBias$log_mx[which(Lifetab_MMY_sub_report
Bias$test_class==4 & Lifetab_MMY_sub_reportBias$age>=3)]), by =

Lifetab_MMY_sub_reportBias$age[which(Lifetab_MMY_sub_reportBias$test
_class==4 & Lifetab_MMY_sub_reportBias$age>=3)]))

par(mfrow = c(2,1), mar = c(5.1,4.1,0,2.1), bty = "n")

dev.off()

rm(Lifetab_MMY_sub_reportBias)
gc()

################################################################
##########

## The same treatment for motorcycles (pooled) and 3-3.5T vans, by
MMY

################################################################
##########

#tiff(filename  = "Publication_docs/
Fig_Supp_MMY_survival_motos_vans.tiff",
#     height = 9, width = 7.5, units = "in", res  =300)

cairo_ps("Publication_docs/Fig_Supp_MMY_survival_motos_vans.eps",
height = 9, width = 7.5)

par(bty = "n", mfrow = c(2,1), mar = c(5.1,4.1,3.1,2.1))

boxplot(Lifetab_MMY_sub$log_mx[which(Lifetab_MMY_sub$test_class<=2 &
Lifetab_MMY_sub$age>=1 & Lifetab_MMY_sub$age<=22)]~
```

```
            Lifetab_MMY_sub$age[which(Lifetab_MMY_sub$test_class<=2 &
Lifetab_MMY_sub$age>=1 & Lifetab_MMY_sub$age<=22)],
        col = "#3498db", notch = F, varwidth = F,
        ylim = c(-4,0),
        xlim = c(1,25),
        pch = 20, outcex =  0.5, outcol = rgb(0,0,1,0.7),
        xlab = "Age in Years", ylab = "log10 Probability of Death",
        axes = F)

abline(h = seq(-4,0, by = 1), lty = 3, col = "darkgrey")

abline(v = seq(1,40, by = 2), lty = 3, col = "darkgrey")

par(new = T)

boxplot(Lifetab_MMY_sub$log_mx[which(Lifetab_MMY_sub$test_class<=2 &
Lifetab_MMY_sub$age>=1 & Lifetab_MMY_sub$age<=22)]~
            Lifetab_MMY_sub$age[which(Lifetab_MMY_sub$test_class<=2 &
Lifetab_MMY_sub$age>=1 & Lifetab_MMY_sub$age<=22)],
        col = "#3498db", notch = F, varwidth = F,
        ylim = c(-4,0),
        xlim = c(1,25),
        pch = 20, outcex =  0.5, outcol = rgb(0,0,1,0.7),
        xlab = "", ylab = "",
        axes = F)

## for ages where N<=20, ages 24 and over, plots actual data
subvec_jit<-
jitter(Lifetab_MMY_sub$age[which(Lifetab_MMY_sub$test_class<=2 &
Lifetab_MMY_sub$age>22)]-2)

points(Lifetab_MMY_sub$log_mx[which(Lifetab_MMY_sub$test_class<=2 &
Lifetab_MMY_sub$age>22)]~
            subvec_jit,
        pch = 20, cex = 0.5, col = rgb(0,0,1,0.7))

axis(side = 3, at = c(1:22), las = 2,
     cex.axis = 0.8,
     labels = table(!
is.na(Lifetab_MMY_sub$log_mx[which(Lifetab_MMY_sub$test_class<=2 &
Lifetab_MMY_sub$age>=1)]), by =

Lifetab_MMY_sub$age[which(Lifetab_MMY_sub$test_class<=2 &
Lifetab_MMY_sub$age>=1)]))

axis(side = 1, at = c(seq(3,25, by = 5)), labels= c(seq(5,25, by =
5)),
     cex.axis = 1)
axis(side = 2)

text(x = 10, y = -3.5, labels = "Motorbikes and Mopeds")

mtext("a   \n", adj = 0, cex = 1.6)
#par(new = T)
```

```
boxplot(Lifetab_MMY_sub$log_mx[which(Lifetab_MMY_sub$test_class==7 &
Lifetab_MMY_sub$age>=3 & Lifetab_MMY_sub$age<=21)]~
          Lifetab_MMY_sub$age[which(Lifetab_MMY_sub$test_class==7 &
Lifetab_MMY_sub$age>=3& Lifetab_MMY_sub$age<=21)],
        col = "lightgreen", notch = F, varwidth = F,
        ylim = c(-4,0),
        xlim = c(1,40),
        pch = 20, outcex =  0.5, outcol = rgb(0,0.5,0,0.7),
        xlab = "Age in Years", ylab = "log10 Probability of Death",
        axes = F)

abline(h = seq(-4,0, by = 1), lty = 3, col = "darkgrey")

abline(v = seq(1,40, by = 2), lty = 3, col = "darkgrey")

par(new = T)
boxplot(Lifetab_MMY_sub$log_mx[which(Lifetab_MMY_sub$test_class==7 &
Lifetab_MMY_sub$age>=3 & Lifetab_MMY_sub$age<=21)]~
          Lifetab_MMY_sub$age[which(Lifetab_MMY_sub$test_class==7 &
Lifetab_MMY_sub$age>=3 & Lifetab_MMY_sub$age<=21)],
        col = "lightgreen", notch = F, varwidth = F,
        ylim = c(-4,0),
        xlim = c(1,40),
        pch = 20, outcex =  0.5, outcol = rgb(0,0.5,0,0.7),
        xlab = "", ylab = "",
        axes = F)

## for ages where N<=20, plots actual data
subvec_jit<-
jitter(Lifetab_MMY_sub$age[which(Lifetab_MMY_sub$test_class==7 &
Lifetab_MMY_sub$age>21)]-2)

points(Lifetab_MMY_sub$log_mx[which(Lifetab_MMY_sub$test_class==7 &
Lifetab_MMY_sub$age>21)]~
          subvec_jit,
        pch = 20, cex = 0.5, col = rgb(0,0.5,0,0.7))

axis(side = 3, at = c(c(1:38)), las = 2,
     cex.axis = 0.8,
     labels = c(table(!
is.na(Lifetab_MMY_sub$log_mx[which(Lifetab_MMY_sub$test_class==7 &
Lifetab_MMY_sub$age>=3)]), by =

Lifetab_MMY_sub$age[which(Lifetab_MMY_sub$test_class==7 &
Lifetab_MMY_sub$age>=3)])[1:20],
                rep(0, times = 13), rep(1, times = 5)))

## direct-labels the awesomely long-lived 1971 Ford Transit van
text(x = 35, y = -1.5, labels = "1971 Ford\n Transit Vans", col =
"darkgreen")

axis(side = 1, at = c(seq(3,40, by = 5)), labels= c(seq(5,40, by =
5)),
```

```
      cex.axis = 1)
axis(side = 2)

text(x = 10, y = -3.5, labels = "Vans 3000-3500kg")

mtext("b    \n", adj = 0, cex = 1.6)

dev.off()

################################################################
####

## Deaths per mile, instead of per year

################################################################
####

Lifetab_MMY_miles_meta<-readRDS("Lifetab_MMY_miles_meta.rds")

Lifetab_MMY_miles_sub<-
Lifetab_MMY_miles_meta[which(Lifetab_MMY_miles_meta$expo_vec>=100),]
Lifetab_MMY_miles_sub$log_mx<-log10(Lifetab_MMY_miles_sub$mx_vec)

Lifetab_MMY_miles_sub<-
Lifetab_MMY_miles_sub[which(Lifetab_MMY_miles_sub$log_mx !=-Inf),]
rm(Lifetab_MMY_miles_meta)

qplot(Lifetab_MMY_miles_sub$age[which(Lifetab_MMY_miles_sub$Tclass==
4)],

Lifetab_MMY_miles_sub$log_mx[which(Lifetab_MMY_miles_sub$Tclass==4)]
,
     alpha = I(0))+
  geom_hex(bins = 60)+
  theme_minimal()

## traces medians
med_vals<-
aggregate(Lifetab_MMY_miles_sub$log_mx[which(Lifetab_MMY_miles_sub$T
class==4)],
                  by =
list(Lifetab_MMY_miles_sub$age[which(Lifetab_MMY_miles_sub$Tclass==4
)]),
                  median, na.rm = T)

upQ_vals<-
aggregate(Lifetab_MMY_miles_sub$log_mx[which(Lifetab_MMY_miles_sub$T
class==4)],
                  by =
list(Lifetab_MMY_miles_sub$age[which(Lifetab_MMY_miles_sub$Tclass==4
)]),
                  quantile,0.75, na.rm = T)
```

```
lowQ_vals<-
aggregate(Lifetab_MMY_miles_sub$log_mx[which(Lifetab_MMY_miles_sub$T
class==4)],
                      by =
list(Lifetab_MMY_miles_sub$age[which(Lifetab_MMY_miles_sub$Tclass==4
)]),
                      quantile,0.25, na.rm = T)

## counts MMY combinations observed at each mileage (table() ignores
zeros)
MMY_obs<-NULL
for(i in 0:1000){
  MMY_obs<-c(MMY_obs,sum(!
is.na(Lifetab_MMY_miles_sub$log_mx[which(Lifetab_MMY_miles_sub$Tclas
s==4
                                                                    &
Lifetab_MMY_miles_sub$age==(i*1000))])))
}

## counts total vehicles observed at each mileage (table() ignores
zeros)
MMY_N_obs<-NULL
for(i in 0:1000){
  MMY_N_obs<-
c(MMY_N_obs,sum(Lifetab_MMY_miles_sub$expo_vec[which(Lifetab_MMY_mil
es_sub$Tclass==4
                                                                    &
Lifetab_MMY_miles_sub$age==(i*1000))]))
}

MMY_obs_smallN<-c(0:1000)[which(MMY_obs<=20)]
MMY_obs_N20<-c(0:1000)[which(MMY_obs>20)]

### Generates a figure with over-plotting, switches to plotting raw
data when N<=20 MMY combinations

#tiff(filename  = "Publication_docs/Fig_1c.tiff",
#     height = 6.5, width = 11, units = "in", res  =300)

cairo_ps("Publication_docs/Fig_1c.eps", height = 6.5, width = 11)

par(mfrow = c(2,1), mar = c(5.1,4.1,0,2.1), bty = "n")

#tiff(filename  = "Publication_docs/Fig_1c.tiff",
#     height = 5, width = 11, units = "in", res  =300)

layout(mat = matrix(c(2,1),
                    nrow = 2,
```

```
                          ncol = 1),
        heights = c(2, 5),     # Heights of the two rows
        widths = c(6, 6))

set.seed(233)

## plots empty frame
plot(med_vals, col = "white",
     ylim = c(-6,0), pch = 20, cex = 0.5,
     xlim = c(0,800), axes = F,
     xlab = "", ylab = "")

## lays down gridlines
abline(h = seq(-6,0, by = 1), lty = 3, col = "darkgrey")
abline(v = seq(0,800, by = 100), lty = 3, col = "darkgrey")

par(new = T)

## box-whisker plot
boxplot(Lifetab_MMY_miles_sub$log_mx[which(Lifetab_MMY_miles_sub$Tcl
ass==4 & Lifetab_MMY_miles_sub$age<=418000)]~ # %in%
c(MMY_obs_N20*1000))]~

c(Lifetab_MMY_miles_sub$age[which(Lifetab_MMY_miles_sub$Tclass==4 &
Lifetab_MMY_miles_sub$age<=418000)]),  #Lifetab_MMY_miles_sub$age
%in% c(MMY_obs_N20*1000))]*1000),
        #col = rgb(0.6,0.7,1,1),
        #border = rgb(0.6,0.7,1,1), #  border = "lightblue",
        col = "white", border = "white",
        notch = F, varwidth = F,
        whisklty = 1,whiskcol = "black", #whiskcol =
c(c("black","white")[as.numeric(MMY_obs<=20)]),
        whisklwd = 0.5,
        ylim = c(-6,0),
        xlim = c(0,800),
        pch = 20, outcex =  0.2, outcol =  rgb(1,0.5,0,0.2),
        xlab = "Age in Miles", ylab = "log10 Probability of Death",
        axes = F)

## manually adds points for boxplots, N>20 samples only
for(i in 1:length(MMY_obs_N20)){

  ## IQR lines (instead of boxes)
  points(x = c((MMY_obs_N20[i]*1000)+1000,
(MMY_obs_N20[i]*1000)+1000)/1000,
         y = c(upQ_vals$x[[MMY_obs_N20[i]+1]],
lowQ_vals$x[[MMY_obs_N20[i]+1]]),
         type = "l", col = "orange")
}

## for ages where N<=20, plots actual data
subvec_jit<-
```

```r
jitter(Lifetab_MMY_miles_sub$age[which(Lifetab_MMY_miles_sub$Tclass=
=4 &

Lifetab_MMY_miles_sub$age %in% c(MMY_obs_smallN*1000))]/1000)

points(Lifetab_MMY_miles_sub$log_mx[which(Lifetab_MMY_miles_sub$Tcla
ss==4 & Lifetab_MMY_miles_sub$age %in% c(MMY_obs_smallN*1000))]~
        c(subvec_jit+1),
      pch = 20, cex = 0.1, col = rgb(1,0.5,0,0.7))

## adds the median values manually over the top
for(i in (MMY_obs_N20+1)){
  ## Medians
  points(x =  (i),
        y = c(med_vals$x[[i]]),
        pch = 20, cex = 0.2, col = rgb(0,0,0,1))

  ## upper and lower quartiles
  #  points(x =  c((i*1000)-1000),
  #         y = c(upQ_vals$x[[i]]),
  #         pch = 20, cex = 0.2, col = rgb(0,0,0,0.5))
  #  points(x =  c((i*1000)-1000),
  #         y = c(lowQ_vals$x[[i]]),
  #         pch = 20, cex = 0.2, col = rgb(0,0,0,0.5))
}

## adds the axes
axis(side = 1, at = seq(0,800, by = 200),
     labels = c("0", "200k", "400k", "600k", "800k"),
     cex.axis = 1.3)
#labels = seq(0,800000, by = 200000), las = 2)
axis(side = 2, cex.axis = 1.2)

par(mar = c(0.1,4.1,0.1,2.1))

## plots this
plot(log10(MMY_N_obs)~c(c(0:1000)*1000),
     xlab = "", ylab = "",
     ylim = c(2,8),
     xlim = c(0,800000),
     type = "l", axes = F)

abline(h = seq(2,8, by = 2), lty = 3, col = "darkgrey")
abline(v = seq(0,1000000, by = 100000), lty = 3, col = "darkgrey")

points(log10(MMY_N_obs)~c(c(0:1000)*1000), pch = 20, cex = 0.2, col
= "red")

axis(side = 2, at = seq(2,8, by = 1), labels = c("100", "1k", "10k",
"100k", "1M", "10M", "100M"),
     las = 2, cex.axis = 1.2)
```

```
#text(x = 500000, y = 6.5, "Common Make-Models\n Exposed to Risk")

dev.off()

par(mfrow = c(1,1), mar = c(5.1,4.1,4.1,2.1), bty = "n")

cairo_ps("Publication_docs/Fig_1c_variant.eps", height = 6.5, width
= 11)

set.seed(233)

## plots empty frame
plot(med_vals, col = "white",
     ylim = c(-6,0), pch = 20, cex = 0.5,
     xlim = c(0,800), axes = F,
     xlab = "", ylab = "")

## lays down gridlines
abline(h = seq(-6,0, by = 1), lty = 3, col = "darkgrey")
abline(v = seq(0,800, by = 100), lty = 3, col = "darkgrey")

par(new = T)

## box-whisker plot
boxplot(Lifetab_MMY_miles_sub$log_mx[which(Lifetab_MMY_miles_sub$Tcl
ass==4 & Lifetab_MMY_miles_sub$age<=418000)])~ # %in%
c(MMY_obs_N20*1000))]~

c(Lifetab_MMY_miles_sub$age[which(Lifetab_MMY_miles_sub$Tclass==4 &
Lifetab_MMY_miles_sub$age<=418000)]),  #Lifetab_MMY_miles_sub$age
%in% c(MMY_obs_N20*1000))]*1000),
        #col = rgb(0.6,0.7,1,1),
        #border = rgb(0.6,0.7,1,1), #  border = "lightblue",
        col = "white", border = "white",
        notch = F, varwidth = F,
        whisklty = 1,whiskcol = "black", #whiskcol =
c(c("black","white")[as.numeric(MMY_obs<=20)]),
        whisklwd = 0.5,
        ylim = c(-6,0),
        xlim = c(0,800),
        pch = 20, outcex =  0.2, outcol =  rgb(1,0.5,0,0.2),
        xlab = "Age in Miles", ylab = "log10 Probability of Death",
        axes = F)

## manually adds points for boxplots, N>20 samples only
for(i in 1:length(MMY_obs_N20)){

  ## IQR lines (instead of boxes)
```

```
  points(x = c((MMY_obs_N20[i]*1000)+1000,
(MMY_obs_N20[i]*1000)+1000)/1000,
        y = c(upQ_vals$x[[MMY_obs_N20[i]+1]],
lowQ_vals$x[[MMY_obs_N20[i]+1]]),
        type = "l", col = "orange")
}

## for ages where N<=20, plots actual data
subvec_jit<-
jitter(Lifetab_MMY_miles_sub$age[which(Lifetab_MMY_miles_sub$Tclass=
=4 &

Lifetab_MMY_miles_sub$age %in% c(MMY_obs_smallN*1000))]/1000)

points(Lifetab_MMY_miles_sub$log_mx[which(Lifetab_MMY_miles_sub$Tcla
ss==4 & Lifetab_MMY_miles_sub$age %in% c(MMY_obs_smallN*1000))]~
        c(subvec_jit+1),
      pch = 20, cex = 0.1, col = rgb(1,0.5,0,0.7))

## adds the median values manually over the top
for(i in (MMY_obs_N20+1)){
  ## Medians
  points(x =  (i),
         y = c(med_vals$x[[i]]),
         pch = 20, cex = 0.2, col = rgb(0,0,0,1))

  ## upper and lower quartiles
  #  points(x =  c((i*1000)-1000),
  #         y = c(upQ_vals$x[[i]]),
  #         pch = 20, cex = 0.2, col = rgb(0,0,0,0.5))
  #  points(x =  c((i*1000)-1000),
  #         y = c(lowQ_vals$x[[i]]),
  #         pch = 20, cex = 0.2, col = rgb(0,0,0,0.5))
}

## adds the axes
axis(side = 1, at = seq(0,800, by = 200),
     labels = c("0", "200k", "400k", "600k", "800k"),
     cex.axis = 1.3)
#labels = seq(0,800000, by = 200000), las = 2)
axis(side = 2, cex.axis = 1.2)

#axis(side = 3, at = seq(0,800, by = 100), labels =
MMY_N_obs[seq(1,801, by = 100)], las= 2)

axis(side = 3, at = seq(0,800, by = 100),
     labels = c("40.6 M", "20.3 M", "11.0M", "55K", "7.5K", "1732",
"678", "127", "0"), las= 2)

dev.off()
```

```
########################################################
####

## Figure 2 — the relationship between wear-and-tear and survival

########################################################
####

## switches to lifetables with metadata
Lifetab_MMY<-readRDS("Lifetab_MMY_meta.rds")
Lifetab_MMY_miles<-readRDS("Lifetab_MMY_miles_meta.rds")

FPT_MMY_miles<-readRDS("data/FPT_MMY_miles_clean.rds")
FPT_MMY_age<-readRDS("data/FPT_MMY_age_clean.rds")

## Joins the failure rates per test to the survival patterns
Lifetab_MMY_miles_combined<-data.frame(FPT_MMY_miles,

Lifetab_MMY_miles[match(FPT_MMY_miles$Var1,

paste(Lifetab_MMY_miles$MMY,Lifetab_MMY_miles$age, sep = "__")),])

## The same for chronological age
Lifetab_MMY_combined<-data.frame(FPT_MMY_age,
                                 Lifetab_MMY[match(FPT_MMY_age$Var1,

paste(Lifetab_MMY$MMY,Lifetab_MMY$age, sep = "__")),])

## filters a few non-matches due to rounding (e.g. '96 Alfa Rom. 146
that is cut off because apparent obs age is in 2004 due to rounding)
Lifetab_MMY_combined<-Lifetab_MMY_combined[!
is.na(Lifetab_MMY_combined$MMY.1),]

######################################

##  Mechanical failure rates by year, for x-year-old vehicles

######################################

## choose the age you want
age_select<-5

Lifetab_subextra<-
Lifetab_MMY_combined[which(Lifetab_MMY_combined$age==age_select &

Lifetab_MMY_combined$expo_vec>100 &

Lifetab_MMY_combined$test_class==4),]

Lifetab_subextra2<-
Lifetab_MMY_combined[which(Lifetab_MMY_combined$age==10 &
```

```
Lifetab_MMY_combined$expo_vec>100 &

Lifetab_MMY_combined$test_class==4),]

cairo_ps("Publication_docs/Fig_LongRates.eps", height = 8.5, width =
8.5)

par(mfrow = c(2,2), mar = c(4.1,4.1,3.1,1.1), bty = "n")

boxplot(100*(Lifetab_subextra$N_failed/Lifetab_subextra$N_tests)~
        c(Lifetab_subextra$YOB+age_select),
        ylim = c(0,100), xlab = "", ylab = "Percent Tests Failed",
        outcol = "#FFD12A", pch = "-", cex = 1.5,
        col = "#FFD12A",las = 2) ## banana crayon

abline(h = seq(0,100, by = 20), lty = 3, col = "grey")
abline(v = seq(1,20, by = 2), lty = 3, col = "grey")

## adds a line at the 2005 median
abline(h = median(100*(Lifetab_subextra$N_failed/
Lifetab_subextra$N_tests)
[c(Lifetab_subextra$YOB+age_select)==2005]),
        col = "hotpink")

axis(side = 3, labels = table(c(Lifetab_subextra$YOB+age_select)),
at = 1:17, las = 2)

boxplot(c(Lifetab_subextra$mx_vec)~
        c(Lifetab_subextra$YOB+age_select),
        xlab = "", ylab = "Mortality Rate",
        outcol = "#E30B5C", pch = "-", cex = 1.5,
        col = "#E30B5C",las = 2) ## Razzamatazz

abline(h = seq(0,1, by = 0.02), lty = 3, col = "grey")
abline(v = seq(1,20, by = 2), lty = 3, col = "grey")

axis(side = 3, labels = table(c(Lifetab_subextra$YOB+age_select)),
at = 1:17, las = 2)

#abline(h = median(c(Lifetab_subextra$mx_vec)
[c(Lifetab_subextra$YOB+age_select)==2005]),
#       col = "black")

boxplot(100*(Lifetab_subextra2$N_failed/Lifetab_subextra2$N_tests)~
        c(Lifetab_subextra2$YOB+10),
        ylim = c(0,100), xlab = "Year", ylab = "Percent Tests
Failed",
        outcol = "#84DE02", pch = "-", cex = 1.5,
        col = "#84DE02",las = 2) ## alien armpit

abline(h = seq(0,100, by = 20), lty = 3, col = "grey")
```

```r
abline(v = seq(1,20, by = 2), lty = 3, col = "grey")

axis(side = 3, labels = table(c(Lifetab_subextra2$YOB+10)), at =
1:17, las = 2)

## adds a line at the 2005 median
abline(h = median(100*(Lifetab_subextra2$N_failed/
Lifetab_subextra2$N_tests)[c(Lifetab_subextra2$YOB+10)==2005]),
       col = "hotpink")

boxplot(c(Lifetab_subextra2$mx_vec)~
          c(Lifetab_subextra2$YOB+10),
       xlab = "Year", ylab = "Mortality Rate",
       outcol = "#4570E6", pch = "-",  cex = 1.5,
       col = "#4570E6",las = 2) ## Blue (II)

abline(h = seq(0,1, by = 0.2), lty = 3, col = "grey")
abline(v = seq(1,20, by = 2), lty = 3, col = "grey")

axis(side = 3, labels = table(c(Lifetab_subextra2$YOB+10)), at =
1:17, las = 2)

dev.off()

#######################

par(mfrow = c(1,1), mar = c(5.1,4.1,4.1,2.1))

boxplot(100*(Lifetab_subextra$N_failed/Lifetab_subextra$N_tests)~
          c(Lifetab_subextra$YOB+age_select),
       ylim = c(0,30), xlab = "Year Tested", ylab = "Percent MOTs
failed",
       outcol = "orange", pch = "-", col = "orange",las = 2)

####### NOTE TWO are cut off above y-axis to show trends!

abline(h = seq(0,50, by  =10), lty = 3, col = "darkgrey")
abline(v = seq(0,40, by  =1), lty = 3, col = "darkgrey")

## the Nissan Leaf, really the first large electric car
Leaf<-Lifetab_subextra[grep("__LEAF__", Lifetab_subextra$MMY),]

## whacks the Leaf in there as dark green
#points(100*(Leaf$N_failed[which(Leaf$age==age_select)]/
Leaf$N_tests[which(Leaf$age==age_select)])~
#         c(Leaf$YOB[which(Leaf$age==age_select)]-(2004.25-
```

```
age_select)),
#        pch =3, cex = 1.4, col = "darkgreen")

#points(100*(Leaf$N_failed[which(Leaf$age==age_select)]/
Leaf$N_tests[which(Leaf$age==age_select)])~
#          c(Leaf$YOB[which(Leaf$age==age_select)]-(2004.25-
age_select)),
#        type = "l", col = "darkgreen")

## the Tesla
Tesla<-Lifetab_MMY_combined[grep("^TESLA__",
Lifetab_MMY_combined$MMY),]

## Teslas as light blue
points(100*(Tesla$N_failed[which(Tesla$age==age_select)]/
Tesla$N_tests[which(Tesla$age==age_select)])~
        c(Tesla$YOB[which(Tesla$age==age_select)]-(2003.75-
age_select)),
      pch =20, cex = 1.4, col = "lightblue")

## adds mortality rates for contrast

boxplot(log10(Lifetab_MMY_combined$mx_vec[which(Lifetab_MMY_combined
$age==age_select &

Lifetab_MMY_combined$expo_vec>100 &

Lifetab_MMY_combined$test_class==4)])~

c(Lifetab_MMY_combined$YOB[which(Lifetab_MMY_combined$age==age_selec
t &

Lifetab_MMY_combined$expo_vec>100 &

Lifetab_MMY_combined$test_class==4)]+age_select),
        xlab = "Year Tested", ylab = "log10 Mortality Rate",
        outcol = "orange", pch = 20, col = "orange",las = 2)

abline(h = seq(0,50, by  =10), lty = 3, col = "darkgrey")
abline(v = seq(0,40, by  =1), lty = 3, col = "darkgrey")

## the Nissan Leaf, really the first large electric car
Leaf<-Lifetab_MMY_combined[grep("__LEAF__",
Lifetab_MMY_combined$MMY),]

## whacks the Leaf in there as dark green
points(100*(Leaf$N_failed[which(Leaf$age==age_select)]/
Leaf$N_tests[which(Leaf$age==age_select)])~
        c(Leaf$YOB[which(Leaf$age==age_select)]-(2004.25-
age_select)),
      pch =20, cex = 1.4, col = "darkgreen")

## the Tesla
```

```
Tesla<-Lifetab_MMY_combined[grep("^TESLA__",
Lifetab_MMY_combined$MMY),]

## Teslas as light blue
points(100*(Tesla$N_failed[which(Tesla$age==age_select)]/
Tesla$N_tests[which(Tesla$age==age_select)])~
        c(Tesla$YOB[which(Tesla$age==age_select)]-(2003.75-
age_select)),
        pch =20, cex = 1.4, col = "lightblue")

cairo_ps("Publication_docs/Fig_MechPassFail.eps", height = 5, width
= 9.5)

par(mfrow = c(1,2), bty = "n")

### Mechanical pass/fails per mile
Lifetab_subextra<-
Lifetab_MMY_miles_combined[which(Lifetab_MMY_miles_combined$N_tests>
100 &

Lifetab_MMY_miles_combined$Tclass==4),]

Lifetab_subextra$pct_Fail<-c(100*(Lifetab_subextra$N_failed/
Lifetab_subextra$N_tests))

boxplot(Lifetab_subextra$pct_Fail[Lifetab_subextra$miles<=370000]~Li
fetab_subextra$miles[Lifetab_subextra$miles<=370000],
        xlab = "Miles", ylab = "Percent Tests Failed",
        xlim = c(1,58),
        pch ="-", outcol = rgb(0.2,0.5,0.9,0.3), col =
rgb(0.2,0.5,0.9,0.5), axes = F)

abline(h = seq(0,100, by = 10), lty = 3, col = "grey")
abline(v = seq(1,60, by = 5), lty = 3, col = "grey")

## adds points for N<20 makes and models
points(Lifetab_subextra$pct_Fail[Lifetab_subextra$miles>370000]~

jitter(c(Lifetab_subextra$miles[Lifetab_subextra$miles>370000]/
10000)+1),
        pch = 20, cex= 0.5, col = rgb(0.2,0.5,0.9,0.3))

Leaf<-Lifetab_subextra[grep("__LEAF__", Lifetab_subextra$MMY),]
Tesla<-Lifetab_subextra[grep("TESLA__", Lifetab_subextra$MMY),]
FF_02<-Lifetab_subextra[grep("FORD__FOCUS__2002",
Lifetab_subextra$MMY),]

Taxi<-Lifetab_subextra[grep("INT__TX4", Lifetab_subextra$MMY),]

#points(100-Leaf$pct_Fail~c(c(Leaf$miles/10000)+0.9),
#       pch =20, cex = 0.5,  col = "#FFD12A")
```

```r
#points(100-Tesla$pct_Fail~c(c(Tesla$miles/10000)+1.1),
#       pch =20, cex = 0.5,  col = "#FF007C")

#points(100-FF_02$pct_Fail~c(c(FF_02$miles/10000)+1.1),
#       pch =20, cex = 0.5,  col = "purple")

axis(side = 1, at = seq(1, 51, by = 10), labels = c(0,"100k",
"200k", "300k","400k", "500k"))
axis(side = 2)
axis(side = 3, labels = c(table(Lifetab_subextra$miles)[seq(1,58, by
= 5)]), at = seq(1,58, by = 5), las = 2)

## REMEMBER this is just fig 1d with a linear x-axis

boxplot(Lifetab_subextra$mx_vec[Lifetab_subextra$miles<=370000]~Life
tab_subextra$miles[Lifetab_subextra$miles<=370000],
        xlab = "Miles", ylab = "Mortality Rate", ylim =c(0,0.2),xlim
= c(1,58),
        pch ="-", outcol = rgb(0.2,0.5,0.9,0.3), col =
rgb(0.2,0.5,0.9,0.5), axes = F)

abline(h = seq(0,1, by = 0.05), lty = 3, col = "grey")
abline(v = seq(1,60, by = 5), lty = 3, col = "grey")

## adds points for N<20 makes and models
points(Lifetab_subextra$mx_vec[Lifetab_subextra$miles>370000]~

jitter(c(Lifetab_subextra$miles[Lifetab_subextra$miles>370000]/
10000)+1),
        pch = 20, cex= 0.5, col = rgb(0.2,0.5,0.9,0.3))

Leaf<-Lifetab_subextra[grep("__LEAF__", Lifetab_subextra$MMY),]
Tesla<-Lifetab_subextra[grep("TESLA__", Lifetab_subextra$MMY),]
FF_02<-Lifetab_subextra[grep("FORD__FOCUS__2002",
Lifetab_subextra$MMY),]

Taxi<-Lifetab_subextra[grep("INT__TX4", Lifetab_subextra$MMY),]

#points(Leaf$mx_vec~c(c(Leaf$miles/10000)+0.9),
#       pch =20, cex = 0.5,  col = "#FFD12A")

#points(Tesla$mx_vec~c(c(Tesla$miles/10000)+1.1),
#       pch =20, cex = 0.5,  col = "#FF007C")

#points(100-FF_02$pct_Fail~c(c(FF_02$miles/10000)+1.1),
#       pch =20, cex = 0.5,  col = "purple")

axis(side = 1, at = seq(1, 51, by = 10), labels = c(0,"100k",
"200k", "300k","400k", "500k"))
axis(side = 2)
axis(side = 3, labels = c(table(Lifetab_subextra$miles)[seq(1,58, by
= 5)]), at = seq(1,58, by = 5), las = 2)
```

```
dev.off()

#####################################

## How does the relationship between breakdowns and survival change
with mileage?
cor_vec3<-NULL
for(i in 1:52){

  xobj<-cor.test(c(Lifetab_subextra$N_failed/
Lifetab_subextra$N_tests)
[Lifetab_subextra$miles==c(seq(10000,600000, by = 10000)[i])],

c(Lifetab_subextra$mx_vec[Lifetab_subextra$miles==c(seq(10000,600000
, by = 10000)[i])]))
  if(sum(Lifetab_subextra$miles==c(seq(10000,600000, by = 10000)
[i]))>=20){
    cor_vec3[[i]]<-c(xobj$est, xobj$conf.int, xobj$p.value,
sum(Lifetab_subextra$miles==c(seq(10000,600000, by = 10000)[i])))
  }

  if(sum(Lifetab_subextra$miles==c(seq(10000,600000, by = 10000)
[i]))<20){
    cor_vec3[[i]]<-c(NA,NA,NA,NA,NA)
  }
}

cor_vec3<-matrix(unlist(cor_vec3), ncol = 5, byrow = T)
cor_vec3<-ldply(unlist(cor_vec3), as.data.frame)

#####################################

##  Compares the survival patterns of several notable vehicles,
combines
##  with previous figure on reliability

#####################################

## what is the most common van and motorcycle?
Motos<-
Lifetab_MMY_miles_combined[which(Lifetab_MMY_miles_combined$Tclass<=
2),]
## the Vespa 2003 of course PIAGGIO__VESPA__2003
Motos[which.max(Motos$N_tests),]

Vans<-
Lifetab_MMY_miles_combined[which(Lifetab_MMY_miles_combined$Tclass==
7),]
## the FORD__TRANSIT__2001
Vans[which.max(Vans$N_tests),]
```

```r
Vespa<-Lifetab_MMY_miles_combined[grep("PIAGGIO__VESPA__2003",
Lifetab_MMY_miles_combined$MMY),]
Transit<-Lifetab_MMY_miles_combined[grep("FORD__TRANSIT__2001",
Lifetab_MMY_miles_combined$MMY),]

## Pulls the London Taxis TX II 2002 model (common long-lived Black
London Cab)
Taxis<-Lifetab_MMY_miles_combined[grep("LONDON",
Lifetab_MMY_miles_combined$MMY),]
#Taxis[which.max(Taxis$expo_vec),]
#Taxis<-Taxis[grep("TX II__2002",Taxis$MMY),]
## gets one that doesn't break the 12-year maximum legal lifespan
Taxis<-Taxis[grep("TX4__2010",Taxis$MMY),]

## gets the 2002 Ford Focus, most common vehicle we have
Focus<-Lifetab_MMY_miles_combined[grep("FORD__FOCUS__2002",
Lifetab_MMY_miles_combined$MMY),]

## Gets the 2004 Piaggio NRG (absolutely the worst. Just really
REALLY bad.)
NRG<-Lifetab_MMY_miles_combined[grep("__NRG__2004",
Lifetab_MMY_miles_combined$MMY),]

## and the supposedly good 2015 Nissan X-trail (which is actually
just not driven very far)
XTRAIL<-Lifetab_MMY_miles_combined[grep("__X-TRAIL__2015",
Lifetab_MMY_miles_combined$MMY),]

## The electric cars....

## the 2013 Leaf, really the first large electric cohort
Leaf<-Lifetab_MMY_miles_combined[grep("__LEAF__2013",
Lifetab_MMY_miles_combined$MMY),]

## the 2015 Tesla, the first large electric Tesla cohort
Tesla<-Lifetab_MMY_miles_combined[grep("TESLA__MODEL S__2015",
Lifetab_MMY_miles_combined$MMY),]

###################################################

expo_size<-99

## who has the worst survival at 100k miles for a car?
temp_table<-
Lifetab_MMY_miles_combined[which(Lifetab_MMY_miles_combined$Tclass==
4 &

Lifetab_MMY_miles_combined$miles==100000 &

Lifetab_MMY_miles_combined$expo_vec>=100),]

head(temp_table[order(temp_table$mx_vec, decreasing = T),])
```

```
## best? LONDON TAXIS INT__TX II__2002 then LONDON TAXIS INT__TX1
BRONZE__2001 then
tail(temp_table[order(temp_table$mx_vec, decreasing = T),], 30)

## the worst? FORD__KA__1999
KaKa<-Lifetab_MMY_miles_combined[grep("FORD__KA__1999",
Lifetab_MMY_miles_combined$MMY),]

## worst at 50k? VAUXHALL__ADAM__2017
Adam<-Lifetab_MMY_miles_combined[grep("VAUXHALL__ADAM__2017",
Lifetab_MMY_miles_combined$MMY),]

Taxis<-Taxis[order(Taxis$miles),]
Focus<-Focus[order(Focus$miles),]
Tesla<-Tesla[order(Tesla$miles),]
Leaf<-Leaf[order(Leaf$miles),]
NRG<-NRG[order(NRG$miles),]
Vespa<-Vespa[order(Vespa$miles),]
Transit<-Transit[order(Transit$miles),]
KaKa<-KaKa[order(KaKa$miles),]
Adam<-Adam[order(Adam$miles),]

#cairo_ps("Publication_docs/Fig_S_notables.eps", height = 5, width =
8)

cairo_ps("Publication_docs/Fig_MechPassFail_compound.eps", height =
8.5, width = 8.5)

par(mfrow = c(2,2), bty = "n", mar =c(4.1,4.1,4.1,1.1))

### Mechanical pass/fails per year

Lifetab_subextra<-
Lifetab_MMY_combined[which(Lifetab_MMY_miles_combined$N_tests>100 &

Lifetab_MMY_miles_combined$Tclass==4),]

Lifetab_subextra$pct_Fail<-c(100*(Lifetab_subextra$N_failed/
Lifetab_subextra$N_tests))

Lifetab_subextra<-Lifetab_subextra[which(Lifetab_subextra$age>=3),]

boxplot(Lifetab_subextra$pct_Fail[Lifetab_subextra$age<=40]~Lifetab_
subextra$age[Lifetab_subextra$age<=40],
        xlab = "Age in Years", ylab = "Percent Tests Failed",
        xlim = c(1,58),
        pch ="-", outcol = rgb(0.2,0.5,0.9,0.5), col =
rgb(0.2,0.5,0.9,0.7), axes = F)

abline(h = seq(0,100, by = 5), lty = 3, col = "grey")
abline(v = seq(3,63, by = 5), lty = 3, col = "grey")
```

```
## adds points for N<20 makes and models
points(Lifetab_subextra$pct_Fail[Lifetab_subextra$age>40]~
        jitter(c(Lifetab_subextra$age[Lifetab_subextra$age>40])-2),
        pch = 20, cex = 0.5, col = rgb(0.2,0.5,0.9,0.5))

axis(side = 1, at = seq(3, 53, by = 10), labels = seq(5,55, by =10))
axis(side = 2)
axis(side = 3, labels = c(table(Lifetab_subextra$age)[seq(1,55, by =
5)]), at = seq(3,57, by = 5), las = 2)

### Mechanical pass/fails per mile
Lifetab_subextra<-
Lifetab_MMY_miles_combined[which(Lifetab_MMY_miles_combined$N_tests>
100 &

Lifetab_MMY_miles_combined$Tclass==4),]

Lifetab_subextra$pct_Fail<-c(100*(Lifetab_subextra$N_failed/
Lifetab_subextra$N_tests))

boxplot(Lifetab_subextra$pct_Fail[Lifetab_subextra$miles<=370000]~Li
fetab_subextra$miles[Lifetab_subextra$miles<=370000],
        xlab = "Age in Miles", ylab = "Percent Tests Failed",
        xlim = c(1,58),
        pch ="-", outcol = rgb(0.2,0.5,0.9,0.5), col =
rgb(0.2,0.5,0.9,0.7), axes = F)

abline(h = seq(1,100, by = 10), lty = 3, col = "grey")
abline(v = seq(1,60, by = 5), lty = 3, col = "grey")

## adds points for N<20 makes and models
points(Lifetab_subextra$pct_Fail[Lifetab_subextra$miles>370000]~

jitter(c(Lifetab_subextra$miles[Lifetab_subextra$miles>370000]/
10000)-2),
        pch = 20, cex = 0.5, col = rgb(0.2,0.5,0.9,0.5))

axis(side = 1, at = seq(1, 51, by = 10), labels = c(0,"100k",
"200k", "300k","400k", "500k"))
axis(side = 2)
axis(side = 3, labels = c(table(Lifetab_subextra$miles)[seq(1,53, by
= 5)]), at = seq(1,53, by = 5), las = 2)

plot(c(100*(Taxis$N_failed/Taxis$N_tests)
[which(Taxis$expo_vec>expo_size)])~
        Taxis$age[which(Taxis$expo_vec>expo_size)],
        cex = 0.4, #cex = log(Taxis$expo_vec)*0.5,
        pch = 20, col = "black", #rgb(0.1,0.1,0.1,0.2),
        xlim = c(0,500000),ylim = c(0,50),
        xlab = "Age in Miles", ylab = "Percent Tests Failed",
```

```
      axes =F)

abline(h = seq(0,50, by = 10), col ="darkgrey", lty = 3)
abline(v = seq(0,1000000, by = 100000), col ="darkgrey", lty = 3)

axis(side = 2)
axis(side = 1, labels = c(0, "100k","200k", "300k", "400k", "500k"),
     at = seq(0,500000,by = 100000))

points(c(100*(Taxis$N_failed/Taxis$N_tests)
[which(Taxis$expo_vec>expo_size)])~
        Taxis$age[which(Taxis$expo_vec>expo_size)], type = "l")

## Most common car
points(c(100*(Focus$N_failed/Focus$N_tests)
[which(Focus$expo_vec>expo_size)])~
        Focus$age[which(Focus$expo_vec>expo_size)], pch = 20, cex =
0.4, col = "orange")
points(c(100*(Focus$N_failed/Focus$N_tests)
[which(Focus$expo_vec>expo_size)])~
        Focus$age[which(Focus$expo_vec>expo_size)], type = "l", col
= "orange")

## most common van
points(c(100*(Transit$N_failed/Transit$N_tests)
[which(Transit$expo_vec>expo_size)])~
        Transit$age[which(Transit$expo_vec>expo_size)], pch = 20,
cex = 0.4, col = "lightblue")
points(c(100*(Transit$N_failed/Transit$N_tests)
[which(Transit$expo_vec>expo_size)])~
        Transit$age[which(Transit$expo_vec>expo_size)], type = "l",
col = "lightblue")

## Most common moto
points(c(100*(Vespa$N_failed/Vespa$N_tests)
[which(Vespa$expo_vec>expo_size)])~
        Vespa$age[which(Vespa$expo_vec>expo_size)], pch = 4, cex =
0.6, col = "hotpink")
points(c(100*(Vespa$N_failed/Vespa$N_tests)
[which(Vespa$expo_vec>expo_size)])~
        Vespa$age[which(Vespa$expo_vec>expo_size)], type = "l", col
= "hotpink")

## First common electric car
points(c(100*(Leaf$N_failed/Leaf$N_tests)
[which(Leaf$expo_vec>expo_size)])~
        Leaf$age[which(Leaf$expo_vec>expo_size)], pch = 3, cex =
0.6, col = "green")
points(c(100*(Leaf$N_failed/Leaf$N_tests)
[which(Leaf$expo_vec>expo_size)])~
        Leaf$age[which(Leaf$expo_vec>expo_size)], type = "l", col =
"green")

## The worst car (at 100k)
```

```
points(c(100*(KaKa$N_failed/KaKa$N_tests)
[which(KaKa$expo_vec>expo_size)])~
        KaKa$age[which(KaKa$expo_vec>expo_size)], type = "l", col =
"purple")
points(c(100*(KaKa$N_failed/KaKa$N_tests)
[which(KaKa$expo_vec>expo_size)])~
        KaKa$age[which(KaKa$expo_vec>expo_size)], pch = 4, cex =
0.6, col = "purple")

#text(x = 300000,y = 17.5, labels = "2013 Nissan Leaf", col =
"green")
#text(x = 300000,y = 15, labels = "2002 Ford Focus", col = "orange")
#text(x = 300000,y = 12.5, labels = "2002 London Taxi TX II", col =
"black")
#text(x = 300000,y = 10, labels = "2003 Piaggio Vespa", col =
"hotpink")
#text(x = 300000,y = 7.5, labels = "2001 Ford Transit", col =
"lightblue")
#text(x = 300000,y = 5, labels = "1999 Ford Ka", col = "purple")

plot(c(Taxis$mx_vec[which(Taxis$expo_vec>expo_size)])~
        Taxis$age[which(Taxis$expo_vec>expo_size)],
      cex = 0.4, #cex = log(Taxis$expo_vec)*0.5,
      pch = 20,  col = "black", #rgb(0.1,0.1,0.1,0.2),
      xlim = c(0,500000),ylim = c(0,0.12),
      xlab = "Age in Miles", ylab = "Mortality Rate", axes = F)

axis(side = 2)
axis(side = 1, labels = c(0, "100k","200k", "300k", "400k", "500k"),
      at = seq(0,500000,by = 100000))

abline(h = seq(0,0.2, by = 0.02), col ="darkgrey", lty = 3)
abline(v = seq(0,1000000, by = 100000), col ="darkgrey", lty = 3)

points(c(Taxis$mx_vec[which(Taxis$expo_vec>expo_size)])~
        Taxis$age[which(Taxis$expo_vec>expo_size)], type = "l")

## Most common car
points(c(Focus$mx_vec[which(Focus$expo_vec>expo_size)])~
        Focus$age[which(Focus$expo_vec>expo_size)], pch = 20, cex =
0.4, col = "orange")
points(c(Focus$mx_vec[which(Focus$expo_vec>expo_size)])~
        Focus$age[which(Focus$expo_vec>expo_size)], type = "l", col
= "orange")

## most common van
points(c(Transit$mx_vec[which(Transit$expo_vec>expo_size)])~
        Transit$age[which(Transit$expo_vec>expo_size)], pch = 20,
cex = 0.4, col = "lightblue")
points(c(Transit$mx_vec[which(Transit$expo_vec>expo_size)])~
        Transit$age[which(Transit$expo_vec>expo_size)], type = "l",
col = "lightblue")
```

```r
## Most common moto
points(c(Vespa$mx_vec[which(Vespa$expo_vec>expo_size)])~
        Vespa$age[which(Vespa$expo_vec>expo_size)], pch = 20, cex =
0.4, col = "hotpink")
points(c(Vespa$mx_vec[which(Vespa$expo_vec>expo_size)])~
        Vespa$age[which(Vespa$expo_vec>expo_size)], type = "l", col
= "hotpink")

## First common electric car
points(c(Leaf$mx_vec[which(Leaf$expo_vec>expo_size)])~
        Leaf$age[which(Leaf$expo_vec>expo_size)], pch = 3, cex =
0.4, col = "green")
points(c(Leaf$mx_vec[which(Leaf$expo_vec>expo_size)])~
        Leaf$age[which(Leaf$expo_vec>expo_size)], type = "l", col =
"green")

## The worst car (at 100k)
points(c(KaKa$mx_vec[which(KaKa$expo_vec>expo_size)])~
        KaKa$age[which(KaKa$expo_vec>expo_size)], type = "l", col =
"purple")
points(c(KaKa$mx_vec[which(KaKa$expo_vec>expo_size)])~
        KaKa$age[which(KaKa$expo_vec>expo_size)], pch = 4, cex =
0.6, col = "purple")

#text(x = 150000,y = 18, labels = "2004 Piaggio NRG Moped", col =
"brown")
#text(x = 210000,y = 16, labels = "2015 Nissan X-Trail", col =
"hotpink")
#text(x = 210000,y = 16, labels = "2015 Tesla Model S", col =
"darkgreen")
text(x = 300000,y = 0.115, labels = "2013 Nissan Leaf", col =
"green")
text(x = 300000,y = 0.105, labels = "2002 Ford Focus", col =
"orange")
text(x = 300000,y = 0.095, labels = "2002 London Taxi TX II", col =
"black")
text(x = 300000,y = 0.085, labels = "2003 Piaggio Vespa", col =
"hotpink")
text(x = 300000,y = 0.075, labels = "2001 Ford Transit", col =
"lightblue")
text(x = 300000,y = 0.065, labels = "1999 Ford Ka", col = "purple")

dev.off()

#####################################

##  using the same style as before, shows MOT failure rates per mile
and per year

#####################################
```

```
FailRate<-Lifetab_MMY_combined$N_failed/Lifetab_MMY_combined$N_tests
FailRate_miles<-Lifetab_MMY_miles_combined$N_failed/
Lifetab_MMY_miles_combined$N_tests

cairo_ps("Publication_docs/Fig_2ad.eps", height = 9.5, width = 9.5)

#tiff(filename  = "Publication_docs/Fig_2ab.tiff",
#     height = 4.5, width = 11, units = "in", res  =300)

par(mfrow = c(2,2), mar = c(5.1,4.1,4.1,0.1), bty = "n")

set.seed(383)

boxplot(100*(FailRate_miles[which(Lifetab_MMY_miles_combined$N_tests
>100 &

Lifetab_MMY_miles_combined$Tclass==4)])~

Lifetab_MMY_miles_combined$miles[which(Lifetab_MMY_miles_combined$N_
tests>100 &

Lifetab_MMY_miles_combined$Tclass==4)],
        col = "white", border = "white",
        notch = F, varwidth = F,
        #   whisklty = 1,whiskcol = "black", #whiskcol =
c(c("black","white")[as.numeric(MMY_obs<=20)]),
        whisklwd = 0.5,
        xlim = c(0,61),
        ylim = c(0,60),outcol =  rgb(0,0.5,1,0),
        xlab = "Age in Miles", ylab = "Percent Inspections Failed",
        axes = F)

abline(h = seq(0,100, by = 10), lty = 3, col = "darkgrey")
abline(v = seq(1,81, by = 10), lty = 3, col = "darkgrey")

par(new = T)

boxplot(100*(FailRate_miles[which(Lifetab_MMY_miles_combined$N_tests
>100 &

Lifetab_MMY_miles_combined$Tclass==4 &

Lifetab_MMY_miles_combined$miles<380000)])~

Lifetab_MMY_miles_combined$miles[which(Lifetab_MMY_miles_combined$N_
tests>100 &

Lifetab_MMY_miles_combined$Tclass==4 &

Lifetab_MMY_miles_combined$miles<380000)],
        col = "steelblue1",# rgb(0,0.5,1,1),
        border = "black",
        whisklwd = 0.5,
        xlim = c(0,61),
```

```
        ylim = c(0,60),
        # ylim = c(-2.5,0),
        # xlim = c(0,800),
        pch = "-",  outcol =  rgb(0,0.5,1,0.3),# outcex =  0.5,
        xlab = "", ylab = "", axes = F)

## for ages where N<=20, plots actual data
subvec_jit<-
jitter(Lifetab_MMY_miles_combined$miles[which(Lifetab_MMY_miles_comb
ined$Tclass==4 &

Lifetab_MMY_miles_combined$N_tests>100 &

Lifetab_MMY_miles_combined$miles>=380000)])

points(c(100*FailRate_miles[which(Lifetab_MMY_miles_combined$Tclass=
=4 &

Lifetab_MMY_miles_combined$N_tests>100 &

Lifetab_MMY_miles_combined$miles>=380000)])~
        c(c(subvec_jit/10000)+1),
     pch = 20, cex =  0.5, col =  rgb(0,0.5,1,0.5))

axis(side = 1, at = seq(1,61, by = 20),
     labels = c("0", "200k", "400k", "600k"),
     cex.axis = 1.2)
#labels = seq(0,800000, by = 200000), las = 2
axis(side = 2, cex.axis = 1.2)

axis(side = 3, at = c(1:58), las = 2,
     cex.axis = 1.2,
     labels = table(!
is.na(Lifetab_MMY_miles_combined$mx_vec[which(Lifetab_MMY_miles_comb
ined$Tclass==4 &

Lifetab_MMY_miles_combined$N_tests>100)]),
                   by =
Lifetab_MMY_miles_combined$miles[which(Lifetab_MMY_miles_combined$Tc
lass==4 &

Lifetab_MMY_miles_combined$N_tests>100)]))

###############

boxplot(100*(FailRate[which(Lifetab_MMY_combined$N_tests>100 &
                            Lifetab_MMY_combined$test_class==4 &
                            Lifetab_MMY_combined$age>=47)])~

Lifetab_MMY_combined$age[which(Lifetab_MMY_combined$N_tests>100 &

Lifetab_MMY_combined$test_class==4 &
```

```
Lifetab_MMY_combined$age>=47)],
        col = "white", border = "white",
        notch = F, varwidth = F,
        #   whisklty = 1,whiskcol = "black", #whiskcol =
c(c("black","white")[as.numeric(MMY_obs<=20)]),
        whisklwd = 0.5,
        xlim = c(0,53),
        ylim = c(0,60),outcol =  rgb(0,0.5,1,0),
        xlab = "Age in Years", ylab = "Percent Inspections Failed",
        axes = F)

abline(h = seq(0,100, by = 10), lty = 3, col = "darkgrey")
abline(v = seq(0,80, by = 5), lty = 3, col = "darkgrey")

par(new = T)

boxplot(100*(FailRate[which(Lifetab_MMY_combined$N_tests>100 &
                            Lifetab_MMY_combined$test_class==4 &
                            Lifetab_MMY_combined$age>=3 &
                            Lifetab_MMY_combined$age<47)])~

Lifetab_MMY_combined$age[which(Lifetab_MMY_combined$N_tests>100 &

Lifetab_MMY_combined$test_class==4 &

Lifetab_MMY_combined$age>=3 &

Lifetab_MMY_combined$age<47)],
        col = "steelblue1",# rgb(0,0.5,1,1),  rgb(0,0.5,1,1), border
= "black",
        whisklwd = 0.5,
        xlim = c(0,53),
        ylim = c(0,60),
        #  ylim = c(-2.5,0),
        # xlim = c(0,800),
        pch = "-", outcol = rgb(0,0.5,1,0.3), #outcex =  0.5,
        xlab = "", ylab = "", axes = F)

## for ages where N<=20, plots actual data
subvec_jit<-
jitter(Lifetab_MMY_combined$age[which(Lifetab_MMY_combined$test_clas
s==4 &

Lifetab_MMY_combined$age>=3 &

Lifetab_MMY_combined$N_tests>100 &

Lifetab_MMY_combined$age>=47)])

points(c(100*FailRate[which(Lifetab_MMY_combined$test_class==4 &
                            Lifetab_MMY_combined$age>=3 &
                            Lifetab_MMY_combined$N_tests>100 &
                            Lifetab_MMY_combined$age>=47)])~
```

```r
          c(subvec_jit-2),
        pch = 20, cex = 0.5, col = rgb(0,0.5,1,0.5))

axis(side = 1,  at = seq(3,49, by = 5),
     labels = seq(5,50, by = 5),
     cex.axis = 1.2)
#labels = seq(0,800000, by = 200000), las = 2)
axis(side = 2, cex.axis = 1.2)

axis(side = 3, at = c(1:52), las = 2,
     cex.axis = 1.2,
     labels = table(!
is.na(Lifetab_MMY_combined$mx_vec[which(Lifetab_MMY_combined$test_cl
ass==4 &

Lifetab_MMY_combined$age>=3 &

Lifetab_MMY_combined$N_tests>100)]),
                     by =
Lifetab_MMY_combined$age[which(Lifetab_MMY_combined$test_class==4 &

Lifetab_MMY_combined$age>=3 &

Lifetab_MMY_combined$N_tests>100)]))

par(mar = c(5.1,4.1,1.1,0.1))

plot(c(100*(Taxis$N_failed/Taxis$N_tests)
[which(Taxis$expo_vec>expo_size)])~
     Taxis$age[which(Taxis$expo_vec>expo_size)],
     cex = 0.4, #cex = log(Taxis$expo_vec)*0.5,
     pch = 20,  col = "black", #rgb(0.1,0.1,0.1,0.2),
     xlim = c(0,610000),ylim = c(0,60),
     xlab = "Miles", ylab = "Percent Inspections Failed",
     axes =F)

abline(h = seq(0,50, by = 10), col ="darkgrey", lty = 3)
abline(v = seq(0,1000000, by = 100000), col ="darkgrey", lty = 3)

axis(side = 2)
axis(side = 1, labels = c(0, "200k",  "400k",  "600k"),
     at = seq(0,610000,by = 200000))

points(c(100*(Taxis$N_failed/Taxis$N_tests)
[which(Taxis$expo_vec>expo_size)])~
       Taxis$age[which(Taxis$expo_vec>expo_size)], type = "l")

## Most common car
points(c(100*(Focus$N_failed/Focus$N_tests)
[which(Focus$expo_vec>expo_size)])~
       Focus$age[which(Focus$expo_vec>expo_size)], pch = 20, cex =
0.4, col = "orange")
```

```
points(c(100*(Focus$N_failed/Focus$N_tests)
[which(Focus$expo_vec>expo_size)])~
        Focus$age[which(Focus$expo_vec>expo_size)], type = "l", col
= "orange")

## most common van
points(c(100*(Transit$N_failed/Transit$N_tests)
[which(Transit$expo_vec>expo_size)])~
        Transit$age[which(Transit$expo_vec>expo_size)], pch = 20,
cex = 0.4, col = "lightblue")
points(c(100*(Transit$N_failed/Transit$N_tests)
[which(Transit$expo_vec>expo_size)])~
        Transit$age[which(Transit$expo_vec>expo_size)], type = "l",
col = "lightblue")

## Most common moto
points(c(100*(Vespa$N_failed/Vespa$N_tests)
[which(Vespa$expo_vec>expo_size)])~
        Vespa$age[which(Vespa$expo_vec>expo_size)], pch = 4, cex =
0.6, col = "hotpink")
points(c(100*(Vespa$N_failed/Vespa$N_tests)
[which(Vespa$expo_vec>expo_size)])~
        Vespa$age[which(Vespa$expo_vec>expo_size)], type = "l", col
= "hotpink")

## First common electric car
points(c(100*(Leaf$N_failed/Leaf$N_tests)
[which(Leaf$expo_vec>expo_size)])~
        Leaf$age[which(Leaf$expo_vec>expo_size)], pch = 3, cex =
0.6, col = "green")
points(c(100*(Leaf$N_failed/Leaf$N_tests)
[which(Leaf$expo_vec>expo_size)])~
        Leaf$age[which(Leaf$expo_vec>expo_size)], type = "l", col =
"green")

## The worst car (at 100k)
points(c(100*(KaKa$N_failed/KaKa$N_tests)
[which(KaKa$expo_vec>expo_size)])~
        KaKa$age[which(KaKa$expo_vec>expo_size)], type = "l", col =
"purple")
points(c(100*(KaKa$N_failed/KaKa$N_tests)
[which(KaKa$expo_vec>expo_size)])~
        KaKa$age[which(KaKa$expo_vec>expo_size)], pch = 4, cex =
0.6, col = "purple")

#text(x = 300000,y = 17.5, labels = "2013 Nissan Leaf", col =
"green")
#text(x = 300000,y = 15, labels = "2002 Ford Focus", col = "orange")
#text(x = 300000,y = 12.5, labels = "2002 London Taxi TX II", col =
"black")
#text(x = 300000,y = 10, labels = "2003 Piaggio Vespa", col =
"hotpink")
#text(x = 300000,y = 7.5, labels = "2001 Ford Transit", col =
"lightblue")
```

```r
#text(x = 300000,y = 5, labels = "1999 Ford Ka", col = "purple")

plot(c(Taxis$mx_vec[which(Taxis$expo_vec>expo_size)])~
        Taxis$age[which(Taxis$expo_vec>expo_size)],
      cex = 0.4, #cex = log(Taxis$expo_vec)*0.5,
      pch = 20,  col = "black", #rgb(0.1,0.1,0.1,0.2),
      xlim = c(0,610000),ylim = c(0,0.12),
      xlab = "Miles", ylab = "Mortality Rate", axes = F)

axis(side = 2)
axis(side = 1, labels = c(0, "200k", "400k","600k"),
      at = seq(0,610000,by = 200000))

abline(h = seq(0,0.2, by = 0.02), col ="darkgrey", lty = 3)
abline(v = seq(0,1000000, by = 100000), col ="darkgrey", lty = 3)

points(c(Taxis$mx_vec[which(Taxis$expo_vec>expo_size)])~
        Taxis$age[which(Taxis$expo_vec>expo_size)], type = "l")

## Most common car
points(c(Focus$mx_vec[which(Focus$expo_vec>expo_size)])~
        Focus$age[which(Focus$expo_vec>expo_size)], pch = 20, cex =
0.4, col = "orange")
points(c(Focus$mx_vec[which(Focus$expo_vec>expo_size)])~
        Focus$age[which(Focus$expo_vec>expo_size)], type = "l", col
= "orange")

## most common van
points(c(Transit$mx_vec[which(Transit$expo_vec>expo_size)])~
        Transit$age[which(Transit$expo_vec>expo_size)], pch = 20,
cex = 0.4, col = "lightblue")
points(c(Transit$mx_vec[which(Transit$expo_vec>expo_size)])~
        Transit$age[which(Transit$expo_vec>expo_size)], type = "l",
col = "lightblue")

## Most common moto
points(c(Vespa$mx_vec[which(Vespa$expo_vec>expo_size)])~
        Vespa$age[which(Vespa$expo_vec>expo_size)], pch = 20, cex =
0.4, col = "hotpink")
points(c(Vespa$mx_vec[which(Vespa$expo_vec>expo_size)])~
        Vespa$age[which(Vespa$expo_vec>expo_size)], type = "l", col
= "hotpink")

## First common electric car
points(c(Leaf$mx_vec[which(Leaf$expo_vec>expo_size)])~
        Leaf$age[which(Leaf$expo_vec>expo_size)], pch = 3, cex =
0.4, col = "green")
points(c(Leaf$mx_vec[which(Leaf$expo_vec>expo_size)])~
        Leaf$age[which(Leaf$expo_vec>expo_size)], type = "l", col =
"green")

## The worst car (at 100k)
points(c(KaKa$mx_vec[which(KaKa$expo_vec>expo_size)])~
```

```
        KaKa$age[which(KaKa$expo_vec>expo_size)], type = "l", col =
"purple")
points(c(KaKa$mx_vec[which(KaKa$expo_vec>expo_size)])~
        KaKa$age[which(KaKa$expo_vec>expo_size)], pch = 4, cex =
0.6, col = "purple")

#text(x = 150000,y = 18, labels = "2004 Piaggio NRG Moped", col =
"brown")
#text(x = 210000,y = 16, labels = "2015 Nissan X-Trail", col =
"hotpink")
#text(x = 210000,y = 16, labels = "2015 Tesla Model S", col =
"darkgreen")
text(x = 400000,y = 0.115, labels = "2002 Ford Focus", col =
"orange")
text(x = 400000,y = 0.105, labels = "2003 Piaggio Vespa", col =
"hotpink")
text(x = 400000,y = 0.095, labels = "2001 Ford Transit", col =
"lightblue")
text(x = 400000,y = 0.085, labels = "2013 Nissan Leaf", col =
"green")
text(x = 400000,y = 0.075, labels = "2002 London Taxi TX II", col =
"black")
text(x = 400000,y = 0.065, labels = "1999 Ford Ka", col = "purple")

dev.off()

## Plots the relationship between failures per test at a given
mileage, and
## the mortality rate at that mileage, for all surviving cohorts
N>100

## Cars
qplot(100*(Lifetab_MMY_miles_combined$N_failed/
c(Lifetab_MMY_miles_combined$N_tests))
[which(Lifetab_MMY_miles_combined$Tclass==4 &
Lifetab_MMY_miles_combined$expo_vec>100)],

log(Lifetab_MMY_miles_combined$mx_vec[which(Lifetab_MMY_miles_combin
ed$Tclass==4  & Lifetab_MMY_miles_combined$expo_vec>100)]),
     alpha = I(0), xlab = "Percent MOTs Failed", ylab = "log mx")+
  geom_hex(bins = 70)+
  geom_smooth(col = "orange")+
  theme_minimal()

cor.test(c(Lifetab_MMY_miles_combined$N_failed/
c(Lifetab_MMY_miles_combined$N_tests))
[which(Lifetab_MMY_miles_combined$Tclass==4 &
Lifetab_MMY_miles_combined$expo_vec>100)],

c(Lifetab_MMY_miles_combined$qx_vec[which(Lifetab_MMY_miles_combined
```

```
$Tclass==4  & Lifetab_MMY_miles_combined$expo_vec>100)]))

## Vans
qplot(100*(Lifetab_MMY_miles_combined$N_failed/
c(Lifetab_MMY_miles_combined$N_tests))
[which(Lifetab_MMY_miles_combined$Tclass==7 &

Lifetab_MMY_miles_combined$expo_vec>100)],

log(Lifetab_MMY_miles_combined$mx_vec[which(Lifetab_MMY_miles_combin
ed$Tclass==7
                                            &
Lifetab_MMY_miles_combined$expo_vec>100)]), alpha = I(0))+
  geom_hex(bins = 70)+
  geom_smooth(col = "orange")+
  theme_minimal()

cor.test(c(Lifetab_MMY_miles_combined$N_failed/
c(Lifetab_MMY_miles_combined$N_tests))
[which(Lifetab_MMY_miles_combined$Tclass==7 &
Lifetab_MMY_miles_combined$expo_vec>100)],

c(Lifetab_MMY_miles_combined$mx_vec[which(Lifetab_MMY_miles_combined
$Tclass==7  & Lifetab_MMY_miles_combined$expo_vec>100)]))

## However, this is a cross-sectional sample, and both target
variables (failure rates, mortality rates)
## are correlated with age. They are confounded because older
vehicles both break down more and die more,
## because they share (chronological or physical) age as a causal
factor.
## What about comparing populations with the same age?

## Say 100,000 miles? N=4752
Age10_plot<-qplot(100*(Lifetab_MMY_miles_combined$N_failed/
c(Lifetab_MMY_miles_combined$N_tests))
[which(Lifetab_MMY_miles_combined$Tclass==4 &

Lifetab_MMY_miles_combined$expo_vec>100 &

Lifetab_MMY_miles_combined$age==100000)],

log10(Lifetab_MMY_miles_combined$mx_vec[which(Lifetab_MMY_miles_comb
ined$Tclass==4
                                            &
Lifetab_MMY_miles_combined$expo_vec>100
                                            &
Lifetab_MMY_miles_combined$age==100000)]),
                 alpha = I(0), xlab = "Percent Inspections Failed
at 100k Miles", ylab = "Mortality Rate Above 100k miles")+
  geom_hex(bins = 70)+
  scale_fill_viridis(option  ="magma", name = "Observed\n
Populations")+
```

```
  geom_smooth(col = "lightblue",bg = "lightblue", method = "lm")+
  theme_minimal()+
  annotate("text", label = "R^2 = -0.006; p = 0.68", x = 80, y =
-1.2)+
  theme(legend.position = c(0.75, 0.3))

## the R2?
cor.test(c(Lifetab_MMY_miles_combined$N_failed/
c(Lifetab_MMY_miles_combined$N_tests))
[which(Lifetab_MMY_miles_combined$Tclass==4 &

Lifetab_MMY_miles_combined$expo_vec>100 &

Lifetab_MMY_miles_combined$age==100000)],

c(Lifetab_MMY_miles_combined$mx_vec[which(Lifetab_MMY_miles_combined
$Tclass==4
                                               &
Lifetab_MMY_miles_combined$expo_vec>100
                                               &
Lifetab_MMY_miles_combined$age==100000)]))

## Nonexistent.

## The total number of observed vehicles in here?
sum(Lifetab_MMY_miles_combined$expo_vec[which(Lifetab_MMY_miles_comb
ined$Tclass==4
                                               &
Lifetab_MMY_miles_combined$expo_vec>100
                                               &
Lifetab_MMY_miles_combined$age==100000)])
## 19,801,729

## what about chronological age?

## Say 10 years?
Age10_plot2<-qplot(100*(Lifetab_MMY_combined$N_failed/
c(Lifetab_MMY_combined$N_tests))
[which(Lifetab_MMY_combined$test_class==4 &

Lifetab_MMY_combined$expo_vec>100 &

Lifetab_MMY_combined$age==10)],

log10(Lifetab_MMY_combined$mx_vec[which(Lifetab_MMY_combined$test_cl
ass==4 &

Lifetab_MMY_combined$expo_vec>100 &

Lifetab_MMY_combined$age==10)]),
                    alpha = I(0), xlab = "Percent Failed MOTs Age
10", ylab = "Mortality Rate")+
```

```
  geom_hex(bins = 70)+
  scale_fill_viridis(option  ="magma", name = "Observed\n
Populations")+
  geom_smooth(col = "lightblue",bg = "lightblue", method = "lm")+
  theme_minimal()+
  annotate("text", label = "R2 = 0.61; p < 2.2e-16", x = 80, y =
-1.2)+
  theme(legend.position = c(0.75, 0.3))

## the R2?
cor.test(c(Lifetab_MMY_combined$N_failed/
c(Lifetab_MMY_combined$N_tests))
[which(Lifetab_MMY_combined$test_class==4 &

Lifetab_MMY_combined$expo_vec>100 &

Lifetab_MMY_combined$age==10)],

c(Lifetab_MMY_combined$mx_vec[which(Lifetab_MMY_combined$test_class=
=4 &

Lifetab_MMY_combined$expo_vec>100 &

Lifetab_MMY_combined$age==10)]))

## What is the correlation between breakdown rates at a given age,
## and survival rates by years of age?

cor_vec1<-NULL
cor_vec2<-NULL

for(i in seq(0,50, by = 1)){

  vec1<-c(c(Lifetab_MMY_combined$N_failed/
c(Lifetab_MMY_combined$N_tests))
[which(Lifetab_MMY_combined$test_class==4 &

Lifetab_MMY_combined$expo_vec>100 &

Lifetab_MMY_combined$age==i)])
  vec2<-log(c(Lifetab_MMY_combined$mx_vec)
[which(Lifetab_MMY_combined$test_class==4
                                          &
Lifetab_MMY_combined$expo_vec>100
                                          &
Lifetab_MMY_combined$age==i)])

  vec2[vec2==-Inf]<-NA

  if(sum(!is.na(vec1+vec2))<100){
    cor_vec1<-c(cor_vec1,NA)
    cor_vec2<-rbind(cor_vec2,c(sum(!is.na(vec1+vec2)),NA,NA))
  }
```

```
  if(sum(!is.na(vec1+vec2))>=100){
    cor_vec1<-c(cor_vec1,cor.test(vec1,vec2)$estimate)
    cor_vec2<-rbind(cor_vec2, c(sum(!
is.na(vec1+vec2)),cor.test(vec1,vec2)$conf.int[1:2]))
  }
}

plot(cor_vec1~seq(0,50, by = 1),
     type = "l", ylim = c(-1,1), xlab = "Years of Age", ylab = "Rsq
breakdown ~ mortality rate")

abline(h = seq(-1,1, by = 0.2), col = "grey")
abline(h = 0, lty = 3, col= "red")

points(c(cor_vec2[,2],cor_vec2[,3])~c(seq(0,50, by = 1),seq(0,50, by
= 1)),
       type = "l", col = "lightblue")
points(c(cor_vec2[,2],cor_vec2[,3])~c(seq(0,50, by = 1),seq(0,50, by
= 1)),
       pch = 20, cex = 0.3)

points(cor_vec1~seq(0,50, by = 1),  pch = 20, col = "darkblue")

#points(cor_vec2[,2]~seq(0,1000000, by = 10000),  pch = "-")
#points(cor_vec2[,3]~seq(0,1000000, by = 10000),  pch = "-")

## makes a marginal boxplot to match

## What is the correlation between breakdown rates at a given
mileage,
## and survival rates over the next 10,000 miles?

cor_vec1<-NULL
cor_vec2<-NULL

for(i in seq(0,1000000, by = 10000)){

  vec1<-c(c(Lifetab_MMY_miles_combined$N_failed/
c(Lifetab_MMY_miles_combined$N_tests))
[which(Lifetab_MMY_miles_combined$Tclass==4 &

Lifetab_MMY_miles_combined$expo_vec>100 &

Lifetab_MMY_miles_combined$age==i)])
  vec2<-log(c(Lifetab_MMY_miles_combined$mx_vec)
[which(Lifetab_MMY_miles_combined$Tclass==4
                                             &
Lifetab_MMY_miles_combined$expo_vec>100
                                             &
Lifetab_MMY_miles_combined$age==i)])
```

```
vec2[vec2==-Inf]<-NA

if(sum(!is.na(vec1+vec2))<100){
  cor_vec1<-c(cor_vec1,NA)
  cor_vec2<-rbind(cor_vec2,c(sum(!is.na(vec1+vec2)),NA,NA))
}

if(sum(!is.na(vec1+vec2))>=100){
  cor_vec1<-c(cor_vec1,cor.test(vec1,vec2)$estimate)
  cor_vec2<-rbind(cor_vec2, c(sum(!
is.na(vec1+vec2)),cor.test(vec1,vec2)$conf.int[1:2]))
  }
}

plot(cor_vec1~seq(0,1000000, by = 10000),
     type = "l", ylim = c(-1,1), xlab = "Miles accumulated", ylab =
"Rsq breakdown ~ mortality rate")

abline(h = seq(-1,1, by = 0.2), col = "grey")
abline(h = 0, lty = 3, col= "red")

points(c(cor_vec2[,2],cor_vec2[,3])~c(seq(0,1000000, by =
10000),seq(0,1000000, by = 10000)),
       type = "l", col = "lightblue")
points(c(cor_vec2[,2],cor_vec2[,3])~c(seq(0,1000000, by =
10000),seq(0,1000000, by = 10000)),
       pch = 20, cex = 0.3)

points(cor_vec1~seq(0,1000000, by = 10000),  pch = 20, col =
"darkblue")

#points(cor_vec2[,2]~seq(0,1000000, by = 10000),  pch = "-")
#points(cor_vec2[,3]~seq(0,1000000, by = 10000),  pch = "-")

#text(y= cor_vec1+0.3, x = seq(0,1000000, by = 10000),labels
=cor_vec2[,1])

###############################################################
#########

## Figure SXX - comparison of three otherwise identical models

###############################################################
#########

tiff(filename  = "Publication_docs/Fig_S_survival_Octavia.tiff",
     height = 8.5, width = 9.5, units = "in", res  =300)

## Plots comparison of all three, for Expo > 100
```

```
A3<-Lifetab_MMY_combined[grep("__A3__2004",
Lifetab_MMY_combined$MMY),]
Octavia<-Lifetab_MMY_combined[grep("__OCTAVIA__2004",
Lifetab_MMY_combined$MMY),]
Golf<-Lifetab_MMY_combined[grep("__GOLF__2004",
Lifetab_MMY_combined$MMY),]

## Plots comparison of all three, for Expo > 100

par(mfrow = c(2,2), bty = "n", mar = c(5.1,4.1,3.1,2.1))

plot(log10(A3$qx_vec[which(A3$expo_vec>100)])~A3$age[which(A3$expo_v
ec>100)],
     type = "l", xlim = c(0,18), ylim = c(-2.5,0),
     xlab = "Age in Years", ylab = "log10 Probability of Death")

abline(v = seq(0,15, by =5), lty = 3, col = "grey")
abline(h = seq(-5,0, by =0.5), lty = 3, col = "grey")

points(log10(Golf$qx_vec)~Golf$age, type = "l", col = "orange")
points(log10(Octavia$qx_vec)~Octavia$age, type = "l", col =
"lightblue")

points(log10(A3$qx_vec[which(A3$expo_vec>100)])~A3$age[which(A3$expo
_vec>100)],
       cex = c(log10(A3$expo_vec[which(A3$expo_vec>100)])/5), pch =
20)
points(log10(Golf$qx_vec[which(Golf$expo_vec>100)])~Golf$age[which(G
olf$expo_vec>100)],
       cex = c(log10(Golf$expo_vec[which(Golf$expo_vec>100)])/5),
pch = 20,
       col = "orange")
points(log10(Octavia$qx_vec[which(Octavia$expo_vec>100)])~Octavia$ag
e[which(Octavia$expo_vec>100)],
       cex = c(log10(Octavia$expo_vec[which(Octavia$expo_vec>100)])/
5), pch = 20,
       col = "lightblue")

## attaches sample sizes and names
#text(12.5,-1.625, paste0("A3 N =",max(A3$expo_vec, na.rm = T)))
#text(12.5,-1.875, paste0("Golf N =",max(Golf$expo_vec, na.rm = T)),
col = "orange")
#text(12.5,-2.125, paste0("Octavia N =",max(Octavia$expo_vec, na.rm
= T)), col = "lightblue")

text(2.5,-0.625, paste0("A3 N = ",max(A3$expo_vec, na.rm = T)))
text(2.5,-0.875, paste0("Golf N = ",max(Golf$expo_vec, na.rm = T)),
col = "orange")
text(2.5,-1.125, paste0("Octavia N = ",max(Octavia$expo_vec, na.rm =
T)), col = "lightblue")

text(7.5,-0.25, "2004 Models", cex = 1.2)

mtext("a  \n", side = 3, adj = 0, cex = 1.5)
```

```
## then 2010

A3<-Lifetab_MMY_combined[grep("__A3__2010",
Lifetab_MMY_combined$MMY),]
Octavia<-Lifetab_MMY_combined[grep("__OCTAVIA__2010",
Lifetab_MMY_combined$MMY),]
Golf<-Lifetab_MMY_combined[grep("__GOLF__2010",
Lifetab_MMY_combined$MMY),]

## Plots comparison of all three, for Expo > 100

plot(log10(A3$qx_vec[which(A3$expo_vec>100)])~A3$age[which(A3$expo_v
ec>100)],
     type = "l", xlim = c(0,18), ylim = c(-2.5,0),
     xlab = "Age in Years", ylab = "log10 Probability of Death")

abline(v = seq(0,15, by =5), lty = 3, col = "grey")
abline(h = seq(-5,0, by =0.5), lty = 3, col = "grey")

points(log10(Golf$qx_vec)~Golf$age, type = "l", col = "orange")
points(log10(Octavia$qx_vec)~Octavia$age, type = "l", col =
"lightblue")

points(log10(A3$qx_vec[which(A3$expo_vec>100)])~A3$age[which(A3$expo
_vec>100)],
       cex = c(log10(A3$expo_vec[which(A3$expo_vec>100)])/5), pch =
20)
points(log10(Golf$qx_vec[which(Golf$expo_vec>100)])~Golf$age[which(G
olf$expo_vec>100)],
       cex = c(log10(Golf$expo_vec[which(Golf$expo_vec>100)])/5),
pch = 20,
       col = "orange")
points(log10(Octavia$qx_vec[which(Octavia$expo_vec>100)])~Octavia$ag
e[which(Octavia$expo_vec>100)],
       cex = c(log10(Octavia$expo_vec[which(Octavia$expo_vec>100)])/
5), pch = 20,
       col = "lightblue")

## attaches sample sizes and names
text(2.5,-0.625, paste0("A3 N = ",max(A3$expo_vec, na.rm = T)))
text(2.5,-0.875, paste0("Golf N = ",max(Golf$expo_vec, na.rm = T)),
col = "orange")
text(2.5,-1.125, paste0("Octavia N = ",max(Octavia$expo_vec, na.rm =
T)), col = "lightblue")

text(7.5,-0.25, "2010 Models", cex = 1.2)

mtext("b  \n", side = 3, adj = 0, cex = 1.5)

#######################
#######################
```

```
### Now per mile

A3<-Lifetab_MMY_miles_combined[grep("__A3__2004",
Lifetab_MMY_miles_combined$MMY),]
Octavia<-Lifetab_MMY_miles_combined[grep("__OCTAVIA__2004",
Lifetab_MMY_miles_combined$MMY),]
Golf<-Lifetab_MMY_miles_combined[grep("__GOLF__2004",
Lifetab_MMY_miles_combined$MMY),]

## reorders
A3<-A3[order(A3$age, decreasing = F),]
Octavia<-Octavia[order(Octavia$age, decreasing = F),]
Golf<-Golf[order(Golf$age, decreasing = F),]

## Plots comparison of all three, for Expo > 100

plot(log10(A3$qx_vec[which(A3$expo_vec>100)])~A3$age[which(A3$expo_v
ec>100)],
     type = "l",
     xlim = c(0,400000), ylim = c(-5,0),
     xlab = "Age in Miles", ylab = "log10 Probability of Death")

abline(v = seq(0,500000, by =100000), lty = 3, col = "grey")
abline(h = seq(-10,0, by =1), lty = 3, col = "grey")

points(log10(A3$qx_vec[which(A3$expo_vec>100)])~A3$age[which(A3$expo
_vec>100)],
       cex = c(log10(A3$expo_vec[which(A3$expo_vec>100)])/5), pch =
20)
points(log10(Golf$qx_vec[which(Golf$expo_vec>100)])~Golf$age[which(G
olf$expo_vec>100)],
       cex = c(log10(Golf$expo_vec[which(Golf$expo_vec>100)])/5),
pch = 20,
       col = "orange")
points(log10(Octavia$qx_vec[which(Octavia$expo_vec>100)])~Octavia$ag
e[which(Octavia$expo_vec>100)],
       cex = c(log10(Octavia$expo_vec[which(Octavia$expo_vec>100)])/
5), pch = 20,
       col = "lightblue")

points(log10(A3$qx_vec)~A3$age, type = "l")
points(log10(Golf$qx_vec)~Golf$age, type = "l", col = "orange")
points(log10(Octavia$qx_vec)~Octavia$age, type = "l", col =
"lightblue")

## attaches sample sizes and names
#text(200000,-2.75, paste0("A3 N = ",max(A3$expo_vec, na.rm = T)))
#text(200000,-3.25, paste0("Golf N = ",max(Golf$expo_vec, na.rm =
T)), col = "orange")
```

```
#text(200000,-3.75, paste0("Octavia N = ",max(Octavia$expo_vec,
na.rm = T)), col = "lightblue")

text(200000,-0.25, "2004 Models", cex = 1.2)

mtext("c  \n", side = 3, adj = 0, cex = 1.5)

########################
########################

A3<-Lifetab_MMY_miles_combined[grep("__A3__2010",
Lifetab_MMY_miles_combined$MMY),]
Octavia<-Lifetab_MMY_miles_combined[grep("__OCTAVIA__2010",
Lifetab_MMY_miles_combined$MMY),]
Golf<-Lifetab_MMY_miles_combined[grep("__GOLF__2010",
Lifetab_MMY_miles_combined$MMY),]

## reorders
A3<-A3[order(A3$age, decreasing = F),]
Octavia<-Octavia[order(Octavia$age, decreasing = F),]
Golf<-Golf[order(Golf$age, decreasing = F),]

## Plots comparison of all three, for Expo > 100

plot(log10(A3$qx_vec[which(A3$expo_vec>100)])~A3$age[which(A3$expo_v
ec>100)],
     type = "l",
     xlim = c(0,400000), ylim = c(-5,0),
     xlab = "Age in Miles", ylab = "log10 Probability of Death")

abline(v = seq(0,500000, by =100000), lty = 3, col = "grey")
abline(h = seq(-10,0, by =1), lty = 3, col = "grey")

points(log10(A3$qx_vec[which(A3$expo_vec>100)])~A3$age[which(A3$expo
_vec>100)],
       cex = c(log10(A3$expo_vec[which(A3$expo_vec>100)])/5), pch =
20)
points(log10(Golf$qx_vec[which(Golf$expo_vec>100)])~Golf$age[which(G
olf$expo_vec>100)],
       cex = c(log10(Golf$expo_vec[which(Golf$expo_vec>100)])/5),
pch = 20,
       col = "orange")
points(log10(Octavia$qx_vec[which(Octavia$expo_vec>100)])~Octavia$ag
e[which(Octavia$expo_vec>100)],
       cex = c(log10(Octavia$expo_vec[which(Octavia$expo_vec>100)])/
5), pch = 20,
       col = "lightblue")

points(log10(A3$qx_vec)~A3$age, type = "l")
points(log10(Golf$qx_vec)~Golf$age, type = "l", col = "orange")
```

```
points(log10(Octavia$qx_vec)~Octavia$age, type = "l", col =
"lightblue")

## attaches sample sizes and names
#text(200000,-2.75, paste0("A3 N = ",max(A3$expo_vec, na.rm = T)))
#text(200000,-3.25, paste0("Golf N = ",max(Golf$expo_vec, na.rm =
T)), col = "orange")
#text(200000,-3.75, paste0("Octavia N = ",max(Octavia$expo_vec,
na.rm = T)), col = "lightblue")

text(200000,-0.25, "2010 Models", cex = 1.2)

mtext("d  \n", side = 3, adj = 0, cex = 1.5)

dev.off()

####################################################################
################

##  Figure 3: Variable importance and marginal plots of survival
effects

####################################################################
################

require(randomForestSRC)

table(sub_frameRF_train$test_class_4)

Survival_rf_all_lge<-readRDS("Survival_rf_all_lge.rds")

cairo_ps("Publication_docs/Fig_3ab_eps.eps", height = 4.5, width =
7.5)

par(bty = "n", mar = c(5.1,4.1,2.1,2.1))
plot.variable(Survival_rf_all_lge,
              xvar.names = c("age_at_test","corrected_odometer"),
              partial = TRUE, npts = 200)

dev.off()

tiff(filename  = "Publication_docs/Fig_3ab_tiff.tiff",
     height = 4.5, width = 7.5, units = "in", res  =300)

par(bty = "n", mar = c(5.1,4.1,2.1,2.1))
plot.variable(Survival_rf_all_lge,
              xvar.names = c("age_at_test","corrected_odometer"),
              partial = TRUE, npts = 200)

dev.off()

par(mfrow =c(3,2), bty = "n")
```

```
cairo_ps("Publication_docs/Fig_3cd_eps.eps", height = 6.5, width =
7.5)

par(bty = "n", mar = c(5.1,4.1,2.1,2.1))

## makes a partial plot of survival effects by e.g fuel type
plot.variable(Survival_rf_all_lge,
              xvar.names =
c("fuel_name_Petrol","fuel_name_Electric"),
              ## non significant marginal effects -
"fuel_name_Hybrid.Electric..Clean.","fuel_name_Diesel",
              partial = TRUE)

dev.off()

tiff(filename  = "Publication_docs/Fig_3cd_tiff.tiff",
     height = 7, width = 8.5, units = "in", res  =300)

par(bty = "n", mar = c(5.1,4.1,2.1,2.1))

## makes a partial plot of survival effects by e.g fuel type
plot.variable(Survival_rf_all_lge,
              xvar.names =
c("fuel_name_Petrol","fuel_name_Electric"),
              partial = TRUE)

dev.off()

par(mfrow =c(3,2), bty = "n")

## makes a barplot of variable importance scores
vimp_rf_lge<-readRDS("vimp_rf_lge.rds")

## knocks out a narrow bar plot
cairo_ps("Publication_docs/Fig_3_barplot_eps.eps", height = 6.5,
width = 3.5)

par(mfrow =c(1,1), bty = "n")

barplot(tail(vimp_rf_lge$importance[order(vimp_rf_lge$importance)],
15),
        las =2, col = "#004225", horiz = T) ## in British racing
green
abline(v = seq(0,0.025, by = 0.005), lty =3)

#text(y = 0.5, x = 0.001, label = "here")

dev.off()
```

```r
#text(x =
c(tail(vimp_rf_lge$importance[order(vimp_rf_lge$importance)],
10)+0.001),
#      y = c(seq(1, 12, by = 1.3)+0.5),
#      labels =
names(tail(vimp_rf_lge$importance[order(vimp_rf_lge$importance)],
10)))

## labels the bars directly, ignores the axis
text(at = c(1,20), labels =
names(tail(vimp_rf_lge$importance[order(vimp_rf_lge$importance)],
20)))

tiff(filename  = "Publication_docs/Fig_S_ML_tiff.tiff",
     height = 4, width = 8.5, units = "in", res  =300)

par(bty = "n", mar = c(5.1,4.1,2.1,2.1))

## makes a partial plot of survival effects by e.g fuel type
plot.variable(Survival_rf_all_lge,
              xvar.names = c("first_use_date",
"N_failed","vote_PC"),
              partial = TRUE, npts = 100)

dev.off()

par(mfrow =c(3,2), bty = "n")

## loads up oversampling for higher mileages, plots the same

Survival_rf_highMiles<-readRDS("Survival_rf_highMiles.rds")

Survival_rf_highYears<-readRDS("Survival_rf_highYears.rds")

tiff(filename  = "Publication_docs/Fig_3_Highmiles_tiff.tiff",
     height = 4.5, width = 7.5, units = "in", res  =300)

par(bty = "n", mar = c(5.1,4.1,2.1,2.1))
plot.variable(Survival_rf_highMiles,
              xvar.names = c("age_at_test","corrected_odometer"),
              partial = TRUE, npts = 200)

dev.off()

cairo_ps("Publication_docs/Fig_3_Highmiles_eps.eps", height = 4.5,
width = 7.5)

par(bty = "n", mar = c(5.1,4.1,2.1,2.1))
```

```
plot.variable(Survival_rf_highMiles,
              xvar.names = c("age_at_test","corrected_odometer"),
              partial = TRUE, npts = 200)

dev.off()

tiff(filename  = "Publication_docs/Fig_3_HighYears_tiff.tiff",
     height = 4.5, width = 7.5, units = "in", res  =300)

par(bty = "n", mar = c(5.1,4.1,2.1,2.1))

plot.variable(Survival_rf_highYears,
              xvar.names = c("age_at_test","corrected_odometer"),
              partial = TRUE, npts = 200)

dev.off()

cairo_ps("Publication_docs/Fig_3_HighYears_eps.eps", height = 4.5,
width = 7.5)

par(bty = "n", mar = c(5.1,4.1,2.1,2.1))
plot.variable(Survival_rf_highYears,
              xvar.names = c("age_at_test","corrected_odometer"),
              partial = TRUE, npts = 200)

dev.off()

plot.variable(Survival_rf_all_lge,
              xvar.names = c("N_prs_fail","N_failed", "N_tests"),
              partial = TRUE, npts = 20)

plot.variable(Survival_rf_all_lge,
              xvar.names = c("first_use_date","age_at_test",
"first_odo"),
              partial = TRUE, npts = 20)

plot.variable(Survival_rf_all_lge,
              xvar.names = c("age_at_test","corrected_odometer"),
              partial = TRUE, npts = 100)

barplot(tail(vimp_rf_lge$importance[order(vimp_rf_lge$importance)],
20), las =2)

par(mfrow =c(1,1))

rm(Survival_rf_all_lge,Survival_rf_highYears,Survival_rf_highMiles)
gc()
```

```
###################################################################
##############

## Plots the brier score of these models as a small supp figure

###################################################################
################

Survival_rf_all_lge<-readRDS("Survival_rf_all_lge.rds")
bs1<-get.brier.survival(Survival_rf_all_lge)
rm(Survival_rf_all_lge)
gc()

Survival_rf_highMiles<-readRDS("Survival_rf_highMiles.rds")
bs2<-get.brier.survival(Survival_rf_highMiles)

rm(Survival_rf_highMiles)
gc()

Survival_rf_highYears<-readRDS("Survival_rf_highYears.rds")
bs3<-get.brier.survival(Survival_rf_highYears)

rm(Survival_rf_highYears)
gc()

Survival_rf_eqN<-readRDS("Survival_rf_eqN.rds")
bs4<-get.brier.survival(Survival_rf_eqN)

rm(Survival_rf_eqN)
gc()

## Gets the variable importance scores

vimp_rf_lge<-readRDS("vimp_rf_lge.rds")
vimp_rf_highYears<-readRDS("vimp_rf_highYears.rds")
vimp_rf_highMiles<-readRDS("vimp_rf_highMiles.rds")
vimp_rf_eqN<-readRDS("vimp_rf_eqN.rds")

cairo_ps("Publication_docs/Fig_S_Brier_scores.eps", height = 9.5,
width = 8.5)

par(mfrow = c(2,2), bty = "n", mar = c(5.1,9.6,3.1,1.1))

plot(bs1$brier.score, type = "s",
     ylab = "Brier Score", xlab = "Time in Years", ylim = c(0,0.25))
points(bs2$brier.score, type = "s", col = "orange")
points(bs3$brier.score, type = "s", col = "#004225")
points(bs4$brier.score, type = "s", col = "hotpink")
```

```
abline(h = seq(0,0.25, by = 0.05), lty = 3, col = "grey")
abline(v = seq(0,10, by = 2), lty = 3, col = "grey")

## makes colours correspond to key in brier score

importance_set<-
tail(vimp_rf_highYears$importance[order(vimp_rf_highYears$importance
)], 15)
names(importance_set)<-gsub("_", " ", names(importance_set))
names(importance_set)<-gsub("  ", " ", names(importance_set))
names(importance_set)<-gsub("odo$", "odometer",
names(importance_set))
names(importance_set)<-gsub("^N ", "N roadworthy ",
names(importance_set))

barplot(importance_set,
        las =2, col = "#004225", horiz = T, main = "Vehicles Over 18
Years") ## in British racing green
abline(v = seq(0,0.025, by = 0.005), lty =3)

importance_set<-
tail(vimp_rf_highMiles$importance[order(vimp_rf_highMiles$importance
)], 15)
names(importance_set)<-gsub("_", " ", names(importance_set))
names(importance_set)<-gsub("  ", " ", names(importance_set))
names(importance_set)<-gsub("odo$", "odometer",
names(importance_set))
names(importance_set)<-gsub("^N ", "N roadworthy ",
names(importance_set))
names(importance_set)<-gsub("LONDON\\.TAXIS\\.INT", "LONDON TAXI",
names(importance_set))
names(importance_set)<-gsub("LAND\\.ROVER", "LAND ROVER",
names(importance_set))

barplot(importance_set,
        las =2, col = "orange", horiz = T, main = "Vehicles Over
200k Miles")
abline(v = seq(0,0.025, by = 0.005), lty =3)

importance_set<-
tail(vimp_rf_eqN$importance[order(vimp_rf_eqN$importance)], 15)
names(importance_set)<-gsub("_", " ", names(importance_set))
names(importance_set)<-gsub("  ", " ", names(importance_set))
names(importance_set)<-gsub("odo$", "odometer",
names(importance_set))
names(importance_set)<-gsub("^N ", "N roadworthy ",
names(importance_set))

barplot(importance_set,
        las =2, col = "hotpink", horiz = T, main = "Balanced by
Vehicle Type")
abline(v = seq(0,0.01, by = 0.002), lty =3)

dev.off()
```

```
par(mfrow = c(1,1), bty = "n", mar = c(5.1,4.1,4.1,2.1))
```

## That's all! Contact me at saul.j.newman@gmail.com with any
questions.